\documentclass[12pt,twoside,a4paper]{cernyrep} 
\usepackage[paperheight=29.5cm,paperwidth=21.cm,margin=0.6in,heightrounded]{geometry}
\usepackage{chngcntr}
\counterwithin{equation}{section}
\counterwithin{figure}{section}
\counterwithin{table}{section}
\counterwithin{footnote}{section}
\usepackage{bibunits}  % bibunits.sty
\usepackage[utf8]{inputenc}

\usepackage{texnames}
\usepackage[T1]{fontenc}
\usepackage{hyperref}
\hypersetup{
linktocpage=true,
colorlinks=true,
linkcolor=blue,          % color of internal links (change box color with linkbordercolor)
citecolor=mygreen,        % color of links to bibliography
filecolor=blue,      % magenta color of file links
urlcolor=blue           % color of external links
}  %YR
\usepackage{url}   
\usepackage{graphicx}
\usepackage{caption}
\usepackage{subcaption}
\usepackage{hepnames}
\usepackage{color}
\definecolor{mygreen}{rgb}{0.0,0.75,0.0}
\usepackage{timestamp}
\usepackage{times}
\usepackage{comment}
\usepackage{tikz}
\usetikzlibrary{arrows,decorations.pathmorphing,backgrounds,positioning,fit,petri}
\usepackage{mathtools}
\usepackage{fancyhdr}  
\usepackage{etoolbox}
\usepackage{placeins}  % 2020-03-27

\definecolor{amber}{rgb}{1.0, 0.75, 0.0}
\usepackage{pagecolor}
\usepackage{afterpage}  %TR YR for use with pagecolor

\usepackage{amssymb,amsmath,amsfonts,mathrsfs,wrapfig,enumerate}
\usepackage{lmodern}
\usepackage{mmacells}
\usepackage{latexsym}
\usepackage{multirow}
\usepackage{textcomp}
\usepackage{framed}
\usepackage[]{hyperref}
\usepackage{colortbl}
\usepackage{makecell}
\usepackage{color}
\usepackage{pifont}
\usepackage{appendix}
\usepackage{cancel}
\usepackage{import}
\usepackage{subfiles}
\usepackage{filecontents}
\usepackage{threeparttable,tablefootnote}
\usepackage{slashed}
\usepackage{mathtools}
\usepackage{pbox}
\renewcommand{\headrulewidth}{0.4pt}   % the default
\newlength\FHoffset
\fancyheadoffset{\FHoffset}
\newlength\FHleft
\newlength\FHright
\newbox\FHline
\setbox\FHline=\hbox{\hsize=\paperwidth%
  \hspace*{\FHleft}%
  \rule{\dimexpr\headwidth-\FHleft-\FHright\relax}{\headrulewidth}\hspace*{\FHright}%
}

\newcommand{\order}[1]{${\cal O}(#1)$}

\newcommand{\AmpE}[2]{{\cal A}^{{\rm #1}}_{#2}}
\newcommand{\Amp}[3]{{\cal A}^{{\rm #1}}_{#2}\ifthenelse{\isempty{#3}}{}{{[#3]}}}
\newcommand{\tAmp}[3]{\tilde{\cal A}^{{\rm #1}}_{#2}\ifthenelse{\isempty{#3}}{}{{[#3]}}}
\newcommand{\cp}{\ensuremath{{\cal CP}}}
\newcommand{\IE}{i.e.,}
\newcommand{\EG}{e.g.,}
\newcommand{\FH}{\texttt{FeynHiggs}}
\newcommand{\I}{\mathrm{i}}
\newcommand{\Un}[1]{(U_n)_{#1}}

\newcommand{\Bnull}[1]{B_{0}{\left(#1\right)}}

\newcommand{\mueff}{\mu_{\rm eff}}

\newcommand{\MSbar}{\ensuremath{\overline{\mathrm{MS}}}}

\newcommand{\simord}{\sim}
\newcommand{\gsim}{\gtrsim}

\newcommand{\half}{\mbox{\small{$\frac{1}{2}$}}}

\providecommand{\gaga}{\upgamma\,\upgamma}
\providecommand{\epem}{\rm e^{+}e^{-}}
\providecommand{\alphas}{\alpha_{\rm s}}

\newcommand{\sqrts}{\sqrt{s}}
\newcommand{\ttbar}{\mathrm{t}\overline{\mathrm{t}}}

\def\mathswitch#1{\relax\ifmmode#1\else$#1$\fi}
\def\mathswitchr#1{\relax\ifmmode{\mathrm{#1}}\else$\mathrm{#1}$\fi}

\newcommand{\Pt}{\mathswitchr t}

\def\order#1{\ensuremath{{\cal O}(#1)}}

\newcommand{\bhlumi}{{\tt BHLUMI}}

\newcommand{\veps}{{\varepsilon}}

\newcommand{\MW}{\mathswitch {M_{\PW}}}
\newcommand{\MZ}{\mathswitch {M_{\PZ}}}
\newcommand{\GZ}{\mathswitch {\Gamma_{\PZ}}}

\newcommand{\me}{\mathswitch {m_{\Pe}}}

\newcommand{\mt}{\mathswitch {m_{\Pt}}}

\newcommand{\scrs}{\scriptscriptstyle}
\newcommand{\sw}{\mathswitch {s_{\scrs\PW}}}

\newcommand{\mw}{\mathswitch {\overline{M}_\PW}} 
\newcommand{\mz}{\mathswitch {\overline{M}_\PZ}}

\newcommand{\as}{\alpha_{\mathrm s}}

\newcommand{\gev}{\,\mathrm{GeV}}
\newcommand{\mev}{\,\mathrm{MeV}}

\newcommand{\nn}{\nonumber}

\newcommand*{\kira}{\texttt{Kira}}

\newcommand*{\firefly}{\texttt{FireFly}}
\newcommand*{\firesix}{\texttt{FIRE\,6}}
\newcommand*{\fermat}{\texttt{Fermat}}

\newcommand{\bea}{\begin{eqnarray}}
\newcommand{\eea}{\end{eqnarray}}

\newcommand{\nl}{\nonumber \\ }

\newcommand{\Fign}[1]{Fig.~\ref{#1}}

\newcommand{\babayaga}{\texttt{{BabaYaga}}}
\newcommand{\bhwide}{\texttt{{Bhwide}}}

\renewcommand{\thechapter}{\Alph{chapter}}
\usepackage{varwidth}

\usepackage{xcolor}

\usepackage[normalem]{ulem}

\begin{document}

% \frontmatter
\pagenumbering{roman}  
\pagestyle{empty}

\renewcommand{\chaptermark}[1]{\markboth{ #1}{}}
\renewcommand{\sectionmark}[1]{\markright{\thesection.\ #1}}

%=================================================================
%BKniehl
\providecommand{\gaga}{\gamma\,\gamma}
\providecommand{\epem}{\rm e^{+}e^{-}}
\providecommand{\alphas}{\alpha_{\rm s}}
%\newcommand{\Lumi}{\mathcal{L}}
%\newcommand{\Lunits}{cm$^{-2}$s$^{-1}$}
%\newcommand{\sqrts}{\sqrt{s}}
%\newcommand{\ttbar}{t\overline{t}}
%AKardos
\newcommand{\fig}[1]{Fig.~\ref{#1}}
\newcommand{\RefAK}[1]{Ref.~\cite{#1}}

%% pagecolortest -- begin
%\pagecolor{amber}  % \newpagecolor{yellow} \afterpage{\restorepagecolor}
%%%%%%%%%\pagecolor{yellow}  %\afterpage{\nopagecolor}

%%Here comes text a la YR standards, yellow title page:

\pagestyle{empty}

\renewcommand{\chaptermark}[1]{\markboth{ #1}{}}
\renewcommand{\sectionmark}[1]{\markright{\thesection.\ #1}}

\thispagestyle{empty}
\setlength{\unitlength}{1mm}
\begin{picture}(0.001,0.001)
\put(-8,8){\large CERN Yellow Reports: Monographs}
\put(120,8){\large CERN-2020-003}

\put(-5,-60){\LARGE\bfseries
                                              Theory for the FCC-ee}
\put(-5,-68){\small Report on the 11th FCC-ee Workshop}
\put(-5,-73){\small Theory and Experiments}
\put(-5,-78){\small CERN, Geneva, 8--11 January 2019}

\put(-5,-94){\Large Editors:}

\put(0,-105){\Large A. Blondel}
\put(5,-109){\small DPNC University of Geneva, Switzerland}

\put(0,-120){\Large J. Gluza}
\put(5,-124){\small Institute  of Physics, University of Silesia, %40-007 
Katowice, Poland}
\put(5,-128){\small Faculty of Science, University of Hradec Kr\'alov\'e, Czech Republic}

\put(0,-139){\Large S. Jadach}
\put(5,-143){\small Institute of Nuclear Physics, PAN, 31-342 Krak\'ow, Poland}

\put(0,-154){\Large P. Janot}
\put(5,-158){\small CERN, CH-1211 Geneva 23, Switzerland}

\put(0,-169){\Large T. Riemann}
\put(5,-173){\small Institute  of Physics, University of Silesia, %40-007 
Katowice, Poland}
\put(5,-177){\small Deutsches Elektronen-Synchrotron, DESY, 15738 Zeuthen, Germany}

\put(65,-250){\includegraphics{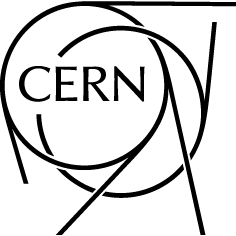}}

\end{picture}

\clearpage
%%%%%%%%%%%%%%%%%%%%%%%%%%%%%%%%%%%%%%%%%%%%%%%%%%%%%%%%%%%%%%%%%%%%%%%%%%%%%%%%%%%%%%%
%     second yellow page, left-side
\thispagestyle{empty}
\mbox{}
\vfill

\begin{flushleft}%\large
CERN Yellow Reports: Monographs\\
Published by CERN, CH-1211 Geneva 23, Switzerland\\[3mm]

\begin{tabular}{@{}l@{~}l}
  ISBN & 978-92-9083-560-8 (paperback) \\
  ISBN & 978-92-9083-559-2 (PDF) \\
  ISSN & 2519-8068 (Print)\\ %for CERN Yellow Reports: Monographs
  ISSN & 2519-8076 (Online)\\ %for CERN Yellow Reports: Monographs
  DOI & \url{http://dx.doi.org/10.23731/CYRM-2020-003}\\%for CERN Yellow Reports: Monographs
\end{tabular}\\[3mm]
Accepted for publication by the CERN Reports Editorial Board (CREB) on 20 March 2020\\[1mm]
Available online at \url{http://publishing.cern.ch/} and \url{http://cds.cern.ch/}\\[3mm]

Copyright \copyright{} CERN, 2020\\[1mm]
\raisebox{-1mm}{\includegraphics[height=12pt]{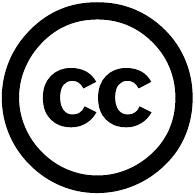}}
Creative Commons Attribution 4.0\\[1mm]
Knowledge transfer is an integral part of CERN's mission.\\[1mm]
CERN publishes this volume Open Access under the Creative Commons Attribution 4.0 license
(\url{http://creativecommons.org/licenses/by/4.0/}) in order to permit its wide dissemination and use.\\
The submission of a contribution to a CERN Yellow Report series shall be deemed to constitute the contributor's agreement to this copyright and license statement. Contributors are requested to obtain any clearances that may be necessary for this purpose.\\[5mm]

This volume is indexed in: INSPIRE and CERN Document Server (CDS)\\[5mm]

This volume should be cited as:\\[1mm]

Theory for the FCC-ee : Report on the 11th FCC-ee Workshop\\Theory and Experiments, CERN, Geneva, 8--11 January 2019\\ 
Eds. A.~Blondel, J.~Gluza, S.~Jadach, P.~Janot and T.~Riemann\\  CERN Yellow Reports: Monographs, CERN-2020-003 (CERN, Geneva, 2020), \url{http://dx.doi.org/10.23731/CYRM-2020-003}\\[3mm]

\end{flushleft}

\clearpage
%\cleardoublepage

%%%%%%%%%%%%%%%%%%%%%%%%%%%%%%%%%%%%%%%%%%%%%%%%%%%%%%%%%%%%%%%%%%%%%%%%%%
%
%                            Title page
%
%%%%%%%%%%%%%%%%%%%%%%%%%%%%%%%%%%%%%%%%%%%%%%%%%%%%%%%%%%%%%%%%%%%%%%%%%%
% now come white pages
\pagecolor{white}

{\flushright{
%{\small
%{\tt  
%BU-HEPP-19-03, %BWard
%CERN-TH-2019-061, %SJones
%CP3-19-22, %who?
%DESY~19-072, %JReuter
%DESY~19-074, %FJegerlehner
%FR-PHENO-2019-005, %BPage
%HU-EP-19/12, %FJegerlehner
%IFIC/19-23,  %GRodrigoGSborlini
%IFT-UAM/CSIC-19-058, %SHeinemeyer
%IPhT-19-050, %who?  
%IPPP/19/32, %BPage
%KW~19-003, %TRiemann
%LTH~1203, %JAGracey, 
%MPP-2019-84, %BPage
%TTK-19-19, %CSchwinn
%TTP19-008, %JZurita
%TUM-HEP-1200/19, %RSzafron
%ZU-TH 22/19 %BPage 
}
%\\
%\hspace*{75.5mm}{\tt  MITP/18-052, MPP-2018-143, SI-HEP-2018-21}
}

%JG 
\vspace{-2mm}
%\begin{center}
\noindent
{\bf\LARGE
%\\[5mm] 
%\large
Theory for the FCC-ee \\
\begin{normalsize}Report on the 11th FCC-ee Workshop\footnote{\url{https://indico.cern.ch/event/766859/}}, CERN, Geneva, 8--11 January 2019 
\end{normalsize}
%Report on the theory part of the 
%\\
%11$^{th}$ FCC-ee workshop: Theory and Experiments \\
% \\[4mm]
% \url{https://indico.cern.ch/event/766859/}
 }
 %\end{center}
 
\vspace{7mm}

%\begin{center}
\noindent
{\bf A.~Blondel$^{1}$, ~J.~Gluza\footnote{Corresponding editor, email: janusz.gluza@cern.ch}$^{,2,3}$,  ~S.~Jadach$^4$, ~P.~Janot$^{5}$, ~T.~Riemann$^{2,6}$ (editors)}
\\
S.~Abreu$^{7}$, ~J.J.~Aguilera-Verdugo$^{8}$,
~A.B.~Arbuzov$^{9}$,
~J.~Baglio$^{10}$, ~S.D.~Bakshi$^{11}$, ~S.~Banerjee$^{12}$,
~M.~Beneke$^{13}$,
~C.~Bobeth$^{13}$, ~C.~Bogner$^{14}$,
~S.G.~Bondarenko$^{9}$,
~S.~Borowka$^{5}$, ~S.~Bra{\ss}$^{15}$, ~C.M.~Carloni~Calame$^{16}$,
~J.~Chakrabortty$^{11}$ ,
~M.~Chiesa$^{17}$,
~M.~Chrzaszcz$^{4}$, ~D.~d'Enterria$^{5}$,
~F.~Domingo$^{18}$, ~J.~Dormans$^{19}$,
~F.~Driencourt-Mangin$^{8}$,
~Y.V.~Dydyshka$^{20}$,  ~J.~Erler$^{21,22}$,\\
~F.~Febres~Cordero$^{19,23}$, ~J.A.~Gracey$^{24}$, ~Z.-G.~He$^{25}$, ~S.~Heinemeyer$^{26,27,28}$, ~G.~Heinrich$^{29}$, ~I.~H\"onemann$^{14}$, ~H.~Ita$^{19}$, ~S.~Jahn$^{29}$, ~F.~Jegerlehner$^{6,30}$, ~S.P.~Jones$^{5}$,
~L.V.~Kalinovskaya$^{20}$,
~A.~Kardos$^{31}$, ~M.~Kerner$^{32}$, ~W.~Kilian$^{15}$,
~S.~Kluth$^{26}$, ~B.A.~Kniehl$^{25}$, ~A.~Maier$^{6}$,
~P.~Maierh\"ofer$^{19}$,
~G.~Montagna$^{16,33}$, ~O.~Nicrosini$^{16}$,
~T.~Ohl$^{17}$, ~B.~Page$^{34}$, ~S.~Pa{\ss}ehr$^{35}$,
~S.K.~Patra$^{11}$, ~F.~Piccinini$^{16}$,  ~R.~Pittau$^{36}$,
~W.~Placzek$^{37}$, ~J.~Plenter$^{8}$, ~S.~Ram\'{\i}rez-Uribe$^{8}$, 
~J.~Reuter$^{38}$, ~G.~Rodrigo$^{8}$, ~V.~Rothe$^{38}$,
~L.A.~Rumyantsev$^{20,39}$,
~R.R.~Sadykov$^{20}$,
 ~G.F.R.~Sborlini$^{8}$,~J.~Schlenk$^{40,44}$, ~M.~Schott$^{21}$, ~A.~Schweitzer$^{41}$, ~C.~Schwinn$^{42}$, ~M.~Skrzypek$^{4}$, ~G.~Somogyi$^{43}$, ~M.~Spira$^{44}$, ~P.~Stienemeier$^{38}$, ~R.~Szafron$^{13}$, 
~K.~Tempest$^{45}$, ~W.J.~Torres Bobadilla$^{8}$, ~S.~Tracz$^{8}$, ~Z.~Tr\'ocs\'anyi$^{43,46}$, ~Z.~Tulip\'ant$^{43}$,
~J.~Usovitsch$^{47}$, ~A.~Verbytskyi$^{26}$, ~B.F.L.~Ward$^{48}$,
~Z.~Was$^{4}$, ~G.~Weiglein$^{38}$, ~C.~Weiland$^{49}$, ~S.~Weinzierl$^{21}$,
~V.L.~Yermolchyk$^{20}$
~S.A.~Yost$^{50}$, ~J.~Zurita$^{51,52}$

\vspace*{2mm}  %tr 14 mm

\begin{flushleft}
{\em\small
{$^{1}$ 
DPNC, University of Geneva, Switzerland}
%DPNC, University of Geneva
\\
{$^2$ Institute  of Physics, University of Silesia, Katowice, Poland}
\\
{$^{3}$ Faculty of Science, University of Hradec Kr\'alov\'e, Czech Republic}
\\
{$^4$ Institute of Nuclear Physics, PAN, 
%ul. Radzikowskiego 152, 
31-342 Krak\'ow, Poland}
\\
{$^{5}$ CERN, CH-1211 Geneva 23, Switzerland}
\\
{$^{6}$ Deutsches Elektronen-Synchrotron, DESY, 15738 Zeuthen, Germany}
\\
{$^{7}$ Center for Cosmology, 
%Particle Physics and Phenomenology (CP3), 
CP3, 
Universit\'e Catholique de Louvain, 1348 Louvain-La-Neuve, Belgium}
\\
{$^{8}$ 
IFIC,
%Instituto de F\'{\i}sica Corpuscular, 
Universitat de Val\`{e}ncia,
% -- Consejo Superior de Investigaciones Cient\'{\i}ficas, 
%Parc Cient\'{\i}fic, 
E-46980 Paterna, Valencia, Spain}
\\
{$^{9}$ Bogoliubov Laboratory of Theoretical Physics, JINR, Dubna, 141980 Russia}
\\
{$^{10}$ Institut f\"ur Theoretische Physik, Eberhard Karls Universit\"at, 
%T\"ubingen, 
%Auf der Morgenstelle 14, 
72076 T\"ubingen, Germany}
\\
{$^{11}$ Indian Institute of Technology, Kanpur, India}
\\
{$^{12}$  University of Louisville, Louisville, KY 40292, USA}
 %        University of Louisville, Louisville, KY 40292 USA
\\
{$^{13}$ Physik Department T31, Technische Universit\"at M\"unchen, Garching, Germany}
\\
{$^{14}$ Institut f{\"u}r Physik, Johannes Gutenberg-Universit{\"a}t Mainz, D-55099 Mainz, Germany}
\\
{$^{15}$ Department Physik, Universit\"at Siegen, 
%Walter-Flex-Str. 3, 
57068 Siegen, Germany}
\\
{$^{16}$ Istituto Nazionale di Fisica Nucleare, Sezione di Pavia, 
% via A. Bassi 6, 
27100 Pavia, Italy}
\\
{$^{17}$ Fakult\"at f\"ur Physik und Astronomie, Universit\"at W\"urzburg, 
% Emil-Hilb-Weg 22, 
97074 W\"urzburg, Germany}
\\
{$^{18}$ Bethe Center for Theoretical Physics \& Physikalisches Institut der Universit\"at Bonn, 
%Nu{\ss}allee 12, 
53115 Bonn, Germany}
\\
{$^{19}$ Physikalisches Institut, Albert-Ludwigs-Universit\"at Freiburg, % Hermann-Herder-Str. 3, D-
79104 Freiburg, Germany}
\\
{$^{20}$  Dzhelepov Laboratory of Nuclear Problems, JINR, Dubna, 141980 Russia}
\\
{$^{21}$ PRISMA Cluster of Excellence, Institut f{\"u}r Physik, Johannes Gutenberg-Universit{\"a}t, 55099 Mainz, Germany}
\\
{$^{22}$ 
        Departamento de F\'isica Te\'orica, Instituto de F\'isica, Universidad Nacional Aut\'onoma de M\'exico, 04510 CDMX, Mexico}
\\
{$^{23}$ Physics Department, Florida State University, 
% 77 Chieftan Way, 
Tallahassee, FL 32306, USA}
\newpage
{$^{24}$ Theoretical Physics Division, Department of Mathematical Sciences, University of Liverpool,
        % P.O. Box 147, 
        Liverpool, L69 3BX, United Kingdom}
\\
{$^{25}$Institut f\"ur Theoretische Physik, Universit\"at Hamburg,
        % Luruper Chaussee 149, 
        22761 Hamburg, Germany}
\\
{$^{26}$ Instituto de F\'isica Te\'orica (UAM/CSIC), Universidad Aut\'onoma de Madrid, Cantoblanco, 28049 Madrid, Spain}
\\
{$^{27}$ Instituto de F\'isica de Cantabria (CSIC-UC), 39005 Santander, Spain}
\\
{$^{28}$ Campus of International Excellence UAM+CSIC, Cantoblanco, 28049, Madrid, Spain}
\\
{$^{29}$ Max Planck Institute for Physics, 
% F\"ohringer Ring 6, 
80805 M\"unchen, Germany}
\\
{$^{30}$ Institut f\"ur Physik, Humboldt-Universit\"at zu Berlin, 
% Newtonstrasse 15, D--
12489 Berlin, Germany}
\\
{$^{31}$Institute of Physics, University of Debrecen, 4010 Debrecen, 
% PO Box 105, 
Hungary}
\\
{$^{32}$ Physik-Institut, Universit{\"a}t Z{\"u}rich, % Winterthurerstrasse 190, 
8057 Z{\"u}rich, Switzerland}
\\
{$^{33}$ Dipartimento di Fisica, Universit\`a di Pavia, 
% via A. Bassi 6, 
27100 Pavia, Italy}
\\
{$^{34}$ Institut de Physique Th\'eorique, CEA, CNRS, Universit\'e Paris-Saclay, F-91191 Gif-sur-Yvette, France}
\\
{$^{35}$ Laboratoire de Physique Th\'eorique et Hautes \'Energies (LPTHE), Sorbonne Universit\'e, CNRS, 
% 4 Place Jussieu, 
75252 Paris CEDEX 05, France}
\\
{$^{36}$ Departamento de F\'{i}sica Te\'orica y del Cosmos and CAFPE, Universidad de Granada, 
%Campus Fuentenueva s.n., E-
18071 Granada, Spain}
\\
{$^{37}$ Marian Smoluchowski Institute of Physics, Jagiellonian University, Krak\'ow, Poland}
\\
{$^{38}$ Deutsches Elektronen-Synchrotron, DESY, 
% Notkestra{\ss}e 85, 
22607 Hamburg, Germany}
\\
{$^{39}$ Institute of Physics, Southern Federal University, Rostov-on-Don, 344090 Russia} 
\\
{$^{40}$ Institute for Particle Physics Phenomenology, University of Durham, Durham DH1 3LE, UK}
\\
{$^{41}$ Institute for Theoretical Physics, ETH Z\"urich, 
% Wolfgang-Pauli-Str. 27, 
8093 Z\"urich, Switzerland}
\\
{$^{42}$ Institut f\"ur Theoretische Teilchenphysik und   
        Kosmologie,   RWTH Aachen University, 
        % Sommerfeldstra\ss e 16,  
        52056 Aachen, Germany}
\\
{$^{43}$ MTA-DE Particle Physics Research Group, University of Debrecen, 4010 Debrecen, 
% PO Box 105, 
Hungary}
\\
{$^{44}$ Paul Scherrer Institut, CH-5232 Villigen PSI, Switzerland}
\\
{$^{45}$ Department of Physics, University of Toronto, 
% 60 St George St., 
Toronto, Ontario, M5S 1A7, Canada}
\\
{$^{46}$ Institute for Theoretical Physics, E\"otv\"os Lor\'and University, 
% P\'azm\'any P\'eter 1/A, H-
1117 Budapest, Hungary}
\\
{$^{47}$  School of Mathematics,
        Trinity College Dublin,  University of Dublin, Ireland}
\\
{$^{48}$ Baylor University, Waco, TX, USA}
\\
{$^{49}$ Pittsburgh Particle Physics, Astrophysics, and Cosmology Center, Department of Physics and Astronomy, University of Pittsburgh,
% 3941 O'Hara St., 
Pittsburgh, PA 15260, USA}
\\
{$^{50}$ The Citadel, Charleston, SC, USA}
\\
{$^{51}$ Institute for Nuclear Physics (IKP), Karlsruhe Institute of Technology,
  % Hermann-von-Helmholtz-Platz 1, D-
  76344 Eggenstein-Leopoldshafen, Germany}
\\
{$^{52}$ Institute for Theoretical Particle Physics (TTP), Karlsruhe Institute of Technology, 
% Engesserstra\ss e 7, D-
76128 Karlsruhe, Germany}
}
\end{flushleft}

\clearpage
\pagestyle{plain}
\pagenumbering{roman}
\setcounter{page}{5}

%################################################
%\chapter*{Copyright statement}
%{
%Theory report on the 11th FCC-ee workshop,  8--11 January 2019, CERN, Geneva, Switzerland,\\
%\url{https://indico.cern.ch/event/766859/}
%
%\vspace*{1cm}
%
%\noindent
%\copyright{} Copyright 2019 under the terms of the Creative Commons
%Attribution 4.0 International License  \href{http://creativecommons.org/licenses/by/4.0/}{CC BY 4.0}, %held by
%\begin{flushleft} 
%S.~Abreu, ~J.J.~Aguilera-Verdugo, ~A.B.~Arbuzov, ~J.~Baglio, ~S.D.~Bakshi, ~S.~Banerjee, ~M.~Beneke, %A.~Blondel, ~C.~Bobeth, ~C.~Bogner, S.~Bondarenko, ~S.~Borowka, ~S.~Bra{\ss}, ~C.M.~Carloni~Calame, %~J.~Chakrabortty, ~M.~Chiesa, ~M.~Chrzaszcz, ~D.~d'Enterria, ~F.~Domingo, ~J.~Dormans, %~F.~Driencourt-Mangin, ~Y.~Dydyshka, ~J.~Erler, ~F.~Febres~Cordero, ~J.~Gluza, ~J.A.~Gracey, ~Z.-G.~He, %~S.~Heinemeyer, ~G.~Heinrich, ~I.~H\"onemann, ~H.~Ita, ~S.~Jadach, ~S.~Jahn, ~F.~Jegerlehner, %~S.P.~Jones, L.~Kalinovskaya, ~A.~Kardos, ~M.~Kerner, ~W.~Kilian,
%~S.~Kluth, ~B.A.~Kniehl, ~A.~Maier, ~P.~Maierh\"ofer, ~G.~Montagna, ~O.~Nicrosini,  ~T.~Ohl, ~B.~Page, %~S.~Pa{\ss}ehr, ~S.K.~Patra, ~F.~Piccinini, ~R.~Pittau, ~W.~P\l{}aczek, ~J.~Plenter, %~S.~Ram\'{\i}rez-Uribe, 
%~J.~Reuter, ~T.~Riemann, ~G.~Rodrigo, ~V.~Rothe, L.~Rumyantsev, R.~Sadykov,  ~G.F.R.~Sborlini, ~%J.~Schlenk,
%~M.~Schott, ~A.~Schweitzer, ~C.~Schwinn, ~M.~Skrzypek, ~G.~Somogyi, ~M.~Spira, ~P.~Stienemeier, %~R.~Szafron, 
%~K.~Tempest, ~W.J.~Torres Bobadilla, ~S.~Tracz, ~Z.~Tr\'ocs\'anyi, ~Z.~Tulip\'ant,
%~J.~Usovitsch, ~A.~Verbytskyi, ~B.F.L.~Ward,
%~Z.~Was, ~G.~Weiglein, ~C.~Weiland, ~S.~Weinzierl, V.~Yermolchyk,  S.A.~Yost, ~J.~Zurita
%\end{flushleft}
%}
%}
\pagenumbering{roman} 
\setcounter{page}{5}
% also: \mainmatter
%\tableofcontents
%%%%%%%%%%%%%%%%%%%%%%%%%%%%%%%%%%%%%%%%%%%%%%%%%
\clearpage
%################################################
\chapter*{Abstract}
\addcontentsline{toc}{chapter}{Abstract}

%\vspace*{20mm}

%\centerline{\bf \Large Abstract}
\vspace*{12mm}
\noindent

{\small

\noindent The Future Circular Collider (FCC) at CERN, a proposed $100\Ukm$ circular facility with several colliders in succession, culminates in a $100\UTeV$ proton--proton collider. It offers a vast new domain of exploration in particle physics, with orders-of-magnitude advances in terms of precision, sensitivity, and energy. The implementation plan published in 2018 foresees, as a first step, an electroweak factory electron--positron collider. This high-luminosity facility, operating at centre-of-mass energies between 90 and $365\UGeV$, will study the heavy particles of the Standard Model (SM), Z, W, and Higgs bosons, and top quarks with unprecedented accuracy. 
The physics programme offers great discovery potential:\\
(i) through precision measurements, 
(ii) through sensitive searches for symmetry violations, forbidden, or extremely rare decays,  
and (iii) through the search for direct observation of new particles with extremely small couplings. 
The electroweak factory $\mathrm{e}^+\mathrm{e}^-$ collider
constitutes a real challenge to the theory and to precision calculations,  triggering the need for the development of new mathematical methods and software tools. 
A first workshop in 2018  focused on the first FCC-ee stage, the \mbox{Tera-Z}, and confronted the theoretical status of precision 
Standard Model calculations on the Z boson resonance to the experimental demands.   

The second workshop, in January 2019, extended the scope to the  next stages, with the production of W bosons (FCC-ee-W), the Higgs boson (FCC-ee-H), and top quarks (FCC-ee-tt). 
In particular, the  theoretical precision in the  determination of the crucial input parameters,  $\alpha_{\mathrm{QED}}$,  $\alpha_{\mathrm{QCD}}$, $M_\mathrm{W}$,  and $m_\mathrm{t}$, at the level of FCC-ee requirements was thoroughly discussed. 
The requirements on Standard Model theory calculations were spelt out, so as to meet the demanding accuracy of the \mbox{FCC-ee} experimental potential.  
The discussion of innovative methods and tools for multiloop calculations was deepened.
Furthermore, phenomenological analyses beyond the Standard Model  were discussed, including effective theory approaches.
%{\color{blue} 
The reports of 2018 and 2019 serve as white papers of the workshop results and subsequent developments. 
}
%}

%%%%%%%%%%%%%%%%%%%%%%%%%%%%%%%%%%%%%%%%%%%%%%%%

%################################################
\chapter*{Editors' note} \label{ch:editorsNote}
\vspace*{1.cm}
\addcontentsline{toc}{chapter}{Editors' note}

\pagestyle{fancy}
\fancyhead[CO]{\thechapter \hspace{1mm} Editors' note}
\fancyhead[RO]{}
\fancyhead[LO]{}
\fancyhead[LE]{}
\fancyhead[CE]{}
\fancyhead[RE]{}
\fancyhead[CE]{A.~Blondel,~J.~Gluza,~S.~Jadach,~P.~Janot,~T.~Riemann}
\lfoot[]{}
\cfoot{  \thepage \hspace*{0.075mm} }
\rfoot[]{}

\begin{bibunit}[elsarticle-num]  
\let\stdthebibliography\thebibliography
\renewcommand{\thebibliography}{%
\let\section\subsection
\stdthebibliography}
      
\noindent
{\em Understanding the origins of the Universe and how it works and evolves} is the present mission of a large community of physicists of many nations and specialities. It calls for a large-scale vision, involving general relativity, astrophysics, and cosmology, together with the detailed basic understanding provided by particle physics; these disciplines work  hand in hand, with the help of several other research fields. 
Presently, particle physics stands at an important moment in its history. With the discovery of the Higgs boson, the matrix of interactions and elementary particles that is called the `Standard Model' (SM), is complete. Yet the Higgs boson itself, and how it breaks the electroweak symmetry, remains a fascinating subject requiring verification at the next order of precision, typically at percent, or even % https://www.linguee.de/deutsch-englisch/uebersetzung/2+promille.html
{per-mille, }%permil 
accuracy. Furthermore,  several experimental facts are not accounted for by the SM; let us mention: (i) the baryon asymmetry of the Universe, (ii) the nature and origin of dark matter, and (iii) the origin of neutrino masses. These have no unique, if any, explanation in the SM and yet will require answers from particle physics. 

Particle physics exploration must continue\dots\  but we no longer have a guiding scale.

How can this exploration be carried out? Which next tool is needed?  Going to higher and higher energies is an obvious idea. It has worked well for the Standard Model particles so far, because they all have roughly the same strong and electroweak couplings.  
It is far from evident, however, that the new phenomena or particles, required to explore these questions, will behave in the same way---the opportunity to explore much smaller couplings or much higher scales must be kept in mind. Here, the role of precision measurements,  the search for extremely rare decays of known particles, for small violations of the SM symmetries, and for direct production of super-weakly coupled objects is in order. A broad search strategy  is thus needed.  

With this in mind, and armed with the recommendation of the European Strategy in 2013 that Europe should be in a position to ``propose an ambitious post-LHC accelerator project at CERN'',  the FCC collaboration has elaborated a strategy of circular colliders fitting in a new facility of $100\Ukm$ circumference. It will start with a high-luminosity $\rm e^+ e^-$ electroweak factory, FCC-ee, and culminate in a proton collider, FCC-hh, of more than $100\UTeV$ collision energy.  Additional options of heavy-ion collisions and e--p scattering are foreseen and, possibly, muon collisions.  This strategy offers, by way of synergy and complementarity,  a thorough study of the Higgs 
boson, as well as unmatched capabilities of high-energy exploration, precision measurements, and sensitive rare process searches~\cite{fcc}. The FCC Conceptual Design Report (CDR) has been prepared and released~\cite{FCC-CDR,Abada:2019zxq,Abada:2019lih}. This powerful exploratory project will, right from its first step as a Z
factory, explore completely uncharted territory in terms of precision and sensitivity. Moreover, it constitutes an extraordinary challenge for theory. The theoretical community has responded with enthusiasm to the challenge; already several workshops have gathered an increasing number of contributions.  

In this report, we collect theory contributions to the 11th FCC-ee meeting held in January 2019 at CERN \cite{mini2}, completed by a few invited guest contributions.
% start here insertion on white book 2019-07-01 
%\textcolor{red}{
The report is a kind of community white paper, rather than a conventional conference report.
It collects, coherently, the contributions from 86 scientists, representing the state of the art in 2019 and envisioning the additional needs of future lepton colliders. 
% 2020April08
We are grateful to Jens Vigen from CERN. Due to his efforts in the final productions, the document meets the highest editorial standards. 
The collective interactions of all of us, in one way or another, at the meeting in January 2019 and for several months after, make the backbone of the final write-up. 
Nevertheless, for the convenience of the reader, we decided to retain  a sectional structure for the bulk of the document, with individual bibliographies for the sections. 
%}
% end here insertion on white book 2019-07-01 

The volume follows the report \cite{Blondel:2018mad} 
on the FCC-ee workshop in January 2018 \cite{mini}, which focused on 
the
theory needs for the Tera-Z, the first stage of the FCC-ee, working in the Z boson mass range.
The purpose is to document existing studies and also to motivate future theoretical studies, enabling by their predictions a full exploration of the experimental potential of the FCC-ee.

It has become evident that a significant work must be accomplished, both in multiloop calculations in the Standard Model and also in projects
beyond the Standard Model. Documentation of these requirements became  highly desirable to complement the submitted Conceptual Design Report.
The present report exemplifies both the well-advanced status of phenomenology for the FCC-ee and, at the same time, the need for further mathematically well-founded deepening of the technologies for precision measurements. In this respect, it is a necessary addition to the FCC CDR.

From a scientific point of view, the FCC is the most challenging collider project for the next few decades \cite{Blondel:2019qlh}. We see it as our duty and pleasure to prepare such a frontier project and to sustain CERN's leading role in basic research worldwide. 
The goals must be set as high as possible, \ie at the level of the statistical uncertainties, because {\it this precision genuinely equates discovery potential.}

We thank all participants of the workshop for their engagement with presentations and in the discussions during the workshop, and the authors of the report for writing such excellent contributions. The exploratory potential of the FCC-ee can be fully exploited only if the talent and efforts of accelerator builders and experimenters is met by theory. 
The message is: we are working on it. 

From this quest for the unknown, driven by curiosity, history shows that
there is a return for all of us, scientists or not \cite{bscience,bscience1,elsenbordry}.

\bigskip

   The editors.
\begin{flushleft}

\end{flushleft}

\end{bibunit}

%################################################
\clearpage \pagestyle{empty} 
\cleardoublepage

\pagestyle{plain}
\fancyhead[LO]{}
\fancyhead[RO]{}
\fancyhead[CO]{\thechapter ~ \leftmark}
\fancyhead[LE]{}
\fancyhead[CE]{}
\fancyhead[RE]{} 

\tableofcontents
\clearpage \pagestyle{empty}
\cleardoublepage

\pagestyle{plain}
\pagenumbering{arabic} % also: \mainmatter
\pagestyle{empty}
%\lfoot[]{}
\lfoot[\thepage]{-  \thepage \hspace*{0.075mm} -}
\cfoot{-  \thepage \hspace*{0.075mm} -}
\rfoot[\thepage]{-  \thepage \hspace*{0.075mm} -}
%\rfoot[]{}
%%%%%%%%%%%%%%%%%%%%%%%%%%%%%%%%%%%%%%%%%%%%%%%%
%%%%%%%%%%%%%%%%%%%%%%%%%%%%%%%%%%%%%%%%%%%%%%%
\chapter*{Executive summary} \label{ch:exsum}
\addcontentsline{toc}{chapter}{Executive summary}
%============================================
%

\pagestyle{fancy}
\chead[]{}
%\lhead[\thechapter.~~ \leftmark]{\thechapter.~~ \leftmark}
\lhead[\thechapter~~ \leftmark]{\thechapter~~ \leftmark}
% []left pages,{}righimprovetimprove pages
\rhead[\rightmark]{\rightmark}
% []left pages, {}right pages 
\lfoot[]{}
%TR20190506 \cfoot{-  \thepage \hspace*{0.075mm} -}
\cfoot[\thepage]{\thepage}
\rfoot[]{}

%\fancyhead[LE,RO]{\slshape \rightmark}
%\fancyhead[LO,RE]{\slshape \leftmark}

\begin{bibunit}[elsarticle-num]  
\let\stdthebibliography\thebibliography
\renewcommand{\thebibliography}{%
\let\section\subsection
\stdthebibliography}

\newcommand{\mbo}[1]{$#1$}

\noindent
The main theoretical issues {of the} FCC-ee studies discussed in this report may be summarised as follows.
\begin{enumerate}
\item {To} adjust {the precision of theory predictions} to the experimental demands from the \mbox{FCC-ee}, an update of existing {software} and {the development of new, independent software} will be needed. This should include, in the first instance, solutions to the following issues: 
\begin{enumerate}
\item factorisation to infinite order of multiphoton {soft-virtual QED} contributions; 
 \item
resummations in Monte Carlo generators; 
\item
 disentangling {of} QED and EW corrections beyond one loop, with soft-photon
factorisation or resummation;
\item {proper implementation} of higher-loop effects, {such as} Laurent series around {the} Z peak; 
\item further progress in methods and tools for multiloop calculations and Monte Carlo generators.
\end{enumerate}
Some discussions have been initiated in the {2018} report \cite{Blondel:2018mad}; here, {they are extended} in {the} Introduction and  Chapters B and C.
\item 
To meet the experimental precision of the FCC-ee Tera-Z for electroweak precision
observables (EWPOs), even three-loop {EW} calculations of the $\mathrm{Zf{\bar f}}$ vertex will be needed, 
comprising the loop orders 
${\cal{O}}(\alpha \alpha_\mathrm{s}^2), 
{\cal{O}}(N_\mathrm{f} \alpha_{}^2 \alpha_\mathrm{s}),
{\cal{O}}(N_\mathrm{f}^2 \alpha_{}^3)$,
and {also the} corresponding  QCD four-loop terms. 
This was mainly a subject of the {2018} report \cite{Blondel:2018mad}.
\item To decrease the $\alpha_{\mathrm{QED}}$ uncertainty by a factor of  five to ten, to the level (3--5$) \times 10^{-5}$, will require improvements in low-energy experiments.  Alongside this, the perturbative QCD (pQCD) prediction of the Adler function must be  improved by a factor of two, accomplished with better uncertainty estimates for $m_\mathrm{c}$ and $m_\mathrm{b}$. The
next mandatory improvements required are: 
\begin{enumerate}
\item four-loop massive pQCD calculation of {the} Adler function;
\item {improved} $ \alpha_\mathrm{s}$ in {the} low $ Q^2$ region above the $\tau$ mass;
\item a better control and understanding of $\Delta\alpha^{(5)}_{\rm had}(M_\mathrm{Z}^2)$, in terms of {\mbo{R} data;}
\item different methods  for directly accessing \mbo{\alpha(M_\mathrm{Z}^2)}, \eg the muon forward--backward asymmetry, or for calculating $\alpha_\mathrm{QED}$, either based on a radiative return experiment, \eg at the FCC-ee Tera-Z, or using lattice QCD methods.
\end{enumerate}
This is discussed in Chapter B.
\item FCC-ee precision measurements require many improvements on the theoretical QCD side. These include: (i) higher-order pQCD fixed-order calculations; (ii) higher-order logarithmic resummations; (iii) per-mille-precision extractions of the $\alpha_\mathrm{s}$ coupling; and (iv) an accurate control of non-perturbative QCD effects (such as, \eg\ colour reconnection, hadronization), both analytically and as implemented in the Monte Carlo generators.
These issues are discussed in Chapter B. 
\item 
The reduction of the theoretical uncertainty of the total W pair
production cross-section to the level of  $\sim 0.01\%$ at the FCC-ee-W
requires at least the calculation of ${\cal{O}}(\alpha^2)$ and dominant ${\cal{O}}(\alpha^3)$ corrections to double-resonant diagrams. Estimates within an effective field theory (EFT) approach show  that the theory-induced systematic uncertainty of the mass measurement from a threshold scan can be at the level of $\Delta \MW=(0.15-0.60)\UMeV$. The lower value results from assuming that the non-resonant corrections are under control. In addition,  it is also essential to reduce the uncertainty from initial-state radiation (ISR) corrections and QCD corrections for hadronic final states to the required accuracy.
This is discussed in {Chapter} B. 
\item
 Predictions for H decay widths and branching ratios are known with sufficient accuracy for the LHC. At the FCC-ee, the Higgs mass can be measured with a precision below $0.05\UGeV$. The dependence of {EWPOs}  on $M_\mathrm{H}$ is mild, $\propto \alpha  \log (M_\mathrm{H}/M_\mathrm{W})$, and an accuracy of $0.05\UGeV$ of $M_\mathrm{H}$ will not affect their determination. The main  improvements in Higgs boson studies will be connected with a better determination of branching ratios and self-couplings. 
More on related issues is discussed in the Introduction and {in} Chapter B.  
\item  The {top pair} line shape for centre-of-mass energies close to the $\mathrm{t \bar t}$  production threshold is highly sensitive to the mass of the
top quark,  allowing its determination with unprecedented precision. The statistical uncertainty of the measurement ($\sim 20\UMeV$) is projected to be significantly less than the current theoretical uncertainty. It is crucial to continuously improve the theoretical prediction. The most sensitive observable is the total production cross-section for $\mathrm{b\bar{b}W}^{+}\mathrm{W}^{-}\mathrm{X}$ final states near the top pair production threshold. A very precise knowledge of the strong coupling constant from other sources will be crucial {in order} to meaningfully
constrain the top Yukawa coupling. 
These issues are discussed in Chapter B.
\item  
Proper truncation of the ultraviolet scale $\Lambda$ depends on the experimental precision of the observables and Standard Model effective field theories  (SMEFTs) must be adjusted to FCC-ee experimental conditions, \eg in construction of appropriate complete operator bases and Wilson coefficients (WCs) for Beyond the Standard Model (BSM) theories. 
This issue is discussed in Chapter D. 
\item The FCC-ee and the FCC-hh will both be sensitive to BSM physics and exotic massive states reaching tens of TeV or very weak couplings. It is proposed to use the SMEFT framework and constrain the Higgs triple coupling by analysing precision measurements. For these studies, but also exotic Higgs decays, it will be important  to combine the LHC and HL-LHC data with an analysis at the FCC-ee.\\ 
These issues are discussed in Chapter E.
\end{enumerate}
\begin{flushleft}
\renewcommand\bibname{Reference}

\end{flushleft}
\end{bibunit}

%%%%%%%%%%%%%%%%%%%%%%%%%%%%%%%%%%%%%%%%%%%%%%%
%%%%%%%%%%%%%%%%%%%%%%%%%%%%%%%%%%%%%%%%%%%%%%%%%
\clearpage \pagestyle{empty} 
\cleardoublepage
%============================================
%============================================
\chapter[{Introduction and overview\\ {\it A. Blondel, J. Gluza, S. Jadach, P. Janot, T. Riemann}}]{Introduction and overview}
\label{ch-s1}
%\input{Introduction/bibunit-intro.tex}
%============================================

\pagestyle{fancy}
\fancyhead[CO]{\thechapter \hspace{1mm} Introduction and overview}
\fancyhead[RO]{}
\fancyhead[LO]{}
\fancyhead[LE]{}
\fancyhead[CE]{}
\fancyhead[RE]{}
\fancyhead[CE]{A.~Blondel,~J.~Gluza,~S.~Jadach,~P.~Janot,~T.~Riemann}
\lfoot[]{}
\cfoot{-  \thepage \hspace*{0.075mm} -}
\rfoot[]{}

\fancyfoot[LO]{}
\fancyfoot[LE]{}

\begin{bibunit}[elsarticle-num]  
\let\stdthebibliography\thebibliography
\renewcommand{\thebibliography}{%
\let\section\subsection
\stdthebibliography}
      
\noindent
%04
{\bf Contribution\footnote{This contribution should be cited as:\\
A.~Blondel,~J.~Gluza,~S.~Jadach,~P.~Janot,~T.~Riemann,
Introduction and overview,  
%04 DOI:10.23731/CYRM-2020-XXX.\thepage, in:
%04 \url{http://dx.doi.org/10.23731/CYRM-2020-XXX.\thepage}, in:
DOI: \href{http://dx.doi.org/10.23731/CYRM-2020-003.\thepage}{10.23731/CYRM-2020-003.\thepage}, in:
Theory for the FCC-ee, Eds. A. Blondel, J. Gluza, S. Jadach, P. Janot and T. Riemann,
\\CERN Yellow Reports: Monographs, CERN-2020-003,
%04 \url{http://dx.doi.org/10.23731/CYRM-2020-XXX}, p. \thepage.} 
DOI: \href{http://dx.doi.org/10.23731/CYRM-2020-003}{10.23731/CYRM-2020-003},
p. \thepage.
\\ \copyright\space CERN, 2020. Published by CERN under the 
%04-2
\href{http://creativecommons.org/licenses/by/4.0/}{Creative Commons Attribution 4.0 license}.
%04-2 Creative Common Attribution CC BY 4.0 Licence.
} by: A.~Blondel,~ J.~Gluza,~S.~Jadach,~ P.~Janot,~T.~Riemann}
\\ 
Corresponding author: 
J. Gluza [janusz.gluza@cern.ch]\\
\vspace*{.5cm}

\noindent This report includes a collection of {studies} devoted to a discussion of 
{(i)} the status of theoretical efforts 
towards the calculation of higher-order Standard Model (SM) corrections needed for the FCC-ee precision measurement programme, 
{(ii)} the possibility of making  discoveries in physics by means of these precision measurements, and 
{(iii)} methods and tools that must be developed to guarantee precision calculations of the observables to be measured. 
This report originates from presentations at the 11th FCC-ee Workshop: Theory and Experiments,
8--11 January 2019, CERN, Geneva \cite{mini2}, with 117 registered participants and 42  talks on theory.

\section[The FCC-ee electroweak  factory]{The FCC-ee electroweak  factory}
%%% Electroweak Factory!!
In the {2018} %previous 
report \cite{Blondel:2018mad}, we focused on  {theoretical} issues {of the} FCC-ee Tera-Z, which will be a $\mathrm{e}^+\mathrm{e}^-$ collider working at the Z resonance energy region. However, {the FCC-ee} collider project will work in several energy regions, making it  a complete \emph{electroweak factory},  {covering the} direct production of all massive bosons {of the SM} and the top {quark}.  {This} plan is summarised in Table~\ref{tab:FCC-ee-runplan}. 

\begin{table}[h!]
\center
\caption{Run plan for FCC-ee in its baseline configuration with two experiments. The WW event numbers are given for the entirety of the FCC-ee running at and above the WW threshold. \label{tab:FCC-ee-runplan}}
\begin{tabular}{l l l l l} \hline \hline 
Phase & Run duration  & Centre-of-mass  &  Integrated   & Event   \\ 
& (years)  & energies & luminosity &  statistics \\ 
&   & (GeV) & (ab$^{-1}$) &   \\ 
\hline 
FCC-ee-Z & 4  & \phantom{1}88--95   & 150   & $3 \times 10^{12}$ visible Z decays   \\ 
FCC-ee-W & 2  & 158--162 &  \phantom{1}12   & \phantom{$3\ \times\,$}$10^8$ WW events               \\ 
FCC-ee-H & 3  & 240     &  \phantom{15}5    & \phantom{$3\ \times\,$}$10^6$ ZH events               \\ 
FCC-ee-tt & 5  & 345--365 &  \phantom{15}1.7 & \phantom{$3\ \times\,$}$10^6$ $\mathrm{t\bar{t}}$ events       \\
 \hline \hline
\end{tabular} 
\end{table}

The exceptional precision of the FCC-ee comes from several features of the programme.
\begin{enumerate}
    \item Extremely high statistics of $5\times10^{12}$ Z decays, $10^8$ WW, $10^6$ ZH, and $10^6$ ${\rm t\bar t}$ events.       
    \item  High-precision (better than $100\UkeV$) absolute determination of the centre-of mass energies at the Z pole and WW threshold, thanks to the availability of transverse polarisation and the resonant depolarisation. This is a unique feature of the circular lepton colliders, ${\rm e^+ e^-}$ and $\upmu^+ \upmu^-$. At higher energies, WW, ZZ, and Z$\upgamma$ production can be used to constrain the centre-of-mass energy with precisions of 2 and $5\UMeV$, at the ZH cross-section maximum and at the ${\rm t\bar t}$ threshold, respectively. At all energies, ${\rm e^+e^-}\to \upmu^+\upmu^-$ events, which occur at a rate in excess of $3\UkHz$ at the Z pole, provide, by themselves, in a matter of minutes, the determination of the centre-of mass energy spread, the residual difference between the energies of ${\rm e^+}$ and ${\rm e^-}$ beams and (relative) centre-of-mass energy monitoring with a precision that is more than sufficient for the precision needs of the programme.   
    \item The clean environmental conditions and an optimised run plan allow a complete programme of ancillary measurements of currently precision-limiting input quantities for the precision EW tests. This is the case for the top quark mass from the scan of the t\={t} production threshold; of the unique, direct, measurement of the QED running coupling constant at the Z mass from the Z--$\upgamma$ interference; of the strong coupling constant by measurements of the hadronic-to-leptonic branching fractions of the Z, the W, and the $\uptau$ lepton; and, of course, of the Higgs and Z masses themselves. 
\end{enumerate}

For {the reader's} convenience, we also include Table~\ref{tab:FCC-observables}  from the CDR, showing 
some of the most significant FCC-ee experimental accuracies compared with those of the current measurements.  More on {the} experimental precision of {the} FCC-ee can be found in volumes 1 and 2 of the CDR documents \cite{Abada:2019zxq,Abada:2019lih}. The experimenters are working hard to reduce systematic uncertainties by devising dedicated methods and ancillary measurements; the task of the theoretical community will be  to ensure that the SM predictions will be precise enough so as not to spoil the best foreseeable experimental accuracies, \ie the statistical uncertainties.   
% \cite{Benedikt:2018qee}

\setlength{\tabcolsep}{2pt}
\begin{table}
\centering
\caption{Measurement of selected electroweak precision observables (EWPOs) at the \mbox{FCC-ee}, compared with the current precision. The systematic uncertainties are initial estimates and might improve on further examination.
This set of measurements, together with those of the Higgs properties, achieves indirect sensitivity to new physics up to a scale $\Lambda$ 
of $70\UTeV$ in a description with dimension-6 operators, and possibly much higher in some specific new physics models. \label{tab:FCC-observables}}
\begin{tabular}{lllllll}
\hline\hline
Observable  & Current &  &          &  FCC-ee  &  FCC-ee  &  Comment,   \\ 
            & value  &$\pm$& Error  &  stat. &   syst.     &  dominant experimental error \\ 
\hline  
$ m_\mathrm{_Z}$ (keV)    &  91186700   & $\pm$ &  2200    & 4  & 100  & From Z line shape scan,  \\ 
$  $  &  & &    &   &  &  beam energy calibration  \\

$ \mathrm{  \Gamma_Z  ~(keV) } $  & 2495200   & $\pm$ &  2300    & 7  & 100  & From Z line shape scan,  \\
$  $  &  & &    &   &   &  beam energy calibration  \\

$  R_{\ell}^\mathrm{Z} ~(\times 10^3) $  & 20767 & $\pm$ &  25   & 0.06   & 0.2--1   &  Ratio of hadrons to leptons, \\
$  $  &  & &    &   &   &  acceptance for leptons  \\

$  \alpha_\mathrm{s} (m_\mathrm{Z})  ~(\times 10^4) $  & 
 1196 & $\pm$ &  30  &  0.1  &  0.4--1.6  &   
From $\mathrm{  R_{\ell}^{Z}}$ \\

$   R_\mathrm{b} ~(\times 10^6) $  & 216290 & $\pm$ &  660   & 0.3   &  <60  &  Ratio of $\rm{ b\bar{b}}$  to hadrons,  \\
$  $  &  & &    &   &   &  stat. extrapolated from SLD
\\
$ \mathrm{\sigma_{had}^0} ~(\times 10^3)$ (nb) & 41541 & $\pm$ &  37   & 0.1  &  4  &  Peak hadronic cross-section,  \\
$  $  &  & &    &   &   &  luminosity measurement  \\
$   N_{\upnu}  (\times 10^3) $  & 2991  & $\pm$ &  7   & 0.005   &  1  &  Z peak cross-sections, \\
$  $  &  & &    &   &  &   luminosity measurement \\
$ \mathrm{ sin^2{\theta_{W}^{\rm eff}}} (\times 10^6) $  & 231480   & $\pm$ &  160   & 3   &  2--5  &   
From $ A_\mathrm{FB}^{\upmu \upmu} $ 
from $  A_\mathrm{FB}^{{\upmu} {\upmu}}$  at Z peak,\\
$  $  &  & &    &   &   &  beam energy calibration  \\
$  1/\alpha_\mathrm{QED} (m_\mathrm{Z})  (\times10^3) $  & 128952 
  & $\pm$ &  14   & 4   &  Small  &   
From $  A_\mathrm{FB}^{{\upmu} {\upmu}}$ off peak\\
$   A_\mathrm{FB}^\mathrm{b},0 ~(\times 10^4) $  & 992 & $\pm$ &  16   & 0.02   &  1-3  &  b quark asymmetry at Z pole,  \\
$  $  &  & &    &   &   &  from jet charge \\
$ A_\mathrm{FB}^{\mathrm{pol},\tau} ~(\times 10^4) $  & 1498 & $\pm$ &  49   & 0.15   &  <2  &  $\uptau$ polarisation and charge asymmetry,  \\
$  $  &  & &    &   &   &  $\uptau$ decay physics \\
$  m_\mathrm{W}$  (MeV)    &  80350   & $\pm$ &  15    & 0.5  & 0.3  & From WW threshold scan, \\ 
$  $  &  & &    &   &  &  beam energy calibration  \\
$ \mathrm{  \Gamma_W  ~(MeV) } $  & 2085   & $\pm$ &  42    & 1.2  & 0.3  & From WW threshold scan, \\
$  $  &  & &    &   &   &  beam energy calibration  \\
$  \alpha_\mathrm{s} (m_\mathrm{W})   (\times 10^4)$  & 
 1170   & $\pm$ & 420 &  3  & Small  &   
  From $ R_{\ell}^\mathrm{W} $\\
$   N_{\upnu}  (\times 10^3) $  & 2920 & $\pm$ &  50   & 0.8   & Small   &   Ratio of invisible to leptonic, \\
$  $  &  & &    &   &  & in radiative Z returns  \\
$  m_\mathrm{top}$  (MeV/$c^2$)    &  172740   & $\pm$ &  500    & 17  & Small  & From $\mathrm {t\bar{t}}$ threshold scan, \\ 
$  $  &  & &    &   &  &  QCD errors dominate  \\
$ \Gamma_\mathrm{top}$  (MeV/$c^2$)    &  1410   & $\pm$ &  190    & 45  & Small  & From $\mathrm {t\bar{t}}$ threshold scan, \\ 
$  $  &  & &    &   &  &  QCD errors dominate \\
$ \mathrm{ \lambda_{top}/\lambda_{top}^{SM}   } $  &   1.2 
    & $\pm$ &  0.3    & 0.10  & Small  & From $\mathrm {t\bar{t}}$ threshold scan, \\ 
$  $  &  & &    &   &  &  QCD errors dominate \\
$ \mathrm{ ttZ ~couplings   } $  &   
   & $\pm$ &  30\%   & 0.5 -- 1.5\%  & Small  & From $ E_\mathrm{CM}=365 \UGeV$ run  \\ 
\hline\hline
\end{tabular} 
\end{table}
\setlength{\tabcolsep}{2pt}

If future theory uncertainties match the FCC-ee experimental precision, the many different measurements from the FCC-ee will provide the capability of exhibiting and deciphering signs of new physics. Here are two examples: the EFT analysis searching for signs of heavy particles physics with SM couplings shows the potential to exhibit signs of new particles up to around $70\UTeV$; with a very different but characteristic pattern, observables involving neutrinos would show a significant deviation if these neutrinos were mixed with a heavy counterpart at the level of one part in $100\,000$, even if those were too heavy to be directly produced. 

%\renewcommand{\arraystretch}{1.07}   %  \rule{0pt}{25pt}
%\begin{table}[h!] %Table Foreword 2
%       \begin{center}
%%              \begin{tabular}{||l|rcl|c|c|r||}

Table~\ref{tab:FCC-observables} shows that the FCC-ee has the potential to achieve (at least) {a} 20--100 {times} higher precision or better in electroweak precision measurements over the present state-of-the-art  situation. This includes such input quantities  as the Z, Higgs, and top masses, and the strong and QED coupling constants at the Z scale. This extremely favourable situation will require leap-jumps in the precision of {the} theoretical computations for Standard Model phenomena,  for  all quantities given in Table~\ref{tab:FCC-observables}. The theory calculation must also be able to include the improved input parameters    \cite{ALEPH:2005ab,Blondel:2018mad}, which, in the particular case of
the FCC-ee, will be measured within the experimental programme.

The quantities listed in  Table \ref{tab:FCC-observables} are called {\em electroweak precision observables} (EWPO) and encapsulate experimental data after extraction of {well-known} and controllable QED and QCD effects, in a model-independent manner.
They provide a convenient bridge between real data and the predictions of the SM, or {of the SM plus new physics}.
%{\cmg The important advantage of EWPOs is that,} 
Contrary to raw experimental data (like differential cross-sections),
EWPOs are also well-suited for archiving and long-term use. Archived EWPOs can be exploited over long periods of time
for comparisons with steadily improving theoretical calculations of the SM predictions,
and for validations of the new physics models beyond the SM.
They are also useful for the comparison and combination of results from different experiments.
However, removing trivial but sizeable QED or QCD effects from EWPOs might induce additional sources of  uncertainty. The work needed is  well-known  concerning QED, more significant conceptual work may need to be done for QCD. 

Let us summarise briefly the mandatory improvements of the calculations of 
QED effects in EWPOs according to recent work \cite{Jadach:2019bye}:
\begin{enumerate}
\item%[(i)] 
improved calculation of the additional light
fermion pair emissions (for Z boson mass and width); 
\item%[(ii)] 
better calculation of the final-state radiation effects in the presence of cut-offs ({for} $R_\ell^\mathrm{Z}$); 
\item%[(iii)] 
implementation of a new QED matrix element in the Monte Carlo (MC) event generator for low-angle Bhabha processes
(for the luminosity determination in view of the measurement of $\sigma_\mathrm{had}^0$ 
and other cross-sections); 
\item%[(iv)] 
${\cal O}(\alpha^2)$ calculation for $\mathrm{e}^+\mathrm{e}^-\to \mathrm{Z}\upgamma$ (for the determination of $N_\upnu$); 
\item%[(v)] 
improved MC simulation of $\uptau$ decays 
(for the effective weak mixing angle and tau branching ratio measurements); 
\item%[(vi)] 
QED effects at the W pair production threshold (for  measurement of the W mass and width); 
\item%[(vii)]
initial--final-state interference (\eg for the forward--backward charge asymmetry of lepton pairs around the Z peak). 
\end{enumerate}

For more on the related subject of the separation of QED effects 
from weak quantities at {the} FCC-ee precision 
and generally on the improvements in the definition of EWPOs,
see recent discussions in Ref. \cite{Blondel:2018mad}.
A similar systematic discussion of the QCD effects in EWPOs is in progress, 
see Ref.~\cite{Blondel:2018mad} and Section~B.2 in this report.

For {the} FCC-ee data analysis, owing to the rise of non-factorisable QED effects above the experimental uncertainties, direct {use of} MC programs might become the standard for fitting EWPOs to {the} data, even at the Tera-Z stage \cite{Blondel:2018mad,Blondel:2019qlh,Jadach:2019bye}. New MC event generators will have to provide built-in provisions for {an}
efficient direct fitting of EWPOs to data, which are not present in the LEP legacy MCs.
Section C.3 of Ref.~\cite{Blondel:2018mad} describes
possible forms of future EWPOs at FCC-ee experiments 
and {specifies the} new required MC software.
It is emphasized there that, owing to non-factorisable QED contributions,  the multiphoton QED effects 
will have to be factorised at the amplitude level. Additional quantities available in tau and heavy flavour physics will reach  $10^{-5}$ precision and are likely to need similar attention. 

Very precise determinations of $M_\mathrm{W}$
at the FCC-ee will rely on the precise measurement
of the cross-section of the $\mathrm{e}^+\mathrm{e}^- \to \mathrm{W}^+\mathrm{W}^-$ process near the threshold.
A statistical precision {of} $0.04\%$ of this cross-section translates
into 0.6\,MeV experimental uncertainty on $M_\mathrm{W}$, compared with the current 3\,MeV theoretical uncertainty  for  $M_\mathrm{W}$.
Therefore, improved theoretical calculations are required 
for the generic $\mathrm{e}^+\mathrm{e}^- \to 4\mathrm{f}$ process near the WW threshold with an improvement of one order of magnitude. The most economical solution will be to combine the \order{\alpha^1}
calculation for the $\mathrm{e}^+\mathrm{e}^- \to 4\mathrm{f}$ process with the \order{\alpha^2}
calculation for the doubly resonant
$\mathrm{e}^+\mathrm{e}^- \to \mathrm{W}^+\mathrm{W}^-$ subprocess.
The former calculation is already available~\cite{Denner:2005fg}.
The latter will need to be developed; inclusion of the resummed QED corrections will be mandatory.
For details, see Chapter B and Ref. \cite{Skrzypek:Jan2019}.

In the case of the FCC-ee-H, $M_\mathrm{H}$ will be obtained from the $\mathrm{e}^+\mathrm{e}^- \to \mathrm{HZ}$ process
with a precision better than 10\,MeV \cite{FCC-CDR,Abada:2019zxq}. 
Theory uncertainties 
(mainly owing to final-state radiation effects) will be subdominant. The main focus will be on calculations of Higgs boson branching ratios and self-couplings. See Chapters B and E.

The anticipated experimental uncertainty
on the $m_\mathrm{t}$ measurement at FCC-ee-tt~\cite{Blondel:2018mad}
is $  \order{20}\mev$.
On the theory side, there are several sources of uncertainties: 
%\cite{Vos:2016til}:
(i) 
the perturbative uncertainty
for the calculation of the threshold shape with higher-order QCD corrections;  
(ii) 
the threshold mass definition translated into the
$\overline{\text{MS}}$ scheme; and
(iii) the precision of $\as$.  
Combining these three sources of uncertainty,
a theoretical uncertainty close to the experimental one and less than $50\UMeV$ for $\mt$  appears feasible.\footnote{Examples show that estimations of higher-order corrections can differ from actual calculations by factors of three to five \cite{Dubovyk:2018rlg,Blondel:2019qlh}.} 
In addition, a very accurate determination of the efficiency of experimental acceptances and selection cuts is needed. 
This task will require the inclusion of higher-order corrections 
and resummation results in a Monte Carlo event generator; 
next-to-leading-order (NLO) QCD corrections for off-shell $\mathrm{t\bar{t}}$ production, and matching between these contributions,   complement previous semi-analytic results. 

In this report, we are especially interested in {the} discussion of input parameters and {of} EWPOs connected with W, H, and top production physics. These are masses of  heavy {SM} particles, their couplings, and also $\alpha_{\mathrm{QED}}$ and $\alpha_{\mathrm{QCD}}$, which, as running quantities, must be adjusted carefully at {the} considered high-energy regions. These issues will be discussed in this report.

\section[What this theory report brings: an overview]{What this theory report brings: an overview} 

The report is divided into four basic chapters. Both the workshop and this report 
are mainly devoted to precision theoretical calculations. 
It is a most important subject because the value
of most of the FCC-ee experimental analyses relies on the precision of the Standard Model and BSM predictions. 

In Chapter B, the status and prospects for measurements and determination of $\alpha_\mathrm{QED}$ and $\alpha_\mathrm{s}$ at the FCC-ee are given, but also issues of QED and QCD resummations, 
an EFT  radiative correction approach to W boson production, 
heavy quarkonia, 
analysis of the weak mixing angle from data (important, as it definitely has non-perturbative effects different from those in $\alpha$), 
QCD vertex functions beyond two loops, 
EFT and QED in flavour physics, 
top pair production and mass determination, 
and a summary of SM precision predictions for partial Higgs decay widths.

In Chapter C, numerical and analytical methods for precision multiloop calculations are presented and recent advances in the field are discussed.
The chapter is an addition to the 2018 report \cite{Blondel:2018mad}. We mentioned already that Monte Carlo generators are very important, as they link pure experimental data with theory. 
Generators for precision $\mathrm{e}^+\mathrm{e}^-$ simulations, $\tau$, top, and W boson physics,  heritage projects, and the need for proper software preservation with Monte Carlo generators are also discussed in Chapter C.

Chapter D consists of only one contribution.
SMEFT theory is a bridge between SM physics and the analysis of extended gauge models. The chapter is connected with this issue and a specific code is presented. For another discussion, see the 
talk by J.~de Blas  \cite{Blas:Jan2019}.

In  Chapter E, finally, three contributions are collected, about Higgs models that go beyond the Standard Model theory.

\end{bibunit}

\clearpage
%============================================
%============================================
\chapter{Precision calculations in the Standard Model}
\label{ch-sm}
%============================================

\pagestyle{fancy}
\fancyhead[CO]{\thechapter.\thesection \hspace{1mm}  $\alpha_{{\rm QED,}\,{\rm eff}}(s)$ for precision physics at the FCC-ee/ILC}
\fancyhead[RO]{}
\fancyhead[LO]{}
\fancyhead[LE]{}
\fancyhead[CE]{}
\fancyhead[RE]{}
\fancyhead[CE]{F. Jegerlehner}
\lfoot[]{}
\cfoot{-  \thepage \hspace*{0.075mm} -}
\rfoot[]{}
    
 \begin{bibunit}[elsarticle-num] % define the bib-style for the unit: elsarticle-num.bst
%  text-1; this is the corresponding section
%\putbib[2loops] % the *.bib
%\end{bibunit}
% go-on
%--- from: bibunits.sty, adapts the font size of ``References'' to section
\begin{flushleft}
\let\stdthebibliography\thebibliography
\renewcommand{\thebibliography}{%
\let\section\subsection
\stdthebibliography}
\end{flushleft}
%===================================================

\def\dahs{\Delta\alpha^{(5)}_{\rm had}(s)}
\def\mz{M_\mathrm{Z}^2}
\def\MZ{M_\mathrm{Z}}
\def\mw{M_\mathrm{W}^2}
\def\MW{M_\mathrm{W}}
\def\dah{\Delta\alpha^{(5)}_{\rm had}}
\newcommand{\noi}{\noindent}
\newcommand{\dal}{\Delta \alpha}
\def\dahs{\Delta\alpha^{(5)}_{\rm had}(s)}
\def\dghs{\Delta\alpha^{(5)}_{2\,{\rm had}}(s)}
\def\dahz{\Delta\alpha^{(5)}_{\rm had}\left(\MZ^2\right)}
\def\dah0{\Delta\alpha^{(5)}_{\rm had}(-s_0)}
\def\damu{\delta \amu}
\newcommand{\SU}{{SU}}
\newcommand{\xnull}[1]{\stackrel{\circ}{#1}}
\newcommand{\cin}[1]{~\cite{#1}}
\newcommand{\pvint}{{\cal P}\!\!\!\!\!\!\! \int}
\newcommand{\labelitemv}{$\bullet$~~}
\newcommand{\labelitemx}{$\mathbf{-}$~~}
\newcommand{\gapprox}{\raisebox{-.2ex}{$\stackrel{\textstyle>}{\raisebox{-.6ex}[0ex][0ex]{$\sim$}}$}}
\newcommand{\lapprox}{\raisebox{-.2ex}{$\stackrel{\textstyle<}{\raisebox{-.6ex}[0ex][0ex]{$\sim$}}$}}
\newcommand{\mytexttilde}{{\raise.17ex\hbox{$\scriptstyle\mathtt{\sim}$}}}
\newcommand{\LPl}{\Lambda_{\rm Pl}}
\newcommand{\lpl}{\Lambda_{\rm Pl}}
\newcommand{\MSb}{$\overline{\mathrm{MS}}$ }
\newcommand{\be}{\begin{equation}}
\newcommand{\ee}{\end{equation}}
\newcommand{\Gmu}{G_\upmu}
\newcommand{\dalh}{\Delta \alpha_{\rm had}}
\newcommand{\al}{\alpha }
\newcommand{\gv}{\mbox{GeV}}
\newcommand{\mv}{\mbox{MeV}}
\newcommand{\ppm}{\pi^+ \pi^-}
\newcommand{\eepp}{e^+e^- \rightarrow \pi^+\pi^-}
\newcommand{\x}{\phantom{x}}
\newcommand{\bl}{\phantom{-}}
\renewcommand{\veps}{\varepsilon}
\newcommand{\mbo}[1]{$#1$}
\newcommand{\epm}{\mathrm{e}^+\mathrm{e}^- }
\newcommand{\mumu}{\upmu^+\upmu^- }
\newcommand{\power}[1]{\times 10^{#1} }
\newcommand{\D}{\mathrm{d}}
\newcommand{\E}{\mathrm{e}}
\newcommand{\Impa}{{\rm Im}\,}
\newcommand{\Repa}{{\rm Re}\,}
\newcommand{\semis}{\;;\;\; }
\newcommand{\comas}{\;,\;\; }
\newcommand{\wz}{\sqrt{2}}
\newcommand{\amu}{a_\upmu}
\newcommand{\amuh}{a_\upmu^{\rm had} }
\renewcommand{\bea}{\begin{eqnarray}}
\renewcommand{\eea}{\end{eqnarray}}
\newcommand{\epo}{\;. }
\newcommand{\cA}{{\cal A} }
\newcommand{\cJ}{{\cal J} }
\newcommand{\cL}{{\cal L} }
\newcommand{\crn}{\nonumber \\}
\newcommand{\bra}[1]{\langle{#1}|}
\newcommand{\braket}[1]{\langle{#1}\rangle}
\newcommand{\ket}[1]{|{#1}\rangle}
\newcommand{\FF}{{\cal F}_{\pi^{0*}\gamma^*\gamma^*}}
\newcommand{\FFa}{{\cal F}_{\pi^0\gamma^*\gamma^*}}
\newcommand{\FFb}{{\cal F}_{\pi^{0*}\gamma\gamma^*}}
\newcommand{\FFc}{{\cal F}_{\pi^{0*}\gamma^*\gamma}}
\newcommand{\FFbc}{{\cal F}_{\pi^{0*}\gamma\gamma}}
\newcommand{\FFac}{{\cal F}_{\pi^0\gamma^*\gamma}}
\newcommand{\FFab}{{\cal F}_{\pi^0\gamma\gamma^*}}
\newcommand{\FFabc}{{\cal F}_{\pi^0\gamma\gamma}}
\renewcommand{\nn}{\nonumber}
\newcommand{\bary}{\begin{array}}
\newcommand{\eary}{\end{array}}
\newcommand{\ttcb}[1]{\multicolumn{2}{c|}{#1}}
\newcommand{\ttc}[1]{\multicolumn{2}{c}{#1}}
\newcommand{\ba}{\begin{eqnarray}}
\newcommand{\ea}{\end{eqnarray}}
\newcommand{\cS}{{\cal S}}
\newcommand{\cP}{{\cal P}}
\newcommand{\cQ}{{\cal Q}}
\newcommand{\cO}{{\cal O}}
\newcommand{\cT}{{\cal T}}
\newcommand{\cH}{{\cal H}}
\newcommand{\cC}{{\cal C}}
\newcommand{\cR}{{\cal R}}
\newcommand{\cD}{{\cal D}}
\newcommand{\cI}{{\cal I}}
\newcommand{\cZ}{{\cal Z}}
\newcommand{\cG}{{\cal G}}
\newcommand{\bit}{\begin{itemize}}
\newcommand{\eit}{\end{itemize}}
\newcommand{\cosi}{\cos^2 \Theta_i}
\newcommand{\sini}{\sin^2 \Theta_i}
\newcommand{\sinW}{\sin^2 \Theta_\mathrm{W}}
\newcommand{\cosW}{\cos^2 \Theta_\mathrm{W}}
\newcommand{\sing}{\sin^2 \Theta_\mathrm{g}}
\newcommand{\cosg}{\cos^2 \Theta_g}
\newcommand{\sinf}{\sin^2 \Theta_\mathrm{f}}
\newcommand{\cosf}{\cos^2 \Theta_\mathrm{f}}
\newcommand{\dro}{\Delta \rho}
\newcommand{\ha}{\frac12 }

%----------------------

\section
[ $\alpha_{{\rm QED,}\,{\rm eff}}(s)$ for precision physics at the
FCC-ee/ILC \\ {\it F. Jegerlehner}]
%-----------------------
{$\alpha_{{\rm QED,}\,{\rm eff}}(s)$ for precision physics at the
FCC-ee/ILC}
\label{contr:SM_Jegerlehner}
\noindent
{\bf Contribution\footnote{This contribution should be cited as:\\
F.~Jegerlehner,
$\alpha_{{\rm QED,}\,{\rm eff}}(s)$ for precision physics at the
FCC-ee/ILC,  
%04 DOI:10.23731/CYRM-2020-003.\thepage, in:
%04 \url{http://dx.doi.org/10.23731/CYRM-2020-003.\thepage}, in:
DOI: \href{http://dx.doi.org/10.23731/CYRM-2020-003.\thepage}{10.23731/CYRM-2020-003.\thepage}, in:
Theory for the FCC-ee, Eds. A. Blondel, J. Gluza, S. Jadach, P. Janot and T. Riemann,
\\CERN Yellow Reports: Monographs, CERN-2020-003,
%04 \url{http://dx.doi.org/10.23731/CYRM-2020-003}, p. \thepage.} 
DOI: \href{http://dx.doi.org/10.23731/CYRM-2020-003}{10.23731/CYRM-2020-003},
p. \thepage.
\\ \copyright\space CERN, 2020. Published by CERN under the 
%04-2
\href{http://creativecommons.org/licenses/by/4.0/}{Creative Commons Attribution 4.0 license}.}
by: F. Jegerlehner 
%\\ Corresponding Author: F. Jegerlehner 
{[fjeger@physik.hu-berlin.de]}}
\vspace*{.5cm}

%\begin{abstract}
\noindent Discovering the `physics behind precision' at future linear or circular
colliders (ILC or FCC projects) requires improved SM predictions based
on more precise input parameters. I will review the role of
\mbo{\alpha_{\rm QED,~eff}} at future collider energies and report on
possible progress based on results from low-energy machines.
%\end{abstract}

%################################################
%\clearpage
%%%%%%%%%%%%%%%%%%%%%%%%%%%%%%%%%%%%%%%%%%%%%%%%%

\subsection{$\alpha(M^2_\mathrm{Z})$ in precision physics (precision physics limitations)}
\begin{figure}
\centering
\includegraphics[height=9cm]{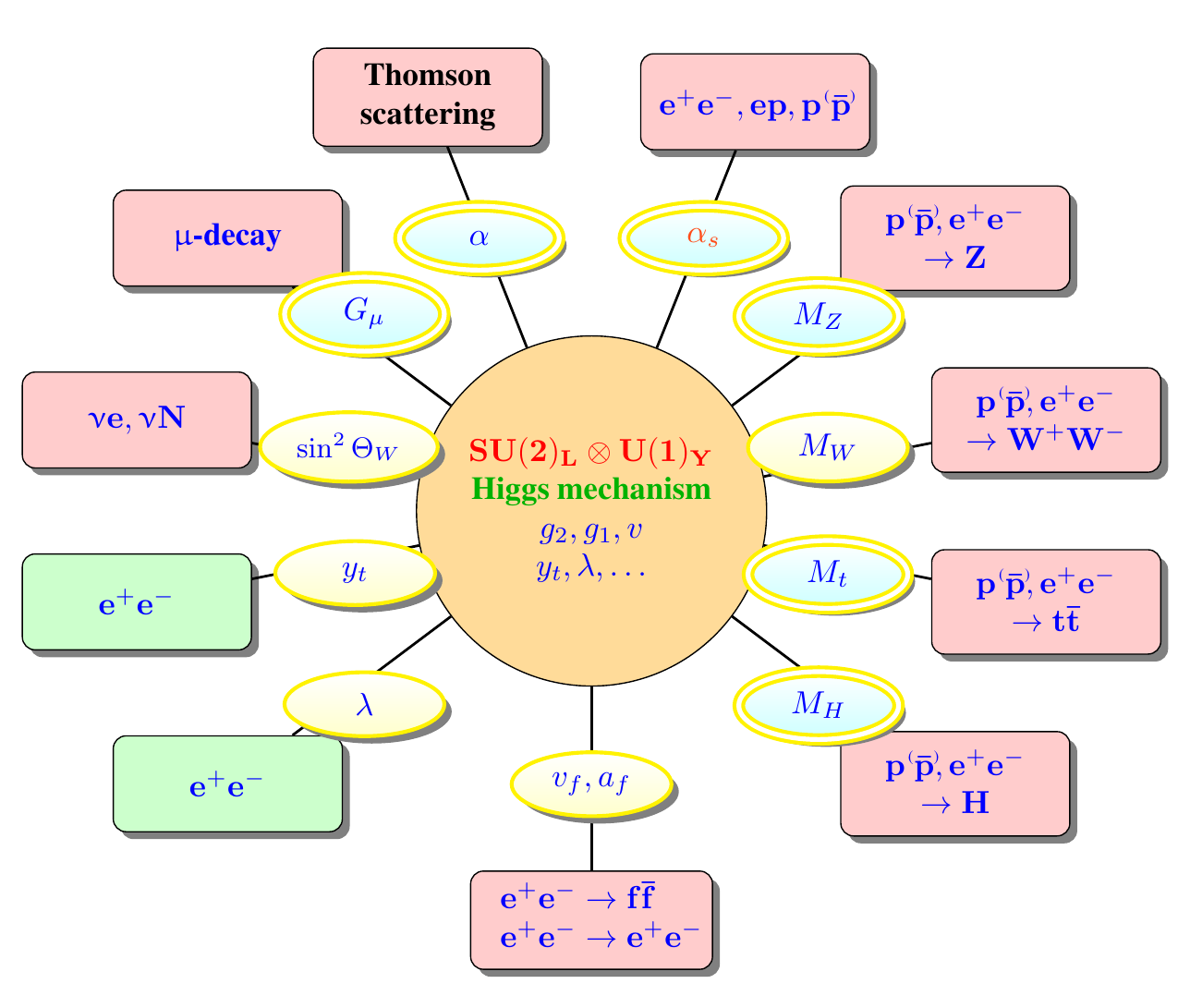}
\caption{Many precisely measurable pseudo-observables associated with
scattering-, production-, and decay processes are interrelated
and predictable in terms of a few independent input parameters.}
\label{fig:SMEWparams}
%\query{Please correct the figure labels of \Fref{fig:SMEWparams}. Put all
%particle names (W, H, t, \=t, f, $\upmu$ and all text labels in roman font,
%using italic font only for variables.}
\end{figure}

Uncertainties of hadronic contributions to the effective fine
structure constant $\alpha \equiv \alpha_{\mathrm{QED}}$ are a problem
for electroweak (EW) precision physics. Presently, we have $\alpha$,   $G_\upmu$,
and $M_\mathrm{Z}$ as the most precise input parameters, which, together with the top
Yukawa coupling \mbo{y_\mathrm{t}},  the Higgs self-coupling $\lambda$, and the strong
interaction coupling $\alpha_\mathrm{s}$ allow us to make precision predictions
for the particle reaction cross-sections encompassed by the Standard
Model (SM). The cross-section data unfolded form detector and photon
radiation resolution effects  are often conveniently representable in terms
of so-called pseudo-observables, such as $\sinf$, $v_\mathrm{f}$, $a_\mathrm{f}$, $M_\mathrm{W}$, $\Gamma_\mathrm{Z}$,
$\Gamma_\mathrm{W}, \dots$, as illustrated in \Fref{fig:SMEWparams}. 

Because of the
large 6\% relative correction between $\alpha$ in the classical limit
and the effective value $\alpha(M^2_\mathrm{Z})$ at the Z mass scale, where
50\% of the shift is due to non-perturbative hadronic effects, one is
losing about a factor of five orders of magnitude in
precision. Nevertheless, for the vector boson Z and W, top quark, and
Higgs boson precision physics possible at future $\epm$ colliders, the
best effective input parameters are given by $\alpha(M_\mathrm{Z}), G_\upmu$,
and 
$M_\mathrm{Z}$. The effective $\alpha(s)$ at a process scale $\sqrt{s}$ is given
in terms of the photon vacuum polarisation (VP) self-energy correction
$\Delta \alpha (s)$ by
\bea
\alpha(s)=\frac{\alpha}{1-\Delta
\alpha (s)} \semis \quad  \Delta \alpha(s)=\Delta \alpha_{\rm lep}(s)+\dahs +\Delta \alpha_{\rm top}(s)\,.
\eea
To be included are the perturbative lepton and top quark contributions, in
addition to the non-perturbative hadronic VP shift $\dahs$ from the
five light quarks and the hadrons they form.

The current accuracies of the corresponding SM input parameter
are:
\be \bary{cccccc}
\frac{\updelta \alpha}{\alpha} &\sim & 3.6 &\times& 10^{-9}\,, & \\[2mm]
\frac{\updelta G_\upmu}{G_\upmu} &\sim& 8.6 &\times& 10^{-6}\,, & \\[2mm]
\frac{\updelta M_\mathrm{Z}}{M_\mathrm{Z}} &\sim& 2.4 &\times& 10^{-5}\,, &\\[2mm]
\frac{\updelta \alpha(M_\mathrm{Z})}{\alpha(M_\mathrm{Z})} &\sim& 0.9 \div 1.6 &\times&
10^{-4} & {\rm (present: {lost\ }{ 10^{5}}{  \ in \ precision! })\,,}\\[2mm]
{ \frac{\updelta \alpha(M_\mathrm{Z})}{\alpha(M_\mathrm{Z})}} &{ \sim}& { 5 }& {
\times}& { 10^{-5}}
& \text{(FCC-ee/ILC requirement)\epo}
\eary
\ee
We further note that ${\updelta M_\mathrm{W}} / {M_\mathrm{W}}\sim 1.5 \times 10^{-4}\,,\,\,
   {\updelta M_\mathrm{H}} / {M_\mathrm{H}}\sim 1.3 \times 10^{-3}\,,\,\,
  {\updelta M_\mathrm{t}} / {M_\mathrm{t}}\sim 2.3 \times 10^{-3}\,,$ at present.
Evidently, $\alpha(M_\mathrm{Z})$ is the least precise among the basic input
parameters $\alpha(M_\mathrm{Z})$, $G_\upmu$, and $M_\mathrm{Z}$, and requires a major effort of
improvement. As an example, one of the most precisely
measured derived observables, the leptonic weak mixing parameter
$\sin^2 \Theta_{\ell\,{\rm eff}}=(1-v_{\ell}/a_{\ell})/4 =0.231\,48 \pm
0.000\,17$ and also the related W mass $M_\mathrm{W}=80.379\pm0.012\,\gv$ are
affected by the present hadronic uncdertainty $\updelta \Delta
\alpha(M_\mathrm{Z})=0.000\,20$ in predictions by $\updelta \sin^2
\Theta_{\ell\,{\rm eff}}=0.000\,07$ and $\updelta M_\mathrm{W}/M_\mathrm{W} \sim
4.3\power{-5}$, respectively. 

Here, one has to keep in mind that, besides $\Delta \alpha$, there is a
second substantial leading one-loop correction, which enters the neutral
to charged current effective Fermi-couplings ratio $\rho= G_{\rm
NC}(0)/G_{\rm CC}(0)=1+\Delta \rho\,,$ where $\Delta
\rho= {3\sqrt{2}M_\mathrm{t}^2\,G_\upmu} / {16 \pi^2}$ is quadratic in the
top quark mass. The mentioned ${\updelta M_\mathrm{t}} / {M_\mathrm{t}}$ uncertainty
affects the $M_\mathrm{W}$ and $\sin^2 \Theta_{\ell\,{\rm eff}}$
predictions, as given by
\begin{align}
\frac{\updelta M_\mathrm{W}}{M_\mathrm{W}} &\sim 
M_\mathrm{W}^2/(2 M_\mathrm{W}^2-M_\mathrm{Z}^2) \cdot \Delta \rho
\:\frac{\updelta M_\mathrm{t}}{M_\mathrm{t}}\sim 1.3 \power{-2}\,\frac{\updelta
M_\mathrm{t}}{M_\mathrm{t}}\simeq 3.0\power{-5}\,, \\
\frac{\updelta \sinf}{\sinf} &\sim \frac{2\,\cosf}{\cosf-\sinf}
\;\Delta \rho \:\frac{\updelta M_\mathrm{t}}{M_\mathrm{t}} \sim
2.7\power{-2}\,\frac{\updelta M_\mathrm{t}}{M_\mathrm{t}} \simeq 6.2\power{-5}\;,
\end{align}
which are comparable to the current
uncertainties from $\updelta \Delta \alpha$. Thus, an improvement of
$\updelta M_\mathrm{t}$
by a factor of five appears to be as important as an improvement of
$\alpha(M_\mathrm{Z})$. We are reminded that the dependence on $M_\mathrm{H}$ is very much
weaker because of the custodial symmetry, which implies the absence of $M_\mathrm{H}^2$ corrections, such that only relatively weak $\log M_H$ effects are remaining.

%\query{Should that be $\log H_M$ or $\log M_H$?}

The input parameter uncertainties affect most future
precision tests and may obscure new physics searches!  To
reduce hadronic uncertainties for perturbative QCD (pQCD)
contributions, last but not least, it is also very crucial  to improve
the precision of QCD parameters $\alpha_\mathrm{s},\; m_\mathrm{c}, \; m_\mathrm{b},\; m_\mathrm{t}$, which
is also a big challenge  for lattice QCD.

\subsubsection{The relevance of $\alpha(M^2_\mathrm{Z})$}
Understanding precisely even the simplest four-fermion, vector boson,
and Higgs boson production and decay processes, requires very precise
input parameters.

Unlike in QED and QCD in the SM, a spontaneously
broken non-Abelian gauge theory, there are intricate parameter
inter-dependences, all masses are related to couplings, and only six
quantities (besides $\mathrm{f} \neq \mathrm{t}$ fermion masses and mixing parameters), $\alpha$,
$G_\upmu$, and $M_\mathrm{Z}$, in addition to the QCD coupling $\alpha_\mathrm{s}$, the
top quark Yukawa coupling $y_\mathrm{}$, and the Higgs boson
self-coupling $\lambda_\mathrm{H}$, are independent. The effective
$\alpha(M_\mathrm{Z}^2)$
exhibits large hadronic correction that affects prediction-like
versions of the weak mixing parameter via
\bea
\sini\,\cosi\,
=\frac{\pi\,\alpha}{\sqrt{2}\,G_\upmu\,M_\mathrm{Z}^2}\,
\frac{1}{1-\Delta r_i}\semis \quad \Delta r_i =\Delta r_i(\alpha , \Gmu ,
M_\mathrm{Z} , m_\mathrm{H}, m_{\mathrm{f}\neq \mathrm{t}},m_\mathrm{t})\,,
\eea
with quantum corrections from gauge-boson self-energies and vertex and
box corrections, where
$\Delta
r_i$ depends on the definition of $\sini$. The various definitions
coincide at tree level and hence only differ by quantum effects. From
the weak gauge-boson masses, the electroweak gauge couplings, and the
neutral current couplings of the charged fermions, we obtain
\begin{align}
\sinW &= 1-\frac{M_\mathrm{W}^2}{M_\mathrm{Z}^2}\,,\\
\sing &= e^2/g^2=\frac{\uppi \al}{\sqrt{2}\:G_\upmu\:M_\mathrm{W}^2}\,,\\
\sinf &=
\frac{1}{4|Q_\mathrm{f}|}\;\left(1-\frac{v_\mathrm{f}}{a_\mathrm{f}} \right)\;,\;\;\mathrm{f}\neq \upnu\;,
\end{align}
for the most important cases and the general form of $\Delta r_i$ reads
\ba
\Delta r_i &=& \dal - f_i(\sini)\:\dro + \Delta r_{i\:\mathrm{reminder}}\,,
\label{derstruct}
\ea
with a universal term $\dal$, which affects the predictions of
$M_\mathrm{W}$, $A_\mathrm{LR}$, $A^\mathrm{f}_\mathrm{FB}$, $\Gamma_\mathrm{f}$, etc. 
The leading corrections are $\Delta
\alpha (M_\mathrm{Z}^2)=\Pi_\upgamma'(0)-\Repa \Pi'_\upgamma (M_\mathrm{Z}^2)$ from the
running fine structure constant and
\[
 \Delta \rho
=\frac{\Pi_\mathrm{Z}(0)}{M_\mathrm{Z}^2}-\frac{\Pi_\mathrm{W}(0)}{M_\mathrm{W}^2}+2\,\frac{\sin
\Theta_\mathrm{W}}{\cos \Theta_\mathrm{W}}\,\frac{\Pi_{\gamma \mathrm{Z}}(0)}{M_\mathrm{Z}^2}
,
\]
 which is
proportional to $G_\upmu\,M_\mathrm{t}^2$ and therefore large, dominated by the
heavy top quark mass effect, or by the large top Yukawa coupling.

The uncertainty $\updelta
\Delta \alpha$ implies uncertainties $\updelta M_\mathrm{W}$, $\updelta \sini$
given by
\begin{align}
\frac{\updelta M_\mathrm{W}}{M_\mathrm{W}} &\sim \ha \frac{\sinW}{\cosW-\sinW}
\;\updelta \dal  \sim  0.23 \;\updelta \dal\,, \\
\frac{\updelta \sinf}{\sinf} &\sim \frac{\cosf}{\cosf-\sinf}
\;\updelta \dal  \sim  1.54 \;\updelta \dal\;.
\end{align}
Also affected are the important relationships between couplings and
masses, such as
\bea
\lambda =3\,\sqrt{2}G_\upmu\,
M_\mathrm{H}^2\,(1+\updelta_\mathrm{H}(\al,\dots))\semis \quad  y_\mathrm{t}^2=2\,\sqrt{2} G_\upmu\,
M_\mathrm{t}^2\,(1+\updelta_\mathrm{t}(\al,\dots)\,,
\eea
which currently offer the only way to determine $\lambda$ and $y_\mathrm{t}$ via
the experimentally accessible masses $M_\mathrm{H}$ and $M_\mathrm{t}$. Direct
measurement of $\lambda$ and $y_\mathrm{t}$  will probably be possible only at
future lepton colliders, such as the FCC-ee.

The \emph{parameter relationships} between very precisely
measurable quantities provide stringent precision tests and, at high
enough precision, would reveal the physics missing within the SM. Currently,
the non-perturbative hadronic contribution
$\dalh^{(5)}(M_\mathrm{Z}^2)$  limits the precision predictions. Concerning
the relevance of quantum corrections and their precision,
one should keep in mind that a 30\,SD disagreement between some SM
prediction and experiment is obtained when subleading SM corrections
are neglected, and only the leading corrections $\Delta
\alpha (M_\mathrm{Z}^2)$ and $\Delta \rho$ in \Eref{derstruct} are
accounted for.

{\footnotesize Calculate, for example,
the W and Z mass from $\alpha(M_\mathrm{Z}),\,G_\upmu$ and $\sin^2
\Theta_{\ell\,{\rm eff}}$:
first $\sin^2 \Theta_\mathrm{W}=1-M_\mathrm{W}^2/M_\mathrm{Z}^2$ is related to $\sin^2 \theta_{\ell,
{\rm eff}}(M_\mathrm{Z})$ via
$$\sin^2 \theta_{\ell, {\rm eff}}(M_\mathrm{Z})=\left( 1+ \frac{\cos^2
\Theta_\mathrm{W}}{\sin^2 \Theta_\mathrm{W}}\,\Delta \rho\right)\,\sin^2 \Theta_\mathrm{W}\,,$$
where the leading top quark mass square correction is 
$$\Delta \rho = \frac{3\, M_\mathrm{t}^2\,\sqrt{2} G_\upmu}{16\,\uppi^2}\;;\qquad
M_\mathrm{t} =173\pm0.4\,\gv \, .$$
The iterative solution with input $\sin^2 \theta_{\ell, {\rm eff}}(M_\mathrm{Z}) = 0.23148$ is
$\sin^2 \Theta_\mathrm{W}= 0.224\,26$ while $1-M_\mathrm{W}^2/M_\mathrm{Z}^2=0.222\,63$ is what one gets using
PDG:
 $$M_\mathrm{W}^{\rm exp}=80.379\pm 0.012\,\gv \;;\qquad  M_\mathrm{Z}^{\rm exp}=91.1876\pm 0.0021\,\gv\,.$$
Predicting, then, the masses, we have
$$M_\mathrm{W}=\frac{A_0}{\sin^2\Theta_\mathrm{W}}\;;\qquad A_0=\sqrt{\frac{\uppi
\alpha}{\sqrt{2} G_\upmu}}\;;\qquad 
M_\mathrm{Z}=\frac{M_\mathrm{W}}{\cos \Theta_\mathrm{W}}$$ where, including photon VP correction
$\alpha^{-1}(M_\mathrm{Z})=128.953\pm 0.016$. For the W and Z masses, we then get
$$M_\mathrm{W}^{\rm the}= 81.1636 \pm 0.0346\,\gv \;;\qquad M_\mathrm{Z}^{\rm the}=92.1484 \pm 0.0264\,\gv\,.$$
This gives the following SD values:
$$ \mathrm{W}:~~ 23 \,\sigma  \;;\qquad  \mathrm{Z}:~~36\, \sigma $$
Uncertainties from $\sin^2 \theta$, $\alpha(M_\mathrm{Z})$,  and  $M_\mathrm{t}$, as well as  experimental uncertainties,
are added in quadrature. The result is, of course, scheme-dependent, but illustrates well the
sensitivity to taking into account the proper radiative
corrections. Actually, including full one-loop and leading two-loop
corrections reduces the disagreement below the $2\sigma$ level.}

\subsection{The ultimate motivation for high-precision SM parameters}
\begin{figure}
\includegraphics[width=0.52\textwidth]{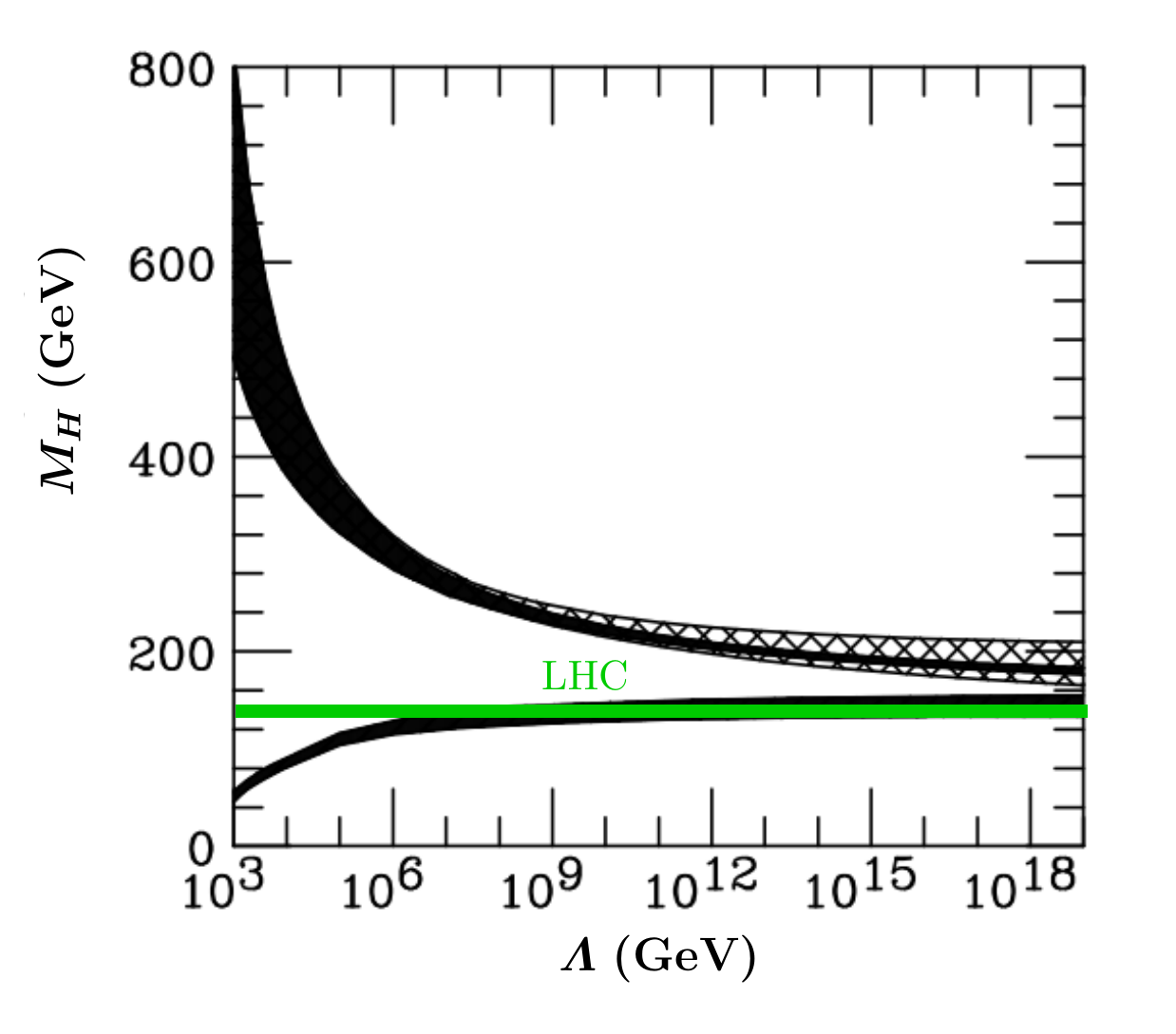}
\includegraphics[width=0.46\textwidth]{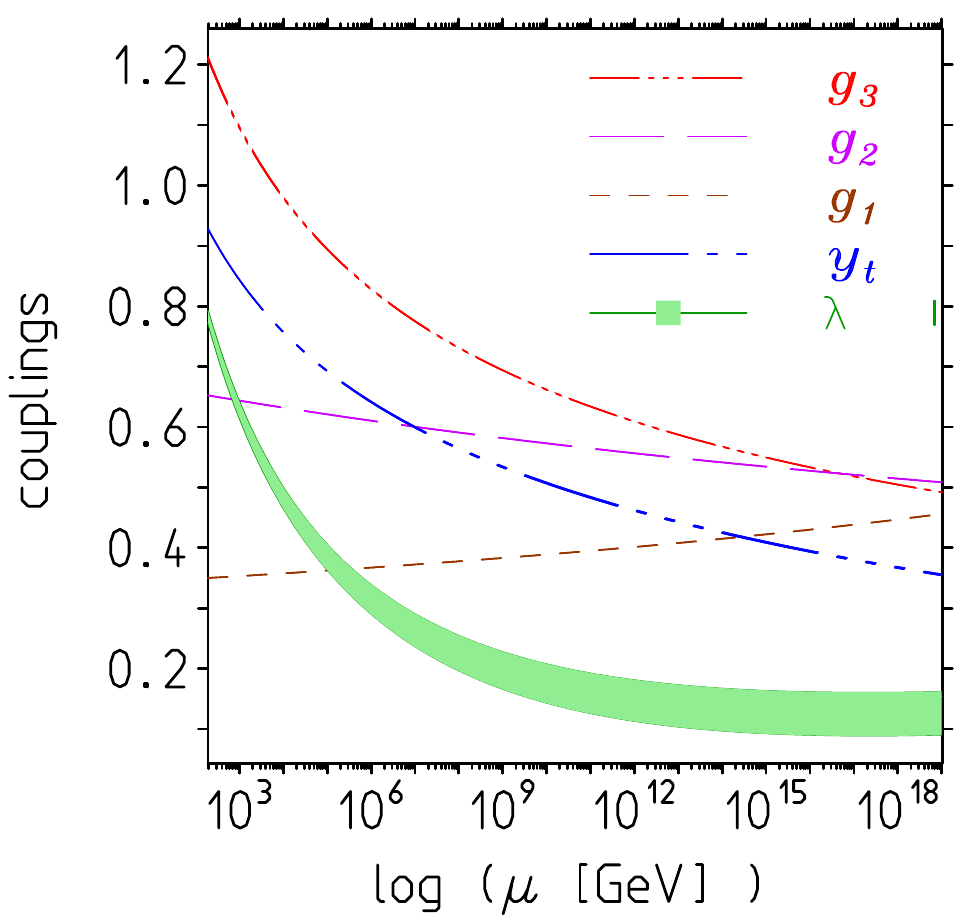}
\caption{Left: Plot by Riesselmann and Hambye in 1996, the first two-loop
analysis after knowing \mbo{M_\mathrm{t}} from CDF~\cite{Hambye:1996wb}. Right: the SM dimensionless
couplings in the \MSb scheme as a function of the renormalization
scale for $M_\mathrm{H} = 124$--126\,$\gv$, which were obtained in Refs.
\cite{Jegerlehner:2012kn,Jegerlehner:2013cta,Jegerlehner:2014mua,Jegerlehner:2014xxa}.}
\label{fig:runparI}
%\query{Please correct the figure labels in \Fref{fig:runparI}. Set all variables
%in italic font. (Do not set particle names in italic font.)}
\end{figure}

\begin{figure}
\centering
\includegraphics[width=0.46\textwidth]{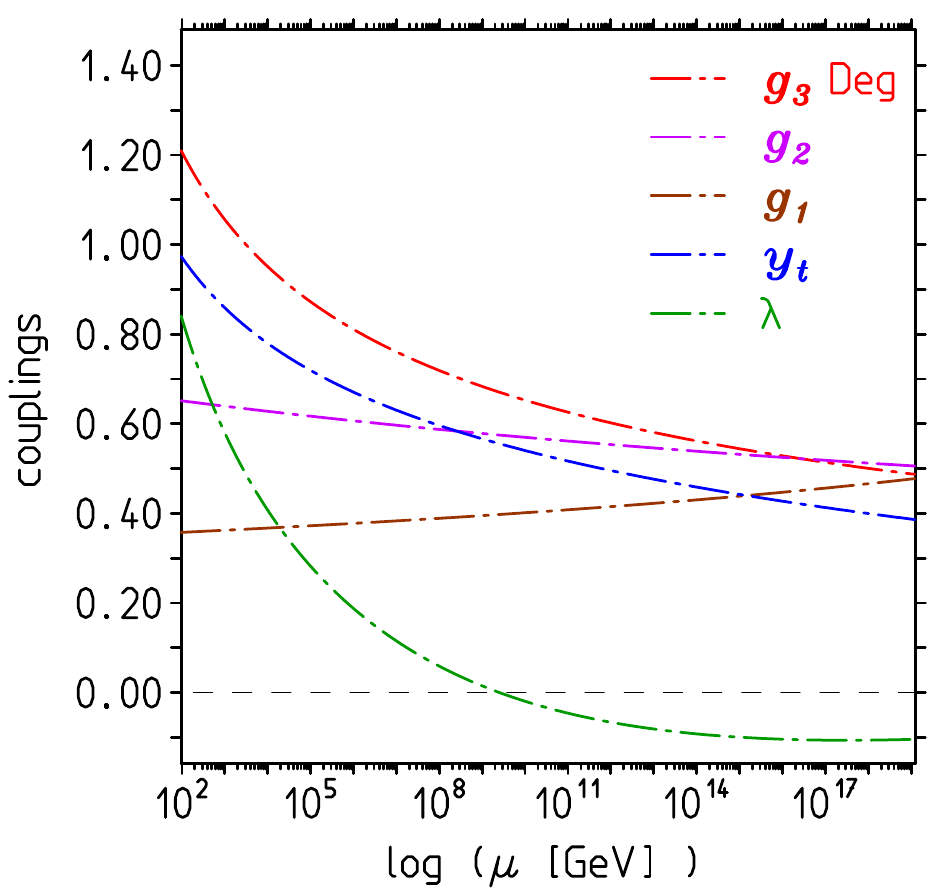}
\includegraphics[width=0.46\textwidth]{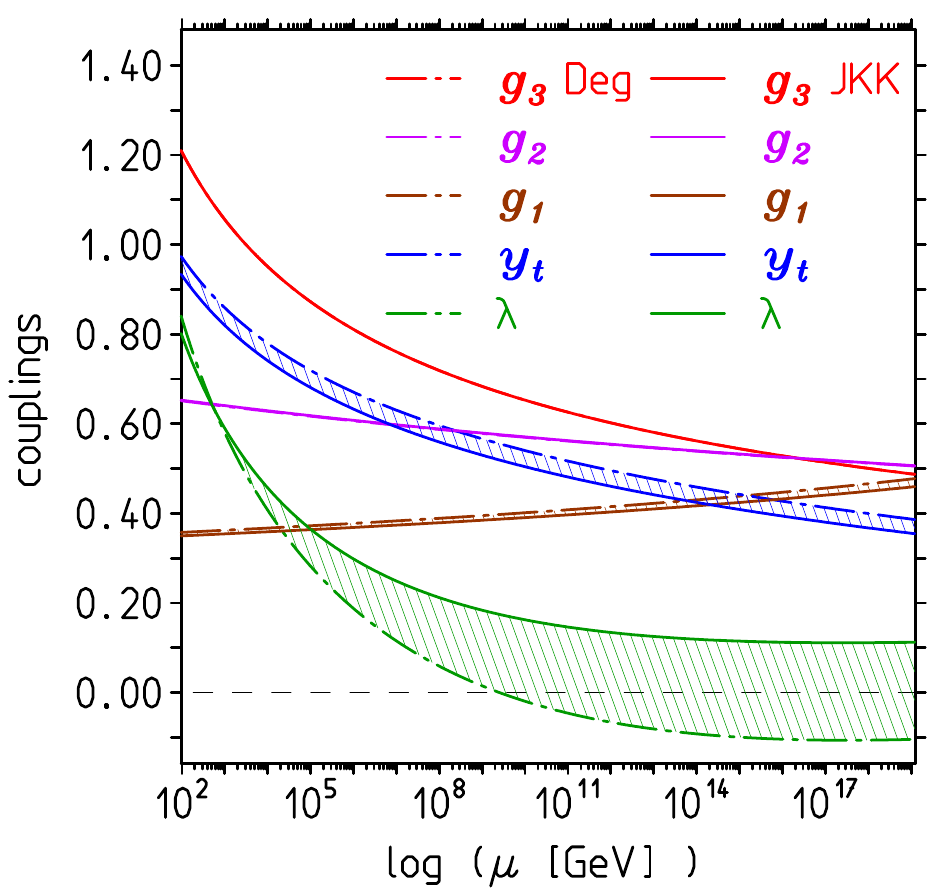}
\caption{Left: Shaposhnikov \textit{et al.} and Degrassi \textit{et al.}  matching~\cite{Bezrukov:2012sa,Degrassi:2012ry}.
Right: The shaded bands show the difference in the SM parameter
extrapolation using the central values of the \MSb parameters obtained
from differences in the matching procedures.}
\label{fig:runparII}
%\query{Please correct the figure labels in \Fref{fig:runparII}. Set all variables
%in italic font. (Do not set particle names in italic font.)}
\end{figure}

After the ATLAS and CMS Higgs discovery at the LHC, the Higgs vacuum
stability issue is one of the most interesting to be clarified at
future $\epm$ facilities. Much more surprising than the discovery of
its true existence is the fact that the Higgs boson turned out to
exhibit a mass very close to what has been expected from vacuum stability
extending up to the Planck scale $\lpl$ (see \Fref{fig:runparI}). There
appears to be a very tricky conspiracy with other couplings to achieve
this `purpose'. Related is the question of whether the SM allows us
to extrapolate  up to the Planck scale.  Thus, the central issue for the
future is the very delicate `acting together' between SM couplings,
which makes the precision determination of SM parameters more important
than ever. Therefore, higher-precision SM parameters
$g',g,g_\mathrm{s},y_\mathrm{t}$, and $\lambda$ are mandatory for progress in this direction. Actually, the vacuum stability is controversial at present at the
 $1.5\sigma$ level between a metastable and a stable EW vacuum,
which depends on whether $\lambda$ stays positive up to $\lpl$ or not. This
is illustrated in \Fref{fig:runparII}. If the SM extrapolates stable to $\lpl$, obviously the resulting
effective parameters affect early cosmology, Higgs inflation, Higgs
reheating, \etc \cite{Jegerlehner:2013cta}. The sharp dependence of
the Higgs vacuum stability on the SM input parameters, as well as  on possible
SM extensions and the vastly different scenarios that can result as a consequence of minor shifts in parameter space, makes the stable vacuum
case a particularly interesting one and it could reveal the Higgs particle as `the master of the Universe'. After all, it is commonly
accepted that dark energy provided by some scalar field is the
`stuff' shaping the Universe both at very early (inflation) as well
as at  late times (accelerated expansion).

It is highly conceivable that perturbation expansion works up to
the Planck scale without a Landau pole or other singularities and
that the Higgs potential remains (meta)stable! The discovery of the Higgs boson
 has supplied us, for the first time, with the complete set of SM
parameters and, for the peculiar SM configuration, revealed that all SM
couplings, with the exception of the hypercharge $g_1$, are decreasing
with energy. Very surprisingly, this implies that perturbative SM
predictions improve at higher energies. More specifically,
the pattern now looks as follows: the gauge coupling related to \mbo{U(1)_\mathrm{Y}} is
screening (IR-free), the couplings associated with
\mbo{SU(2)_\mathrm{L}} and \mbo{SU(3)_\mathrm{c}} are antiscreening (UV-free). Thus $g_1$, $g_2$, and $g_3$
behave as expected (standard wisdom). By contrast, the top Yukawa
coupling {\mbo{y_\mathrm{t}}} and Higgs self-coupling {\mbo{\lambda}}, while
screening if stand-alone (IR-free, like QED), as part of the SM 
are transmuted from IR-free to UV-free. The SM reveals an amazing
parameter conspiracy, which reminds us of phenomena often observed in
condensed matter systems:
\textit{``There is a sudden rapid passage to a totally new and more
comprehensive type of order or organisation, with quite new emergent
properties''}~\cite{HuxleyHuxley1947}, \ie there must be reasons
that couplings are as they are. This manifests itself in the QCD
dominance within the renormalization group (RG) of the top Yukawa
coupling, which requires $g_3> {3} \,y_\mathrm{t} / 4$, and in the top Yukawa
dominance within the RG of the  Higgs boson coupling, which requires $
\lambda<  {3\,(\sqrt{5}-1)} \,y_\mathrm{t}^2 / 2$ in the 
gaugeless (\mbo{g_1,g_2=0}) limit.  Under focus is the Higgs
self-coupling. Does it stay positive
\mbo{\lambda > 0} up to \mbo{\lpl}? A zero-valued \mbo{\lambda} would be
an essential singularity. The key problem concerns
the precise size of the top Yukawa coupling \mbo{y_\mathrm{t}}, which decides
 the stability of our world! The metastability vs. stability controversy
will be decided by obtaining more precise input parameters and by better-established EW matching conditions. Most important in this context is
the direct measurement of \mbo{y_\mathrm{t}} and \mbo{\lambda} at future
$\epm$ colliders, but also the important role that the running gauge
couplings are playing requires substantial progress in obtaining more
precise hadronic cross-sections in order to reduce hadronic uncertainties in
\mbo{\alpha(M_\mathrm{Z})} and \mbo{\alpha_2(M_\mathrm{Z})}. This is a big challenge for low-energy hadron facilities. Complementary, progress in lattice QCD simulations of
two-point correlators will be important to pin down hadronic effects
from first principles. Such improvement in SM precision physics could
open a new gateway to precision cosmology of the early Universe!

\subsection{$R$ data evaluation of $\alpha(M^2_\mathrm{Z})$
\label{sec:R-data}}
\begin{figure}
\centering
\includegraphics[height=2.0cm]{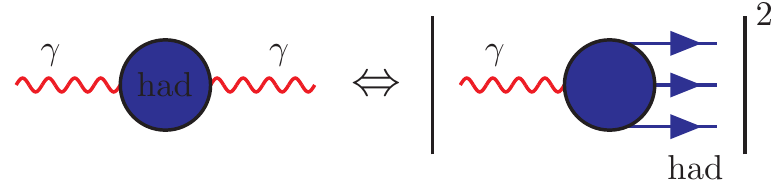}
\caption{The master equation (\ref{DRalpZ}), relating
$\Pi^{'\:\mathrm{had}}_{\gamma}(q^2)$
and $\sigma_{\mathrm{tot}}^{\mathrm{had}}(q^2)$, is based on analyticity and the optical theorem.}
\label{fig:Fred}
%\query{All figures are figures and must be treated as such. Please check
%that the caption for \Fref{fig:Fred} is appropriate, and replace if necessary.}
\end{figure}

What we need is a precise calculation of the hadronic photon vacuum
polarisation function. The non-perturbative hadronic piece from the five light quarks $
\dahs=- (\Pi'_\upgamma(s)-\Pi'_\upgamma(0) )_{\rm had}^{(5)}$ can be evaluated in
terms of $\sigma(\mathrm{e}^+\mathrm{e}^- \to {\rm hadrons})$ data via the dispersion
integral
\bea
\dahs = - \frac{\alpha\, s}{3\uppi}\;\bigg(\;\;
 \pvint\limits_{m_{\uppi_0}^2}^{E^2_{\rm cut}}\! \D s'\;
\frac{{ R^{\mathrm{data}}_\upgamma(s')}}{s'(s'-s)}
\;+\;\;
\pvint\limits_{E^2_{\rm cut}}^\infty\! \D s'\;
\frac{{ R^{\mathrm{pQCD}}_\upgamma(s')}}{s'(s'-s)}\,\,
\bigg)\,,
\label{DRalpZ}
\eea
where $R_\upgamma(s) \equiv \sigma^{(0)}(\mathrm{e}^+\mathrm{e}^- \rightarrow \gamma^*
\rightarrow {\rm hadrons})/ ( {4\uppi \alpha^2} / {3s} )$
measures the hadronic cross-section in units of the tree-level $\epm \to \mumu$ cross-section
sufficiently above the muon pair production threshold ($s\gg
4m_\upmu^2$).  The master equation (\Eref{DRalpZ}) is based on
analyticity and the optical theorem, as shown in \Fref{fig:Fred}.

\begin{figure}
\centering
\includegraphics[width=0.85\textwidth]{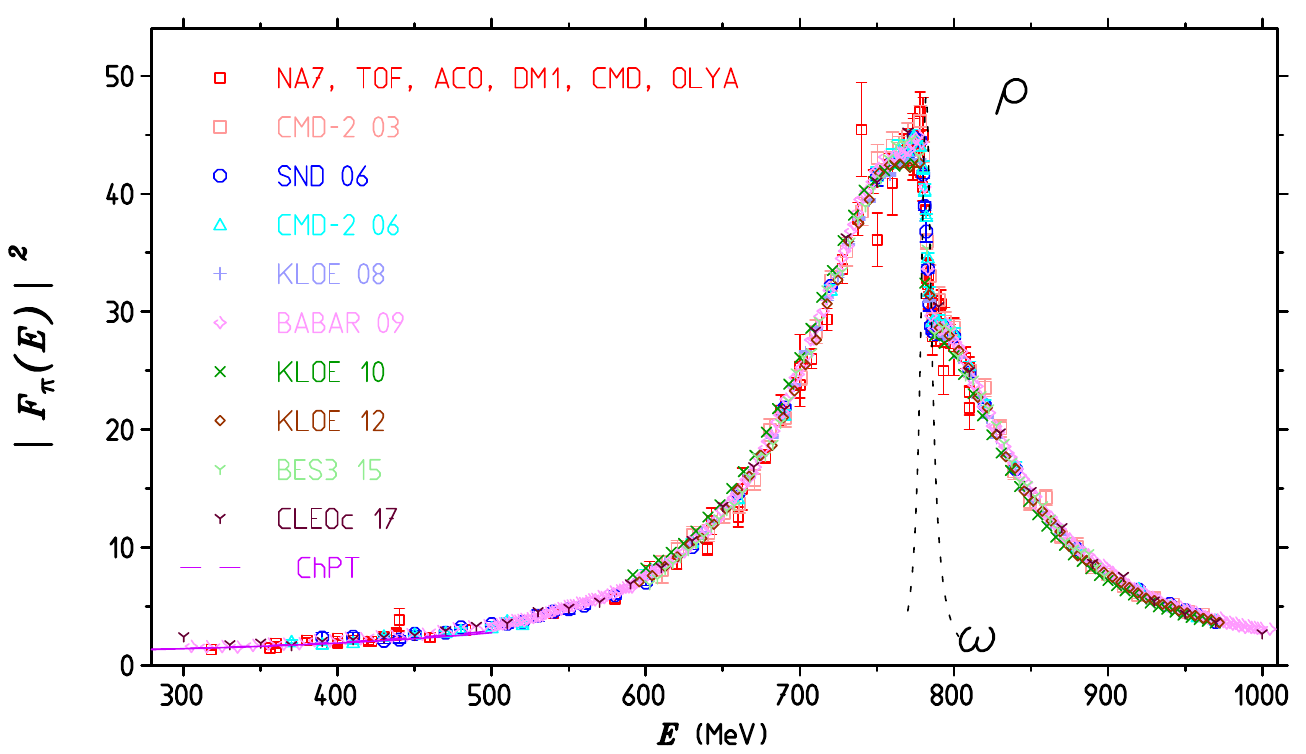}
\caption{The low-energy tail of $R$ is provided by $\uppi^+\uppi^-$
production data. Shown is a compilation of the modulus square of the
pion form factor in the $\rho$ meson region. The corresponding $R(s)$
is given by
$R(s)=\frac14\,\beta_\uppi^3\,|F_\uppi^{(0)}(s)|^2\,,\,\,\beta_\uppi=(1-4m^2_\uppi/s)^{1/2}$
is the pion velocity ($s=E^2$). Data from CMD-2, SND, KLOE, BaBar,
BESIII, and
CLEOc~\cite{Akhmetshin:2003zn,Aulchenko:2006na,Akhmetshin:2006wh,Akhmetshin:2006bx,
Achasov:2006vp,Aloisio:2004bu,Ambrosino:2008aa,Ambrosino:2010bv,Babusci:2012rp,Anastasi:2017eio,Venanzoni:2017ggn,
Aubert:2009ad,Lees:2012cj,Ablikim:2015orh,Xiao:2017dqv}
besides some older sets.}
\label{fig:Rpipi}
%\query{Please correct the figure labels in \Fref{fig:Rpipi}. Set all variables
%in italic font. (Do not set particle names in italic font.) Please check
%that all the colours are easy to read.}
\end{figure}

\begin{figure}
\centering
\includegraphics[]{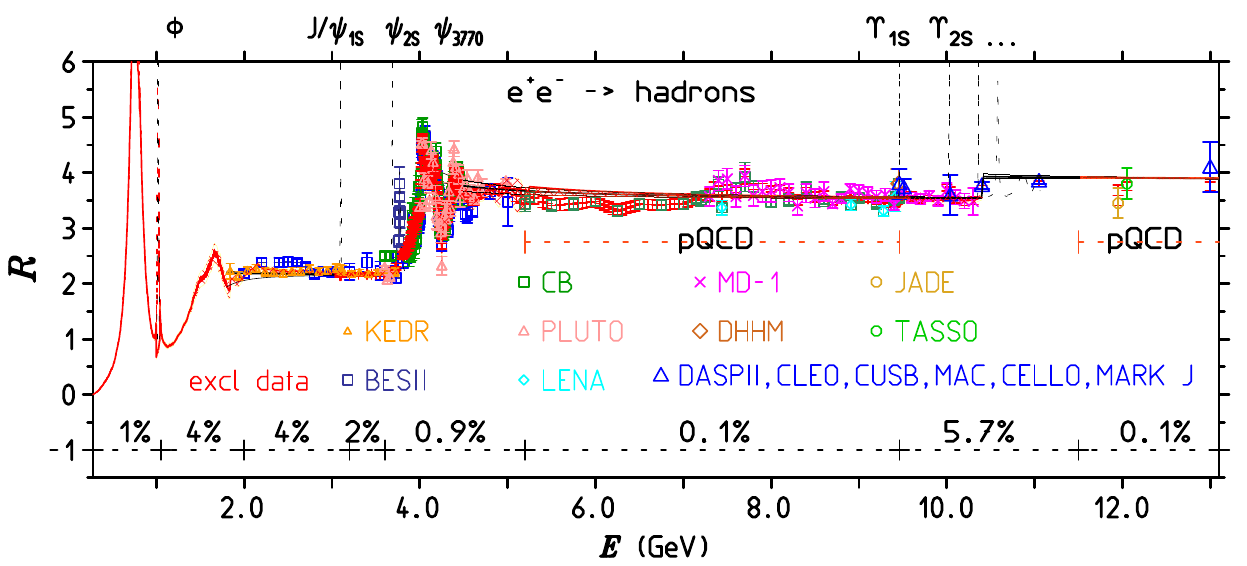}
\caption{The compilation of $R(s)$ data utilised in the evaluation of $\dalh$. The
bottom line shows the relative systematic uncertainties within the split
regions. Different regions are assumed to have uncorrelated
systematics. Data
from Refs.~\cite{Aubert:2004kj,Aubert:2005eg,Aubert:2005cb,Aubert:2006jq,Aubert:2007ef,Aubert:2007ym,
Lees:2012cr,Lees:2011zi,Lees:2013ebn,Lees:2013gzt,Lees:2014xsh,Lees:2018dnv,Davier:2015bka,Davier:2016udg,
Akhmetshin:2013xc,Akhmetshin:2015ifg,Kozyrev:2016raz,Achasov:2014ncd,Aulchenko:2014vkn,
Achasov:2016bfr,Achasov:2016eyg,Achasov:2016qvd,Achasov:2016lbc,Achasov:2016zvn,
Bai:1999pk,Bai:2001ct,Ablikim:2009ad,Anashin:2015woa,Anashin:2016hmv}
and others. We apply pQCD from 5.2\,GeV
to 9.46\,GeV and above 11.5\,GeV using the
code of Ref.~\cite{Harlander:2002ur}.}
\label{fig:Rdata} 
%\query{Please correct the figure labels in \Fref{fig:Rdata}. Set all variables
%in italic font. (Do not set particle names in italic font.)  Please check
%that all the colours are easy to read.}
\end{figure}

\begin{figure}
\centering
\includegraphics[width=0.98\textwidth]{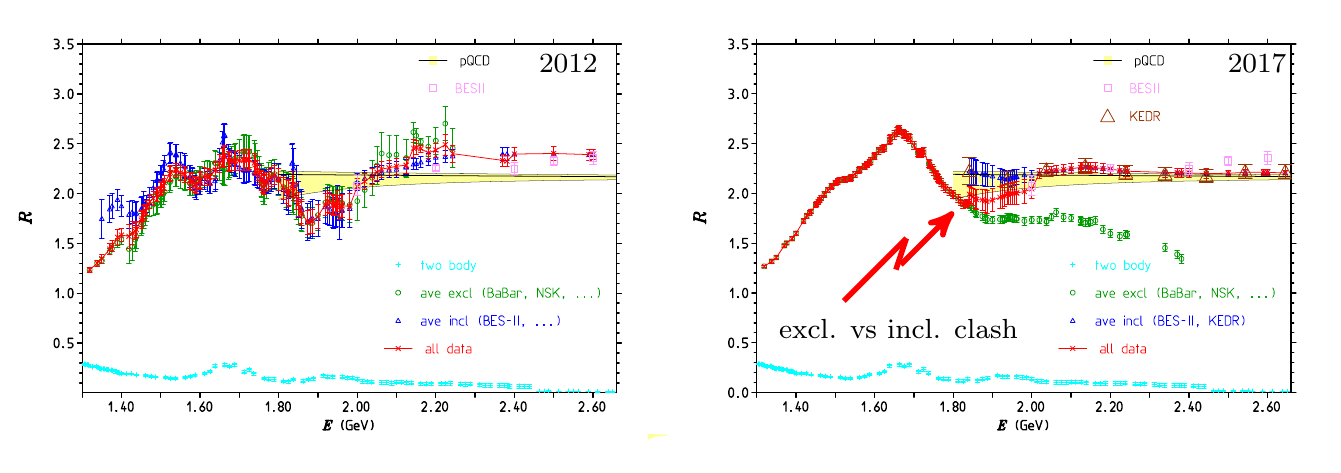}
\caption{Illustrating progress by BaBar and NSK exclusive channel data
vs. new inclusive data by KEDR. Why is the point at 1.84\,GeV so high?}
\label{fig:KEDR}
%\query{Please correct the figure labels in \Fref{fig:KEDR}. Set all variables
%in italic font. (Do not set particle names in italic font.)  Please check
%that all the colours are easy to read. Please also
%check that this figure is cited correctly in the text.}
\end{figure}

A compilation of the available $R$ data is shown in
\Fref{fig:Rpipi} for the low-energy $\uppi\uppi$ channel and in
\Fref{fig:Rdata} for $R(s)$ above the $\rho$ resonance peak. Since the mid
1990s~\cite{Eidelman:1995ny}, enormous progress has been achieved, also
because the new initial-state radiation (ISR) radiative return
approach\footnote{This was pioneered by the KLOE Collaboration, followed by
BaBar and BESIII experiments.} provided good statistics data from $\upphi$ and B meson
factories
(see Refs.~\cite{Akhmetshin:2003zn,Aulchenko:2006na,Akhmetshin:2006wh,Akhmetshin:2006bx,
Achasov:2006vp,Aloisio:2004bu,Ambrosino:2008aa,Ambrosino:2010bv,Babusci:2012rp,Anastasi:2017eio,Venanzoni:2017ggn,
Aubert:2009ad,Lees:2012cj,Ablikim:2015orh,Xiao:2017dqv,
Aubert:2004kj,Aubert:2005eg,Aubert:2005cb,Aubert:2006jq,Aubert:2007ef,Aubert:2007ym,
Lees:2012cr,Lees:2011zi,Lees:2013ebn,Lees:2013gzt,Lees:2014xsh,Lees:2018dnv,Davier:2015bka,Davier:2016udg,
Akhmetshin:2013xc,Akhmetshin:2015ifg,Kozyrev:2016raz,Achasov:2014ncd,Aulchenko:2014vkn,
Achasov:2016bfr,Achasov:2016eyg,Achasov:2016qvd,Achasov:2016lbc,Achasov:2016zvn,
Bai:1999pk,Bai:2001ct,Ablikim:2009ad,Anashin:2015woa,Anashin:2016hmv}).
Still, an issue in hadronic vacuum polarisation (HVP) is the region 1.2--2\,GeV, where we have a test ground for exclusive (more than 30
channels) versus inclusive
\mbo{R}  measurements, where data taking or data analysis is
ongoing with CMD-3 and SND detectors (scan) and BaBar and BESIII
detector data (radiative return). The region still contributes about
50\% to the uncertainty of the hadronic contribution to the muon
$g-2$, as we may learn from Fig.~\ref{fig:distamuvsalpZ}, in the next section. Above
2\,GeV, fairly accurate BES II data~\cite{Bai:1999pk,Bai:2001ct,Ablikim:2009ad} are
available. Recently, a new inclusive determination of $R_\upgamma(s)$ in
the range 1.84--3.72~GeV has been obtained with the KEDR detector at
Novosibirsk~\cite{Anashin:2015woa,Anashin:2016hmv} (see Fig.~\ref{fig:KEDR}).
At present, the results from the direct and the Adler function improved
approach, to be discussed in \Sref{sec:Adler}, reads
\begin{align}
\Delta \al _{\rm hadrons}^{(5)}(\mz) &  =   0.0277\,56 \pm 0.000\,157 & \notag
\\
& \ \quad  { 0.027563 \pm 0.000120} &{\rm  Adler  } \notag \\
\alpha^{-1}(\mz) &  =   128.916 \pm 0.022 & \notag \\
&  \ \quad { 128.953 \pm 0.016 }  &{\rm Adler }
\label{dhadr5alpha}
\end{align}

\begin{figure}
\centering
\includegraphics{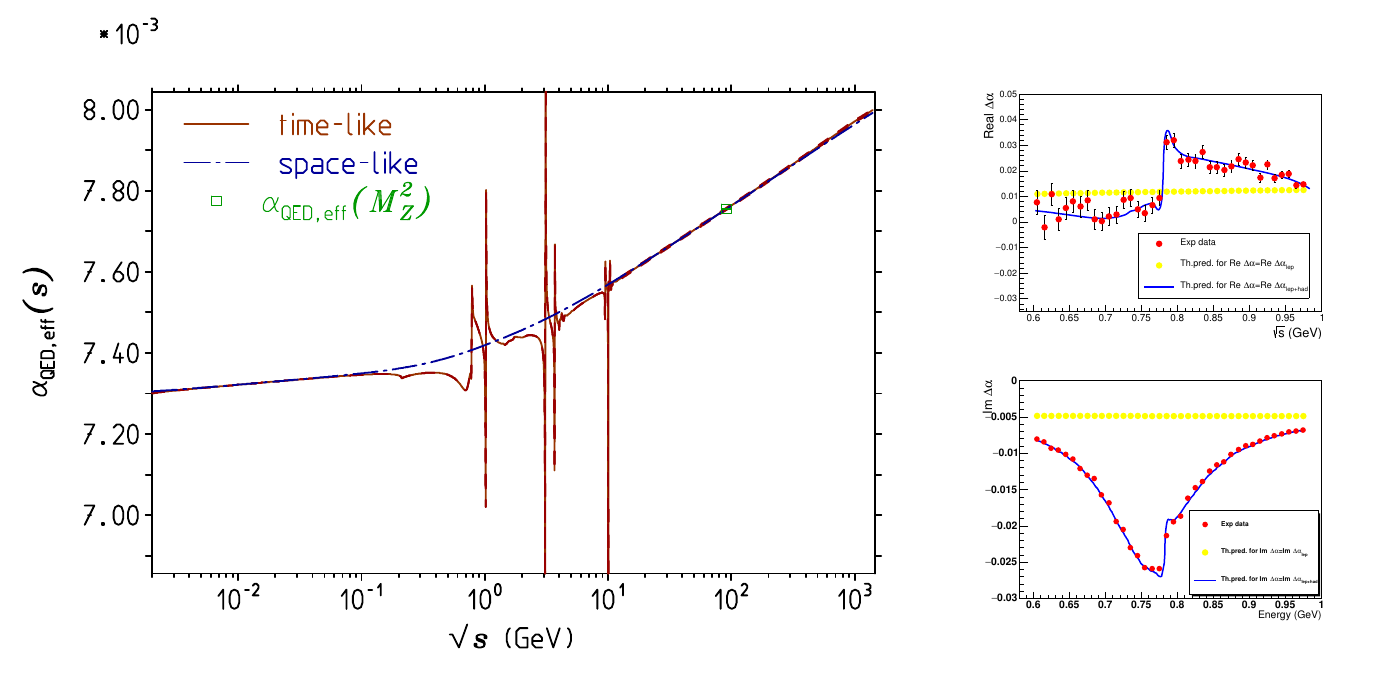}
\caption{Left: The effective \mbo{\alpha(s)} at time-like vs.
space-like momentum transfer, showing quark--hadron duality at work. In
the time-like region, the effective charge  varies dramatically near
resonances but agrees quite well on average with the space-like
version. Locally, it is ill-defined near OZI suppressed meson decays
\mbo{\mathrm{J}/\uppsi,\uppsi_1,\Upsilon_{1,2,3}} where Dyson series of self-energy
insertions do not converge (see Section~5 of Ref.~\cite{Jegerlehner:2015stw}). 
Right: A first experimental determination of the effective charge in
the $\rho$ resonance region by KLOE-2\cite{KLOE-2:2016mgi}, which demonstrates
the pronounced variation of the vacuum polarisation (charge screening)
across a resonance.}
\label{fig:alphatands}
%\query{Please correct the figure labels in \Fref{fig:alphatands}. Set all variables
%in italic font. (Do not set particle names in italic font.)  Please check
%that all the colours are easy to read.}
\end{figure}

In \Fref{fig:alphatands}, we show the effective fine structure
constant as a function of the c.m. energy $E=\sqrt{s}$, for the
time-like and  space-like regions. The question now is: what are the possible
improvements? 
\begin{enumerate}
\item Evidently, 
%\begin{itemize}
%    \item[(i)]
%    \end{itemize}
%$\bullet$~
a direct improvement of the
dispersion integral involves reducing the uncertainty of $R(s)$ to 1\% up to
above the $\Upsilon$ resonances; probably, nobody will do that. One may  rely on pQCD above 1.8\,GeV and refer to quark--hadron duality, as
in Ref.~\cite{Davier:2017zfy}. Then experimental input above 1.8\,GeV is not
required. But then we are left with questions about where precisely to
assume thresholds and what are the mass effects near
thresholds. Commonly, pQCD is applied, taking into account
uncertainties in $\alpha_\mathrm{s}$ only.  This certainly does not provide a
result that can be fully trusted, although the $R$ data integral in
this range is much less precise at present. The problem is that, in
this theory-driven approach, 70\% of $\dahz$ comes from pQCD. Thereby,
one has to assume that, in the time-like region above 1.8\,GeV, pQCD,  on
average, is  as precise as the usually adopted \MSb parametrization
suggests. Locally, pQCD does not work near thresholds and resonances, obviously.

\item The more promising approach discussed in the following relies on the
%$\bullet$~ 
%\begin{itemize}
%    \item[(ii)]
%    \end{itemize}
 Euclidean split method (Adler function controlled pQCD),
which only requires improved $R$ measurements in the exclusive region
from 1 to 2\,GeV. Here, NSK, BESIII, and Belle II can top what BaBar has
achieved. However, in this rearrangement,  a
substantially more precise calculation of the pQCD
Adler function is as important. Required is an essentially exact massive four-loop
result, which is equivalent to sufficiently high-order low- and
high-energy expansions, of which a few terms are available already
\cite{Maier:2017ypu}.
\end{enumerate}

Because of the high
sensitivity to the precise charm and bottom quark values, one also needs better
parameters $m_\mathrm{c}$ and $m_\mathrm{b}$ besides $\alpha_\mathrm{s}$. Here one can profit from activities
going on anyway and the FCC-ee and ILC projects pose further strong
motivation to attempt to reach higher precision for QCD parameters.

\subsubsection{$\dalh (M_\mathrm{Z}^2)$ results from ranges}
Table~\ref{tab:alpZranges} shows the contributions and uncertainties to
\mbo{\Delta \alpha^{(5)}_{\rm had}(M_\mathrm{Z})} for $M_\mathrm{Z}=91.1876\,\gv$ 
in units $10^{-4}$ from different regions.  Typically, depending on
cuts applied, the direct evaluation of the dispersion integral of $R$
yields 43\% from data and 57\% from perturbative QCD. Here, pQCD is
used between 5.2\,GeV and 9.5\,GeV and above 11.5\,GeV. Systematic uncertainties
are taken to be correlated within the different ranges, but taken as
independent between the different ranges.

\begin{table}
\centering
\caption{\mbo{\Delta \alpha^{(5)}_{\rm had}(M_\mathrm{Z})} in terms of \mbo{\epm}
data and pQCD.
The last two columns list the relative accuracy and the percentage contribution
of the total. The systematic uncertainties (syst) are assumed to be
independent among the different energy ranges listed in the table.}
\label{tab:alpZranges}
\begin{tabular}{llllll}
\hline \hline
Final state &  Range & $\Delta \alpha^{(5)}_{\rm had}\power{4}$ (stat) (syst) [tot] & Rel & Abs \\
&  (GeV)& & (\%)  & (\%) \\
\hline
   $\rho  $     & (0.28, 1.05) &    \phantom{1}34.14 (0.03) (0.28) [0.28]&  0.8 &  \phantom{10}3.1 \\
   $\omega $    & (0.42, 0.81) &      \phantom{13}3.10 (0.03) (0.06) [0.07]&  2.1 &  \phantom{10}0.2 \\
   $\phi   $    & (1.00, 1.04) &      \phantom{13}4.76 (0.04) (0.05) [0.06]&  1.4 &  \phantom{10}0.2 \\
   $\mathrm{J}/\uppsi $    &             &     \phantom{1}12.38 (0.60) (0.67) [0.90]&  7.2 & \phantom{1}31.9 \\
   $\Upsilon$   &             &      \phantom{13}1.30 (0.05) (0.07) [0.09]&  6.9 &  \phantom{10}0.3 \\
     Had        & (1.05, 2.00) &     \phantom{1}16.91 (0.04) (0.82) [0.82]&  4.9 & \phantom{1}26.7 \\
     Had        & (2.00, 3.20) &     \phantom{1}15.34 (0.08) (0.61) [0.62]&  4.0 & \phantom{1}15.1 \\
     Had        & (3.20, 3.60) &      \phantom{13}4.98 (0.03) (0.09) [0.10]&  1.9 &  \phantom{10}0.4 \\
     Had        & (3.60, 5.20) &     \phantom{1}16.84 (0.12) (0.21) [0.25]&  0.0 &  \phantom{10}2.4 \\
    pQCD        & (5.20, 9.46) &     \phantom{1}33.84 (0.12) (0.25) [0.03]&  0.1 &  \phantom{10}0.0 \\
     Had        & (9.46,11.50) &     \phantom{1}11.12 (0.07) (0.69) [0.69]&  6.2 & \phantom{1}19.1 \\
    pQCD        & (11.50, 0.00) &    123.29 (0.00) (0.05) [0.05]&  0.0 &  \phantom{10}0.1 \\
    Data        & (0.3,$\infty$)&    120.85 (0.63) (1.46) [1.58]&  1.0 &  \phantom{10}0.0 \\
    Total       &             &    277.99 (0.63) (1.46) [1.59]&  0.6 &100.0 \\
\hline \hline
\end{tabular}
\end{table}

\begin{figure}
\centering
\includegraphics[width=0.5\textwidth]{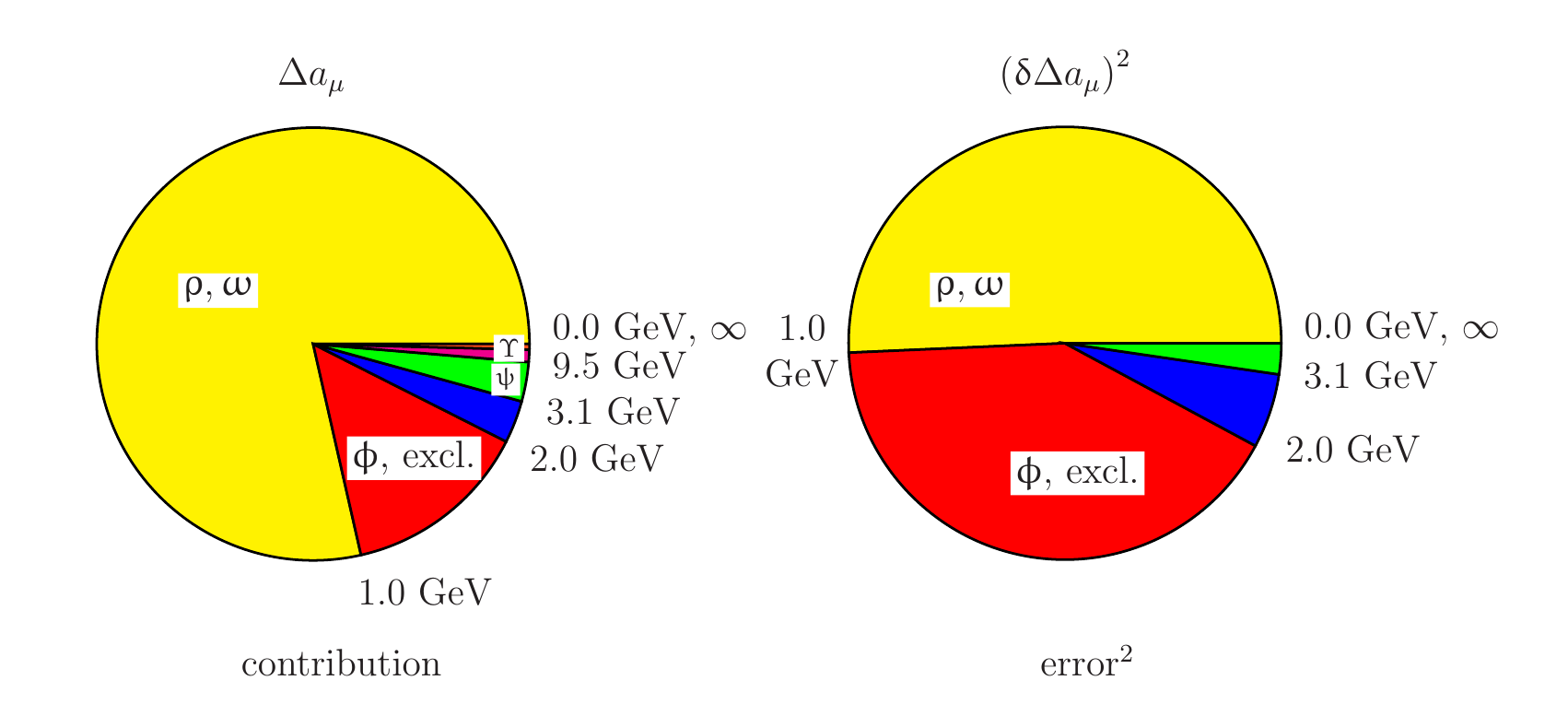}
\includegraphics[width=0.5\textwidth]{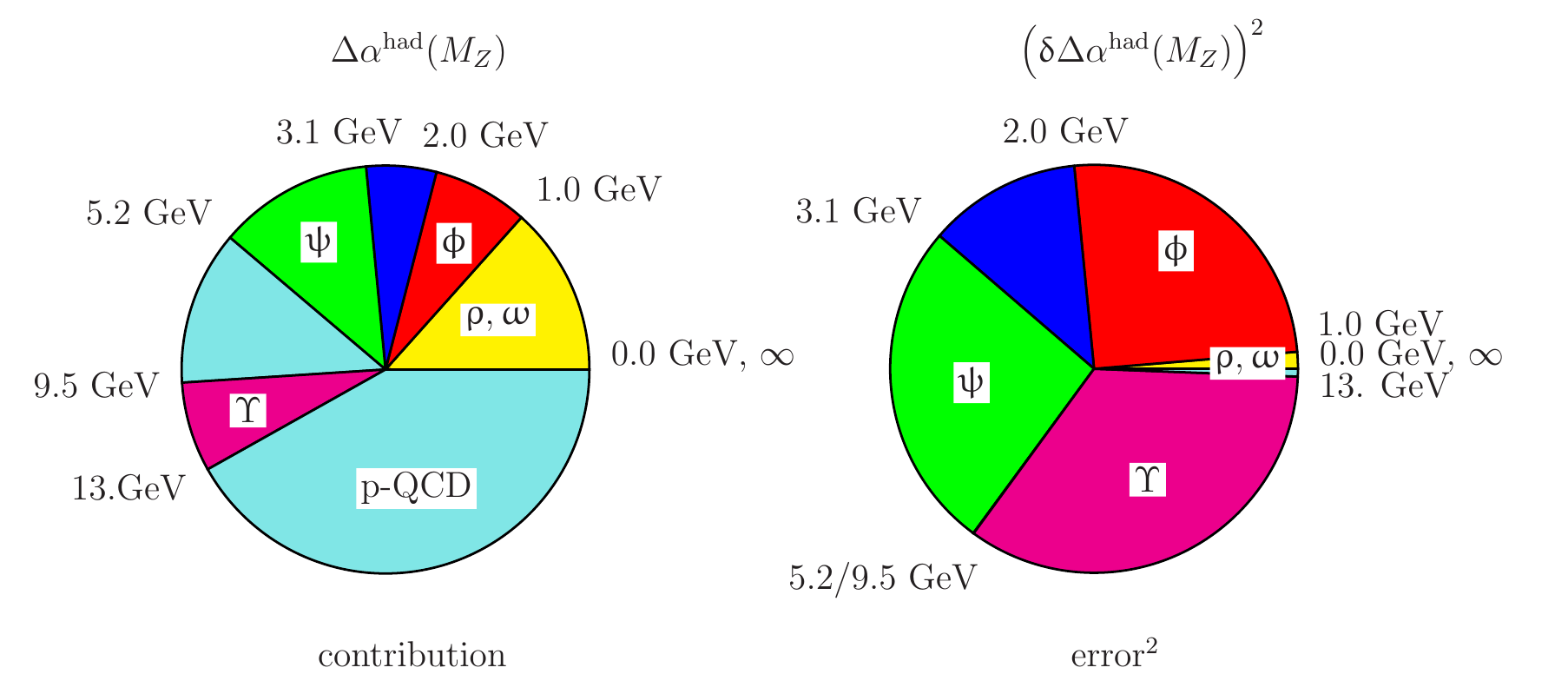}
\caption{A comparison of the weights and square uncertainties between
$a_\upmu^{\rm had}$ and $\dahz$ of
contributions from different regions. It reveals the importance of the
different energy regions. In contrast to the low-energy
dominated $a_\upmu^{\rm had}$, $\dahz$ is
sensitive to data from much higher energies.}
\label{fig:distamuvsalpZ}
%\query{Please correct the figure labels in \Fref{fig:distamuvsalpZ}. Set
%particle names and the $\updelta$ in roman (upright) font.}
\end{figure}

In \Fref{fig:distamuvsalpZ}, we illustrate the relevance of
different energy ranges by comparing the hadronic contribution to the
muon $g-2$ with that to the hadronic shift of the effective charge
at $M_\mathrm{Z}$. The point is that the new muon $g-2$ experiments strongly motivate
efforts the measure $R(s)$ in the low-energy region more
precisely. From \Fref{fig:distamuvsalpZ}, we learn that low-energy
data alone are not able to substantially improve a direct evaluation
of the dispersion integral (\Eref{DRalpZ}). Therefore,  to
achieve the required factor of five improvement, alternative methods
to determine $\Delta \alpha^{(5)}_{\rm had}(s)$ at high energies
must be developed.

\subsection{Reducing uncertainties via the Euclidean split trick:
Adler function controlled pQCD}
\label{sec:Adler}
As we learn from \Fref{fig:Rdata}, it is difficult, if not
impossible, to tell at what precision pQCD can replace data. This
especially concerns resonance and threshold effects and to what extent
quark--hadron duality can be made precise. This is much simpler to
accommodate by comparison in the Euclidean (space-like) region, as
 suggested by Adler~\cite{Adler:1974gd} a long time ago and
 successfully tested ~\cite{Eidelman:1998vc}. As the data pool has
been improving greatly, the `experimental' Adler function is
now known with remarkable precision. Actually, on the experimental side,
new more precise measurements of
\mbo{R(s)} are being made, primarily in the low-energy range. On the
theory side, pQCD calculations for Euclidean two-point current
correlators are expected to be pushed further. Advances are also
expected from lattice QCD, which  can also produce data for the Adler
function. As suggested in Refs.~\cite{Jegerlehner:1999hg,Jegerlehner:2003ip,Jegerlehner:2008rs}, in the Euclidean region, a
split into a non-perturbative and a pQCD part is self-evident. One may
write
\bea
\alpha \left (\mz \right )=\alpha^{\mathrm{data}} \left (-M_0^2 \right )+\left[\alpha
\left (-\mz \right )-\alpha \left (-M_0^2 \right )\right]^{\mathrm{pQCD}}+\left[\alpha
\left (\mz \right )-\alpha \left (-\mz \right )\right]^{\mathrm{pQCD}}\,,
\label{EsplitalphaZ}
\eea
where the space-like offset \mbo{M_0} is chosen such that pQCD is
well under control for \mbo{-s<-M_0^2}. The non-perturbative offset $
\alpha^{\mathrm{data}}(-M_0^2)$ may be obtained by integrating \mbo{R(s)}
data, by choosing $s = -M_0^2$ in \Eref{DRalpZ}.

The crucial point is that the contribution from different energy
ranges to $\alpha^{\mathrm{data}}(-M_0^2)$ is very different from those
to $ \alpha^{\mathrm{data}}(\mz)$. Table~\ref{tab:alpZranges} now
is replaced by Table~\ref{tab:alp0ranges}, where
$\alpha^{\mathrm{data}}(-M_0^2)$ is listed for $M_0=2\,\gv$ in units $10^{-4}$.
Here 94\% results using data and only 6\% pQCD, applied again
between 5.2\,GeV and 9.5\,GeV and above 11.5\,GeV. Of $\Delta \alpha^{(5)}_{\rm had}(M_\mathrm{Z}^2)$ 22\% data, 78\% pQCD! The
split point, $M_0$, may be shifted to optimise the uncertainty
contributed from the pQCD part and the data based offset value. A
reliable estimate of the latter is mandatory and we have
also crosschecked its evaluation using the phenomenological effective
Lagrangian global fit approach~\cite{Benayoun:2015gxa,Benayoun:2019zwh}, specifically,
within the broken hidden local symmetry implementation.

\begin{table}
\centering
\caption{\mbo{\Delta \alpha^{(5)}_{\rm had}(-M_0^2)} at $M_0=2\,\gv$ in terms of \mbo{\epm} data and pQCD.
Labels as in Table~\ref{tab:alpZranges}.}
\label{tab:alp0ranges}
\begin{tabular}{llllll}
\hline \hline
Final state &  Range  & $\Delta \alpha^{(5)}_{\rm had}(-M_0^2)\power{4}$ (stat) (syst) [tot] & Rel & Abs \\
 &  (GeV) & & (\%) & (\%)\\
\hline
   $\rho  $     & (0.28, 1.05) &     29.97 (0.03) (0.24) [0.24]&  0.8 &  \phantom{6}14.3 \\
   $\omega $    & (0.42, 0.81) &      \phantom{6}2.69 (0.02) (0.05) [0.06]&  2.1 &   \phantom{61}0.8 \\
   $\phi   $    & (1.00, 1.04) &       \phantom{6}3.78 (0.03) (0.04) [0.05]&  1.4 &   \phantom{61}0.6 \\
   $\mathrm{J}/\uppsi $    &             &       \phantom{6}3.21 (0.15) (0.15) [0.21]&  6.7 &  \phantom{6}11.2 \\
   $\Upsilon$   &             &       \phantom{6}0.05 (0.00) (0.00) [0.00]&  6.8 &   \phantom{61}0.0 \\
     Had        & (1.05, 2.00) &     10.56 (0.02) (0.48) [0.48]&  4.6 &  \phantom{6}56.9 \\
     Had        & (2.00, 3.20) &       \phantom{6}6.06 (0.03) (0.25) [0.25]&  4.2 &  \phantom{6}15.7 \\
     Had        & (3.20, 3.60) &       \phantom{6}1.31 (0.01) (0.02) [0.03]&  1.9 &   \phantom{61}0.2 \\
     Had        & (3.60, 5.20) &       \phantom{6}2.90 (0.02) (0.02) [0.03]&  0.0 &   \phantom{61}0.2 \\
    pQCD        & (5.20, 9.46) &       \phantom{6}2.66 (0.02) (0.02) [0.00]&  0.1 &   \phantom{61}0.0 \\
     Had        & (9.46, 11.50) &       \phantom{6}0.39 (0.00) (0.02) [0.02]&  5.7 &   \phantom{61}0.1 \\
    pQCD        & (1.50, 0.00) &       \phantom{6}0.90 (0.00) (0.00) [0.00]&  0.0 &   \phantom{61}0.0 \\
    Data        &(0.3, $\infty$)&     60.92 (0.16) (0.62) [0.64]&  1.0 &   \phantom{61}0.0 \\
    Total       &             &     64.47 (0.16) (0.62) [0.64]&  1.0 &100.0 \\
\hline \hline
\end{tabular}
\end{table}

\begin{figure}
\centering
\includegraphics[width=0.5\textwidth]{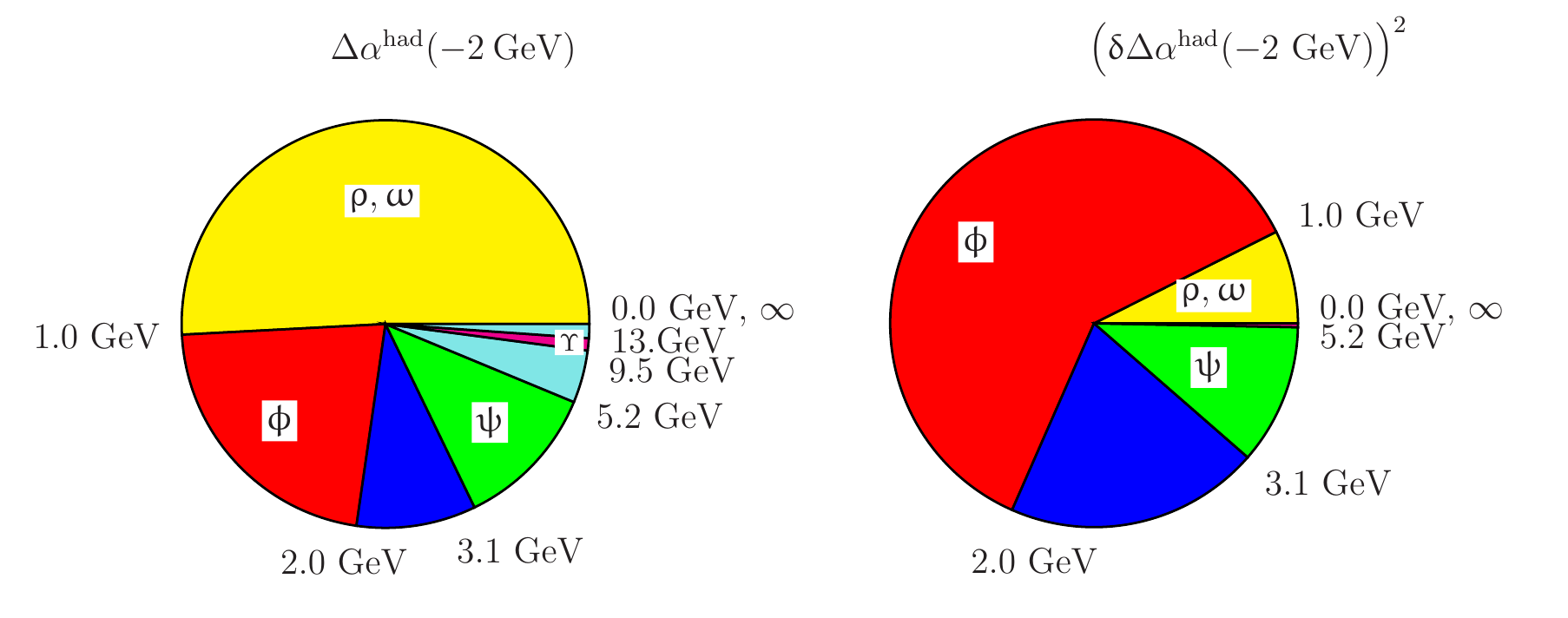}
\includegraphics[width=0.5\textwidth]{SM_FJegerlehner/aldistRc}
\caption{Contributions and square errors from $\mathrm{e}^+\mathrm{e}^-$ data ranges and from pQCD to
$\Delta \alpha_{\rm had}^{(5)}(-M_0^2)$ vs. $\Delta \alpha_{\rm had}^{(5)}(M_\mathrm{Z}^2)$.}
\label{fig:distalp0vsalpZ}
%\query{Please correct the figure labels in \Fref{fig:distalp0vsalpZ}. Set
%particle names and the $\updelta$ in roman (upright) font.}
\end{figure}

In Fig.~\ref{fig:distalp0vsalpZ}, we illustrate the relevance of
different energy ranges by comparing the hadronic shift of the
effective charge as evaluated at  space-like low-energy scale
$M_0=2~\gv$ with those at the time-like $M_\mathrm{Z}$ scale. The crucial
point is that the profile of the offset $\alpha$ at $M_0$ much more closely
resembles the profile found for the hadronic contribution to $a_\upmu$
and improving $a_\upmu^{\rm had}$ automatically leads to an improvement
of $\Delta \alpha_{\rm had}^{(5)}(-M_0^2)$; this is the profit gained
from the Euclidean split trick.

What does this have to do with the Adler function?  (i)
The Adler function is the monitor to control the
applicability of pQCD and  (ii) the pQCD part
$\left[\alpha(-\mz)-\alpha(-M_0^2)\right]^{\mathrm{pQCD}}$
is favourably calculated by integrating the Adler function
\mbo{D(Q^2)}. The small remainder \newline
$\left[\alpha(\mz)-\alpha(-\mz)\right]^{\mathrm{pQCD}}$ can be obtained
in terms of the VP function \mbo{\Pi'_\upgamma(s)}. In fact, the Adler
function is the ideal monitor for comparing theory and data.
The Adler function is defined as the derivative of the VP function:
\ba
D(-s) \doteq
\frac{3\uppi}{\alpha}\:s\:\frac{\D}{\D s} \Delta \alpha_{\mathrm{had}}(s)
= -\left( 12 \uppi^2 \right)\:s\: \frac{\D\Pi'_{\upgamma}(s)}{\D s}
\ea
and can be evaluated in terms of $\epm$ annihilation data by the dispersion integral
\ba
D(Q^2) =  Q^2\;\left (\int\limits_{4
m_\uppi^2}^{E^2_\mathrm{cut}} \D s\,
\frac{R(s)^\mathrm{data}}{\left( s+Q^2 \right)^2}
+ \int_{E^2_\mathrm{cut}}^{\infty} \D s\, \frac{R^\mathrm{pQCD}(s)}{(s+Q^2)^2}\right)\;.
\ea
It is a finite object not subject to renormalization and it tends to a
constant in the high-energy limit, where it is perfectly
perturbative. Comparing the direct $R(s)$-based and the
$D(Q^2)$-based methods
\begin{center}
\begin{tabular}{ccc}
 pQCD $\leftrightarrow$ $R(s)$ &$\qquad$&  pQCD $\leftrightarrow$ $D(Q^2)$\\
Very difficult to obtain &$\qquad$& Smooth simple function  \\
in theory &$\qquad$&  in \emph{Euclidean} region\\
\end{tabular}
\end{center}
we note that in the time-like approach pQCD only works well in
`perturbative windows' roughly in the  ranges 3.00--3.73\,GeV, 5.00--10.52\,GeV and 11.50\,GeV to $\infty$  ~\cite{Harlander:2002ur}, while
in the space-like approach pQCD works well for $Q > 2.0\,\gv$, a clear advantage.

\begin{figure}
\centering
\includegraphics[width=0.95\textwidth]{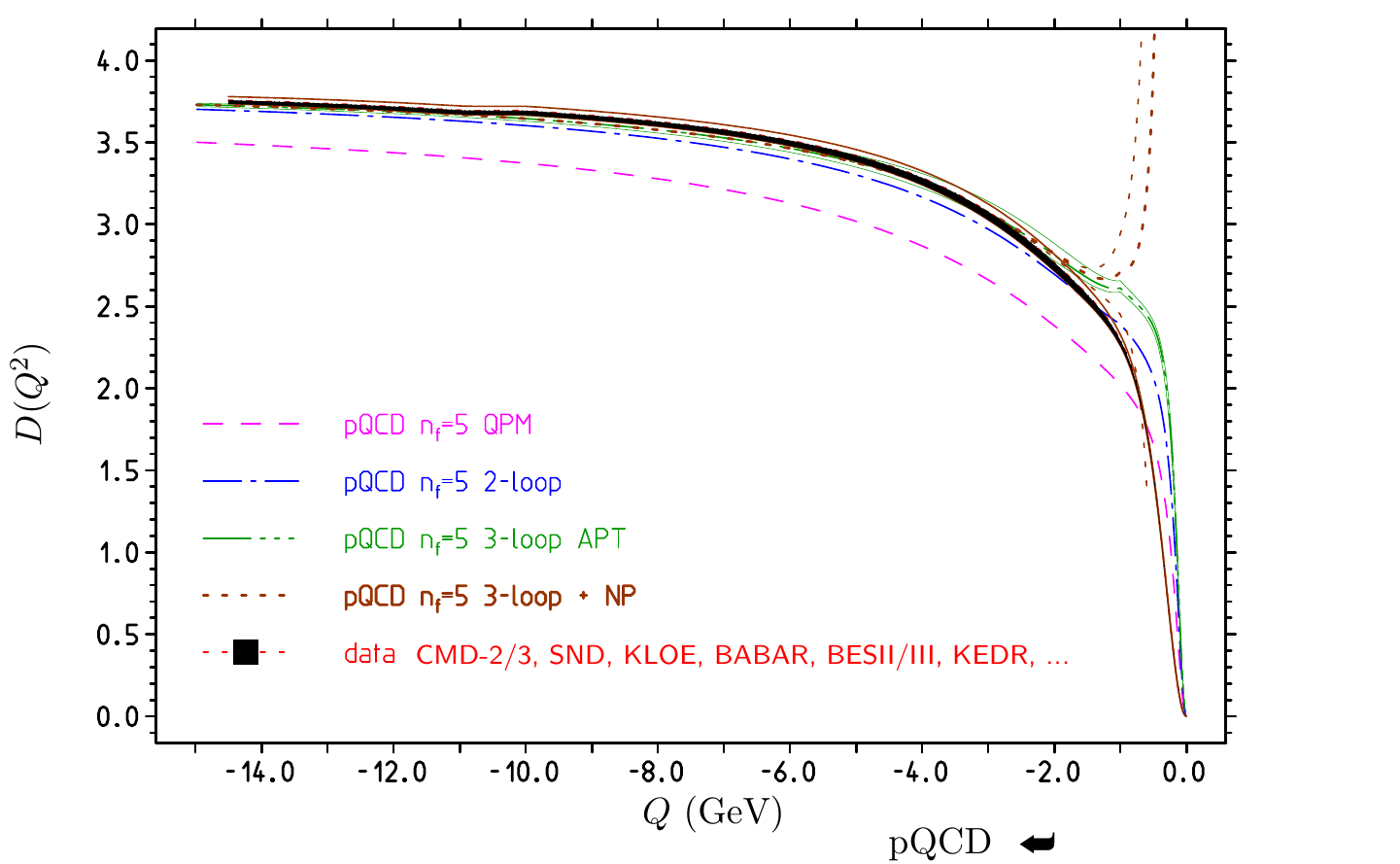}
\caption{Monitoring pQCD vs. data: the pQCD prediction of $D(Q^2)$
works well down to
$M_0= 2.0\, \gv$, provided full massive QCD at three-loop order or higher
 is employed.}
\label{fig:adlerapr17}
%\query{Please correct the figure labels in \Fref{fig:adlerapr17}. Set variables
%in italic font. Set
%particle names in roman font.}
\end{figure}

In \Fref{fig:adlerapr17}, the `experimental' Adler function is
confronted with theory (pQCD + NP). Note that, in contrast to most
$xfR$ plots, like \Fref{fig:Rdata},  showing statistical errors only,
in \Fref{fig:adlerapr17}. the total error is displayed as the
shaded band. We see that while one-loop and two-loop predictions clearly
fail  to follow
the data band, a full massive three-loop QCD prediction in the gauge-invariant background field MOM scheme~\cite{Jegerlehner:1998zg}
reproduces the experimental Adler function surprisingly well. This has
been worked out \cite{Eidelman:1998vc} by Pad\'e improvement of the moment
expansions provided in Refs.~\cite{Chetyrkin:1996cf,Chetyrkin:1997qi,Chetyrkin:1997mb}. Figure \ref{fig:adlerapr17} also shows that non-perturbative (NP) contributions from the
quark and gluon condensates~\cite{Shifman:1978bx,Shifman:1978by}\footnote{These are evaluated
by means of operator product expansions; the explicit expressions
may be found in Ref.~\cite{Eidelman:1998vc}.} start to contribute substantially only
at energies where pQCD fails to converge because one is approaching
the Landau pole in \MSb parametrized QCD. Strong coupling constant
freezing, as in analytic perturbation theory, advocated
in Ref.~\cite{Shirkov:1997wi} or similar schemes,  is not actually able to improve
the agreement in the low-energy regime. Coupling constant freezing
also contradicts lattice QCD results~\cite{Bruno:2017gxd}.

From the three terms of \Eref{EsplitalphaZ}, we already
know the low-energy offset $\Delta \alpha_{\mathrm{had}}(-M_0^2)$ for
$M_0=2.0\,\gv$. We obtain the second term by
integrating the pQCD predicted Adler function
\ba
\Delta_1=\Delta \alpha_{\mathrm{had}} \left (-M_\mathrm{Z}^2 \right )-\Delta \alpha_{\mathrm{had}} \left (-M_0^2 \right )=  \frac{\alpha}{3\uppi} \int_{M_0^2}^{M_\mathrm{Z}^2}
\D {Q'}^{2} \frac{D \left ({Q'}^{2} \right )}{{Q'}^{2}}\,,
\ea
based on a complete three-loop massive QCD analysis.
The QCD parameters used are $\alpha_\mathrm{s}(M_\mathrm{Z})=0.1189(20)$,
$m_\mathrm{c}(m_\mathrm{c})=1.286(13)[M_\mathrm{c}=1.666(17)]\,\gv\,,$
$m_\mathrm{b}(m_\mathrm{c})=4.164(25)[M_\mathrm{b}=4.800(29)]\,\gv\,.$
The result obtained is
$$\Delta_1=\Delta \alpha_{\mathrm{had}} \left (-M_\mathrm{Z}^2 \right )-\Delta \alpha_{\mathrm{had}} \left (-M_0^2 \right )=0.021\,074\pm 0.000\,100\epo$$
This includes a shift { $+0.000\,008$} from the massless four-loop
contribution included in the high-energy tail.
The error $\pm 0.000\,100$ will be added in quadrature.
Up to three loops, all contributions
have the same sign and are substantial. Four-loop and higher orders
could still add up to non-negligible contributions. An error for
missing higher-order terms is not included.

The remaining term concerns the link between the space-like and the
time-like region at the Z boson mass scale and is given by the
difference
 $$
 \Delta_2=\dahz-\Delta\alpha^{(5)}_{\rm had} \left(-M_\mathrm{Z}^2 \right
 )
=0.000\,045 \pm 0.000\,002\,,
 $$
 which can be calculated in pQCD. It
accounts for the $\I\uppi$-terms from the logs $\ln
(-q^2/\mu^2)=\ln(|q^2/\mu^2|)+\I\uppi.$
Since the term is small, we can
also get it  from direct data integration based on our data
compilation. We obtain $\dalh \left(-M_\mathrm{Z}^2 \right )=276.44\pm 0.64\pm1.78$ and
$\dalh \left (+M_\mathrm{Z}^2 \right )=276.84\pm 0.64\pm1.90$, and taking into
account that errors are almost 100\% correlated, we have
$\dalh (M_\mathrm{Z}^2 )-\dalh  (-M_\mathrm{Z}^2 
)=0.40 \pm 0.12$ less precise but in
agreement with the pQCD result. We then have
\begin{align*}
\Delta\alpha^{(5)}_{\rm had} \left (-M_0^2 \right )^{\mathrm{data}} &= 0.006\,409 \pm
0.000\,063 \\
\Delta\alpha^{(5)}_{\rm had} \left (-M_\mathrm{Z}^2 \right ) &=  0.027\,483 \pm 0.000\,118  \\
\Delta\alpha^{(5)}_{\rm had} \left ( M_\mathrm{Z}^2 \right ) &=  0.027\,523 \pm 0.000\,119  \epo
\end{align*}

\begin{figure}
\centering
\includegraphics[width=0.67\textwidth]{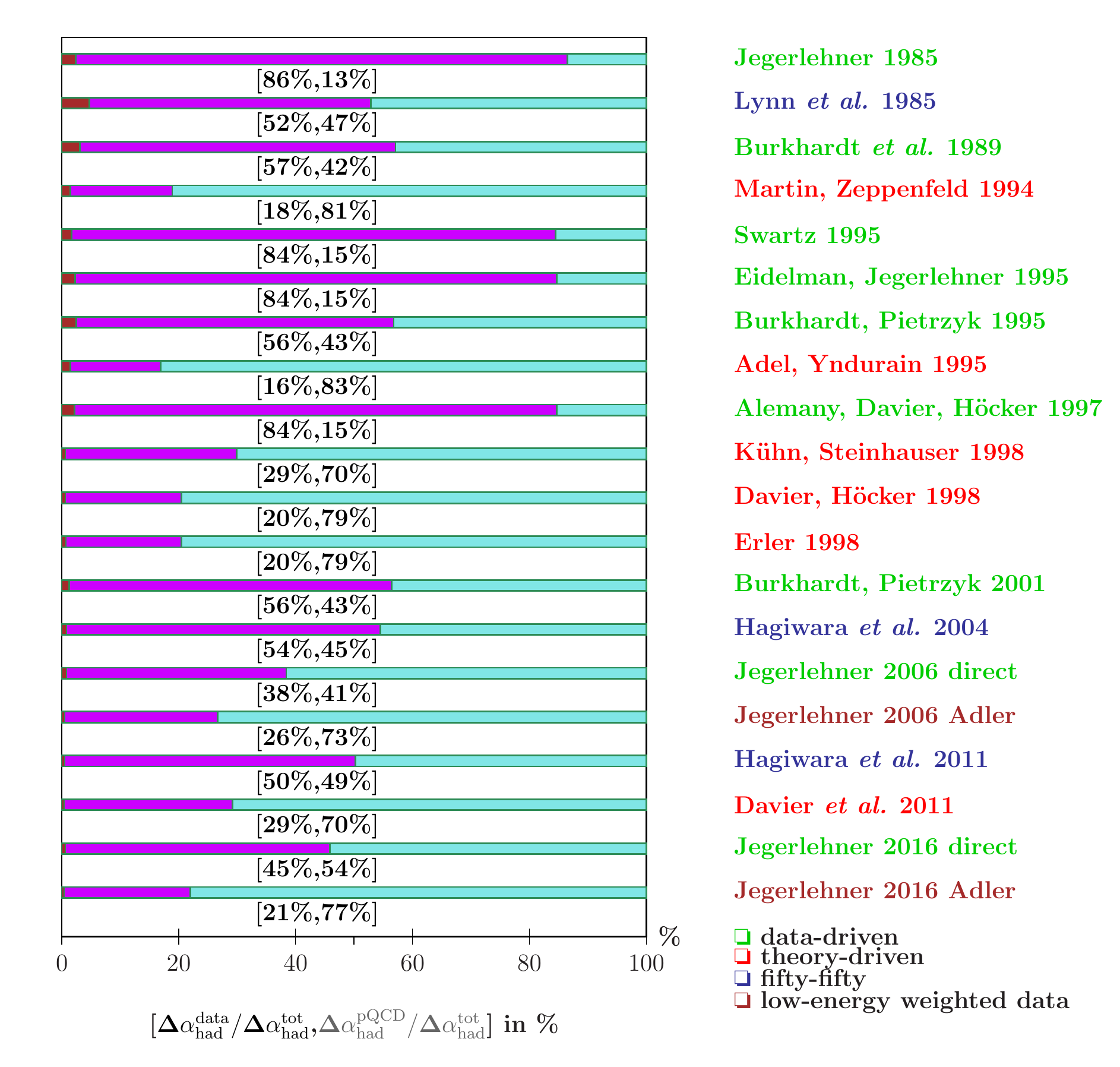}
\caption{How much pQCD? Here a history of results by different
authors. It shows that the Adler function controlled approach to $\dahz$
is barely more pQCD-driven than many of the standard evaluations. The
pQCD piece is 70\% in Davier \textit{et al.}~\cite{Davier:2017zfy} and 77\% in our
Adler-driven case, with an important difference:
in the Adler-controlled case, the major part of 71\% is based on pQCD in the
space-like region and only 6\% contributing to the non-perturbative offset value is evaluated in the time-like region, while in the
standard theory-driven, as well as in the more data-driven approaches,
pQCD is applied in the time-like region, where it is much harder to be
tested against data.}
\label{fig:howmuchQCD}
%\query{Please correct the figure labels in \Fref{fig:howmuchQCD}. Set `\textit{et
%al.}' in italic font each time. Correct `low energy weighted data' to `low-energy
%weighted data'.}
\end{figure}

To get $\alpha^{-1}(M_\mathrm{Z}^2)$, we also have to include
 the leptonic piece~\cite{Steinhauser:1998rq}
\bea
\Delta \al _{\rm lep} \left (M_\mathrm{Z}^2 \right)\simeq 0.031\,419\,187\,418\,,
\eea
and the top quark contribution.
A very heavy top quark decouples as
\bea
\Delta \al _{\rm top}\simeq -\frac{\al}{3\uppi}\frac{4}{15} \frac{s}{m_\mathrm{t}^2} \to 0
\nn
\eea
when { $m_\mathrm{t} \gg s$}. At $s=M^2_\mathrm{Z}$, the top quark contributes
\bea
\Delta \alpha_{\rm top} \left(\mz \right ) = -0.76 \times 10^{-4}\;\;.
\eea
Collecting terms, this leads to the result presented in \Eref{dhadr5alpha}.
One should note that the Adler function controlled Euclidean data
vs. pQCD split approach is only moderately more pQCD-driven than the
time-like approach adopted by Davier \textit{et al.}~\cite{Davier:2017zfy} and
others, as follows from the collection of results shown in
\Fref{fig:howmuchQCD}. The point is that the Adler function driven
method only uses pQCD where reliable predictions are possible and
direct cross checks against lattice QCD data may be carried
out. Similarly, possible future direct measurements of $\alpha(-Q^2)$
in $\upmu$-e scattering~\cite{Abbiendi:2016xup} can provide Euclidean
HVP data, in particular, also for the offset $\dalh(-M_0^2)\epo$

\subsection{Prospects for future improvements}
The new muon $g-2$ experiments at Fermilab and at JPARC in Japan (expected to
go into operation later) trigger the continuation of $\epm \to $
hadrons cross-section measurements in the low-energy region by CMD-3
and SND at BINP Novosibirsk, by BES III at IHEP Beijing and soon by
Belle II at KEK Tsukuba. This automatically helps to improve $ \Delta
\alpha (-M_0^2)$ and hence $\alpha (M_\mathrm{Z}^2)$ via the Adler function
controlled split-trick approach. Equally  important are the results from
lattice QCD, which come closer to being competitive with the data-driven
dispersive method.

The improvement by a factor of five to ten in this case largely relies on
improving the QCD prediction of the two-point vector correlator above the
2\,GeV scale, which is a well-defined and comparably simple task. The
mandatory pQCD improvements required are as follows.
\begin{itemize}
%\item[$bullet$]
\item[(a)] Four-loop massive
pQCD calculation of Adler function. In practice, this requires the
calculation of a sufficient number of terms in the low- and high-momentum
series expansions, such that an accurate Pad\'e
improvement is possible.
%\item[$\bullet$] 
\item[(b)] \mbo{m_\mathrm{c}}, \mbo{m_\mathrm{b}} improvements by sum
rule or lattice QCD evaluation.
%\item[$\bullet$] 
\item[(c)] Improved $ \alpha_\mathrm{s}$
in low $ Q^2$ region above the $\uptau$ mass.
\end{itemize}

Note that the direct dispersion relations (DR) approach requires
precise data up to much higher energies or a heavy reliance on the
pQCD calculation of the time-like $R(s)$! The virtues of the
Adler function approach are obvious:
\begin{itemize}
\item[(a)] no problems with physical threshold and resonances;
\item[(b)] pQCD is used only where we can check it to work accurately
      (Euclidean $Q\:\gapprox\:$ 2.0\,GeV);
\item[(c)] no manipulation of data, no assumptions about global or local
duality;
\item[(d)] the non-perturbative `remainder' $\Delta \alpha (-M_0^2)$ is mainly
 sensitive to low-energy data;
\item[(e)]  $ \Delta \alpha (-M_0^2)$ would be directly accessible in a
       MUonE experiment (project)~\cite{Abbiendi:2016xup} or in lattice QCD.
\end{itemize}
In the direct approach, \eg Davier \textit{et al.}~\cite{Davier:2017zfy} use
pQCD above $ 1.8\,\gv$, which means that no error reduction follows
from remeasuring cross-sections above $1.8\,\gv$. Also, there is no
proof that pQCD is valid at 0.04\% precision as adopted. This is a
general problem when utilising pQCD at time-like momenta exhibiting
non-perturbative features.

\begin{figure}
\centering
\includegraphics[height=10cm]{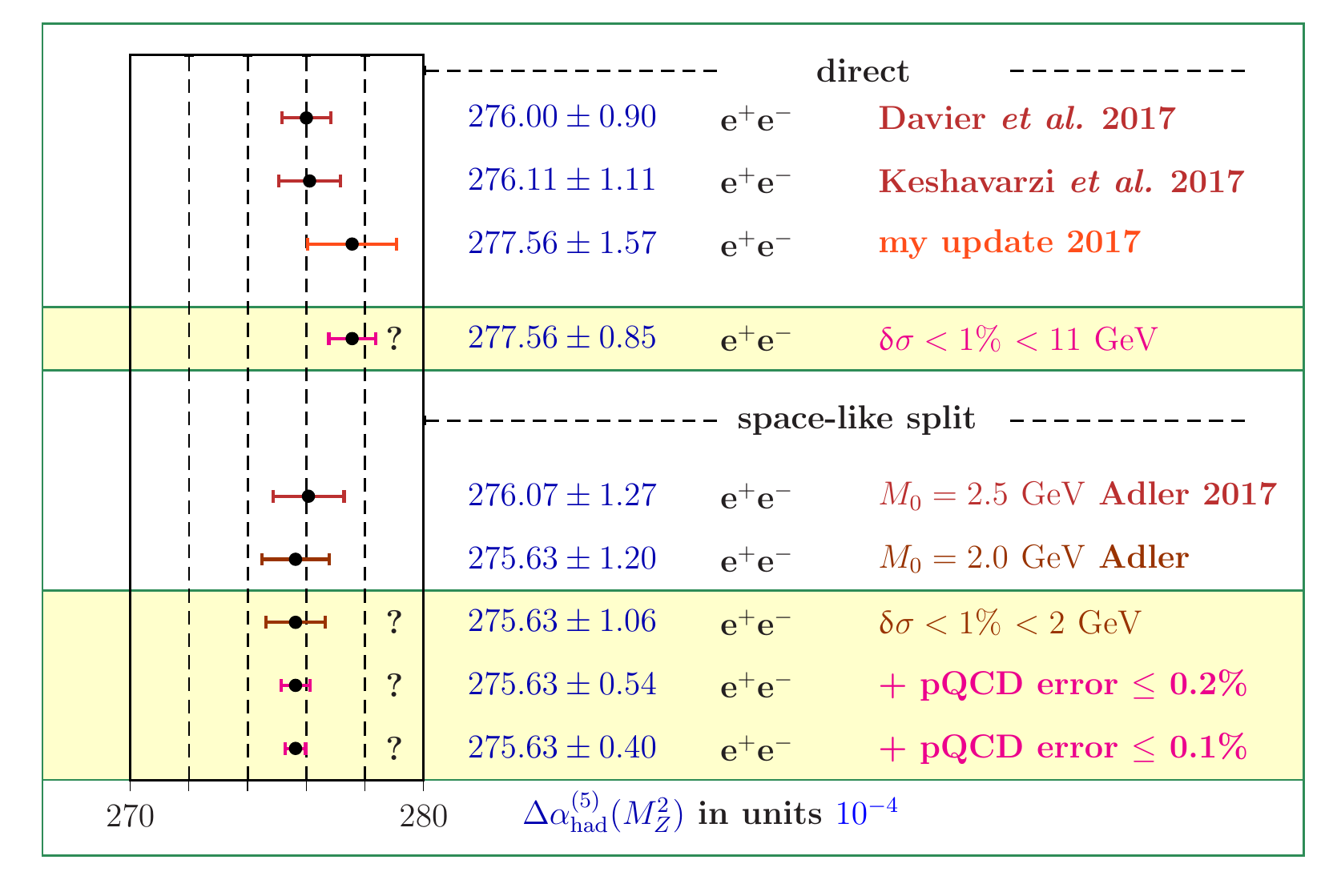}
\caption{Comparison of possible improvements. 
My `direct' analysis is data-driven, adopting pQCD in the window 5.2--9.5\,\gv\
 and above $11.5\,\gv\,.$ The Adler-driven results under `space-like
split' show the current status for the two offset energies,
$M_0=2.5\,\gv$ and $2\,\gv$. The improvement potential is displayed for
three options: reducing the error of the data offset by a factor of two,
improving pQCD to a 0.2\% precision Adler function in addition and the same by
improving pQCD to a 0.1\% precision Adler function. The direct results are from
Refs.~\cite{Davier:2017zfy,Keshavarzi:2018mgv,Jegerlehner:2017zsb}.}
\label{fig:future}
%\query{Please correct the figure labels in \Fref{fig:future}. Set
%particle names and the $\updelta$ in roman (upright) font.}
\end{figure}

What we can achieve is illustrated in Fig.~\ref{fig:future} and  \Tref{table:Fred}.
Our analysis shows that the Adler function inspired method is
competitive with Patrick Janot's~\cite{Janot:2015gjr} direct near-Z
pole determination via a measurement of the forward--backward asymmetry
\mbo{A_{\rm FB}^{\upmu\upmu}} in $\epm \to \upmu^+\upmu^-$.  The modulus
square of the sum of the two tree-level diagrams has three terms: the
Z exchange alone, \mbo{\cZ\propto (\mz
G_\upmu)^2}, the \mbo{\upgamma}--Z interference, \mbo{\cI\propto
\alpha(s)\,\mz G_\upmu}, and the $\upgamma$-exchange only, 
\mbo{\cG\propto \alpha^2(s)}. The
interference term determines the forward--backward (FB) asymmetry, which
is linear in $\alpha(s)$;
\mbo{v} denotes the vector \mbo{\mathrm{Z}\upmu\upmu} coupling that depends on
\mbo{\sin^2\Theta_{\ell\,{\rm eff}}(s)}, while \mbo{a} denotes the axial \mbo{\mathrm{Z}\upmu\upmu}
coupling that is sensitive to the \mbo{\rho}-parameter (strong \mbo{M_\mathrm{t}}
dependence). In extracting $\alpha(\mz)$, one is using the \mbo{v} and \mbo{a}
couplings as measured at the {Z} peak directly. At tree level, one then has
\bea
 A_{\rm FB}^{\upmu\upmu}=A_{\rm FB,0}^{\upmu\upmu}+
\frac{3\,a^2}{4\,v^2}\,\frac{\cI}{\cZ+\cG}
\semis \quad A_{\rm FB,0}^{\upmu\upmu} =
\frac{3}{4}\,\frac{4v^2 a^2}{(v^2+a^2)^2}\,,
\eea
where
\begin{gather*}
\cG=\frac{c^2_\upgamma}{s}\comas \quad  \cI=\frac{2c_\upgamma
c_\mathrm{Z}\,v^2\, \left (s-\mz \right )}{\left (s-\mz \right )^2+\mz \Gamma_\mathrm{Z}^2}\comas \quad
\cZ=\frac{c^2_\mathrm{Z}\, \left (v^2+a^2 \right )\,s}{\left (s-\mz \right
)^2+\mz \Gamma_\mathrm{Z}^2}\\
c_\upgamma=\sqrt{\frac{4\uppi}{3}}\, \alpha(s)\comas \quad 
c_\mathrm{Z}=\sqrt{\frac{4\uppi}{3}}\,\frac{\mz}{2\uppi}\,\frac{G_\upmu}{\wz}\comas \quad
v= \left (1-4\,\sin^2\Theta_\ell \right )\,a\comas \quad a=-\frac12 \epo
\end{gather*}

\begin{table}
\begin{center}
\caption{Precision in \mbo{\alpha(M^2_\mathrm{Z})\,}}
\label{table:Fred}
\begin{tabular}{llll}
\hline
\hline
 Present & Direct & $1.7\power{-4}$ &\\
        & Adler  & $1.2\power{-4}$ &\\
Future  & Adler QCD 0.2\% & $5.4\power{-5}$ &\\
        & Adler QCD 0.1\% & $3.9\power{-5}$ &\\
Future & Via \mbo{A_{\rm FB}^{\upmu\upmu}} off-Z & $3\phantom{.9}\power{-5}$&~\cite{Janot:2015gjr}\\
\hline\hline
\end{tabular}
\end{center}
%\query{All tables are tables and must be treated as such. Please check
%that the caption for \Tref{table:Fred} is appropriate, and replace if necessary.}
\end{table}

 Note that $\mz
G_\upmu=\mw
G_\upmu/\cos^2\Theta_\mathrm{W}= {\uppi}\, {(\alpha_2(s))} / {\wz}{(\cos^2\Theta_\mathrm{g}(s))}$
and $\sin^2\Theta_\mathrm{g}(s)=\alpha(s)/\alpha_2(s)$. \ie all parameters
vary more or less with energy, depending on the renormalization scheme
utilised. The challenges for this direct measurement are precise
radiative corrections (see Refs.~\cite{Blondel:2018mad,Dubovyk:2018rlg} and references
therein) and the required dedicated off-{Z} peak running. Short
accounts of the methods proposed for improving $\alpha(M^2_\mathrm{Z})$
may be found in Sections~8 and 9 of~Ref. \cite{Azzi:2017iih}.

The Adler function based method is much cheaper, I think, and
does not depend on understanding the Z peak region with
unprecedented precision. Another very crucial point may be that the
dispersive method and the Adler function modified version provide the
effective $\alpha(s)$ for arbitrary c.m. energies, not at $s=M_\mathrm{Z}^2$
only; although,  given a very precise $\alpha(\mz)$, one
can reliably calculate $\alpha(s)-\alpha(\mz)$ via pQCD for
values of $s$ in the perturbative regime,
\ie especially  going to higher energies. In any case, the
requirements specified here that must be satisfied in order to reach a
factor of five improvement appears to be achievable.

\subsection{The need for a space-like effective $\alpha(t)$}

\begin{figure}
\centering
\includegraphics[height=4cm]{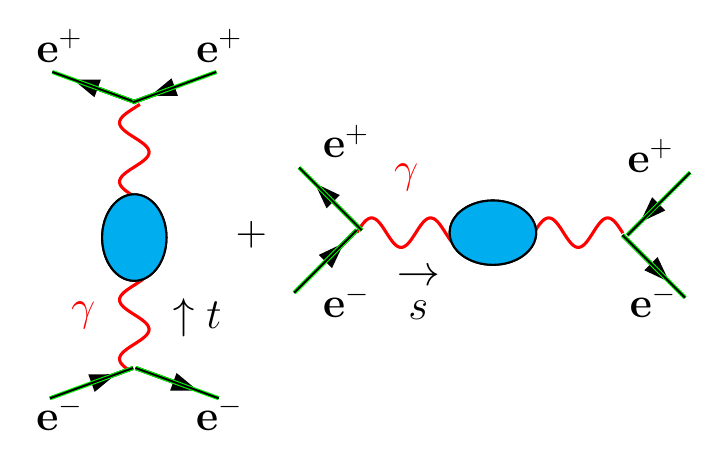}~~\raisebox{2cm}{;}~~
\includegraphics[height=4cm]{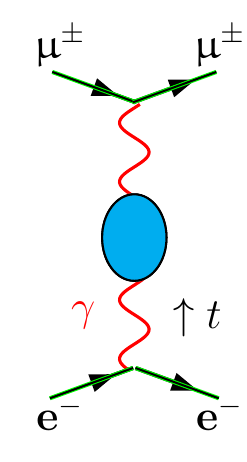}
\caption{t-channel dominated QED processes. Left: VP dressed tree-level Bhabha scattering at small scattering angles. Right:
the leading VP effect in $\upmu$e scattering.}
\label{fig:EuclideanVP}
%\query{Please correct the figure labels in \Fref{fig:EuclideanVP}.  Set
%particle names in roman font.}
\end{figure}

\begin{figure}
\centering
\includegraphics[width=0.43\textwidth]{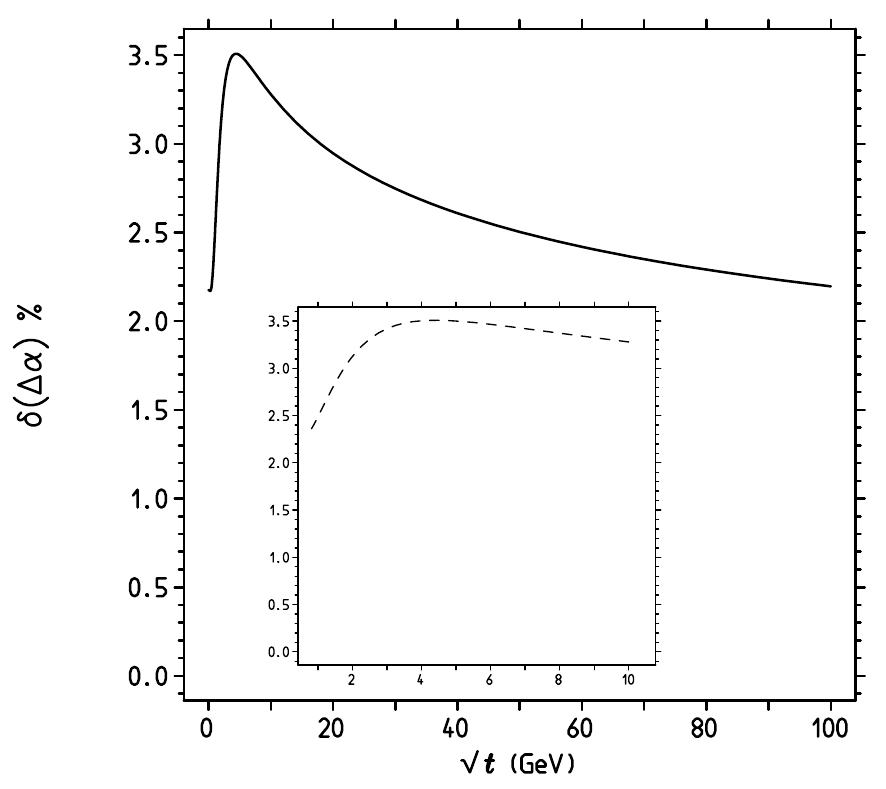}
\includegraphics[width=0.45\textwidth]{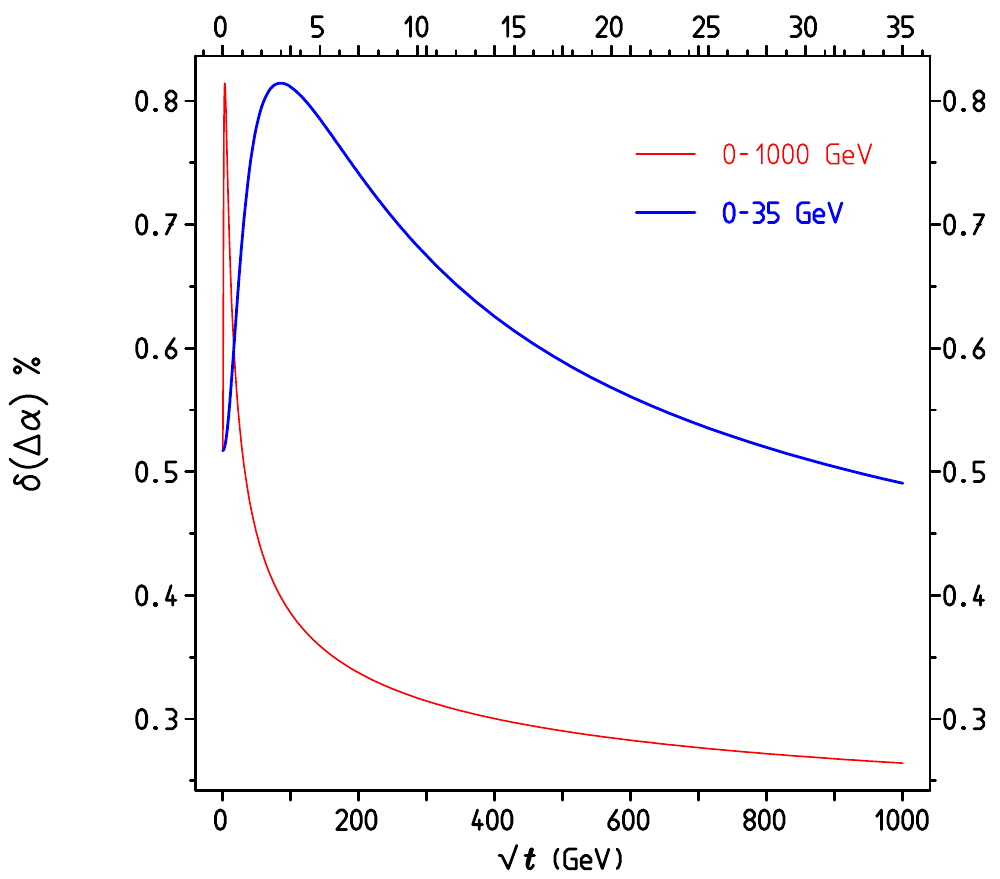}
\caption{Hadronic uncertainty $ \updelta
\Delta \alpha_{\rm had} (\sqrt{t})$. The progress since LEP times, from 1996
(left) to now (right) is remarkable. A great deal of much more precise  low-energy data,
$\uppi\uppi$, \etc, are now available.}
\label{fig:Bhabhaalpha}
%\query{Please correct the figure labels in \Fref{fig:Bhabhaalpha}. Set variables
%in italic font.}
\end{figure}

As a normalization in measurements of cross-sections in $\epm$ collider
experiments, small-angle Bhabha scattering is the standard
choice. This reference process is
dominated by the t-channel diagram of the Bhabha scattering process
shown in the left of Fig.~\ref{fig:EuclideanVP}. In small-angle Bhabha scattering, we have
${\delta_{\rm HVP} \sigma}/{\sigma}=2\, {\updelta \alpha (\bar{t})}/{\alpha
(\bar{t})}$, and for the FCC-ee luminometer $\sqrt{\bar{t}}\simeq 3.5\,\gv$ near the
Z peak and $\simeq 13\,\gv$ at 350\,GeV~\cite{Jadach:2018jjo}.
The progress achieved after LEP times is displayed in
Fig.~\ref{fig:Bhabhaalpha}. What can be achieved for the FCC-ee project
is listed in \Tref{table:Tom}. The estimates are based on expected improvements possible for
$\dalh(-Q^2)$ in the appropriate energy ranges, centred at $\sqrt{\bar{t}}$.

\begin{table}
\begin{center}
\caption{Possible achievements for the FCC-ee project}
\label{table:Tom}
\begin{tabular}{lllll}
\hline \hline
$\sqrt{s}$ & $\sqrt{\bar{t}}$ & 1996 \cite{Jadach:1996gu,Arbuzov:1996eq} & Present & FCC-ee expected \cite{Jadach:2018jjo}\\
& (GeV)\\
\hline
 $M_\mathrm{Z}$ & \phantom{$1$}$3.5$ &0.040\% & 0.013\% &
 $0.6\power{-4}$ \\
 $350~\gv$ &$ 13$ &   & $1.2\power{-4}$ & $2.4\power{-4}$\\
 \hline \hline
\end{tabular}
\end{center}
%\query{All tables are tables and must be treated as such. Please check
%that the caption for \Tref{table:Tom} is appropriate, and replace if necessary.}
\end{table}

\subsubsection{A new project: measuring  the low-energy \mbo{\alpha(t)}
directly}
The possible direct measurement of $\dalh(-Q^2)$ follows a very
different strategy of evaluating the HVP contribution to the muon
$g-2$. There is no VP subtraction issue, there is no exclusive channel
separation and recombination, no issue of combining data from very
different experiments and controlling correlations. Even a 1\% level
measurement can provide invaluable independent information. The recent
proposal~\cite{Abbiendi:2016xup} to measure $\alpha(-Q^2)$ via $\upmu^-$e$^-$-scattering (see right part of Fig.~\ref{fig:Bhabhaalpha})
in the {\tt MUonE} projects at CERN is very important for future
precision physics. It is based on a cross-section measurement
\bea
\frac{\D \sigma^{\rm unpol.}_{\upmu^-\mathrm{e}^- \to \upmu^-\mathrm{e}^-}}{\D
t}=4\uppi\,{ \alpha(t)^2}\,\frac{1}{\lambda(s,m_\mathrm{e}^2,m_\upmu^2) }\,\left\{\frac{\left(s-m_\upmu^2-m_\mathrm{e}^2\right)^2}{t^2}+\frac{s}{t}+\frac12\right\}\epo
\eea
The primary goal of the project concerns the determination of
\mbo{\amuh} in an alternative way
\bea
\amuh=\frac{\alpha}{\uppi}\int\limits_0^1 \D x\:(1-x)\: \dalh
\left(-Q^2(x)\right)\,,
\eea
where $ Q^2(x)\equiv  {x^2} m_\upmu^2 / (1-x) $ is the space-like
square momentum transfer and
\bea
\dalh(-Q^2)=\frac{\alpha}{\alpha(-Q^2)}+\Delta
\alpha_{\rm lep}(-Q^2)-1
\eea
directly compares with lattice QCD data and the offset
$\alpha(-M_0^2)$ discussed before. We propose to determine, very
accurately, $\dalh\left(-Q^2\right)$ at $Q \approx 2.5\,\gv$ by this
method (one single number!) as the non-perturbative part of \mbo{\dalh
\left(M_\mathrm{Z}^2\right)}, as needed in the `Adler function approach'.
It  would also be of direct use for a precise small-angle Bhabha
luminometer! Because of the high precision required, accurate radiative
corrections are mandatory and corresponding calculations are in
progress~\cite{Mastrolia:2017pfy,DiVita:2018nnh,Fael:2018dmz,Alacevich:2018vez}.

\subsection{Conclusions}
Reducing the muon \mbo{g-2} prediction uncertainty remains the key issue of high-precision physics and
strongly motivates  more precise measurements of low-energy
$\epm \to \mathrm{hadrons}$ cross-sections. Progress is
expected from Novosibirsk (VEPP 2000/CMD3,SND), Beijing
(BEPCII/BESIII), and Tsukuba (SuperKEKB/BelleII). This helps to improve
\mbo{\alpha(t)} in the region relevant
for small-angle Bhabha scattering and in calculating \mbo{\alpha(s)}
at FCC-ee/ILC energies using the Euclidean split-trick method.  The
latter method requires pQCD prediction of the Adler function to
improve by a factor  of two. This also means that we need improved
parameters, in particular, $m_\mathrm{c}$ and $m_\mathrm{b}$.

One question remains to be asked. Are presently estimated and
essentially agreed-on evaluations of $\dahz$ in terms of \mbo{R} data
reliable? One has to keep in mind that the handling of systematic
errors is rather an art than a science. Therefore, alternative methods
are very important and, fortunately, are under consideration.

Patrick Janot's approach is certainly  an important
alternative method, directly accessing \mbo{\alpha(M_\mathrm{Z}^2)} with
very different systematics. This is a challenging project.

Another interesting option is an improved radiative return
measurement of  $\sigma(\epm \to \mathrm{hadrons})$ at
the GigaZ, allowing for directly improved dispersion integral input,
which would include all resonances and thresholds in one experiment!

In any case, on paper, \mbo{\mathrm{e}^- \upmu^+ \to \mathrm{e}^- \upmu^+} appears to be the
ideal process to perform an  unambiguous measurement of
\mbo{\alpha(-Q^2)}, which determines the leading-order (LO) HVP to \mbo{\amu},  as well
as the non-perturbative part of \mbo{\alpha(s)}! 

Lattice QCD results are very close to becoming competitive here as
well. Thus, in the end, we will have alternatives, allowing
for important improvements and crosschecks.

The improvement obtained by reducing the experimental error to 1\% in
the range from $\phi$ to 3\,GeV would allow one to choose a higher cut
point, \eg for $\sqrt{M_0}=3.0\,\gv$. One  can then balance the
importance of data against pQCD differently. This would provide further
important consolidation of results. For a 3\,GeV cut, one gets
$\dalh(-M_0^2)=82.21\pm 0.88[0.38]$ in $10^{-4}$. The QCD contribution
is then smaller, as well as safer, because the mass effects that are
responsible for the larger uncertainty of the pQCD prediction are also substantially reduced. Taking the view that a massive four-loop QCD
calculation is a challenge, the possibility of optimising the choice of
split scale $M_0$ would be very useful. Therefore, the ILC/FCC-ee
community should actively support these activities as an integral part
of the $\epm$-collider precision physics programme!

\subsection{Addendum: the coupling $ \alpha_2$, $M_\mathrm{W}$,  and $\sinf$}
Besides $\alpha$,  the $SU(2)$ gauge coupling
$\alpha_2=g^2/(4\uppi)$ is also running and thereby affected by
non-perturbative hadronic
effects~\cite{Jegerlehner:1985gq,Jegerlehner:2011mw,Jegerlehner:2017zsb}. Related with
the $U_\mathrm{Y}(1)\otimes SU_\mathrm{L}(2)$ gauge couplings is the running of the
weak mixing parameter $ \sinf$, which is actually defined by the ratio
$\alpha/\alpha_2$. In Refs.~\cite{Jegerlehner:1985gq,Jegerlehner:2011mw,Jegerlehner:2017zsb}, the hadronic effects
have been evaluated by means of DRs in terms of
$\epm$ data with appropriate flavour separation and
reweighting. Commonly, a much simpler approach is adopted in studies
of the running of $ \sinf$, namely  using pQCD with effective quark
masses~\cite{Czarnecki:1995fw,Czarnecki:2000ic,Erler:2004in,Erler:2017knj}, which have
been determined elsewhere.

\begin{figure}
\centering
\includegraphics[height=7.6cm]{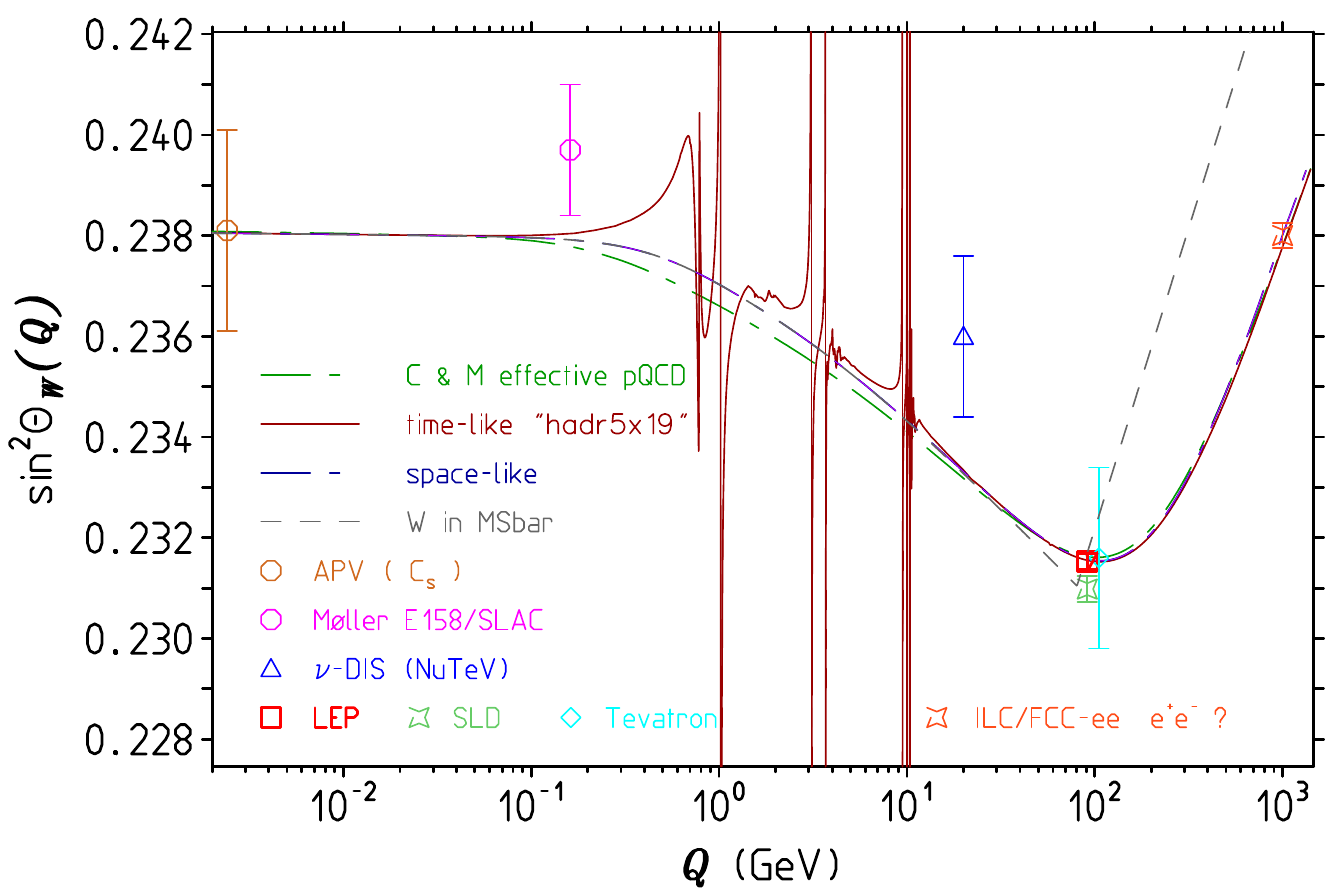}
\caption{$\sin^2 \Theta_W(Q)$ as a function of $Q$ in the time-like
and space-like region. Hadronic uncertainties are included but barely
visible in this plot. Uncertainties from the input parameter $\sin^2
\Theta_W(0)=0.238\,22(100)$ or $\sin^2 \Theta_W(M_\mathrm{Z}^2)=0.231\,53(16)$ are
not shown. Note the substantial difference from applying pQCD with
effective quark masses. Future FCC-ee/ILC measurements at 1\,TeV would be sensitive to
Z$'$, H$^{--}$, etc.}
\label{fig:sin2theta}
%\query{Please correct the figure labels in \Fref{fig:sin2theta}. Set variables
%in italic font. Set
%particle names in roman font.}
\end{figure}

Given $ g \equiv g_2$ and the Higgs vacuum expectation value (VEV) $v$, then
%5.04.2019 JG I should come to this later \varv not defined
\[
M_\mathrm{W}^2=\frac{g^2\,v^2}{4 }=\frac{\uppi\, { \alpha_2}}{\sqrt{2}\,G_\upmu} \, .
\]
The running $\sinf(s)$ relates electromagnetic to weak  neutral channel
mixing at the LEP scale to low-energy $ \upnu_\mathrm{e} \mathrm{e}$ scattering as
\bea
\sin^{2}\Theta_{\rm lep}(\mz)=\left\{ \frac{1-\dal_2}{1-\dal}
+\Delta_{\upnu_{\upmu}\mathrm{e},\mathrm{vertex+box}}
+\Delta \kappa_{\mathrm{e},\mathrm{vertex}} \right\}
\sin^{2}\Theta_{\upnu_{\upmu}\mathrm{e}}(0)\epo
\eea
The first correction from the running coupling ratio is largely
compensated for by the \mbo{\upnu_\upmu} charge radius, which dominates the
second term. The ratio
\mbo{\sin^{2}\Theta_{\upnu_{\upmu}\mathrm{e}}/\sin^{2}\Theta_{\rm lep}} is close to 1.002,
independent of the top and Higgs masses. Note that errors in the ratio
\mbo{{(1-\dal_2)} / {(1-\dal})} can be taken to be 100\% correlated and
thus largely cancel. A similar relation between $\sin^{2}\Theta_{\rm
lep}(\mz)$ and the weak mixing angle appearing in polarised M{\o}ller
scattering asymmetries has been worked out ~\cite{Czarnecki:1995fw,Czarnecki:2000ic}.
It includes specific bosonic contribution $\Delta \kappa_\mathrm{b}(Q^2)$, such that
\ba
\kappa(s=-Q^2)=\frac{1-\Delta \alpha_2(s)}{1-\Delta \alpha(s)}+\Delta \kappa_\mathrm{b}(Q^2)-\Delta \kappa_\mathrm{b}(0) \, ,
\ea
where, in our low-energy scheme, we require $\kappa(Q^2)=1$ at $Q^2=0$.
Explicitly~\cite{Czarnecki:1995fw,Czarnecki:2000ic}, at one-loop order
\begin{align}
\Delta
\kappa_\mathrm{b}(Q^2)&=-\frac{\alpha}{2\uppi\,s_\mathrm{W}}\,\biggl\{-\frac{42\,c_\mathrm{W}+1}{12}\,\ln c_\mathrm{W}
+\frac{1}{18}-\left(\frac{r}{2}\,\ln
\xi-1\right)\,\biggl[(7-4z)\,c_\mathrm{W}  
 +\frac16\,(1+4z)\biggr] \nn \\
 & \qquad  \qquad  \qquad  \qquad -z\,\biggl[\frac34-z+\left(z-\frac23\right)\,r\,\ln
\xi+z\,(2-z)\,\ln^2\xi\biggr]\biggr\}\comas \\
\Delta
\kappa_\mathrm{b}(0)&= -\frac{\alpha}{2\uppi\,s_\mathrm{W}}\,\left \{-\frac{42\,c_\mathrm{W}+1}{12}\,\ln c_\mathrm{W} +\frac{1}{18}
+\frac{6\,c_\mathrm{W}+7}{18}\right \}\,,
\end{align}
with $z=M_\mathrm{W}^2/Q^2$, $r=\sqrt{1+4z}$, $\xi= {(r+1)} / {(r-1)}$, $s_\mathrm{W}=
\sin^2\Theta_\mathrm{W}$, and $c_\mathrm{W}=\cos^2\Theta_\mathrm{W}$. Results
obtained in Refs.~\cite{Czarnecki:1995fw,Czarnecki:2000ic} based on one-loop perturbation
theory using light quark masses $m_\mathrm{u} =m_\mathrm{d} =m_\mathrm{s}
=100\,\mv$ are compared with
results obtained in our non-perturbative approach
in Fig.~\ref{fig:sin2theta}.

How can we evaluate the leading non-perturbative hadronic corrections 
to $\alpha_2$? As in the case of $\alpha$, they are related to
quark-loop contributions to gauge-boson self-energies (SE) 
\mbo{\upgamma \upgamma}, \mbo{\upgamma}Z, {ZZ}, and {WW}, in
particular, those involving the photon, which exhibit large leading
logarithms. To disentangle the leading corrections, decompose
the self-energy functions as follows ($s^2_\Theta=e^2/g^2\semis c^2_\Theta=1-s^2_\Theta$):
\bea
\bary{lcclcclccl}
\Pi^{\upgamma \upgamma} &=& e^2&\hat{\Pi}^{\upgamma \upgamma}\,, &&&&&& \\[2mm]
\Pi^{\mathrm{Z} \upgamma} &=& \frac{eg}{c_\Theta}&\hat{\Pi}^{3 \upgamma}_V
&-&\frac{e^2\,s_\Theta}{c_\Theta}&\hat{\Pi}^{\upgamma \upgamma}_V\,, &&&\\[2mm]
\Pi^{\mathrm{Z} \mathrm{Z}} &=& \frac{g^2}{c^2_\Theta}&\hat{\Pi}^{33}_{V-A}
&-&2\,\frac{e^2}{c^2_\Theta}&\hat{\Pi}^{3 \upgamma}_V
 &+&\frac{e^2\,s^2_\Theta}{c^2_\Theta}&\hat{\Pi}^{\upgamma \upgamma}_V\,,\\[2mm]
\Pi^{\mathrm{W}\mathrm{W}} &=& g^2 & \hat{\Pi}^{+-}_{V-A}\,. &&&&&&
\eary
\eea

\begin{figure}
\centering
\includegraphics[width=0.47\textwidth]{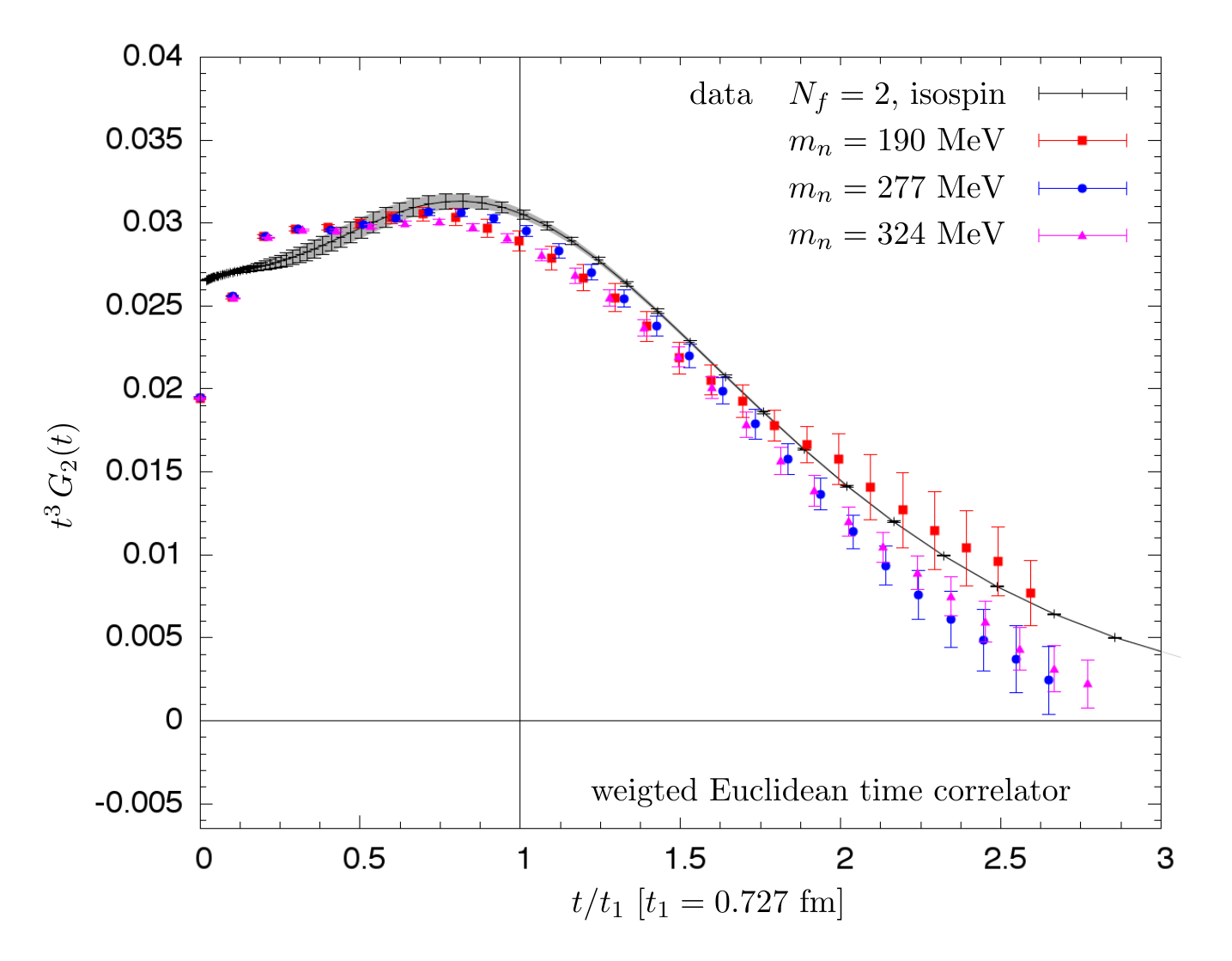}
\includegraphics[width=0.47\textwidth]{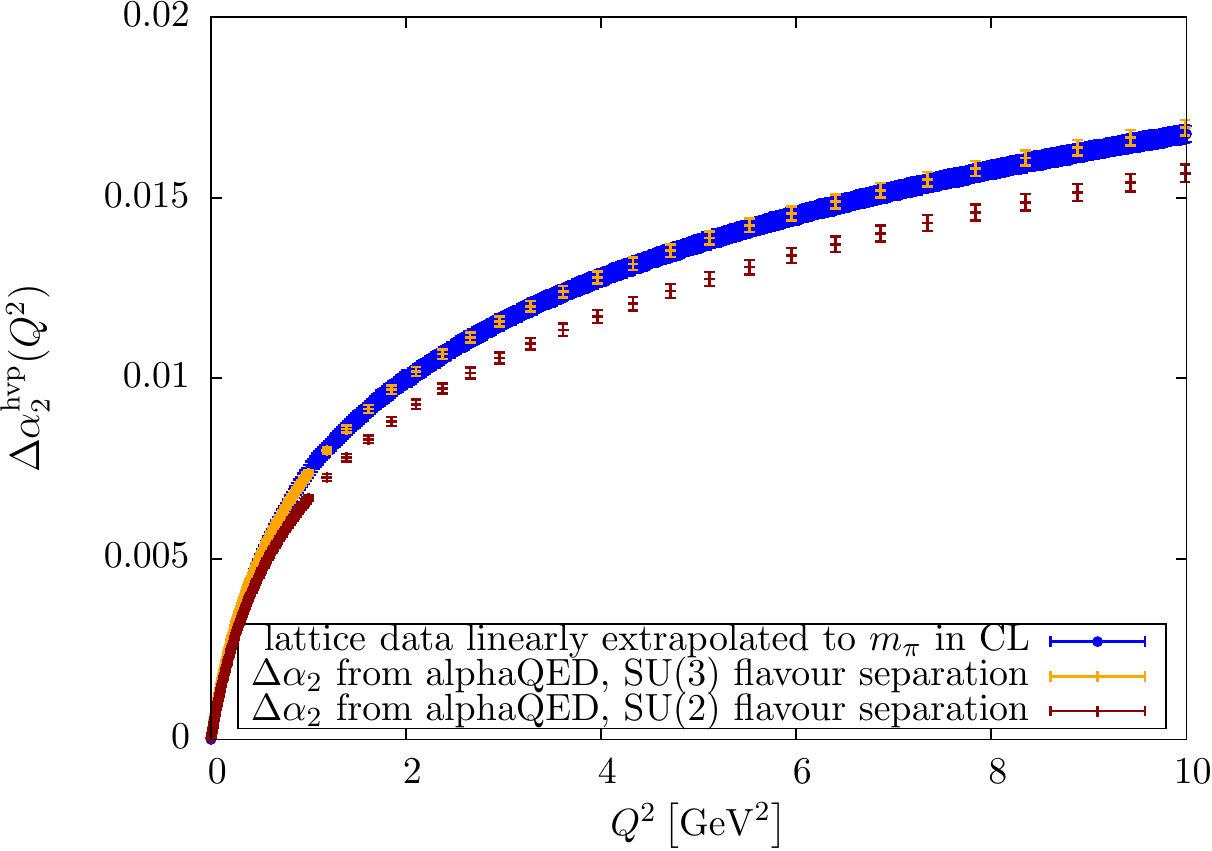}
\caption{Testing flavour separation in lattice QCD. Left: a rough test
by checking the Euclidean time correlators clearly favours the flavour
separation of
\Eref{flasep}~\cite{Bernecker:2011gh,Francis:2013jfa,Ce:2018ziv},
while the pQCD reweighting (not displayed) badly fails. Right: the
renormalised photon self-energy at Euclidean $Q^2$~\cite{Burger:2015lqa}
is in good agreement with the flavour $SU(3)$
limit, while again it fails with the $SU(2)$ case, which coincides with
perturbative reweighting.}
\label{fig:flavorseptest}
%\query{Please correct the figure labels in \Fref{fig:flavorseptest}. Set variables
%in italic font. Set
%labels in roman font.}
\end{figure}

With \mbo{\hat{\Pi}(s)=\hat{\Pi}(0)+s\,{\Pi'}(s)}, we find the
leading hadronic corrections
\begin{align}
\dahs &= -e^2\, \left[\Repa {\Pi'}^{\upgamma \upgamma}(s)
-{\Pi'}^{\upgamma \upgamma}(0)\right]\,,\\
\dghs &= -\frac{e^2}{s^2_\Theta}\, \left[\Repa {\Pi'}^{3 \upgamma}(s)
-{\Pi'}^{3 \upgamma}(0)\right]\,,
\end{align}
which exhibit the leading hadronic non-perturbative parts, \ie the
ones involving the photon field via mixing. Besides \mbo{\dahs},
\mbo{\dghs} can also then be obtained in terms of \mbo{\epm} data, together with
isospin flavour separation of (u, d) and s components
\bea
\Pi^{3\upgamma}_\mathrm{ud}=\frac{1}{2}\, \Pi^{\upgamma \upgamma}_\mathrm{ud}\semis
\quad \Pi^{3\upgamma}_\mathrm{s}=\frac{3}{4}\, \Pi^{\upgamma \upgamma}_\mathrm{s} 
\eea
and for resonance contributions
\begin{align}
\Pi^{\upgamma\upgamma} & =\Pi^{(\rho)}+\Pi^{(\omega)}+\Pi^{(\phi)}+
\cdots \nn \\
\Pi^{3\upgamma} &=\frac12\,\Pi^{(\rho)}+\frac34\,\Pi^{(\phi)}+ \cdots
\label{flasep}
\end{align}
We are reminded that gauge-boson self-energies are potentially very sensitive to new
physics (oblique corrections) and the discovery of what is missing in
the SM may be obscured by non-perturbative hadronic effects. Therefore,
it is important to reduce the related uncertainties.  Interestingly,
flavour separation assuming OZI violating terms to be small implies a
perturbative reweighting, which, however, has been shown to disagree
with lattice QCD results
\cite{Bernecker:2011gh,Francis:2013jfa,Ce:2018ziv,Burger:2015lqa}!
Indeed, the `wrong' perturbative flavour weighting
$$
\Pi^{3\upgamma}_\mathrm{ud}=\frac{9}{20}\, \Pi^{\upgamma \upgamma}_\mathrm{ud}\semis
\quad 
\Pi^{3\upgamma}_\mathrm{s}=\frac{3}{4}\, \Pi^{\upgamma \upgamma}_\mathrm{s}
$$ 
clearly mismatch lattice results, while the replacement
${9} / {20}\Rightarrow  {10} / {20}$ is in good agreement. This
also means that the OZI suppressed contributions should be at the 5\% level
and not negligibly small. Actually, if we assume flavour $SU(3)$
symmetry to be an acceptable approximation, we obtain
$$
\Pi^{3\upgamma}_\mathrm{uds}=\frac{1}{2}\, \Pi^{\upgamma \upgamma}_\mathrm{uds}\,,
$$
which does not require any flavour separation in the uds-sector, \ie
up to the charm threshold at about $3.1\,\gv$. 
Figure \ref{fig:flavorseptest} shows a lattice QCD test of two flavour
separation schemes. One, labelled `$SU(2)$', denotes the perturbative
reweighting advocated in Refs. \cite{Czarnecki:1995fw,Czarnecki:2000ic,Erler:2004in,Erler:2017knj} and the
other, labelled `$SU(3)$', represents that proposed
in Ref.~\cite{Jegerlehner:1985gq}. Lattice data clearly disprove pQCD reweighting for the uds-sector! 
This also shows that pQCD-type predictions based on effective quark
masses cannot be accurate. This criticism also applies in cases where
the effective quark masses have been obtained by fitting $\dahs$, even
more so, when constituent quark masses are used.

\begin{figure}
\centering
\includegraphics[width=0.5\textwidth]{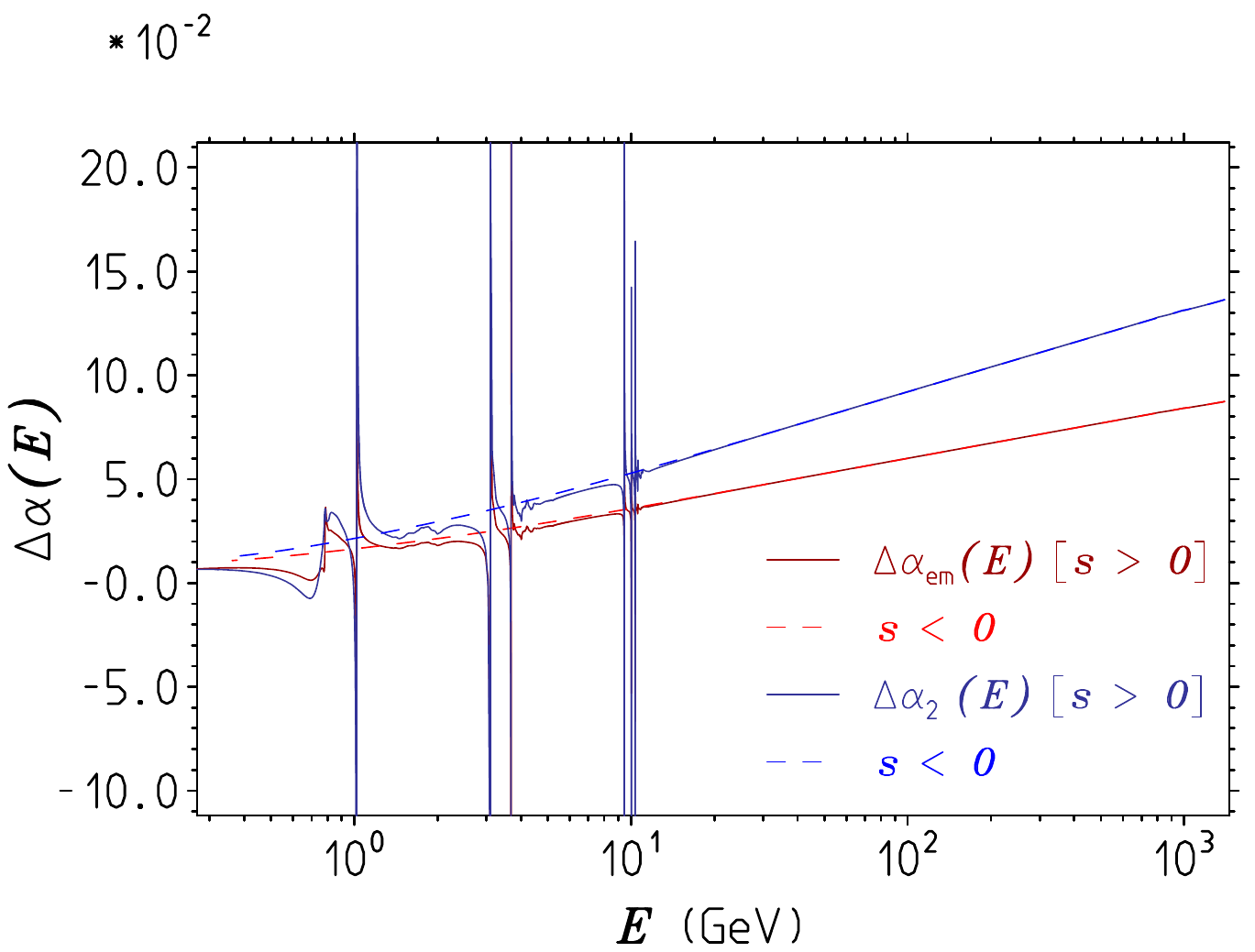}
\caption{$ \Delta \alpha_{\rm QED}(E)$ and $ \Delta \alpha_{2}(E)$ as functions of energy
$E$ in the time-like and space-like domain.
The smooth space-like correction (dashed line) agrees rather well
with the non-resonant `background' above the $ \phi$ resonance (a kind
of duality). In resonance regions, as expected, `agreement' is
observed in the mean, with huge local deviations.}
%\label{fig:Tom}
%\query{All figures must be cited in the text. Please cite %\Fref{fig:Tom}
%in the text.}
% \query{Please correct the figure labels in \Fref{fig:Tom}. Set variables
%in italic font. Set
%labels in roman font.}
%\end{figure}
\label{fig:dalphaem}
\end{figure}

The effective $SU(2)$ coupling $\Delta \alpha_{2}(E)$ in comparison with $ \Delta \alpha_{\rm QED}(E)$
is displayed in Fig.~\ref{fig:dalphaem}, and
the updated $\sin^2 \Theta_W(s)$ is shown in Fig.~\ref{fig:sin2theta} for time-like as
well as for space-like momentum transfer.
Note that \mbo{\sin^2 \Theta_\mathrm{W}(0)/\sin^2 \Theta_\mathrm{W}(M_\mathrm{Z}^2)=1.028\,76};
 a
3\% correction is established at $ 6.5\sigma$.
Except for the LEP and SLD points (which deviate by {$1.8 \sigma$}), all existing measurements are of rather
limited accuracy, unfortunately. Upcoming experiments will substantially improve
results at low space-like $Q$.
We are reminded that $ \sin^{2}\Theta_{\ell\,{\rm eff}}$,  exhibiting a specific dependence on the
gauge-boson self-energies, is an excellent monitor for new physics. At
pre-LHC times, it was  the predestined monitor for virtual Higgs
particle effects and a corresponding limiter for the Higgs boson mass.

%%%%%%%%%%%%%%%%%%%%%%%%%%%%%%%%%%%%%%%%%%%%%%%%%
%\input{BSM_your1/bsm_your1.tex} \label{sec-bsm-your1}
%\clearpage \pagestyle{empty}  \cleardoublepage

%################################################
%\clearpage
%%%%%%%%%%%%%%%%%%%%%%%%%%%%%%%%%%%%%%%%%%%%%%%%%

%\noindent
\subsection*{Acknowledgements}
%Acknowledgments \\ 
I would like to thank Janusz Gluza and the
organising committee for the invitation to this workshop and for support.
Thanks also to Maurice Benayoun, Simon Eidelman, and Graziano Venanzoni
for collaboration and many useful discussions on the topics presented
here.

\end{bibunit}

\label{sec-sm-fjeg}  
\clearpage \pagestyle{empty}  \clearpage
%============================================
%\setcounter{figure}{0} %\renewcommand{\thefigure}{\arabic{figure}} 
%\setcounter{table}{0} %\renewcommand{\thefigure}{\arabic{table}} 

\pagestyle{fancy}
\fancyhead[CO]{\thechapter.\thesection \hspace{1mm} Precision quantum chromodynamics}
\fancyhead[RO]{}
\fancyhead[LO]{}
\fancyhead[LE]{}
\fancyhead[CE]{}
\fancyhead[RE]{}
\fancyhead[CE]{D. d'Enterria}
\lfoot[]{}
\cfoot{-  \thepage \hspace*{0.075mm} -}
\rfoot[]{}

%=================================================================
 
\begin{bibunit}[elsarticle-num] % define the bib-style for the unit: elsarticle-num.bst
%  text-1; this is the corresponding section
%\putbib[2loops] % the *.bib
%\end{bibunit}
% go-on
%--- from: bibunits.sty, adapts the font size of ``References'' to section
\begin{flushleft}
\let\stdthebibliography\thebibliography
\renewcommand{\thebibliography}{%
\let\section\subsection
\stdthebibliography}
\end{flushleft}
\newcommand{\bbbar}{\rm {b\bar{b}}}
\newcommand{\ccbar}{\rm {c\bar{c}}}
\newcommand{\qqbar}{\rm {q\bar{q}}}
\newcommand{\qbar}{\rm \bar{q}}
\newcommand{\alphaqed}{\alpha_{_{\rm QED}}}
\newcommand{\pt}{p_{_\perp}}
\newcommand{\alphasmZ}{\alphas(m_{_{\rm Z}})}

\newcommand{\AFBb}{A_{_{\textsc{fb}}}^{0,\mathrm{b}}}
\newcommand{\AFBbb}{A_{_{\textsc{fb}}}^{b}}
\newcommand{\AFBobs}{(A_{_{\textsc{fb}}}^{b})_{_{\rm obs}}}
\newcommand{\weakang}{\sin^2\theta_{\rm W}}
\newcommand{\weakeff}{\sin^2\theta_{_{\rm eff}}^\mathrm{f}}
\newcommand{\weakeffb}{\sin^2\theta_{_{\rm eff}}^b}

\providecommand{\GWhexp}{\Gamma^{\rm W}_{\rm had,exp}}
\providecommand{\GWhBisexp}{\Gamma^{\rm W}_{\rm had',exp}}
\providecommand{\GWl}{\Gamma^{\rm W}_{\rm lep}}
\providecommand{\GWhl}{\Gamma^{\rm W}_{\rm had,lep}}
\providecommand{\GWlexp}{\Gamma^{\rm W}_{\rm lep,exp}}
\providecommand{\GWtot}{\Gamma^{\rm W}_{\rm tot}}
\providecommand{\GWtotexp}{\Gamma^{\rm W}_{\rm tot,exp}}
\providecommand{\GWtotewk}{\Gamma^{\rm W}_{\rm tot,fit}}
\providecommand{\BRWh}{\rm {\cal B}^{\rm W}_{\rm had}}
\providecommand{\BRWl}{\rm {\cal B}^{\rm W}_{\rm lep}}
\providecommand{\BRWhl}{\rm {\cal B}^{\rm W}_{\rm had,lep}}
\providecommand{\BRWhexp}{\rm {\cal B}^{\rm W}_{\rm had,exp}}
\providecommand{\BRWhBisexp}{\rm {\cal B}^{\rm W}_{\rm had',exp}}
\providecommand{\BRWlexp}{\rm {\cal B}^{\rm W}_{\rm lep,exp}}
\providecommand{\RW}{\rm R_\mrm{W}}
\providecommand{\RWexp}{\rm R_\mrm{W,exp}}

\newcommand{\Rlz}   {R^0_\ell}
\newcommand {\so}   {\rm \sigma_0^{had}}
\newcommand{\Ghad}  {\rm\Gamma_{\mathrm{had}}}

\section[Precision quantum chromodynamics \\ {\it D. d'Enterria}]
{Precision quantum chromodynamics
\label{contr:QCD}}
\noindent
{\bf Contribution\footnote{This contribution should be cited as:\\
D. d'Enterria, Precision quantum chromodynamics,  
%04 DOI:10.23731/CYRM-2020-XXX.\thepage, in:
%04 \url{http://dx.doi.org/10.23731/CYRM-2020-XXX.\thepage}, in:
DOI: \href{http://dx.doi.org/10.23731/CYRM-2020-003.\thepage}{10.23731/CYRM-2020-003.\thepage}, in:
Theory for the FCC-ee, Eds. A. Blondel, J. Gluza, S. Jadach, P. Janot and T. Riemann,
\\CERN Yellow Reports: Monographs, CERN-2020-003,
%04 \url{http://dx.doi.org/10.23731/CYRM-2020-003}, p. \thepage.} 
DOI: \href{http://dx.doi.org/10.23731/CYRM-2020-003}{10.23731/CYRM-2020-003},
p. \thepage.
\\ \copyright\space CERN, 2020. Published by CERN under the 
%04-2
\href{http://creativecommons.org/licenses/by/4.0/}{Creative Commons Attribution 4.0 license}.} by: D. d'Enterria 
%\\ Corresponding Author: David d'Enterria 
{[dde@cern.ch]}}
\vspace*{.5cm}

\noindent The unprecedentedly small experimental uncertainties expected in the electron--positron measurements at the FCC-ee, key to searches for physics beyond the SM up to $\Lambda\approx 50$\,TeV, 
%through ``stress tests'' of the SM, 
impose precise calculations for the corresponding theoretical observables. At the level of theoretical precision required to match that of the FCC-ee experimental measurements, the current relevant QCD uncertainties have to be reduced at at least  four different levels.
\begin{enumerate}
    \item Purely theoretical perturbative uncertainties from missing higher-order (HO) corrections in perturbative QCD (pQCD) calculations of $\epem$ scattering amplitudes and decay processes involving multiple real emissions or virtual exchanges of quarks and gluons. 
    Such fixed-order (FO) corrections include pure QCD and mixed QCD--QED or QCD--weak terms.
    %Such corrections can be fixed-order (FO), including pure-QCD and mixed QCD-QED and QCD-EWK terms and soft and/or collinear logarithmic resummations. The latter can be analytical or properly incorporated into matched parton shower Monte Carlo generators. 
    Reducing such uncertainties requires pQCD calculations beyond the current
state
of the art, often given  by next-to-next-to-leading-order (NNLO) accuracy, pure, or mixed with higher-order electroweak terms. %$\mathcal{O}(\alphaqed\alphas^2)$, $\mathcal{O}(N_{f}\alphaqed^2\alphas)$, $\mathcal{O}(N_{f}\alphas^3)$.
    %, and next-to-next-to-leading-log (NNLL) accuracies.
    \item Theoretical uncertainties due to incomplete logarithmic resummations of different energy scales potentially appearing in the theoretical calculations. Examples include resummations of (i) soft and collinear logs in final states dominated by jets---either in analytical calculations or (only partially) incorporated into matched parton shower Monte Carlo generators---and (ii) logarithmic terms in the velocity of the produced top quarks in $\epem\to\ttbar$ cross-sections. Reducing such uncertainties requires calculations beyond the current state of the art, often given  by the next-to-next-to-leading-log (NNLL) accuracy.
    \item Parametric uncertainties propagated into the final theoretical result, owing to the dependence of the calculation on the input values of (i) the QCD coupling at the Z pole scale, $\alphasmZ$, known today with a relatively poor $\pm0.9\%$ precision, and (ii) the heavy quark (charm and bottom) masses $m_\mathrm{c}$ and $m_\mathrm{b}$. Theoretical progress in lattice QCD determinations of $\alphas$ and $m_\mathrm{c,b}$ is needed, complemented with much more precise experimental measurements. A per-mille extraction of $\alphasmZ$ is thereby also a key axis of the FCC-ee physics programme~\cite{dEnterria:2015kmd}.
    \item Non-perturbative uncertainties from final-state hadronic effects linked to power-suppressed infrared phenomena, such as colour reconnection, hadronization, and multiparticle correlations (in spin, colour, space, momenta), that cannot be currently computed from first-principles QCD theory, and that often rely on phenomenological Monte Carlo models. The high-precision study of parton hadronization and other non-pQCD phenomena is also an intrinsic part of the FCC-ee physics programme~\cite{Anderle:2017qwx}.
\end{enumerate}

Examples of key observables where such four sources of QCD uncertainty will have an impact at the FCC-ee are numerous.
\begin{enumerate}
\item Uncertainties from missing HO terms are non-negligible in theoretical predictions for electroweak precision observables (EWPOs) at the Z pole, WW and $\ttbar$ cross-sections, (N)MSSM Higgs cross-sections and decays, etc. 
\item Uncertainties from missing soft and collinear log resummations, in analytical calculations or in parton shower MC generators, impact all $\epem$ final states with jets---\eg\ the accurate extraction of forward--backward quark asymmetries at the Z pole---as well as precision flavour physics studies via B meson decays. Similarly, the size of the NNLL corrections (in the $\ln v$ top quark velocity) appears to be as large as that from the FO N$^3$LO terms in $\epem\to\ttbar$ cross-section calculations.
\item The $\alphasmZ$ parametric uncertainty has a significant effect on the determination of all top properties ($m_{\rm top}$, $\lambda_{\rm top}$ , $\Gamma_{\rm top}$), all hadronic Higgs decay widths ($\rm H \to \ccbar, \bbbar, \qqbar, g\,g$) and associated Yukawa couplings, as well as on the extraction of other similarly crucial SM parameters ($m_\mathrm{c}, m_\mathrm{b}, \alphaqed$).
\item Non-perturbative uncertainties, in particular colour reconnection and hadronization effects, impact hadronic final states in $\epem \to WW$ and $\epem \to \ttbar$, and forward--backward angular asymmetries of quarks at the Z pole.
\end{enumerate}

In the following sections, the current status and FCC-ee prospects for these four axes of QCD studies are summarised.

%=================================================================
\subsection[Higher fixed-order pQCD corrections]{Higher fixed-order pQCD corrections}

Computations of pQCD corrections beyond the N$^{2,3}$LO accuracy are required for many theoretical FCC-ee observables, in order to match their expected experimental precision. New analytical, algorithmic, and numerical concepts and tools are needed to be able to compute HO QCD and mixed QCD+electroweak  multiloop, -legs, and -scales corrections for processes involving the heaviest SM particles (W, Z, H, t) to be carefully scrutinised at the FCC-ee. Concrete developments are covered in more detail in various other sections of this report, and are summarised here.
\begin{enumerate}
\item %\item[$\bullet$]
EWPOs: Mixed QCD-electroweak calculations of the $\mathrm{Zf{\bar f}}$ vertex will be needed at the FCC-ee at higher order than known today, including the ${\cal{O}}(\alpha \alpha_\mathrm{s}^2), {\cal{O}}(N_\mathrm{f} \alpha_{}^2 \alpha_\mathrm{s}), {\cal{O}}(N_\mathrm{f}^2 \alpha_{}^3)$ loop orders, where $N_\mathrm{f}^n$ denotes $n$ or more closed internal fermion loops, plus the corresponding QCD four-loop terms~\cite{Blondel:2018mad}. The number of QCD diagrams for $\rm Z \to \bbbar$ decays at two (three) loops is 98 (10\,386)~\cite{Blondel:2018mad}. Section~\ref{contr:jgracey} provides, \eg\ details on the extension of calculations beyond the two-loop QCD off-shell vertex functions, noting that for the triple-gluon vertex there are 2382  (63\,992) three- (four-) loop graphs to evaluate. 
%For both the other 3-point vertices, the numbers of graphs in each case are the %same and are 688 and 17\,311 respectively at three and four loops. 
Including massive quarks in three- and four-point functions is a further requirement in order to reduce the FO theoretical uncertainties.
%\item[$\bullet$] 
\item W bosons (Section~\ref{contr:schwinn}): The resonant $\epem \to \PW\PW$ cross-section contains soft corrections to the Coulomb function, analogous to ultrasoft  ($m_{\rm top}v^2$) QCD corrections in $\ttbar$ production~\cite{Beneke:2008cr}. For the W hadronic decay modes, QCD corrections to the partial decay widths have to be included beyond NNLO to match the corresponding theoretical QED precision given by the counting $\mathcal{O}(\alphas^2) \sim \mathcal{O}(\alphaqed)$. QCD corrections to $\PW$ self-energies and decay widths up to $ {\cal{O}}(\alphaqed\,\alphas^2)$ and ${\cal{O}}(\alphas^4)$ are required. Currently, $\mathcal{O}(\alphas^4)$ corrections for inclusive hadronic vector boson decays are known~\cite{Baikov:2008jh}, while mixed QCD-EW corrections are known up to $\mathcal{O}(\alphaqed\,\alpha_\mathrm{s})$~\cite{Kara:2013dua}.
%\item[$\bullet$]
\item Higgs bosons (Section~\ref{contr:mspira}): The pure QCD corrections to Higgs boson decays into quarks, gluons, and photons are known up to N$^4$LO (no mass effects), N$^3$LO (heavy top limit), and NLO, respectively. Those translate into approximately 0.2\%, 1\%, and <3\% scale uncertainties from missing HO corrections. In the case of the (N)MSSM Higgs sector (Section~\ref{contr:HEINEMEYER}), HO pQCD corrections to the Higgs bosons decays are mostly known at NLO accuracy; thereby, their uncertainty is larger than for the SM Higgs case.
%\item[$\bullet$] 
\item Top quarks (Section~\ref{contr:ttbar_prod}): The total cross-section for inclusive $\epem\to\bbbar\rm W^+W^-X$ production can be computed in a non-relativistic effective field theory with local effective vertices and matching corrections known up to N$^3$LO in pQCD~\cite{Beneke:2015kwa}. Those translate into about 3\% theoretical scale uncertainties of the threshold $\ttbar$ cross-sections that propagate into an uncertainty of $\pm$60\,MeV in the position of the resonant peak. Although the uncertainty has been reduced by a factor of two going from NNLO to N$^3$LO, perturbative progress is still needed, in particular in the threshold top mass definition translated into the $\overline{\rm MS}$ scheme.
%\item[$\bullet$] 
\item The extraction of $\alphaqed$ from the $R$ ratio  requires the calculation of the four-loop massive pQCD calculation of the Adler function (together with better estimates of $\alphas$ in the low-$Q^2$ region above the $\uptau$ mass, as well as of the $m_\mathrm{c}$ and $m_\mathrm{b}$ masses).
\end{enumerate}

%=================================================================
\subsection[Higher-order logarithmic resummations]{Higher-order logarithmic resummations}

Improvements in the resummations of all-order logarithmic terms from different energy scales, appearing in the theoretical calculations for certain processes, are needed in various directions.
\begin{enumerate}
%\item[$\bullet$] 
\item Soft and collinear parton radiation impacts many $\epem$ observables with jets in the final state. Such uncertainties enter through incomplete NNLL resummations in analytical calculations (\eg\ based on soft-collinear effective theory, SCET), or through approximate models of the coherent branching implemented in the parton shower MC generators used to unfold and interpret the experimental data. Among those experimental observables, the measured forward--backward (FB) angular asymmetries of charm and bottom quarks in $\epem$ collisions around the Z pole, directly connected to the weak mixing angle, will need a careful study. The asymmetry value measured at LEP, $(\AFBb)_{_{\rm exp}} = 0.0992\pm0.0016$, remains today the electroweak precision observable with the largest disagreement (2.9$\sigma$) with respect to the SM prediction, $(\AFBb)_{_{\rm th}} = 0.1038$~\cite{ALEPH:2005ab,ALEPH:2010aa}. Consequently, so also does  the effective weak mixing angle derived from it, $\weakeff = 0.232\,21\pm0.000\,29$, compared with the $\weakeff = 0.231\,54\pm0.000\,03$ world-average \cite{PDG}. The dominant systematic uncertainties on $(\AFBb)_{_{\rm exp}}$ arise from angular decorrelations induced in the thrust axis by soft and collinear parton radiation or parton-to-hadron b quark hadronization, and were estimated using MC simulations 20 years ago \cite{Abbaneo:1998xt}. A recent reanalysis of the QCD corrections to $\AFBb$ \cite{dEnterria:2018jsx}, with different modern parton shower models~\cite{Sjostrand:2014zea,Giele:2011cb,Fischer:2016vfv}, indicates propagated uncertainties of about 1\% (0.4\%) for the lepton (jet) charge-based measurements, slightly smaller but still consistent with the original measurements derived at LEP. The measurement of $\AFBb$ at the FCC-ee will feature insignificant statistical uncertainties, and improvements in the modelling of parton radiation will be required for any high-precision extraction of the associated $\weakeff$ value.

%\item[$\bullet$] 
\item Another field of $\epem$ measurements where progress in logarithmic resummations is needed is in the studies of event shapes---such as the thrust $T$, $C$ parameter, and jet broadening. All those observables are commonly used to extract the QCD coupling~\cite{dEnterria:2015kmd}. Theoretical studies of event shapes supplement FO perturbation theory with the resummation of enhanced logarithmic contributions, specifically accounting for terms ranging from $\alphas^n \ln^{n+1}$ down to $\alphas^n \ln^{n-2}$, \ie\ N$^3$LL~\cite{Salam:2017qdl}. However, the $\alphasmZ$ values derived from the $T$ and $C$ measurements differ and their combination has thereby a final 2.9\% systematic uncertainty~\cite{PDG}. This result points to limits in the resummation formalism that (i) hold only for $C, 1-T \ll 1$, where every emission is so soft and collinear that one can effectively neglect the kinematic cross-talk (\eg\ energy--momentum conservation) that arises when there are a number of emissions, and (ii) use a power correction valid only in the two-jet limit, $1-T\ll 1$~\cite{Salam:2017qdl}.

%\item[$\bullet$] 
\item High-precision studies of $n$-jet rates at the FCC-ee will also benefit from a reduction of resummation uncertainties. Jet rates in $\epem$ rely on an algorithm to reconstruct them that comes with a parameter ($y_\text{cut}=k_{\rm T}^2/s$, in the $k_{\rm T}$ Durham~\cite{Catani:1991hj} and Cambridge~\cite{Dokshitzer:1997in} cases) to define how energetic the emission should be in order to be considered a jet. For $\ln y_\text{cut} > -4$, the extracted $\alphas$ value from three-jet rates is fairly independent of $y_\text{cut}$, whereas the result depends substantially on the choice of $y_\text{cut}$ below that~\cite{Dissertori:2009qa}. This feature points to a breakdown of FO perturbation theory, owing to logarithmically enhanced $(\alphas \ln^2 y_\text{cut})^n$ terms. Jet rates at the one-in-a-million level in $\epem$ at the Z pole will be available at the FCC-ee, including: four-jet events up to $k_{\rm T} \approx$\,30\,GeV (corresponding to $|\ln{y_\text{cut}}| \approx 2$), five-jet events at $k_{\rm T} \approx$\,20\,GeV ($|\ln{y_\text{cut}}| \approx 3$), six-jet events at $k_{\rm T} \approx$\,12\,GeV ($|\ln{y_\text{cut}}| \approx 4$), and seven-jet events at $k_{\rm T} \approx$\,7.5\,GeV ($|\ln{y_\text{cut}}| \approx 5$). Such results will be compared with theoretical calculations with an accuracy beyond the NNLO\,+\,NNLL provided today by the {\sc eerad3}~\cite{Ridder:2014wza}, {\sc mercutio}~2~\cite{Weinzierl:2010cw}, and CoLoRFulNNLO~\cite{DelDuca:2016csb} (NNLO), and {\sc ARES}~\cite{Banfi:2014sua} (NNLL) codes, thereby leading to $\alphas$ extractions with uncertainties well below the current few-percent level. In general, with the envisioned FCC-ee luminosities, jet measurements will extend along the six axes of higher accuracy, finer binning, higher jet resolution scales, larger numbers of resolved final-state objects, more differential distributions, and possibility  placing stringent additional cuts to isolate specific interesting regions of the $n$-jet phase spaces not strongly constrained by LEP measurements~\cite{Fischer:2015pqa}.

%\item[$\bullet$]
\item In top physics studies, the size of the NNLL corrections (in top quark velocity, $\ln v$) in $\epem\to\ttbar$ cross-section calculations appears to be as large as that from the FO N$^3$LO terms~\cite{Beneke:2015kwa}, calling for improved resummation studies for such an observable.

%\item[$\bullet$]
\item In the sector of flavour physics (Section~\ref{contr:szafron}), new tools based on SCET, developed to study processes with energetic quarks and gluons, can be applied after certain modifications to improve the accuracy of theoretical corrections for B-physics studies at the FCC-ee, in particular for regions of phase space where the perturbative approach breaks down, owing to the presence of large logarithmic enhancements, and where the next-to-soft effects become more important.
\end{enumerate}

%=================================================================
\subsection[Per-mille-precision $\alphas$ extraction]{\textbf{Per-mille-precision $\alphas$ extraction}}

The strong coupling, $\alphas$, is one of the fundamental parameters of the Standard Model, and its value not only directly affects the stability of the electroweak vacuum~\cite{Buttazzo:2013uya} but it chiefly impacts all theoretical calculations of $\epem$ scattering and decay processes involving real or virtual quarks and gluons~\cite{dEnterria:2015kmd}. Known today with a 0.9\% precision, making  it the worst known of all fundamental interaction couplings in nature~\cite{PDG}, the input value of $\alphasmZ$ propagates as a parametric uncertainty into many of the FCC-ee physics observables, chiefly in the Z, Higgs, and top quark sectors.
\begin{enumerate}
%\item[$\bullet$] 
\item The leading source of uncertainty in the calculation of crucial EWPOs  at the Z pole, such as $\Gamma_{\rm Z}$, $\sigma^{0}_{\rm had}$, and $R_{\ell}$, is the propagated $\updelta\alphas$ parametric source~\cite{Blondel:2018mad}.
%\item[$\bullet$] 
\item In the Higgs sector (Section~\ref{contr:mspira}), the current $\alphasmZ$ parametric uncertainty (combined with uncertainties arising from our imperfect knowledge of $m_\mathrm{c}$ and $m_\mathrm{b}$) propagates into total final uncertainties of $\sim$2\% for the BR(H\,$\to$WW, ZZ) and BR(H\,$\to \uptau^+\uptau^-,\upmu^+\upmu^-$) branching ratios, of $\sim$6--7\% for BR(H\,$\to$ gg) and BR(H\,$\to \mathrm{c\bar c})$, $\sim$3\% for BR(H\,$\to\upgamma\upgamma$), and $\sim$7\% for BR(H\,$\to \mathrm{Z}\upgamma$). %The total decay width of $\sim 4.1$ MeV can be predicted with $\sim 2\%$ total uncertainty.
%\item[$\bullet$] 
\item Precise studies of the $\epem\to\ttbar$ cross-section (Section~\ref{contr:ttbar_prod}) indicate that it should be possible to extract the top quark width and mass with an uncertainty of around 50\,MeV, provided that a precise independent extraction of the strong coupling is available. Such a requirement is, in particular, crucial to meaningfully constrain the top Yukawa coupling.
\end{enumerate}

 The current world-average value, $\alphasmZ = 0.1181~\pm~0.0011$, is derived from a combination of six subclasses of approximately independent observables~\cite{PDG} measured in $\epem$ (hadronic Z boson and $\uptau$ decays, and event shapes and jet rates), DIS (structure functions and global fits of parton distributions functions), and p--p collisions (top pair cross-sections), as well as from lattice QCD computations constrained by the empirical values of hadron masses and decay constants. To enter into the $\alphasmZ$ world-average, the experimental (or lattice) results need to have a counterpart pQCD theoretical prediction at NNLO (or beyond) accuracy.

Of the current six $\alphasmZ$ extractions entering in the PDG average, that
 derived from comparisons of NNLO pQCD predictions with lattice QCD results (Wilson loops, $\qqbar$ potentials, hadronic vacuum polarisation, QCD static energy)~\cite{Aoki:2019cca} today provides  the most precise result: $\alphasmZ = 0.1188 \pm 0.0011$. The current $\sim$0.9\% uncertainty is dominated by finite lattice spacing, truncations of the pQCD expansion up to NNLO, and hadron extrapolations. Over the next 10 years, reduction of the statistical uncertainties, at least by a factor of two, can be anticipated with increased computing power, while reaching the $\sim$0.1\% uncertainty level will also require the computation of fourth-order pQCD corrections~\cite{dEnterria:2015kmd}.

After the lattice result, the most theoretically and experimentally `clean' extractions of $\alphas$ are those based on the hadronic decays of the $\uptau$ lepton, and W and Z bosons that will be measured with unparalleled accuracies at the FCC-ee. To derive $\alphasmZ$, the experimental ratios of hadronic-to-leptonic decays are compared with the corresponding pQCD theoretical prediction, known today up to $\order{\alphas^4}$~\cite{Baikov:2008jh,Baikov:2012zn}:
\begin{align}
  \label{eq:R_alphas}
  R^{\rm \uptau,W,Z}_{\ell}(Q=m_\uptau,m_{\rm W},m_{\rm Z})  & =  \frac{\sigma(\rm \epem \to (\uptau,W,Z) \to {\rm hadrons})}{\sigma(\rm \epem \to (\uptau,W,Z) \to \ell^+\ell^-)} \nonumber\\
    &  =  R_{\rm EW}(Q) \left(1 + \sum^{N=4}_{i=1} c_n(Q)\left(\frac{\alphas(Q)}{\pi}\right)^n+\mathcal{O}(\alphas^5)+\updelta_\mathrm{m}+\updelta_\mathrm{np}\right).
\end{align} 
In this equation, $Q$ is the typical momentum transfer in the process used for measuring $R_\ell$, $c_n$ are coefficients of the perturbative series that can, in practice, be calculated up to some finite order $n = N$, %, and depend on the choice of renormalization scale $Q$, as does the coupling itself.
and the terms $\delta_{\rm m}$ and $\delta_{\rm np}$ correspond, respectively, to mixed QCD+EW higher-order and power-suppressed $\order{\Lambda^p/Q^p}$ non-perturbative corrections, which affect, differently, the tau lepton and electroweak boson decays. For $\alphasmZ = 0.118$, the size of the QCD term in \Eref{eq:R_alphas} amounts to a $4\%$ effect, so at least per-mille measurement accuracies for the $R_{\ell}$ ratios are required for a competitive $\alphasmZ$ determination~\cite{ALEPH:2005ab}. Such an experimental precision has been reached in measurements of $\uptau$ and Z boson decays, but not for the W boson and that is why the latter still does not provide a precise $\alphas$ extraction~\cite{dEnterria:2016rbf}. Reaching per-mille uncertainties in $\alphas$ determinations based on \Eref{eq:R_alphas} requires 100 times smaller uncertainties in the experimental $\uptau$, W, and Z measurements, a situation only reachable at the FCC-ee.

The ratio of hadronic to leptonic tau decays, known experimentally to within $\pm0.23\%$ ($R^{\rm \uptau,exp}_{\ell} = 3.4697 \pm 0.0080$), compared with next-to-NNLO (N$^3$LO) calculations, yields $\alphasmZ = 0.1192 \pm 0.0018$, \ie\ a 1.5\% uncertainty, through a combination of results from different theoretical approaches (contour-improved perturbation theory (CIPT) and fixed-order perturbation theory (FOPT)) with different treatments of non-pQCD corrections~\cite{Pich:2016bdg,Boito:2016oam}. 
The current $\alphas$ uncertainty is shared roughly equally between experimental and theoretical systematics. The latter are driven by differences in the CIPT and FOPT results, although the power-suppressed non-perturbative $\delta_{\rm np}$ term in \Eref{eq:R_alphas}, which is of $\order{\Lambda^2/m_\uptau^2}\approx 10^{-2}$, is not negligible for the tau, at variance with the much heavier W and Z bosons. High-statistics $\uptau$ spectral functions (\eg from B factories now, and the FCC-ee in the future), and solving CIPT--FOPT discrepancies (extending the calculations to N$^4$LO accuracy and controlling the non-pQCD uncertainties) are required to reduce the relative $\alphas$ uncertainty below the $\sim$1\% level. 

\begin{figure}
\centering
\includegraphics[width=0.46\columnwidth]{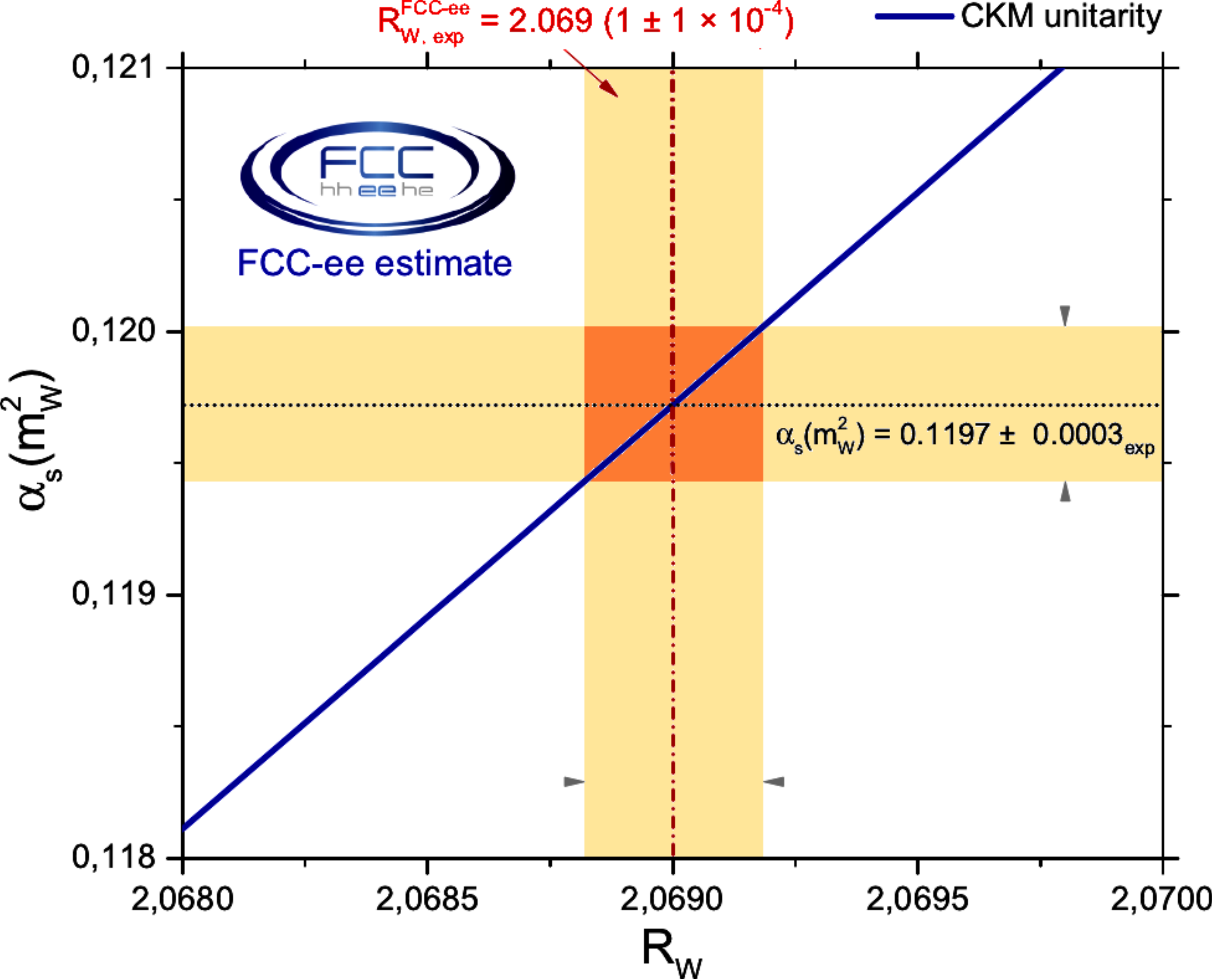}
\includegraphics[width=0.53\columnwidth,height=6.2cm]{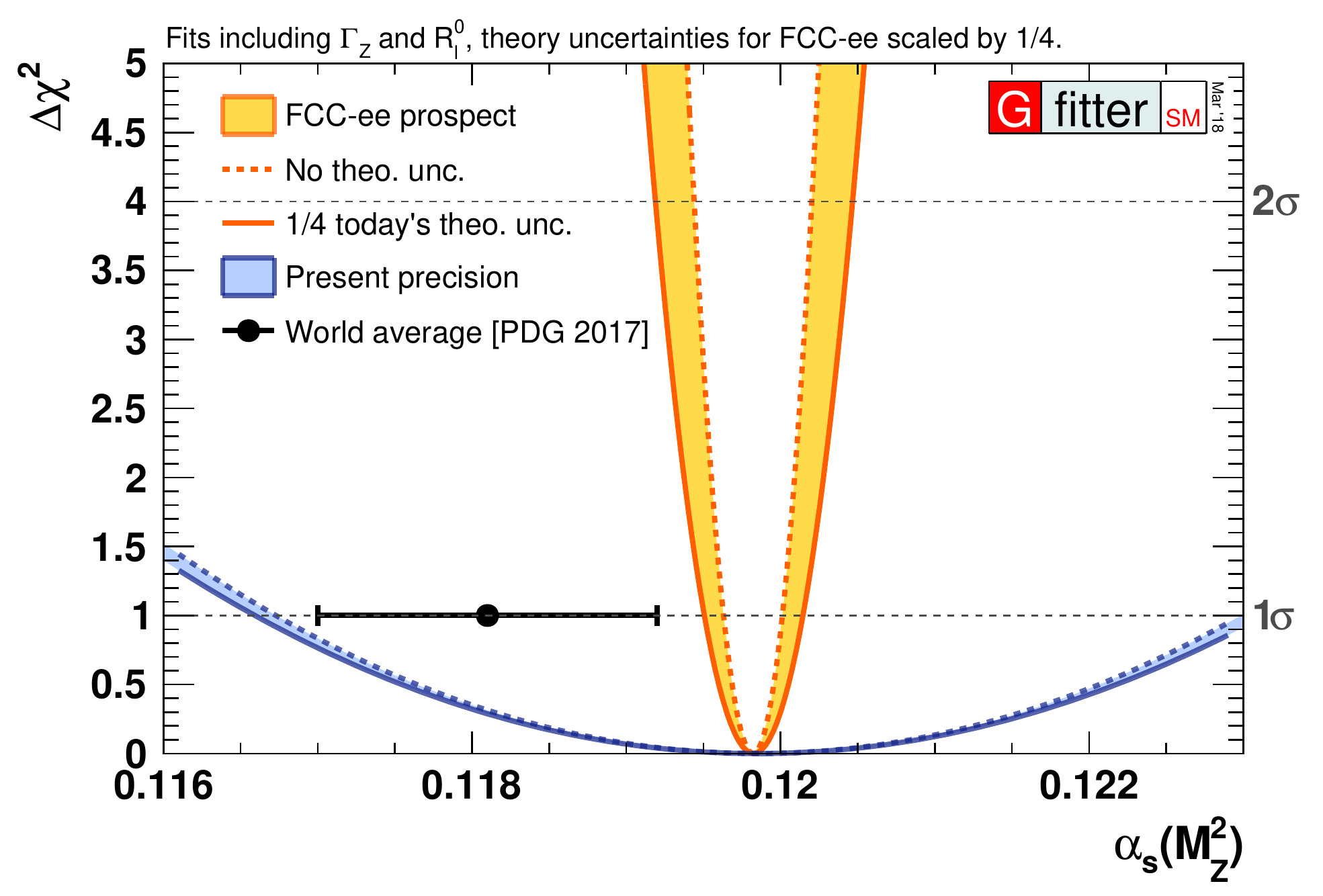}
\caption{Left: Expected $\alphas$ determination from the W hadronic-to-leptonic decay ratio ($R^{\rm W}_{\ell}$) at the FCC-ee (the diagonal blue line assumes CKM matrix unitarity)~\cite{dEnterria:2016rbf}.
Right: Precision on $\alphas$ derived from the electroweak fit today (blue band)~\cite{Haller:2018nnx} and expected at the FCC-ee (yellow band, without theoretical uncertainties and with the current theoretical uncertainties divided by a factor of four).}
\label{fig:alphas_fccee}
% \query{Please correct the figure labels in \Fref{fig:alphas_fccee}. Set variables
%in italic font.}
\end{figure}

The current state-of-the-art N$^3$LO calculations of W boson decays~\cite{Kara:2013dua} would allow a theoretical extraction of $\alphas$ with a $\sim$0.7\% uncertainty, provided that one would have experimental measurements of sufficient precision. Unfortunately, the relevant LEP W$^+$W$^-$ data are poor, based on $5\times 10^4$ W bosons alone, and result in a QCD coupling extraction, $\alphasmZ = 0.117 \pm 0.040$, with a huge $\sim$37\% uncertainty today~\cite{dEnterria:2016rbf}. A determination of $\alphas$ with per-mille uncertainty from W boson decays can only be achieved through the combination of two developments: (i) data samples commensurate with those expected at the FCC-ee (10$^8$ W bosons) %(ii) %improved N$^4$LO corrections, and (iii) 
and (ii) a significantly reduced uncertainty of the V$_\mathrm{cs}$ CKM element, which directly enters into the leading $R_{\rm EW}(Q)$ prefactor of \Eref{eq:R_alphas} and propagates into a significant parametric uncertainty on the extracted $\alphas$. Figure \ref{fig:alphas_fccee} (left) shows the expected $\alphasmZ$ value derived from the $R^{\rm W}_{\ell}$ ratio with 10$^8$ W bosons at the FCC-ee, assuming that V$_\mathrm{cs}$ has a negligible uncertainty (or, identical, assuming
Cabibbo--Kobayashi--Maskawa (CKM) matrix unitarity). The extracted QCD coupling would have $\sim$0.2\% propagated experimental uncertainties.

The current QCD coupling extraction based on Z boson hadron decays uses three closely related pseudo-observables measured at the LEP: $\Rlz = \Ghad/\Gamma_\ell$, $\sigma^{0}_{\rm had} = 12 \uppi/m_Z \cdot \Gamma_e\Ghad/\Gamma_Z^2$, and $\GZ$, combined with N$^3$LO calculations, to give $\alphasmZ = 0.1203 \pm 0.0028$ with a 2.5\% uncertainty~\cite{PDG}. Alternatively, fixing all SM parameters to their measured values and letting free $\alphas$ in the electroweak fit yields $\alphas = 0.1194 \pm 0.0029$ ($\sim$2.4\% uncertainty, shallow blue curve in Fig.~\ref{fig:alphas_fccee} (right)) \cite{Haller:2018nnx}. At the FCC-ee, with 10$^{12}$ Z bosons providing high-precision measurements with $\Delta m_{\rm Z} = 0.1\UMeV$, and $\Delta \Gamma_{Z} = 0.1$\,MeV, $\Delta R^{0}_{\ell}  = 10^{-3}$ (achievable thanks to the possibility of performing a threshold scan including energy self-calibration with resonant depolarisation) reduces the uncertainty on $\alphasmZ$ to $\sim$0.15\%. Figure \ref{fig:alphas_fccee} (right) shows the expected $\alphas$ extractions from $R^{\rm Z}_{\ell}$ and $\Gamma^{\rm Z}$ %, and $ \sigma_0^{\rm had}$ 
at the FCC-ee (yellow band) with the experimental uncertainties listed in Table~(\ref{tab:FCC-observables}), without theoretical uncertainties (dotted red curve) and with the theoretical uncertainties reduced to one-quarter of their current values (solid red curve)~\cite{Haller:2018nnx}. 

\begin{figure}
\centering
\includegraphics[width=0.65\linewidth]{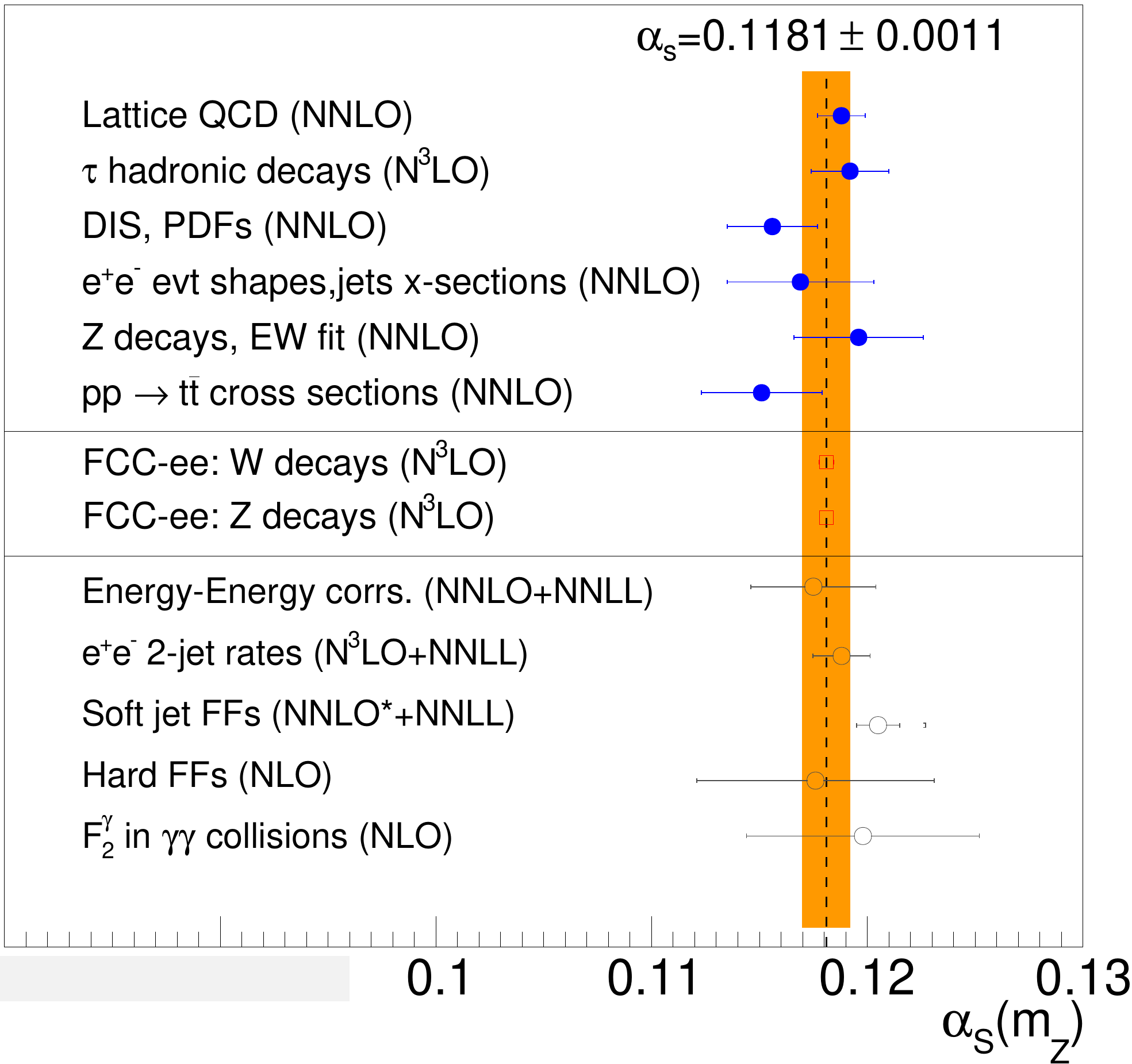}
\caption[]{Summary of the $\alphasmZ$ determinations discussed here. Top: Subclasses entering in the current PDG world-average (solid dots, orange band) whose numerical value is listed on top~\cite{PDG}. Middle: Expected FCC-ee values via W, Z hadronic decays (open squares). Bottom: Other methods based on $\epem$ data not (yet) in the $\alphasmZ$ world-average: recent EEC~\cite{Kardos:2018kqj} and two-jet rates~\cite{Verbytskyi:2019zhh}, plus other extractions at a (currently) lower level of theoretical accuracy.
}
\label{fig:alphas}
%\query{Please correct the figure labels in \Fref{fig:alphas}. Set variables
%in italic font. Do not use a hyphen (-) where a minus sign ($-$) or an en
%dash (--) is wanted.}
\end{figure}

The FCC-ee will not only provide an unprecedented amount of electroweak boson data, but also many orders of magnitude more jets than collected at LEP. The large and clean set of accurately reconstructed (and flavour-tagged) $\epem$ hadronic final states will provide additional high-precision $\alphas$ determinations from studies of event shapes, jet rates, and parton-to-hadron fragmentation functions (FFs)~\cite{dEnterria:2015kmd}. 
The existing measurements of $\epem$ event shapes (thrust $T$, $C$ parameter)~\cite{Bethke:2008hf,Abbate:2012jh,Banfi:2014sua,Hoang:2015hka} and $n$-jet rates~\cite{Weinzierl:2008iv,Dissertori:2009qa,Schieck:2012mp}, analysed with N$^{2,3}$LO calculations matched, in some cases, to soft and collinear N$^{(2)}$LL resummations, yield $\alphasmZ$~=~0.1169~$\pm$~0.0034, with a 2.9\% uncertainty~\cite{PDG}. This relatively large uncertainty is mostly driven by the span of individual extractions that use different (Monte Carlo or analytical) approaches to account for soft and collinear radiation as well as to correct for hadronization effects. Modern jet substructure techniques~\cite{Bendavid:2018nar} can help mitigate the latter corrections. In terms of event shapes, the recent combination of the CoLoRFulNNLO subtraction method~\cite{DelDuca:2016ily} with NNLL corrections in the back-to-back region~\cite{deFlorian:2004mp} has led to a precise calculation of the energy--energy correlation (EEC) observable in electron--positron collisions, and thereby an accurate NNLO+NNLL extraction of 
$\alphasmZ = 0.1175 \pm 0.0029$ ($\sim$2.5\% uncertainty)~\cite{Kardos:2018kqj}, as discussed in detail in Section~\ref{contr:SM_Kardos}.
%Reduction of the non-pQCD uncertainties, \eg\ through new $\epem$ jet data at lower (higher) $\sqrts$ for the event shapes 
%(jet rates), plus jet cross-sections with improved resummation (beyond NLL), are needed to reach $\alphas$ uncertainties below~1\%.
Moreover, a very recent analysis of two-jet rates in $\epem$ collisions at N$^3$LO+NNLL accuracy~\cite{Verbytskyi:2019zhh} has provided a new QCD coupling determination with  $\sim$1\% uncertainty: $\alphasmZ = 0.118\,81 \pm 0.001\,32$.
In addition, other sets of observables computed today with a lower degree of accuracy (NLO, or approximately NNLO, bottom part of Fig.~\ref{fig:alphas}), and thereby not now included  in the PDG average, will provide additional constraints~\cite{dEnterria:2015kmd}. The energy dependence of the low-$z$ FFs today provides  $\alphasmZ = 0.1205 \pm 0.0022$ ($\sim$2\% uncertainty) at NNLO*+NNLL~\cite{Perez-Ramos:2013eba,dEnterria:2015ljg}, whereas NLO scaling violations of the high-$z$ FFs yield $\alphasmZ = 0.1176 \pm 0.0055$ ($\sim$5\% uncertainty, mostly of experimental origin)~\cite{Albino:2005me}. In addition, measurements of the photon structure function F$_2^\upgamma(x,Q^2)$, via $\epem \to \gaga\to$\,hadrons, have been employed to derive $\alphasmZ$ = 0.1198~$\pm$~0.0054 ($\sim$4.5\% uncertainty) at NLO~\cite{Albino:2002ck}. Extension to full-NNLO accuracy of the FFs and F$_2^\upgamma(x,Q^2)$ fits using the much larger $\epem$ datasets available at various centre-of-mass energies at the FCC-ee will enable  subpercentage precision in $\alphasmZ$
to be attained. Figure~\ref{fig:alphas} presents a comparison of the current $\alphasmZ$ results (top), the expected FCC-ee extractions (middle), and the other aforementioned methods based on $\epem$ data not currently included in the world-average.

%=================================================================
\subsection[High-precision non-perturbative QCD]{\textbf{High-precision non-perturbative QCD}}

All $\epem$ processes with quarks and gluons in the final state have an intrinsic uncertainty linked to the final non-perturbative conversion of the partons, present in the last stage of the QCD shower, into hadrons. Such a process cannot be computed using first-principles QCD calculations and is described using phenomenological models, such as the Lund string \cite{Andersson:1983jt}, as implemented in the {\sc pythia} MC generator \cite{Sjostrand:2014zea}, or the cluster hadronization approach \cite{Webber:1983if} typical of the {\sc herwig} event generator \cite{Corcella:2000bw}. The analysis and unfolding of any $\epem$ experimental measurement of hadronic final states relies on these very same Monte Carlo generators; therefore, the final results are sensitive to their particular implementation of soft and collinear parton radiation (whose MC modelling is equivalent to an approximate next-to-leading-log
(NLL) accuracy~\cite{Dasgupta:2018nvj}) and of the hadronization process. Examples of such propagated uncertainties have been discussed already in the context of $\alphas$ extractions from various experimental $\epem$ observables. An improved MC reproduction of the experimental hadron data can, \eg\ help in enabling advanced light quark and gluon jet tagging in constraints of the Higgs Yukawa couplings to the first and second family of quarks. Controlling the uncertainties linked to hadronization and other final-state partonic effects, such as colour reconnection and multiparticle (spin, momenta, space, etc.) correlations, is, therefore, basic for many high-precision SM studies. Such effects are optimally studied in the clean environment provided by $\epem$ collisions, without coloured objects in the initial state. An  FCC-ee goal, therefore, is to produce truly precise QCD measurements to constrain many aspects of non-perturbative dynamics to the 1\% level or better, leaving an important legacy for MC generators for the FCC-eh and FCC-hh physics programme, much as those from LEP proved crucial for the parton shower models used today at the LHC~\cite{Anderle:2017qwx}. In particular, the FCC-ee operating at different c.m.\ energies will enormously help to control resummation and hadronization effects in event shape distributions, reducing, in particular, non-perturbative uncertainties from a 9\% effect at $\sqrts = 91.2$\,GeV to a 2\% at 400\,GeV \cite{Anderle:2017qwx,Bell:2018gce}.

The modelling of parton hadronization in the current MC event generators has achieved a moderate success, and the LHC data have only further complicated the situation. First, the production of baryons (in particular containing strange quarks) remains poorly understood and is hard to measure in the complicated hadron--hadron environment. Second, and most importantly, the LHC measurements have challenged the standard assumption of parton hadronization universality, \ie\ that models developed from $\epem$ data can be directly applied to hadron--hadron collisions. Strong final-state effects, more commonly associated with heavy-ion physics and quark--gluon--plasma formation, such as the `ridge'~\cite{Khachatryan:2010gv} or the increase of strangeness production in high-multiplicity pp events~\cite{ALICE:2017jyt}, cannot be accommodated within the standard MC generators. The large statistical samples available at the FCC-ee will allow  parton hadronization to be controlled in the QCD vacuum with subpercentage uncertainties, and thereby provide a better understanding of any collective final-state effects present in hadron--hadron collisions, starting with multistrange baryons, whose total production rates could only be determined with 5--20\% accuracy at the LEP~\cite{Abreu:1995qx,Alexander:1996qj}, and going further to excited~\cite{Alexander:1996qj,Akers:1995wx}, exotic, or multiple heavy hadrons, with implications for more advanced fragmentation models. For $\Lambda$--$\Lambda$ correlation distributions, where MC generator programs today fail to describe the LEP~\cite{Abbiendi:1998ux} and LHC data, the huge FCC-ee samples of hadronic Z decays will have statistical uncertainties matching the best LEP systematic uncertainties, corresponding to a total errors reduction by a factor of ten or more.

In $\epem \to \ttbar$, when the top and antitop quarks decay and hadronize close to each other, interactions and interferences between them, the decay bottoms, and any radiated gluons affect the rearrangement of the colour flow and thereby the kinematic distributions of the final hadronic state. Whereas the perturbative radiation in the process can, in principle, be theoretically controlled, there is a `cross-talk' among the produced hadronic strings, also known as colour reconnection (CR), that can only be modelled phenomenologically~\cite{Khoze:1994fu}.
In the pp case, such CR effects can decrease the precision that can be achieved in the extraction of the top mass, and constitute 20--40\% of its uncertainty~\cite{Argyropoulos:2014zoa}. Colour reconnection can also impact limits for CP-violation searches in $\rm H \to W^+ W^- \to q_1\qbar_2 q_3\qbar_4$ decays~\cite{Christiansen:2015yca}. Searches for such effects can be optimally studied in the process $\epem \to \rm W^+W^- \to q_1\qbar_2 q_3\qbar_4$~\cite{Christiansen:2015yca}, where CR could lead to the formation of alternative `flipped' singlets $\rm q_1\qbar_4$ and $\rm q_3\qbar_2$, and correspondingly more complicated string topologies~\cite{Sjostrand:1993hi}. The combination of results from all four LEP collaborations excluded the no-CR null hypothesis at 99.5\% CL~\cite{Schael:2013ita}, but the size of the WW data sample was too small for any quantitative studies. At the FCC-ee, with the W mass determined to better than 1\,MeV by a threshold scan, the semileptonic WW measurements (unaffected by CR) can be used to probe the impact of CR in the hadronic WW events~\cite{Anderle:2017qwx,Mangano:2018mur}. Alternative CR constraints at the FCC-ee have been proposed through the study of event shape observables sensitive to string overlap, such as sphericity for different hadron flavours, as described in `rope hadronization' approaches~\cite{Bierlich:2016vgw,Bierlich:2017sxk}.

\end{bibunit}

\label{sec-sm-denterria}  
\clearpage \pagestyle{empty}  \cleardoublepage
%============================================

\pagestyle{fancy}
\fancyhead[CO]{\thechapter.\thesection \hspace{1mm} Inclusion of mixed QCD--QED resummation effects at higher orders}
\fancyhead[RO]{}
\fancyhead[LO]{}
\fancyhead[LE]{}
\fancyhead[CE]{}
\fancyhead[RE]{}
\fancyhead[CE]{G.F.R. Sborlini}
\lfoot[]{}
\cfoot{-  \thepage \hspace*{0.075mm} -}
\rfoot[]{}

%%% definitions %%%%%%%%%%%%%%%%%%%%%%%%%%
\def\beqn{\begin{eqnarray}} \def\eeqn{\end{eqnarray}}
\def\beeq{\begin{eqnarray}}
\def\eeeq{\end{eqnarray}}
\def\nn{\nonumber}
\def\alphas{\alpha_{\rm S}}
    
\begin{bibunit}[elsarticle-num] % define the bib-style for the unit: elsarticle-num.bst
%  text-1; this is the corresponding section
%\putbib[2loops] % the *.bib
%\end{bibunit}
% go-on
%--- from: bibunits.sty, adapts the font size of ``References'' to section
\let\stdthebibliography\thebibliography
\renewcommand{\thebibliography}{%
\let\section\subsection
\stdthebibliography}
%---
    
\section
[Inclusion of mixed QCD--QED resummation effects at higher orders \\ {\it G.F.R. Sborlini}]
{Inclusion of mixed QCD--QED resummation effects at higher orders
\label{contr:SBORLINI}}
\noindent
{\bf Contribution\footnote{This contribution should be cited as:\\
G.F.R. Sborlini, Inclusion of mixed QCD--QED resummation effects at higher orders,  
%04 DOI:10.23731/CYRM-2020-003.\thepage, in:
%04 \url{http://dx.doi.org/10.23731/CYRM-2020-003.\thepage}, in:
DOI: \href{http://dx.doi.org/10.23731/CYRM-2020-003.\thepage}{10.23731/CYRM-2020-003.\thepage}, in:
Theory for the FCC-ee, Eds. A. Blondel, J. Gluza, S. Jadach, P. Janot and T. Riemann,
\\CERN Yellow Reports: Monographs, CERN-2020-003,
%04 \url{http://dx.doi.org/10.23731/CYRM-2020-003}, p. \thepage.} 
DOI: \href{http://dx.doi.org/10.23731/CYRM-2020-003}{10.23731/CYRM-2020-003},
p. \thepage.
\\ \copyright\space CERN, 2020. Published by CERN under the 
%04-2
\href{http://creativecommons.org/licenses/by/4.0/}{Creative Commons Attribution 4.0 license}.} by: G.F.R. Sborlini {[german.sborlini@ific.uv.es]}}
\vspace*{.5cm}

\noindent In this section, we review some recent results concerning the inclusion of mixed QCD--QED corrections in the computation of physical observables. First, we comment on the extension of the Dokshitzer--Gribov--Lipatov--Altarelli--Parisi (DGLAP) equations to deal with the presence of mixed QCD--QED interactions. We describe the calculation of the full set of higher-order corrections to the splitting kernels, through the Abelianization algorithm. This procedure allows us to build the functional form of the QCD--QED corrections, starting from pure QCD terms. As a practical application of this technique, we also explore the computation of fixed-order corrections to diphoton production, and the inclusion of higher-order mixed QCD--QED resummation effects to Z production. In both cases, we directly apply the Abelianization to the $q_T$ subtraction or resummation formalism, obtaining the universal ingredients that allow us to compute the aforementioned corrections to any process involving colourless and neutral particles in the final state.

%%%%%%%%%%%%%%%%%%%%%%%%%%%%%%%%%%%%%%%%%%%%%%%%%%%%%%%%%%%%%%%%%%%%%%%%%%%%%%%%%%%%%%%%%%%%%%%%%%%%%%%%%%%%%%%%%%%%%%%%%%%
%%%%%%%%%%%%%%%%%%%%%%%%%%%%%%%%%%%%%%%%%%%%%%%%%%%%%%%%%%%%%%%%%%%%%%%%%%%%%%%%%%%%%%%%%%%%%%%%%%%%%%%%%%%%%%%%%%%%%%%%%%%

\subsection[Introduction and motivation]{Introduction and motivation}
\label{sec:Introduction:sborlini}
The large amount of data that high-energy experiments are collecting allows  the precision of several measurements to be  increased. In consequence, theoretical predictions must be pushed forward by including previously neglected small effects. This is the case for electroweak (EW) or QED corrections, which are subdominating for collider physics. However, from  na\"ive power
counting, it is easy to notice that ${\cal O}(\alpha) \approx {\cal O}(\alphas^2)$. In addition, QED interactions (as well as the full set of EW calculations) lead to novel effects that could interfere with the well-known QCD signals. Moreover, these effects might play a crucial role in the context of future lepton colliders, such as the FCC-ee. For these reasons, EW and QED higher-order corrections must be seriously studied in a fully consistent framework.

The aim of this brief section is to present some results related to the impact of QED corrections on the calculation of physical observables for colliders. In  \Sref{sec:DGLAP}, we recall the computation of the full set of QCD--QED splitting functions at ${\cal O}(\alpha \, \alphas)$ and ${\cal O}(\alpha^2)$, centring into the Abelianization algorithm and the relevance of the corrections to achieve a better determination of the photon PDF. Then, we apply the Abelianization to the well-established $q_T$ subtraction or resummation \cite{Catani:2007vq,Catani:2013tia} framework. In  \Sref{sec:difotones}, we show the impact of the NLO QED corrections to diphoton production. After that, we characterize the mixed QCD--QED resummation of soft gluons or photons for Z boson production in  \Sref{sec:Zproduction}. Conclusions are drawn and future research directions are discussed in  \Sref{sec:conclusions:sborlini}.

%20190204: Hecho!!

\subsection[Splittings and PDF evolution]{Splittings and PDF evolution}
\label{sec:DGLAP}
Splitting functions are crucial in describing the singular collinear behaviour of scattering amplitudes. On the one hand, they are used to build the counterterms to subtract infrared (IR) singularities from cross-sections. On the other hand, they are the evolution kernels of the integro-differential DGLAP equations \cite{Altarelli:1977zs}, which govern the perturbative evolution of PDFs. When taking  QCD and EW or QED interactions into account, it is necessary to include photon and lepton PDFs, and this will lead to the presence of new splitting functions. In Refs. \cite{deFlorian:2015ujt,deFlorian:2016gvk}, we computed the ${\cal O}(\alpha \, \alphas)$ and ${\cal O}(\alpha^2)$ corrections to the DGLAP equations, as well as the associated kernels. The strategy that we adopted was based on the implementation of a universal algorithm, called \textit{Abelianization}, which aims to explode previously known pure QCD results to obtain the corresponding QCD--QED or QED expressions. Roughly speaking, the key idea behind this method is that of \textit{transforming gluons into photons}: colour factors are replaced by suitable electric charges, as well as symmetry or counting factors.

\begin{figure}
\begin{center}
\includegraphics[width=0.98\textwidth]{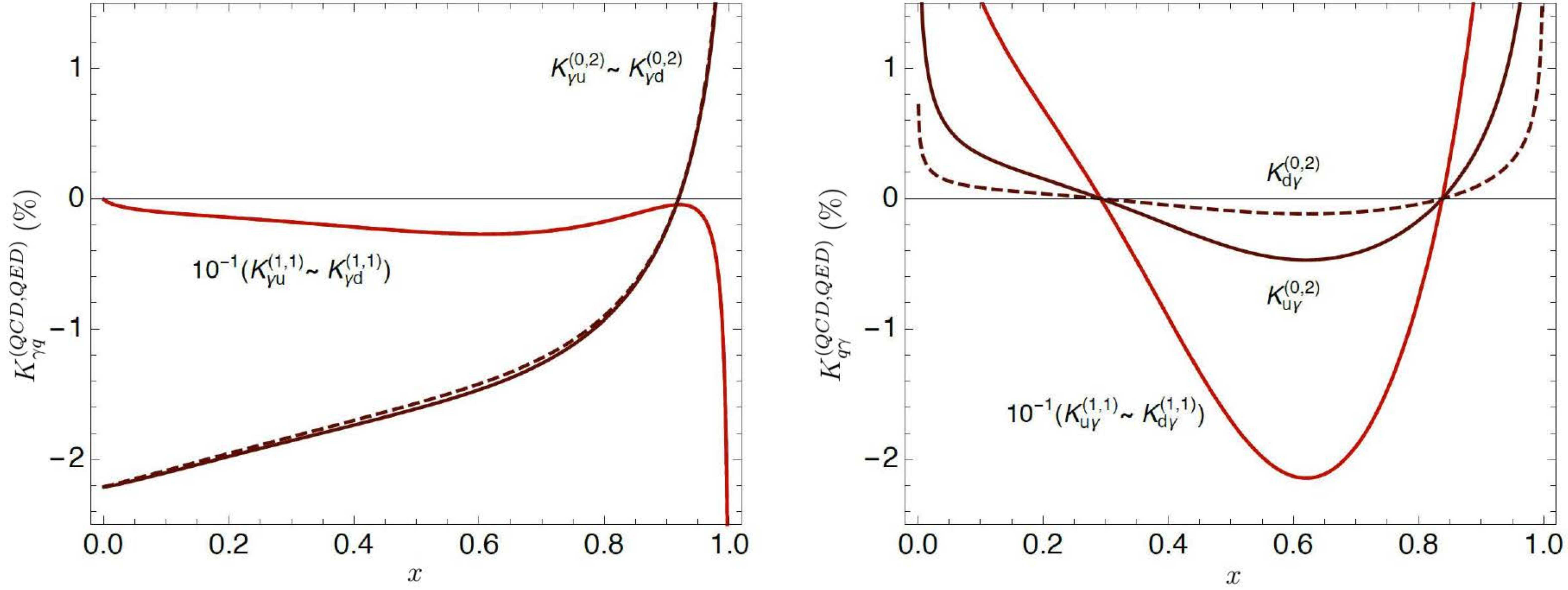}
\caption{Corrections due to the inclusion of QED contributions in the $P_\mathrm{q\upgamma}$ (right) and $P_\mathrm{\upgamma q}$ (left) splitting kernels. We include both ${\cal O}(\alpha^2)$ (brown) and ${\cal O}(\alpha\,  \alphas)$ (red) terms. The $K$ ratio is defined using the leading order as normalization. To ease the visual presentation, we rescaled the ${\cal O}(\alpha \, \alphas)$ terms by a factor of  $0.1$.} 
\label{fig:Kernelsqgamma}
%\query{Please correct the figure labels in \Fref{fig:Kernelsqgamma}. Set variables
%in italic font.}
\end{center}
\end{figure}

With the purpose of exhibiting the quantitative effects that mixed QCD--QED or ${\cal O}(\alpha^2)$ corrections might have, we plot the $K$ ratio for quark--photon and photon--quark splitting functions in \Fref{fig:Kernelsqgamma}. It is important to notice that these contributions are not present in pure QCD, which implies that the evolution of photon PDF is noticeably affected by  ${\cal O}(\alpha \, \alphas)$ splittings or even higher orders in the mixed QCD--QED perturbative expansion. We would like to point out that a precise determination of photon distributions is crucial to obtaining more accurate predictions for several physical observables.  

%********************************************************************%
%********************************************************************%

%20190205: Hecho!!

\subsection[Fixed-order effects: application to diphoton production]{Fixed-order effects: application to diphoton production}
\label{sec:difotones}
The $q_T$ subtraction or resummation formalism \cite{Catani:2007vq,Catani:2013tia} is a powerful approach to computing higher-order corrections to physical observables. This formalism has been mainly applied to QCD calculations, and relies on the colour neutrality of the final-state particles.\footnote{An extension to deal with massive or coloured particles in the final state is presented in Refs. \cite{Catani:2014qha,Bonciani:2015sha}.} Thus, we used the Abelianization algorithm to compute the universal coefficients required to implement NLO QED corrections to any process involving only neutral particles in the final state. In this way, we demonstrate that this extension can deal consistently with the cancellation of IR divergences in the limit $q_T\to 0$.

\begin{figure}
\begin{center}
\includegraphics[width=0.99\textwidth]{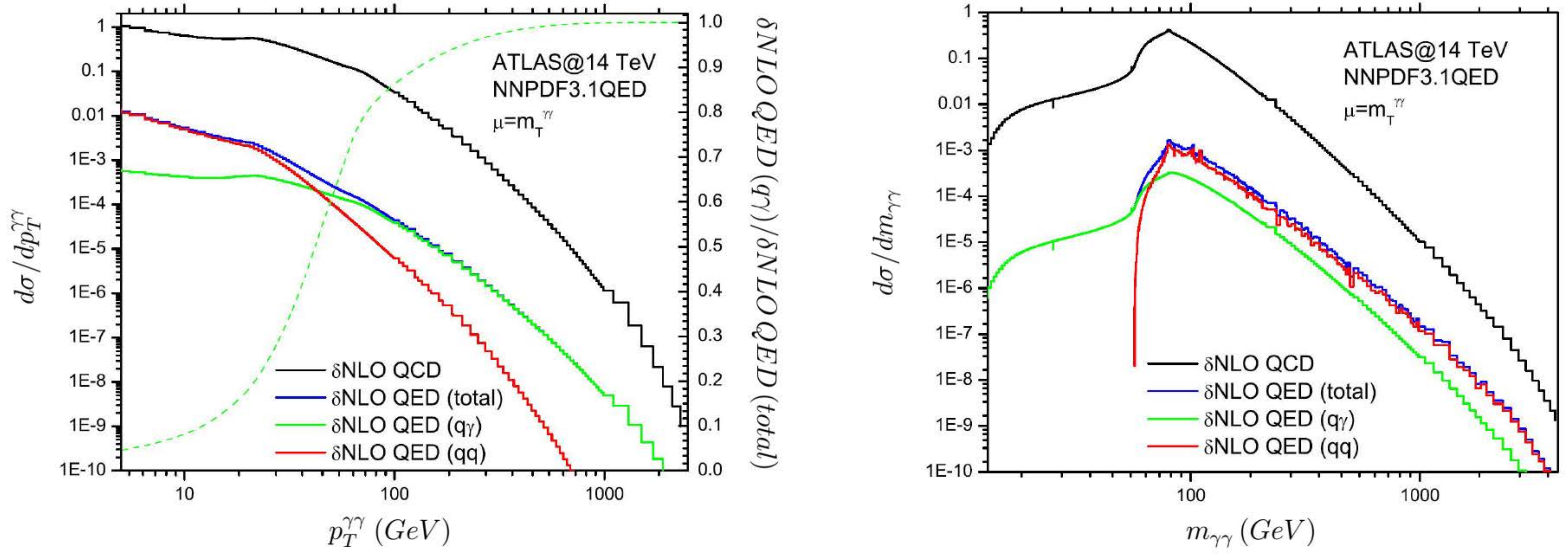}  
\caption{Impact of higher-order QED corrections on the transverse momentum (left) and invariant mass (right) distributions for diphoton production. The black (blue) curve shows the total NLO QCD (QED) prediction, without including the LO contribution. The dashed green line indicates the relative contribution of the $\mathrm{q \upgamma}$-channel to the total NLO QED correction.}
\label{fig:Difotones}
%\query{Please correct the figure labels in \Fref{fig:Difotones}. Set variables
%in italic font.}
\end{center}
\end{figure}

As a practical example, we used the public code \texttt{2gNNLO} \cite{Catani:2011qz,Catani:2018krb}, which provides up to NNLO QCD corrections to diphoton production, and we implemented the corresponding NLO QED corrections\cite{Sborlini:2017gpl,Sborlini:2018fhr}. We applied the default ATLAS cuts, with 14\,TeV centre-of-mass energy, and the \texttt{NNPDF3.1QED} \cite{Ball:2017nwa,Bertone:2017bme} PDF set. The transverse momentum and invariant mass spectra are shown in  \Fref{fig:Difotones}. It is interesting to note that, even if the corrections are small compared with the QCD contributions, the QED interactions lead to novel features, such as a dynamic cut in the invariant mass spectrum. This is because real radiation in the $\mathrm{q \bar q}$ channel contains three final-state photons, which must be ordered according to their transverse momenta before imposing the selection cuts. Moreover, introducing the QED corrections (or, even better, mixed NLO QCD--QED corrections) will allow us to reduce the scale uncertainties and produce more reliable theoretical predictions. 

%********************************************************************%
%********************************************************************%

% 20190205: Hecho!!

\subsection[Mixed resummation effects: Z boson production]{Mixed resummation effects: Z boson production}
\label{sec:Zproduction}
Finally, we studied the impact of including mixed QCD--QED terms within the $q_T$ resummation formalism. This is equivalent to considering the simultaneous emission of soft or collinear gluons and photons. A detailed description of the formalism is presented in Ref. \cite{Cieri:2018sfk}, which gives
the computation of the modified Sudakov form factors as well as all the required universal coefficients to reach mixed NLL$'$+NLO accuracy in the double expansion in $\alpha$ and $\alphas$. Explicitly, we obtained 
\begin{multline}
\label{eqG2}
{\cal G}_{N}'(\alphas,\alpha, L) = {\cal G}_{N}(\alphas, L) + L \; g'^{(1)}(\alpha L)+ g_N'^{(2)}(\alpha L) 
+\sum_{n=3}^\infty \left(\frac{\alpha}{\uppi}\right)^{n-2} g_N'^{(n)}(\alpha L) 
\\ + g'^{(1,1)}(\alphas L,\alpha L) + \sum_{n,m=1 \atop n+m\neq 2}^\infty \left(\frac{\alphas}{\uppi}\right)^{n-1}\left(\frac{\alpha}{\uppi}\right)^{m-1} g_N'^{(n,m)}(\alphas L,\alpha L) \, 
\end{multline}
and
\begin{equation}
\label{eqH2}
{\cal H}_N'^{F}(\alphas,\alpha) = {\cal H}_N^{F}(\alphas) + \frac{\alpha}{\uppi} \,{\cal H}_N'^{F \,(1)} +\sum_{n=2}^\infty \left(\frac{\alpha}{\uppi}\right)^{n}\,{\cal H}_N'^{F \,(n)} 
 + \sum_{n,m=1}^\infty \left(\frac{\alphas}{\uppi}\right)^{n}\left(\frac{\alpha}{\uppi}\right)^{m}\,{\cal H}_N'^{F \,(n,m)}\,
\end{equation}
for the expansion of the Sudakov exponents and the hard-virtual coefficients, respectively. A similar expansion is available for the soft-collinear coefficients $C_{a b}$. Other important ingredients of the formalism are the mixed QCD--QED renormalization group equations, which include a double expansion of the corresponding $\beta$ functions \cite{Cieri:2018sfk}.

\begin{figure}
\begin{center}
\includegraphics[width=0.95\textwidth]{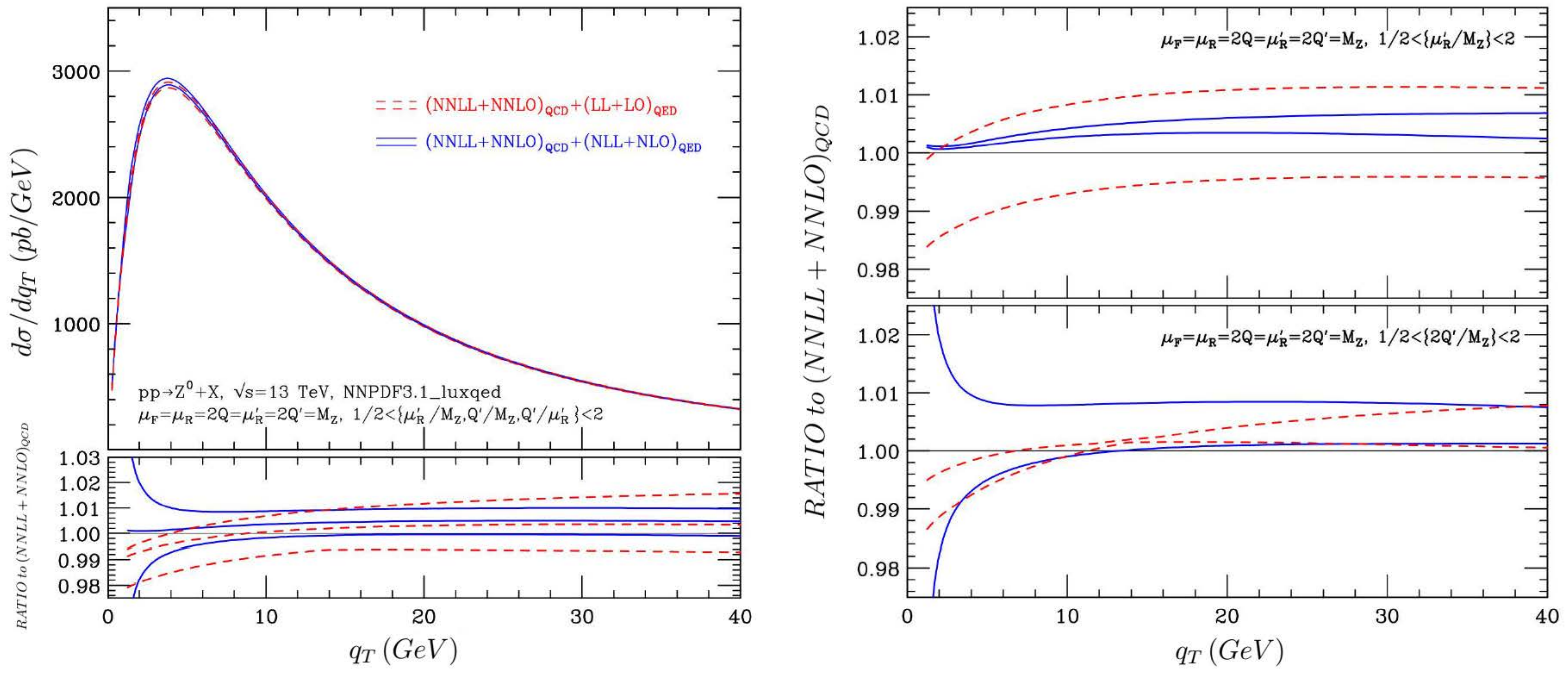}
\caption{The $q_T$ spectrum for Z boson production at the LHC with 13\,TeV centre-of-mass energy. In the left panel, we show the combination of NNLL+NNLO QCD contributions together with the LL (red dashed curve) and NLL$'$+NLO (blue solid curve) QED effects. We include the uncertainty bands that result from the full scale variation by a factor of two (up and down). More details about scale uncertainties are shown in the right panel, where we independently modify  the resummation (upper plot) and renormalization (lower plot) scales.}
\label{fig:Zproduction}
%\query{Please correct the figure labels in \Fref{fig:Zproduction}. Set variables
%in italic font.}
\end{center}
\end{figure}

To test our formalism, we used Z boson production as a benchmark process. We started from the code \texttt{DYqT}\cite{Bozzi:2008bb} to compute the next-to-next-to-leading logarithmic QCD (NNLL) corrections properly matched to the fixed-order contribution (\ie NNLO QCD in this case). In  \Fref{fig:Zproduction}, we show the combination of NNLL+NNLO QCD predictions for the $q_T$ spectrum of the produced Z (in the narrow width approximation), together with the LL (red dashed curve) and mixed NLL$'$+NLO QED contributions (blue solid
curve). The effects introduced by mixed QCD--QED terms reach the percentage
level for $q_T \approx 20\,{\rm GeV}$, when considering LHC kinematics at 13\,TeV centre-of-mass energy. However, the most noticeable consequence of introducing these corrections is the scale-dependence reduction. This means that our predictions are more stable when varying the electroweak parameters or the factorisation, renormalization, or resummation scales. 

%********************************************************************%
%********************************************************************%

%20190206: Hecho!!!

\subsection[Conclusions]{Conclusions}
\label{sec:conclusions:sborlini}
In this brief section, we reviewed some of our recent efforts towards more precise phenomenological predictions for colliders. We centred the discussion on the inclusion of QED and mixed QCD--QED corrections to the evolution of PDFs (through the computation of novel splitting functions), QED fixed-order computations (using diphoton production as a benchmark), and mixed QCD--QED $q_T$ resummation (applied to Z boson production). In all these cases, the corrections constitute percentage-level deviation from the dominant QCD correction, but this could still be detected through an increased precision of the forthcoming experimental measurements (such as those provided by the FCC-ee). Thus, understanding how to extend the exposed frameworks to deal with even higher perturbative orders is crucial to match the quality of the experimental data, allowing us to detect any possible deviation from the Standard Model and discover new physical phenomena.

\subsection*{Acknowledgements}

The work was done in collaboration with D. de Florian, G. Rodrigo, L. Cieri, and G. Ferrera.

\end{bibunit}

\label{sec-sm-Sborlini} 

\clearpage \pagestyle{empty}  \cleardoublepage
%============================================

\pagestyle{fancy}
\fancyhead[CO]{\thechapter.\thesection \hspace{1mm} CoLoRFulNNLO at work: a determination of $\alpha_{\rm S}$}
\fancyhead[RO]{}
\fancyhead[LO]{}
\fancyhead[LE]{}
\fancyhead[CE]{}
\fancyhead[RE]{}
\fancyhead[CE]{A. Kardos, S. Kluth, G.~Somogyi, Z.~Tr\'ocs\'anyi,
Z.~Tulip\'ant, A.~Verbytskyi}
\lfoot[]{}
\cfoot{-  \thepage \hspace*{0.075mm} -}
\rfoot[]{}

%%% definitions %%%%%%%%%%%%%%%%%%%%%%%%%%
\def\beqn{\begin{eqnarray}} \def\eeqn{\end{eqnarray}}
\def\beeq{\begin{eqnarray}}
\def\eeeq{\end{eqnarray}}
\def\nn{\nonumber}
\def\alphas{\alpha_{\rm S}}

\begin{bibunit}[elsarticle-num]
\let\stdthebibliography\thebibliography
\renewcommand{\thebibliography}{%
\let\section\subsection
\stdthebibliography}

\section
[CoLoRFulNNLO at work: a determination of $\alpha_{\rm S}$\\
{\it A. Kardos, S. Kluth, G. Somogyi, Z. Tr\'ocs\'anyi, Z. Tulip\'ant,  A. Verbytskyi}]
{CoLoRFulNNLO at work: a determination of $\alpha_{\rm S}$
\label{contr:SM_Kardos}}
\noindent
{\bf Contribution\footnote{This contribution should be cited as:\\
A. Kardos, S. Kluth, G. Somogyi, Z. Tr\'ocs\'anyi, Z. Tulip\'ant,  A. Verbytskyi, CoLoRFulNNLO at work: a determination of $\alpha_{\rm S}$,  
%04 DOI:10.23731/CYRM-2020-XXX.\thepage, in:
%04 \url{http://dx.doi.org/10.23731/CYRM-2020-XXX.\thepage}, in:
DOI: \href{http://dx.doi.org/10.23731/CYRM-2020-003.\thepage}{10.23731/CYRM-2020-003.\thepage}, in:
Theory for the FCC-ee, Eds. A. Blondel, J. Gluza, S. Jadach, P. Janot and T. Riemann,
\\CERN Yellow Reports: Monographs, CERN-2020-003,
%04 \url{http://dx.doi.org/10.23731/CYRM-2020-003}, p. \thepage.} 
DOI: \href{http://dx.doi.org/10.23731/CYRM-2020-003}{10.23731/CYRM-2020-003},
p. \thepage.
\\ \copyright\space CERN, 2020. Published by CERN under the 
%04-2
\href{http://creativecommons.org/licenses/by/4.0/}{Creative Commons Attribution 4.0 license}.} by: A. Kardos, S. Kluth, G. Somogyi, Z. Tr\'ocs\'anyi,
Z. Tulip\'ant,\\ A.~Verbytskyi\\
Corresponding author: A. Kardos {[kardos.adam@science.unideb.hu]}}
\vspace*{.5cm}

\subsection{Introduction}

The most precise determination of fundamental parameters of the Standard Model 
is very important.
One such fundamental parameter is the strong coupling of QCD. Its importance can 
be gauged by taking a look at the various experiments and configurations where 
it was measured; for an up-to-date summary, see Ref.~\cite{Bethke:2017uli}. The 
precise measurement of such a parameter is  difficult for two reasons. First, 
high-quality data with small and well-controlled uncertainties are needed. 
Second, high-precision calculations are needed from the 
theory side, such that theoretical uncertainties are small as well.

In a theoretical prediction based on calculation in perturbation theory, the
uncertainty has two main sources: the omission of higher-order 
terms, which are estimated by the renormalization scale, and the numerical 
stability of the integrations. While the dependence on unphysical scales can, in principle, 
be decreased by including more and more higher-order contributions in the prediction, 
the numerical uncertainty can be intrinsic to the method used to obtain the
theoretical prediction.  Moreover, the method of comparing experiment with 
theory is also affected by another uncertainty. While an experiment measures colour singlet 
objects, hadrons, the predictions are made in QCD for colourful ones, partons. 
The assumption of local parton--hadron duality ensures a correspondence between
these 
two up to non-perturbative effects. Non-perturbative effects are power 
corrections in nature, going with some negative power of the collision energy. 
This means that, for an accurate comparison,  either (i) these effects should be 
estimated and taken into 
account, or (ii)  the experiment should have a high enough energy that these 
contributions become negligible compared with other effects, or (iii)  an observable 
must be chosen that is not sensitive to these effects. 

To take these non-perturbative effects into account, we must choose from 
phenomenological \cite{Webber:1983if,Andersson:1983ia} or analytical models 
\cite{Dokshitzer:1999sh}. It is worth noting that none of these models is 
derived from first principles; hence, there is still room for improvement. 
Non-perturbative effects derived from first principles would also be favoured 
because these corrections are to be used in comparisons of predictions 
with experimental measurements. Currently, phenomenological 
models use several parameters fitted to experimental data; thus, bias is 
introduced in the measurement of physical parameters. The calculation of 
non-perturbative corrections from first principles is also advocated because 
the only available analytical model seems to be ill-suited for the current 
precision of theoretical calculations, as  shown in Ref.~\cite{Tulipant:2017ybb}.

In this report, we show two approaches to  how the measurement of a physical 
parameter, the strong coupling, can be carried out with high precision. Because 
the used observables allow for such measurements, these can be considered as 
interesting subjects to study in a future electron--positron collider.

\subsection{Precision through higher orders}

\begin{figure}
\centering
\includegraphics[width=8cm]{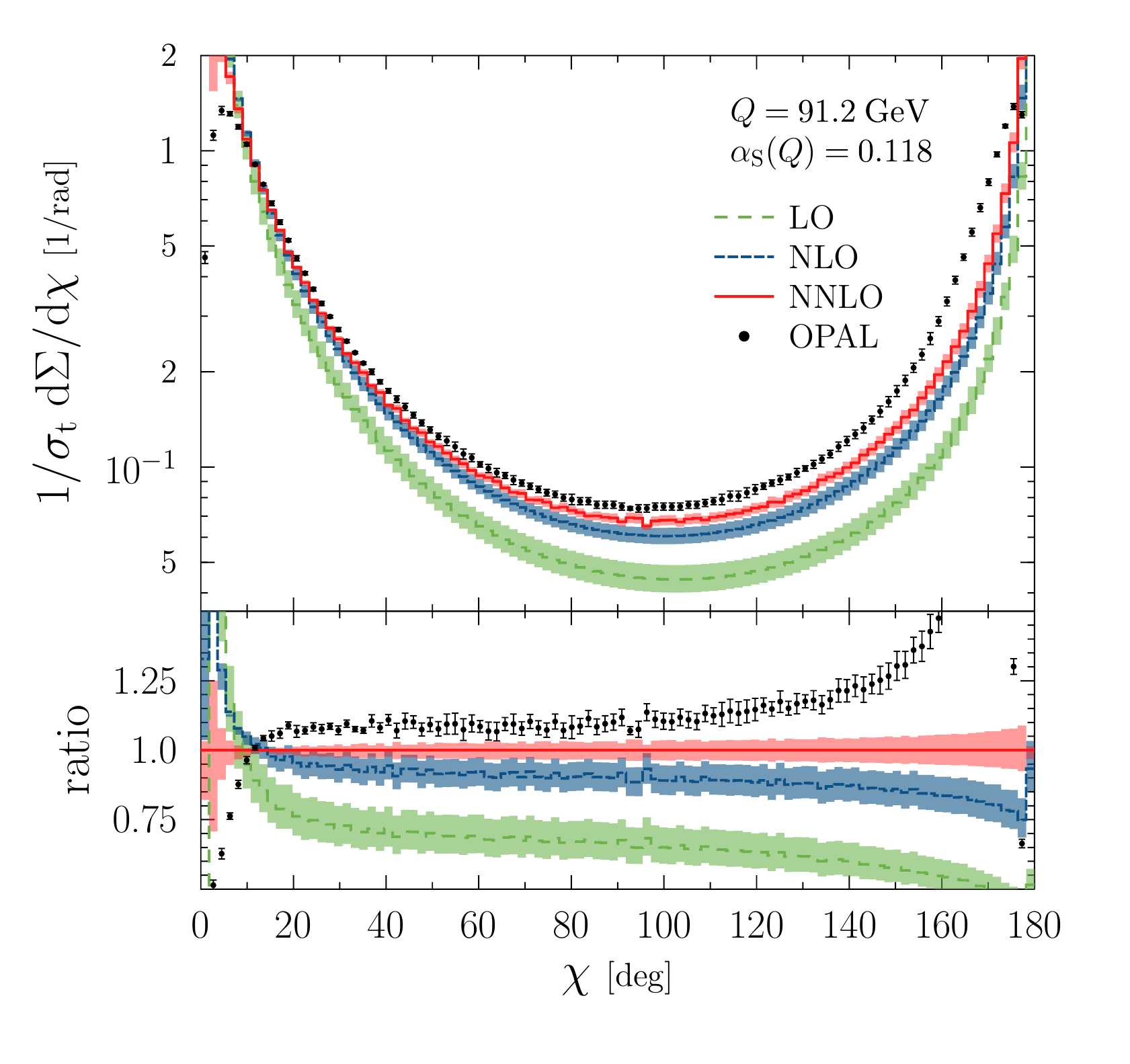}
\caption{{\label{fig:EECfo}} Top: Fixed-order prediction for EEC
in the first three orders of perturbation theory with theoretical
uncertainties. The dots show the measurement by the OPAL collaboration
\cite{Acton:1993zh}.
Bottom: Comparison of the predictions and the measurement
with the NNLO result.}
\end{figure}

A possible approach to increasing the precision of a measurement from the 
theoretical perspective is to select an observable and refine its prediction
by including higher-order contributions in fixed-order perturbation theory or
by means of resummation. With the completion of the CoLoRFulNNLO subtraction 
method \cite{Somogyi:2006da,Somogyi:2006db,DelDuca:2016ily} for electron--positron 
collisions,
the next-to-next-to-leading-order (NNLO) QCD prediction for energy--energy 
correlation (EEC) recently  became available 
\cite{DelDuca:2016csb} for the first time. Matching this with predictions
obtained by resumming leading (LL), next-to-leading (NLL), and next-to-next-to-leading logarithms (NNLL) in the back-to-back region \cite{deFlorian:2004mp}, 
it was possible,
by matching the two calculations, to arrive at the most precise theoretical
prediction for this observable at NNLO+NNLL accuracy in QCD 
\cite{Tulipant:2017ybb}.
The energy--energy correlation is defined as a normalised energy-weighted sum of
two-particle correlations:
\begin{align}
\frac{1}{\sigma_{\rm t}}\frac{{\rm d}\Sigma(\chi)}{{\rm d}\cos\chi} \equiv
\frac{1}{\sigma_{\rm t}}\int\!\sum_{i,j}\frac{E_i E_j}{Q^2}{\rm d}
\sigma_{{\rm e}^+{\rm e}^-\to ij+X}\delta(\cos\chi + \cos\theta_{ij})
\,,
\end{align}
where $Q$ is the centre-of mass energy of the collision, $\sigma_{\rm t}$ is the 
corresponding total cross-section, $E_i$ is the energy of the $i$th
particle, and $\cos\theta_{ij}$ is the enclosed angle between  particles
$i$ and $j$. The theoretical prediction for EEC in the first three
orders of perturbation theory is depicted in \Fref{fig:EECfo}. The
theoretical uncertainties were obtained by varying the renormalization scale
between $m_\mathrm{Z}/2$ and $2m_\mathrm{Z}$. As can be seen from the lower 
panel, even when using the highest-precision prediction, the difference between
measurement and theory is sizeable. This can be attributed to missing higher-order terms becoming important at the edge of phase space and missing 
hadronization corrections.

The behaviour near $\chi=0$ can be improved by including all-order results
through resummation. As described in Ref.~\cite{Kardos:2018kqj}, we used modern Monte Carlo (MC)
tools to extract such corrections for EEC. To do this, we generated event
samples  at both the hadron and  the parton level and the ratio of these
provided the hadron-to-parton ratio or $H/P$. Using this ratio and multiplying 
our parton-level predictions bin by bin, we obtained our theoretical 
prediction at the hadron level. As MC tools, we used
\texttt{SHERPA2.2.4} \cite{Gleisberg:2008ta} and \texttt{Herwig7.1.1}
\cite{Bellm:2015jjp}. The exact set-up of the MC tools is presented in
Ref.~\cite{Kardos:2018kqj}.

\begin{figure}
\centering
\begin{tabular}{cc}
\includegraphics[width=0.47\textwidth]{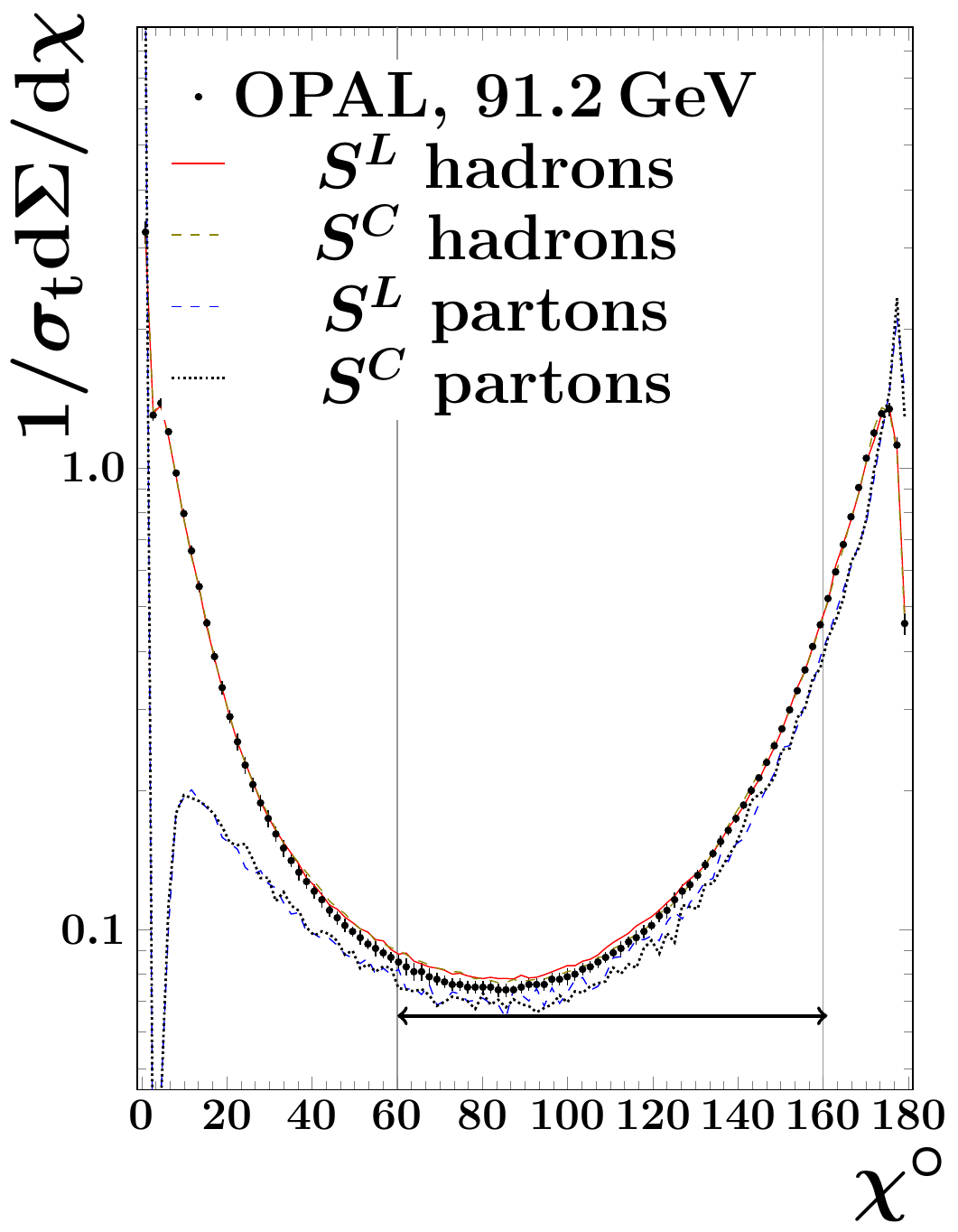} &
\includegraphics[width=0.47\textwidth]{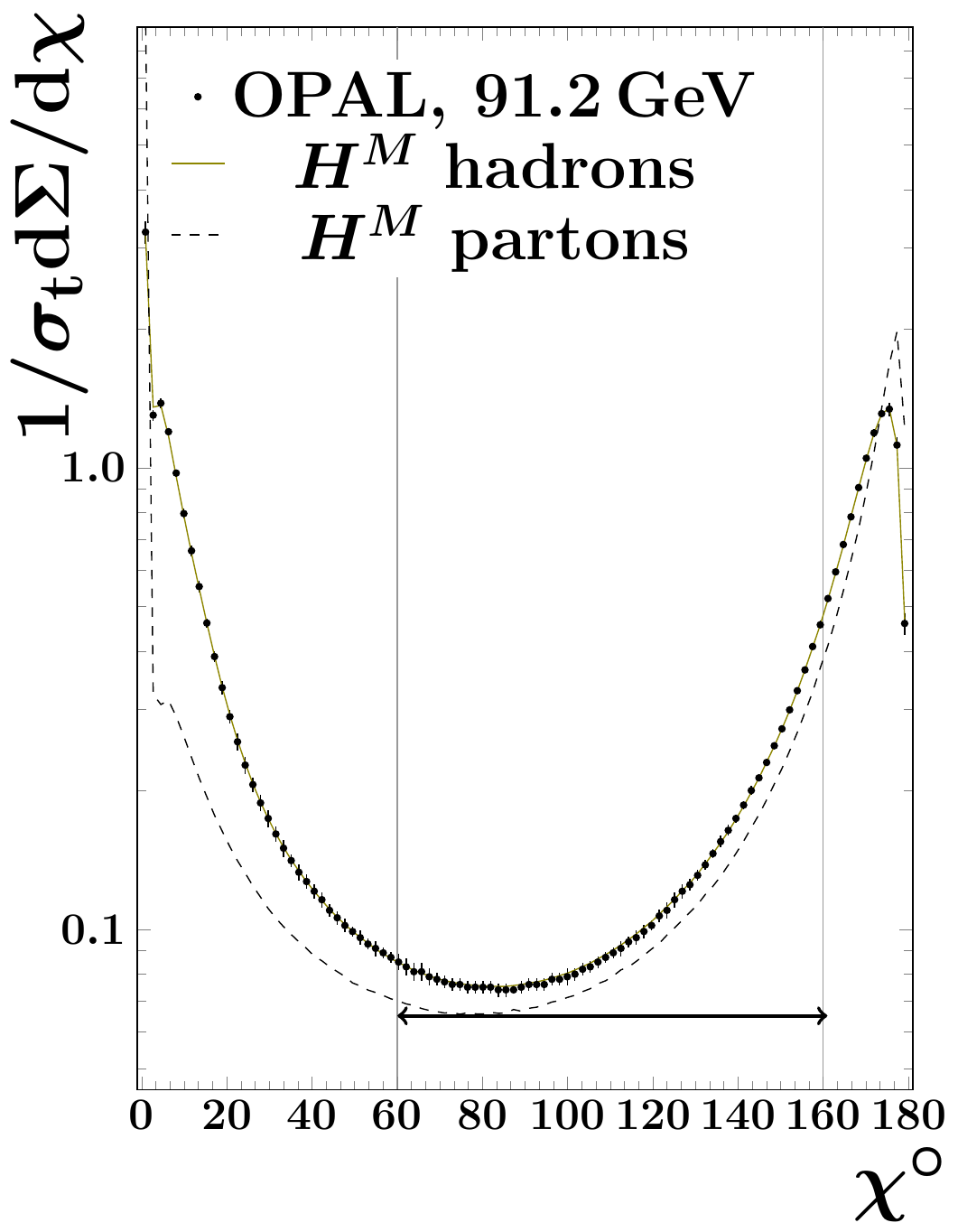}
\end{tabular}
\caption{{\label{fig:SherpaHerwig}} EEC distributions obtained with the
two MC tools at the parton and hadron level at $91.2\,{\rm GeV}$, with 
corresponding OPAL data. Note that for these two plots a
different definition of $\chi$  was used: this time, the back-to-back region 
corresponds to $\chi\to 180^\circ$.}
%\query{Please correct the figure labels in \Fref{fig:SherpaHerwig}. Set text
%labels (\eg `t' for `total') in roman font.}
\end{figure}

The value for the strong coupling was determined by fitting the predictions 
to 20 different datasets (for details, see Table 1 of Ref.~\cite{Kardos:2018kqj}).
For illustrative purposes,  \Fref{fig:SherpaHerwig}  shows the predictions
obtained with \texttt{SHERPA} and \texttt{Herwig} at the parton and hadron 
level. For \texttt{SHERPA}, we used both the Lund ($S^\mathrm{L}$) 
\cite{Andersson:1983ia} and cluster ($S^\mathrm{C}$) \cite{Webber:1983if} hadronization
models, while in \texttt{Herwig} we used the built-in cluster model. The 
figure  also indicates the range  used in the actual fitting 
procedure.

For the fitting, the \texttt{MINUIT2} program \cite{James:1975dr}
was used to minimize the quantity:
\begin{align}
\chi^2(\alpha_{\rm S}) = \sum_{\rm datasets}\chi^2_{\rm dataset}(\alpha_{\rm S})
\end{align}
with the $\chi^2(\alpha_{\rm S})$\footnote{Not to be confused with the angle 
$\chi$.} quantity calculated as:
\begin{align}
\chi^2(\alpha_{\rm S}) &=
(\boldsymbol{D} - \boldsymbol{P}(\alpha_{\rm S}))^{\rm T}
\underline{\underline{V}}^{-1}
(\boldsymbol{D} - \boldsymbol{P}(\alpha_{\rm S}))
\,,
\end{align}
where $\boldsymbol{D}$ is the vector of data points, $\boldsymbol{P}$ is 
the vector of predictions as functions of $\alpha_{\rm S}$ and 
$\underline{\underline{V}}$ is the covariance matrix. 

With the fitting procedure performed in the range between $60^\circ$ and 
$160^\circ$. The resulting strong coupling NNLL+NNLO prediction is
\begin{align}
\alpha_{\rm S}(m_\mathrm{Z}) &= 
0.117\,50
\pm 0.002\,87
\end{align}
and at NNLL+NLO accuracy is
\begin{align}
\alpha_{\rm S}(m_\mathrm{Z}) &= 
0.122\,00
\pm 0.005\,35
\,.
\end{align}
Notice the reduction in uncertainty as we go from NNLL+NLO to NNLL+NNLO.

\subsection{Precision through small power corrections}

As outlined in the introduction,
the current methods of taking the effect of non-perturbative
contributions into account can raise concerns, mainly
because only phenomenological models are present for them. The other big
concern is that these models rely on experimental results through
tuned parameters. The best option, without any model derived from first 
principles, is to decrease
these effects as much as possible. The idea is simple: if the 
non-perturbative contribution can be shrunk, 
its large uncertainties will make a smaller contribution to the final
uncertainty of the extracted value of the strong coupling.

In this section, we focus on altering the definitions of existing observables
to decrease the non-perturbative corrections. The most basic and most used 
observables in electron--positron collisions are the thrust ($T$) and the 
various jet masses. In their original definitions, these all  incorporate 
all registered hadronic objects of the event or a given, well-defined region. 
Hence, a
natural way to modify them is to filter the tracks contributing to their
value in an event. One possible way to remove tracks is by means of grooming
\cite{Butterworth:2008iy,Krohn:2009th,Ellis:2009me,Ellis:2009su,
Dasgupta:2013ihk,Larkoski:2014wba}. In particular, the
soft drop \cite{Larkoski:2014wba} is a grooming when a part of the soft content 
of the event is removed according to some criteria.

In Ref.~\cite{Baron:2018nfz}, soft-drop variants were defined for thrust, 
$\tau'_{\rm SD} = 1 - T'_{\rm SD}$, hemisphere jet mass, $e^{(2)}_2$ and 
narrow jet mass, $\rho$.
As  showed in that paper, the non-perturbative corrections can be 
drastically decreased if soft drop is applied. The effect of soft drop
turns out to be the most significant in the peak region of the 
distributions, where the contribution from all-order resummation and 
non-perturbative
effects is the greatest. This makes these observables promising candidates 
for strong coupling measurements at a future electron--positron collider.
The application of these observables---although very interesting---is
limited at LEP measurements, owing to the
limited amount of data taken and because the soft-drop procedure inherently
results in a decrease of cross-section.

In our recent paper \cite{Kardos:2018kth}, we analysed the proposed 
observables from the standpoint of perturbative behaviour by calculating
the NNLO QCD corrections to the observables and analysing their dependence on
the non-physical renormalization scale as an indicator of the size of 
neglected
higher-order terms. The soft-drop versions of the observables listed  
have two parameters related to soft drop: $z_\mathrm{c}$ and $\beta$ \cite{Baron:2018nfz}.
This allows for optimisation in order to minimize the decrease in cross-section when the soft-drop procedure is applied. 

\begin{figure}
\centering
\begin{tabular}{cc}
\includegraphics[width=0.47\textwidth]{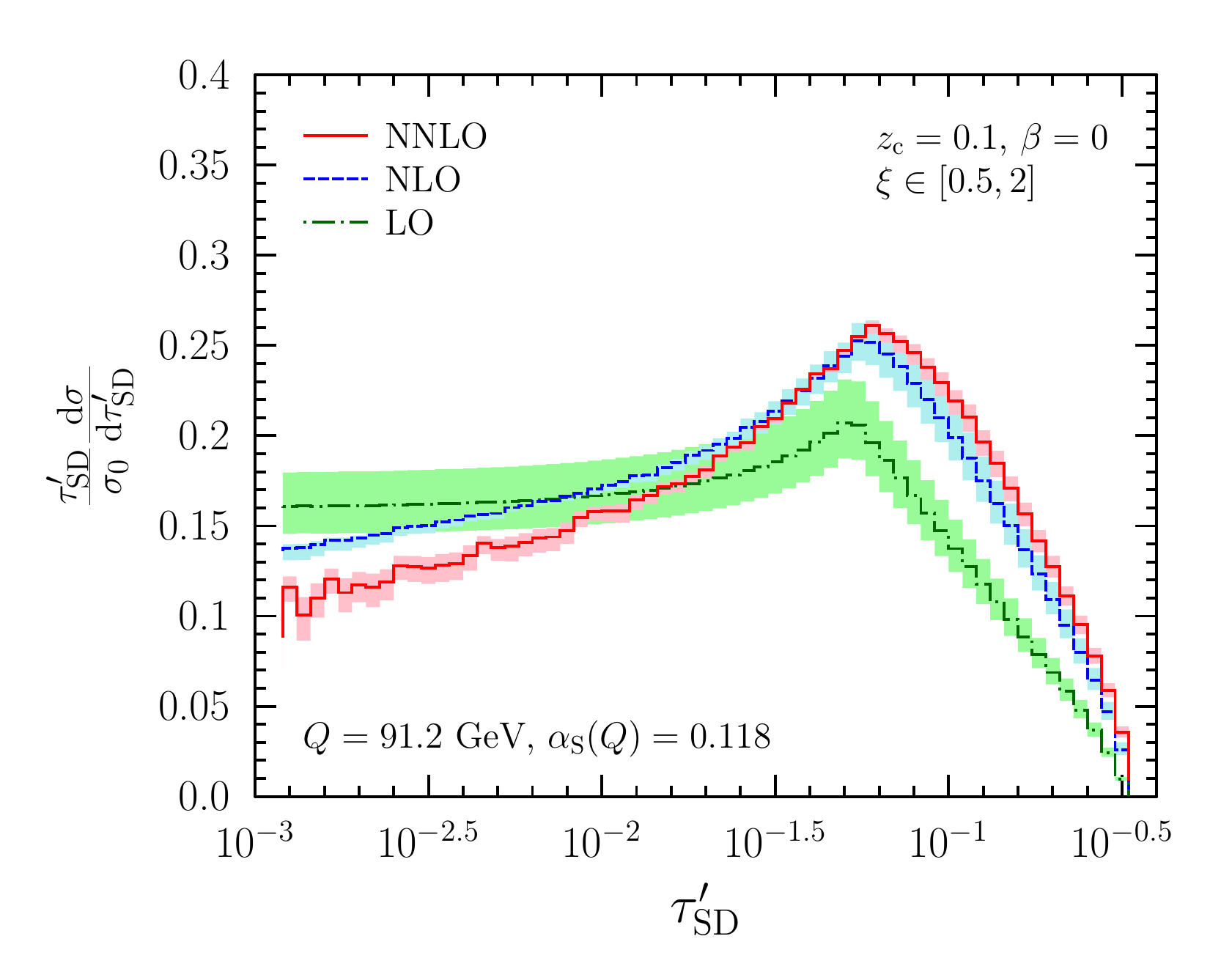} &
\includegraphics[width=0.47\textwidth]{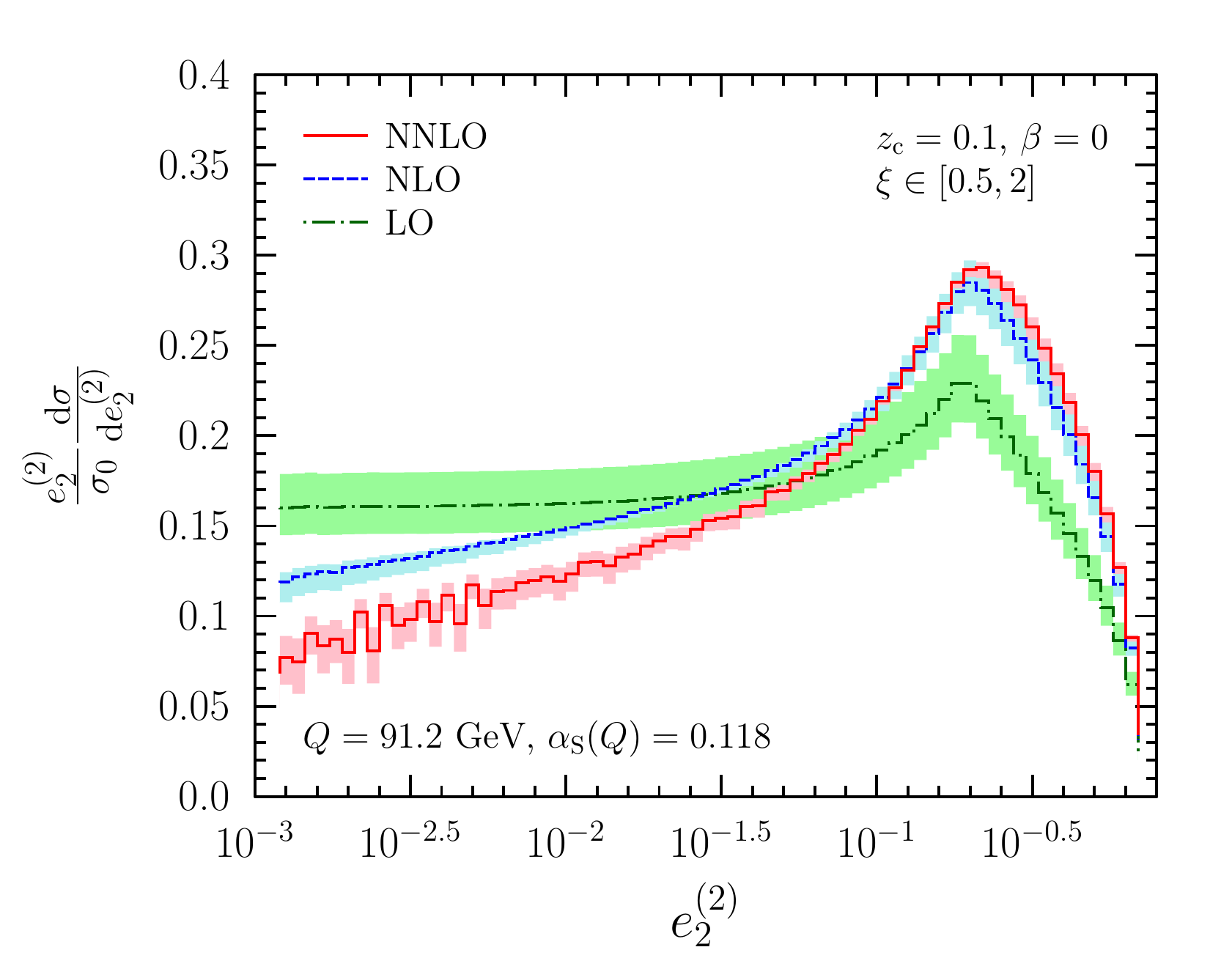}
\end{tabular}
\caption{{\label{fig:tauSD}} Left: Soft-dropped thrust distribution at the Z peak
in the first three orders of perturbation theory; the bands represent the 
uncertainty coming from the variation of the renormalization scale between
$Q/2$ and $2Q$. Right: The $K$ factors for the soft-dropped thrust distribution for
various choices of the soft-drop parameters.}
\end{figure}

Figure \ref{fig:tauSD} shows the soft-dropped thrust distribution  in the 
first three orders of QCD perturbation theory for a specific choice of the two
soft-drop related parameters. On the right-hand side of the  figure, the
$K$ factors are depicted for various parameter choices to illustrate the stability
of the result. We found that the most stable perturbation prediction and
moderate drop in cross-section can be achieved when $(z_\mathrm{c},\beta) = (0.1, 0)$.

\begin{figure}
\centering
\begin{tabular}{cc}
\includegraphics[width=0.47\textwidth]{SM_Kardos/e22SD-beta0p0-zcut0p1-logx-large} &
\includegraphics[width=0.47\textwidth]{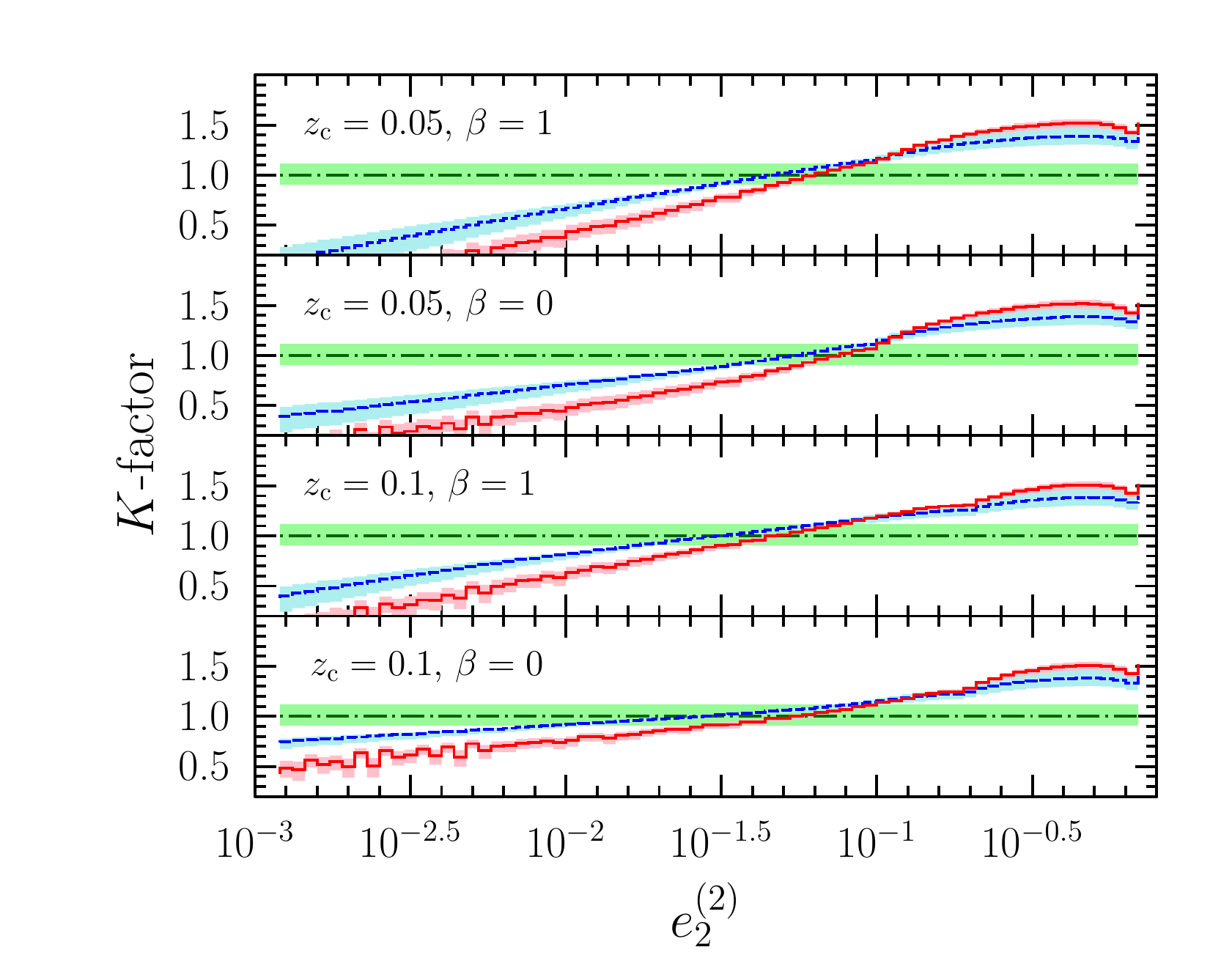}
\end{tabular}
\caption{{\label{fig:e22SD}} The same as \fig{fig:tauSD} but for the 
hemisphere jet mass}
\end{figure}

Figure \ref{fig:e22SD} depicts the soft-dropped hemisphere jet mass  in 
exactly the same way as the soft-dropped thrust shown in \Fref{fig:tauSD}. In this
case, it can be seen once more that the perturbative behaviour stabilizes
on going to higher orders in perturbation theory. This is most pronounced
at the left-hand side of the peak, where the NLO and NNLO predictions 
coincide. 
For this observable, we found that the best choice for the soft-drop 
parameters is also $(z_\mathrm{c},\beta) = (0.1, 0)$. For the traditional versions of
these observables, the peak region is the one where the all-order resummed
results and non-perturbative corrections must have agreement with
the experiment, but for the soft-dropped versions neither the 
higher-order contributions nor the non-perturbative corrections are drastic.
The minimal role of higher orders in perturbation theory can be seen from the 
perturbative
stability of our results, while the small size of non-perturbative 
corrections has been shown in Ref.~\cite{Baron:2018nfz}. These properties make the
soft-dropped event shapes attractive observables for the extraction of the
strong coupling.

\subsection{Conclusions}

A future electron--positron collider would be considered a dream machine
for many reasons. A machine of this type would allow for a precise tuning of
collision energy; it would have no annoying underlying event and it would
have coloured partons in the initial state. Several possible measurements could be 
envisioned 
at such a machine but from the QCD point of view, determination of the strong
coupling stands out. The strong coupling is a fundamental parameter of the
Standard Model of particle physics, so knowing its value is of key importance.

In this report, we showed two possible ways to conduct such a measurement.
First, it can be achieved by including higher-order corrections in the theoretical prediction
and comparing this with the experimental result modelling non-perturbative
effects with modern MC tools. Second, we showed modified
versions of well-known observables defined in electron--positron collisions 
where non-perturbative corrections can be minimized, hence diminishing the
effects of their uncertainties on theoretical predictions. These observables 
seem to be promising
candidates, not just for strong coupling measurements but also for the 
purpose of testing the Standard Model further. Thus, they should be seriously 
considered as important measurements at a future electron--positron facility.

\subsection*{Acknowledgements}
A.K. is grateful to the organisers of the CERN FCC-ee 2019 workshop %held at CERN
for the invitation to give a talk about recent advancements in QCD related to $\mathrm{e}^+\mathrm{e}^-$
%electron-positron 
colliders and for the warm atmosphere they created.

\end{bibunit}

\label{sec-sm-Kardos} 

\clearpage \pagestyle{empty}  \cleardoublepage
%============================================

\pagestyle{fancy}
\fancyhead[CO]{\thechapter.\thesection \hspace{1mm} Theoretical luminosity precision for the FCC-ee: overview of the path to 0.01\%}
\fancyhead[RO]{}
\fancyhead[LO]{}
\fancyhead[LE]{}
\fancyhead[CE]{}
\fancyhead[RE]{}
\fancyhead[CE]{B.F.L. Ward, S. Jadach,~W. P\l{}aczek,~M. Skrzypek, S.A. Yost}
\lfoot[]{}
\cfoot{-  \thepage \hspace*{0.075mm} -}
\rfoot[]{}

%=================================================================

%\vspace*{5.cm}

\begin{bibunit}[elsarticle-num] % utphys_spires define the bib-style for the unit: elsarticle-num.bst
%  text-1; this is the corresponding section
%\putbib[2loops] % the *.bib
%\end{bibunit}
% go-on
%--- from: bibunits.sty, adapts the font size of ``References'' to section
\let\stdthebibliography\thebibliography
\renewcommand{\thebibliography}{%
\let\section\subsection
\stdthebibliography}

\section[Theoretical luminosity precision for the FCC-ee: overview of the path to 0.01\% \\ 
{\it %B.F.L. Ward,~S. Jadach,~W. Placzek,~M. Skrzypek, and S.A. Yost
B.F.L. Ward,~S.~Jadach,~W. P\l{}aczek,~M. Skrzypek, S.A.~Yost}]
{Theoretical luminosity precision for the FCC-ee: overview of the path to 0.01\%
\label{contr:bhlumi}}
\noindent
{\bf Contribution\footnote{This contribution should be cited as:\\
B.F.L. Ward,~S.~Jadach,~W. P\l{}aczek,~M. Skrzypek, S.A.~Yost, Theoretical luminosity precision for the FCC-ee: overview of the path to 0.01\%,  
%04 DOI:10.23731/CYRM-2020-XXX.\thepage, in:
%04 \url{http://dx.doi.org/10.23731/CYRM-2020-XXX.\thepage}, in:
DOI: \href{http://dx.doi.org/10.23731/CYRM-2020-003.\thepage}{10.23731/CYRM-2020-003.\thepage}, in:
Theory for the FCC-ee, Eds. A. Blondel, J. Gluza, S. Jadach, P. Janot and T. Riemann,
\\CERN Yellow Reports: Monographs, CERN-2020-003,
%04 \url{http://dx.doi.org/10.23731/CYRM-2020-XXX}, p. \thepage.} 
DOI: \href{http://dx.doi.org/10.23731/CYRM-2020-003}{10.23731/CYRM-2020-003},
p. \thepage.
\\ \copyright\space CERN, 2020. Published by CERN under the 
%04-2
\href{http://creativecommons.org/licenses/by/4.0/}{Creative Commons Attribution 4.0 license}.} by: B.F.L. Ward,~S.~Jadach,~W. P\l{}aczek,~M. Skrzypek,  S.A.~Yost \\
Corresponding author: B.F.L.~Ward {[BFL\_Ward@baylor.edu]}}

\vspace*{.5cm}
%\centerline{\bf Abstract}
%% Text of abstract
\noindent We present an overview of the pathways to the required theoretical precision for the luminosity targeted by the FCC-ee precision studies. We put the discussion in context with a
brief review of the situation at the time of the LEP. We then present the current status and an overview of routes to the desired 0.01\% targeted by the FCC-ee (as well as by the ILC). 

 We use the situation that existed at the end of the LEP as our starting point. At the end of the LEP, the error budget for the \bhlumi4.04 MC used by all LEP collaborations to
simulate the luminosity process was that calculated in Ref.~\cite{Ward:1998ht}. For reference, we reproduce this result here in \Tref{tab:error99}.
In this table, we show the published works on which the various error estimates are based, as  discussed in Ref.~\cite{Ward:1998ht}.

%%%%%%%%%%%%%%%%%%%%%%%%%%%%%%%%%%%%%%%%%%%%%%%%%%%%%%%%%%%%%%%%%%%%%%%%%%%%%%%
%%%%%%%%%%%%%%%%%%%%%%%%%%%%%%%%%%%%%%%%%%%%%%%%%%%%%%%%%%%%%%%%%%%%%%%%%%%%%%%
\begin{table}[h!]
\centering
\caption{%\sf
Summary of the total (physical + technical) theoretical uncertainty
for a typical calorimetric detector.
For LEP1, this estimate is valid for a generic angular range
of   $1^{\circ}$--$3^{\circ}$ ($18$--$52$ mrad), and
for  LEP2 it is valid for energies up to $176$\,GeV and an
angular range of $3^{\circ}$--$6^{\circ}$.
Total uncertainty is taken in quadrature.
Technical precision is included in (a).
}
\begin{tabular}{llllll}
\hline \hline %%%%%%%%%%%%%%%%%%%%%%%%%%%%%%%%%%%%%%%%%%%%%%%%%%%%%%%%%%%%%%
    & \multicolumn{2}{l}{LEP1} 
              & \multicolumn{2}{l}{LEP2}
\\  %%%%%%%%%%%%%%%%%%%%%%%%%%%%%%%%%%%%%%%%%%%%%%%%%%%%%%%%%%%
Type of correction or error
    & 1996
         & 1999
              & 1996
                   & 1999 \\
                   & (\%)  & (\%)  & (\%)  & (\%) 
\\  \hline %%%%%%%%%%%%%%%%%%%%%%%%%%%%%%%%%%%%%%%%%%%%%%%%%%%%%%%%%%
%Technical Precision& -- & --& --& (0.03)\% 
%\\ %%%%%%%%%%%%%%%%%%%%%%%%%%%%%%%%%%%%%%%%%%%%%%%%%%%%%%%%%%%%%%%%%%
(a) Missing photonic ${\cal O}(\alpha^2 )$~\cite{Jadach:1995hy,Jadach:1999pf} 
    & 0.10%      
        & 0.027    
            & 0.20  
                & 0.04
\\ %%%%%%%%%%%%%%%%%%%%%%%%%%%%%%%%%%%%%%%%%%%%%%%%%%%%%%%%%%%%%%%%%%
(b) Missing photonic ${\cal O}(\alpha^3 L_e^3)$~\cite{Jadach:1996ir} 
    & 0.015%     
        & 0.015    
            & 0.03  
                & 0.03 
\\ %%%%%%%%%%%%%%%%%%%%%%%%%%%%%%%%%%%%%%%%%%%%%%%%%%%%%%%%%%%%%%%%%%
(c) Vacuum polarisation~\cite{Burkhardt:1995tt,Eidelman:1995ny} 
    & 0.04%     
        & 0.04    
           & 0.10  
                & 0.10 
\\ %%%%%%%%%%%%%%%%%%%%%%%%%%%%%%%%%%%%%%%%%%%%%%%%%%%%%%%%%%%%%%%%%%
(d) Light pairs~\cite{Jadach:1992nk,Jadach:1996ca} 
    & 0.03      
        & 0.03    
            & 0.05  
                & 0.05 
\\ %%%%%%%%%%%%%%%%%%%%%%%%%%%%%%%%%%%%%%%%%%%%%%%%%%%%%%%%%%%%%%%%%%
(e) Z and s-channel $\upgamma$~\cite{Jadach:1995hv,Arbuzov:1996eq}
    & 0.015      
        & 0.015   
            &  0.0  
                & 0.0 
\\  %%%%%%%%%%%%%%%%%%%%%%%%%%%%%%%%%%%%%%%%%%%%%%%%%%%%%%%%%%%
Total  
    & 0.11~\cite{Arbuzov:1996eq}
        & 0.061~\cite{Ward:1998ht}
            & 0.25~\cite{Arbuzov:1996eq}
                & 0.12~\cite{Ward:1998ht}
\\ \hline \hline %%%%%%%%%%%%%%%%%%%%%%%%%%%%%%%%%%%%%%%%%%%%%%%%%%%%%%%%%%%
\end{tabular}
\label{tab:error99}
\end{table}
%%%%%%%%%%%%%%%%%%%%%%%%%%%%%%%%%%%%%%%%%%%%%%%%%%%%%%%%%%%%%%%%%%%%%%%%%%%%%%%

One way to address the 0.01\% precision tag needed for the luminosity theory error for the FCC-ee is to develop the corresponding improved version
of the \bhlumi. This problem is addressed  in Ref.~\cite{Jadach:2018jjo}, wherein the path to 0.01\% theory precision for the FCC-ee luminosity is 
presented in some detail. The results of this latter reference are shown in \Tref{tab:lep2fcc}, wherein we also present the current state of the art for completeness, as  discussed in more detail in Ref.~\cite{Jadach:2018jjo}.

\begin{table}
\centering
\caption{%\sf
Anticipated total (physical + technical) theoretical uncertainty 
for a FCC-ee luminosity calorimetric detector with
 angular range  $64$--$86\,$mrad (narrow), near the Z peak.
Description of photonic corrections in square brackets is related to 
the second column.
The total error is summed in quadrature.
}
\label{tab:lep2fcc}
\begin{tabular}{llll}
\hline \hline
Type of correction or error
        & Update 2018
                &  FCC-ee forecast
\\ 
& (\%)\\
\hline %%%%%%%%%%%%%%%%%%%%%%%%%%%%%%%%%%%%%%%%%%%%%%%%%%%%%%%%%%
(a) Photonic $[{\cal O}(L_e\alpha^2 )]\; {\cal O}(L_e^2\alpha^3)$
        & \phantom{(}0.027
                &  $ 0.1 \times 10^{-4} $
\\ %%%%%%%%%%%%%%%%%%%%%%%%%%%%%%%%%%%%%%%%%%%%%%%%%%%%%%%%%%%%%%%%%
(b) Photonic $[{\cal O}(L_e^3\alpha^3)]\; {\cal O}(L_e^4\alpha^4)$
        &  \phantom{(}0.015
                & $ 0.6 \times 10^{-5} $
\\%%%%%%%%%%%%%%%%%%%%%%%%%%%%%%%%%%%%%%%%%%%%%%%%%%%%%%%%%%%%%%%%%
(c) Vacuum polarisation
        &  \phantom{(}0.014 \cite{JegerlehnerCERN:2016}
                & $ 0.6 \times 10^{-4} $
\\%%%%%%%%%%%%%%%%%%%%%%%%%%%%%%%%%%%%%%%%%%%%%%%%%%%%%%%%%%%%%%%%%%
(d) Light pairs
        &  \phantom{(}0.010 \cite{Montagna:1998vb,Montagna:1999eu}
                & $ 0.5 \times 10^{-4} $
\\%%%%%%%%%%%%%%%%%%%%%%%%%%%%%%%%%%%%%%%%%%%%%%%%%%%%%%%%%%%%%%%%%%
(e) Z and s-channel $\upgamma$ exchange
        &  \phantom{(}0.090 ~\cite{Jadach:1995hv}
                & $ 0.1 \times 10^{-4} $
\\ %%%%%%%%%%%%%%%%%%%%%%%%%%%%%%%%%%%%%%%%%%%%%%%%%%%%%%%%%%%%%%%%%%
(f) Up--down interference
    & \phantom{(}0.009 ~\cite{Jadach:1990zf}
        & $ 0.1 \times 10^{-4} $
\\%%%%%%%%%%%%%%%%%%%%%%%%%%%%%%%%%%%%%%%%%%%%%%%%%%%%%%%%%%%%%%%%%
(f) Technical precision & (0.027)  
                & $ 0.1 \times 10^{-4} $
\\  %%%%%%%%%%%%%%%%%%%%%%%%%%%%%%%%%%%%%%%%%%%%%%%%%%%%%%%%%
Total
        &  \phantom{(}0.097 
                & $ 1.0 \times 10^{-4} $
\\ \hline \hline %%%%%%%%%%%%%%%%%%%%%%%%%%%%%%%%%%%%%%%%%%%%%%%%%%%%%%%%%%
\end{tabular}
\end{table}
%%%%%%%%%%%%%%%%%%%%%%%%%%%%%%%%%%%%%%%%%%%%%%%%%%%%%%%%%%%%%%%%%%%
%%%%%%%%%%%%%%%%%%%%%%%%%%%%%%%%%%%%%%%%%%%%%%%%%%%%%%%%%%%%%%%%%%%
%%%%%%%%%%%%%%%%%%%%%%%%%%%%%%%%%%%%%%%%%%%%%%%%%%%%%%%%%%%%%%%%%%%

The key steps in arriving at \Tref{tab:lep2fcc} are as follows. The errors associated with the photonic corrections in lines (a) and (b) in the LEP results in \Tref{tab:error99} are due to effects that are known from Refs.~\cite{Jadach:1995hy,Jadach:1999pf,Jadach:1996ir} but  were not implemented into \bhlumi. In \Tref{tab:lep2fcc}, we show what these errors will become after these known results are included in \bhlumi,\ as discussed in Ref.~\cite{Jadach:2018jjo}. Similarly, in line (c) of \Tref{tab:error99}, the error is due to the uncertainty at the time of LEP on the hadronic contribution to the vacuum polarisation for the photon at the respective momentum transfers for the luminosity process; in \Tref{tab:lep2fcc}, we show the improvement of this error that is expected for the FCC-ee, as discussed in Refs.~\cite{JegerlehnerCERN:2016,Jegerlehner:2017zsb}.
\par
Continuing in this way, in line (d) in \Tref{tab:lep2fcc}, we show the expected  improvement \cite{Jadach:2018jjo}, with reference to the LEP time for \Tref{tab:error99}, in the light pairs error for the FCC-ee. As explained in Ref.~\cite{Jadach:2018jjo}, the complete matrix element for the additional real $\mathrm{e}^{+} \mathrm{e}^{-} $ pair radiation should be
used, because non-photonic graphs can contribute as much as $0.01\%$ for the cut-off, $z_{\rm cut} \sim 0.7$. This can be done with the MC generators developed for the $\mathrm{e}^+ \mathrm{e}^-\rightarrow 4\mathrm{f}$ processes for  LEP2 physics---see Ref.~\cite{Jadach:2018jjo} for further discussion. With known methods~\cite{Jadach:2018jjo}, the contributions of light quark pairs, muon pairs, and non-leading, non-soft additional $\mathrm{e}^+\mathrm{e}^- +n\upgamma$ corrections can be controlled such that the error on the pairs contribution is as given in line (d) for the FCC-ee. As noted, we also show the current state of the art~\cite{Jadach:2018jjo} for this error in line
(d) of \Tref{tab:lep2fcc}.

Turning to line (e) in \Tref{tab:lep2fcc},  we show the improvement of the error for the Z and s-channel $\upgamma$ exchange for the FCC-ee as well as its current state of the art.
In Ref.~\cite{Jadach:2018jjo}, a detailed discussion is presented of all of the six interference and three additional squared modulus terms  that result from the s-channel $\upgamma$,  s-channel Z, and t-channel Z exchange contributions to the amplitude for the luminosity process. It is shown that, if the predictions of \bhlumi\ for the luminosity
measurement at FCC-ee are combined with those  from \bhwide\ \cite{Jadach:1995nk} for
this Z and s-channel $\upgamma$ exchange contribution, then the error in the second column of  line (e) of
\Tref{tab:lep2fcc} could be reduced to $0.01\%$. To reduce the uncertainty of this contribution
practically to zero we would include these Z and $\upgamma_\mathrm{s}$ exchanges
within the CEEX-type matrix element at \order{\alpha^1} in \bhlumi~\cite{Jadach:2000ir}. Here, CEEX stands for coherent exclusive exponentiation, which acts at the level of the amplitudes,
as compared with the original Yennie--Frautschi--Suura~\cite{Yennie:1961ad}
(YFS)  exclusive exponentiation (EEX), which is used in \bhlumi4.04 and which acts at the level of the squared amplitudes.
It is expected to be enough to add the EW corrections
to the  large angle Bhabha (LABH) process in the form of effective couplings in the Born amplitudes. This leads to the error estimate shown in \Tref{tab:lep2fcc} in line
(e) for the FCC-ee.

For completeness,  we note that for our discussion of the Z- and s-channel $\upgamma$ exchanges we made \cite{Jadach:2018jjo} a numerical study using \bhwide\
for the the calorimetric LCAL-type
detector, as described in Ref.~\cite{Jadach:1991cg},
for the symmetric angular range $64$--$86\,$mrad without
any cut in acoplanarity. The pure weak corrections were calculated with the  {\tt ALIBABA} 
EW library~\cite{Beenakker:1990mb,Beenakker:1990es}. The results, shown in \Tref{tab:Zsgam},
were obtained for three values of the centre-of-mass (CM)
energy: $E_{\rm CM} = M_\mathrm{Z},\, M_\mathrm{Z}\pm 1\,$GeV, where the latter two values have Z contributions that are close to maximal in size. The results in the second column for the total size of the  Z and $\upgamma_\mathrm{s}$ exchanges are consistent with our expectations, as explained in Ref.~\cite{Jadach:2018jjo}: the contribution is positive below the Z peak, where it reaches a size $\sim$$0.64\%$, is close to zero near the peak, and changes sign above the peak, where it reaches a size $\sim$$-0.72\%$.
The third column features the fixed-order (non-exponentiated) ${\cal O}(\alpha)$ QED correction and shows that it is sizeable and up to a half of the size of the Born level effect,
with a sign that is opposite to that of the latter effect. The fourth column shows the size of the higher-order QED effects from YFS exponentiation, which also change their sign near the Z peak, in opposition to the corresponding change of the ${\cal O}(\alpha)$ corrections. We see that the size of the former effects is about
a quarter of that of the latter. The effects in the fourth column allow us to make a conservative estimate of the size of the missing higher-order QED effects in \bhwide\ using the big log factor $\gamma= {\alpha} \ln
({|\bar{t}|} / {m_\mathrm{e}^2}) / \uppi =0.042$ from Section 4 of Ref.~\cite{Jadach:2018jjo} and the safety factor of 2 from Ref.~\cite{Jadach:1995hv}, together with the largest higher-order effect  in \Tref{tab:Zsgam}, $0.081\%$, as $0.081\%\times \gamma \times 2 \simeq 0.007\%$. The last column shows that the size of the pure weak corrections, as implemented within the ${\cal O}(\alpha)$ YFS exponentiation scheme, is at a level of $0.01\%$ below and at $M_\mathrm{Z}$ and increases up to $\sim$$0.04\%$ above $M_\mathrm{Z}$. We may use the same factor as we did for the higher-order corrections to estimate the size of the missing higher-order pure weak corrections in \bhwide\ as $\sim$$0.003\%$. Altogether, by adding the two estimates of its massing effects, we obtain a conservative estimate of $0.01\%$ for the physical precision of \bhwide\ to justify our remarks  concerning the error in line (e) of \Tref{tab:lep2fcc} that would result from the combination of the prediction of \bhlumi\ and that of \bhwide\ for this contribution. 

\begin{table}
\centering
\caption{%\sf
Results from \bhwide\ for the Z and $\upgamma_\mathrm{s}$ exchange
contributions to the FCC-ee luminosity with respect to the
$\upgamma_\mathrm{t}\otimes\upgamma_\mathrm{t}$ process for the calorimetric
LCAL-type detector \cite{Jadach:1991cg} 
with the symmetric angular range $64$--$86\,$mrad; 
no acoplanarity cut was applied.
MC errors are marked in brackets.
}
\label{tab:Zsgam}
\begin{tabular}{lllll}
\hline \hline %%%%%%%%%%%%%%%%%%%%%%%%%%%%%%%%%%%%%%%%%%%%%%%%%%%%%%%%%%%
$E_{\rm CM}$
    & $\Delta_{\rm tot}$
         & $\updelta_{{\cal O}(\alpha)}^{\rm QED}$ 
              & $\updelta_{\rm h.o.}^{\rm QED}$ 
                   & $\updelta_{\rm tot}^{\rm weak}$ 
\\ 
(GeV) & (\%) & (\%)  & (\%)  & (\%) \\
 \hline %%%%%%%%%%%%%%%%%%%%%%%%%%%%%%%%%%%%%%%%%%%%%%%%%%%%%%%%%%
%\\ %%%%%%%%%%%%%%%%%%%%%%%%%%%%%%%%%%%%%%%%%%%%%%%%%%%%%%%%%%%%%%%%%%
$90.1876$       
  & $+0.642\,(12)$ 
    & $-0.152\,(59)$
      & $+0.034\,(38)$
        & $-0.005\,(12)$
\\ %%%%%%%%%%%%%%%%%%%%%%%%%%%%%%%%%%%%%%%%%%%%%%%%%%%%%%%%%%%%%%%%%%
%\\ %%%%%%%%%%%%%%%%%%%%%%%%%%%%%%%%%%%%%%%%%%%%%%%%%%%%%%%%%%%%%%%%%%
$91.1876$       
  & $+0.041\,(11)$ 
    & $+0.148\,(59)$
      & $-0.035\,(38)$
        & $+0.009\,(12)$
\\ %%%%%%%%%%%%%%%%%%%%%%%%%%%%%%%%%%%%%%%%%%%%%%%%%%%%%%%%%%%%%%%%%%
%\\ %%%%%%%%%%%%%%%%%%%%%%%%%%%%%%%%%%%%%%%%%%%%%%%%%%%%%%%%%%%%%%%%%%
$92.1876$       
  & $-0.719\,(13)$ 
    & $+0.348\,(59)$
      & $-0.081\,(38)$
        & $+0.039\,(13)$
\\ \hline \hline %%%%%%%%%%%%%%%%%%%%%%%%%%%%%%%%%%%%%%%%%%%%%%%%%%%%%%%%%%%
\end{tabular}
%30Aug \setcounter{table}{0}
\end{table}

In line (f) in \Tref{tab:lep2fcc}, we show the estimate of the error in the up--down interference between radiation from the $\mathrm{e}^-$ and $\mathrm{e}^+$  lines. Unlike in LEP1, where it was negligible, for the FCC-ee, this effect, calculated in Ref.~\cite{Jadach:1990zf} at ${\cal O}(\alpha^1)$, is ten times larger and must be included in the upgraded \bhlumi. Once this is done, the error estimate shown in line (f) for the FCC-ee is obtained~\cite{Jadach:2018jjo}.

This brings us to the issue of the technical precision. In an ideal situation,
 to get the upgraded \bhlumi's technical precision
at a level of $10^{-5}$ for the total cross-section and $10^{-4}$
for single differential distributions, one would need to compare
it with another MC program developed independently,
which properly implements the soft-photon resummation,
LO corrections up to \order{\alpha^3 L_e^3}, and
the second-order corrections with the complete \order{\alpha^2 L_e}.
In principle, an extension of a program like \babayaga ~\cite{CarloniCalame:2000pz,CarloniCalame:2001ny,Balossini:2006wc}, which is currently exact at NLO with a matched QED shower,
to the level of NNLO for the hard process, while
keeping the correct soft-photon resummation,
would provide the best comparison with the upgraded \bhlumi\ to establish the
technical precision of both programs at the $10^{-5}$
precision level.%
\footnote{The upgrade of the \bhlumi\ distributions will be
  relatively straightforward because its multiphoton 
  phase space is exact~\cite{Jadach:1999vf} for any number of photons.}
During the intervening time period, a very good test of the technical precision of the upgraded \bhlumi\ would follow from the comparison of its results with EEX and CEEX matrix elements; 
the basic multiphoton phase space integration module of \bhlumi\
was already well tested in Ref.~\cite{Jadach:1996bx}
and such a test can be repeated at an even higher precision level.

In summary,  we conclude that, with the appropriate resources, the path to $0.01\%$ precision for the FCC-ee luminosity (and the ILC luminosity) at the Z peak is open
via an upgraded version of \bhlumi.

\end{bibunit}

\label{sec-sm-bward} 

\clearpage \pagestyle{empty}  \cleardoublepage
%============================================

\pagestyle{fancy}
\fancyhead[CO]{\thechapter.\thesection \hspace{1mm} ${ \mathrm{e}^+  \mathrm{e}^-  \to  \upgamma  \upgamma}$ at large angles for FCC-ee luminometry}
\fancyhead[RO]{}
\fancyhead[LO]{}
\fancyhead[LE]{}
\fancyhead[CE]{}
\fancyhead[RE]{}
\fancyhead[CE]{C.M. Carloni, M.~Chiesa, G.~Montagna, O.~Nicrosini, F.~Piccinini}
\lfoot[]{}
\cfoot{-  \thepage \hspace*{0.075mm} -}
\rfoot[]{}

%=================================================================

%\vspace*{5.cm}

\begin{bibunit}[elsarticle-num] % define the bib-style for the unit: elsarticle-num.bst
%  text-1; this is the corresponding section
%\putbib[2loops] % the *.bib
%\end{bibunit}
% go-on
%--- from: bibunits.sty, adapts the font size of ``References'' to section
\let\stdthebibliography\thebibliography
\renewcommand{\thebibliography}{%
\let\section\subsection
\stdthebibliography}

\newcommand{\di}{\mathrm{d}}
\newcommand{\modulo}[1]{\left |#1\right |}
\newcommand{\sqmodulo}[1]{{\modulo{#1}}^{2}}
\newcommand{\M}{{\cal{M}}}

\def\bit{\begin{itemize}}    \def\eit{\end{itemize}}
\def\ben{\begin{enumerate}}  \def\een{\end{enumerate}}
\def\bce{\begin{center}}    \def\ece{\end{center}}

\section[$\mathrm{e}^+  \mathrm{e}^-  \to  \upgamma  \upgamma$ at large angles for FCC-ee luminometry \\ {\it C.M.~Carloni, M.~Chiesa, G.~Montagna, O.~Nicrosini, F.~Piccinini}]
{$\mathrm{e}^+ \mathrm{e}^-  \to  \upgamma  \upgamma$ 
at large angles for FCC-ee luminometry
\label{contr:gg}
}
\noindent
{\bf Contribution\footnote{This contribution should be cited as:\\
C.M.~Carloni, M.~Chiesa, G.~Montagna, O.~Nicrosini, F.~Piccinini, $\mathrm{e}^+ \mathrm{e}^-  \to  \upgamma  \upgamma$ 
at large angles for FCC-ee luminometry,  
%04 DOI:10.23731/CYRM-2020-XXX.\thepage, in:
%04 \url{http://dx.doi.org/10.23731/CYRM-2020-XXX.\thepage}, in:
DOI: \href{http://dx.doi.org/10.23731/CYRM-2020-003.\thepage}{10.23731/CYRM-2020-003.\thepage}, in:
Theory for the FCC-ee, Eds. A. Blondel, J. Gluza, S. Jadach, P. Janot and T. Riemann,
\\CERN Yellow Reports: Monographs, CERN-2020-003,
%04 \url{http://dx.doi.org/10.23731/CYRM-2020-XXX}, p. \thepage.} 
DOI: \href{http://dx.doi.org/10.23731/CYRM-2020-003}{10.23731/CYRM-2020-003},
p. \thepage.
\\ \copyright\space CERN, 2020. Published by CERN under the 
%04-2
\href{http://creativecommons.org/licenses/by/4.0/}{Creative Commons Attribution 4.0 license}.} by: C.M.~Carloni, M.~Chiesa, G.~Montagna, O.~Nicrosini, F.~Piccinini
\\ Corresponding author: C.M. Carloni 
{[carlo.carloni.calame@pv.infn.it]}}
\vspace*{.5cm}
  
\section*{Abstract}
We examine large-angle two-photon production in $\mathrm{e}^+ \mathrm{e}^-$ annihilation as a possible process 
to monitor the luminosity of the FCC-ee. We review the current status of the theoretical predictions
and perform an exploratory phenomenological study of the next-to-leading and higher-order QED 
corrections using the Monte Carlo event generator BabaYaga@NLO. 
We also consider the one-loop weak corrections, which are necessary to meet
the high-precision requirements of the FCC-ee. Possible ways to 
approach the target theoretical accuracy are sketched.

\subsection{Introduction}

The successful accomplishment of the FCC-ee physics goals requires a 
detailed knowledge of the collider luminosity. 
The ambitious FCC-ee target is a luminosity measurement with a total error 
of the order of $10^{-4}$ (or even better) 
and calls for a major effort by both the experimental and theoretical community. 

At the FCC-ee, the standard luminosity process is expected to be 
small-angle Bhabha scattering, likewise at the LEP. However,  the process of large-angle 
two-photon production, \ie $\mathrm{e}^+ \mathrm{e}^- \to \upgamma\upgamma$, has also been recently 
proposed as a possible alternative normalization process for FCC-ee 
operation~\cite{Janot:2015,Dam:2016,Carloni:2019}.  
Actually, this is a purely QED process at leading order 
at any energy; it receives QED corrections from the initial state only 
and does not contain at order $\alpha$ the contribution due to the vacuum polarisation 
(in particular, hadronic loops), which enters at 
next-to-next-to-leading-order (NNLO) only. Conversely, 
the cross-section of $\mathrm{e}^+ \mathrm{e}^- \to \upgamma\upgamma$ is significantly smaller than that of
small-angle Bhabha scattering but adequate everywhere at the FCC-ee, 
with the exception of the running at the Z resonance. Moreover, the process 
is affected by a large background, owing to large-angle Bhabha scattering.

In spite of these limitations, the possibility of using photon-pair production as 
a luminosity process at the FCC-ee is an interesting option to be pursued. 
Contrarily to Bhabha scattering, which received a lot of attention over the past decades, 
there is  rather scant theoretical literature about $\mathrm{e}^+ \mathrm{e}^- \to \upgamma\upgamma$ annihilation and the most recent 
phenomenological results refer to $\mathrm{e}^+ \mathrm{e}^-$ colliders of 
moderate energies~\cite{Arbuzov:1997pj,Balossini:2008xr,Actis:2010gg,Eidelman:2010fu}. 
Moreover, the few available Monte Carlo (MC) generators~\cite{Eidelman:2010fu,Balossini:2008xr}
are tailored for low-energy accelerators and need to be improved
for the high-energy, high-precision requirements of the FCC-ee.

In this contribution, we provide a first assessment of 
the current status of the theoretical accuracy for large-angle 
two-photon production at FCC-ee energies. 
For this purpose, we use the MC program 
\textsc{BabaYaga@nlo}~\cite{CarloniCalame:2000pz,CarloniCalame:2001ny,CarloniCalame:2003yt,Balossini:2006wc,Balossini:2008xr},
 which includes next-to-leading-order (NLO) QED corrections matched to a QED parton shower, 
 and compute the one-loop weak corrections from heavy boson exchange. The
QED corrections to 
$\mathrm{e}^+ \mathrm{e}^- \to \upgamma\upgamma$ at order $\alpha$ were previously 
calculated some time ago~\cite{Berends:1973tm,Eidelman:1978rw,Berends:1980px}
and NLO electroweak corrections are reported in Refs.~\cite{CapdequiPeyranere:1978ce,Bohm:1986dn,Fujimoto:1986xb}. 
A generator based on Ref. \cite{Berends:1980px} was used at LEP for the 
analysis of photon-pair production at energies above the Z~\cite{Alcaraz:2006mx}.
Here, we perform an exploratory phenomenological study of the QED corrections at 
NLO and evaluate the impact of higher-order contributions due to multiple photon emission, 
by considering typical values for the c.m. energies of the FCC-ee.
Possible perspectives to achieve the target theoretical accuracy are briefly outlined.

\subsection{Theoretical approach and numerical results}

According to the theoretical formulation implemented in \textsc{BabaYaga@nlo}, the photonic corrections
are computed  using a fully exclusive QED parton shower matched to QED 
contributions at NLO. The matching of the parton shower ingredients with the NLO QED corrections 
is realised in such a way that its $O(\alpha)$ expansion reproduces the NLO cross-section, and 
exponentiation of the leading contributions owing to soft and collinear radiation is preserved, as in a pure parton shower algorithm. 
Various studies and comparisons with independent calculations \cite{Balossini:2006wc,Actis:2010gg,CarloniCalame:2011zq} 
showed that this formulation enables a theoretical accuracy at a level of 0.1\% (or slightly better)
for the calculation of integrated cross-sections.

To meet the high-precision requirements of FCC-ee, we 
also computed the one-loop weak corrections due to 
heavy boson exchange. The calculation was 
performed by treating the ultraviolet divergencies in dimensional regularisation and using the 
computer program \textsc{Recola}~\cite{Actis:2016mpe}, which internally adopts the \textsc{Collier}~\cite{Denner:2016kdg} 
library for the evaluation of 
one-loop scalar and tensor integrals. In our calculation, we used  the on-shell 
renormalization scheme, with complex mass values for the heavy boson masses~\cite{Denner:2006ic}.

In the following, we show a sample of 
numerical results obtained using the code\\ \textsc{BabaYaga@nlo}. 
They refer to four canonical c.m. energy values, which are representative of the expected FCC-ee operation 
programme (Z pole, WW, ZH, and $\mathrm{t \bar{t}}$ thresholds)
\begin{equation}
\sqrt{s}=91,\ 160,\ 240,\ 365\,\UGeV
\end{equation}

To study the effects due to the QED corrections, we consider 
a simulation set-up, in which we require at least two photons within the angular acceptance 
$20^\circ \leq \theta_\upgamma \leq 160^\circ$ with energy $E_\upgamma\ge 0.25\times\sqrt{s}$. 
In \Tref{Tab:Tab2}, we examine the impact of the QED radiative 
corrections on the integrated cross-sections, when considering these kinematic cuts.

\begin{table}
\caption{Two-photon production cross-section at LO, NLO, and  higher-order QED corrections  
for four FCC-ee c.m. energies. Numbers in parentheses are the 
relative contributions of NLO and higher-order QED corrections.}
\label{Tab:Tab2}
\bce
\begin{tabular}[c]{llll}
\hline \hline
$\sqrt{s}$  & LO  & NLO  & Higher-order \\
(GeV) & (pb) & (pb) & (pb) \\
\hline
     \phantom{3}$91$  & $39.821$ & $41.043$  $[+3.07\%]$ & $40.868(3)$    $[-0.44\%]$\\ 
$160$ & $12.881$ & $13.291$  $[+3.18\%]$ & $13.228(1)$    $[-0.49\%]$\\ 
$240$ & $\phantom{3}5.7250$ & $\phantom{3}5.9120$  $[+3.26\%]$ & $\phantom{3}5.884(2)$  \; $[-0.49\%]$\\ 
$365$ & $\phantom{3}2.4752$ & $\phantom{3}2.5582$  $[+3.35\%]$ & $\phantom{3}2.5436(2)$    $[-0.59\%]$\\ 
\hline \hline
\end{tabular}
\ece
\end{table}

The photon-pair production cross-section is shown 
for different accuracy levels, \ie at LO, NLO QED, and including
higher-order contributions due to multiphoton radiation. The numbers in parentheses are the 
relative contributions due to NLO and higher-order QED corrections, respectively. 
It can be observed that the NLO corrections are at the level of a few percent, while 
the higher-order contributions amount to about 5\textperthousand\   and 
reduce the effect due to $O(\alpha)$ corrections.

\begin{figure}
\centering
\includegraphics[width=9.cm]{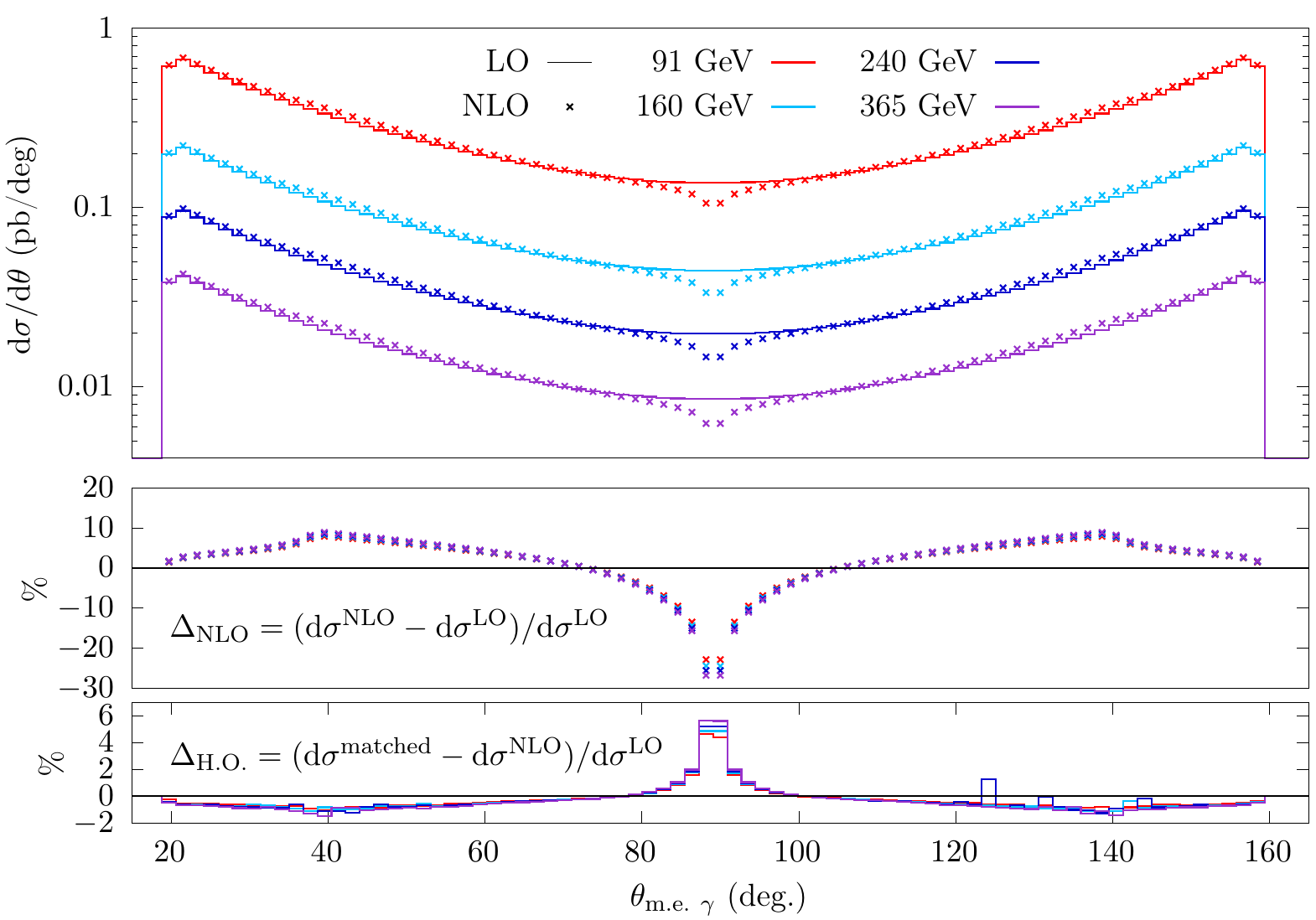}
\caption{Top:  Angular distribution of the most energetic photon, 
for four FCC-ee c.m. energies. Bottom: 
Relative contributions of NLO and higher-order QED corrections.}
\label{Fig:Fig5}
%\query{Please correct the figure labels in \Fref{Fig:Fig5}. Set variables
%in italic font, but the differential d and  text
%labels  in roman font.}
\end{figure}

A representative example of the effects due to QED corrections on the differential cross-sections is given in \Fref{Fig:Fig5}, 
which shows the angular distribution of the most energetic photon for the four energy points. 
One can see that the NLO corrections are particularly important in the central region, where they 
reach the 20--30\% level, being mainly due to soft-photon radiation. This effect is partially compensated for by the higher-order
corrections, which amount to some percent in the same region. 

We also preliminarily explored the contribution of one-loop weak corrections, to conclude that 
their size is at the percentage level, \ie roughly as large as QED contributions beyond NLO. 
A more detailed investigation of their effects is being made.
          
\subsection{Summary and outlook}

We have examined large-angle two-photon production in $\mathrm{e}^+ \mathrm{e}^-$ annihilation as a possible process 
to monitor the luminosity at the FCC-ee. We have assessed the present 
status of the theoretical accuracy through an exploratory phenomenological study of the radiative 
corrections to $\mathrm{e}^+ \mathrm{e}^- \to \upgamma\upgamma$ annihilation at the c.m. energies of main interest. 
To this end, we have improved the theoretical content of the code \textsc{BabaYaga@nlo}, which 
includes exact NLO QED corrections matched to parton shower, by computing the weak corrections due to the presence 
of heavy bosons in the internal loops.

The accuracy of the present calculation can be estimated to be at the 0.1\% level or 
slightly better. A first way to improve it is given by the calculation of NNLO fermion 
loop contributions, accompanied by the computation of the same-order real pair 
corrections, along the lines described in Refs.~\cite{Montagna:1998vb,CarloniCalame:2011zq}. This 
should be sufficient to get close to an accuracy at the $10^{-4}$ level. 
Beyond that, a full calculation of NNLO QED corrections and, eventually, of two-loop
weak contributions will  ultimately be needed to reach the challenging frontier of 
the 10\,ppm theoretical accuracy. These developments are  now under consideration.

\end{bibunit}

\label{sec-sm-gg} 

\clearpage \pagestyle{empty}  \cleardoublepage
%============================================

\pagestyle{fancy}
\fancyhead[CO]{\thechapter.\thesection \hspace{1mm} Prospects for higher-order corrections to $\PW$ pair
production near threshold}
\fancyhead[RO]{}
\fancyhead[LO]{}
\fancyhead[LE]{}
\fancyhead[CE]{}
\fancyhead[RE]{}
\fancyhead[CE]{C. Schwinn}
\lfoot[]{}
\cfoot{-  \thepage \hspace*{0.075mm} -}
\rfoot[]{}
    \begin{bibunit}[elsarticle-num] % define the bib-style for the unit: elsarticle-num.bst
%  text-1; this is the corresponding section
%\putbib[2loops] % the *.bib
%\end{bibunit}
% go-on
%--- from: bibunits.sty, adapts the font size of ``References'' to section
\let\stdthebibliography\thebibliography
\renewcommand{\thebibliography}{%
\let\section\subsection
\stdthebibliography}

\section
[Prospects for higher-order corrections to $\PW$ pair
production near threshold in the EFT approach \\
{\it C. Schwinn}]
{Prospects for higher-order corrections to $\PW$ pair
  production near threshold in the EFT approach
\label{contr:schwinn}}
\noindent
{\bf Contribution\footnote{This contribution should be cited as:\\
C. Schwinn, Prospects for higher-order corrections to $\PW$ pair
  production near threshold in the EFT approach,  
%04 DOI:10.23731/CYRM-2020-XXX.\thepage, in:
%04 \url{http://dx.doi.org/10.23731/CYRM-2020-XXX.\thepage}, in:
DOI: \href{http://dx.doi.org/10.23731/CYRM-2020-003.\thepage}{10.23731/CYRM-2020-003.\thepage}, in:
Theory for the FCC-ee, Eds. A. Blondel, J. Gluza, S. Jadach, P. Janot and T. Riemann,
\\CERN Yellow Reports: Monographs, CERN-2020-003,
%04 \url{http://dx.doi.org/10.23731/CYRM-2020-XXX}, p. \thepage.} 
DOI: \href{http://dx.doi.org/10.23731/CYRM-2020-003}{10.23731/CYRM-2020-003},
p. \thepage.
\\ \copyright\space CERN, 2020. Published by CERN under the 
%04-2
\href{http://creativecommons.org/licenses/by/4.0/}{Creative Commons Attribution 4.0 license}.} by: C. Schwinn 
%\\ Corresponding Author: Christian Schwinn 
{[schwinn@physik.rwth-aachen.de]}}
\vspace*{.5cm}

\noindent The precise measurement of the mass of the $\PW$ boson plays an
essential role for precision tests of the Standard Model (SM) and indirect
searches for new physics through global fits to electroweak observables. Cross-section measurements near the $\PW$ pair production threshold at a possible
future $\Pem\Pep$ collider promise to reduce the experimental uncertainty to
the level of $3\UMeV$ at an International Linear
Collider~(ILC)~\cite{Baer:2013cma,Baak:2013fwa}, while a high-luminosity
circular collider offers a potential improvement  to $0.5\UMeV$ in
%Mangano:2018mur->Abada:2019lih
the case of the FCC-ee~\cite{Azzi:2017iih,Abada:2019lih} or $1\UMeV$ at the
CEPC~\cite{CEPCStudyGroup:2018ghi}.  At the point of highest
sensitivity, an uncertainty in the cross-section measurement of $0.1\%$
translates to an uncertainty of $\sim$$1.5\UMeV$ on
$\MW$~\cite{Azzi:2017iih}. Therefore, a theoretical prediction for the cross-section with an accuracy of $\Delta\sigma\sim 0.01\%$ at threshold is required
to fully exploit the potential of a future circular $\Pem\Pep$ collider.
Theory predictions using the double-pole
approximation~(DPA)~\cite{Aeppli:1993rs} at next-to-leading
order~(NLO)~\cite{Jadach:1996hi,Jadach:2001uu,Beenakker:1998gr,Denner:1999kn,Denner:2000bj}
successfully described LEP2 results with an accuracy of better than $1\%$
above threshold. An extension of the DPA to NNLO appears to be appropriate for
a future $\Pem\Pep$  collider operating above the $\PW$ pair
threshold, \eg for the interpretation of anomalous triple-gauge-coupling
measurements at $\sqrt s=240\UGeV$.  However, the accuracy of the DPA at NLO
degrades to $2$--$3\%$ near the threshold. In this region, the combination of a full
NLO calculation of four-fermion production~\cite{Denner:2005es,Denner:2005fg} with leading
NNLO effects obtained using effective field theory~(EFT)
methods~\cite{Beneke:2007zg,Actis:2008rb} reduces the theory uncertainty of
the total cross-section to below
$0.3\%$; sufficient for the ILC target uncertainty but far above that of
the FCC-ee.  This raises the question of the methods required to reach a theory
accuracy $\sim$$0.01\%$. In this contribution, this issue is addressed from the EFT point of
view.  The discussion is limited to the total cross-section, where the EFT
approach is best developed so far, although cuts on the $\PW$ decay products
can also be incorporated~\cite{Actis:2008rb}.  To reach the target accuracy,
it will also be essential to have theoretical control of effects beyond the
pure electroweak effects considered here. In particular, it is assumed that
next-to-leading logarithmic corrections $(\alpha/\uppi)^2 \ln(\me^2/s)$ from
collinear initial-state photon radiation~(ISR), which have been estimated to
be $\lesssim$$0.1\%$~\cite{Denner:2005es}, will be resummed to all orders.  The QCD
effects, which are particularly important for the fully hadronic decay modes,
are only briefly considered.  In Section~\ref{sec:eft:schwinn}, aspects of the EFT
approach are reviewed
from an updated perspective using insight into the
factorisation of soft, hard, and Coulomb corrections~\cite{Beneke:2010da}.
The NLO and leading NNLO results are summarised and compared with the
NLO$^{\text{ee4f}}$ calculation~\cite{Denner:2005es}.
In
Section~\ref{sec:nnlo}, the structure of the EFT expansion and 
calculations  of subsets of corrections are used to estimate
the magnitude of the NNLO and leading N$^3$LO corrections and to determine
whether such
calculations are sufficient to meet the FCC-ee target accuracy.
%=================================================================
\subsection{Effective theory approach to $\PW$ pair production}
\label{sec:eft:schwinn}

In the EFT approach to four-fermion production near the
$\PW$ pair production 
threshold~\cite{Beneke:2007zg}, the cross-section is expanded simultaneously
in the coupling, the $\PW$ decay width, and the energy relative to the
production threshold, which are taken to be of similar order and are denoted
collectively by
\begin{equation}
  \label{eq:counting}
\delta\sim v^2\equiv \frac{(s-4\MW^2)}{\MW^2}\sim\frac{\Gamma_{\PW}}{\MW}\sim \alpha.
\end{equation}
An NNLO$^{\text{EFT}}$ calculation includes corrections up to
$\mathcal{O}(\delta^n)$, whereas, as usual, NNLO refers to 
the $\mathcal{O}(\alpha^n)$ corrections.
 As discussed in Sections 7.1.1 and 7.1.2, non-resonant and Coulomb
corrections lead to odd powers of $v$, so that the expansion proceeds in
half-integer powers of $\delta$. 
The current state of the art in the EFT is the calculation of the total cross-section for the semi-leptonic
final state $\upmu^- \overline{\upnu}_\upmu \Pqu  \Paqd$ up to  NLO$^{\text{EFT}}$~\cite{Beneke:2007zg},
which includes  corrections of the order 
\begin{align}
\label{eq:nlo-power}
  \text{NLO}^{\text{EFT}}&: v^2, \alpha, \alpha^2/v^2,
\end{align}
supplemented with the genuine
$\mathcal{O}(\alpha^2,\alpha^3)$ corrections at the next order, $\delta^{3/2}$,  in the 
 $\delta$-expansion~\cite{Actis:2008rb},
 \begin{equation}
\label{eq:three-half-power}
 \mbox{N}^{3/2}\mbox{LO}^{\text{EFT}}: \quad \alpha v,\quad \alpha^2/v, \quad \alpha^3/v^3.
\end{equation}
In the following, aspects of these results and the EFT method are reviewed that are useful for
the estimate of NNLO$^{\text{EFT}}$ corrections and the remaining uncertainty.

%=================================================================
\subsubsection{Expansion of the Born cross-section}
\label{sec:born}

The total cross-section $\Pem \Pep\to 4\mathrm{f}$ can be obtained
from the imaginary part of the forward-scattering amplitude $\Pem\Pep\to \Pem
\Pep$, where the Cutkosky cuts are restricted to those with  four-fermion final
states. Flavour-specific final states can be selected accordingly.
The expansion of the forward-scattering amplitude in $\delta$ can be
formulated in terms of an EFT~\cite{Beneke:2003xh,Beneke:2004km,Beneke:2007zg},
where the initial-state leptons are described by
soft-collinear effective theory~\cite{Bauer:2000yr}, and the $\PW$ bosons by a
non-relativistic EFT. 
Similarly to the DPA~\cite{Aeppli:1993rs}, the
cross-section is decomposed into resonant and non-resonant contributions:
\begin{equation}
  \label{eq:sigma-pole}
  \sigma^{\text{ee4f}}(s\approx 4\MW^2)= \sigma_{\text{res}}(s)+\sigma_{\text{non-res}}(s).
\end{equation}
The EFT method enables computation of the Born
cross-section as an expansion according to the
counting (\Eref{eq:counting}),
$\sigma_{\text{Born}}^{\text{ee4f}}=\sigma_{\text{Born}}^{(0)}+\sigma_{\text{Born}}^{(1/2)}+\dots$
This is not necessary in practice since the full  $\Pem\Pep\to 4\mathrm{f}$ Born cross-section  for arbitrary kinematics can be
computed using automated Monte Carlo programs. However, the expansion serves as a
test-case of  the EFT method and provides useful input for estimating the accuracy of  a future
NNLO$^{\text{EFT}}$ calculation.
The leading-order resonant contribution to the cross-section is given by the
imaginary part of a one-loop EFT diagram with non-relativistic
W propagators, denoted by dashed lines,
\begin{equation}
\label{eq:sigmaLO}
\sigma_{\text{Born}}^{(0)}(s)=\sigma^{(0)}_{\text{res}}(s)
=\frac{1}{s}\text{Im}\left[
\parbox{22mm}{\includegraphics{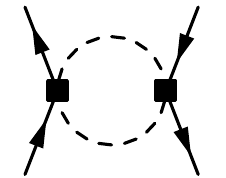}} 
\right]
=\frac{\uppi \alpha^2}{\sw^4 s}\text{Im}\left[\,-\sqrt{-\frac{\mathcal{E}_{\PW}}{\MW}}\;\right] .
\end{equation}
Here, the complex energy variable 
${\cal E}_{\PW}\equiv  \sqrt{s}-2\MW+\mathrm{i} \Gamma_{\PW}\sim\MW v^2$ has been introduced and $\sw=\sin\theta_{\PW}$
with the weak mixing angle $\theta_{\PW}$. A specific final
state is selected by multiplying~\Eref{eq:sigmaLO}  by the LO
branching ratios,
\begin{equation}
\label{eq:decay-lo}
\sigma^{(0)}_{\mathrm{f}_1 \overline{\mathrm{f}}_2 {\mathrm{f}_3  \overline{\mathrm{f}}_4}}
=\frac{\Gamma^{(0)}_{\PWm\to \mathrm{f}_1\overline{\mathrm{f}}_2} \Gamma^{(0)}_{\PWp \to \mathrm{f}_3 \overline{\mathrm{f}}_4}}{\Gamma_{\PW}^2} \sigma^{(0)}_{\text{res}}.
\end{equation}
The  non-resonant contribution to the cross-section  arises from 
local four-electron operators,
\begin{equation}\label{eq:nonres}
\sigma_{\text{non-res}}(s)  
=\frac{1}{s}\mathrm{Im}\left[
  \parbox{17mm}{
\includegraphics{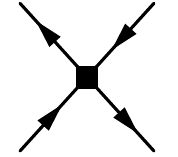} 
 }
\right]=\frac{\alpha^3}{\sw^6 s} \,\mathcal{K},
\end{equation}
where the dimensionless constant
$\mathcal{K}=\mathcal{K}^{(0)}+ {\alpha} 
\mathcal{K}^{(1)} / \sw^2 +\dots$ is
 computed from the forward-scattering amplitude in the
full SM without  self-energy resummation in the
$\PW$ propagators. The first contribution is of   order $\alpha^3$ and arises from cut two-loop
diagrams corresponding to squared tree diagrams of the $\Pem\Pep\to
\PWpm \mathrm{f} \overline{\mathrm{f}}$  processes.
Hence, the leading
non-resonant contribution $\sigma_{\text{non-res}}^{(1/2)}\equiv
\sigma_{\text{Born}}^{(1/2)}$  is 
 suppressed by $\alpha/v\sim\delta^{1/2}$ compared with the resonant
LO cross-section (\Eref{eq:sigmaLO}).  
For the final state, $\upmu^- \overline{\upnu}_\upmu \Pqu  \Paqd$, the explicit \setcounter{footnote}{0}%
result is~\cite{Beneke:2007zg}\footnote{Equation \eqref{eq:4fsqrt} is obtained by setting $s=4\MW^2$ in Eq.~(37) in Ref. \cite{Beneke:2007zg}, where an
  additional $s$-dependence of $\mathcal{K}$ has been kept.} 
\begin{equation}
  \label{eq:4fsqrt}
  \mathcal{K}^{(0)}=-4.25698.
\end{equation}
The $\mathcal{O}(v^2)$ corrections in~\Eref{eq:nlo-power} originate
from higher-order terms in the EFT expansion of the resonant
Born cross-section, $\sigma_{\text{Born}}^{(1)}$, 
and depend strongly on the centre-of-mass
 energy~\cite{Beneke:2007zg},
 \begin{equation}
    \sigma_{\text{Born}}^{(1)}(\sqrt s=161\UGeV)=8\%\times
    \sigma_{\text{Born}}^{\text{ee4f}} ,\qquad 
   \sigma_{\text{Born}}^{(1)}(\sqrt s=170\UGeV)=-8\%\times
    \sigma_{\text{Born}}^{\text{ee4f}}    .
 \end{equation}
%=================================================================
\subsubsection{Radiative corrections}

Including radiative corrections, the resonant cross-section factorises into  hard, soft,
 and Coulomb functions~\cite{Beneke:2010da}. ({This formula holds for the leading term in
   the expansion in $v$. Subleading terms
   result in a sum over Wilson coefficients and Green functions related to
higher partial waves.
In higher orders, there are also soft corrections to the Coulomb
function analogous to ultrasoft  QCD corrections
in $\Pqt\Paqt$ production~\cite{Beneke:2008cr}.)
}
 \begin{equation}
   \label{eq:fact}
   \sigma_{\text{res}}(s)=\mathrm{Im}\left[\,
  \parbox{22mm}{
\includegraphics{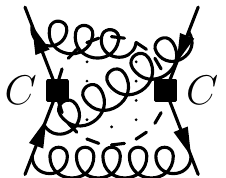}
}\right]=\frac{4\uppi^2 \alpha^2}{ s\MW^2 \sw^4} 
   \text{Im}\left[C^2\int \mathrm{d}\omega W(\omega) G_\mathrm{C}(0,0,\mathcal{E}_{\PW}-\omega) \right].
 \end{equation}
 Here, curly lines depict soft photons with momenta
 $(q^0,\vec q)\sim (\delta,\delta)$, while dotted lines denote potential
 (Coulomb) photons with $(q^0,\vec q)\sim (\delta,\sqrt \delta)$.  The Wilson
 coefficient $C=1+ {\alpha} C^{(1)} / {2\uppi} \dots $ is related to
 contributions of hard loop momenta $q\sim \MW$ to the on-shell amplitudes
 $\Pem\Pep\to \PWm\PWp$ evaluated at the production threshold.  For the input
 parameters used in Ref.~\cite{Beneke:2007zg}, the explicit value of the one-loop
 coefficient is
\begin{equation}
  \label{eq:hard-one}
  C^{(1)}=\text{Re}\, c_{p,\mathrm{LR}}^{(1,{\rm fin})} =-10.076.
\end{equation}
The function $W(\omega)$ includes soft-photon  effects, which decouple
from the W bosons~\cite{Fadin:1993dz,Melnikov:1993np} for the total cross-section, since soft
radiation is only sensitive to the total (\ie vanishing)
electric charge of the produced system. This function is the
QED analogue of the soft function for Drell--Yan production near
the partonic threshold~\cite{Belitsky:1998tc,Becher:2007ty}.
The leading Coulomb Green function at the origin,
\begin{equation}
  \label{eq:greenfunc}
   G_{\rm C}^{(0)}(0,0;\mathcal{E}_{\PW})= -\frac{\MW^2}{4\uppi} \Bigg\{\! \sqrt{-\frac{\mathcal{E}_{\PW}}{\MW}} + 
    \frac{\alpha}{2} \ln \left( - \frac{\mathcal{E}_{\PW}}{\MW}\right)
    - \frac{\alpha^2\uppi^2}{12}\sqrt{-\frac{\MW}{\mathcal{E}_{\PW}}}+\alpha^3 \frac{\zeta(3)}{4}\,
     \frac{\MW}{\mathcal{E}_{\PW}}+\cdots
    \Bigg\},  
\end{equation}
 sums Coulomb exchange and is known to all orders (see, \eg~Ref. \cite{Beneke:2007zg}).
At each order, the Coulomb
corrections  $\sim$$(\alpha/v)^n\sim \delta^{n/2}$ are
parametrically enhanced over the remaining  $\mathcal{O}(\alpha^n)$ corrections but
do not have to be resummed to all orders, owing to the screening of the Coulomb
singularity by  $\Gamma_{\PW}$~\cite{Fadin:1994pm}.
  The convolution of the soft and Coulomb functions results
in logarithms of $\mathcal{E}_{\PW}\sim \MW v^2 $, which can be resummed
 in analogy to threshold resummation at hadron
colliders~\cite{Sterman:1986aj,Catani:1989ne,Becher:2007ty}. However, for QED corrections,  $\alpha\log v$ is
not enhanced, so this resummation is formally not necessary.\footnote{An initial study obtained NLL effects of
 $0.1\%$~\cite{Falgari:2009zz}, so the relevance for the FCC-ee may have to be
 revisited.} Higher-order corrections to the non-resonant cross-section~(\Eref{eq:nonres})  only arise through  hard corrections to 
 $\mathcal{K}$, while loop corrections in the EFT vanish.

These ingredients provide results for massless initial-state electrons and could be used, in analogy to QCD predictions at hadron colliders, to
 define appropriate `partonic' cross-sections
that are convoluted with corresponding electron
structure functions resumming  large mass logarithms. 
Structure functions in such a scheme are known up to NNLO~\cite{Blumlein:2011mi}. In the NLO$^{\text{EFT}}$ calculation
 of Ref.~\cite{Beneke:2007zg}, however, electron mass effects have been treated by
 including collinear corrections and matching to the commonly used resummed
 structure functions~\cite{Skrzypek:1992vk} by subtracting double-counting
 contributions.\footnote{In the
   process of finalizing this report, we have noted that NLL
   contributions arising from the combination of numerator factors of
   $\me$ and integrals with negative powers of $\me$ have been
   inadvertently omitted in the computation of the collinear
   corrections. The expressions and numerical
   predictions in this report are preliminary results including the
   missing contributions. A more complete discussion
   will be given elsewhere.}

A useful result~\cite{Actis:2008rb} for computing a class of higher-order effects
of the form $\alpha^{n+1}/v^n$ is obtained from~\Eref{eq:fact}
 by combining the all-order Coulomb Green function with one-loop hard and soft corrections and matching to
ISR structure functions, as in the  NLO$^{\text{EFT}}$
calculation:
\begin{equation}
\label{eq:csh}
  \Delta{\sigma}^{\rm C\times[S+H]_1} (s) =\frac{4\pi^2\alpha^2}{s\MW^2\sw^4}\frac{\alpha}{\pi}\bigg\{ \bigg( \frac{7}{2}+\frac{\pi^2}{4}+ C^{(1)} \bigg)\,\mbox{Im}\, G_{\rm C}(0,0;\mathcal{E}_{\PW}).
\end{equation}
Corrections of the same order, $\alpha^{n+1}/v^n$, result from the NLO Green
function~\cite{Beneke:1999qg}~$G_{\rm C}^{(1)}$, which includes the $\mathcal{O}(\alpha)$
correction to the Coulomb potential. In the $G_\upmu$ input
parameter scheme, the $\mathcal{O}(\alpha^{2}/v)$ correction
reads~\cite{Actis:2008rb}
   \begin{equation}
\label{eq:nlo-coulomb}
     \Delta G_{\rm C}^{(1)}(0,0,\mathcal{E}_\mathrm{W})=   -\frac{M_\mathrm{W}^2}{4\uppi}
   \frac{\alpha^2}{8\uppi} \ln\!\left( -\frac{{\cal E}_\mathrm{W}}{M_\mathrm{W}}\right)\left\{
    - \frac{\beta_0}{2}
     \left[
     \ln\!\left( -\frac{{\cal E}_\mathrm{W}}{M_\mathrm{W}}\right)\right]
                                                      +\Delta_{G_\upmu}  \right\}+\mathcal{O}(\alpha^3)
   \end{equation}
   with the QED beta function with five quark flavours,
   $ \beta_0=- {4} (\sum_{\mathrm{f}\neq \Pt} N_{C_\mathrm{f}} Q_\mathrm{f}^2)
   / 3=-80/9$, and where
   the scheme-dependent constant $\Delta_{G_\upmu}=61.634$ is related to 
   the quantity
\[
   \delta_{\alpha(\MZ)\to
     G_\upmu}=\frac{\alpha}{4\uppi}\left(\Delta_{G_\mu}+2\beta_0\ln\!\left(\frac{2M_\mathrm{W}}{M_\mathrm{Z}}\right)\right)
 \]
   used in Ref.~\cite{Actis:2008rb}. 
  Equations~\eqref{eq:csh} and~\eqref{eq:nlo-coulomb} are the basis for
  computing 
  examples of leading N$^3$LO corrections in Section~\ref{sec:nnlo}.

%=================================================================
\subsubsection{NLO$^{\text{EFT}}$ result}
\label{sec:nloeft}

The genuine radiative corrections at NLO$^{\text{EFT}}$ can be
obtained by expanding \Eref{eq:csh} to
$\mathcal{O}(\alpha)$ relative to the leading order and adding the
second-order Coulomb correction from~\Eref{eq:greenfunc}.
A specific four-fermion final state is selected by multiplying the NLO
correction with the LO branching ratios~(\Eref{eq:decay-lo}) and adding
NLO decay corrections, 
\begin{equation}
  \Delta\sigma^{(1)}_{\text{decay}}=
\left(\frac{\Gamma^{(1,\mathrm{ew})}_{\mathrm{f}_1\overline{\mathrm{f}}_2}}{
\Gamma^{(0)}_{\mathrm{f}_1\overline{\mathrm{f}}_2} }+
\frac{\Gamma_{\mathrm{f}_3\overline{\mathrm{f}}_4}^{(1,\mathrm{ew})}}{\Gamma_{\mathrm{f}_3\overline{\mathrm{f}}_4}^{(0)} }
\right) \sigma^{(0)}_{\text{res}},
\label{eq:delta-decay}
\end{equation}
with the one-loop electroweak corrections to the partial decay widths,
$\Gamma^{(1,\mathrm{ew})}_{\mathrm{f}_i\overline{\mathrm{f}_j}}$.  For hadronic decay modes,  QCD
corrections to the partial decay widths must  also be included up to NNLO, using
the counting $\alpha_\mathrm{s}^2\sim \alpha$.
In \Tref{tab:eftvs4f}, the
$\mathcal{O}(\alpha)$-contributions of the NLO$^{\text{EFT}}$ result are
compared with  the NLO$^{\text{ee4f}}$ calculation in the full
SM~\cite{Denner:2005es}.\footnote{Note that here the updated results in the
  erratum to~Ref. \cite{Denner:2005es} are used.  The EFT results here
  and in Table~\ref{tab:tabnnlo} differ from  those of Refs.~\cite{Beneke:2007zg, Actis:2008rb} because of the
  corrected collinear contributions.}  The differences are of the order
\begin{equation}
   \Delta\sigma^{(1)}_{\text{4f}}(s)\equiv\sigma^{\text{ee4f}}_{\text{NLO}}(s)
-\sigma^{(1)}_{\text{EFT}}(s)= \sigma_{\text{Born}}^{\text{ee4f}}(s)\times (0.7-0.1)\%
\end{equation}
for $\sqrt s=161$--$170\UGeV$. Near the threshold, the dominant source of this discrepancy is expected to be the
$\mathcal{O}(\delta^{3/2})$ contribution from the
$\mathcal{O}(\alpha)$ correction to the non-resonant  cross-section~(\Eref{eq:nonres}),
 which has not been computed in the EFT
approach.\footnote{For $\Pem\Pep\to \Pqt\Paqt$, a related calculation has been
  performed recently~\cite{Beneke:2017rdn}.}
 Attributing the difference
 at $\sqrt{s}=161\UGeV$ to this correction, one obtains 
\begin{equation}
  \mathcal{K}^{(1)}\approx 1.4,
\end{equation}
indicating that the $\mathcal{O}(\alpha)$ corrections to the non-resonant
contribution~(\Eref{eq:4fsqrt}) are moderate,
$ |\mathcal{K}^{(1)}/\mathcal{K}^{(0)}|\approx
0.3$. Above the threshold, $\mathcal{O}(\delta^{3/2})$ and $\mathcal{O}(\delta^{2})$
corrections to the resonant cross-section are expected to be important; these
arise 
from the combination of $\mathcal{O}(\alpha/v,\alpha)$ corrections in the EFT
with $\mathcal{O}(v^2)$ kinematic corrections and from $\mathcal{O}(\alpha)$
corrections to the Wilson coefficients of subleading production operators.
Naive estimates using the $\mathcal{O}(v^2)$ expansion of the Born amplitude
and the first Coulomb correction,
\begin{equation}
    \sigma_{\alpha v}^{(3/2)}(s) \sim
   |\sigma^{(1)}_{\text{Born}}(s)|\sigma^{(1/2)}_\mathrm{C}(s)/\sigma^{(0)}_{\text{Born}}(s),
   \qquad 
   \sigma_{\alpha v^2}^{(2)}(s)  \sim
   \frac{\alpha}{\sw^2}|\sigma^{(1)}_{\text{Born}}(s)|,
\label{eq:nlo-kin}
   \end{equation}
indicate that both corrections are $\sim 0.3\%\times
    \sigma_{\text{Born}}^{\text{ee4f}}$ at $\sqrt s=170\UGeV$,
 overestimating the discrepancy to the
NLO$^{\text{ee4f}}$ calculation.
To assess the accuracy of the EFT expansion, it would be interesting to
calculate these corrections exactly and investigate whether the difference to the NLO$^{\text{ee4f}}$ calculation  could be reduced, \eg
by resumming relativistic corrections to the $\PW$ propagators.

\begin{table}
\caption{Comparison of the strict electroweak NLO results (without
QCD corrections, second Coulomb correction and ISR resummation) in the EFT approach to the full NLO$^{\text{ee4f}}$ calculation and the DPA implementation of Ref.~\cite{Denner:2000bj}.}
\label{tab:eftvs4f}
\begin{center}
\begin{tabular}{lllll}
\hline \hline 
& \multicolumn{3}{l}{
$\sigma(\Pem\Pep\to \upmu^- \overline{\upnu}_\upmu \Pqu  \Paqd \,\mathrm{X})$(fb)}&
\\
$\sqrt{s}$ & Born & NLO(EFT)~\cite{Beneke:2007zg} & ee4f~\cite{Denner:2005es}
& DPA~\cite{Denner:2005es} \\
(GeV)
\\
\hline
161& 150.05(6) &107.34(6)  & 106.33(7) & 103.15(7)\\\hline
170 & 481.2(2) & 379.03(2)  & 379.5(2) & 376.9(2)  
\\\hline \hline
\end{tabular}
\end{center}
\end{table}

%=================================================================
\subsubsection{Leading NNLO corrections}
\label{sec:threehalf}

In Ref. \cite{Actis:2008rb}, those $\mathcal{O}(\delta^{3/2})$ corrections according to~\Eref{eq:three-half-power} have been computed that originate from
genuine NNLO corrections in the usual counting in $\alpha$.  These consist of
several classes: (a) interference of one-loop Coulomb corrections with soft and
hard corrections~(\Eref{eq:csh}); (b) interference of one-loop Coulomb
corrections with corrections to $\PW$ decay, obtained
from \Eref{eq:delta-decay} by replacing the LO cross-section with the first
Coulomb correction; (c) interference of one-loop Coulomb corrections with NLO
corrections to residues of $\PW$ propagators; and (d) radiative NLO corrections to
the Coulomb potential (\Eref{eq:nlo-coulomb}). The third Coulomb
correction from \Eref{eq:greenfunc} contributes at the same order,
$\delta^{3/2}$.  Care has been taken to avoid double-counting corrections
already included in the NLO$^{\text{ee4f}}$ calculation, so the two results can
be added to obtain the current best prediction for the total cross-section near the  threshold.
The numerical results are reproduced in \Tref{tab:tabnnlo}, together with
the second Coulomb correction included in the
NLO$^{\text{EFT}}$ calculation.  The results show that the leading Coulomb-enhanced two-loop corrections are of the order of $0.3\%$.
% relative to the LO cross-section.
The uncertainty due to the remaining non-Coulomb-enhanced
NNLO corrections was estimated to be below the ILC target accuracy of
$\Delta \MW=3\UMeV$~\cite{Actis:2008rb} but not sufficient for
the FCC-ee.

  \begin{table}[t]
    \caption{Leading $\mathcal{O}(\alpha^2)$ corrections~\cite{Actis:2008rb}
    (second and third column) and contributions to leading
    $\mathcal{O}(\alpha^3)$ corrections from triple-Coulomb
    exchange~\cite{Actis:2008rb} (fourth column), interference of
    double-Coulomb exchange with soft and hard corrections~(\Eref{eq:c2-sh})
    (fifth column), and double-Coulomb exchange with the NLO Coulomb
    potential~(\Eref{eq:nloc2}) (sixth column).  The relative correction is
    given with respect to the Born cross-section without ISR improvement, as
    quoted in Ref.~\cite{Actis:2008rb}. }
  \label{tab:tabnnlo}
  \begin{center}
  \begin{tabular}{llllll}
    \hline \hline
    &
    \multicolumn{4}{l}{ $\sigma(\Pem\Pep\to \upmu^- \overline{\upnu}_\upmu \Pqu  \Paqd\,\mathrm{X})$(fb)}&
    \\
    $\sqrt{s}$& $\mathcal{O}(\alpha^2/v^2)$&
                                                           $\mathcal{O}(\alpha^2/v)$&  $\mathcal{O}(\alpha^3/v^3)$ &  $\mathcal{O}(\alpha^3/v^2)|_{\rm C_2\times[S+H]_1}$
&  $\mathcal{O}(\alpha^3/v^2)|_{\rm C_2^{NLO}}$
    \\
  (GeV)  \\
    \hline
      $  158$ &$ 0.151$ &$0.061$  & $3.82\times 10^{-3}$ & $-1.50\times 10^{-3}$ & $5.38\times 10^{-3}$\\
 &[$+0.245\%$]&[$+0.099\%$] &[$+0.006\%$]& [$-0.002\%$]& [$+0.009\%$]
    \\\hline
    %%%%%%%%%%%%%%%%%%%%%%%%%%%%%%%%%%%%%
    $161$ & $0.437$ & $0.331$ & $9.92\times 10^{-3}$ &$-0.433\times 10^{-2}$ &   $1.52\times 10^{-2}$\\
&[$+0.284\%$]&[$+0.215\%$] &[$+0.006\%$ ]& [$-0.003\%$]& [$+0.010\%$]    \\\hline
%%%%%%%%%%%%%%%%%%%%%%%%%%%%%%%%%%%%%%%%%%%%%%%%%%%%%%%%%%    
$ 164$ & $0.399$ & $1.038$ & $2.84\times 10^{-3}$ &$ -3.95\times
   10^{-3}$ & $1.97\times 10^{-2}$  \\
& [$+0.132\%$]&  [$+0.342\%$]&[$+0.001\%$]&[$-0.001\%$] &[$+0.007\%$] \\\hline
%%%%%%%%%%%%%%%%%%%%%%%%%%%%%%%%%%%%%%%%%%%%%%%%%%%%%%%%%
  $167$ & $0.303$ & $1.479$ & $9.43\times 10^{-4}$ & $-3.00\times 10^{-3} $&  $1.77\times 10^{-2}$ \\
&[$+0.074\%$]&[$+0.362\%$]&[$+0.000\%$]&[$-0.001\%$]&[$+0.004\%$]
    \\\hline
%%%%%%%%%%%%%%%%%%%%%%%%%%%%%%%%%%%%%%%%%%%%%%%%%%%%%%%%%%%%    
 $170$ &$ 0.246$ &$ 1.734$ & $4.39\times 10^{-4}$ &$ -2.43\times
                                              10^{-3}$ & $1.56\times 10^{-2}$  \\
 &[$+0.051\%$]&[$+0.360\%$]&[$+0.000\%$]& [$-0.001\%$]&[$+0.003\%$]
    \\\hline \hline
  \end{tabular}
  \end{center}
  \end{table}
 %=================================================================

%=================================================================
\subsection{Estimate of NNLO$^{\text{EFT}}$ corrections and beyond}
\label{sec:nnlo}

In this section, the structure of the EFT expansion of the cross-section
and the ingredients for higher-order corrections reviewed in \Sref{sec:eft:schwinn} are used to estimate the possible effects of a future NNLO$^{\text{EFT}}$
calculation. Owing to the counting~(\Eref{eq:counting}), this also includes  leading 
 corrections beyond NNLO in the conventional perturbative expansion:
 \begin{equation}
   \label{eq:nnlo-count}
\text{NNLO}^{\text{EFT}}:\quad   v^4,\quad \alpha v^2,\quad  \alpha^2\, \quad \alpha^3/v^2, \quad \alpha^4/v^4.
\end{equation}
The contributions of $\mathcal{O}(v^4,\alpha v^2)$ in~\Eref{eq:nnlo-count}
arise from kinematic corrections to the Born and NLO cross-section in the full
SM, as discussed in Sections~\ref{sec:born} and~\ref{sec:nloeft}, respectively.
The genuine $\mathcal{O}(\alpha^2)$ corrections are estimated in
Section~\ref{sec:hard-nnlo}.  A representative subset of the
$\mathcal{O}(\alpha^3/v^2)$ corrections is computed in
Section~\ref{sec:mix-c-nnlo} and serves as an estimate of effects beyond a
conventional NNLO calculation. The quadruple-Coulomb correction $\alpha^4/v^4$
follows from the expansion of the known Coulomb Green function and is smaller
than $0.001\%$ and therefore negligible.  Counting $\alpha\sim \alpha_\mathrm{s}^2$,
 QCD corrections to $\PW$ self-energies and decay widths up to
\begin{equation}
 \alpha \alpha_\mathrm{s}^2 ,\quad   \alpha_\mathrm{s}^4
\end{equation}
are also required.  Currently, the required $\mathcal{O}( \alpha_\mathrm{s}^4)$ corrections
for inclusive hadronic vector boson decays are known~\cite{Baikov:2008jh},
while mixed QCD-EW corrections are known up to
$\mathcal{O}(\alpha\alpha_\mathrm{s})$~\cite{Kara:2013dua}.  The uncertainty of a
future NNLO$^{\text{EFT}}$ calculation can be estimated by considering the
impact of corrections at the next order in the $\delta$-expansion, \ie
\begin{equation}
  \label{eq:5half}
\mathrm{N}^{5/2}\mathrm{LO}^{\text{EFT}}:\quad \alpha v^3,\quad \alpha^2v,\quad \alpha^3/v,  \quad \alpha^4/v^3,\quad \alpha^5/v^5.
\end{equation}
The  contributions $\sim$$\alpha v^3$ are already included in the
NLO$^{\rm{ee4f}}$ calculation.
The fifth Coulomb correction $\sim$$\alpha^5/v^5$ is known but
negligibly small. The corrections $\sim$$\alpha^4/v^3$ arise from the
combination of $\mathcal{O}(\alpha)$
corrections with triple-Coulomb exchange and are also expected to be
negligible, since the latter is <$0.01\%$.  Therefore, the dominant genuine radiative corrections
beyond NNLO$^{\text{EFT}}$ are expected to be of order $\alpha^3/v$. These arise from
a combination of single Coulomb exchange and various sources of  $\mathcal{O}(\alpha^2)$
corrections and are estimated in Section~\ref{sec:fivehalf}. Further contributions
   from triple-Coulomb exchange  combined with $\sim$$v^2$
   kinematic corrections are again expected to be negligible. 
The $\mathcal{O}(\alpha^2)$ corrections to the non-resonant cross-section~(\Eref{eq:nonres}) also provide  $\sim$$\alpha^3/v$ corrections
 relative to the LO cross-section, while
 corrections $\sim$$\alpha^2 v$ arise from a combination of single
Coulomb exchange with kinematic
 corrections $\sim$$\alpha v^2$.  Such
non-resonant and kinematic corrections are estimated in
Section~\ref{sec:nonresNNLO}.  It is assumed throughout that large logarithms
of $\me$ are absorbed in electron structure functions and only the uncertainty
due to non-universal $\mathcal{O}(\alpha^{2},\alpha^3)$ corrections is
considered.

%=================================================================
\subsubsection{ $\mathcal{O}(\alpha^2)$  corrections in the EFT}
\label{sec:hard-nnlo}
The most involved corrections of order $\alpha^2$ in the EFT arise from 
hard two-loop corrections to
the Wilson coefficients of production operators and to decay rates and from soft
two-loop corrections to the forward-scattering amplitude. Additional
corrections from 
higher-order potentials  or 
 the combination of
double-Coulomb exchange with kinematic
corrections  $\sim$$v^2$ are anticipated to be
subdominant.  The soft corrections for massless initial-state electrons can be
extracted from the two-loop Drell--Yan soft
function~\cite{Belitsky:1998tc,Becher:2007ty} and converted to the
electron mass regulator scheme using the NNLO structure functions
computed in Ref. \cite{Blumlein:2011mi}.   We make no attempt here to
estimate these soft corrections, which are formally of the same order as the
hard corrections. This is supported by the NLO result, where hard
corrections alone provide a reasonable order-of-magnitude estimate and soft
corrections contribute less than $50\%$ of the NLO corrections for
$\sqrt s=158$--$170\UGeV$.  
The contribution of the NNLO  Wilson coefficient of the
production operator to the cross-section reads
\begin{equation}
  \sigma^{(2)}_{\text{hard}}(s)=\frac{\uppi \alpha^2}{\sw^4 s}\text{Im}\left[\,-\sqrt{-\frac{\mathcal{E}_{\PW}}{\MW}}
\left(\frac{\alpha}{\uppi}\right)^2 C^{(2)}\;\right],
\end{equation}
where the NNLO hard coefficient is defined in terms of the squared Wilson
coefficient,
\[ 
 C^2=1+\frac{\alpha}{\uppi}C^{(1)}+\left(\frac{\alpha}{\uppi}\right)^2 C^{(2)}+\cdots
\]
The computation of $C^{(2)}$ involves the two-loop
amplitude for $\Pem\Pep\to \PWm\PWp$, evaluated directly at the threshold.
Such a computation is beyond the
current state of the art, which includes two-loop EW
corrections to three-point
functions~\cite{Actis:2008ts,Freitas:2014hra,Dubovyk:2018rlg},  but will
presumably be
feasible before the
operation of the FCC-ee.
A naive estimate of the NNLO coefficient in terms of the the one-loop
result~(\Eref{eq:hard-one}),
\begin{equation}
  \label{eq:hard-est}
C^{(2)}\sim (C^{(1)})^2  ,
\end{equation}
suggests an
effect on the cross-section of
\begin{equation}
\label{eq:nnlo-hard}
  \Delta\sigma^{(2)}_{\text{hard}}\approx \sigma^{(0)}_{\text{res}}  \times 0.06\% .
\end{equation}
The NNLO corrections to $\PW$ boson decay give rise to the
correction
\begin{equation}
  \Delta\sigma^{(2)}_{\text{decay}}=
\left(\frac{\Gamma^{(2,\mathrm{ew})}_{\upmu^- \overline{\upnu}_\upmu}}{{
\Gamma^{(0)}_{\upmu^- \overline{\upnu}_\upmu} }}+
\frac{\Gamma_{\Pqu\Paqd}^{(2,\mathrm{ew})}}{\Gamma_{\Pqu\Paqd}^{(0)}}+
\frac{\Gamma^{(1,\mathrm{ew})}_{\upmu^- \overline{\upnu}_\upmu}\Gamma_{\Pqu\Paqd}^{(1,\mathrm{ew})}}{{
\Gamma^{(0)}_{\upmu^- \overline{\upnu}_\upmu} \Gamma_{\Pqu\Paqd}^{(0,\mathrm{ew})}}}
\right) \sigma^{(0)}_{\text{res}}.
\label{eq:delta-decay-2}
\end{equation}
The product of NLO corrections in the last term  contributes a negligible $0.001\%$ to the $G_\upmu$ input parameter scheme.
A naive estimate of the currently unknown $\mathcal{O}(\alpha^2)$ corrections to $\PW$ decay suggests
\[
\Gamma_{\mathrm{f}_i\overline{\mathrm{f}}_j}^{(2,\mathrm{ew})} \approx \frac{\alpha}{\sw^2}
\Gamma_{\mathrm{f}_i\overline{\mathrm{f}}_j}^{(1,\mathrm{ew})} \sim 0.01\%\times
\Gamma_{\mathrm{f}_i\overline{\mathrm{f}}_j}^{(0)} \, ,
\]
consistent with the size of the $\mathcal{O}(\alpha^2)$ corrections to
$\PZ$ decay~\cite{Freitas:2014hra,Dubovyk:2018rlg}.
The estimates given in this subsection indicate that the combined non-Coulomb-enhanced corrections
of $\mathcal{O}(\alpha^2)$ are of the order of $ 0.1\%$ and are therefore
mandatory to reduce the uncertainty below $\Delta\MW\lesssim 1.5\UMeV$.

%=================================================================

\subsubsection{Corrections of $\mathcal{O}(\alpha^3/v^2)$}
\label{sec:mix-c-nnlo}
The corrections of $\mathcal{O}(\alpha^3/v^2)$ involve a double-Coulomb
exchange in combination with an $\mathcal{O}(\alpha)$ correction and arise
from similar sources to those of the $\mathcal{O}(\alpha^2/v)$ corrections discussed in
Section~\ref{sec:threehalf}.   The subclass of contributions arising from the
combination of double-Coulomb exchange with soft and hard corrections is
obtained by inserting the $\mathcal{O}(\alpha^2)$ term in
the expansion of the Coulomb Green function~(\Eref{eq:greenfunc}) into~\Eref{eq:csh}, resulting in the contribution to
the cross-section
  \begin{equation}
  \label{eq:c2-sh}
   \Delta \sigma^{\rm C_2\times[S+H]_1}
    = \,\frac{\alpha^2}{\sw^4s}\,\frac{\alpha^3\uppi^2}{12}
    \mbox{Im}\,
 \Bigg\{ \sqrt{-\frac{\MW}{\mathcal{E}_{\PW}}} \left[ \bigg( \frac{7}{2}+\frac{\pi^2}{4}+C^{(1)}
    \bigg)  \right]
  \end{equation}
Corrections from the  NLO Coulomb potential to double-Coulomb exchange  can be obtained by expanding the
expression for the NLO Coulomb Green function~\cite{Beneke:1999qg} quoted
in Ref.~\cite{Beneke:2011mq} and using the result for the Coulomb potential in the
$G_\upmu$ input parameter scheme~\cite{Actis:2008rb}, resulting in
 \begin{align}
\label{eq:nloc2}
   \Delta{\sigma}^{\rm C_2^{NLO}} (s)=\frac{\alpha^2}{\sw^4s} \frac{\alpha^3}{24}
                                   \text{Im}\Biggl\{\sqrt{ -\frac{M_\mathrm{W}}{{\cal E}_\mathrm{W}}}
                                  & \Biggl[
    \uppi^2\left(-\beta_0
      \ln\!\left( -\frac{{\cal E}_\mathrm{W}}{M_\mathrm{W}}\right)+\Delta_{G_\upmu} \right)
-12\beta_0\zeta_3 \Biggr]\Biggr\}.
 \end{align}
 The combination of double-Coulomb exchange with NLO corrections to
 $\PW$ decay is obtained from \Eref{eq:delta-decay} by replacing
 $\sigma^{(0)}_{\text{res}}$ with the second Coulomb correction. The resulting
 effect is, at most, $0.002\%$.  Further corrections arise from corrections to
 the propagator residues and can be computed with current methods, but are
 beyond the scope of the present simple estimates. At
 $\mathcal{O}(\alpha^2/v)$, the corresponding corrections are of a similar
 size to the mixed soft+hard Coulomb corrections~\cite{Actis:2008rb}.
 Therefore, the  predictions from Eqs.~\eqref{eq:c2-sh}
 and~\eqref{eq:nloc2}, which are shown in  \Tref{tab:tabnnlo} together
 with the known two- and three-loop corrections~\cite{Actis:2008rb}, are
 expected to be representative  of the the $\mathcal{O}(\alpha^3/v^2)$
 corrections. They are of a similar order as the third Coulomb correction, and
 individually of the order $\lesssim$$0.01\%$ near the threshold.  The sum of all
 $\mathcal{O}(\alpha^3/v^2)$ corrections may, therefore, be of the order
 of  
 $0.01\%$, indicating the need to go beyond a strict
 $\mathcal{O}(\alpha^2)$ calculation to reach the FCC-ee accuracy goal.
%%%

\subsubsection{Radiative corrections  of $\mathcal{O}(\alpha^3/v)$}
\label{sec:fivehalf}
Genuine three-loop corrections at $\mathcal{O}(\alpha^3/v)$ can arise from a
combination of the first Coulomb correction and soft or hard
$\mathcal{O}(\alpha^2)$ corrections, corrections from higher-order potentials
to the Coulomb Green function or a combination of $\mathcal{O}(\alpha)$
hard or soft and potential corrections. One contribution in the latter class can
be computed by inserting the NLO Green function~(\Eref{eq:nlo-coulomb}) into
the product with the $\mathcal{O}(\alpha)$ hard and soft
corrections~(\Eref{eq:csh}),
\begin{equation}
   \hat{\sigma}^{\rm C^{NLO}\times[S+H]_1} (s) = \frac{\alpha^2}{s \sw^4}\frac{\alpha^3}{8\uppi} 
\mbox{Im}  \Biggl\{ 
 \bigg( \frac{7}{2}+\frac{\pi^2}{4}+C^{(1)}\bigg)\left(-\frac{\beta_0}{2} \ln\!\left( -\frac{{\cal E}_W}{M_W}\right) +\Delta_{ G_\mu}
  \right) \ln\!\left( -\frac{{\cal E}_W}{M_W}\right)    \Biggr\}.
  \end{equation}
 The corrections to the cross-section for  $\sqrt s=161$--$170\UGeV$ are given by
\begin{equation}
     \Delta{\sigma}^{\rm C^{NLO}\times[S+H]_1}
     =- 0.001\%\times \sigma_{\text{LO}}.
   \end{equation}
A further indication for the magnitude of corrections at this order can be obtained
from  the combination of the NNLO hard coefficient with the first Coulomb
correction,
\begin{equation}
\Delta{\sigma}^{\rm{C}_1\times \rm{H}_2} =
    -\frac{\uppi \alpha^2}{\sw^4 s}
  \frac{\alpha^3 C^{(2)}}{2\uppi}
   \mathrm{Im}\bigg[\ln\left(-\frac{{\cal E}_{\PW}}{\MW} \right)
    \bigg],
\end{equation}
and using the estimate~(\Eref{eq:hard-est}) for the hard two-loop coefficient,
which results in
\begin{equation}
     \Delta{\sigma}^{\rm{C}_1\times \rm{H}_2}
     (161\UGeV) \approx 0.005\%\times \sigma_{\text{LO}} ,\qquad
     \Delta{\sigma}^{\rm{C}_1\times \rm{H}_2}
     (170\UGeV) \approx 0.002\%\times \sigma_{\text{LO}} .
\end{equation}
These results indicate that the $\mathcal{O}(\alpha^3)$ corrections beyond 
NNLO$^{\text{EFT}}$ are $\lesssim$$0.01\%$.  It is expected that the
factorisation~(\Eref{eq:fact})  and the N$^3$LO Coulomb Green
function~\cite{Beneke:2015kwa} enable the computation of all $\mathcal{O}(\alpha^3/v)$ corrections  once the
NNLO$^{\text{EFT}}$ result is known, as for a related
calculation for hadronic
$\Pqt\Paqt$ production~\cite{Piclum:2018ndt}.

\subsubsection{Non-resonant and kinematic $\mathcal{O}(\alpha^2)$ corrections}
\label{sec:nonresNNLO}
Kinematic $\mathcal{O}(\alpha^2v)$ corrections and $\mathcal{O}(\alpha^2)$
corrections to the non-resonant cross-section in~\Eref{eq:5half} would be
included in a full NNLO$^{\text{ee4f}}$ calculation, which is far beyond
current calculational methods.  The comparison of the NLO$^{\text{EFT}}$ and
NLO$^{\text{ee4f}}$ results in \Sref{sec:nloeft} indicate a
well-behaved perturbative expansion of the non-resonant
corrections~(\Eref{eq:nonres}), with coefficients $\mathcal{K}^{(i)} $ of order
one.
This suggests that the non-resonant and kinematic NNLO corrections are
reasonably estimated by scaling the corresponding NLO corrections,
\begin{equation}
  \Delta\sigma^{(2)}_{\text{4f}}(s)=\sigma^{\text{ee4f}}_{\text{NNLO}}(s)-\sigma^{(2)}_{\text{EFT}}(s) \approx
  \frac{\alpha}{\sw^2}
  \left(\sigma^{\text{ee4f}}_{\text{NLO}}(s)-\sigma^{(1)}_{\text{EFT}}(s)\right)
=\sigma_{\text{Born}}^{\text{ee4f}}(s)\times 0.02\%
\end{equation}
for $\sqrt s=161$--$170\UGeV$.  Therefore, these effects must be under
control to reach the desired accuracy for the FCC-ee.  A calculation of the
$\mathcal{O}(\alpha^2)$ non-resonant correction in the EFT involves a
combination of $\mathcal{O}(\alpha^2)$ corrections to the processes
$\Pem\Pep\to \PWpm \mathrm{f} \overline{\mathrm{f}}$ with $\mathcal{O}(\alpha)$ corrections for
$\Pem\Pep\to 4 \mathrm{f}$.  Such a computation is beyond current capabilities, but
may be possible before a full NNLO$^{\text{ee4f}}$ calculation is available.
A comparison of future NNLO calculations in the EFT and the conventional DPA
may also enable  these corrections to be constrained.

%=================================================================
\subsection{Summary and outlook}

The prospects of reducing the theoretical uncertainty of the total $\PW$
pair production cross-section near the threshold to the level of $\sim$$0.01\%$
required to fully exploit the high statistics at a future circular $\Pem\Pep$
collider have been investigated within the EFT approach, building on
results for the NLO and dominant NNLO corrections.  The estimates in
Section~\ref{sec:hard-nnlo} suggest that $\mathcal{O}(\alpha^2)$ corrections
beyond the leading Coulomb effects~\cite{Actis:2008rb} are of the order
\begin{equation}
  \Delta\sigma_{\text{NNLO}}\approx %\text{few}\times
  0.1\%\times \sigma_{\text{Born}}
\end{equation}
at the threshold and are therefore mandatory to reach FCC-ee precision.
In Sections~\ref{sec:mix-c-nnlo} and~\ref{sec:fivehalf}, the dominant, Coulomb-enhanced three-loop effects have been estimated to be 
of the order
\begin{equation}
      \Delta\sigma_{\text{N}^3\text{LO}}\approx \text{few}\times  0.01 \%\times \sigma_{\text{Born}} \, ,
\end{equation}
based on computations or estimates of representative examples  of
$\mathcal{O}(\alpha^3/v^2,\alpha^3/v)$ effects.  These corrections are either part of the NNLO$^{\text{EFT}}$ result or can be computed
once this result is available. The effect of the remaining
$\mathcal{O}(\alpha^3)$ corrections without Coulomb enhancement is expected to
be below the FCC-ee target accuracy.  However, the accuracy of the
NNLO$^{\text{EFT}}$ calculation is limited by non-resonant and kinematic
corrections. An extrapolation of the difference of the NLO$^{\text{EFT}}$ and
NLO$^{\text{ee4f}}$ calculations suggests the magnitude
\begin{equation}
   \Delta\sigma^{(2)}_{\text{4f}}\approx 0.02\%\times \sigma_{\text{Born}}.
 \end{equation}
 Related estimates, $\Delta\sigma_{\text{N}^3\text{LO}}\approx 0.02\%$ and
 $\Delta\sigma_{\text{NNLO}}^{(\text{non-res})}\approx 0.016\%$, have been
 obtained using scaling arguments and an extrapolation of the accuracy of the
 DPA~\cite{Jadach:2019bye}.  Our results suggest that a theory-induced
 systematic error of the mass measurement from a threshold scan of
 \begin{equation}
   \Delta \MW=(0.15-0.45)\UMeV
 \end{equation}
 should be achievable, where the lower value results from assuming that the
 non-resonant corrections are under control.  In addition to the corrections
 considered here, it is also essential to reduce the uncertainty from ISR
 corrections and QCD corrections for hadronic final states to the required
 accuracy. It would also be desirable to bring the precision for differential
 cross-sections to a similar level to that of the total cross-section.

\subsection*{Acknowledgements}

I would like to thank M.~Beneke for useful comments.

\end{bibunit}

\label{sec-sm-schwinn} 
\clearpage \pagestyle{empty}  \cleardoublepage
%============================================

\pagestyle{fancy}
\fancyhead[CO]{\thechapter.\thesection \hspace{1mm}  Perspectives of heavy quarkonium
production at the FCC-ee}
\fancyhead[RO]{}
\fancyhead[LO]{}
\fancyhead[LE]{}
\fancyhead[CE]{}
\fancyhead[RE]{}
\fancyhead[CE]{Z.-G. He and B.A. Kniehl}
\lfoot[]{}
\cfoot{-  \thepage \hspace*{0.075mm} -}
\rfoot[]{}
    
 \begin{bibunit}[elsarticle-num] % define the bib-style for the unit: elsarticle-num.bst
%  text-1; this is the corresponding section
%\putbib[2loops] % the *.bib
%\end{bibunit}
% go-on
%--- from: bibunits.sty, adapts the font size of ``References'' to section
\let\stdthebibliography\thebibliography
\renewcommand{\thebibliography}{%
\let\section\subsection
\stdthebibliography}

\section
[Perspectives of heavy quarkonium
production at the FCC-ee \\ {\it Z.-G. He, B.A. Kniehl}]
{Perspectives of heavy quarkonium
production at the FCC-ee
\label{contr:SM_He_Kniehl}}
\noindent
{\bf Contribution\footnote{This contribution should be cited as:\\
Z.-G. He, B.A. Kniehl, Perspectives of heavy quarkonium
production at the FCC-ee,  
%04 DOI:10.23731/CYRM-2020-XXX.\thepage, in:
%04 \url{http://dx.doi.org/10.23731/CYRM-2020-XXX.\thepage}, in:
DOI: \href{http://dx.doi.org/10.23731/CYRM-2020-003.\thepage}{10.23731/CYRM-2020-003.\thepage}, in:
Theory for the FCC-ee, Eds. A. Blondel, J. Gluza, S. Jadach, P. Janot and T. Riemann,
\\CERN Yellow Reports: Monographs, CERN-2020-003,
%04 \url{http://dx.doi.org/10.23731/CYRM-2020-XXX}, p. \thepage.} 
DOI: \href{http://dx.doi.org/10.23731/CYRM-2020-003}{10.23731/CYRM-2020-003},
p. \thepage.
\\ \copyright\space CERN, 2020. Published by CERN under the 
%04-2
\href{http://creativecommons.org/licenses/by/4.0/}{Creative Commons Attribution 4.0 license}.} by: Z.-G. He, B.A. Kniehl \\
Corresponding author: B.A. Kniehl {[kniehl@desy.de]}}
\vspace*{.5cm}

\noindent Owing to its non-relativistic nature, heavy quarkonium, constituting heavy quark--antiquark
pairs ($\mathrm{Q\bar{Q}}=\mathrm{b\bar{b}}\; \mathrm{or}\; \mathrm{c\bar{c}}$) is an ideal object to 
investigate both perturbative and non-perturbative aspects of QCD. The non-relativistic
QCD factorisation formalism \cite{Bodwin:1994jh}, built on  rigorous effective field 
theory \cite{Caswell:1985ui}, provides a powerful tool to calculate heavy quarkonium
production and decay systematically. In this formalism, the production of heavy quarkonium is factorised 
into the process-dependent short-distance coefficients (SDCs) multiplied
by supposedly universal 
long-distance matrix elements (LDMEs). The SDC describing the production of a $\mathrm{Q\bar{Q}}$ pair 
in Fock state $n=~^{2S+1}L_{J}^{[a]}$ with total spin $S$, orbital angular momentum $L$, 
and total angular momentum $J$ can be calculated perturbatively as an expansion in
$\alpha_\mathrm{s}$. The LDMEs related to the probability that Fock state $n$ will evolve into the 
heavy meson are organised by the velocity scaling rules~\cite{Lepage:1992tx} of non-relativistic QCD (NRQCD), 
and their values can be determined by fitting to experimental data. Here, the velocity, 
$v_\mathrm{Q}$, refers to the motion of a heavy quark, Q, in the rest frame of the heavy meson. 
Although NRQCD has greatly improved our understanding of the heavy quarkonium production 
mechanism, the long-standing `$\mathrm{J}/\uppsi$ polarisation puzzle' has not yet been resolved.
The SDCs for the relevant colour singlet (CS) channel $(^3S_1^{[1]})$ and the three colour octet (CO) $(^3S_{1}^{[8]},
~^1S_{0}^{[8]},~ ^3P_{J}^{[8]})$ channels have been obtained by three groups independently, 
while the corresponding LDMEs were fitted to different sets of experimental data, based on 
different considerations~\cite{Butenschoen:2011yh,Chao:2012iv,Gong:2012ug}. However, none 
of these predictions can explain both the $\mathrm{J}/\uppsi$ yield and polarisation data at hadron 
colliders simultaneously. Recently, the universality of the NRQCD LDMEs was challenged 
by $\upeta_\mathrm{c}$ hadroproduction data~\cite{Butenschoen:2014dra}. 

Compared with hadron colliders, in $\mathrm{e}^{+}\mathrm{e}^{-}$ colliders, the production mechanism is 
simpler, the uncertainties in the theoretical calculations are smaller, and the convergence 
of perturbative calculations is faster. Moreover, on the experimental side, the much cleaner background 
makes it possible to study the production of other heavy quarkonia besides the $\mathrm{J}/\uppsi$ and $\Upsilon$ mesons, 
such as $\upeta_\mathrm{c,b}$ and $\upchi_\mathrm{c,b}$, and to study more production processes, such as the associated 
production of heavy quarkonium with a photon or a heavy quark pair, in detail. Therefore, heavy quarkonium
production in $\mathrm{e}^{+}\mathrm{e}^{-}$ colliders plays an important role in testing NRQCD 
factorisation, so as to help resolving the `$\mathrm{J}/\uppsi$ polarisation puzzle'. There are two 
ways to produce heavy quarkonium \setcounter{footnote}{0} directly.\footnote{Here, we mean production other than  through the decay of other heavy particles, 
like the Z boson, Higgs boson, or top quark.} One is in $\mathrm{e}^{+}\mathrm{e}^{-}$ annihilation and the other is
in $\upgamma\upgamma$ collisions. We review heavy quarkonium production, concentrating on the $\mathrm{J}/\uppsi$
case, by these two processes in Sections \ref{ee} and \ref{rr}, respectively, and discuss the prospects of heavy quarkonium production at the FCC-ee beyond the current measurements made at 
B factories and CERN LEP-II. Section \ref{sum} contains a summary and an outlook.

%=================================================================
\subsection[Heavy quarkonium production through $\mathrm{e}^{+}\mathrm{e}^{-}$ annihilation]{Heavy quarkonium 
production through $\mathrm{e}^{+}\mathrm{e}^{-}$ annihilation}\label{ee}

The total cross-section for inclusive $\mathrm{J}/\uppsi$ production in $\mathrm{e}^{+}\mathrm{e}^{-}$ annihilation was 
measured by the Babar~\cite{Aubert:2001pd}, Belle~\cite{Abe:2001za}, and CLEO~\cite{Briere:2004ug} 
collaborations at $\sqrt{s}=10.6\UGeV$, yielding
\begin{equation}
 \sigma(\mathrm{e}^{+}\mathrm{e}^{-}\to \mathrm{J}/\uppsi +\mathrm{X})=
\begin{cases}
 2.5\pm 0.21 \pm 0.21 \;\rm{pb} & \rm{Babar} \nonumber \\
 1.47\pm 0.10 \pm 0.13 \;\rm{pb} & \rm{Belle} \nonumber \\
 1.9 \pm 0.2 \; \rm{pb} &  \rm{CLEO}\nonumber
\end{cases}
\end{equation}
The NRQCD prediction at leading order (LO) is in the wide range of
0.8-1.7\,pb~\cite{Yuan:1996ep,Yuan:1997sn,Baek:1998yf,Schuler:1998az}, including 0.3\,pb 
from the CS mechanism. The Belle collaboration further managed to discriminate the contributions due to the final states  
$\mathrm{J}/\uppsi~+\mathrm{c\bar{c}}+\mathrm{X}$ and $\mathrm{J}/\uppsi +\mathrm{X}_{{\rm non-c\bar{c}}}$, and found that $\sigma(\mathrm{e}^{+}\mathrm{e}^{-}\to 
\mathrm{J}/\uppsi +\mathrm{c\bar{c}}+\mathrm{X})=0.74\pm0.08^{+0.09}_{-0.08}~\rm{pb}$ and $\sigma(\mathrm{e}^+\mathrm{e}^-\to \mathrm{J}/\uppsi +\mathrm{X}_{{\rm non-c\bar{c}}})
=0.43\pm0.09\pm0.09\,\mathrm{pb}$~\cite{Pakhlov:2009nj}. Neither of these results is compatible with LO NRQCD
predictions.

The LO NRQCD prediction for $\sigma(\mathrm{e}^+\mathrm{e}^-\to \mathrm{J}/\uppsi + \mathrm{c\bar{c}}+ \mathrm{X})$ is about $0.15\,\mathrm{pb}$, in which the
CO contribution is negligible~\cite{Liu:2003jj}. To solve the problem, both 
the next-to-leading-order (NLO) QCD~\cite{Zhang:2006ay} and relativistic 
corrections~\cite{He:2007te} were calculated. The relativistic 
correction was found to be less than one percent of the LO contribution. The effect of the NLO QCD 
correction is large. Its $K$ factor is about $1.8$ for $m_\mathrm{c}=1.5\UGeV$ and $\alpha_\mathrm{s}=0.26$.
After including the feed-down contribution from $\psi(2S)$, the NRQCD prediction at NLO becomes 
$0.53^{+0.59}_{-0.23}\,\rm{pb}$ and largely removes the discrepancy~\cite{Zhang:2006ay}. However, 
the theoretical prediction depends strongly on the chosen values of $m_\mathrm{c}$ and $\alpha_\mathrm{s}$. According 
to the design~\cite{Benedikt:2018qee}, the FCC-ee will run at several beam energies. Measuring
$\mathrm{J}/\uppsi +\mathrm{c\bar{c}}$ production at different energies will definitely help to improve our understanding of the 
parameter setting in the theoretical calculation.

At high energies, the predominant contribution to $\mathrm{J}/\uppsi +\mathrm{c\bar{c}}$ production comes from the
fragmentation process. For heavy quarkonium production, it is found that there are two types of 
fragmentation~\cite{Kang:2011mg}, (1) single-parton fragmentation (SPF) and (2) double-parton 
fragmentation (DPF). At hadron colliders, experimentally, the $\mathrm{J}/\uppsi
+ \mathrm{c\bar{c}}$ final state is hard to
detect and, theoretically, both SPF and DPF contribute, so that it is very difficult to study their properties separately.   

In the $\mathrm{e}^{+}\mathrm{e}^{-}$ annihilation process, only SPF contributes. Thus, the differential cross-section
in the fragmentation limit can be expressed as
\begin{equation}
\mathrm{d}\sigma(\mathrm{e}^{+}\mathrm{e}^{-}\to \mathrm{J}/\uppsi + \mathrm{c\bar{c}})=2\int \mathrm{d}\sigma(\mathrm{e}^{+}\mathrm{e}^{-}\to \mathrm{c\bar{c}}) D_{\mathrm{c}\to 
\mathrm{J}/\uppsi}(z) \mathrm{d}z,
\end{equation}
where 
\begin{equation}
D_{\mathrm{c}\to \mathrm{J}/\uppsi}(z)=\frac{8\alpha_\mathrm{s}^2}{27\uppi}\,\frac{z(1-z)^2(5z^4-32z^3+72z^2-32z+16)}{(2-z)^6}\,
\frac{|R(0)|^2}{m_\mathrm{c}^3},
\end{equation}
with $z=E_{\mathrm{J}/\uppsi}/\sqrt{s}$, where $|R(0)|$ is the wave function of $\mathrm{J}/\uppsi$ at the origin~\cite{Braaten:1993mp}.
 
\begin{figure}
\centering
\includegraphics[scale=0.5]{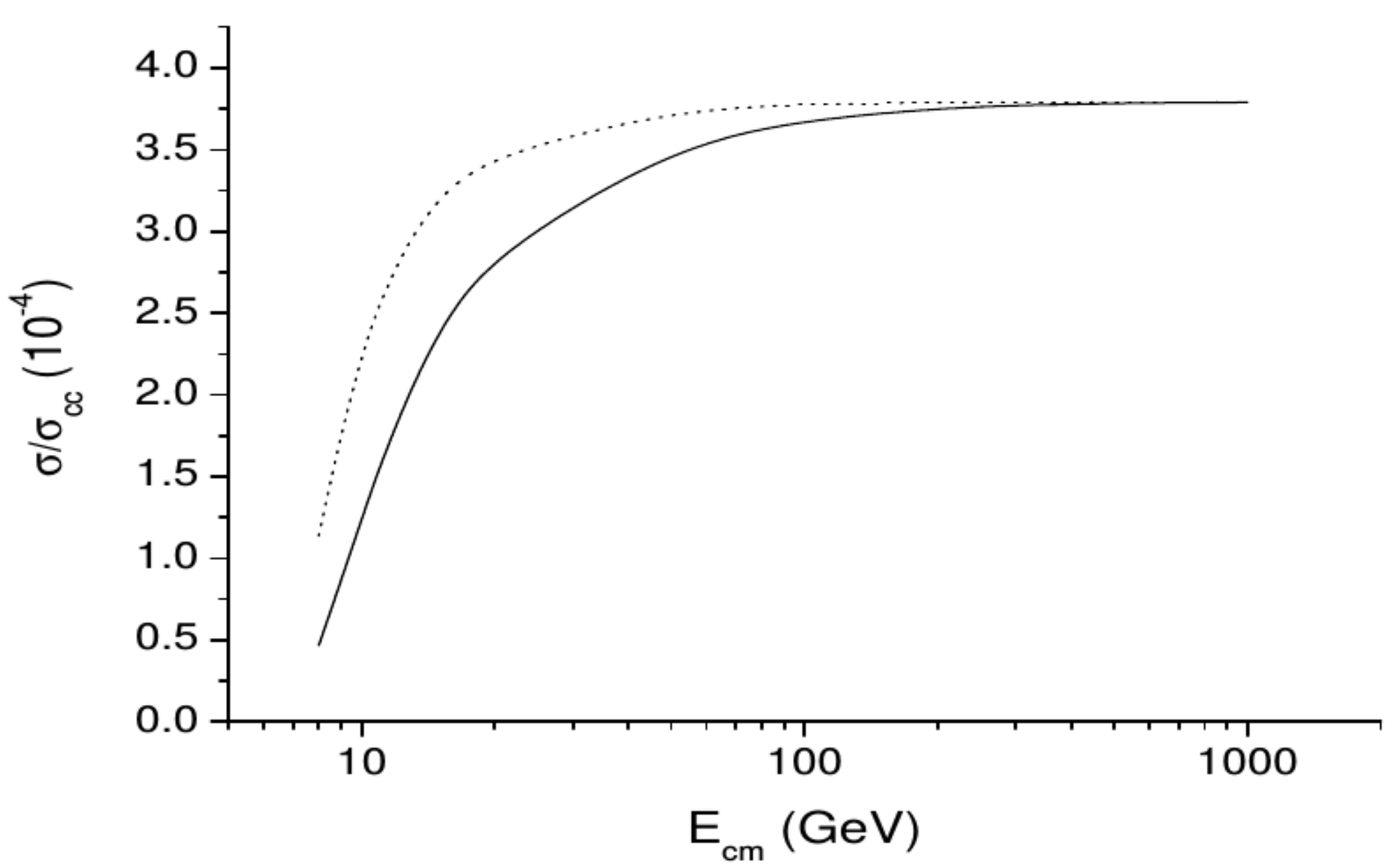}
\caption{Cross-section of $\sigma(\mathrm{e}^{+}\mathrm{e}^{-}\to \mathrm{J}/\uppsi +\mathrm{c\bar{c}}+\mathrm{X})$ normalised to $\sigma(
\mathrm{e}^{+}\mathrm{e}^{-}\to \mathrm{c\bar{c}}+\mathrm{X})$ at LO in NRQCD as a function of the centre-of-mass energy. The dotted line denotes 
the complete result, and the solid line denotes the fragmentation calculation. Figure courtesy ref.~\cite{Liu:2003jj}.}
\label{frag}
%\query{Please correct the figure labels in \Fref{frag}. Set variables
%in italic font. Be sure that minus signs are clearly minus signs.}
\end{figure}

At $\sqrt{s}=10.6\UGeV$, the fragmentation contribution can only account for $58\%$ of the 
complete calculation~\cite{Liu:2003jj}. The comparison between the complete calculation and the
fragmentation approximation is shown in \Fref{frag}. We observe that, only in the 
energy range of the FCC-ee or even beyond, the fragmentation contribution provides a good approximation. 
Conversely, the differential cross-section of $\mathrm{e}^{+}\mathrm{e}^{-}\to \mathrm{Q\bar{Q}}$ is known at 
$\mathcal{O}(\alpha_\mathrm{s}^2$)~\cite{Gao:2014nva,Gao:2014eea}. By comparing experimental measurements 
with higher-order theoretical calculations, the fragmentation function at higher orders can also 
be extracted.

For $\mathrm{J}/\uppsi +X_{\text{non-}\mathrm{c\bar{c}}}$ production, in the CS contribution, the NLO QCD corrections~\cite{Ma:2008gq} 
and relativistic corrections~\cite{He:2009uf} are equally important. Their $K$ factors are 
both around 1.2~\cite{Ma:2008gq,He:2009uf}, and the cross-section through NLO in QCD and $v^2$ becomes
$\sigma(\mathrm{e}^{+}\mathrm{e}^{-}\to \mathrm{J}/\uppsi + \mathrm{gg}) \simeq 437\,\rm{fb}$ for $\mu=\sqrt{s}/2$ and $m_\mathrm{c}=1.5\UGeV$, 
which almost saturates the Belle measurement and leaves little room for the CO
contribution~\cite{He:2009uf}. 
The NLO QCD corrections to the CO channels ${}^1S^{[8]}_{0}$ and ${}^3P^{[8]}_{J}$ were 
also computed~\cite{Zhang:2009ym}. A lower bound on the CO contribution is obtained by using the 
LDMEs from Ref.~\cite{Butenschoen:2011yh},  yielding $0.3\,\rm{pb}$. Therefore, the total NRQCD 
prediction is larger than the Belle measurements, but does not conflict with the Babar and CLEO measurements 
if we assume that $\sigma(\mathrm{e}^+\mathrm{e}^-\to \mathrm{J}/\uppsi +
\mathrm{c\bar{c}}+ \mathrm{X})$ is similar in these three experiments. To understand the CO 
mechanism in $\mathrm{e}^{+}\mathrm{e}^{-}$ annihilation, further analysis of $\mathrm{J}/\uppsi
+\mathrm{X}_{\text{non-}\mathrm{c\bar{c}}}$ production 
at 10.6\,GeV and in the future at the FCC-ee is necessary.

Besides charmonium, the production of bottomonium in $\mathrm{e}^{+}\mathrm{e}^{-}$ annihilation is also of great interest. 
However, the collision energy at B factories is so close to the $\Upsilon$ production threshold
that perturbative calculations are no longer reliable. Moreover, such a low energy is not sufficient to
enable $\Upsilon+ \mathrm{b\bar{b}}$ production. At the FCC-ee, the collision energy is of the order of $10^2\UGeV$ and,
therefore, provides a unique opportunity to study $\Upsilon+\mathrm{X}_{\text{non-}\mathrm{b\bar{b}}}$ and 
$\Upsilon+ \mathrm{b\bar{b}}$ production in $\mathrm{e}^{+}\mathrm{e}^{-}$ annihilation. Theoretically, the NRQCD prediction through 
NLO can  easily be obtained from the known $\mathrm{J}/\uppsi$ calculation by changing the 
value of $\sqrt{s}$ and replacing $m_\mathrm{c}$ with $m_\mathrm{b}$ and the LDMEs of $\mathrm{J}/\uppsi$ by those of $\Upsilon$.  

\subsection[Heavy quarkonium production in $\upgamma\upgamma$ collisions]{Heavy
quarkonium production
in $\upgamma\upgamma$ collisions}
\label{rr}

$\mathrm{J}/\uppsi$ photoproduction in $\upgamma\upgamma$ collisions ($\mathrm{e}^{+}\mathrm{e}^{-}\to \mathrm{e}^{+}\mathrm{e}^{-}\mathrm{J}/\uppsi + \mathrm{X}$) 
was measured by the DELPHI collaboration at  LEP-II~\cite{TodorovaNova:2001pt,Abdallah:2003du}. The total cross-section was found to be $\sigma(\mathrm{e}^{+}\mathrm{e}^{-}\to
\mathrm{e}^{+}\mathrm{e}^{-}\mathrm{J}/\uppsi +X)=(45\pm9\pm17)\,\rm{pb}$~\cite{Abdallah:2003du}. The DELPHI collaboration also measured the 
transverse momentum ($p_\mathrm{T}$) distribution of the cross-section. Since the higher excited states $\upchi_{\mathrm{c}J}$ and $\uppsi^{\prime}$ can decay into $\mathrm{J}/\uppsi$ via radiative decays or 
hadronic transitions, their feed-down contributions should also be considered. In such processes, 
the $\mathrm{c\bar{c}}$ pair can either be produced by photons directly (direct photoproduction) or via 
the light quark and gluon content of the photons (resolved photoproduction), so that there are three 
channels: direct, single resolved, and double resolved, all of which contribute formally at the 
same order in the perturbative expansion and should be included.

Working in the Weizs\"acker--Williams approximation to describe the bremsstrahlung photons radiated
off the $\mathrm{e}^{\pm}$ beams and using the factorisation theorems of the QCD parton model and NRQCD, the general formula for the differential cross-section for the production of the heavy quarkonium state H can be written as 
\begin{equation}
\frac{\mathrm{d}\sigma (\mathrm{e}^{+}\mathrm{e}^{-}\to \mathrm{e}^{+}\mathrm{e}^{-} \mathrm{H}+\mathrm{X})}{\mathrm{d}x_1 \mathrm{d}x_2 \mathrm{d}x_a \mathrm{d}x_b} = \sum_{a,b,n} f_{\upgamma}(x_1)
f_{\upgamma}(x_2)f_{a/\upgamma}(x_a)f_{b/\upgamma}(x_b) 
 \times\mathrm{d}\hat{\sigma}(a+b\to \mathrm{Q\bar{Q}}(n)+\mathrm{X})
\langle\mathcal{O}^\mathrm{H}(n)\rangle,
\end{equation}
where $f_{\upgamma}(x)$ is the flux function of the photon in the $\mathrm{e}^{\pm}$ beam, $f_{j/\upgamma}(x)$ is 
$\delta(1-x)$ if $j=\upgamma$ and otherwise the parton distribution function
of parton $j$ in the resolved photon, $\mathrm{d}\hat{\sigma}(a+b\to \mathrm{Q\bar{Q}}(n)+\mathrm{X})$ is the partonic cross-section, and $\langle\mathcal{O}^\mathrm{H}(n)\rangle$ is the NRQCD LDME. 

\begin{figure}
\centering
\begin{tabular}{cc}
\includegraphics[scale=0.37]{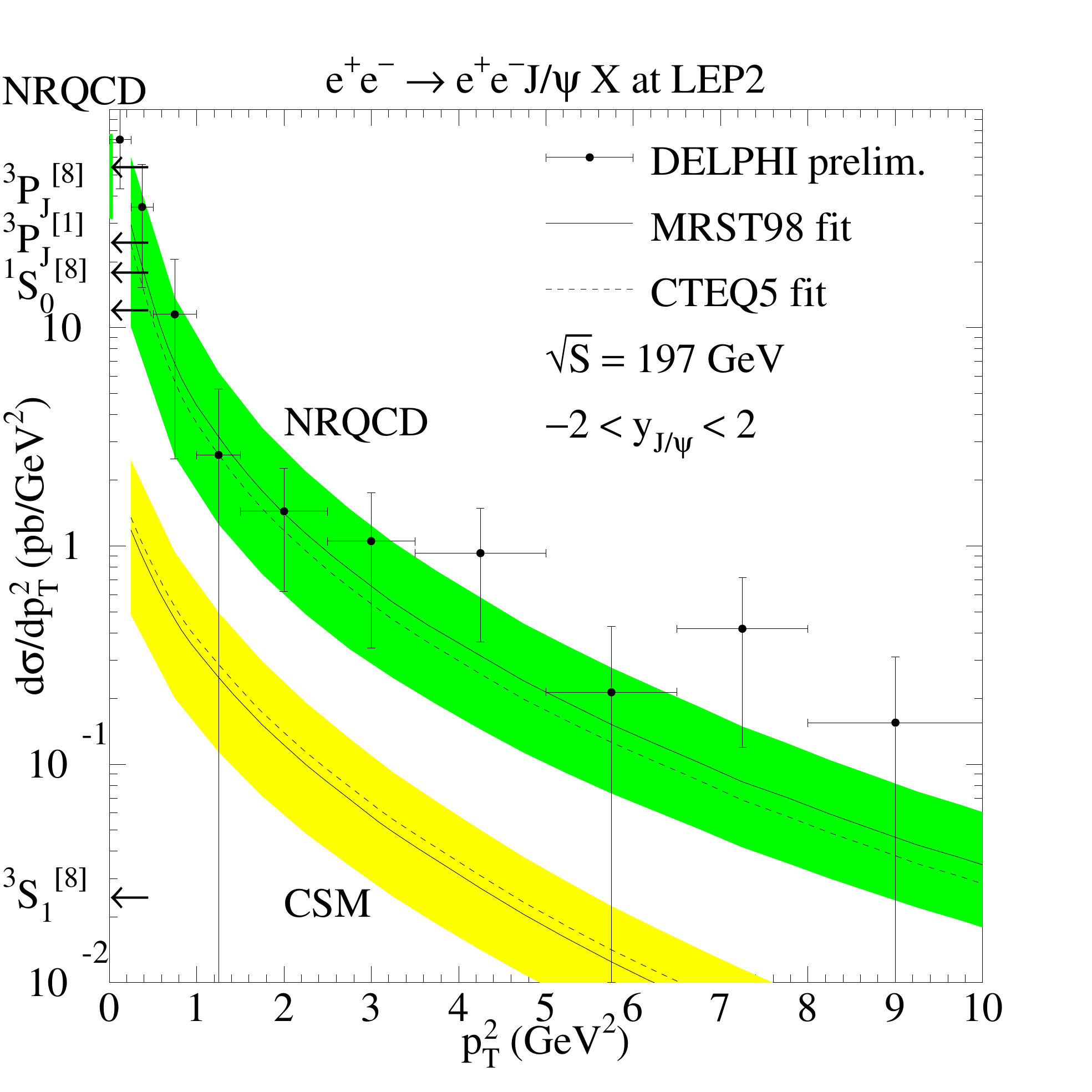}&
\end{tabular}
\caption{Comparison between NRQCD and CS model predictions of $\mathrm{d}\sigma/\mathrm{d} p_\mathrm{T}^2$ as functions of $p_\mathrm{T}^2$ for $\upgamma\upgamma\to \mathrm{J}/\uppsi$ at LO and DELPHI 
measurement at LEP-II. The solid and dashed lines are calculated with the MRST98 LO and CTEQ5 parton distribution functions, respectively. The bands indicate the theoretical uncertainties. Figure courtesy ref.~\cite{Klasen:2001cu}.}
\label{lep-LO}
%\query{Please correct the figure labels in \Fref{lep-LO}. Set particle names in roman font but variables
%in italic font. Check that labels are clear to read.}
\end{figure}

In the LO calculation, both direct $\mathrm{J}/\uppsi$ production and the feed-down from $\chi_{\mathrm{c}J}$ for
$J=0,1,2$ and $\uppsi^{\prime}$ are included~\cite{Klasen:2001cu}. For $\mathrm{J}/\uppsi$ ($\uppsi^{\prime}$) 
production through relative order $\mathcal{O}(v^4)$, the Fock states include $n={}^3S_1^{[1,8]},{}^1S_0^{[8]},{}^3P_J^{[8]}$, 
and for $\chi_{\mathrm{c}J}$ production at LO in $v^2$ one needs $n={}^3P_J^{[1]},{}^3S_1^{[8]}$. As shown in
\Fref{lep-LO}, the LO NRQCD prediction of $\mathrm{d}\sigma/\mathrm{d}p_\mathrm{T}^2$, evaluated with the LDMEs from the
LO fit to Tevatron data~\cite{Braaten:1999qk}, agree very well with the DELPHI data, while the CS
contribution itself lies far below the data, as the central values are about 16 
times smaller. The total cross-section in the range $1\leq p_\mathrm{T}^2\leq 10\UGeV^2$ measured by
DELPHI is $6.4\pm2.0\,\rm{pb}$~\cite{TodorovaNova:2001pt}. The NRQCD prediction is
$4.7^{+1.9}_{-1.2}\,\rm{pb}$~\cite{Klasen:2001cu}, which is also consistent with the DELPHI result,
within errors. However, the CS contribution is only $0.39^{+0.16}_{-0.09}\,\rm{pb}$~\cite{Klasen:2001cu}. 
The nice agreement between the NRQCD calculation and the experimental measurement for $\mathrm{J}/\uppsi$ photoproduction is one of the earliest pieces of evidence for the CO mechanism predicted by NRQCD.
 
\begin{figure}
\centering
\begin{tabular}{cc}
\includegraphics[scale=0.48]{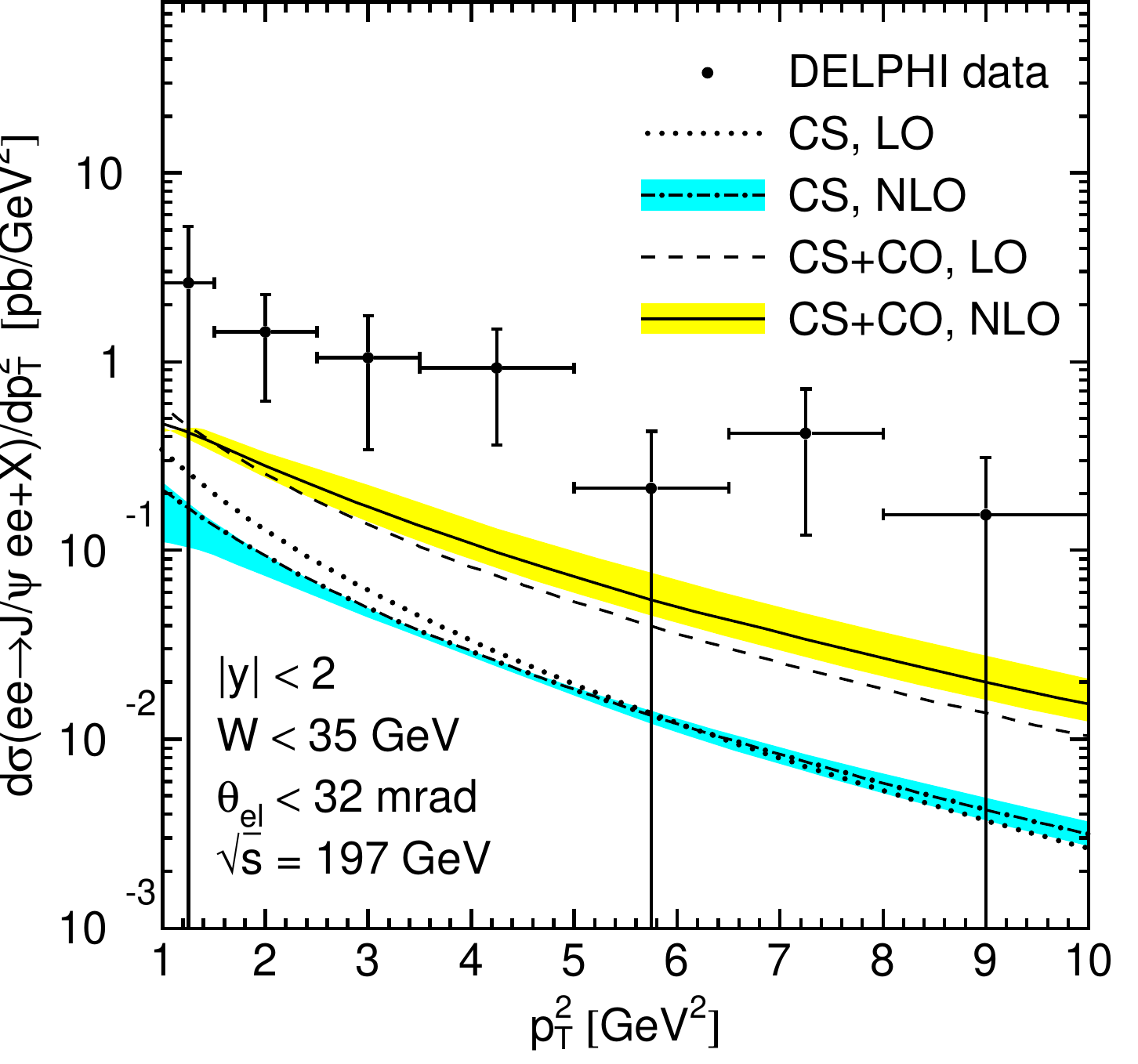}
\end{tabular}
\caption{Comparison of LEP-II data on $\upgamma\upgamma\to \mathrm{J}/\uppsi$ with NLO NRQCD predictions evaluated with
LDMEs obtained via a global data analysis. Figure courtesy ref.~\cite{Butenschoen:2011yh}.}
\label{lep-NLO}
%\query{Please correct the figure labels in \Fref{lep-NLO}. Set particle names
%in roman font but variables
%in italic font. Check that labels are clear to read.}
\end{figure}

In 2011, two groups independently obtained  complete NLO QCD corrections to $\mathrm{J}/\uppsi$ direct
hadroproduction for the first time \cite{Butenschoen:2010rq,Ma:2010yw}. However, their LDMEs are 
different because they fitted to data in different $p_\mathrm{T}$ ranges. To eliminate such problems and 
further check the universality of the NRQCD LDMEs at NLO, a global analysis to worldwide data
including $\upgamma\upgamma$ collisions was conducted. The resulting three CO LDMEs,
$\langle{\cal O}^{\mathrm{J}/\uppsi}({}^1S_0^{[8]}) \rangle=(4.97\pm0.44)\times10^{-2}\UGeV^{3}$,
$\langle{\cal O}^{\mathrm{J}/\uppsi}({}^3S_1^{[8]}) \rangle=(2.24\pm0.59)\times10^{-3}\UGeV^{3}$, and
$\langle{\cal O}^{\mathrm{J}/\uppsi}({}^3P_0^{[8]}) \rangle=(-1.61\pm0.20)\times10^{-2}\UGeV^{5}$,
which obey the velocity scaling rules, were found to explain all the $\mathrm{J}/\uppsi$ yield data fairly well, except for the case of $\upgamma\upgamma$ collisions \cite{Butenschoen:2011yh}. In contrast to the situation at LO, the DELPHI data systematically overshoot the NLO NRQCD prediction, as may be seen in \Fref{lep-NLO}. However, Figs.~\ref{lep-LO} and \ref{lep-NLO} indicate that the uncertainties in the experimental measurements are very large. There are only $36\pm7$ $\mathrm{J}/\uppsi\to \upmu^{+}\upmu^{-}$ events in total (and 16 thereof in the region
$p_\mathrm{T}>1\UGeV$), collected with an integrated luminosity of 617\,$\rm{pb}^{-1}$. The integrated luminosity at the FCC-ee will reach the $\rm{ab}^{-1}$ level, which is more than three orders of magnitude larger than that of LEP-II.
Measuring $\mathrm{J}/\uppsi$ production in $\upgamma\upgamma$ collisions at the FCC-ee would not only serve as a cross-check of the LEP-II results, but also provide results with high accuracy. Such a study could
surely clarify the current conflict and deepen our understanding of the heavy
quarkonium production
mechanism in $\upgamma\upgamma$ collisions.

Unlike the case of $\mathrm{e}^{+}\mathrm{e}^{-}$ annihilation, $\mathrm{J}/\uppsi+\mathrm{c\bar{c}}+\mathrm{X}$ production in $\upgamma\upgamma$
collisions is predicted to have a smaller cross-section than $\mathrm{J}/\uppsi+X_{\text{non-}\mathrm{c\bar{c}}}$ production. While $\upgamma\upgamma\to \mathrm{J}/\uppsi+\mathrm{X}_{\text{non-}\mathrm{c\bar{c}}}$ proceeds dominantly via
single resolved photoproduction, $\upgamma\upgamma\to \mathrm{J}/\uppsi+\mathrm{c\bar{c}}+\mathrm{X}$ proceeds dominantly via
direct photoproduction~\cite{Li:2009fd}. The total cross-section in the region $p_\mathrm{T}^{\mathrm{J}/\uppsi}>1\UGeV$
is predicted to be about 0.16--$0.20\,\rm{pb}$, depending on the chosen values of $\alpha_\mathrm{s}$ and the CS LDME~\cite{Li:2009fd,Chen:2016hju}. Its NLO NRQCD correction has also been calculated, and the $K$ factor is found to be 1.46, enhancing the total cross-section in the region $p_\mathrm{T}^{\mathrm{J}/\uppsi}>1\UGeV$ to become
around 0.23--$0.29\,\rm{pb}$, which is too small to be analysed at LEP-II~\cite{Chen:2016hju}. The cross-section becomes larger as the $\mathrm{e}^{+}\mathrm{e}^{-}$ collision energy increases. Based on the results given in 
Ref.~\cite{Chen:2016hju}, we estimate the numbers of $\mathrm{J}\uppsi+\mathrm{c\bar{c}}$ events accumulated with the FCC-ee at the ZZ and ZH thresholds  to be around $2\times 10^{6}$ each, assuming the kinematic-cut conditions
for the FCC-ee to be the same as for LEP-II. Such large data samples should be enough to usefully study $\mathrm{J}/\uppsi+ \mathrm{c\bar{c}}+ \mathrm{X}$ production in $\upgamma\upgamma$ collisions.    

\subsection[Summary and outlook]{Summary and outlook}\label{sum}

The production mechanisms of heavy quarkonium, especially of the $\mathrm{J}/\uppsi$ meson, have not yet been fully understood within the framework of NRQCD factorisation. We have discussed here two modes of $\mathrm{J}/\uppsi$ production at $\mathrm{e}^{+}\mathrm{e}^{-}$ colliders, through $\mathrm{e}^{+}\mathrm{e}^{-} $ annihilation and $\upgamma\upgamma$ collisions. In
the $\mathrm{e}^{+}\mathrm{e}^{-}$ annihilation case, for $\mathrm{J}/\uppsi+
\mathrm{c\bar{c}}
+ \mathrm{X}$ production, the NRQCD prediction 
and the Belle measurement agree within errors; however, for $\mathrm{J}/\uppsi+\mathrm{X}_{\text{non-}
\mathrm{c\bar{c}}}$ production, the
Belle result favours the CS model prediction and is overshot by NRQCD predictions evaluated using any of the available LDME sets, although the latter are mutually inconsistent. We note that the NRQCD predictions seem
to be compatible with the Babar and CLEO results. As for $\mathrm{J}/\uppsi$ production in $\upgamma\upgamma$ collisions, the NRQCD prediction can explain the LEP-II data, whose uncertainties are large, at LO, but fails once
the NLO correction is included. 

The FCC-ee will run at different energy points with considerable integrated luminosity, of 
$\mathcal{O}(\rm{ab}^{-1})$ or even $\mathcal{O}(10^2~\rm{ab}^{-1})$ at the Z boson
peak~\cite{Benedikt:2018qee}, which will provide a perfect environment to judge the
disagreements independently. Moreover, it can significantly enrich our knowledge of heavy quarkonium production in $\mathrm{e}^{+}\mathrm{e}^{-}$ collisions, especially by studying bottomonium production, the
fragmentation function of $\mathrm{c}\to \mathrm{J}/\uppsi$, and $\mathrm{J}/\uppsi+
\mathrm{c\bar{c}}$ production in $\upgamma\upgamma$ collisions.
 
%=================================================================

\end{bibunit}

\label{sec-sm-He-Kniehl}
\clearpage \pagestyle{empty}  \cleardoublepage
%============================================

 \pagestyle{fancy}
 \fancyhead[CO]{\thechapter.\thesection \hspace{1mm} Vertex functions in QCD---preparation for beyond two loops}
 \fancyhead[RO]{}
 \fancyhead[LO]{}
 \fancyhead[LE]{}
 \fancyhead[CE]{}
 \fancyhead[RE]{}
 \fancyhead[CE]{J.A.~Gracey}
 \lfoot[]{}
 \cfoot{-  \thepage \hspace*{0.075mm} -}
 \rfoot[]{}
 
 \begin{bibunit}[elsarticle-num] % define the bib-style for the unit: elsarticle-num.bst
        %  text-1; this is the corresponding section
        %\putbib[2loops] % the *.bib
        %\end{bibunit}
        % go-on
        %--- from: bibunits.sty, adapts the font size of ``References'' to section
        \let\stdthebibliography\thebibliography
        \renewcommand{\thebibliography}{%
                \let\section\subsection
                \stdthebibliography}
        %---
        
        \section
        [Vertex functions in QCD---preparation for beyond two loops \\ {\it J.A. Gracey }]
        {Vertex functions in QCD---preparation for beyond two loops
                \label{contr:jgracey}}
        \noindent
        {\bf Contribution\footnote{This contribution should be cited as:\\
J.A. Gracey, Vertex functions in QCD---preparation for beyond two loops,  
%04 DOI:10.23731/CYRM-2020-XXX.\thepage, in:
%04 \url{http://dx.doi.org/10.23731/CYRM-2020-XXX.\thepage}, in:
DOI: \href{http://dx.doi.org/10.23731/CYRM-2020-003.\thepage}{10.23731/CYRM-2020-003.\thepage}, in:
Theory for the FCC-ee, Eds. A. Blondel, J. Gluza, S. Jadach, P. Janot and T. Riemann,
\\CERN Yellow Reports: Monographs, CERN-2020-003,
%04 \url{http://dx.doi.org/10.23731/CYRM-2020-XXX}, p. \thepage.} 
DOI: \href{http://dx.doi.org/10.23731/CYRM-2020-003}{10.23731/CYRM-2020-003},
p. \thepage.
\\ \copyright\space CERN, 2020. Published by CERN under the 
%04-2
\href{http://creativecommons.org/licenses/by/4.0/}{Creative Commons Attribution 4.0 license}.} by: J.A.~Gracey {[gracey@liverpool.ac.uk]}}
        \vspace*{.5cm}
        
%\begin{document}
%\title{Vertex functions in QCD - preparation for beyond two loops}
%\author{J.A. Gracey, \\ Theoretical Physics Division, \\ 
%Department of Mathematical Sciences, \\ University of Liverpool, \\ P.O. Box 
%147, \\ Liverpool, \\ L69 3BX, \\ United Kingdom.} 
%\date{March, 2019.}

%\maketitle 
%\vspace{5cm} 
\section*{Abstract}

We summarise the algorithm to determine the two-loop off-shell
three-point vertex functions of QCD before outlining the steps required to
extend the results to three and more loops.

%\newpage 

\subsection{Introduction}

In our current generation of high-energy particle accelerators involving hadron
collisions, a major source of background is radiation derived from the strong 
sector. As this is governed by quantum chromodynamics (QCD),  to 
quantify the background effects one must carry out high loop order 
computations. There has been remarkable activity and progress in this direction
since around the turn of the millennium. The primary focus has been with the 
evaluation of on-shell $n$-point gluonic and fermionic amplitudes to several 
loop orders, both analytically and numerically. Indeed, such results have been 
crucial in ensuring that the Higgs particle was observed at CERN's LHC. However,
having information on the off-shell Green's functions, such as the three-point 
vertices of QCD, is also important for theory as well as experiment. For 
instance, various articles in this direction have appeared over the years. A 
non-exhaustive literature for this status of three- and four-point functions
at various external momenta configurations is given by Refs. 
\cite{Pascual:1980yu,Davydychev:1996pb,Davydychev:2001uj,Binger:2006sj,
Davydychev:1997vh,Davydychev:1998aw,Davydychev:2000rt,Chetyrkin:2000dq,
Chetyrkin:2000fd,Kellermann:2008iw,Ahmadiniaz:2012xp}. There are various 
theoretical reasons for having such off-shell Green's functions. One is that 
knowing, say, the two-loop off-shell vertex functions enables higher-loop 
$n$-point on-shell amplitudes to be modelled numerically. This could be an
interim position in the absence of the technology to compute them fully 
explicitly. Such an approach is not uncommon.  Equally, in solving QCD beyond 
the perturbative limit analytically to probe deep infrared properties using the
Schwinger--Dyson formalism, approximations must be made in order to solve the
infinite tower of Green's functions. Until recent years, the validity of such 
approximations could not be fully quantified. However, with explicit 
perturbative results, for instance, such error analyses have been made possible. For
instance, one approximation in solving two- and three-point Schwinger--Dyson 
equations is to neglect the summed graphs deriving from the quartic gluon 
vertex. Work in this direction over a period of time 
\cite{Blum:2014gna,Blum:2016fib,Mitter:2014wpa,Alkofer:2008tt,Alkofer:2013bk}
has confirmed that such a step does not affect final results by more than a few 
percent. Equally, the Schwinger--Dyson method has been applied to finding the 
behaviour of the vertex functions. While similar approximations have been made, 
such analyses must be consistent with explicit perturbative results where 
{\em no} approximation is made at a particular loop order to drop a subset of 
contributing graphs. As an aside, lattice gauge theory calculations of vertex 
functions equally have to match on to perturbative results. Therefore, in light 
of these different areas of activity, there is a clear need to compute QCD 
$n$-point, and specifically vertex functions, off-shell as well as on-shell. For 
the former, which is the focus of this article, we will review the status of 
the two-loop evaluation of the three-point vertices as well as outline the
algorithm to extend this to higher-loop order. While the discussion will be 
technical in nature, we will pool together all the necessary ingredients for 
the goal to be obtained at three loops.

While it is not immediately obvious, it is the case that the route to achieve
this will involve higher-level mathematics extracted, for instance, from an 
algebraic geometry approach. Indeed, this also lies at the heart of on-shell 
amplitude
computations. This technology has revolutionalized the programme of loop
calculations. An example of this can be seen in the results for two-loop 
off-shell vertex results of Ref. \cite{Gracey:2014mpa}, where harmonic polylogarithms
based on a specific type of polynomials, known as cyclotomic, 
\cite{Ablinger:2011te}, appeared. One corollary of such results is the 
possibility of effecting renormalization schemes other than the canonical 
$\MSbar$ one, which is universally accepted as the default scheme. Although this 
is the scheme with which one can carry out very high loop order calculations, it
is not a kinematic one and retains no data within the $\beta$ function, for 
instance, of the information on the subtraction point. In  Ref.
\cite{Celmaster:1979km}, the momentum subtraction scheme, denoted  MOM, was 
introduced and the $R$ ratio studied \cite{Celmaster:1980ji}. Extending Ref. 
\cite{Celmaster:1979km} to the next order in Ref. \cite{Gracey:2011vw} produced the 
three-loop MOM renormalization group functions. This allowed for studies of 
physical quantities at a loop order where scheme effects were apparent 
\cite{Gracey:2014pba}. One consequence is that choosing alternative 
renormalization schemes could lead to a different way of estimating theory 
errors in measurements. In other words, similar to an experiment estimating a
measured quantity in different schemes, the average of the result could be a 
more sound way of assessing truncation errors as an alternative to using values
at different scales. With fully off-shell vertex functions, for instance, this 
idea can be extended beyond the symmetric point subtraction of the MOM case to 
have a region bounding the central value.

This section is organised as follows. The method used to evaluate three-point
off-shell vertex functions is discussed  next, with reference to
the triple-gluon vertex. This forms the basis for higher-loop computations,
with the algorithm being outlined in Section $9.3$. Concluding remarks are made
in Section $9.4$.  

\subsection{Current status}

At the outset, it is worth reviewing aspects of the early QCD vertex 
evaluations. By this we mean that our focus will be on cases where there is 
{\em no} nullification of an external momentum. This is important, since, in the
computation of the QCD $\beta$-function to very high loop order, the extraction
of the $\MSbar$ coupling constant renormalization constant can be facilitated
by setting the momentum of one of the external fields of the vertex function to
zero. This is a mathematical shortcut, since the ultraviolet divergence is not
contaminated by any infrared ones. By contrast, this infrared-safe procedure 
does not produce the {\em correct} finite part of the vertex functions, so it
is not an appropriate method for gaining insight into any aspect of the 
kinematic properties of the vertex functions themselves. To be more concrete in
the discussion, we will focus on the triple-gluon vertex function of \Fref{jagfig1}, which represents
\begin{equation}
\left\langle A^a_\mu(p_1) A^b_\nu(p_2) A^c_\sigma(-p_1-p_2) \right\rangle = 
f^{abc} \Sigma^{\mbox{\footnotesize{ggg}}}_{\mu \nu \sigma}(p_1,p_2) = 
f^{abc} \sum_{k=1}^{14}
{\cal P}^{\mbox{\footnotesize{ggg}}}_{(k)   \mu \nu \sigma }(p_1,p_2)  
\Sigma^{\mbox{\footnotesize{ggg}}}_{(k)}(p_1,p_2) \, , 
\label{verdef}
\end{equation}
where $f^{abc}$ are the colour group structure constants. The momenta $p_i$ 
satisfy energy--momentum conservation 
\begin{equation}
\sum_{n=1}^3 p_i ~=~ 0
\end{equation}
and the underlying Lorentz invariants that the three-point functions depend on
are expressed in terms of two dimensionless variables, $x$ and $y$, and one 
mass scale, $\mu$, which are defined by
\begin{equation}
x = \frac{p_1^2}{p_3^2} \, ,~\qquad 
y = \frac{p_2^2}{p_3^2} \, ,~\qquad 
p_3^2 = - \mu^2 \, ,
\label{momconf}
\end{equation}
and we assume that none of $p_i^2$ vanishes. In \Eref{verdef}, we have 
decomposed the vertex into its $14$ scalar amplitudes 
$\Sigma^{\mbox{\footnotesize{ggg}}}_{(k)}(p_1,p_2)$ with respect to a basis of
Lorentz tensors 
${\cal P}^{\mbox{\footnotesize{ggg}}}_{(k) \, \mu \nu \sigma }(p_1,p_2)$. With 
this structure in mind for the other two three-point vertices, the full one-loop 
vertex functions were studied in Ref. \cite{Celmaster:1979km} in the early years 
following the discovery of asymptotic freedom. 

{\begin{figure}
\begin{center}
\includegraphics[width=5.5cm,height=4.0cm]{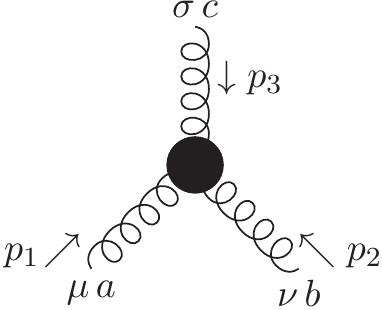}
\end{center}
%\vspace{0.5cm}
\caption{Triple-gluon vertex function}
\label{jagfig1}
\end{figure}}

Two important main early papers that stand out are Refs. 
\cite{Celmaster:1979km,Ball:1980ax}. The former 
focused on the vertex functions at the fully symmetric subtraction point 
defined by $x$~$=$~$y$~$=$~$1$ and introduced the MOM kinematic renormalization 
scheme known as MOM for momentum subtraction. Unlike the $\MSbar$ scheme, the 
renormalization is carried out at this specific symmetric point and the finite 
part of the vertex functions is absorbed into the renormalization constants. 
Therefore, the $\beta$-functions contain kinematic data. The motivation of
Ref. \cite{Celmaster:1979km} was to study whether the convergence of the perturbative 
series could be improved in this new scheme. The other article 
\cite{Ball:1980ax} reported  a systematic study of each fully off-shell three-point 
vertex with a view to writing each in terms of amplitudes dictated by external 
gluons being transverse. As such, it has served as the default vertex function 
convention, where Schwinger--Dyson techniques are used to approximate other 
Green's functions. Consequently, there have been a large number of one-loop 
studies of the three three-point vertices for different external momentum 
configurations, as noted earlier. In some cases, these studies have been at two 
loops, but for the most part one or more external gluon legs were on-shell and 
quarks have been massless, except in the case of  Refs.
\cite{Davydychev:2001uj,Davydychev:2000rt}. In the main, the evaluation has been
by standard quantum field theory techniques via Feynman graphs. However, modern 
string-inspired methods have been used 
\cite{Ahmadiniaz:2012xp,Ahmadiniaz:2013rla} for off-shell one-loop vertex 
functions. The case where a gluon, for example, is on-shell must be treated 
separately from the configuration introduced in \Eref{momconf}, owing to 
potential infrared singularities in taking the on-shell limit from the fully 
off-shell results. 

Studies of the vertex functions for the special cases where one or more
external lines are on-shell has direct applications to experimental set-ups. One
of the reasons why these were computed was, in the main, that the calculational
tools for the off-shell case were not developed until much later. Several main 
components were necessary for this, with the main breakthrough arriving in the
form of the Laporta algorithm \cite{Laporta:2001dd}. This is a procedure of 
relating scalar Feynman integrals of a particular $n$-point function at a 
specified loop order to core or master integrals of $r$-point functions with 
$r$~$\leq$~$n$ and the same loop order, the connection between integrals being 
made via integration by parts. Then, starting with the most complicated integral,
the relations derived from integration by parts could be solved algebraically. 
While  such a large set of equations clearly contains a degree of redundancy, the
whole process can be encoded for a computer to handle and several packages
to do so are publicly available  
\cite{Studerus:2009ye,vonManteuffel:2012np,Smirnov:2008iw,Lee:2012cn,
Lee:2013mka,Anastasiou:2004vj,Maierhoefer:2017hyi}. The second breakthrough
necessary to complete this task was the determination of the master 
integrals. For three-point functions, these had to be constructed by specialised 
methods 
\cite{Davydychev:1992xr,Usyukina:1992wz,Usyukina:1994iw,Birthwright:2004kk}
to two loops, as integration by parts had been exhausted by the Laporta 
algorithm. To give a flavour of the resultant mathematical structure, the one-loop master integral of  \Fref{jagfig2} is, for instance, given by 
\cite{Davydychev:1992xr,Usyukina:1992wz,Usyukina:1994iw}
\begin{equation}
I_1(x,y) = - \frac{1}{\mu^2} \left[ \Phi_1(x,y) + \Psi_1(x,y) \epsilon 
+ \left[ \frac{\zeta(2)}{2} \Phi_1(x,y) + \chi_1(x,y) \right] 
\epsilon^2 + O(\epsilon^3) \right] 
\label{intI}
\end{equation}
in $d = 4 - 2\epsilon$ dimensions, where $\zeta(z)$ is the Riemann zeta
function. Here, the functions are related to polylogarithms $\mbox{Li}_n(z)$. 
For instance
\begin{equation}
\Phi_1(x,y) = \frac{1}{\lambda} \left[ 2 \mbox{Li}_2(-\rho x)
+ 2 \mbox{Li}_2(-\rho y)
+ \ln \left( \frac{y}{x} \right)
\ln \left( \frac{(1+\rho y)}{(1+\rho x)} \right)
+ \ln(\rho x) \ln(\rho y) + \frac{\uppi^2}{3} \right] \, ,
\end{equation}
with
\begin{equation}
\rho(x,y) = \frac{2}{[1-x-y+\lambda(x,y)]} \, , \qquad
\lambda(x,y) = \left  [1-2x-2y+x^2-2xy+y^2 \right ]^{\half} \, ,
\end{equation}
with the other functions of \Eref{intI} given in 
\cite{Davydychev:1992xr,Usyukina:1992wz,Usyukina:1994iw} too. While the 
$O(\epsilon)$ terms may not, at first sight, appear to be necessary, they are
required for various reasons. One is that, at higher loops, these one-loop
expressions are multiplied by the counterterms. Thus, when a pole in $\epsilon$ 
multiplies a term that is $O(\epsilon)$, then that will contribute to the
finite part of the vertex function at the next loop order. Accordingly, one
needs the master integrals to at least $O(\epsilon^2)$ at one loop for a three-loop evaluation. We have indicated this since it could be the case that, in
the reduction using the Laporta algorithm, a spurious pole in $\epsilon$ arises,
which we discuss later. This is not an uncommon occurrence but the latest 
Laporta algorithm packages now have tools to circumvent this possibility. These
technical issues aside, the full off-shell three-point QCD vertex functions are 
available to two loops with more details provided in Ref. \cite{Gracey:2014mpa}.

{\begin{figure}
\begin{center}
\includegraphics[width=6.5cm,height=4.5cm]{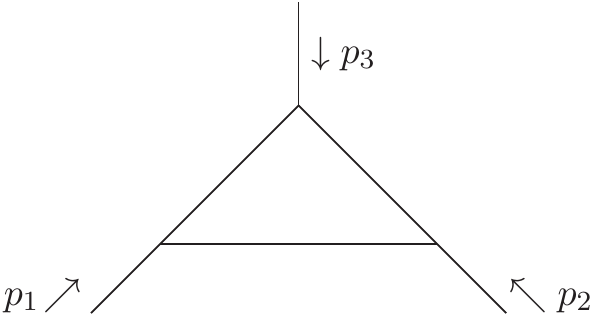}
\end{center}
%\vspace{0.5cm}
\caption{One-loop three-point master integral $I_1(x,y)$}
\label{jagfig2}
\end{figure}}

\subsection{Three-loop strategy}

One reason for detailing the formalism to carry out the two-loop 
computations is that it points the way for higher-loop corrections. On that
basis, we outline the next parts of the jigsaw to construct the three-loop 
extension of Ref. \cite{Gracey:2014mpa}. First, we assume that the procedure of the 
general algorithm for the Green's functions is applied to obtain the three-loop
scalar amplitudes, as illustrated in \Eref{verdef}. From these, the large set of 
scalar Feynman integrals is assembled; these must be reduced to the master
integrals. The Laporta algorithm can, in principle, be applied in the three-loop 
case, using one of the latest packages that have the built-in improvements, such
as the refined algebraic reduction of the {\sc Kira} package 
\cite{Maierhoefer:2017hyi}. However, to speed the integration by parts procedure,
it is not inconceivable that a faster algorithm could be developed. For 
instance, for many years, the {\sc Mincer} package served the multiloop community
well for three-loop massless two-point graphs in four dimensions 
\cite{Gorishnii:1989gt,Larin:1991fz}. It implemented the star--triangle rule to 
produce an efficient code to evaluate even the heaviest fully gluonic three-loop graphs. With the need for more precision experimentally, the four-loop 
{\sc Forcer} package \cite{Ueda:2016yjm,Ruijl:2017cxj} has superseded
{\sc Mincer} in the journey to hit the latest precision benchmark. Each has 
been encoded in the symbolic manipulation language {\sc Form} 
\cite{Vermaseren:2000nd,Tentyukov:2007mu}. With  increasing loop order, the 
evaluation time for a Green's function increases. However, the {\sc Forcer} 
algorithm implements a {\em new} integration rule to handle an internal 
topology that has no three-loop antecedents and hence is a purely four-loop 
feature. We have mentioned this since {\sc Forcer}, like {\sc Mincer}, applies 
only to two-point functions. However, the same new rule should be applicable or 
adaptable to three-loop three-point functions, since such a configuration emerges 
when one slices the vertex off a two-point function, where that vertex contains 
one of the external legs. The remaining graph would retain the internal 
topology of the two-point four-loop case. Therefore, an adaptation of the new 
feature of {\sc Forcer} could, in principle, be transferred to the three-point case
to provide an efficient alternative to the application of the Laporta algorithm
for massless three-point functions. 

While such technology is already in effect in situ, the main obstacle to the
full implementation of a three-loop evaluation is the determination of the
required three-loop master integrals. In recent years, this field has advanced,
with progress  made in understanding the mathematical properties
of high-order Feynman integrals. Examples of such articles include Refs. 
\cite{Brown:2011ik,Schnetz:2016fhy}, which provide novel procedures to compute 
Feynman graphs. The background to this is that there is a wide range of tools 
to evaluate a graph. One method is to introduce the Schwinger parameter representation
of each propagator and convert the $L$-loop $d$-dimensional space--time integral 
into an integral over Schwinger parameters. The resulting integral has a large 
number of parameter integrations to be carried out and there is no guarantee 
that this can be achieved analytically. This is to be preferred over a 
numerical approach, as the latter, if a Monte Carlo approach is used, could 
require a sizeable amount of computer resources to obtain reasonable accuracy. 
In certain instances, an analytical evaluation is possible and, in essence, uses 
algebraic geometry to produce an integration strategy. Such higher mathematics 
is relevant, since the integrand contains polynomials of the parameters, which
represent higher-dimensional geometries. Established mathematical theorems are 
then effected, which determine which parameter integration order is to be used, 
with the guiding principle being linear reducibility. By this, we mean that 
after each parameter integration the polynomial degree reduces but the key to 
achieve this is to have the polynomial factor off a smaller polynomial 
involving only factors linear in the next variable to be integrated. It is this
linearity that is key, as it allows one to use the machinery of hyperlogarithms 
to carry out the integration over that Schwinger parameter. What was not 
immediately evident is whether this procedure could be iterated without obstruction
and that when it terminates the value of the integral is found. It has now been
shown that, if an integral is linearly reducible 
\cite{Brown:2008um,Brown:2009ta}, in this sense there is at least one 
choice of integration order that allows the integral to be determined. While 
this is, in essence, the general current position, it is known that, to three loops, the three-point vertex master integrals are all linearly reducible. Thus, in 
principle, the required master integrals can be determined.

The actual practicalities of this have yet to be carried out. However, several
packages are available to assist with this task. For instance, converting a 
scalar Feynman integral into Schwinger parameter representation via the
underlying graph polynomials is now a standard feature of integration
packages, such as {\sc Hyperint} \cite{Panzer:2014caa}. This package is 
appropriate for an analytical determination, since any evaluation can be written 
in various hyperlogarithm representations. It has features that allow one to 
find the order of integration over the parameter variables to ensure that there
is no obstruction to the linear reducibility. In principle, one can expand to 
several orders in the $\epsilon$ expansion in $d = 4 - 2\epsilon$ 
dimensions. However, for terms beyond the leading few, the parameter integration
can become tedious, especially for high loop orders. Therefore, a more appropriate
strategy would be one where only the first term of the $\epsilon$ expansion of 
a master integral was required, which would then require the Laporta reduction 
to be constrained to producing a basis of masters that is finite. There is a 
caveat with this because one is using dimensional regularisation,
which means that the reduction produces factors of rational polynomials in $d$.
Such functions can include poles in $(d-4),$ which are termed spurious poles. 
This is in the sense that while they correspond to a divergence it is not 
necessarily one due to the divergence of an actual graph. There are now ways to
circumvent this, which work hand in hand with another property of the beauty of 
computing in $d$ dimensions. This was analysed in depth in Refs.
\cite{Tarasov:1996br,Tarasov:1997kx}, where it was shown that $d$-dimensional
integrals can be related to the corresponding topology in $(d+2)$ dimensions
plus a sum of others that have the same core topology but with propagators
missing. Such higher-dimensional integrals can be incorporated in the Laporta
reduction process and have been implemented in version $2.11$ of the 
{\sc Reduze} package \cite{vonManteuffel:2012np}. The advantage is that, with 
the increase in dimensionality in the higher-dimensional integral, it is not as
ultraviolet divergent as its lower-dimensional counterpart. Thereby, in 
principle, one reduces the evaluation of the more difficult master integrals to 
finite higher-dimensional ones, which should therefore be more accessible to the
{\sc Hyperint} package. 

In summarising the algorithm to extend the two-loop QCD off-shell vertex
functions, it is worth noting that, for the triple-gluon vertex, there will be
$2382$ three-loop graphs to evaluate and $63\,992$ at four loops. For both the 
other three-point vertices, the numbers of graphs in each case are the same and
are $688$ and $17\,311$, respectively, at three and four loops. Thus, the evaluation
of even just the three-loop vertex functions will require a substantial 
amount of work and computing time. This would especially be the case at four
loops without access to appropriate computers to build the necessary databases
of integral relations. In the interim, there is a potential alternative to gain
some insight into or estimate of the three-loop contributions. In the period
between the early work of Celmaster and Gonsalves \cite{Celmaster:1979km} and
its extension to the next order in Ref. \cite{Gracey:2011vw}, a method was developed  \cite{Chetyrkin:2000fd} where the vertex functions were computed, at the 
fully symmetric point, numerically at two loops in QCD. The approach was to 
apply a large momentum expansion of the vertex functions to very high order. 
This produced a set of two-point integrals, which were evaluated using 
{\sc Mincer} \cite{Gorishnii:1989gt,Larin:1991fz}. Provided that enough terms were 
computed, the approximate value of the contributing graphs could be accurately 
estimated numerically. The stability and accuracy of the expansion could be 
checked by choosing different external momenta to play the role of the large 
momentum. What was remarkable when the analytic two-loop expressions became 
available\cite{Gracey:2011vw} was how accurate the large-momentum 
{\sc Mincer}-based expansion values were. The only major difference was for a 
colour group Casimir coefficient in one three-loop MOM $\beta$ function, which 
turned out to be of the order $0.01$ \cite{Chetyrkin:2000fd}. The numerical coefficient was small and 
the expansion needed to a higher accuracy than was computationally available at
the time  \cite{Chetyrkin:2000fd}. With  advances in symbolic manipulation,
such as the provision of the {\sc Forcer} program, which is significantly more
efficient than {\sc Mincer}, such an interim numerical evaluation of the 
vertex functions would at least give information on the magnitude of the next-order corrections. As a corollary, it would provide the four-loop MOM
$\beta$ functions numerically.

\subsection{Discussion}

To recap, we have reviewed recent results in the determination of the three-point
vertex functions of QCD at two loops. We have for the most part concentrated on
the off-shell case; it would not have been possible to achieve this without the 
earlier work on different external momentum configurations. While the two-loop 
off-shell results followed a long time after the one-loop case, the main
reason for this was lack of the required computational technology. The last decade has
seen a revolution in this direction with the Laporta algorithm 
\cite{Laporta:2001dd}, as well as a systematic way of computing master 
integrals from high-level mathematics. Consequently, the road to achieve the 
extension to three loops is, in principle, possible. One useful corollary of such
a computation would be the extension of the renormalization group functions to 
four loops in kinematic schemes such as MOM. To go to higher orders beyond 
three, this depends on whether the linear reducibility of four-loop masters can 
be established. One case that we have not touched on is that of the four-point 
functions. The technology to compute the full off-shell one-loop amplitudes is 
already available. However, the current situation is that the relevant two-loop 
off-shell masters have not been computed. Moreover, it has not been established 
whether they are linearly reducible in order that the hyperlogarithm approach can be
applied. This at present appears to be an open question for future work. 
Finally, including massive quarks in three- and four-point functions is another
direction that needs consideration. However, this is not straightforward at two loops, since the three-point
masters with one mass scale and off-shell momentum configuration are not yet
known. 

%\vspace{0.25cm}
%\noindent
%{\bf Acknowledgements.} This work was supported by a DFG Mercator Fellowship.

\end{bibunit}

\label{sec-sm-jgracey}  
\clearpage \pagestyle{empty}  \cleardoublepage
%============================================

\pagestyle{fancy}
\fancyhead[CO]{\thechapter.\thesection \hspace{1mm} Effective field theory approach to QED corrections in flavour physics}
\fancyhead[RO]{}
\fancyhead[LO]{}
\fancyhead[LE]{}
\fancyhead[CE]{}
\fancyhead[RE]{}
\fancyhead[CE]{M. Beneke, C. Bobeth, R. Szafron}
\lfoot[]{}
\cfoot{-  \thepage \hspace*{0.075mm} -}
\rfoot[]{}
    
        \begin{bibunit}[elsarticle-num] % define the bib-style for the unit: elsarticle-num.bst
%  text-1; this is the corresponding section
%\putbib[2loops] % the *.bib
%\end{bibunit}
% go-on
%--- from: bibunits.sty, adapts the font size of ``References'' to section
\let\stdthebibliography\thebibliography
\renewcommand{\thebibliography}{%
\let\section\subsection
\stdthebibliography}
    
\section
[Effective field theory approach to QED corrections in flavour physics \\ {\it M. Beneke, C. Bobeth, R. Szafron}]
{Effective field theory approach to QED corrections in flavour physics
\label{contr:szafron}}
\noindent
{\bf Contribution\footnote{This contribution should be cited as:\\
M. Beneke, C. Bobeth, R. Szafron, Effective field theory approach to QED corrections in flavour physics,  
%04 DOI:10.23731/CYRM-2020-XXX.\thepage, in:
%04 \url{http://dx.doi.org/10.23731/CYRM-2020-XXX.\thepage}, in:
DOI: \href{http://dx.doi.org/10.23731/CYRM-2020-003.\thepage}{10.23731/CYRM-2020-003.\thepage}, in:
Theory for the FCC-ee, Eds. A. Blondel, J. Gluza, S. Jadach, P. Janot and T. Riemann,
\\CERN Yellow Reports: Monographs, CERN-2020-003,
%04 \url{http://dx.doi.org/10.23731/CYRM-2020-XXX}, p. \thepage.} 
DOI: \href{http://dx.doi.org/10.23731/CYRM-2020-003}{10.23731/CYRM-2020-003},
p. \thepage.
\\ \copyright\space CERN, 2020. Published by CERN under the 
%04-2
\href{http://creativecommons.org/licenses/by/4.0/}{Creative Commons Attribution 4.0 license}.} by: M. Beneke, C. Bobeth, R. Szafron \\
Corresponding author: R. Szafron {[robert.szafron@tum.de]}}
\vspace*{.5cm}

%=================================================================
\subsection{Introduction and motivation}

Thanks to the accurate measurements performed at the low-energy facilities \cite{Bevan:2014iga}
and the LHC, flavour physics of light quarks, especially the
bottom quark, emerged on the precision frontier for tests of the Standard
Model (SM) and in searches for new physics effects. On the theoretical side, 
short-distance perturbative higher-order QCD and electroweak corrections 
are under good control for many processes. Moreover, tremendous progress
in lattice computations \cite{Aoki:2019cca} allows  percentage to
even subpercentage accuracy to be achieved for long-distance non-perturbative quantities. This
allows for the prediction of some key observables with unprecedented accuracy
and, in turn, the determination of short-distance parameters, such as the elements
of the quark-mixing matrix (CKM) in the framework of the SM. Given these 
prospects, it is also desirable to improve the understanding and treatment of
QED corrections, which are generally assumed to be small. Unfortunately, not much
new development has taken place in the evaluation of such corrections. 

For the future $\mathrm{e}^{+}\mathrm{e}^{-}$ machines, the proper computation of QED corrections
will be particularly important because the large data samples allow for precision
measurements that require their inclusion in theoretical predictions. We would
like to advocate a framework for a proper and systematic treatment of QED effects
based on the effective field theory~(EFT) approach, which exploits scale hierarchies
present in processes involving mesons. In this spirit, QED corrections to
$\mathrm{B}_\mathrm{s} \to \upmu^+\upmu^-$ have recently been analysed \cite{Beneke:2017vpq}, revealing
an unexpectedly large contribution owing to power enhancement. Such an effect
cannot be found in the standard approach based on soft-photon approximation
\cite{Yennie:1961ad, Weinberg:1965nx, Isidori:2007zt}, as it requires a helicity
flip induced by the photon. Further, the common assumption that hadrons are point-like
objects neglects effects related to the structure of hadrons. It implies implicitly
that the soft-photon approximation itself is performed in the framework of an EFT
in which photons have virtuality below a typical hadronic binding scale $\Lambda_\text{QCD}
\sim \mathcal{O}(100\UMeV)$ of partons in hadrons, below which they do not
resolve the partonic structure of the hadrons. In consequence, this approach cannot address QED corrections, owing to virtualities above the scale $\Lambda_\text{QCD}$. These observations are 
a motivation to scrutinise further QED corrections in flavour physics in the light
of upcoming precise measurements and existing tensions in flavour measurements,
in particular, related to tests of lepton flavour universality.

In addition to a systematic power counting, the EFT treatment offers the possibility 
of the all-order resummation of the corrections. This is particularly important
for the mixed QCD--QED corrections, owing to the size of the QCD coupling constant
and the presence of large logarithmic corrections. While the soft-exponentiation 
theorem allows resumming leading QED effects related to ultrasoft photons that do
not resolve the partonic structure of hadrons, not much is known about the resummation
of the subleading logarithms in QED for photons with larger virtuality. Standard
factorisation theorems derived in QCD cannot be directly translated to QED, for, in
the QCD case, the mass effects related to light degrees of freedom are typically 
neglected. This is not the case in QED, where the lepton mass provides a cut-off
for collinear divergences. Moreover, the fact that in QCD one can observe only
colour singlet states additionally simplifies the computations, while in QED, and 
more generally in the electroweak sector of the SM \cite{Manohar:2018kfx, Fornal:2018znf}, 
it is necessary to account for charged particles  in both the final and initial states.
As a result, the QED factorisation theorems have not been explored intensively in the
literature so far, but this gap should be filled before a precise $\mathrm{e}^{+}\mathrm{e}^{-}$ collider
becomes operational. 

Power corrections to the standard soft approximation may also play an important role
in certain processes. Studies of power corrections in the QCD case  recently gained
much attention \cite{Bonocore:2016awd, Moult:2016fqy, Moult:2017jsg, Beneke:2017ztn,
Beneke:2018rbh, Moult:2018jjd, Beneke:2018gvs}. New tools based on soft-collinear~EFT~(SCET)
developed to study processes with energetic quarks and gluons can, after certain
modifications, be applied to improve the accuracy of electroweak corrections in future
lepton colliders. This is particularly important in collider physics for regions of
phase space where the perturbative approach breaks down, owing to the presence of large
logarithmic enhancements, and the next-to-soft effects become more important. Particularly
interesting are mass-suppressed effects related to soft fermion exchange 
\cite{Liu:2017vkm, Liu:2018czl, Penin:2014msa}, whose consistent treatment in the 
SCET language is not yet fully known. Beyond applications to precision SM physics,
the SCET framework may be necessary after possible discovery of new physics at the
LHC \cite{Alte:2018nbn, Alte:2019iug}. 

%--------+---------+---------+---------+---------+---------+---------+---------+

\subsection{QED corrections in $\mathrm{B}_\mathrm{q} \to \ell^+\ell^-$}

The decay of a neutral meson $\mathrm{B}_\mathrm{q} \to \ell^+\ell^-$ ($\ell = \mathrm{e}, \upmu, \uptau$)
is the first step in an investigation of QED effects in QCD bound states. Its purely
leptonic final state and neutral initial state keep complications related to the
non-perturbative nature of QCD to the necessary minimum. Yet, as we shall see,
even this simple example requires investigation of power corrections in SCET.
The importance of this decay derives from the fact that it depends, at leading
order (LO) in QED, only on the $\mathrm{B}_\mathrm{q}$ meson decay constant, which can  nowadays be
calculated with subpercentage precision on the lattice \cite{Bazavov:2017lyh}, 
necessitating the inclusion of higher-order QED corrections from all scales at
this level. This decay has been observed for $\ell = \upmu$ by LHCb \cite{Aaij:2013aka, 
Aaij:2017vad}, CMS \cite{Chatrchyan:2013bka}, and ATLAS \cite{Aaboud:2016ire}. 
The currently measured branching fraction for $\mathrm{B}_\mathrm{s}$ decays of about $3 \times 10^{-9}$
is compatible with the latest SM predictions \cite{Beneke:2017vpq, Bobeth:2013uxa,
Crivellin:2018gzw} and it is expected that the LHCb experiment will be able to
measure the branching fraction with 5\% accuracy with 50/fb (Run~4) around the
year 2030 \cite{Bediaga:2012py}. The FCC-ee
running on the $\mathrm{Z}$ resonance is expected to provide, with  about $\mathcal{O}(10^3)$
reconstructed events \cite{Mangano:2018mur}, an even higher event yield compared
with LHCb Run 4. This, together with the cleaner hadronic environment at the FCC-ee,
should allow better control of backgrounds and also systematic uncertainties,
such that one can expect improved accuracy. However, the gain in accuracy cannot be quantified without a dedicated study.

On the theory side, electroweak and QCD corrections above the scale $\mu_\mathrm{b} \sim 5\UGeV$
of the order of the b quark mass $m_\mathrm{b}$ are treated in the standard
framework of weak EFT of the SM \cite{Buras:1998raa}. The effective Lagrangian
is a sum of four-fermion and dipole operators
\begin{align}
  {\cal L}_{\Delta B=1} &
  = \mathcal{N}_{\Delta B=1} 
    \left[ \sum_{i=1}^{10} C_i(\mu_\mathrm{b}) \, Q_i \right] + \text{h.c.} \,,
\end{align}
with $\mathcal{N}_{\Delta B=1} \equiv2 \sqrt{2}G_\mathrm{F} V_\mathrm{tb}^{} V_\mathrm{tq}^*$
and covers, in principle, all weak decays of b hadrons. The pertinent operators
relevant for $\mathrm{B}_\mathrm{q} \to \ell^+\ell^-$ ($\mathrm{q} = \mathrm{d,s}$) are 
\begin{align}
  Q_{7} & = \frac{e}{(4\uppi)^{2}}
    \big[\bar{q} \sigma^{\mu\nu} (m_\mathrm{b} P_R + m_\mathrm{q} P_L) b \big] F_{\mu\nu} \,,
\nonumber\\
  Q_{9} & = \frac{\alpha_\mathrm{em}}{4\uppi}
    \big[\bar{q} \gamma^\mu P_L b\big] \sum_\ell \big[\bar{\ell} \gamma_\mu \ell\big] \,, 
\nonumber\\
  Q_{10} & = \frac{\alpha_\mathrm{em}}{4\uppi}
    \big[\bar{q} \gamma^\mu P_L b\big]
    \sum_\ell \big[\bar{\ell} \gamma_\mu \gamma_5 \ell\big] \,.
    \label{eq:operators}
\end{align}
The matching $C_i(\mu_\mathrm{b})$ coefficients are computed at the electroweak scale
$\mu_\mathrm{W} \sim \mathcal{O}(100\UGeV)$ and evolved to the scale of $\mu_\mathrm{b}\sim m_\mathrm{b}$
with the renormalization group equation of the weak EFT.

Because the neutral $\mathrm{B}_\mathrm{q}$ meson is a pseudo-scalar and the SM interactions
are mediated by axial and vector currents, the decay rate must vanish in the limit
$m_\ell \to 0$, and therefore the decay amplitude is proportional to the lepton
mass. The hadronic matrix element at LO in QED is parametrized
by a single decay constant $f_{\mathrm{B}_\mathrm{q}}$, defined by 
$\langle 0 | \bar{q} \gamma^\mu \gamma_5 b | \bar{B}_\mathrm{q}(p) \rangle = \mathrm{i} f_{\mathrm{B}_\mathrm{q}} p^\mu$.
The leading amplitude for $\mathrm{B}_\mathrm{q} \to \ell^+ \ell^-$ is 
\begin{align}
  \mathrm{i} \mathcal{A} & 
  = m_\ell \, f_{\mathrm{B}_\mathrm{q}}\, \mathcal{N} \, C_{10}(\mu_\mathrm{b}) \, 
    \big[\bar{\ell} \gamma_5 \ell\big], &
 \Big( \mathcal{N} &
  \equiv \mathcal{N}_{\Delta B=1} \frac{\alpha_\mathrm{em}}{4\uppi} \Big) 
 \end{align}
and the branching fraction is
\begin{equation}
\label{eq:Fred}
  \text{Br}^{(0)}_{\mathrm{q}\ell} \equiv \text{Br}^{(0)} \big[\mathrm{B}_\mathrm{q} \to \ell^+\ell^- \big]
  = \frac{\tau_{\mathrm{B}_\mathrm{q}} m_{\mathrm{B}_\mathrm{q}}^3 f_{\mathrm{B}_\mathrm{q}}^2}{8\uppi}\, 
    |\mathcal{N}|^2\, \frac{m_\ell^2}{m_{\mathrm{B}_\mathrm{q}}^2}
    \sqrt{1 - \frac{4m_\ell^2}{m_{B_\mathrm{q}}^2}} \, |C_{10}|^2 \,,
\end{equation}
with $m_{\mathrm{B}_\mathrm{q}}$ denoting the mass of the meson and $\tau_{\mathrm{B}_{\mathrm{q}}}$
its total lifetime. For neutral $\mathrm{B}_\mathrm{s}$ mesons, the mixing needs to be accounted for
\cite{DeBruyn:2012wk}, thereby allowing for the
measurement of related CP asymmetries, to be discussed next. In this case, \Eref{eq:Fred} refers to the `instantaneous' branching fraction at time
$t=0$, which  differs from the measured untagged time-integrated branching fraction by the
factor $(1-y_\mathrm{s}^2)/(1 + y_\mathrm{s} \mathcal{A}_{\Delta \Gamma})$, where $y_\mathrm{s} = \Delta 
\Gamma_\mathrm{s}/(2 \Gamma_\mathrm{s})$ is related to the lifetime difference and
$\mathcal{A}_{\Delta \Gamma}$ denotes the mass-eigenstate rate asymmetry.
Concerning QED corrections, this branching fraction refers to the
`non-radiative' one prior to the inclusion of photon bremsstrahlung effects.

If one takes into account soft-photon radiation (both real and virtual) with
energies smaller than the lepton mass, the decay amplitude is dressed by the
standard Yennie--Frautschi--Suura exponent \cite{Yennie:1961ad, Buras:2012ru}
\begin{equation}
  \text{Br}\big[\mathrm{B}_\mathrm{q} \to \ell^+\ell^- + n\upgamma \big]
  = \text{Br}^{(0)}_{\mathrm{q}\ell} \times 
    \left(\frac{2 E_\text{max}}{m_{\mathrm{B}_\mathrm{q}}}\right)^{2\frac{\alpha_{\mathrm{em}}}{\uppi}
       \left(\ln \frac{m_{\mathrm{B}_\mathrm{q}}^2}{m_\ell^2} - 1 \right) + \mathcal{O}(m_\ell)} .
\end{equation}
This `photon-inclusive' branching fraction is based on eikonal approximation,
in the limit when the total energy carried away by the $n$ photons, $E_{\text{max}}$,
is much smaller than the lepton mass. The QED corrections in the initial state are
entirely neglected and photons are assumed to couple to leptons through eikonal
currents 
\begin{equation}
  J^\mu (q)
  = e \sum_i Q_i \eta_i \frac{p_i^\mu}{p_i \cdot q} ,
\end{equation}
where $\eta=-1$ for incoming particles and $\eta=+1$ for outgoing particles.
The sum runs over all charged particles with momenta $p_i$ and charges $Q_i$.
Eikonal currents are spin-independent and thus they do not change the helicity
of the leptons.

From this point, we focus only on the case of muons in the final state, $\ell = \upmu$. 
In the experimental analysis \cite{Aaij:2017vad, Chatrchyan:2013bka,
Aaboud:2016ire}, the signal is simulated fully inclusive of
final-state radiation off the muons by applying PHOTOS \cite{Golonka:2005pn}
corresponding to a convolution of the $E_\text{max}$-dependent exponential
factor in the determination of the signal efficiency.
Conversely, photon emission from the quarks (initial state) vanishes 
in the limit of small photon energies because it is infrared-safe, since the
decaying meson is electrically neutral. Hence, it can be
neglected as long as the signal window is sufficiently small, in practice of 
$\mathcal{O}(60\UMeV$) \cite{Aditya:2012im}, and is effectively treated
as negligible background on both experimental and theory sides. 
In consequence,  the experimental analyses currently provide the non-radiative
branching fraction relying on the simulation with PHOTOS.

The limitations of the conventional approximation had missed the important
effect responsible for the power enhancement of QED corrections to the 
$\mathrm{B}_\mathrm{s} \to \upmu^+\upmu^-$ decay. Indeed, even when the cut on the real photon
emission is much smaller than the muon mass,  virtual photons with 
virtualities of the order of the muon mass or larger can resolve the structure
of the meson, whose typical size is of the order of $1/\Lambda_\text{QCD}$.
In this case, the meson cannot be treated as a point-like object. Moreover,
the eikonal approximation is not suitable for such photons, as they can induce
a helicity flip of the leptons. However, straightforward computation of the QED
corrections is not possible, as it requires the evaluation of non-local
time-ordered products of the $ \mathcal{L}_{\Delta B=1}(0)$ Lagrangian
with the electromagnetic current $j_{\rm QED} = Q_\mathrm{q} \bar q\gamma^\mu q$, 
such as 
\begin{equation}
  \label{eq:NLO:QED:ME}
  \langle 0|\int \mathrm{d}^4 x\,
  T\{j_{\rm QED}(x), \mathcal{L}_{\Delta B=1}(0) \} 
  |\bar{B}_\mathrm{q} \rangle .
\end{equation}
Currently, this object is beyond the reach of lattice QCD, while the SCET approach
allows one to systematically expand this matrix element and reduce the non-perturbative
quantities to universal ones at leading order.  

%--------+---------+---------+---------+---------+---------+---------+---------+

\begin{figure}
\begin{center}
  \includegraphics[width=0.4\textwidth]{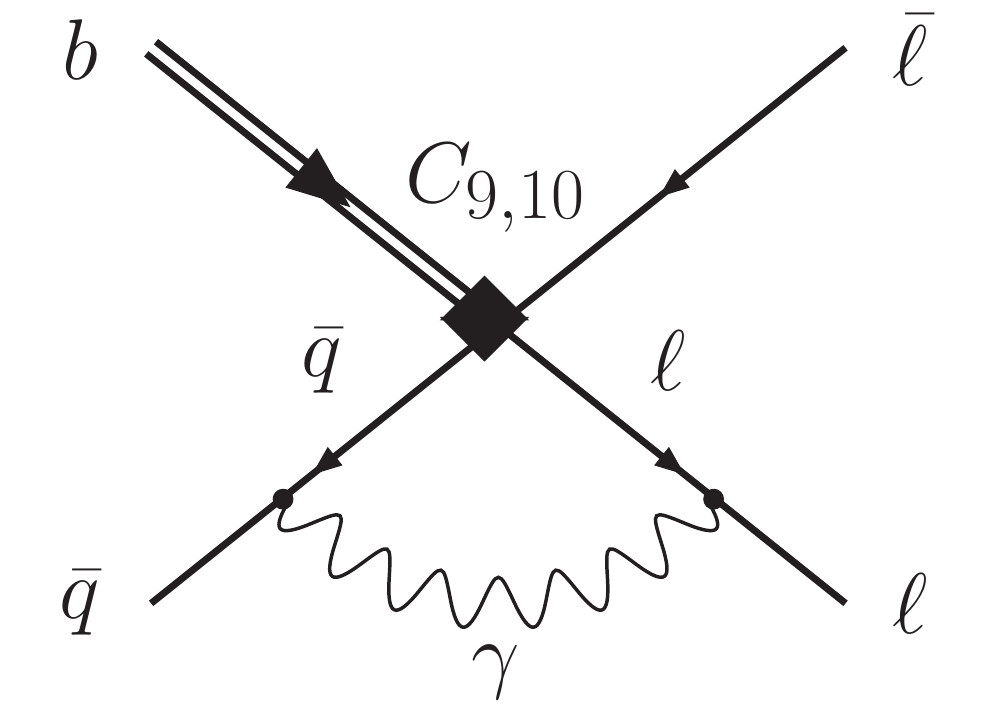}
\end{center}
\caption{\label{fig:diagram}Example diagram that gives rise to 
  the power-enhanced QED correction. A photon can be either collinear with
  virtuality $k^2 \sim m_\upmu^2$ or hard-collinear, $k^2 \sim m_\upmu m_\mathrm{b}$.}
 % \query{Please correct the figure labels in \Fref{fig:diagram}. Set particle
 % names in roman (upright) font.}
\end{figure}

Let us consider  \Fref{fig:diagram}, where the photon is exchanged
between the light quark and the lepton. There are two low-energy scales in the
diagram, set by the external kinematics of the process $\mathrm{B}_\mathrm{q} \to \upmu^+ \upmu^-$. 
One is the muon mass $m_\upmu$, which is related to the collinear scale.
We parametrize the lepton momentum in terms of the light-cone co-ordinates as 
$p_\ell = (n_+ p_\ell,\, n_- p_\ell,\, p_\ell^\perp) \sim m_\mathrm{b} \left(1,\, \lambda_\mathrm{c}^2,\, 
\lambda_\mathrm{c} \right)$, where we introduced the small counting parameter 
$\lambda_\mathrm{c} \sim m_\upmu / m_\mathrm{b}$. The second low-energy scale is related to the typical
size of the soft light quark momentum $l_\mathrm{q} \sim \Lambda_\text{QCD}$ and for 
counting purposes we introduce $\lambda_\mathrm{s} \sim \Lambda_\text{QCD} / m_\mathrm{b}$.
In the case of muons, it happens that numerically $\lambda_\mathrm{c} \approx \lambda_\mathrm{s}$
and in the following we equate them and do not distinguish between them. It turns out that
there also exists a hard-collinear invariant constructed from the lepton and quark
momentum $p_\ell \cdot l_\mathrm{q} \sim \lambda m_\mathrm{b}^2$, thus in addition to the collinear
and soft regions we must also consider a hard-collinear region, where momenta
scale as $k \sim m_\mathrm{b} \left(1,\, \lambda,\, \lambda^{1/2} \right)$.
This non-trivial hierarchy of intermediate scales must be properly accounted
to evaluate the leading QED corrections, which can be done by 
subsequent matching on SCET$_{\rm I}$ and SCET$_{\rm II}$ \cite{Beneke:2003pa} at the hard
($\sim$$m_\mathrm{b}$) and hard-collinear scales, respectively. 

The power enhancement is directly related to the interplay of collinear
and hard-collinear scales. When the hard-collinear or collinear photon
interacts with the soft quark, momentum conservation forces the quark
to become hard-collinear. These modes can be integrated out perturbatively
with the help of the EFT methods. In this case, we must first match the
operators in \Eref{eq:operators} on SCET$_{\rm I}$ currents \cite{Beneke:2019}. 
In SCET$_{\rm I}$, we retain soft, collinear, and hard-collinear modes; only the 
hard modes are integrated out. The leading SCET$_{\rm I}$ operator contains a
hard-collinear quark field, which scales as $\lambda^{1/2}$ instead of
the soft quark field with scaling $\lambda^{3/2}$. When we integrate out
the hard-collinear modes, we must convert the hard-collinear quark field
$\xi_\mathrm{C}(x)$ to the soft quark field $q_\mathrm{s}$. This is done with
the help of power-suppressed Lagrangian \cite{Beneke:2002ni}
\begin{align*}
  \mathcal{L}_{\xi \mathrm{q}}^{(1)} &
  = \bar{q}_\mathrm{s}(x_-) \,W_{\xi \mathrm{C}}^{\dagger} \mathrm{i} \slashed{D}_{\perp} \,\xi_\mathrm{C}(x)
  - \bar{\xi}_\mathrm{C}(x) \,\mathrm{i} \overleftarrow{\slashed D}_{\perp}W_{\xi \mathrm{C}}\, q_\mathrm{s}(x_-),
\end{align*}
where $W_{\xi \mathrm{C}}$ is a collinear Wilson line carrying charge of the
collinear field $\xi_\mathrm{C}$. This Lagrangian insertion costs an additional
power of $\lambda^{1/2},$ but the resulting SCET$_{\rm II}$ operators are
still power-enhanced, as compared with the operators obtained without
an intermediate hard-collinear scale. The power-enhanced correction to the
amplitude is \cite{Beneke:2017vpq}
\begin{multline}
  i\Delta\mathcal{A} 
  = \frac{\alpha_\mathrm{em}}{4\uppi} Q_\ell Q_\mathrm{q}  m_\ell m_{\mathrm{B}_\mathrm{q}} f_{\mathrm{B}_\mathrm{q}}
    \mathcal{N}  \big[\bar{\ell} (1 + \gamma_5) \ell\big]
\\  \times
  \Bigg\{ \int_0^1  \mathrm{d}u (1-u) C_9^\text{eff} (um_\mathrm{b}^2)
          \int_0^\infty \frac{\mathrm{d}\omega}{\omega} \phi_{\mathrm{B}+}(\omega)
          \left[\ln\frac{m_\mathrm{b}\omega}{m_\ell^2} + \ln\frac{u}{1-u} \right]
\\ 
  - Q_{\ell}C_7^\text{eff} \int_0^\infty \frac{\mathrm{d}\omega}{\omega}\, \phi_{\mathrm{B}+}(\omega)
  \left[\ln^2\frac{m_\mathrm{b}\omega}{m_\ell^2} - 2\ln\frac{m_\mathrm{b}\omega}{m_\ell^2}
       + \frac{2\uppi^2}{3} \right] \Bigg\} ,
       \label{eq:correction}
\end{multline}
where $\phi_{\mathrm{B}+}(\omega)$ is the $\mathrm{B}_\mathrm{q}$ meson light-cone distribution
amplitude (LCDA), which contains information about the non-perturbative structure
of the meson. This virtual correction is, by itself, infrared-finite,
as it modifies the exclusive decay rate. The power enhancement manifests
itself in \Eref{eq:correction} as the inverse power of the $\omega$
variable that results from the decoupling of the hard-collinear quark modes 
\begin{equation}
  m_{\mathrm{B}_\mathrm{q}} \int_0^\infty\frac{\mathrm{d} \omega}{\omega}\,\phi_{\mathrm{B}+}(\omega) \, \ln^k\omega 
  \sim \frac{m_{\mathrm{B}_\mathrm{q}}}{\Lambda_\text{QCD}}
  \sim \frac{1}{\lambda} .
\end{equation}
The $\omega$ may be interpreted as a momentum of the soft quark along the
light-cone direction of the lepton, and thus $\omega \sim \Lambda_\text{QCD}$.
The annihilation of the quark into leptons is a non-local process in the
presence of the QED interactions and the virtual leptons with the wrong
helicity can propagate over distances of the order of the meson size.
Thus, the helicity flip costs a factor $m_\ell/\Lambda_\text{QCD}$ instead
of the typical suppression factor of $m_\ell/m_\mathrm{b}$ present in the leading-order amplitude.

The terms proportional to $C_{10}$ cancel after the collinear and anticollinear 
contributions are added, such that only $C_9$ contributes out of the
semileptonic operators. The term $\propto$$C_7$ requires separate treatment
since the convolution integral containing the hard matching coefficient
exhibits an endpoint singularity. In addition, the collinear contribution has 
a rapidity-type divergence. There exists an additional contribution related to
the soft region, which, after a suitable rapidity regularisation, can be 
combined with the collinear contribution. When the convolution integral is
performed in dimensional regularisation before taking the limit $d\to4$,
the total correction is finite and exhibits the double-logarithmic 
enhancement. 

The numerical evaluation \cite{Beneke:2017vpq} of the power-enhanced correction
(\Eref{eq:correction}) shows a partial cancellation of the terms $\propto$$C_9^\text{eff}$
and $\propto$$C_7^\text{eff}$. The final impact on the branching fraction 
$\text{Br}^{(0)}_{\mathrm{q}\upmu}$ is a decrease in the range  (0.3--1.1)\%,
with a central value of 0.7\%.
Despite the cancellation, the overall correction is still sizeable compared
with the natural size of a QED correction of $\alpha_\mathrm{em}/\uppi \sim 0.3$\%. The large
uncertainties of the power-enhanced QED correction are due to the poorly known
inverse moment $\lambda_\mathrm{B}$ and almost unknown inverse-logarithmic moments $\sigma_1$
and $\sigma_2$ of the B meson LCDA.\setcounter{footnote}{0}%
\footnote{Throughout, the same numerical values as in Ref. \cite{Beneke:2017vpq} are used for $B_\mathrm{s}$ and $B_\mathrm{d}$, neglecting $SU(3)$-flavour breaking effects.}
The prediction for the muonic modes for the untagged time-integrated branching
fractions for $B_\mathrm{s} \to \upmu^+\upmu^-$ and $B_\mathrm{d} \to \upmu^+\upmu^-$ are
\begin{footnotesize}\begin{align}
   \overline{\text{Br}}^{(0)}_{\mathrm{s}\upmu}  
 & = \begin{pmatrix} 3.59 \\ 3.65 \end{pmatrix} \left[
    1 
    \pm \begin{pmatrix} 0.032 \\ 0.011 \end{pmatrix}_{f_{\mathrm{B}_\mathrm{s}}}
    \pm 0.031|_\text{CKM}
    \pm 0.011|_{m_\mathrm{t}}
    \pm 0.012|_\text{non-pmr}
    \pm 0.006|_\text{pmr} 
    \pm {}^{+0.003}_{-0.005}|_\text{QED}
    \right] \times 10^{-9},
\\
   \overline{\text{Br}}^{(0)}_{\mathrm{d}\upmu} 
&  = \begin{pmatrix} 1.05 \\ 1.02 \end{pmatrix} \left[
    1 
    \pm \begin{pmatrix} 0.045 \\ 0.014 \end{pmatrix}_{f_{\mathrm{B}_\mathrm{d}}} 
    \pm 0.046|_\text{CKM}
    \pm 0.011|_{m_\mathrm{t}}
     \pm 0.012|_\text{non-pmr}
    \pm 0.003|_\text{pmr} 
    \pm {}^{+0.003}_{-0.005}|_\text{QED}
    \right] \times 10^{-10},
\end{align}\end{footnotesize}%
where we group uncertainties: (i) main parametric long-distance ($f_{\mathrm{B}_\mathrm{q}}$) and
short-distance (CKM and $m_\mathrm{t}$), (ii) remaining non-QED parametric ($\tau_{\mathrm{B}_\mathrm{q}}$, 
$\alpha_\mathrm{s}$) and non-QED non-parametric ($\mu_\mathrm{W}$, $\mu_\mathrm{b}$, higher order, see Ref. 
\cite{Bobeth:2013uxa}), and (iii) from the QED correction ($\lambda_\mathrm{B}$ and 
$\sigma_{1,2}$, see Ref.  \cite{Beneke:2017vpq}).
We provide here two values, depending on the choice of  the lattice calculation
of $f_{\mathrm{B}_\mathrm{q}}$ for $N_f = 2 + 1$ (upper) and $N_f = 2 + 1 + 1$ (lower), with averages
from FLAG 2019 \cite{Aoki:2019cca}. Note that the small uncertainties of the 
$N_f = 2 + 1 + 1$ results are currently dominated by a single group \cite{Bazavov:2017lyh}
and confirmation by other lattice groups in the future is desirable. It can
be observed that, in this case, the largest uncertainties are due to CKM parameters,
such that they can be determined provided the accuracy of the measurements at the FCC-ee is 
at the 1\%\ level. Still fairly large errors are due to the top quark mass
$m_\mathrm{t} = (173.1 \pm 0.6)$\,GeV, here assumed to be in the pole scheme, where an additional
non-parametric uncertainty of 0.2\% is included (in `non-pmr') for the conversion
to the $\overline{\text{MS}}$ scheme. Further `non-pmr' contains a 0.4\% uncertainty
from $\mu_\mathrm{W}$ variation and 0.5\% further higher-order uncertainty, all linearly 
added. For the CKM input, we use  Refs. \cite{Beneke:2017vpq, Crivellin:2018gzw}.

As mentioned, for the $B_\mathrm{s}$ meson, the mixing provides the opportunity
to measure CP asymmetries in a time-dependent analysis
\begin{align}
  \frac{\Gamma[\mathrm{B}_\mathrm{s}(t)\to \upmu^+_\lambda\upmu^-_\lambda] - 
        \Gamma[\bar {\mathrm{B}}_\mathrm{s}(t)\to \upmu^+_\lambda\upmu^-_\lambda]}
       {\Gamma[\mathrm{B}_\mathrm{s}(t)\to \upmu^+_\lambda\upmu^-_\lambda] + 
        \Gamma[\bar {\mathrm{B}}_\mathrm{s}(t)\to \upmu^+_\lambda\upmu^-_\lambda]}
  & = \frac{C_\lambda \cos(\Delta m_{\mathrm{B}_\mathrm{s}}t) + S_\lambda \sin(\Delta m_{\mathrm{B}_\mathrm{s}}t)}
           {\cosh(y_\mathrm{s} t/\tau_{\mathrm{B}_\mathrm{s}}) 
           + \mathcal{A}^\lambda_{\Delta \Gamma} \sinh(y_\mathrm{s} t/\tau_{\mathrm{B}_\mathrm{s}}) }\,,
\end{align}
where all quantities are defined in Ref.~\cite{DeBruyn:2012wk} and 
$|\mathcal{A}^\lambda_{\Delta \Gamma}|^2 + |C_\lambda|^2 + |S_\lambda|^2 = 1$ holds.
For example, the mass-eigenstate rate asymmetry $\mathcal{A}_{\Delta\Gamma} = +1$
in the SM exactly, if only a pseudo-scalar amplitude exists, and
is therefore assumed to be very sensitive to possible new flavour-changing interactions,
with essentially no uncertainty from SM background. We now see that the QED
correction of the SM itself generates small `contamination' of the observable,
given by Ref. \cite{Beneke:2017vpq}
\begin{align}
  \mathcal{A}^\lambda_{\Delta\Gamma} & 
  \approx 1 - 1.0\cdot 10^{-5} \,, &
  S_\lambda &
  \approx -0.1\%\,, &
  C_\lambda & 
  \approx \eta_\lambda\,0.6\%\,,
\end{align}
where $\eta_{L/R} = \pm 1$. Present measurements \cite{Aaij:2017vad} set only very
weak constraints on the deviations of $A_{\Delta\Gamma}^\lambda$ from unity, 
and $C_\lambda$, $S_\lambda$ have not yet been measured,\footnote{Note that $C^\lambda$ requires the measurement of the muon helicity, whereas
$\mathcal{A}^\lambda_{\Delta\Gamma}$ and $S_\lambda$ can also be determined  as averages over the muon helicity; furthermore,  $\mathcal{A}^\lambda_{\Delta\Gamma}$
can be measured without flavour-tagging, whereas it is required for $S_\lambda$ and $C_\lambda$.} but the uncertainty in the B meson LCDA is, in principle, a limiting factor for the precision
with which new physics can be constrained from these observables.
Also, $S_\lambda$ and $C_\lambda$ deviate marginally from the leading-order
SM prediction of zero, but signals from new physics should be substantially
larger to distinguish them from the SM QED correction.

A similar framework can be used to analyse QED corrections to $\mathrm{B}^{\pm} \to
\ell^{\pm}\nu_{\ell}$.
In this case, power enhancement does not arise, owing to the different chirality
structure of the current and the presence of only one charged lepton in
the final state \cite{Beneke:2017vpq}. QED corrections that depend on the meson 
structure are subleading in this case. The leading QED corrections for this
process can be obtained from the usual soft-photon approximation, where the
charged meson is treated as a point-like charge.  

%--------+---------+---------+---------+---------+---------+---------+---------+

\subsection{Summary and outlook}

The proper treatment of QED corrections in theoretical predictions is essential
to the success of future $\mathrm{e}^+ \mathrm{e}^-$ colliders. We have shown 
how this goal could be achieved in flavour physics for the example of a power-enhanced
leading QED correction to the leptonic decays $\mathrm{B}_\mathrm{q} \to \upmu^+\upmu^-$ with $\mathrm{q} = \mathrm{d, s}$
\cite{Beneke:2017vpq} and provided updated predictions. 
A systematic expansion based on the appropriate EFTs must be implemented to cover
dynamics from the hard scale $\mu_\mathrm{b} \sim 5$~GeV over hard-collinear (SCET$_{\rm I}$)
and collinear scales (SCET$_{\rm II}$) down to the ultrasoft scales $\mathcal{O}(10\,\text{MeV})$.
Further, the EFTs allow for a systematic resummation of the leading logarithmic
corrections and they provide a field-theoretical definition of 
non-perturbative objects in the presence of QED, as, for example, generalised
light-cone distribution amplitudes of the B meson dressed by process-dependent
Wilson lines \cite{Beneke:2019}. The consistent evaluation of the QED corrections
is thus a challenging task, but it can be accomplished with the help of effective
field theory.  

In the example at hand, the special numerical value of the muon mass and its proximity
to the typical size of hadronic binding energies $\Lambda_\text{QCD}$ gave rise
to a special tower of EFTs. The application to the cases of electrons and taus
requires additional considerations. Full theoretical control of QED corrections
is also desirable for other decays that will allow future precision determination
of short-distance parameters. For example, an important class is that of exclusive 
$\mathrm{b}\to \mathrm{u} \ell\bar\upnu_\ell$ and $\mathrm{b}\to \mathrm{c} \ell\bar\upnu_\ell$ decays for the determination
of CKM elements $V_\mathrm{ub}$ and $V_\mathrm{cb}$, respectively. Owing to the absence of
resonant hadronic contributions, the only hadronic uncertainties from $\mathrm{B} \to \mathrm{M}$
form factors could become controllable with high accuracy in lattice calculations
for large dilepton invariant masses, \ie energetic leptons, which is also the
preferred kinematic region for the tower of EFTs discussed here. Other interesting
applications are observables that are predicted in the SM to vanish when restricting
to the leading order in the weak operator product expansion but might be sensitive
to non-standard interactions. Then the QED corrections in the SM provide a 
background to the new physics searches, as in the example of $\mathcal{A}_{\Delta \Gamma}$
in $\mathrm{B}_\mathrm{s} \to \upmu^+\upmu^-$ given here. This concerns observables in the
angular distributions of $\mathrm{B}\to \mathrm{K}^{(*)} \ell^+\ell^-$ as, for example, discussed
in Refs. \cite{Bobeth:2007dw, Beaujean:2015gba}.

\end{bibunit}

\label{sec-sm-szafron}  
\clearpage \pagestyle{empty}  \cleardoublepage
%============================================

\pagestyle{fancy}
\fancyhead[CO]{\thechapter.\thesection \hspace{1mm} Top pair production and mass determination}
\fancyhead[RO]{}
\fancyhead[LO]{}
\fancyhead[LE]{}
\fancyhead[CE]{}
\fancyhead[RE]{}
\fancyhead[CE]{A. Maier}
\lfoot[]{}
\cfoot{-  \thepage \hspace*{0.075mm} -}
\rfoot[]{}
    \begin{bibunit}[elsarticle-num] % define the bib-style for the unit: elsarticle-num.bst
%  text-1; this is the corresponding section
%\putbib[2loops] % the *.bib
%\end{bibunit}
% go-on
%--- from: bibunits.sty, adapts the font size of ``References'' to section
\let\stdthebibliography\thebibliography
\renewcommand{\thebibliography}{%
\let\section\subsection
\stdthebibliography} 

%\fancyhead[CE]{Andreas Maier}
%\begin{bibunit}[JHEP]

\section[Top pair production and mass determination\\
{\it A.~Maier}]
{Top pair production and mass determination}
\label{contr:ttbar_prod}
\noindent
{\bf Contribution\footnote{This contribution should be cited as:\\
A.~Maier, Top pair production and mass determination,  
%04 DOI:10.23731/CYRM-2020-XXX.\thepage, in:
%04 \url{http://dx.doi.org/10.23731/CYRM-2020-XXX.\thepage}, in:
DOI: \href{http://dx.doi.org/10.23731/CYRM-2020-003.\thepage}{10.23731/CYRM-2020-003.\thepage}, in:
Theory for the FCC-ee, Eds. A. Blondel, J. Gluza, S. Jadach, P. Janot and T. Riemann,
\\CERN Yellow Reports: Monographs, CERN-2020-003,
%04 \url{http://dx.doi.org/10.23731/CYRM-2020-XXX}, p. \thepage.} 
DOI: \href{http://dx.doi.org/10.23731/CYRM-2020-003}{10.23731/CYRM-2020-003},
p. \thepage.
\\ \copyright\space CERN, 2020. Published by CERN under the 
%04-2
\href{http://creativecommons.org/licenses/by/4.0/}{Creative Commons Attribution 4.0 license}.} by: A. Maier
%\\ Corresponding Author: Andreas Maier 
[andreas.martin.maier@desy]}
\vspace*{0.5cm}

\noindent The mass of the top quark can be measured in a well-defined scheme and
  with unrivalled precision at a future electron--positron collider, like the
  FCC-ee. The most sensitive observable is the total production cross-section for $\mathrm{b}\bar{\mathrm{b}}\mathrm{W}^+\mathrm{W}^-\mathrm{X}$ final states near the top pair production
  threshold. I review the state of the art in theory predictions for
  this quantity.

\subsection{Introduction}
\label{sec:intro:maier}

The total cross-section for inclusive $\mathrm{b}\bar{\mathrm{b}}\mathrm{W}^+\mathrm{W}^-\mathrm{X}$ production can be
measured with very high precision at a future high-energy
electron--positron collider. Owing to the potential for large integrated
luminosity, the FCC-ee is especially well-suited for such a
measurement. The line shape for centre-of-mass energies close to the
top--antitop production threshold is highly sensitive to the mass of the
top quark, which allows its determination with unprecedented
precision. Since the statistical uncertainty of the measurement is projected to be
significantly below the current theory
error~\cite{Seidel:2013sqa,Simon:2016pwp}, it is crucial to continuously
improve the theoretical prediction.

\subsection{Effective theory framework}
\label{sec:eft:maier}

The $\mathrm{\mathrm{b}}\bar{\mathrm{b}}\mathrm{W}^+\mathrm{W}^-\mathrm{X}$ final state is mostly produced through the
creation and decay of non-relativistic top--antitop pairs, interacting
predominantly via a colour Coulomb potential. The dynamics of this
system are described by potential non-relativistic effective field
theory (PNREFT)~\cite{Pineda:1997bj,Beneke:1997zp,Brambilla:1999xf},
combined with unstable particle effective field
theory~\cite{Beneke:2003xh,Beneke:2004km}. Within this framework,
higher-order corrections can be treated systematically through a
simultaneous expansion in the non-relativistic velocity $v$ and the
strong, electromagnetic, and top Yukawa couplings $\alpha_\mathrm{s}, \alpha, $
and $y_\mathrm{t}$. We adopt the power counting $v \sim \alpha_\mathrm{s} \sim
\sqrt{\alpha} \sim y_\mathrm{t}$ with the top quark width $\Gamma_\mathrm{t} \sim
m_\mathrm{t}\alpha$. Powers of $\alpha_\mathrm{s}/v$ from the bound-state interaction are
resummed to all orders in perturbation theory.

At leading order, the PNREFT Lagrangian is given by
\begin{equation}
  \label{eq:L_PNREFT}
    {\cal L}_\text{PNREFT, LO} ={} \psi^\dagger \bigg(\mathrm{i}\partial_0 +
    \frac{\vec{\partial}^2 + \mathrm{i} m_\mathrm{t} \Gamma_\mathrm{t}}{2m_\mathrm{t}}\bigg)
    \psi +{\cal L}_\text{anti-quark}
    - \int \mathrm{d}^3\vec{r}\ \big[\psi^\dagger\psi\big](\vec{x}+\vec{r})
    \frac{C_F\alpha_\mathrm{s}}{r}\big[\chi^\dagger\chi\big](\vec{x})\,,
\end{equation}
where $\psi$ is the quark field and $\chi$ the antiquark field.
The resulting top pair propagator is the Green function of the
Schrödinger equation with the colour Coulomb potential interaction. Its
imaginary part is closely related to the resonant top pair production
cross-section via the optical theorem.

\subsection{Higher-order corrections}
\label{sec:corr:maier}

Higher-order corrections to the PNREFT Lagrangian are obtained by
matching to the full Standard Model. In the first step, hard modes with
large four-momenta $k \sim m_\mathrm{t}$ are integrated out. This gives rise to a
non-relativistic effective field theory with local effective
vertices. These matching corrections are known to NNNLO in the QCD and
Higgs
sector~\cite{Eiras:2006xm,Beneke:2013kia,Marquard:2014pea,Beneke:2015lwa}
and to NNLO in the electroweak
sector~\cite{Grzadkowski:1986pm,Guth:1991ab,Hoang:2004tg,Hoang:2006pd,Beneke:2017rdn}.

In the second matching step,  soft modes $k \sim m_\mathrm{t}v$ and potential modes
$k_0 \sim m_\mathrm{t}v^2, \vec{k} \sim m_\mathrm{t} v$ for gluons and light quarks are also
eliminated from the theory. The most challenging part is the calculation
of the corrections to the static colour Coulomb potential to NNNLO, which
is reported in Refs.~\cite{Anzai:2009tm,Smirnov:2009fh,Lee:2016cgz}.

\subsubsection{Resonant production}
\label{sec:prod_res:maier}

With the matched PNREFT Lagrangian, the resonant top pair production
cross-section can be calculated, including NNNLO QCD and Higgs effects and
NNLO electroweak effects, by computing corrections to the Green function
to this order. The complete result for the QCD corrections was first
presented for the S-wave contribution in Ref.~\cite{Beneke:2015kwa} and for
the P-wave contribution in Ref.~\cite{Beneke:2013kia} (see
also Ref.~\cite{Penin:1998mx}). Schematically, the known contributions to the
top pair production cross-section can be written as
\begin{equation}
  \label{eq:PNRQCD_power_counting}
  \sigma_\text{res} \sim \alpha^2 v \sum_{k=0}^\infty
  \bigg(\frac{\alpha_\mathrm{s}}{v}\bigg)^{k}\times
  \begin{cases}
    1 & \text{LO}\\
    \alpha_\mathrm{s},v,\tfrac{\alpha}{v} & \text{NLO}\\
    \alpha_\mathrm{s}^2,\alpha_\mathrm{s} v,
    v^2,\tfrac{\alpha}{v}\times\left\{\tfrac{\alpha}{v}, \alpha_\mathrm{s}s,
      v\right\}, y_\mathrm{t}\sqrt{\alpha}, y_t^2  & \text{NNLO}\\
    \alpha_\mathrm{s}^{3-i} v^i, y_\mathrm{t}^2\times\left\{\tfrac{\alpha}{v}, \alpha_\mathrm{s},
      v\right\} & \text{NNNLO}
  \end{cases}\,.
\end{equation}

\subsubsection{Non-resonant production}
\label{sec:prod_nonres:maier}

\begin{figure}
  \centering
%  \includegraphics[width=80pt]{SM_maier/figures/nonres_h1}
%  \qquad\includegraphics[width=80pt]{SM_maier/figures/nonres_h2}
%  \qquad\includegraphics[width=80pt]{SM_maier/figures/nonres_h3}
%  \qquad\includegraphics[width=80pt]{SM_maier/figures/nonres_h4}
   \includegraphics[width=80pt]{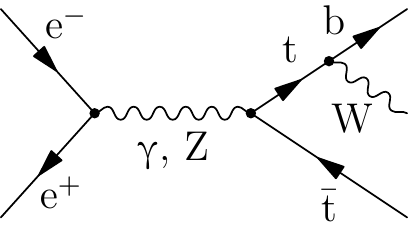}
  \qquad\includegraphics[width=80pt]{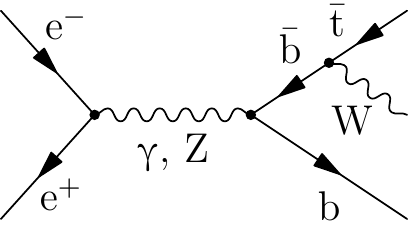}
  \qquad\includegraphics[width=80pt]{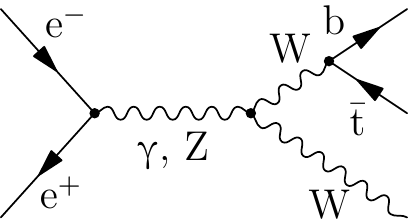}
  \qquad\includegraphics[width=80pt]{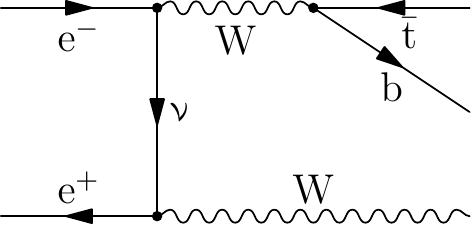}
  \caption{Non-resonant diagrams contributing to the $\mathrm{W}^+\mathrm{b}\bar{\mathrm{t}}$ final
    state at NLO. The final state $\mathrm{W}^-\bar{\mathrm{b}}\mathrm{t}$ follows from charge
    conjugation.}
  \label{fig:NLO_nonres}
 % \query{Please correct the figure labels in \Fref{fig:NLO_nonres}. Set particle
 % names in roman (upright) font.}
\end{figure}

Top quarks are unstable and the final state $\mathrm{b}\bar{\mathrm{b}}\mathrm{W}^+\mathrm{W}^-\mathrm{X}$ can also
be produced in non-resonant channels that do not involve the creation of
a top--antitop pair near the mass shell. According to unstable particle effective field
theory, the full cross-section is given by the sum of the resonant
and non-resonant contributions:
\begin{equation}
  \label{eq:sigma}
  \sigma = \sigma_\text{res} + \sigma_\text{non-res}\,.
\end{equation}
While non-resonant production is suppressed by one power of $\alpha$, it
does not suffer from the same phase space suppression as resonant
production and therefore contributes with a factor of
${\alpha} / {v}$ relative to the leading-order cross-section,
\ie at NLO. The diagrams at this order are shown in \Fref{fig:NLO_nonres}; their
contribution was first detailed in Ref.~\cite{Beneke:2010mp}. The NNLO
non-resonant cross-section was later detailed in Ref.~\cite{Beneke:2017rdn}.

Virtual top quarks in the non-resonant channels are formally far
off-shell with squared momenta $p_\mathrm{t}^2 - m_\mathrm{t}^2 \sim m_\mathrm{t}^2 \gg
m_\mathrm{t}\Gamma_\mathrm{t}$, so the width must not be resummed in the
propagators. Since we integrate over the full phase space, endpoint
divergences occur whenever $p_\mathrm{t}^2 - m_\mathrm{t}^2$ vanishes. At NNLO, this leads
to poles proportional to ${\Gamma_\mathrm{t}} / {\epsilon}$ in $4-2\epsilon$
dimensions. As usual in asymptotic expansions, these cancel against
poles in a different expansion region. In this case, the corresponding
poles appear in the form of finite-width divergences in the resonant
cross-section. A detailed account of the NNLO calculation including the
arrangement of pole cancellations is given in Ref.~\cite{Beneke:2017rdn}.

\subsubsection{Initial-state radiation}
\label{sec:ISR:maier}

Formally, photonic corrections in the initial state are suppressed by
one order in $\alpha$ and therefore contribute at NNLO according to our
power counting. However, it is well-known that these corrections are
enhanced by logarithms of $m_\mathrm{t}$ over $m_\mathrm{e}$, which have to be resummed to
all orders. The resummed cross-section is given by~\cite{Kuraev:1985hb,Fadin:1987wz}
\begin{equation}
  \label{eq:ISR}
  \sigma(s) = \int_0^1 \mathrm{d}x_1\int_0^1 \mathrm{d}x_2 \,\Gamma_\mathrm{ee}^\mathrm{LL}(x_1)
  \Gamma_\mathrm{ee}^\mathrm{LL}(x_2) \hat{\sigma}(x_1x_2 s) + \sigma^{\text{ISR}}_{\text{const}}(s)\,,
\end{equation}
where $\Gamma_\mathrm{ee}^\mathrm{LL}(x)$ is a leading logarithmic structure function,
$\hat{\sigma}$ is the `partonic' cross-section without ISR resummation
and $\sigma^{\text{ISR}}_\text{const}$ accounts for the non-logarithmic
  NNLO contribution.

\subsection{Cross-section predictions}
\label{sec:xs:maier}

\begin{figure}
  \centering
\includegraphics[width=0.45\linewidth]{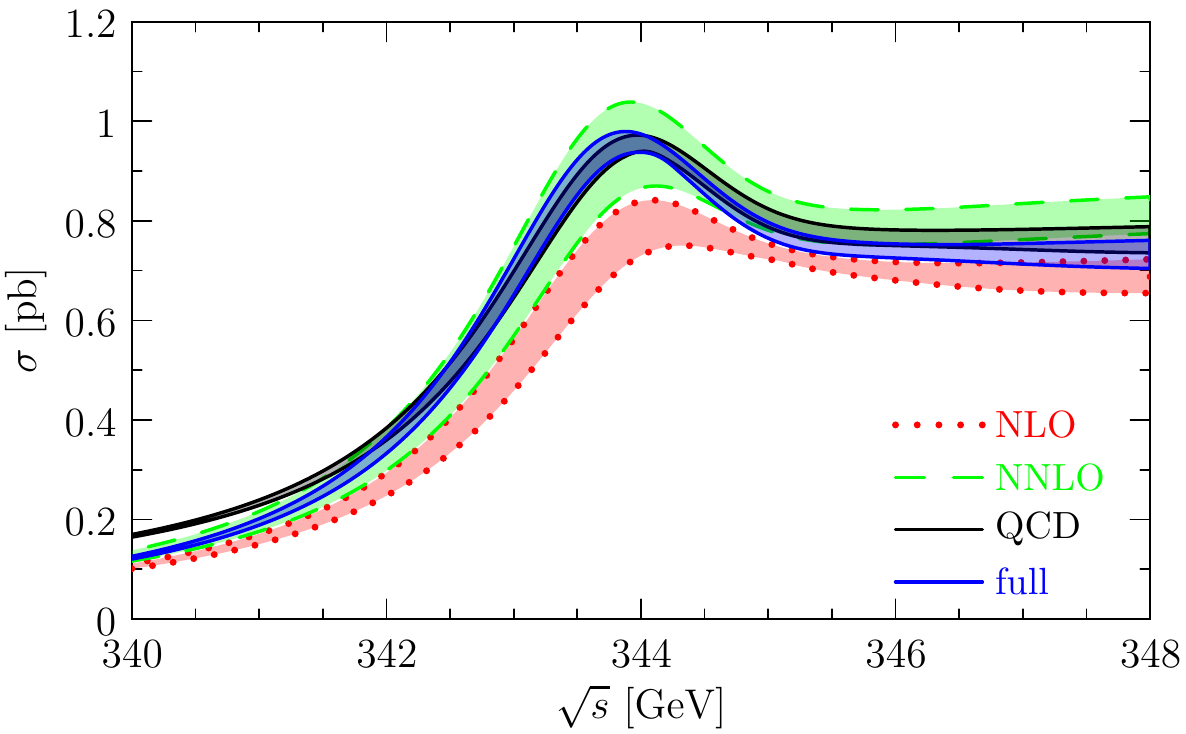} \quad\includegraphics[width=0.45\linewidth]{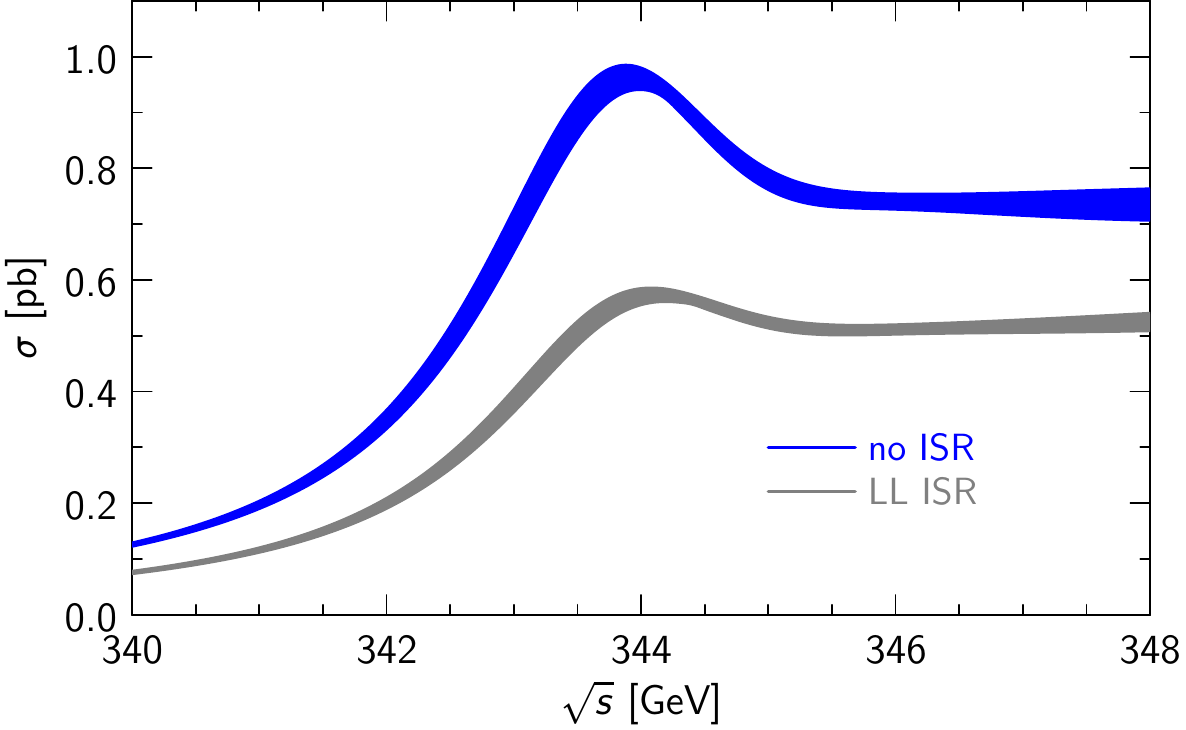}
  \caption{Total cross-section for the process $\mathrm{e}^+\mathrm{e}^-\to$ at various
    orders in perturbation theory. Left: Cross-section without ISR from NLO to NNNLO
    with the pure NNNLO QCD result as comparison. Right: Effect of ISR on
  the cross-section.}
  \label{fig:xs}
\end{figure}

The formulae for the cross-section can be evaluated numerically with the
code\\ \texttt{QQbar\_threshold}~\cite{Beneke:2016kkb}, which includes all
aforementioned corrections. Figure~\ref{fig:xs} shows the behaviour of the total
cross-section near threshold for a top quark mass of
$m_\mathrm{t}(20\UGeV) = 171.5\UGeV$ in the potential-subtracted  scheme~\cite{Beneke:1998rk}
and input parameters $\Gamma_\mathrm{t} = 1.33\UGeV$, $m_\mathrm{H} = 125\UGeV$,
$\alpha_\mathrm{s}(m_\mathrm{Z}) = 0.1177$, $\alpha(m_\mathrm{Z}) = 1/128.944$. The uncertainty
bands originate from a variation of the renormalization scale between
$50$ and $350\UGeV$.

\begin{figure}
  \centering
  \begin{tabular}{cc}
    \includegraphics[width=0.45\linewidth]{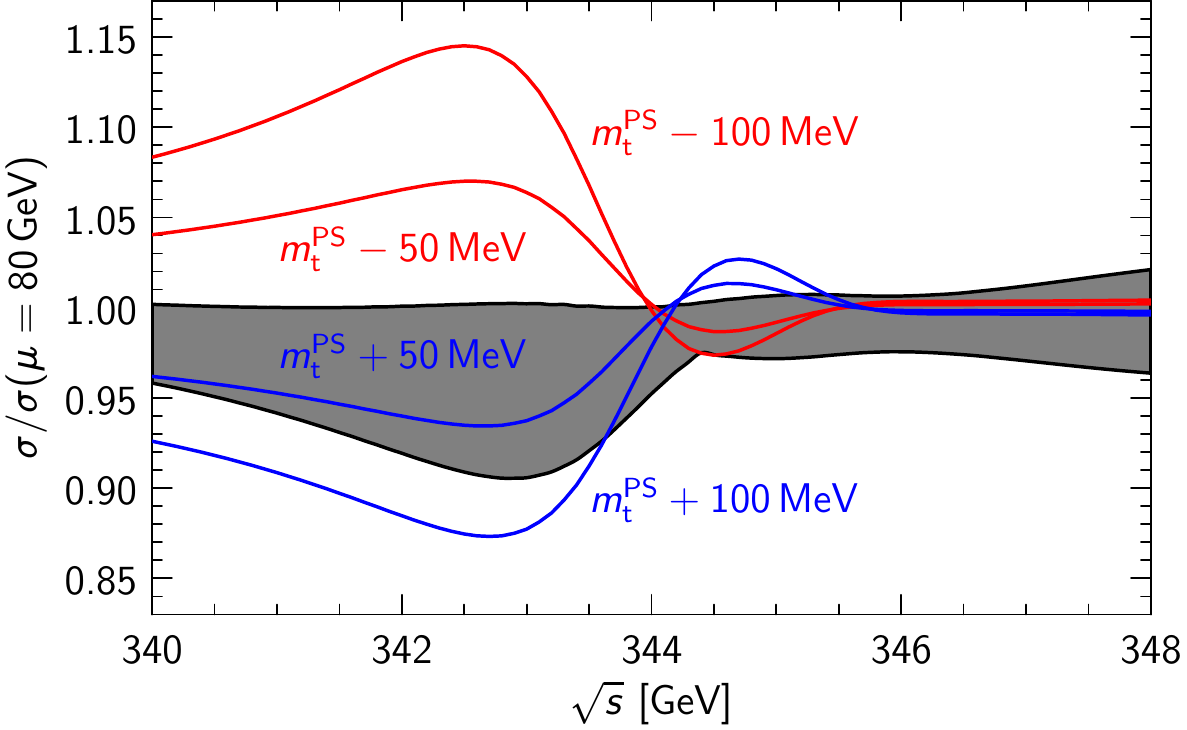}&\includegraphics[width=0.45\linewidth]{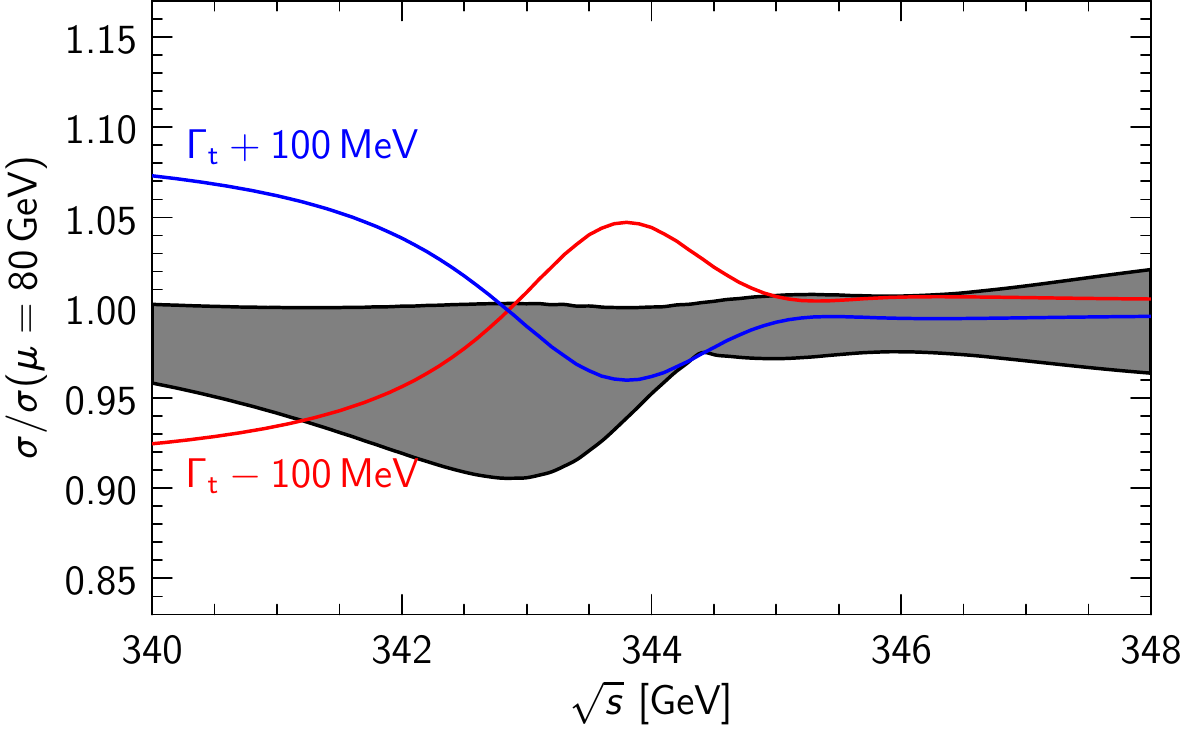}\\
    \includegraphics[width=0.45\linewidth]{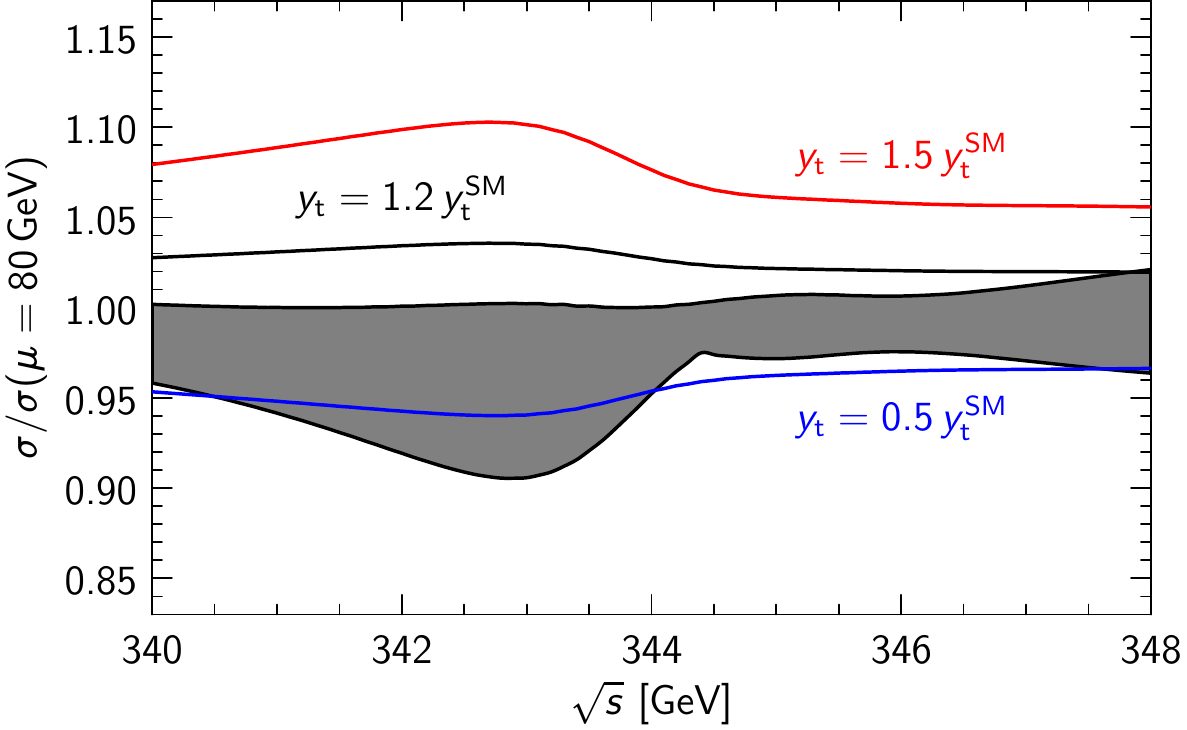}&\includegraphics[width=0.45\linewidth]{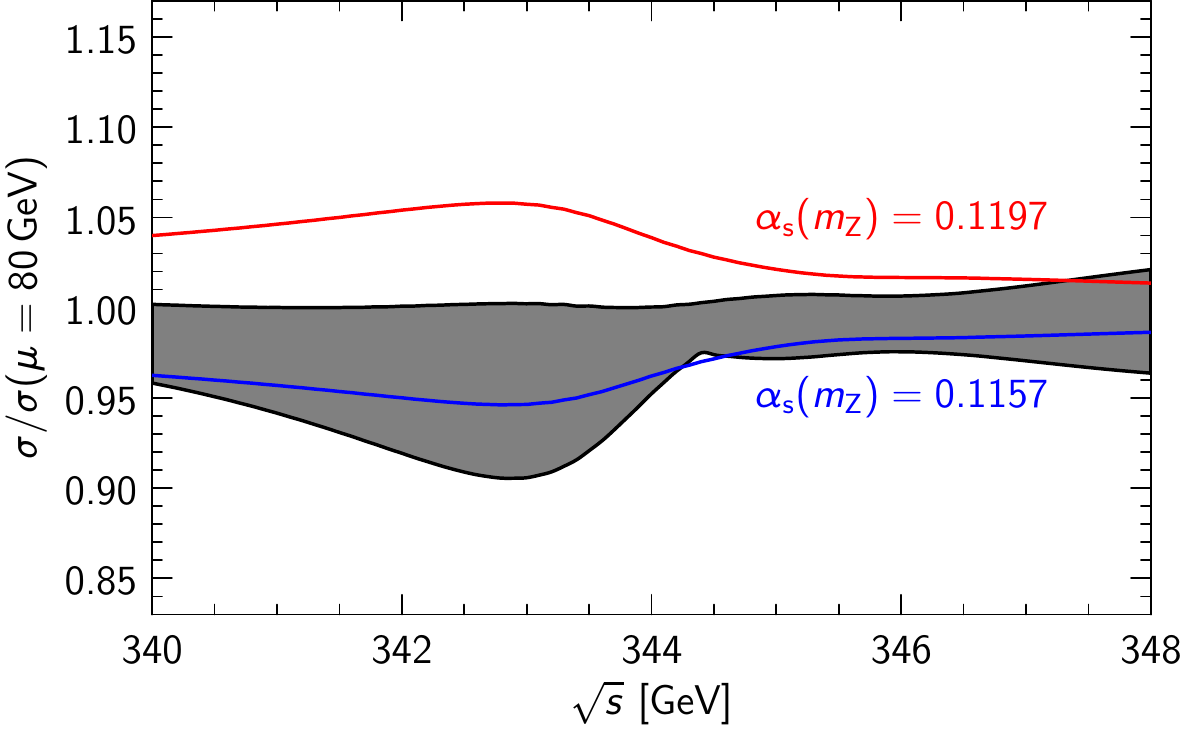}
  \end{tabular}
  \caption{Sensitivity of the cross-section to parameter variation. Top
    left: Variation of the top quark mass by up to $\pm 100\UMeV$. Top
    right: Variation of the top quark width by up $\pm 100\UMeV$. Bottom
  left: Variation of the top Yukawa coupling. Bottom
  right: Variation of the strong coupling constant.}
  \label{fig:xs_var}
  %  \query{Please correct the figure labels in \Fref{fig:xs_var}. Set text
  %  labels and particle
  %names in roman  font.}
\end{figure}

Figure~\ref{fig:xs_var} shows the effect of changing various
parameters. The variation suggests that it should be possible to extract
the top quark width and mass in the potential-subtracted scheme with an uncertainty of
better than $100\UMeV$. The sensitivity to the top Yukawa coupling and
the strong coupling is less pronounced and there is a considerable
degeneracy between the two parameters. A precise knowledge of the strong
coupling constant from other sources will be crucial to meaningfully
constrain the Yukawa coupling. In any case, a dedicated experimental analysis will be
needed to determine the exact precision with which the various top quark
properties can be extracted from a measurement of the cross-section.

\subsection*{Acknowledgements}

I thank the CERN Theory  Department for their hospitality and support during the workshop.

\end{bibunit}

\label{sec-sm-maier}  
\clearpage \pagestyle{empty}  \cleardoublepage
%============================================

\pagestyle{fancy}
\fancyhead[CO]{\thechapter.\thesection \hspace{1mm} Higgs boson decays: theoretical status}
\fancyhead[RO]{}
\fancyhead[LO]{}
\fancyhead[LE]{}
\fancyhead[CE]{}
\fancyhead[RE]{}
\fancyhead[CE]{M. Spira}
\lfoot[]{}
\cfoot{-  \thepage \hspace*{0.075mm} -}
\rfoot[]{}

\newcommand{\tc}[1]{\textcolor{#1}}
\newcommand{\lsim}{\raisebox{-0.13cm}{~\shortstack{$<$ \\[-0.07cm] $\sim$}}~}
%\newcommand{\gsim}{\raisebox{-0.13cm}{~\shortstack{$>$ \\[-0.07cm] $\sim$}}~}

%=================================================================

%\vspace*{5.cm}

\begin{bibunit}[elsarticle-num] % define the bib-style for the unit: elsarticle-num.bst
%  text-1; this is the corresponding section
%\putbib[2loops] % the *.bib
%\end{bibunit}
% go-on
%--- from: bibunits.sty, adapts the font size of ``References'' to section
\let\stdthebibliography\thebibliography
\renewcommand{\thebibliography}{%
\let\section\subsection
\stdthebibliography}

\section[Higgs boson decays: theoretical status \\ {\it M. Spira}]
{Higgs boson decays: theoretical status
\label{contr:mspira}}
\noindent
{\bf 
Contribution\footnote{This contribution should be cited as:\\
M. Spira, Higgs boson decays: theoretical status,  
%04 DOI:10.23731/CYRM-2020-XXX.\thepage, in:
%04 \url{http://dx.doi.org/10.23731/CYRM-2020-XXX.\thepage}, in:
DOI: \href{http://dx.doi.org/10.23731/CYRM-2020-003.\thepage}{10.23731/CYRM-2020-003.\thepage}, in:
Theory for the FCC-ee, Eds. A. Blondel, J. Gluza, S. Jadach, P. Janot and T. Riemann,
\\CERN Yellow Reports: Monographs, CERN-2020-003,
%04 \url{http://dx.doi.org/10.23731/CYRM-2020-XXX}, p. \thepage.} 
DOI: \href{http://dx.doi.org/10.23731/CYRM-2020-003}{10.23731/CYRM-2020-003},
p. \thepage.
\\ \copyright\space CERN, 2020. Published by CERN under the 
%04-2
\href{http://creativecommons.org/licenses/by/4.0/}{Creative Commons Attribution 4.0 license}.} by: M. Spira {[Michael.Spira@psi.ch]}}
\vspace*{.5cm}

\subsection{Introduction}
%           ============
The discovery of a Standard-Model-like Higgs boson at the LHC
\cite{Aad:2012tfa,Chatrchyan:2012xdj} completed the theory of
electroweak and strong interactions.  The measured Higgs mass of
($125.09 \pm 0.24$)\,GeV \cite{Khachatryan:2016vau} ranges at the order
of the weak scale. The existence of the Higgs boson \cite{Higgs:1964ia,
Higgs:1964pj, Englert:1964et, Guralnik:1964eu, Higgs:1966ev,
Kibble:1967sv} allows the Standard Model (SM) particles to be weakly
interacting up to high-energy scales.  This, however, is only possible
for particular Higgs boson couplings to all other particles so that with
the knowledge of the Higgs boson mass all its properties are uniquely
fixed. The massive gauge bosons and fermions acquire mass through their
interaction with the Higgs field, which develops a finite vacuum
expectation value in its ground state.  The minimal model requires the
introduction of one isospin doublet of the Higgs field and leads after
spontaneous symmetry breaking to the existence of one scalar Higgs
boson.

Since all Higgs couplings are fixed within the SM, any meaningful
approach to introduce variations requires the introduction of effects beyond
the SM (BSM). Two major branches are being pursued for this purpose: 
(i) the introduction of higher-dimension operators in terms of a
general effective Lagrangian with dimension-6 operators providing the
leading contributions for energy scales sufficiently below the novel
cut-off scale of these operators and (ii) the introduction of
specific BSM models with extended Higgs, gauge, and fermion sectors. The
extraction of BSM effects, however, strongly relies on the accuracy of
the SM part as, \eg~sketched in the basic decomposition of the SM-like
Higgs boson decay widths as
\begin{equation}
\Gamma = \Gamma_\mathrm{SM} + \Delta\Gamma_\mathrm{BSM}
\end{equation}
Any potential to extract the BSM effects $\Delta\Gamma_\mathrm{BSM}$ is limited
by the uncertainties $\delta \Gamma_\mathrm{SM}$ of the SM part.

\subsection{SM Higgs boson decays}
%           =====================
The determination of the branching ratios of Higgs boson decays thus
necessitates the inclusion of the available higher-order corrections
(for a recent overview see, \eg~Ref. \cite{Spira:2016ztx}) and a
sophisticated estimate of the theoretical and parametric uncertainties.

\subsubsection{$\mathrm{H}\to \mathrm{f\bar f}$}
%              ===================
The Higgs decay $\mathrm{H} \to \mathrm{b\bar b}$ is the dominant Higgs boson decay with a
branching ratio of about 58\%. The subleading fermionic decays $\mathrm{H}
\to
\uptau^+\uptau^-$ and $\mathrm{H}\to \mathrm{c\bar c}$ reach branching ratios of about 6\% and
3\%, respectively. The rare decay $\mathrm{H} \to \upmu^+\upmu^-$ will become visible
at the HL-LHC and happens with about 0.02\% probability
\cite{deFlorian:2016spz}. The present status of the partial decay
widths can be summarised in terms of the (factorised) expression
\begin{equation}
\Gamma (\mathrm{H}\to \mathrm{f}{\overline{\mathrm{f}}}) = \frac{N_\mathrm{c} G_\mathrm{F} M_\mathrm{H} } {4\sqrt{2}\uppi}
\ m_\mathrm{f}^2\ (1 + \updelta_{\rm QCD}+\updelta_\mathrm{t} + \updelta_\mathrm{mixed})
\left(1+\updelta_\mathrm{elw} \right) \, ,
\end{equation}
where $N_\mathrm{c}=3(1)$ for quarks (leptons), $G_\mathrm{F}$ denotes the Fermi constant,
$M_\mathrm{H}$ denotes the Higgs mass, and $m_\mathrm{f}$ denotes the fermion mass. In general, the pure QCD
corrections $\updelta_{\rm QCD}$ to the Higgs boson decays into quarks are
known up to NLO including the full quark mass dependence
\cite{Braaten:1980yq, Sakai:1980fa, Inami:1980qp, Drees:1990dq,
Drees:1989du} and up to N$^4$LO for the leading corrections with the
leading mass effects \cite{Gorishnii:1983cu, Gorishnii:1990zu,
Gorishnii:1991zr, Kataev:1993be, Surguladze:1994gc, Chetyrkin:1996sr,
Melnikov:1995yp}. The dominant part of the QCD corrections can be
absorbed in the running quark mass evaluated at the scale of the Higgs
mass. The top-induced QCD corrections, which are related to interference
effects between $\mathrm{H}\to \mathrm{gg}$ and $\mathrm{H}\to \mathrm{q\bar q}$, are known at NNLO in the
limit of heavy top quarks and light bottom quarks
\cite{Chetyrkin:1995pd, Larin:1995sq, Primo:2018zby}. In the case of
leptons, there are no QCD corrections ($\updelta_{\rm QCD}=\updelta_\mathrm{t} =
\updelta_\mathrm{mixed}=0$). The electroweak corrections $\updelta_\mathrm{elw}$ are known
at NLO exactly \cite{Fleischer:1980ub, Bardin:1990zj, Dabelstein:1991ky,
Kniehl:1991ze}. In addition, the mixed QCD-elw~corrections range at the
per-mille level if the factorised expression with respect to QCD and
elw~corrections is used \cite{Kataev:1997cq, Kniehl:1994ju,
Kwiatkowski:1994cu, Chetyrkin:1996wr, Mihaila:2015lwa, Chaubey:2019lum}.
The public tool {\tt Hdecay} \cite{Djouadi:1997yw, Djouadi:2018xqq}
neglects these mixed QCD-elw~corrections but includes all other
corrections. The partial decay width of $\mathrm{H}\to \mathrm{b\bar b}$ is also known
fully differential at N$^3$LO QCD \cite{Anastasiou:2011qx,
DelDuca:2015zqa, Ferrera:2017zex, Mondini:2019gid}. 

\subsubsection{$\mathrm{H}\to \mathrm{W}^{(*)}\mathrm{W}^{(*)},\mathrm{Z}^{(*)}\mathrm{Z}^{(*)}$}
%              ==========================================
The branching ratios of SM Higgs boson decays into (off-shell) W and
Z bosons amount to about 21\% and 3\%, respectively. Off-shell effects
of the W and Z bosons are important \cite{Rizzo:1980gz, Keung:1984hn,
Cahn:1988ru} and lead to the $\mathrm{H}\to \mathrm{Z}^{*}\mathrm{Z}^{(*)}\to 4\ell^\pm$ decay as
one of the discovery modes of the SM Higgs boson
\cite{Aad:2012tfa,Chatrchyan:2012xdj}. The electroweak corrections to
the full decay modes $\mathrm{H}\to \mathrm{V}^{(*)}\mathrm{V}^{(*)}\to 4\mathrm{f}~(\mathrm{V}=\mathrm{W,Z})$ have been
calculated \cite{Fleischer:1980ub, Kniehl:1990mq, Bardin:1991dp,
Bredenstein:2006rh, Bredenstein:2006ha}. The public tool {\tt
Prophecy4f} \cite{Bredenstein:2006rh, Bredenstein:2006ha}  for calculating
the exclusive decay processes has been used in the experimental
analyses. An improvement beyond the pure elw~corrections has been made
by the proper matching to parton showers at NLO \cite{Boselli:2015aha}.
However, shower effects have not been relevant for the analyses
performed so far. 

\subsubsection{$\mathrm{H}\to \mathrm{gg}$}
%              ================
The loop-induced Higgs decay into gluons reaches a branching ratio of
about 8\%. The decay is dominantly mediated by top and bottom quark
loops, with the latter providing a 10\% contribution. The charm quark
contributes at the level of about 2\%. The two-loop QCD corrections are
known, including the exact quark mass dependences \cite{Inami:1982xt,
Djouadi:1991tka, Spira:1995rr}. They enhance the partial decay width by
about 70\% and thus cannot be neglected in the decay profile of the
Higgs boson.  The NNLO, N$^3$LO, and, recently, the N$^4$LO QCD corrections
have been obtained for the top loops in the limit of heavy top quarks,
\ie~the leading term of a heavy top mass expansion
\cite{Chetyrkin:1997iv, Baikov:2006ch, Herzog:2017dtz}. The QCD
corrections beyond NLO amount to less than 20\% of the NLO QCD-corrected
partial decay width, thus signalling perturbative convergence in spite of
the large NLO corrections. The residual theoretical uncertainties have
been estimated at the level of about 3\% from the scale dependence of
the QCD-corrected partial decay width. The NLO elw~corrections have
been calculated for the top-loop contributions first in the limit of
heavy top quarks \cite{Djouadi:1994ge, Chetyrkin:1996wr,
Chetyrkin:1996ke}, then the electroweak corrections involving light
fermion loops exactly \cite{Aglietti:2004nj, Aglietti:2006yd,
Degrassi:2004mx}, and finally the full electroweak corrections involving
W, Z, and top-loop contributions, including the full virtual mass
dependences, by means of a numerical integration \cite{Actis:2008ug,
Actis:2008ts}.  They amount to about 5\% for the SM Higgs mass value.
The public tool {\tt Hdecay} \cite{Djouadi:1997yw, Djouadi:2018xqq}
includes the NLO QCD results with the full quark mass dependences, the
NNLO and N$^3$LO QCD corrections in the heavy top limit, and the full NLO
elw~corrections in terms of a grid in the Higgs and top masses used for
an interpolation. 

\subsubsection{$\mathrm{H}\to \upgamma\upgamma$}
%              =========================
The rare loop-induced Higgs decay into photons reaches a branching ratio
of about 0.2\%. The decay is dominantly mediated by W and top quark
loops, with the W loops being dominant. The two-loop QCD corrections
are known, including the exact top mass dependences \cite{Spira:1995rr,
Zheng:1990qa, Djouadi:1990aj, Dawson:1992cy, Djouadi:1993ji,
Melnikov:1993tj, Inoue:1994jq, Fleischer:2004vb, Harlander:2005rq,
Anastasiou:2006hc, Aglietti:2006tp}. They correct the partial decay
width by a small amount,  about 2\%. The QCD corrections beyond NLO
have been estimated in the limit of heavy top quarks to be in the
per-mille range \cite{Steinhauser:1996wy, Sturm:2014nva,
Maierhofer:2012vv}. The NLO elw~corrections to the W and top-induced
contributions have been obtained by a numerical integration of the
corresponding two-loop diagrams \cite{Djouadi:1997rj, Degrassi:2005mc,
Passarino:2007fp, Actis:2008ts}. They decrease the partial photonic
branching ratio of the SM Higgs boson by about 2\%, thus nearly cancelling
against the QCD corrections by accident.  The public tool {\tt Hdecay}
\cite{Djouadi:1997yw, Djouadi:2018xqq} includes the NLO QCD results with
the full quark mass dependences and the full NLO elw~corrections in
terms of a grid in the Higgs and top masses used for an interpolation,
but neglects all corrections beyond NLO. 

\subsubsection{$\mathrm{H}\to \mathrm{Z}\upgamma$ and Dalitz decays}
\begin{figure}
% 2020-03 JG  Spira B.12.1
\centering
\includegraphics[width=0.83\linewidth]{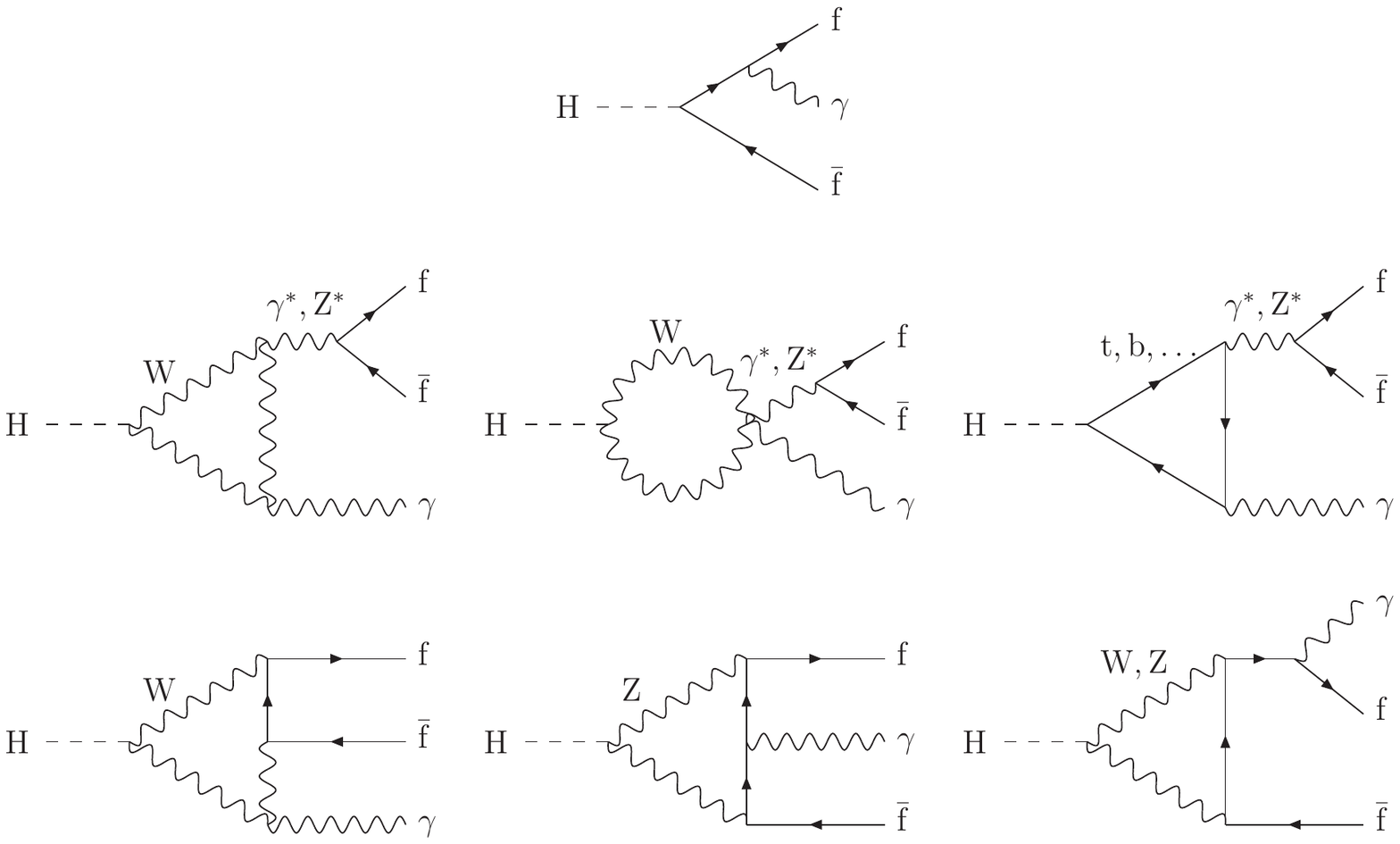}\\ [-11cm]
\caption{\label{fg:dalitzdia} Generic diagrams contributing to
the Dalitz decays $\mathrm{H}\to \upgamma \mathrm{f\bar f}$}
%\query{Please correct the figure labels in \Fref{fg:dalitzdia}. Set  particle
%  names in roman  (upright) font.}
\end{figure}

\begin{figure}[h!]
% 2020-03-22  Spira  B.12.2

%\center 
% 400,500 -> 0,500
\begin{picture}(400,315)(0,0)
%\put(400,450){$H\to \gamma e^+e^-$}
%\put(400,275){$H\to \gamma  \mu^+\mu^-$}
%\put(400,100){$H\to \gamma \tau^+\tau^-$}
\put(125,295){$\mathrm{H}\to \upgamma \; \mathrm{e}^+\mathrm{e}^-$}
\put(125,225){$\mathrm{H}\to \upgamma \; \upmu^+\upmu^-$}
\put(125,155){$\mathrm{H}\to \upgamma \; \uptau^+\uptau^-$}
\put(20,260){\includegraphics[width=1.5\linewidth]{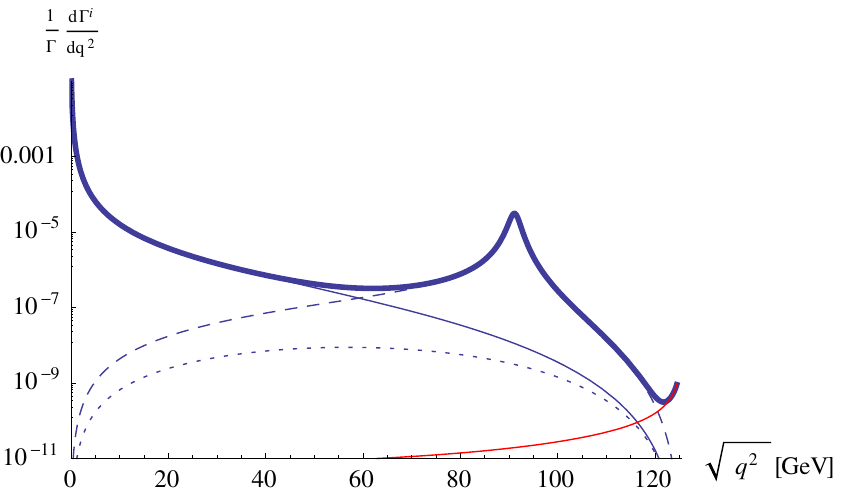}}
\put(20,190){\includegraphics[width=1.5\linewidth]{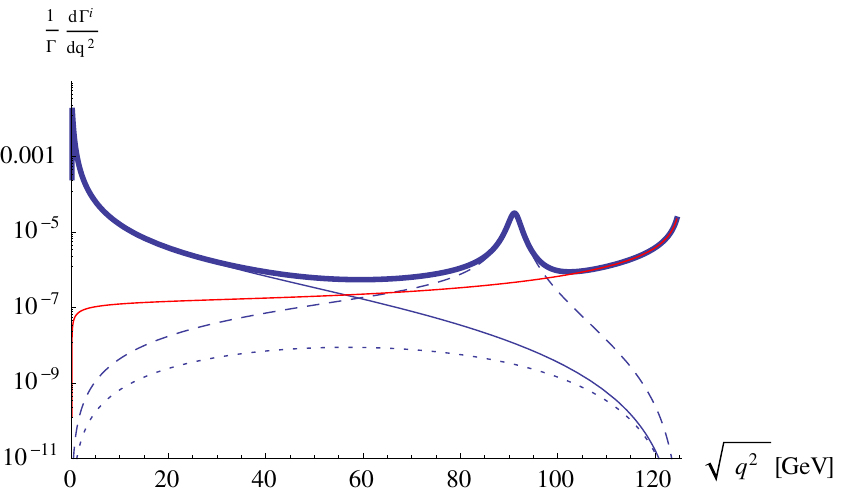}}
\put(20,120){\includegraphics[width=1.5\linewidth]{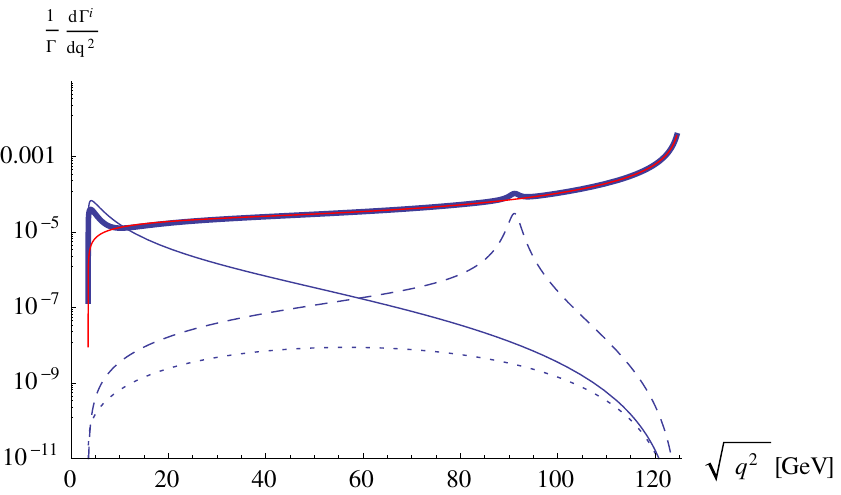}}
\end{picture} \\[-12cm]
%%%%%%%%%%%%%%%%%%%%%%%%%%%%%
%\begin{picture}(0,500)(0,0)
%\put(0,450){$\mathrm{H}\to \upgamma %\mathrm{e}^+\mathrm{e}^-$}
%\put(0,275){$\mathrm{H}\to \upgamma \upmu^+\upmu^-$}
%\put(0,100){$\mathrm{H}\to \upgamma \uptau^+\uptau^-$}
%\put(100,350)
%{\includegraphics[width=1.5\linewidth]{SM_mspira/differen%tialwidthfore2.pdf}} 
%\put(100,175)
%{\includegraphics[width=1.5\linewidth]{SM_mspira/differentialwidthformu.pdf}} 
%\put(100,00)
%{\includegraphics[width=1.5\linewidth]{SM_mspira/differentialwidthfortau.pdf}} 
%\end{picture}

\caption{\label{fg:dalitz}  The invariant mass distributions in
$\sqrt{q^2} = M_{\ell^+\ell^-}$ of the Dalitz decays $\mathrm{H}\rightarrow
\upgamma + \mathrm{e}^+ \mathrm{e}^-/\upmu^+\upmu^-/\uptau^+\uptau^-$ normalised to
$\Gamma(\mathrm{H}\rightarrow \upgamma\upgamma)$ with a cut $E_\upgamma > 1\UGeV$ on the
photon energy. The red lines show the contribution of the tree diagrams,
the thin solid lines denote the contribution of the photon conversion
$\mathrm{H}\to \upgamma\upgamma^*\to \upgamma\ell^+\ell^-$, and the dashed line the
contribution from the $\mathrm{Z}^*$ exchange diagrams, while the thick lines
present the total contributions. The dotted lines denote the
contribution from the box diagrams (in 't Hooft--Feynman gauge). From
Ref.~\cite{Sun:2013rqa}.}
%\query{Please correct the figure labels in %\Fref{fg:dalitz}. Set variables in italic font.}
\end{figure}

%              =====================================
The rare loop-induced Higgs decay into a Z boson and a photon reaches a
branching ratio of less than 0.2\%. The decay is dominantly mediated by W
and top quark loops, with the W loops being dominant. The two-loop QCD
corrections are known, including the exact top mass dependences
\cite{Spira:1991tj, Bonciani:2015eua, Gehrmann:2015dua}.  They correct
the partial decay width by a small amount in the per-mille range and
thus can safely be neglected. The electroweak corrections to this decay
mode are unknown. However, the decay mode $\mathrm{H}\to \mathrm{Z}\upgamma\to \mathrm{f\bar
f}\upgamma$ is one of the more general Dalitz decays $\mathrm{H}\to \mathrm{f\bar f}\upgamma$
\cite{Abbasabadi:1996ze, Abbasabadi:2006dd, Abbasabadi:2004wq,
Dicus:2013ycd, Chen:2012ju, Passarino:2013nka, Sun:2013rqa}. The latter
are described by the diagrams in \Fref{fg:dalitzdia}, where the
Z boson exchange appears in a part of the triangle diagrams. The resonant Z boson exchange corresponds to the $\mathrm{H}\to \mathrm{Z}\upgamma$ decay
mode. The separation of this part, however, depends on the experimental
strategy to reconstruct the Z boson in the final state. A first step
for the reconstruction of the Z boson is to cut on the invariant mass
of the final-state fermion pair. The corresponding distributions of the
Dalitz decays are shown in \Fref{fg:dalitz} for the three
charged lepton final states normalised to the partial width into photons
with a cut $E_\upgamma > 1\UGeV$ on the photon energy. For small invariant
masses, the photon conversion $\mathrm{H}\to\upgamma\upgamma^\ast\to \upgamma
\ell^+\ell^-$ provides the dominant contribution, while for invariant
masses around the Z boson mass the Z boson contribution $\mathrm{H}\to \upgamma
\mathrm{Z}^\ast\to \upgamma \mathrm{f\bar f}$ takes  the dominant role. At the endpoint
$q^2\lsim M_\mathrm{H}^2$ of the spectrum, the direct contribution determines the
distributions. This increases with growing Yukawa coupling, \ie~it is
largest for $\mathrm{H}\to \upgamma \uptau^+\uptau^-$ (where it dominates in the whole
$q^2$ range). (It should be noted that the endpoint in the
$\mathrm{e}^+\mathrm{e}^-\upgamma$ case is four or five orders of magnitude smaller than the photon
and Z exchange contributions, thus making it impossible to determine
the electron Yukawa coupling. The same conclusion is also valid for the
reverse process $\mathrm{e}^+\mathrm{e}^-\to \mathrm{H}\upgamma$, so that the $s$-channel line shape
measurement proposed in Ref.~\cite{Jadach:2015cwa} will not be sensitive
to the electron Yukawa coupling but dominated by the loop-induced
contribution with an additional photon.) For a clean separation of the
$\mathrm{H}\to\upgamma\upgamma$, $\mathrm{H}\to\upgamma\upgamma^\ast\to \upgamma \ell^+\ell^-$,
$\mathrm{H}\to \mathrm{Z}\upgamma$, and $\mathrm{H}\to \ell^+\ell^-$ contributions, appropriate cuts must be implemented for the Dalitz decays. The low-$q^2$ part must
be attributed to $\mathrm{H}\to\upgamma\upgamma$, the $q^2$-part around $M_\mathrm{Z}^2$ to
$\mathrm{H}\to \mathrm{Z}\upgamma$ and the endpoint region close to $M_\mathrm{H}^2$ to the QED
corrections to $\mathrm{H}\to\ell^+\ell^-$. The public code {\tt Hdecay}
\cite{Djouadi:1997yw, Djouadi:2018xqq} does not include the full Dalitz
decays.

%=================================================
%\hspace{-5.0cm}
%=======================  end of fig. B.12.2  ===========

\subsection{Uncertainties}
%           =============
The parametric errors are dominated by the uncertainties in the top,
bottom, and charm quark masses, as well as the strong coupling $\alpha_\mathrm{s}$.
We have used the $\overline{\rm MS}$ masses for the bottom and charm
quarks, $\overline{m}_\mathrm{b} (\overline{m}_\mathrm{b}) = (4.18 \pm 0.03)\UGeV$ and
$\overline{m}_\mathrm{c} (3\UGeV) = (0.986 \pm 0.026)\UGeV$, and the
top quark pole mass $m_\mathrm{t} = (172.5 \pm 1)\UGeV$, according to the
conventions of the LHC Higgs cross-section WG (HXSWG)
\cite{deFlorian:2016spz}. The $\overline{\rm MS}$ bottom and charm
masses are evolved from the input scale to the scale of the decay
process with four-loop accuracy in QCD. The strong coupling $\alpha_\mathrm{s}$ is
fixed by the input value at the Z boson mass scale, $\alpha_\mathrm{s}(M_\mathrm{Z}) =
0.118 \pm 0.0015$.  The total parametric uncertainty for each branching
ratio has been derived from a quadratic sum of the individual impacts of
the input parameters on the decay modes along the lines of the original
analyses in Refs.~\cite{Djouadi:1995gt, Gross:1994fv} and the later analysis
in Ref.
\cite{Denner:2011mq}.

\begin{figure}[h!]
\centering
% did work: \includegraphics[width=0.83\linewidth]{SM_mspira/SMHiggsBR-120-130-crop.pdf}
\includegraphics[width=0.75\linewidth]{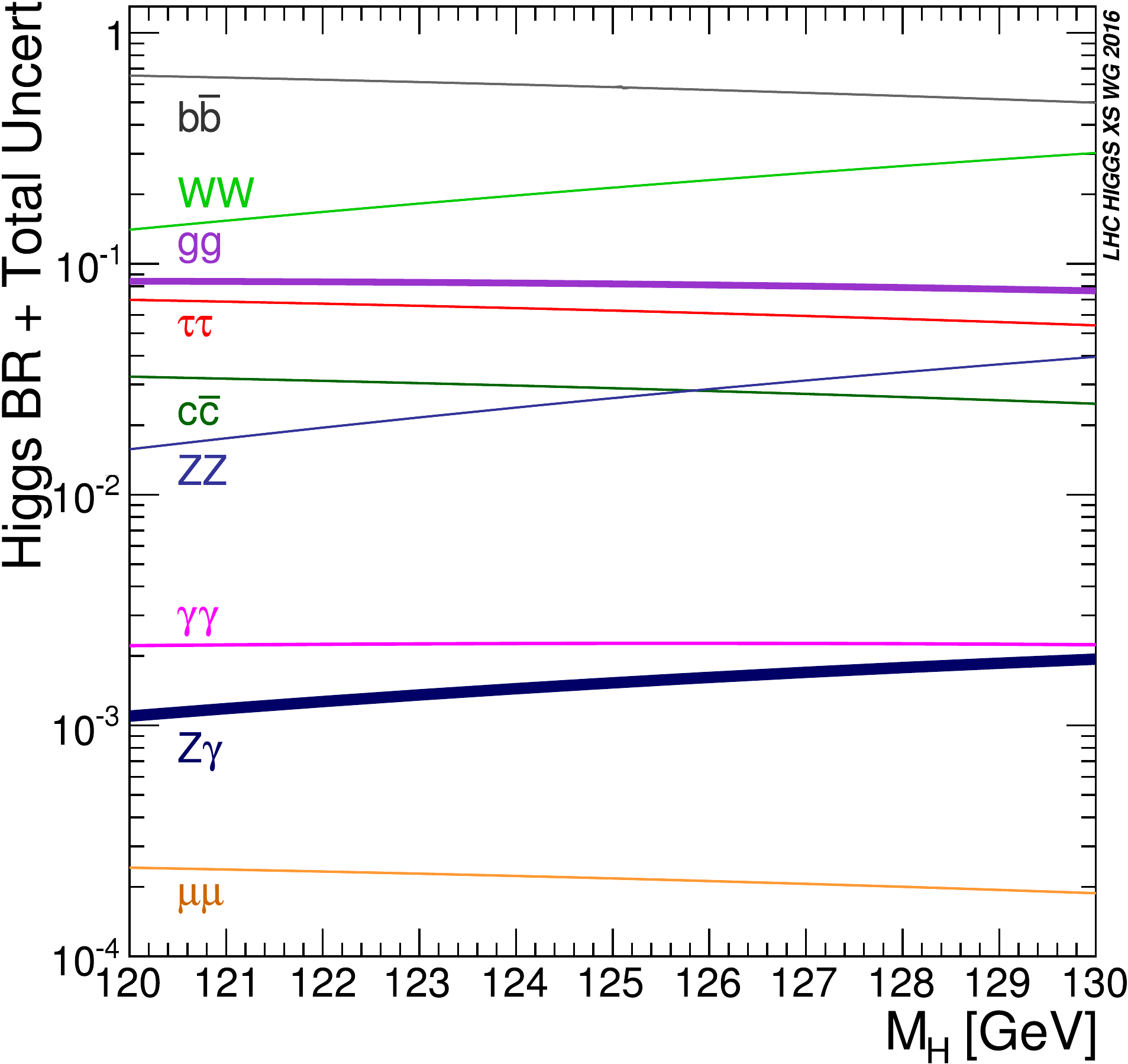}
\caption{Higgs boson branching ratios and their uncertainties for
Higgs masses around 125\,GeV. Figure courtesy ref.~\cite{deFlorian:2016spz}.}
\label{fig:hbr}
%\query{Please correct the figure labels in \Fref{fig:hbr}. Set variables in italic font (but keep particle names in roman (upright) font).}
\end{figure}

The theoretical uncertainties from missing higher orders in the
perturbative expansion are summarised in  \Tref{tab:hbrunc} for the
individual partial decay processes, along with the perturbative orders of
the included QCD or elw~corrections \cite{Spira:2016ztx,
deFlorian:2016spz}. To be conservative, the total parametric
uncertainties are added linearly to the theoretical uncertainties. The
final result for the branching ratios is shown in \Fref{fig:hbr} for
the leading Higgs decay modes with branching ratio larger than $10^{-4}$
for the Higgs mass range between 120 and 130\,GeV. These have been
obtained using {\tt Prophecy4f} \cite{Bredenstein:2006rh,
Bredenstein:2006ha} for the decays $\mathrm{H}\to \mathrm{WW},\mathrm{ZZ}$ and {\tt Hdecay}
\cite{Djouadi:1997yw, Djouadi:2018xqq} for the other decay modes. The
bands represent the total uncertainties of the individual branching
ratios. For a Higgs mass $M_\mathrm{H}=125\UGeV$, the total uncertainty of the
leading decay mode $\mathrm{H}\to \mathrm{b\bar b}$ amounts to less than 2\%, since the
bulk of it cancels out within the branching ratio.  The uncertainty of
$\Gamma(\mathrm{H} \to \mathrm{b\bar b})$, however, generates a significant increase in the
uncertainties of the subleading decay modes. The total uncertainties of
$\mathrm{BR}(\mathrm{H}\to \mathrm{WW}/\mathrm{ZZ})$ and $\mathrm{BR}(\mathrm{H}\to \uptau^+\uptau^-/\upmu^+\upmu^-)$ amount to $\sim$$2\%$, while the uncertainties of $\mathrm{BR}(\mathrm{H}\to \mathrm{gg})$ and $\mathrm{BR}(\mathrm{H}\to \mathrm{c\bar c})$
range at $\sim$6--7\%, of $\mathrm{BR}(\mathrm{H}\to\upgamma\upgamma)$ at $\sim$3\% and of
$\mathrm{BR}(\mathrm{H}\to \mathrm{Z}\upgamma)$ at $\sim$7\%. The total decay width of $\sim$$4.1\UMeV$ can be predicted with $\sim$2\% total uncertainty.

\FloatBarrier 

 \begin{table}[h!]
\caption{Estimated theoretical uncertainties from missing higher
orders and the perturbative orders (QCD/elw) of the results included in
the analysis.}
\label{tab:hbrunc}
%\vspace{0.4cm}
\begin{center}
\renewcommand{\arraystretch}{1.4}
\setlength{\tabcolsep}{1.5ex}
\begin{tabular}{llllll}
\hline \hline
{Partial width} & {QCD} & {Electroweak} &
{Total} & \tc{red}{{On-shell Higgs}} \\
& (\%) & (\%) & (\%)\\
\hline
 $\mathrm{H} \to \mathrm{b\bar b}/\mathrm{c\bar c}$ & $\sim$0.2
& $\sim$0.5 &
$\sim$0.5  & \tc{blue}{N$^4$LO / NLO} \\ 
$\mathrm{H}\to \uptau^+ \uptau^-/\upmu^+\upmu^-$ & --- & $\sim$0.5
& $\sim$0.5 & \tc{blue}{--- / NLO} \\
$\mathrm{H} \to \mathrm{gg}$ & $\sim$3   &
$\sim$1  &   $\sim$3 & \tc{blue}{N$^3$LO / NLO} \\
$\mathrm{H} \to \upgamma \upgamma$  & <1 & <1    &  $\sim$1  &
\tc{blue}{NLO / NLO} \\
$\mathrm{H} \to \mathrm{Z}\upgamma$  & <1 & $\sim$5   &  $\sim$5 &
\tc{blue}{LO / LO} \\ 
$\mathrm{H} \to \mathrm{WW}/\mathrm{ZZ}\to 4\mathrm{f}$ & <0.5 &   $\sim$0.5
&  $\sim$0.5 & \tc{blue}{NLO/NLO} \\
\hline \hline
\end{tabular}
\end{center}
\end{table}
%==========================

\FloatBarrier % 2020-03-27

\vfill

% \FloatBarrier  from package placeins, gibt alle schwebenden Gleitobjekte aus und setzt danach Seite normal fort
\FloatBarrier

\end{bibunit}

\label{sec-sm-mspira}  
\clearpage \pagestyle{empty}  \cleardoublepage
%============================================
%============================================
\chapter
%[Methods and tools]
{Methods and tools} \label{chmt}
%\input{MTools/mt_intro.tex} %\label{sec-mtintro}
%%%%%%%%%%%%%%%%%%%%%%%%%%%%%%%%%%%%%%%%%%%%%%%%%
%============================================

\pagestyle{fancy}
\fancyhead[CO]{\thechapter.\thesection \hspace{1mm}
Heritage projects, preservation, and re-usability concerns  }
\fancyhead[RO]{}
\fancyhead[LO]{}
\fancyhead[LE]{}
\fancyhead[CE]{}
\fancyhead[RE]{}
\fancyhead[CE]{S.~Banerjee, M.~Chrzaszcz, Z.~Was, J.~Zaremba}
\lfoot[]{}
\cfoot{-  \thepage \hspace*{0.075mm} -}
\rfoot[]{}

%%% definitions %%%%%%%%%%%%%%%%%%%%%%%%%%
\def\beqn{\begin{eqnarray}} \def\eeqn{\end{eqnarray}}
\def\beeq{\begin{eqnarray}}
\def\eeeq{\end{eqnarray}}
\def\nn{\nonumber}
\def\alphas{\alpha_{\rm S}}
    
\begin{bibunit}[elsarticle-num]  
\let\stdthebibliography\thebibliography
\renewcommand{\thebibliography}{%
\let\section\subsection
\stdthebibliography}
%---
\section
[Heritage projects, preservation, and re-usability concerns \\
{\it S. Banerjee, M. Chrzaszcz, Z.~Was, J.~Zaremba}]
{Heritage projects, preservation, and re-usability concerns}
\noindent
{\bf 
Contribution\footnote{This contribution should be cited as:\\
S. Banerjee, M. Chrzaszcz, Z.~Was, J.~Zaremba, Heritage projects, preservation, and re-usability concerns,  
%04 DOI:10.23731/CYRM-2020-XXX.\thepage, in:
%04 \url{http://dx.doi.org/10.23731/CYRM-2020-XXX.\thepage}, in:
DOI: \href{http://dx.doi.org/10.23731/CYRM-2020-003.\thepage}{10.23731/CYRM-2020-003.\thepage}, in:
Theory for the FCC-ee, Eds. A. Blondel, J. Gluza, S. Jadach, P. Janot and T. Riemann,
CERN Yellow Reports: Monographs, CERN-2020-003,
%04 \url{http://dx.doi.org/10.23731/CYRM-2020-XXX}, p. \thepage.} 
DOI: \href{http://dx.doi.org/10.23731/CYRM-2020-003}{10.23731/CYRM-2020-003},
p. \thepage.
\\ \copyright\space CERN, 2020. Published by CERN under the 
%04-2
\href{http://creativecommons.org/licenses/by/4.0/}{Creative Commons Attribution 4.0 license}.} by: S.~Banerjee, M.~Chrzaszcz, Z.~Was, J.~Zaremba
\\
{Corresponding author:
 Z.~Was {[z.was@cern.ch]}}}
\vspace*{.5cm}

\noindent The FCC is a long-term project, novel in many respects, and new calculations,
including simulation programs, will be appearing in the forthcoming years. However, many of the approaches developed for  previous experiments, in particular LEP, will be useful, either directly as a tools or as a means to prepare substantial benchmarks. In addition, programs that will be prepared for Belle II, especially in the domain of $\uptau$, B, and D resonance physics, 
will continue to be valuable tools. Such programs and projects will undoubtedly evolve in the meantime, but one can expect that ready-to-use versions will be available when the need  arises. Then only interfaces will need to be archived solely for the FCC. In some cases, the whole projects will require long-term preservation. Before we will explain some attempts on preservation of some example projects, such as $\uptau$ decays, radiative corrections in decays, or electroweak corrections, let us mention possible general  approaches.

There are many helpful tools  for managing software projects, in both development and preservation. However, preservation-development tools become obsolete and code history, necessary for future extensions and validation, may become lost; therefore, it is important to ensure proper migration from one preservation-development
tool to the next. In addition, very stable solutions belong to repositories beyond an author's responsibility, specifically targeting long-term  preservation.

The CPC International Program Library \cite{CPClib}  serves such a purpose;  CERN web pages like that used for TAUOLA~\cite{wwwTauola}
or PHOTOS~\cite{wwwPhotos} may also offer the
necessary facility.

Issues arise if parts of the code are prepared using automated code development
tools.
If those tools (\ie other programs) are not published, the programs
prepared with their help are of limited help for future applications, 
especially if extensions are needed.

The interfaces between segments of the code can be of a different type.
The reassurance offered by solutions based on some tools is indisputable, 
but can be overshadowed if such a tool evolves over an  unsuitable time. 
We have experienced minor, but at inconvenient moments for our project 
evolution, difficulties as a result of ROOT  library~\cite{Brun:1997pa} upgrades
to new versions. 
Manual intervention on our part and changes to the work routine were 
necessary.\footnote{We had to modify the code for our projects, owing to changes in the ROOT library, first in July
  2002 and  again for the changes introduced with ROOT release 6.} Because many software development projects of phenomenology represent a sizeable fraction 
of the total effort and the people  involved  may often not be immediately available, 
this may represent a major inconvenience.

\subsection{Common tools for all FCC design studies}

As already mentioned, it is of crucial importance to have a common software platform with all the repositories. In the FCC, this effort has begun with the creation of a twiki page \cite{twiki}, where different MC generators are available. This collection should be extended with documentation of programs, links to available original \texttt{git} repositories, etc.
Everybody is welcome to link or put there related codes or results to be used in future software.

The twiki page \cite{twiki} currently includes three sections.
\begin{enumerate}
    \item \textit{FccComputing}, in which  the installation procedure of the FCC-ee software is described.
    \item \textit{FccGenerators},  containing different MC generators. Currently, Tauola, Higgsline, KKMC, and Bhabha generators are presented.
    \item  \textit{FccSoftware}, containing various examples of  simulations
run in the FCC framework.
\end{enumerate}

Next, we will briefly describe currently available generators and discuss their preservation.

\subsubsection{Tauola}
The $\uptau$ decay phenomenology relies to a large degree on experimental data.
This is because of the complexity of experimental analyses and
the difficulty in phenomenology modelling decays where intermediate resonances used
in hadronic currents are broad and perturbative QCD description is only  partly
suitable. Background analysis for multidimensional distributions is a problem.
Collaborations are hesitant to enable outside use
of their matrix element parametrizations, because they may be unsuitable
for other, externally studied, distributions. 
Nevertheless, if they become available, it is worthwhile to store them in publicly available repositories.  
In Ref.~\cite{Golonka:2003xt}, parametrizations developed for ALEPH and CLEO were archived,
together with the original parametrization, useful for technical testing of the Tauola algorithm.
In Ref. \cite{Was:2004dg}, thanks to discussions with the BaBar community,
an extension of Tauola with multichannel capacity that is easy to manipulate by users was prepared. 
The resulting default
parametrization equivalent to that work was archived
in Ref. \cite{Chrzaszcz:2016fte}. In that reference, a framework
for work with C++ currents and for Belle II was prepared.
Hopefully, this may provide a means to feed back code at the FCC. A smooth
transition period for evolution from partly Fortran to fully C++ code
is envisaged in this solution.

\subsubsection{Photos and Tauola Universal interface}

The code for these projects is currently in C++.
Preservation of the up-to-date variants is assured, thanks to CERN special
accounts and web pages
\cite{wwwPhotos,wwwTauola}. Some versions are archived in the CPC
~\cite{Davidson:2010ew,Davidson:2010rw}.
The main issue for the project is the fast evolution of event format
HepMC \cite{Dobbs:2001ck}, and especially  how other projects use that
format to write down generated events. In addition, long-term preservation efforts
may be impaired because of the evolution of configuration and make-file arrangements.  

\subsubsection{EvtGen}

The decays of heavy flavoured hadrons provide huge constrains on Beyond Standard Model physics~\cite{Aebischer:2019mlg}. The FCC-ee is due to run on the Z pole, and will also be a heavy flavour factory. The decays of such mesons and hadrons are modelled with the {\tt EvtGen} package~\cite{wwwEvtGen}. The package consists of various models, which are constantly being updated with the theory predictions, such as form factors and amplitude calculations. Currently, the main developers of {\tt EvtGen} are involved in LHCb collaboration; however, the package is made publicly available~\cite{gitEvtGen} via the git repository. It was recently extended to describe the decays of spin $1/2$ particles. The project is written in C++ and interfaces with the HepMC format~\cite{Dobbs:2001ck}. It is also possible to interface it with the {\tt Tauola}, {\tt Pythia}, and {\tt Photos} packages. 

\subsubsection{Electroweak corrections}

The {\tt KKMC} code is published and archived in Ref.~\cite{Jadach:1999vf}.
Its electroweak correction library, in use until today, is also published and
archived: {\tt Dizet} version {\tt 6.21}~\cite{Bardin:1989tq,Bardin:1999yd}.
At present, only {\tt Dizet} version 6.42 ~\cite{Andonov:2008ga, Akhundov:2013ons} is available for the {\tt KKMC} electroweak sector upgrade.
This version of {\tt Dizet} is missing updates, owing to  the photonic vacuum polarisation, \eg as provided in Refs.~\cite{Burkhardt:2005se,Jegerlehner:2017zsb}.
We could implement the updates ourselves because {\tt Dizet} version 6.42 is well-documented.
However, some versions of {\tt Dizet} that still exist may become
unavailable at a later date. In fact, it was difficult for us to obtain access and we
decided to revert to version 6.42.
This indicates the necessity for  code maintenance, even if
authors may at some time become unavailable.

In any case,  {\tt Dizet} version 6.42 ~\cite{Andonov:2008ga, Akhundov:2013ons}
with updates from  Refs.~\cite{Burkhardt:2005se,Jegerlehner:2017zsb}
is prepared as a facility for the electroweak tables used in {\tt KKMC} \cite{Jadach:1999vf}.

Using tables prepared by one program in another program is not only the method to enhance the speed of the calculation. Interpolation of values enable technical
regularisation of the functions. Technical instabilities at the phase space edges can be regulated.

The tables can be used by other programs that understand the format.
In this way, for example, the  {\tt TauSpinner} package~\cite{Davidson:2010rw, Czyczula:2012ny} can be used for graphical presentation for different variants of 
{\tt Dizet} and of its initialization as a natural continuation of work \cite{Richter-Was:2018lld}
for the LHC or similar activities for the FCC.

This limit is a substantial burden for interfaces. Preservation of  projects
is only partly assured by the CPC publications. The most up-to-date versions
are available at user webpages, which sometimes are not available or may become unavailable.

%\subsection*{References of Rodrigo}

\end{bibunit}

\label{sec-mtools-repo}

\clearpage \pagestyle{empty}  \cleardoublepage

%==============================================

\definecolor{myred}{rgb}{0.75,0.0,0.0}

% \varepsilon => d
% \epsilon    => causal
%
%%PHAN edit new commands for Hypergeometric functions
\newcommand{\Fh}[2]{\,{}_#1F_#2}
\newcommand{\Fs}[3]{\!\!\left[\begin{matrix}#1\,;\\#2\,;\end{matrix}#3\right]}
\newcommand{\Fz}[3]{\Fs{#1}{#2}{#3}}
\newcommand{\DD}{\frac{d}{2}}
\newcommand{\DDD}{\frac{d-1}{2}}
\newcommand{\DDDD}{\frac{d-2}{2}}
\newcommand{\bq}{\begin{equation}}
\newcommand{\eq}{\end{equation}}
\newcommand{\ba}{\begin{eqnarray}}
\newcommand{\ea}{\end{eqnarray}}
\newcommand{\ban}{\begin{eqnarray*}}
\newcommand{\ean}{\end{eqnarray*}}
% from cernth6590/92
\newcommand{\z}{$Z$}
\newcommand{\q}{$Q^2\:$}
\newcommand{\swtwo}{$ \sin^{2} \theta_{W} \:$}
\newcommand{\w}{$W^{\pm}$}
\newcommand{\zp}{$Z'$}
\newcommand{\ee}{$e^+e^-\ $}
\newcommand{\oalf}{${\cal O }(\alpha$)$\:$}
\newcommand{\ff}{$f^+f^-\ $}
\newcommand{\mumu}{$\mu^+ \mu^-\:$}
\newcommand{\nobody}{\rule{0ex}{1ex}}
%newcommand{\nn}{\nobody \hfill \\ \noindent }
%newnewcommands
%%\newcommand{\nn}{\nonumber \\}
%\newcommand{\zf}{{\tt ZFITTER}}
%00000000
\newcommand{\az}{A_0}
\newcommand{\ao}{A_1}
\newcommand{\apol}{A_{pol}}
\newcommand{\alr}{A_{LR}}
\newcommand{\swt}{\sin^2 \theta_W}
\newcommand{\rgt}{r_{\gamma}^T}
\newcommand{\rt}{R_T}
\newcommand{\rfb}{R_{FB}}
\newcommand{\rpol}{R_{pol}}
\newcommand{\ra}{R_A}
\newcommand{\rza}{r_0^A}
\newcommand{\rzt}{r_0^T}
\newcommand{\ia}{J_A}
\newcommand{\roa}{r_1^A}
\newcommand{\rot}{r_1^T}
\newcommand{\iT}{J_T}
 
\newcommand{\st}{\sigma_T}
\newcommand{\sfb}{\sigma_{FB}}
\newcommand{\slr}{\sigma_{LR}}
\newcommand{\spol}{\sigma_{pol}}

\pagestyle{fancy}
\fancyhead[CO]{\thechapter.\thesection \hspace{1mm} Scalar one-loop Feynman integrals in arbitrary space--time
dimension $d$---an update}
\fancyhead[RO]{}
\fancyhead[LO]{}
\fancyhead[LE]{}
\fancyhead[CE]{}
\fancyhead[RE]{}
\fancyhead[CE]{T. Riemann, J. Usovitsch}
\lfoot[]{}
\cfoot{-  \thepage \hspace*{0.075mm} -}
\rfoot[]{}
    
\begin{bibunit}[elsarticle-num] % define the bib-style for the unit: elsarticle-num.bst
%  text-1; this is the corresponding section
%\putbib[2loops] % the *.bib
%\end{bibunit}
% go-on
%--- from: bibunits.sty, adapts the font size of ``References'' to section
\let\stdthebibliography\thebibliography
\renewcommand{\thebibliography}{%
\let\section\subsection
\stdthebibliography}
%---
\section
% 2020-03-13 % 2020-03-13
[Scalar one-loop Feynman integrals in arbitrary space--time
dimension $d$ -- an update
\\ {\it T. Riemann, J. Usovitsch}]
{Scalar one-loop Feynman integrals in arbitrary space--time
dimension $d$  {--}  an update
\label{contr:yourcontr}}
\noindent
{\bf Contribution\footnote{This contribution should be cited as:\\
T. Riemann, J. Usovitsch, Scalar one-loop Feynman integrals in arbitrary space-time
dimension $d$ {--} an update,  
%04 DOI:10.23731/CYRM-2020-XXX.\thepage, in:
%04 \url{http://dx.doi.org/10.23731/CYRM-2020-XXX.\thepage}, in:
DOI: \href{http://dx.doi.org/10.23731/CYRM-2020-003.\thepage}{10.23731/CYRM-2020-003.\thepage}, in:
Theory for the FCC-ee, Eds. A. Blondel, J. Gluza, S. Jadach, P. Janot and T. Riemann,\\
CERN Yellow Reports: Monographs, CERN-2020-003,
%04 \url{http://dx.doi.org/10.23731/CYRM-2020-XXX}, p. \thepage.} 
DOI: \href{http://dx.doi.org/10.23731/CYRM-2020-003}{10.23731/CYRM-2020-003},
p. \thepage.
\\ \copyright\space CERN, 2020. Published by CERN under the 
%04-2
\href{http://creativecommons.org/licenses/by/4.0/}{Creative Commons Attribution 4.0 license}.} by: T. Riemann, J. Usovitsch 
\\
Corresponding author: T. Riemann {[tordriemann@gmail.com]}}
\vspace*{.5cm}
  
%=================================================================
\subsection{Introduction}
%=================================================================
The study and use of analyticity of scattering amplitudes was founded by 
R. Eden, P. Landshoff, D. Olive, and J. Polkinghorn in their famous book \textit{The Analytic S-Matrix} in 1966 \cite{Eden:1966AA}.
Indeed, as early as  1969, J. Schwinger quoted:
``One of the most remarkable discoveries in elementary particle physics has been that of the complex plane, [\dots] the theory of functions of complex variables plays the role not of a mathematical tool, but of a fundamental description of nature inseparable from physics.''
\cite{Schwinger:1970xc}.

It took many years to make  use of analyticity and unitarity, together with renormalizability and gauge invariance of quantum field theory, as a practical tool for the calculation of cross-sections at real colliders.
When the analysis of LEP~1 data, around 1989, was prepared, it became evident that the S-matrix language helps to efficiently sort the various perturbative contributions of the Standard Model.

The scattering amplitude for the reaction $\mathrm{e}^+\mathrm{e}^- \to (\mathrm{Z},\upgamma) \to \mathrm{f {\bar f}}$ at LEP energies depends on two variables, $s$ and $\cos\theta$, and the integrated cross-section may be described by an analytical function of $s$ with a simple pole, describing mass and width of the Z resonance:
\begin{equation}\label{eq-zres}
A = \frac{R}{s-M_\mathrm{Z}^2+ \mathrm{i}M_\mathrm{Z} \Gamma_\mathrm{Z}} + \sum_{i=0}^\infty a_i \left (s-M_\mathrm{Z}^2+ \mathrm{i}M_\mathrm{Z} \Gamma_\mathrm{Z}
\right )^i
.
\end{equation}
Here, position $s_0=M_\mathrm{Z}^2- \mathrm{i}M_\mathrm{Z} \Gamma_\mathrm{Z}$ and residue $R$ of the pole, as well as the background expansion, are of interest.
The analytic form of \Eref{eq-zres} must be respected when deriving a Z amplitude at multiloop accuracy; see Ref. \cite{Blondel:2018mad} and the references therein. 

Shortly after the work by Eden \textit{et al.} \cite{Eden:1966AA}, physical amplitudes were also proposed for consideration  as complex functions of space--time dimension $d$ (dimensional regularisation) \cite{Bollini:1972ui,
tHooft:1972tcz}.

\begin{figure}
\center 
\includegraphics[angle=0,scale=0.125]{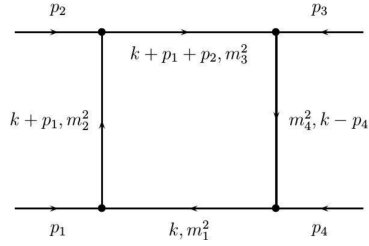}
\caption{\label{fig-1loopFI}
One-loop Feynman integral}
\end{figure}

In perturbative calculations with dimensional regularisation, Feynman integrals $I$ are complex functions of the space--time dimension $d=4-2\varepsilon$.
In fact, they are {meromorphic} functions of $d$ 
and may be expanded in {Laurent series} around poles at, \eg $d_{s}=4+2N_0, N_0\geq0$.
Let $J_n$ be an $n$-point one-loop Feynman integral, as shown in  \Fref{fig-1loopFI}:
\begin{eqnarray}\label{npoint-3}
J_n
\equiv 
J_{n} \left (d; \{p_ip_j\}, \left \{m_i^2 \right \} \right )
 =
    \int \dfrac{\mathrm{d}^d k}{\mathrm{i} \uppi^{d/2}} \dfrac{1}{D_1^{\nu_1} D_2^{\nu_2}\cdots D_n^{\nu_n}}
\end{eqnarray}
with 
\begin{equation}\label{npoint-propagators-1}
 {D_i} = \frac{1}{(k+q_i)^2-m_i^2+ \mathrm{i}\epsilon }
\end{equation}
% for $i, j=1,2,\cdots, N$ which are inverse Feynman propagators;
and
\begin{equation} 
\nu = \sum_{i=1}^n \nu_i,
\qquad 
\sum_{e=1}^n p_e =0
.
\end{equation}

The Feynman integrals are analytical functions of $d$ everywhere with exclusion of isolated singular points $d_s$, where they behave not worse than 
\begin{equation}
 \frac{A_s}{(d-d_s)^{N_s}}
.
 \end{equation}

In physics applications, we need the Feynman integrals at a potentially singular point, $d=4$, so that their general behaviour  at non-singular points is not in the original focus.  
Nevertheless, the question arises:
\begin{quotation} 
{\it Can we determine the general $d$-dependence of a Feynman integral?}
\end{quotation}

For one-loop integrals, the question has been answered  recently, in 
Ref. \cite{Phan:2018cnz}.

At the beginning of systematic cross-section calculations in $d$ dimensions
came two seminal papers on one-loop Feynman integrals in dimensional regularisation \cite{Passarino:1978jh, 'tHooft:1978xw}.
Later, many improvements and generalisations were introduced in various respects.

We see several reasons to study the $d$-dependence of one-loop Feynman integrals and will discuss them briefly in the next subsection.

%%%%%%%%%%%%%%%%%%%%%%%%%%%%%%%%%%%%%%%%%%%%%%%%%%%%%%%%%%%%%%
\subsection{Interests in the $d$-dependence of one-loop Feynman integrals}
%%%%%%%%%%%%%%%%%%%%%%%%%%%%%%%%%%%%%%%%%%%%%%%%%%%%%%%%%%%%%%

%%%%%%%%%%%%%%%%%%%%%%%%%%%%%%%%%%%%%%%%%%%%%%%%%%%%%%%%%%%%%%
\subsubsection{Interest from mathematical physics}
%%%%%%%%%%%%%%%%%%%%%%%%%%%%%%%%%%%%%%%%%%%%%%%%%%%%%%%%%%%%%%
There is a general interest in  the Feynman integrals as meromorphic functions of space--time dimension $d$; the easiest case is that of one loop.
Early attempts, for the massless case, trace back to Boos and Davydychev
\cite{Boos:1987bg}.
The general one-loop integrals were tackled systematically by Tarasov \textit{et al.} since the  1990s; see, \eg\ Refs. 
\cite{Tarasov:1996br,Tarasov:1997kx,Tarasov:1998nx,Fleischer:1999hq}  and references therein.
In 
Refs. \cite{Tarasov:2000sf,Fleischer:2003rm}, the class of generalised hypergeometric functions for massive one-loop Feynman integrals with unit indices was determined and studied with a novel approach based on dimensional difference equations.
\begin{itemize}
\item[(a)]
{ $_2F_1$ Gauss} hypergeometric functions are needed for self-energies.
\item[(b)]
{ $F_1$ Appell} functions are needed for vertices.
\item[(c)] 
{ $F_S$ Lauricella--Saran} functions are needed for boxes.
\end{itemize}
Finally, the correct, general massive one-loop one- to four-point functions with unit indices at arbitrary kinematics were determined by  Phan and Riemann \cite{Phan:2018cnz}, who also calculated the numerics of the generalised hypergeometric functions.

%%%%%%%%%%%%%%%%%%%%%%%%%%%%%%%%%%%%%%%%%%%%%%%%%%%%%%%%%%%%%%
\subsubsection{Interest from tensor reductions of $n$-point functions in higher space--time dimensions}
%%%%%%%%%%%%%%%%%%%%%%%%%%%%%%%%%%%%%%%%%%%%%%%%%%%%%%%%%%%%%%
For many-particle calculations, inverse Gram determinants $1/G(p_i)$ from tensor reductions appear at certain kinematic configurations $p_i$.
These terms {$1/G(p_i)$} may diverge, because 
Gram determinants  can exactly vanish: {$G(p_i)\equiv 0$}.
{One may perform tensor reductions so that no inverse Gram determinants appear.
But then one has to calculate scalar one-loop integrals in higher 
dimensions,  
{$D=4+2n-2\epsilon, n>0$} \cite{Davydychev:1991va,Fleischer:2010sq}}.
In fact, one introduces new scalar integrals 
%in higher dimensions and with higher indices 
\cite{Davydychev:1991va}.
Let us take as an example a rank-5 tensor of an $n$-point function:
\begin{align}
\label{tensor5}
I_{n}^{\mu  \nu  \lambda  \rho  \sigma}
&=
\int \frac{\mathrm{d}^d k}{\mathrm{i}\pi^{d/2}} \frac{ k^{\mu}   k^{\nu}   k^{\lambda} k^{\rho}  k^{\sigma}}
{\prod_{j=1}^{n}   {c_j}}
\nn \\
&=
-    \sum_{i,j,k,l,m=1}^{n}   q_i^{\mu}  q_j^{\nu}  q_k^{\lambda}
 q_l^{\rho}  q_m^{\sigma}   n_{ijklm}   \textcolor{black}{I_{n,ijklm}^{[d+]^5}}
    + \frac{1}{2} \sum_{i,j,k=1}^{n} g^{[\mu \nu} q_i^{\lambda} q_j^{\rho} q_k^{\sigma]}
  n_{ijk}   \textcolor{black}{I_{n,ijk}^{[d+]^4}}
% 2020-03-13
  \nn \\
  &~~~~
 - ~\frac{1}{4} \sum_{i=1}^{n} g^{[\mu \nu} g^{\lambda \rho}  q_i^{\sigma]}   {I_{n,i}^{[d+]^3}}
.
\end{align}
The integrals $I_{n,ab\ldots}^{[d+]^l}$ are special cases of $I_{n,ab\ldots}^{[d+]^l,s}$,   defined in {$[d+]^l=4 -2 \varepsilon + 2 l$} dimensions,  by
shrinking line $s$
and raising the powers of propagators (indices) $a,b,\ldots$

At this step, the tensor integral is represented by scalar integrals with higher space--time dimensions and higher propagator powers.
The publicly available Feynman integral libraries deliver, though, ordinary scalar integrals in $d=4-2\varepsilon$ dimensions, with unit propagator powers.
With the usual integration by parts reduction technique 
\cite{Tkachov:1981wb,Chetyrkin:1981qh}, one may shift indices, \ie reduce  propagator powers to unity:
\begin{equation}\label{eq-007}
%--------------------------------------------------------
 \nu_j   {\bf j^+} I_5
=% &=&
\frac{1}{{\color{black!   80!black}{{0\choose 0}_5}}}
  \sum^{5}_{k=1} {0j\choose 0k}_5
\left[ d - \sum_{i=1}^{5} \nu_i(   {{\bf k^-}}  { {\bf i^+}}+1)
             \right]   I_5
             .
\end{equation}
The operators   ${\bf  i^{\pm}, j^{\pm}, k^{\pm} }$ act by shifting
the indices   {$\nu_i, \nu_j, \nu_k$} by $\pm 1$.

After this step, one has yet to deal with scalar functions in $d=4-2\varepsilon+2l$ dimensions.
This may be further reduced by applying dimensional reduction formulae invented by Tarasov  \cite{Tarasov:1996br,Fleischer:1999hq}:
shift of dimension and index,\begin{equation}
\label{eq:RR1}
 \nu_j    \left ( {\bf j^+} I_5^{{\color{black!   80!black} [d+]}} \right
 )
=%&=&
\frac{1}{{\color{black!   80!black}  \left( \right)_5}}
%&& \nonumber \\ &&
\left[  - {{j \choose 0}_5} +\sum_{k=1}^{5} {j \choose k}_5
   {{\bf k^-} }\right]   {I_5 }
,
\end{equation}
and  shift only of dimension,\begin{equation} % \\ 
%---------------------------------------------------------
\label{eq:RR2}  %   {\color{black!   80!black}  }
 \left  (d-\sum_{i=1}^{5}\nu_i+1 \right )      I_5^{{\color{black!   80!black} [d+]}}
=%  &=&
\frac{1}{ {\color{black!   80!black}{\left( \right)_5}}}
  \left[ {{0 \choose 0}_5}
 - \sum_{k=1}^5 {0 \choose k}_5 {{\bf k^-}} \right]   I_5
 .
\end{equation}
The procedure is elegant, but it introduces inverse powers of potentially vanishing Gram determinants  in both cases.
As a consequence, one has finally to treat  the numerical implications in sophisticated ways.

At this stage, one might try an alternative.
Perform the reductions of tensor functions to scalar functions with unit indices, but allowing for the use of higher space--time dimensions.
This avoids the vanishing inverse Gram problem, but introduces the need of a library of scalar Feynman integrals in higher dimensions.
This idea makes it attractive to derive an algorithm allowing the systematic calculation of scalar one- to $n$-point functions in arbitrary dimensions, and to implement a numerical solution for it.

To be a little more definite, we quote here some unpublished formulae from Refs.
\cite{Fleischer:2010sq,Riemann:september2013}.
The following reduction of a five-point tensor
in terms of tensor coefficients $E_{ijklm}^s$, with line $s$ skipped from the five-point integral, may be used as a starting point: 
\bea\label{compl5a}
I_{5}^{\mu\, \nu\, \lambda \rho \sigma} &=&  
\sum_{s=1}^5 \Biggl[
\sum_{i,j,k,l,m=1}^{5}  q_i^{\mu} q_j^{\nu}  q_k^{\lambda} q_l^{\rho} q_m^{\sigma}
{E_{ijklm}^s}
+
\sum_{i,j,k=1}^5 g^{[\mu \nu} q_i^{\lambda} q_j^{\rho} q_k^{\sigma]} E_{00ijk}^s
%\nonumber \\ &&
+~
\sum_{i=1}^5
g^{[\mu \nu} g^{\lambda \rho} q_i^{\sigma]} E_{0000i}^s \Biggr]
.
\nl
\label{compl5}
\eea

The tensor coefficients  $E_{ijklm}^s$
are expressed in terms of integrals $I_{4,i\cdots}^{[d+]^l,s}$, e.g.: 
% according to (I.4.62):
\bea
\textcolor{black}{E_{ijklm}^s} &=&
-\frac{1}{{0\choose 0}_5}  \Biggl\{ \left[
{0l\choose sm}_5  
{ {n}_{ijk}I_{4,ijk}^{[d+]^4,s}}
+(i \leftrightarrow l)+(j \leftrightarrow l)+
(k \leftrightarrow l)\right]
%\nonumber \\ &&
+~{0s\choose 0m}_5   \textcolor{black}{{n}_{ijkl} I_{4,ijkl}^{[d+]^4,s}} \Biggr\} .
\nl
\label{Ewxyz5}
\eea
No factors $1/G_5 = {\choose }_5$ appear.
Now, in a next step, one may avoid the appearance of inverse sub-Gram determinants $()_4$.
Further, the complete dependence on the indices $i$   of the tensor coefficients can be shifted into the integral's pre-factors with signed minors. 
One can say that \textcolor{black}{the indices \emph{decouple} from the integrals}.
As an example, we reproduce the four-point part of $\textcolor{black}{ I_{4,ijkl}^{[d+]^4}}$:
\begin{align}
\textcolor{black}{{n}_{ijkl} I_{4,ijkl}^{[d+]^4}} &=
\frac{{0\choose i}}{{0\choose 0}}
\frac{{0\choose j}}{{0\choose 0}}\frac{{0\choose k}}{{0\choose 0}}\frac{{0\choose l}}{{0\choose 0}}
d(d+1)(d+2)(d+3)\textcolor{black}{I_4^{[d+]^4} }
\nn 
\\
& \quad + \frac{{0i\choose 0j}{0\choose k}{0\choose l}+{0i\choose 0k}{0\choose j}{0\choose l}+{0j\choose 0k}{0\choose i}{0\choose l}+
{0i\choose 0l}{0\choose j}{0\choose k}+{0j\choose 0l}{0\choose i}{0\choose k}+{0k\choose 0l}{0\choose i}{0\choose j}}
{{0\choose 0}^3}  
% 2020-03-13
\nn \\
& \quad \times ~ d(d+1)\textcolor{black}{I_4^{[d+]^3}}
% 2020-03-13  \nn \\ & \quad 
+ \frac{{0i\choose 0l}{0j\choose 0k}+{0j\choose 0l}{0i\choose 0k}+{0k\choose 0l}{0i\choose 0j}}
{{0\choose 0}^2}\textcolor{black}{I_4^{[d+]^2} }
+ \cdots
\label{fulld4}
\end{align}
In \Eref{fulld4}, one has to understand the four-point integrals to carry the corresponding index $s$ of \Eref{compl5a} and that the
signed minors are ${0\choose k}\to {0s\choose ks}_5$, etc.
We arrived at:
\begin{itemize}
%$\checkmark$ 
\item[(a)] no scalar five-point integrals in higher dimensions;
%\\
%$\checkmark$ 
\item[(b)] no inverse Gram determinants $()_5$;
%\\
%$\checkmark$ 
\item[(c)] \textcolor{black}{ four-point integrals  without indices;}
%\\
%$\dag$ 
\item[(d)] scalar four-point integrals in higher dimensions appearing as \textcolor{black}{$ I_{4}^{[d+]^2,s}$}, etc.;
%\\
%$\dag$ 
\item[(e)] inverse four-point Gram determinants \textcolor{black}{${0\choose 0}_5 \equiv ()_4$}.
\end{itemize}
%%%%%%%%%%%%%%%%%%%%%%%%%%%%%%%%%%%%%%%%%%%%%%%%%%%%%%%%%%%%%%%%%%
\subsubsection{Interest from multiloop calculations}
%%%%%%%%%%%%%%%%%%%%%%%%%%%%%%%%%%%%%%%%%%%%%%%%%%%%%%%%%%%%%%%%%%
Higher-order loop calculations need higher-order contributions from $\epsilon$-expansions of one-loop terms, typically stemming from the expansions
\ba
\frac{1}{d-4} = -\frac{1}{2\epsilon}
\ea
and
\ba
\Gamma(\epsilon) = \frac{a_1}{\epsilon} + a_0 + a_1 \epsilon + \cdots 
\ea
A seminal paper on the $\varepsilon$-terms of one-loop functions is
Ref. \cite{Nierste:1992wg}.
    A general analytical solution of the problem of determining the general 
    $\varepsilon$-expansion of Feynman integrals is unsolved so far, even for the one-loop case, although see Refs. \cite{Passarino:2001wv,Ferroglia:2002yr,Ferroglia:2003wc,Ferroglia:2002mz}.
  The determination of one-loop Feynman integrals as meromorphic functions of $d$ might be a useful preparatory step for determining the pole expansion in $d$ around, \eg $d=4$.

  %%%%%%%%%%%%%%%%%%%%%%%%%%%%%%%%%%%%%%%%%%%%%%%%%%%%%%%%%%%%%%%%%%
  \subsubsection{Interest from Mellin--Barnes representations}
%%%%%%%%%%%%%%%%%%%%%%%%%%%%%%%%%%%%%%%%%%%%%%%%%%%%%%%%%%%%%%%%%%
A powerful approach to arbitrary Feynman integrals is based on Mellin--Barnes representations \cite{Smirnov:1999gc,Tausk:1999vh}.
One-loop integrals with variable, in general non-integer, indices are needed in the context of the {\it loop-by-loop Mellin--Barnes approach} to multiloop integrals.
Details may be found in the literature on the Mathematica package AMBRE 
\cite{Gluza:2007rt,Gluza:2009mj,Gluza:2010v22,Gluza:2010mz,%
Dubovyk:2015yba,Dubovyk:2016ocz,Usovitsch2018Numerical,%
Dubovyk19}, and in references therein.

{A crucial technical problem of the Mellin--Barnes representations arises from the increasing number of dimensions of these representations with an
increasing number of physical scales.}
We will detail this in \Sref{sec-mbdim}.
Thus, there is an unresolved need for low-dimensional one-loop Mellin--Barnes
(MB) integrals, with arbitrary indices. 

%
% %%%%%%%%%%%%%%%%%%%%%%%%%%%%%%%%%%%%%%%%%%%%%%%%%%%%%%%%%%%%%%%%%%%%%%%%%%
%%\begin{frame}

%

%%%%%%%%%%%%%%%%%%%%%%%%%%%%%%%%%%%%%%%%%%%%%%%%%%%%%%%%%%%%%%%%%%%%%%%
\subsection
[Mellin--Barnes representations for one-loop Feynman integrals]
{Mellin--Barnes representations for one-loop Feynman integrals} 
Two numerical MB approaches are advocated. 
\subsubsection
[AMBRE]
{AMBRE \label{sec-mbdim}}
%%%%%%%%%%%%%%%%%%%%%%%%%%%%%%%%%%%%%%%%%%%%%%%%%%%%%%%%%%%%%%%%%%%
There are several ways to take advantage of Mellin--Barnes representations for the calculation of Feynman integrals. 
One approach is the replacement of massive propagators by Mellin--Barnes integrals over massless propagators, invented by Usyukina  \cite{Usyukina:1975yg}.
%,Smirnov:1999gc,Tausk:1999vh}.
Another approach transforms the Feynman parameter representation with Mellin--Barnes representations into a number of complex path integrals, invented in 1999 by Smirnov for planar diagrams \cite{Smirnov:1999gc} and  Tausk for non-planar diagrams \cite{Tausk:1999vh}.
This approach  `automatically' implies a general solution of the infrared problem and has been worked out in the AMBRE approach 
\cite{Gluza:2007rt,Kajda-phd-thesis:2009,Dubovyk:2015yba,Usovitsch2018Numerical,%
Dubovyk19}.

%%%%%%new input copied from MBnumerics contribution%%%%

The general definitions for a multiloop Feynman integral are
\begin{eqnarray}\label{npoint-1}
J_n^L
\equiv 
J_{n}^{L} \left (d; \{p_ip_j\}, \{m_i^2\} \right)
 =
    \int\prod_{j=1}^L \dfrac{\mathrm{d}^d k_{j}}{\mathrm{i} \uppi^{d/2}} \dfrac{1}{D_1^{\nu_1} D_2^{\nu_2}\cdots D_n^{\nu_n}}
\end{eqnarray}
with 
\begin{equation}
D_{i}
% =q_{i}^2-m_{i}^{2}+i\delta
=\left(\sum\limits_{l=1}^{L}a_{il}k_{l}+\sum\limits_{e=1}^{E}b_{ie}p_{e}\right)^{2}-m_{i}^{2}+\mathrm{i}\delta,
\qquad a_{il},\;b_{ie}\in\{-1,0,1\},
\label{eq:propagator}
\end{equation}
where $m_i$ are the masses, $p_e$ the external momenta, $k_l$ the loop momenta,  $\mathrm{i}\delta$ the Feynman prescription, and, finally, $\nu_i$ the complex variables.

With the following Feynman trick, we get a really neat parametric representation:
\begin{equation}
 \frac{(-1)^{\nu}}{\prod\limits_{j=1}^{n} \left (-D_{j}^{\nu_{j}} \right
 )}=\frac{(-1)^{\nu}\Gamma(\nu)
 \left(\prod\limits_{j=1}^{n}
 \int\limits_{\{x_{j}\geq0\}}\frac{\mathrm{d} x_{j}\,x_{j}^{\nu_{j}-1}}{\Gamma(\nu_{j})}
 \right)
 \delta \left (1-{\sum\limits_{j=1}^{n}}x_{j} \right )}{( -k_{l}^{\mu}M_{ll'}k_{l'\mu}+2k_{l}^{\mu}Q_{l\mu}+J-i\delta)^{\nu}},
 \qquad \nu=\sum\limits_{j=1}^{n}\nu_{j},
 \label{eq:FeynmanIntroduction}
\end{equation}
where
\begin{equation}
M_{ll'}=\sum\limits_{j=1}^{n}a_{jl}a_{jl'}x_{j}
\label{eq:matrixM}
\end{equation}
is an $L\times L$ symmetric matrix, 
\begin{equation}
Q_{l}^{\nu}=-\sum\limits_{j=1}^{n}x_{j}a_{jl}\sum\limits_{e=1}^{E}b_{je}p_{e}^{\nu}
\label{eq:vectorQ}
\end{equation}
is a vector with $L$ components, and 
\begin{equation}
J=-\sum\limits_{j=1}^{n}x_{j}\left(\sum\limits_{e=1}^{E}b_{je}p_{e}^{\mu}\sum\limits_{e'=1}^{E}p_{e'}^{\nu}b_{je'}g_{\mu\nu}-m_{j}^{2}\right),
\label{eq:scalarJ}
\end{equation}
where $x_{j}$ are the Feynman parameters introduced using the Feynman trick. The metric tensor is $g_{\mu\nu}= \mathrm{diag}(1,-1,\dots,-1)$.

The Feynman integral can now be written in the Feynman parameter integral representation:
\begin{equation}
J_{n}^{L}=(-1)^{\nu}\Gamma(\nu-LD/2)
% \left(\prod\limits_{j=1}^{N}\int\limits_{x_{j}\geq 0}{\mathrm d} x_{j}\right)
\left(\prod\limits_{j=1}^{n}\int\limits_{\{x_{j}\geq0\}}\frac{\mathrm{d} x_{j}\,x_{j}^{\nu_{j}-1}}{\Gamma(\nu_{j})}\right)
% \int\limits_{\{x_{i}\geq0\}}\left(\prod\limits_{j=1}^{N_{G}}\frac{{\mathrm d} x_{j}\,x_{j}^{\nu_{j}-1}}{\Gamma(\nu_{j})}\right)
\delta \left (1-\sum\limits_{j=1}^{n}x_{j} \right )\frac{U(x)^{\nu-(L+1)D/2}}{F(x)^{\nu-LD/2}},
\label{eq:finalFeynman}
\end{equation}
where
\begin{align}
 U(x)&= \det M,
 \label{eq:Upolynom}\\
 F(x)&= U(x) \left (Q_{l}^{\mu}M_{ll'}^{-1}Q_{l'\mu}+J- \mathrm{i}\delta \right).
 \label{eq:Fpolynom}
\end{align}
From these definitions, it follows that the functions $F(x)$ and $U(x)$ are homogeneous in the Feynman parameters $x_{i}$. The function $U(x)$ is of degree $L$ and the function $F(x)$ is of degree $L+1$. The functions $U(x)$ and $F(x)$ are also known as Symanzik polynomials.
%%%%%%

At one-loop level, the definition of the Feynman integral simplifies drastically and gives many insights straight away, which we will bring to light in this work:
\begin{eqnarray}\label{npoint-2}
J_n 
\equiv 
J_{n} \left (d; \{p_ip_j\}, \{m_i^2\} \right )
 =
    \int \dfrac{\mathrm{d}^d k}{\mathrm{i} \uppi^{d/2}} \dfrac{1}{D_1^{\nu_1} D_2^{\nu_2}\cdots D_n^{\nu_n}}
\end{eqnarray}
with propagators depending only on one-loop momenta:
\begin{eqnarray}\label{npoint-propagators-2}
 {D_i} = \frac{1}{(k+q_i)^2-m_i^2+\mathrm{i}\epsilon },
\end{eqnarray}
with
\begin{eqnarray}\label{npoint-propagators-3}
 q_i=\sum\limits_{e=1}^{i}p_e.
 \end{eqnarray}
We assume here, for brevity,
\begin{eqnarray}%\begin{eqnarray} 
\nu_i = 1,
\qquad 
\sum_{e=1}^n p_e =0
.
\end{eqnarray}%\end{eqnarray}
%The $p_e$ are external momenta and $q_i$ internal ones.
If we take the argument of the Dirac delta function to be $1-\sum_{j=1}^n x_j$, the Feynman parameter representation for one-loop Feynman integrals simplifies to
\begin{eqnarray}\label{bh-1}
J_n
&=& 
(-1)^{n}
{\Gamma\left(n -d/2\right)}
   \int_0^1 \prod_{i=1}^n \mathrm{d}x_i
\delta \left( 1-\sum_{j=1}^n x_j \right)
\frac { 1 } {{F_n(x)}^{n-d/2}}
.
\end{eqnarray}
%The general case for $L$ loops and arbitrary indices $\nu_i$:
%% /riemann-capp-2017-hamburg.pdf
%\begin{eqnarray}\label{bh-1L}
%J_n
%&\to& 
%(-1)^{N_\nu}
%\frac{ {\Gamma\left(N_\nu -dL/2 \right)}} {\prod_{i=1}^n \Gamma(\nu_i)}
%   \int_0^1 \prod_{i=1}^n dx_i
%   x_i^{\nu_i-1}
%\delta \left( 1-\sum_{j=1}^n x_j \right)
%\frac { U_n(x)^{N_\nu-d(L+1)/2} } {{F_n(x)}^{N_\nu-dL/2}}
%,
%\end{eqnarray}
%with $N_\nu = \sum \nu_i$. 
%\\
Here, the $F$ function is the second Symanzik polynomial, which is just of second degree in the Feynman parameters:
\begin{eqnarray}\label{eq-2ndsy}
{F_n(x)} =  
%-(\sum_{i} x_i)~J +Q^2 
%~=~
\frac{1}{2} 
\sum_{i,j} x_i {Y_{ij}} x_j-i\epsilon
.
\end{eqnarray}
The {$Y_{ij}$} are elements of the {Cayley matrix}
$Y=(Y_{ij})$, 
\begin{eqnarray}\label{eq-yij}
{Y_{ij}} = Y_{ji} = m_i^2+m_j^2-(q_i-q_j)^2 . 
\end{eqnarray}
Gram and Cayley determinants were introduced by Melrose \cite{Melrose:1965kb}; see also Ref. \cite{Fleischer:1999hq}. 
The 
$(n-1)\times(n-1)$-dimensional {Gram determinant} $G_n \equiv G_{12\cdots n}$ is
%-----------------------------------------------
\begin{eqnarray} \label{Gram}
G_n % \equiv G_{12 \cdots n}
= 
- 
\left|
\begin{matrix}
  \! (q_1-q_n)^2 & (q_1-q_n)(q_2-q_n)   &\ldots & (q_1-q_n)(q_{n-1}-q_n) \\
  \! (q_1-q_n)(q_2-q_n)  &  (q_2-q_n)^2 &\ldots & (q_2-q_n)(q_{n-1}-q_n) \\
  \vdots  & \vdots  &\ddots   & \vdots \\
  \! (q_1-q_n)(q_{n-1}-q_n)  & (q_2-q_n)(q_{n-1}-q_n)  &\ldots & (q_{n-1}-q_n)^2
\end{matrix}
\right|. 
\end{eqnarray} 
%-----------------------------------------------------------------------------
The $2^n G_n$ equals, notationally, the $G_{n-1}$ of Ref. \cite{Fleischer:1999hq}.
Evidently, the Gram determinant {$G_n$} is independent of the propagator masses.

The Cayley determinant $\Delta_n = \lambda_{12\cdots n}$ is composed of the 
$Y_{ij}$ introduced in \Eref{eq-yij}: 
\begin{eqnarray} \label{npoint-cayley}
{\bf \mathrm{Cayley\;determinant:}}\;\;\;\;\Delta_n = \lambda_n  \equiv \lambda_{12 \cdots n}
=  
\left|
\begin{matrix}
Y_{11}  & Y_{12}  &\ldots & Y_{1n} \\
Y_{12}  & Y_{22}  &\ldots & Y_{2n} \\
\vdots  & \vdots  &\ddots & \vdots \\
Y_{1n}  & Y_{2n}  &\ldots & Y_{nn}
\end{matrix}
         \right|
.
\end{eqnarray}
We also define the 
modified Cayley determinant
\begin{eqnarray} \label{mod-npoint-cayley}
{\bf \mathrm{modified\;Cayley\;determinant:}}\;\;\;\;()_n 
=  
\left|
\begin{matrix}
0 & 1       & 1       &\ldots & 1      \\
1 & Y_{11}  & Y_{12}  &\ldots & Y_{1n} \\
1 & Y_{12}  & Y_{22}  &\ldots & Y_{2n} \\
\vdots  & \vdots  & \vdots  &\ddots & \vdots \\
1 & Y_{1n}  & Y_{2n}  &\ldots & Y_{nn}
\end{matrix}
         \right|
.
\end{eqnarray}
The determinants $\Delta_n$, $()_n$, and $G_n$ are evidently independent of a common shifting of the momenta $q_i$.

One may use Mellin--Barnes integrals \cite{Barnes:1908},
\begin{eqnarray} \label{MB}
\frac{1}{(1 + z)^{\lambda}} 
 =  
\frac{1}{2\uppi \mathrm{i}} \int\limits_{-\mathrm{i}\infty}^{+\mathrm{i}\infty}\mathrm{d}s \;
    \frac{\Gamma(-s)\;\Gamma(\lambda +s)}{\Gamma(\lambda) } 
    \; z^s
%\nonumber \\ &=&  
 = 
%TR 
\Fh21\Fz{\lambda,b}{b}{-z}
,
\end{eqnarray} 
 to split the sum $F_n(x)$ in \Eref{eq-2ndsy} into a product, enabling nested MB integrals  to be calculated. 
For some mathematics behind the derivation, see the corollary at p. 289  in Ref. \cite{Whittaker:1927}.
Equation \eqref{MB} is valid if $|\mathrm{Arg}(z)|<\uppi$.
The 
integration contour must be chosen such that the poles of $\Gamma(- s)$ and $\Gamma(\lambda + s)$ are 
well-separated. 
The right-hand side of \Eref{MB} is identified as Gauss's hypergeometric function.
%%%%%%%%%%%%%%%%%%%%%%%%%%%%%%%%%%%%%%%%%%%%%%%%%%%%%%%%%%%%%%%%%%%%%%%%%
%\begin{frame}%[allowframebreaks=0.9]

There are $N_n= n(n+1) / 2$ different $Y_{ij}$ for $n$-point functions,
leading to $N_n=[n(n+1) / 2-1]$-dimensional Mellin--Barnes integrals when splitting the sum in \Eref{eq-2ndsy} into a product:
\begin{itemize}
 \item 
 $N_3=5$ MB dimensions  for the most general massive vertices;
 \item 
$N_4=9$  MB dimensions for the most general massive box integrals;
 \item 
$N_5=14$  MB dimensions for the most general massive pentagon integrals.
\end{itemize}
The introduction of $N_n$-dimensional MB integrals allows  $x$ integrations
to be calculated. 
The MB integrations must be calculated afterwards, and this raises some mathematical problems with increasing integral dimensions.
This is, for Mellin--Barnes integrals numerical applications, one of the most important limiting factors.

For further details of this approach, we refer to the quoted literature on AMBRE and MBnumerics.

%%%%%%%%%%%%%%%%%%%%%%%%%%%%%%%%%%%%%%%%%%%%%%%%%%%%%%%%%%%%%%%%%%%
\subsubsection
[MBOneLoop]
{MBOneLoop}
%%%%%%%%%%%%%%%%%%%%%%%%%%%%%%%%%%%%%%%%%%%%%%%%%%%%%%%%%%%%%%%%%%%%%%%%%%%%%%%%%%
A completely different approach was initiated in Refs. \cite{Bluemlein:2017rbi,Phan:2018cnz}.
The idea is based on rewriting the $F$ function in \Eref{bh-1} by exploring the factor $\delta(1-\sum x_i)$, 
which makes the $n$-fold  $x$-integration to be an integral over an $(n-1)$-simplex.

The $\delta$ function allows for the elimination of $x_n$, just one of the $x_i$, which creates linear terms in the remaining $x_i$ variables in the $F$ function:
\begin{eqnarray} 
{F_n(x)}=x^\mathrm{T} G_n x +2 H_n^\mathrm{T} x +K_n
.
\end{eqnarray}
The $F_n(x)$ may be recast   into a bilinear form by shifts {$x \to (x-y)$},
\begin{eqnarray}
\label{c4}
{F_n(x)}
 =  {(x - y)^\mathrm{T}} {G_n} {(x-y)}
%\binom{x_1-y_1}{x_2-y_2}
 + r_{n} {  - \mathrm{i} \varepsilon} 
%\nonumber \\ & = & 
~=~ {\Lambda_{n}(x)} + {r_{n}{  - \mathrm{i} \varepsilon}}
%\nonumber \\ &\equiv& 
~=~  \Lambda_{n}(x) + {R_{n}}
.                 
\end{eqnarray}
As a result, there is a separation of $F$ into a homogeneous part $\Lambda_n(x)$,
\begin{eqnarray}
\Lambda_n(x) 
=
(x-y)^\mathrm{T} G_{n} (x-y) ,
%\binom{x_1-y_1}{x_2-y_2} ,
\end{eqnarray} 
and an inhomogeneity $R_{n}$,
\begin{eqnarray}
R_{n} = r_n  {  - \mathrm{i} \varepsilon}
= %\\&=& 
K_n- H_n^\mathrm{T} G_{n}^{-1} H_n {  - \mathrm{i} \varepsilon}
= %\\ \nonumber &=& 
-\frac{\lambda_n}{g_n} {  - \mathrm{i} \varepsilon}
= % \\ \nonumber &=&
-\frac{ 
\begin{pmatrix}
0\\
0\\
\end{pmatrix}_n}
{()_n}
.
\end{eqnarray}
It is only this inhomogeneity  ${R_{n}=r_{n}{ ~ - \mathrm{i} \varepsilon}}$ that carries the $\mathrm{i} \varepsilon$ prescription.
The $(n-1)$ components $y_i$ of the shift vector $y$ appearing here in $F_n(x)$ are 
\begin{eqnarray}
 y_i = -\left(G_{n}^{-1} K_n\right)_i, \qquad i\neq n
 .
\end{eqnarray}
The following relations are also valid:
\begin{eqnarray} \label{eq-def-yi}
y_i = \dfrac{\partial r_{n}}{\partial m_i^2} 
=   - \frac{1}{g_{n}}~\dfrac{\partial \lambda_n}{\partial m_i^2} 
= - \dfrac{\partial_i \lambda_n}{g_{n}}
= \dfrac{2}{g_{n}}  
\begin{pmatrix}
0\\
i\\
\end{pmatrix}
_n
 , 
\qquad i=1, \dots, n.
\end{eqnarray} 
One further notation has been introduced in \Eref{eq-def-yi}, namely that of {\it cofactors of the modified Cayley matrix},
also called {signed minors} in, 
\eg \cite{Melrose:1965kb,Regge:1964}
\begin{eqnarray}\label{gram2}
\begin{pmatrix}
  j_1 & j_2 & \cdots & j_m\\
  k_1 & k_2 & \cdots & k_m\\
\end{pmatrix}_n .
\end{eqnarray}
The signed minors are determinants, labelled by those {rows
% 2020-03-13
$j_1,j_2,\dots, j_m$ and columns $k_1, k_2, \dots,$ $k_m$ that have been
discarded from the definition of the modified Cayley determinant $()_n$}, with a sign convention:
\begin{multline}
\label{eq-modc}
\mathrm{sign}  
\begin{pmatrix}
  j_1 & j_2 & \cdots &  j_m\\
  k_1 & k_2 & \cdots & k_m\\
\end{pmatrix}_n \\
=
(-1)^{j_1+j_2+ \cdots +j_m+k_1+k_2+ \cdots +k_m}
%\\ 
\times \texttt{Signature}[j_1,j_2, \dots, j_m]
\times
\texttt{Signature} [k_1,k_2, \dots, k_m] .
\end{multline}
Here, \texttt{Signature} (defined like the Wolfram Mathematica command) gives the sign of permutations needed to place the indices in increasing 
order.
The Cayley determinant is a signed minor of the modified Cayley determinant,
\begin{eqnarray}
\Delta_n = \lambda_{n} = 
\begin{pmatrix}
  0\\
  0\\
\end{pmatrix}_n.
\end{eqnarray}
For later use, we also introduce
\begin{eqnarray} \label{eq-yn}
y_n = 1-\sum_{i-1}^{n-1} y_i  \equiv   \dfrac{\partial r_{n}}{\partial m_n^2}
.
\end{eqnarray}
The auxiliary condition $\sum_i^n y_i =1$ is fulfilled.
Further, 
the notations for the $F$ function are finally independent of the choice of the variable that was 
eliminated by the use of the 
$\delta$ function in the integrand of \Eref{bh-1}.  
Moreover, the inhomogeneity $R_{n}$ 
is the only variable carrying the causal $\mathrm{i}\epsilon$ prescription, while, \eg $\Lambda(x)$ and  $y_i$ are, by 
definition, real quantities. 
The $R_{n}$
may be expressed by the ratio of the Cayley determinant (\Eref{npoint-cayley}) and the Gram determinant 
(\Eref{Gram}),
\begin{eqnarray}\label{eq-r-ikl}
R_n = r_{1\cdots n}  - \mathrm{i}\epsilon     = - \frac{\lambda_{1\cdots n}}{g_{1\cdots n}} - \mathrm{i}\epsilon
 %                         \quad \text{for} \quad  g_{1\cdots n} \neq 0. 
.
\end{eqnarray}

One may use the Mellin--Barnes relation (\Eref{MB})  to decompose the 
integrand of $J_n$ given in \Eref{bh-1}, as follows:
\begin{align}
\label{c5}
J_n \sim {\int \mathrm{d}x} ~\dfrac{1}{[{F(x)}]^{ n-\frac{d}{2} }} 
&\equiv
{\int \mathrm{d}x}
\dfrac{1}{[{\Lambda_n(x) + R_n}]^{ n-\frac{d}{2} }} 
\equiv 
{\int \mathrm{d}x}
\frac{R_n^{-(n-\frac{d}{2}) }}  {\left [1+\frac{\Lambda_n(x)}{R_n} \right]^{ n-\frac{d}{2} }} 
\nonumber\\
 &=
{ \int \mathrm{d}x~}
\frac{R_n^{-(n- \frac{d}{2})}}{2\uppi \mathrm{i}} 
\int\limits_{-\mathrm{i}\infty}^{+\mathrm{i}\infty}\mathrm{d}s \; 
  \dfrac{\Gamma(-s)\;\Gamma \left ( n-\frac{d}{2} +s \right ) } { \Gamma
  \left (n-\frac{d}{2} \right ) }    
  \left[{\frac{\Lambda_n(x)}{ R_n}} \right]^s,
\end{align}
for 
$|\mathrm{Arg}(\Lambda_n/R_n)|<\uppi$. 
The condition always applies.
\textcolor{black}{ 
Further, the integration path in the complex $s$-plane separates the poles of $\Gamma(-s)$ and $\Gamma( n- {d} / {2} +s) $.}

As a result of \Eref{c5}, the Feynman parameter integral of $J_n$ becomes homogeneous:
\begin{align}
\label{c6}
\kappa_n &= 
{ \int \mathrm{d}x~}
\left[{\frac{\Lambda_n(x)}{ R_n}} \right]^s
\nonumber \\
&=
\prod_{j=1}^{n-1} {\int_0^{1-\sum_{i=j+1}^{n-1} x_i}  \mathrm{d}x_j}
\; 
\left[ 
      \frac{\Lambda_{n}(x)}{ R_n} \right]^s  ~\equiv~
\int \mathrm{d}S_{n-1}
\; 
\left[ 
      \frac{\Lambda_{n}(x)}{ R_n} \right]^s  
.
\end{align}
% from APP  Podlesice2017, modified
 To reformulate this integral, one may introduce the differential 
operator $\hat{P}_n$ \cite{Bernshtein1971-from-springer2,Golubeva:1978},
\begin{eqnarray}
\label{c7}
\frac{\hat{P}_n}{s} \left[\dfrac{\Lambda_n(x)}{R_n} \right]^s
~\equiv~ 
\sum_{i=1}^{n-1} 
   \frac{1}{2s} (x_i-y_i)\frac{\partial}{\partial x_i}
\left[\dfrac{\Lambda_n(x)}{R_n} \right]^s
   =    \;\left[\dfrac{\Lambda_n(x)}{R_n } \right]^s
,
\end{eqnarray}
into \Eref{c6}:
\begin{eqnarray}
\label{c8}
K_n
= \frac{1}{s}
% \prod_{j=1}^{n-1} \int_0^{1-\sum_{i=n-1}^{j+1} x_i}  dx_j
\int \mathrm{d}S_{n-1}\; \hat{P}_n 
    \left[ \dfrac{\Lambda_n(x)}{R_n} \right]^s
~=~ \frac{1}{2s} \sum_{i=1}^{n-1} 
\prod_{k=1}^{n-1}
\int\limits_0^{u_k}  \mathrm{d}x'_k ~ 
   (x_i-y_i)\frac{\partial}{\partial x_i} 
   \left[\dfrac{\Lambda_n(x)}{R_n} \right]^s
.
\end{eqnarray}
\normalsize 
\noindent 
After calculating  one of the $x$ integrations---by partial integration, eliminating in this way the 
corresponding differential, and applying a Barnes relation \cite{Barnes:1908} 
% also: Barnes:1908
(see Ref.~\cite{watson}), 
one  arrives at a  recursion relation in the number of internal lines $n$:
\begin{multline} \label{JNJN1}
J_n(d,\{q_i,m_i^2\})
\\
= \dfrac{{-1}}{2\uppi \mathrm{i}} \int\limits_{-\mathrm{i}\infty}^{+\mathrm{i}\infty}
\mathrm{d}s  
     \dfrac{\Gamma(-s) \Gamma(\frac{d-n+1}{2}+s) \Gamma(s+1) }
           { 2\Gamma(\frac{d-n+1}{2}) } 
     \left(\frac{1}{{R_{n}}}\right)^s
 \times     \sum\limits_{k=1}^n 
     \left( \frac{1}{{R_{n}}} 
            \frac{\partial {r_n}}{\partial m_k^2} 
      \right) \;
     {\bf k}^- J_n \left (d+2s;\{q_i,m_i^2\} \right).
\end{multline}
{The operator {$\bf k^{-}$}}, introduced in \Eref{eq-007}, will reduce an $n$-point Feynman integral $J_n$ to a sum of $(n-1)$-point integrals $J_{n-1}$ 
by shrinking propagators $D_k$ from the original $n$-point integral.
The starting term is the one-point function, or tadpole,
\begin{align} \label{eq-tadpole}
J_{1}(d;m^2) 
& = 
\int \dfrac{\mathrm{d}^d k}{\mathrm{i} \uppi^{d/2}} 
\dfrac{1}{k^2-m^2+\mathrm{i} \varepsilon}
% \nonumber \\
% &=&
=
- \frac{\Gamma( 1 -d/2)}  {R_1^{1-d/2}}
,
\\
R_1 &= m^2-\mathrm{i} \varepsilon
.
\end{align}
The cases $G_n=0$ and $\lambda_n=r_n=0$ are discussed in  \Sref{sec-tr-vanish}.

Equation \eqref{JNJN1}  is the master integral for one-loop $n$-point functions in space--time dimension $d$, representing 
them by $n$ 
integrals over $(n-1)$-point functions with a shifted dimension $d+2s$.
The recursion was first published in Ref. \cite{Bluemlein:2017rbi}.
This implies a series of Mellin--Barnes representations for arbitrary massive one-loop $n$-point integrals with  $(n-1)$ Mellin--Barnes integral dimensions.
This linear increase in  the MB dimension is highly advantageous compared with the number of MB integral dimensions in the AMBRE approach (increasing as $n^2$ with the number $n$ of scales). 

Based on \Eref{JNJN1}, one has now several opportunities to proceed.
\begin{itemize}
 \item[(i)] Evaluate the MB integral in a direct numerical way.
 \item[(ii)] Derive $\varepsilon$-expansions for the Feynman integrals.
 \item[(iii)] Apply the Cauchy theorem for deriving sums and determine analytical expressions in terms of known special functions.
\end{itemize}

The first approach is based on AMBRE/MBOneLoop, the middle one is not yet finished, and the last approach was applied in Ref. \cite{Bluemlein:2017rbi} for massive vertex integrals and in Ref. \cite{Phan:2018cnz}  for massive box integrals.

A few comments are in order.
\begin{enumerate}
\item
Any four-point integral, \eg is, in the recursion, a {three-fold Mellin--Barnes} integral,
whereas, {with AMBRE, one gets for, \eg box integrals up to { nine-fold} MB
integrals}.
\item
Euclidean and Minkowskian integrals converge equally well;
{see Refs. \cite{Usovitsch:April2018slides,%
Usovitsch:2018shx}. }
\item 
There appear to be no numerical problems due to vanishing Gram determinants. For a  few details, see \Tref{tab-d1111} and Ref. \cite{UsovitschRiemannMinirep:2018aa}.
\end{enumerate}

%%%%%%%%%%%%%%%%%%%%%%%%%%%%%%%%%%%%%%%%%%%%%%%%%%%%%%%%%%%%%%%%%%%%%%%%%%%%%%%%%%%
\subsection{The basic scalar one-loop functions}
\subsubsection{Massive two-point functions}
From the recursion relation (\Eref{JNJN1}), taken at $n=2$ and using  
\Eref{eq-tadpole} with $d\to d+2s$ for the one-point 
functions under the 
integral, one  
gets the following Mellin--Barnes representation: 
\begin{multline}%1707
\label{MBJ2}
 J_2(d; q_1, m_1^2,q_2, m_2^2)  = % -1 from generic times -1 from J_1 = 1
\frac{\mathrm{e}^{\epsilon \upgamma_E}}{2\uppi \mathrm{i}} \int\limits_{-\mathrm{i}\infty}^{+\mathrm{i}\infty}\mathrm{d}s \; 
\dfrac{
\Gamma({-s})\;
\Gamma\left(\frac{d-1}{2} +s\right)\;
\Gamma(s+1) }
{2\;\Gamma\left(\frac{d-1}{2} \right) }
R_2^{s} 
 \\
\times
\left[
\frac{1}{r_2}
\dfrac{\partial 
r_2%\overline{m}^2_{12}
}
{\partial m_2^2} 
\frac{\Gamma\left({1-\frac{d+2s}{2}}\right)}{(m_1^2)^{1-\frac{d+2s}{2}}} + (m_1^2 \leftrightarrow m_2^2)
\right].      
\end{multline}
One may {close the integration contour of the MB integral} in \Eref{MBJ2}
to the 
%corrTR left -> right
right, apply the Cauchy theorem, 
and {collect the residua originating from two series of zeros of arguments of 
$\Gamma$ functions} at $s=m$  and $s=m-d/2-1$ for $m \in \mathbb{N}$.
The first series stems from the MB-integration kernel, the other one from the dimensionally shifted one-point functions.
And then one may sum up analytically in terms of Gauss' hypergeometric functions.

{The two-point function, with $R_2 \equiv R_{12}$}, becomes
 \begin{multline}
\label{b99rewritten}
 J_2(d;Q^2, m_1^2,q_2, m_2^2) 
\\
 = 
 % corr. 2018-04-27: \Gamma\left({1- \frac{d}{2}} \right)
- \frac{{\Gamma\left(2- \frac{d}{2}\right)}}{(d-2)}
\dfrac{ \Gamma\left(\frac{d}{2}-1\right)}     {\Gamma\left(\frac{d}{2} \right) }
\dfrac{\partial_2 R_2}{R_{2}} 
\Biggl[
 (m_1^2)^{\frac{d}{2} -1}
 { 
 \Fh21\Fz{1            ,\frac{d}{2}-\frac{1}{2}}{\frac{d}{2}}{\frac{m_1^2}{R_{2 }}}
 }
+
\dfrac{ R_2^{\frac{d}{2} -1} } {\sqrt{1-\frac{m_1^2}{R_2}}}
\sqrt{\uppi} \frac{\Gamma\left(\frac{d}{2}\right)} {\Gamma\left(\frac{d}{2}-\frac{1}{2}\right)}
\Biggr]
\\ 
    + (m_1^2 \leftrightarrow m_2^2)
    .
\end{multline}
Equation \eqref{b99rewritten} is valid for $\left|{m_1^2} / {r_{12}}\right|<1$, 
$\left| {m_2^2} / {r_{12}}\right|<1$ and 
$\mathcal{R}$e$({(d-2)} / {2})>0$. 
The result is in agreement with Eq.\,(53)   of Ref. \cite{Fleischer:2003rm}.

The iterative determination of higher-point functions proceeds analogously.
Closing the integration contours to the right or to the left will cover different kinematic regions in the invariants $R_n$.  

%%%%%%%%%%%%%%%%%%%%%%%%%%%%%%%%%%%%%%%%%%%%%%%%%%%%%%%%%%%%%%%%%%%%%%%%%%%%%%%%%%%
\subsubsection{Massive three-point functions}
%%%%%%%%%%%%%%%%%%%%%%%%%%%%%%%%%%%%%%%%%%%%%%%%%%%%%%%%
The Mellin--Barnes integral for the massive vertex is a sum of three terms \cite{Riemann:calc2018}:
\bq
 J_3= J_{123} + J_{231} + J_{312},
\eq
using the representation for, \eg $J_{123}$,
\begin{equation} \label{JNJN1-2}  % 4.34 of 16feb17, copy from calc2018
{J_{123}}({d},\{q_i,m_i^2\})
= -\dfrac{\mathrm{e}^{\epsilon \upgamma_E}}{2\uppi \mathrm{i}} \int\limits_{-\mathrm{i}\infty}^{+\mathrm{i}\infty} \mathrm{d}s 
     \dfrac{\Gamma({-s})\; \Gamma(\frac{d-2+2s}{2}) \Gamma(s+1) }
           { 2~\Gamma(\frac{d-2}{2}) } 
    R_{3}^{-s} 
\times     
      \frac{1}{r_3}   \frac{\partial r_{3}}{\partial m_3^2} ~
  {J_2} ({d+2s};q_1,m_1^2,q_2,m_2^2).
\end{equation}

After applying the Cauchy theorem and summing up, one gets an analytical representation.
The integrated massive vertex has been published in Ref. \cite{Bluemlein:2017rbi}.
We quote here the representation given in Ref \cite{Phan:2018cnz}:
\begin{align}
\label{J123altern}
{J_{123}} &=  %J_{123} &=&
\Gamma\left(2-\frac{d}{2}\right)  \frac{\partial_3 r_3}{r_3}   
\frac{\partial_2 r_2}{r_2} 
\frac{r_2}{2\sqrt{1-m_1^2/r_2}}
\nonumber\\                    %  ----   OK   ----
& \qquad 
\left[
{-} 
R_2^{d/2-2}  
\frac{\sqrt{\pi}}{2} 
\frac{\Gamma\left(\frac{d}{2}-1\right) }{\Gamma\left(\frac{d}{2}-\frac{1}{2}\right) }
 \Fh21\Fz{ \DDDD, 1}{\DDD}{ \dfrac{R_{2}}{R_{3} } }
{+} ~ R_3^{d/2-2}  
\Fh21\Fz{ 1, 1}{3/2}{ \dfrac{R_{2}}{R_{3} }}
\right]   
\nonumber\\&  \quad                %  ----   OK   ----
%%%%%%%%%===================================================
+ 
\Gamma\left(2-\frac{d}{2}\right)  \frac{\partial_3 r_3}{r_3}   
\frac{\partial_2 r_2}{r_2} 
 \frac{m_1^2}{4\sqrt{1-m_1^2/r_2}} 
\nonumber\\                   %  ----   OK   ----
& \qquad
\left[
{+} 
\frac{2(m_1^2)^{d/2-2}  }{d-2}
F_1 \left( \dfrac{d-2}{2}; 1, \frac{1}{2}; \DD; \dfrac{m_1^2}{R_{3}}, \dfrac{m_1^2}{R_{2}} \right)
{-} 
R_3^{d/2-2}  
F_1 \left( 1; 1, \frac{1}{2}; 2; \dfrac{m_1^2}{R_{3}}, \dfrac{m_1^2}{R_{2}} \right)
\right] 
\nonumber\\                   %  ----   OK   ----
& \quad 
+  (m_1^2 \leftrightarrow m_2^2)  
,
  \nonumber
\end{align}
with the short notation
\bq
R_3=R_{123}, \qquad R_2=R_{12},
\eq
etc.
{For $d \to 4$, the bracket expressions vanish so that their product with the prefactor  }
$  \Gamma(2-{d}/{2})$ stays finite in this limit, as it must come out for a massive vertex function.
For some numerics, see Tables \ref{table-j3table7line1}, \ref{table2-j3table7line1},  \ref{table3-j3table7line1}, and \ref{table4-j3table7line1}.

%%======= table 1 ===============  table 6 line 1 = table 7 line 1 ==============
\begin{table}
\caption[]{\label{table-j3table7line1}%\it 
Numerics for a vertex, $d=4-2\epsilon$. 
Input quantities suggest that, according to Eq. 
(73) in Ref.  \cite{Fleischer:2003rm}, one has to set $b_3=0$.
Although $b_3$ of Ref. \cite{Fleischer:2003rm} deviates from our vanishing value, it has to be set to zero, $b_3\to0$.
 The results of both calculations for $J_3$  agree for this case. }
\begin{center}
\small 
%\begin{tabular}{|l@{\hspace{2.7cm}}|l|}\hline \hline
\begin{tabular}{llll}
\hline \hline
      [$p_i^2$],~~[$m_i^2$] & [{$+100$}, {$+200$}, {$+300$}],~~~[$10$, $20$, $30$]  &
% \\
%   $m_i^2$    & 10, 20, 30 & 
\\ \hline
$G_{123}$ &$-160\,000$
&
\\
$\lambda_{123}$ &$-8\,860\,000$
& 
\\
$m_i^2/r_{123}$  &$-0.180\,587$,
$-0.361\,174$,
$-0.541\,761$
&
\\    
$m_i^2/r_{12} $ &$-0.975\,61$,
$-1.951\,22$,
$-2.926\,83$
& 
\\
$m_i^2/r_{23} $ &$-0.398\,01$,
$-0.796\,02$,
$-1.194\,03$
&
\\
$m_i^2/r_{31} $ &$-0.180\,723$,
$-0.361\,446$,
$-0.542\,169$
&
\\
$\sum J$ terms \cite{Fleischer:2003rm} &
 $(0.019\,223\,879 - 0.007\,987\,267\;\mathrm{i})$& 
%  2017-08-12: for RVZ\[Rule]-1
\\  
$\sum b_3$ terms (TR) &
$0$ & 
\\
 $J_3$ (TR)   &
$(0.019\,223\,879 - 0.007\,987\,267\;\mathrm{i})$ &
% $-(  0.` + 3*20^{-19}\;i)$/eps
%$(-0.0123074 - 0.00930135\;i)$& $(3.03577*10-18 - 1.73472*10-18\;i)$/eps
\\  %================================= Tarasov ======================
$b_3$ term \cite{Fleischer:2003rm} &
$(-0.089\,171\,509 + 0.069\,788\,641\;\mathrm{i})$ && $(0.022\,214\,414)$/eps
% \\ 
% $J$-terms &
% $(0.019223879 - 0.007987267\;i)$$ &  $(
%  5*-18 - 3*-19\;i)$/eps
\\
$b_3+\sum J$ terms &
$(-0.012\,307\,377 - 0.009\,301\,346\;\mathrm{i})$ & 
% $(
%  4*-18 - 6*10^{-18}\;0)$/eps
\\ 
$J_3$ (OT) & 
$\sum J$ terms, {$b_3$ term $\to$ $0$, OK} &
\\ 
MB suite &&
\\ 
$(-1)\times$fiesta3 \cite{Smirnov:2013eza}&
$-(0.012\,307 + 0.009\,301\;\mathrm{i})$&& + $(8 \times 10^{-6} + 0.000\,01\;\mathrm{i})\pm (1+\mathrm{i})10^{-4}$  )
\\ 
LoopTools \cite{Hahn:1998yk1}
% & $0.0192238790286 - 0.00798726725497\;i$
& $0.019\,223\,88 - 0.007\,987\,267\;\mathrm{i}$&
\\ \hline\hline
  \end{tabular}
\end{center}
\end{table}
%%===== end table 1 =======================  table 6 line 1 = table 7 line 1 

% 

%%=====  table 2  ===================  table 6 line 6 = table 7 line 2
\begin{table}
\caption[]{%\it 
\label{table2-j3table7line1}
%\footnotesize%\tiny
Numerics for a vertex, $d=4-2\epsilon$. 
%Causal $\varepsilon =   10^{-20}$.
% Compares to {Table 6, line 
% 6} and {Table 7, line 2} of draft 
% oneloop-16Feb2017.pdf. 
Input quantities suggest that, according to Eq. 
(73) in Ref. \cite{Fleischer:2003rm}, one has to set $b_3=0$.
Further, we have set in the numerics for Eq. (75) of Ref. \cite{Fleischer:2003rm} so that the root of the Gram determinant is
$\sqrt{-g_{123} + \mathrm{i}\varepsilon}$, 
% RVZg123 = -1}, 
which seems counterintuitive for a `momentum'-like function.
{Both results agree if we {\it do not} set Tarasov's $b_3 \to 0$.}
Table courtesy of Ref. \cite{Riemann:2017podlesice}.
}
\begin{center}
\small 
%\begin{tabular}{|l@{\hspace{2.7cm}}|l|}\hline \hline
\begin{tabular}{llll}\hline \hline
      [$p_i^2$],~~[$m_i^2$]    &   [$-100$, {$+200$}, $-300$],~~~[$10$, $20$, $30$]  &
% \\
%   $m_i^2$    & 10, 20, 30 & 
\\ \hline
$G_{123}$ &{$480\,000$}
&
\\
$\lambda_{3}$ &$-19\,300\,000$
& 
\\
$m_i^2/r_{3}$  &$0.248\,705$, $0.497\,409$, $0.746\,114$
&
\\    
$m_i^2/r_{12} $ &$0.248\,447$,
$0.496\,894$,
$0.745\,342$
& 
\\
$m_i^2/r_{23} $ &$-0.398\,01$,
$-0.796\,02$,
$-1.194\,03$
&
\\
$m_i^2/r_{31} $ &$0.104\,895$,
$0.209\,79$,
$0.314\,685$
&
\\
$\sum J$ terms &
$(-0.012\,307\,377 - 0.056\,679\,689\;\mathrm{i} )$ & $(0.012\,825\,498\;\mathrm{i})$/eps
%  2017-08-12: for RVZ\[Rule]-1
\\ 
%\\
$\sum b_3$ terms &
$( 0.047\,378\,343\;\mathrm{i})$ & $(- 0.012\,825\,498\;\mathrm{i})$/eps   
\\
$J_3$(TR)   &
$(-0.012\,307\,377 - 0.009\,301\,346\;\mathrm{i})$& 
% $( 3*-18 + 0 I)$/eps
%$(-0.0123074 - 0.00930135\;i)$& $+ (3.03577*10-18 - 1.73472*10-18\;i)$/eps
\\ 
$b_3$ term &
$( 0.047\,378\,343\;\mathrm{i})$ & $(- 0.012\,825\,498\;\mathrm{i})$/eps
% \\ 
% $J$-terms &
% (-0.012307377 - 0.056679689\;i)& + (
%  5*-18 + 0.012825498\;i)/eps
\\
$b_3+\sum J$ terms &
$(-0.012\,307\,377 - 0.009\,301\,346\;\mathrm{i})$& 
% + (
%  4*-18 - 5*-18 i)/eps
\\ 
$J_3$(OT) &
$\sum J$ terms, {$b_3$ term$\to$0, { gets wrong!}} &
\\
MB suite &&
\\ 
(-1)*fiesta3 &
$-(0.012\,307 + 0.009\,301\;\mathrm{i})$& $(8 \times 10^{-6} + 0.000\,01\;\mathrm{i})\pm (1+\mathrm{i})10^{-4}$  )
\\ 
LoopTools/FF, $\epsilon^0$ &$-0.012\,307\,377\,367\,78 - 0.009\,301\,346\,170\;\mathrm{i}$ &
\\ \hline\hline
\end{tabular}
\end{center}
\end{table}
%%======  end table 2  =================  table 6 line 6 = table 7 line 2  ===============

%%========  table 3   =======table 6 line 2  = table 7 line 3 ===============
\begin{table}
\caption[]{\label{table3-j3table7line1}%\it 
%\footnotesize%\tiny
Numerics for a vertex in space--time dimension  $d=4-2\epsilon$. Causal $\varepsilon =   10^{-20}$.
{Agreement with Ref. \cite{Fleischer:2003rm}.}
% Compares to {Table 6, line 
% 2} and {Table 7, line 3} of draft 
% oneloop-16Feb2017.pdf
% \\
% % {TR 2017-08-25: Check if replacement of r by R=r-I eps changes accuracy in this case.}
Table courtesy Ref. \cite{Riemann:2017podlesice}.
}
\begin{center}
\small 
%\begin{tabular}{|l@{\hspace{2.7cm}}|l|}\hline \hline
\begin{tabular}{lll}
\hline \hline
      $p_i^2$    &   $-100$,$-200$,$-300$  &
\\
  $m_i^2$    & $10$,$20$,$30$ & 
\\ \hline
$G_{123}$ & $-160\,000$  &
\\
$\lambda_{123}$ &
15\,260\,000 
& 
\\
$m_i^2/r_{123}$ & $0.104\,849$, $0.209\,699$, $0.314\,548$  &
\\    
$m_i^2/r_{12} $ & $0.248\,447$, $0.496\,894$, $0.745\,342$  & 
\\
$m_i^2/r_{23} $ & $0.133\,111$, $0.266\,223$, $0.399\,334$ &
\\
$m_i^2/r_{31} $ & $0.104\,895$, $0.209\,79$, $0.314\,685$ &
\\
$\sum J$ terms &$(0.093\,387\,7 - 0\;\mathrm{i})$& $-(0.022\,214\,4 - 0\;\mathrm{i})$/eps
%  2017-08-12: for RVZ\[Rule]-1
\\ 
$\sum b$ terms &$-0.101\,249$ & $+ 0.022\,214\,4$/eps   
\\
$J_3$(TR)   & 
$(-0.007\,861\,55 - 0\;\mathrm{i})$ & 
\\ 
$b_3$ &
$(-0.101\,249 + 0\;\mathrm{i})$
&
 $(0.022\,214\,4 + 0\;\mathrm{i})$/eps
% \\ 
% $J$-terms &
% $(0.0933877 + 0\;i)$
% & 
% $- (0.0222144 + 0\;i)$/eps
\\
$b_3$+$J$ terms &
$(-0.007\,861\,546 + 0\;\mathrm{i})$
&
\\ 
$J_3$(OT) &$b_3+J$ terms $\to$ OK 
 &
\\ 
MB suite &{$-0.007\,862\,014$, {$5.002\,549\,159 \times 10^{-6}$, 0}}&
\\ 
(-1)*fiesta3 & $-0.007\,862$ &$6 \times 10^{-6} + 6 \times 10^{-6}\;\mathrm{i} \pm (1+\mathrm{i})10^{-10}$ 
\\ 
LoopTools/FF, $\epsilon^0$ &
$-0.007\,861\,546\,132\,290\,822\,90$
&
\\ \hline\hline
\end{tabular}
\end{center}
\end{table}
%========== end table 3   %%%%%%%%%%%%%%%%%%%%%%%%%%%%%%%%%%%%%%%%%%

%%========  table 4  ==========  table 6 line 3 = +-+ ============================
\begin{table}
\caption{\label{table4-j3table7line1}%\it 
%\footnotesize%\tiny
Numerics for a vertex in space--time dimension  $d=4-2\epsilon$. Causal $\varepsilon =   10^{-20}$.
% Compares to {Table 6, line 3}  of draft 
% oneloop-16Feb2017.pdf. 
Input quantities suggest that, according to Eq. 
(73) in Ref. \cite{Fleischer:2003rm}, one has to set $b_3=0$.
% Although $b_3$ of \cite{Fleischer:2003rm} deviate from our vanishing value, it has to be set to zero, so that the results of both 
% calculations for $J_3$ agree for this case. 
{Agreement, owing to setting $b_3=0$ there.}
Table courtesy Ref.\cite{Riemann:2017podlesice}.
}
\begin{center}
\small 
%\begin{tabular}{|l@{\hspace{2.7cm}}|l|}\hline \hline
\begin{tabular}{lll}
\hline \hline
      $p_i^2$    &   {$+100$}, {$-200$}, {$+300$}  &
\\
  $m_i^2$    & $10$, $20$, $30$ & 
\\ \hline
$G_{123}$ &{$480\,000$}
&
\\
$\lambda_{123}$  &$4\,900\,000$
& 
\\
$m_i^2/r_{123}$  &
$-0.979\,592$,
$-1.959\,18$,
$-2.938\,78$
&
\\    
$m_i^2/r_{12} $  &
-$0.975\,61$,
$-1.951\,22$,
$-2.926\,83$
& 
\\
$m_i^2/r_{23} $  &
$0.133\,111$,
$0.266\,223$,
$0.399\,334$
&
\\
$m_i^2/r_{31} $  &
$-0.180\,723$,
$-0.361\,446$,
$-0.542\,169$
&
\\
$\sum J$ terms &
$(0.006\,243\,624 - 0.018\,272\,524 \;\mathrm{i})$ &
%  2017-08-12: for RVZ\[Rule]-1
\\  
$\sum b_3$ terms &
0 & 
\\
   $J_3$(TR)   &
$(0.006\,243\,624 - 0.018\,272\,524\;\mathrm{i})$ &
%$(-0.0123074 - 0.00930135\;i)$&  $(3.03577*10-18 - 1.73472*10^{-18} \;i)$/eps
\\ 
%================================= Tarasov ======================
$b_3$ term &
$(0.040\,292\,491 + 0.029\,796\,253\;\mathrm{i})$
&$ (- 0.012\,825\,498\;\mathrm{i})$/eps
% \\ 
% $J$-terms &
% (0.006243624 - 0.018272524\;i)
\\
$b_3+\sum J$ terms &
$(-0.012\,307\,377 - 0.009\,301\,346\;\mathrm{i})$ & $(
 4 \times 10^{-18} - 6 \times 10^{-18}\;\mathrm{i})$/eps
\\ 
$J_3$(OT) &
$\sum J$ terms, {$b_3$ term$\to$0, OK} &
\\ 
MB suite &&
\\ 
$(-1)\times $fiesta3 &
$-(-0.006\,322 + 0.014\,701\;\mathrm{i})$& + $(0.000\,012 + 0.000\,014\;\mathrm{i}) \pm (1+\mathrm{i})10^{-2}$ 
\\ 
LoopTools/FF, $\epsilon^0$&
{\small $0.006\,243\,624\,78 %7277 
- 
        0.018\,272\,524\,0 %487
        \; \mathrm{i}$}
%
%{\small $0.00624362477277 - 
%        0.0182725240487\; i$}
%
&
\\ \hline\hline
\end{tabular}
\end{center}
\end{table}
%%======  end table 4  ========  table 6 line 3 =  +-+   ===============

%%%%%%%%%%%%%%%%%%%%%%%%%%%%%%%%%%%%%%%%%%%%%%%%%%%%%%%%%%%%%%%%%%%%%%%%%%%%%%%%%%%
\subsubsection{Massive four-point functions}  
Finally, we reproduce the box integral, as a three-dimensional Mellin--Barnes representation:
\begin{align} \label{J4J3}
J_4(d;\{p_i^2\},s,t,\{m_i^2\})
&=  %t06
 % {\color{black} C_1\times}  ~ 
 \left(\frac{-1}{4\uppi \mathrm{i}}\right)^4  
% e^{\epsilon \gamma_E} ~
\frac{1}{\Gamma(\frac{d-3}{2})}
\sum_{k_1,k_2,k_3,k_4=1}^4 D_{k_1k_2k_3k_4}
\left( \frac{1}{r_{4}} \frac{\partial r_4}{\partial m_{k_4}^2} \right)
\nonumber \\ 
& \qquad
\left( \frac{1}{r_{k_3k_2k_1}} \frac{\partial r_{k_3k_2k_1}}{\partial 
       m_{k_3}^2} \right)
\left( \frac{1}{r_{k_2k_1}} \frac{\partial r_{k_2k_1}}{\partial m_{k_2}^2} 
      \right)
      (m_{k_1}^2)^{d/2-1}
\\\nonumber 
& \qquad      
\int\limits_{-\mathrm{i}\infty}^{+\mathrm{i}\infty} \mathrm{d}z_4
       \int\limits_{-\mathrm{i}\infty}^{+\mathrm{i}\infty} \mathrm{d}z_3
              \int\limits_{-\mathrm{i}\infty}^{+\mathrm{i}\infty} \mathrm{d}z_2  
          \left(\frac{m_{k_1}^2}{R_{4}}\right)^{z_4}          
       \left(\frac{m_{k_1}^2}{R_{k_3k_2k_1}}\right)^{z_3}
       \left(\frac{m_{k_1}^2}{R_{k_2k_1}}\right)^{z_2}
    \\ \nonumber 
& \qquad   
\Gamma(-z_4) \Gamma(z_4+1) 
\frac  {\Gamma(z_4+\frac{d-3}{2})}  {\Gamma(z_4+\frac{d-2}{2})}  
            \Gamma(-z_3) \Gamma(z_3+1) 
\frac{\Gamma(z_3+z_4+\frac{d-2}{2})}{\Gamma(z_3+z_4+\frac{d-1}{2})} 
% 2020-03-13
\times \cdots
  \\ \nonumber 
& \qquad   
% 2020-03-13
\cdots \times \Gamma \left (z_2+z_3+z_4+\frac{d-1}{2} \right ) 
\Gamma \left (-z_2-z_3-z_4-\frac{d+2}{2} \right )
 \\ \nonumber 
  & \qquad      \Gamma(-z_2) \Gamma(z_2+1)    .   
%        \\\nonumber&&
\end{align}

Equation \eqref{J4J3} can be treated using the 
 Mathematica packages MB and MBnumerics of the MBsuite, replacing AMBRE with a derivative of MBnumerics: {MBOneLoop} 
 \cite{Usovitsch:CERN-Jan-2018,Usovitsch:April2018slides}.
 For numerical examples, see Table \ref{tab-D0-1}.
%      Usovitsch:2018LLslidesJUplb

%==========================================================================
%copy from Phan, Riemann PLB 2019 submission:
\begin{table}                % table 1   d=4-2 eps
\caption{\label{tab-D0-1}%\it 
Comparison of 
the box integral $J_4$ defined in \Eref{J1234altern} with the LoopTools
function 
{\tt D0($p_1^2, p_2^2, p^2_3, p^2_4, (p_1+p_2)^2, (p_2+p_3)^2, m^2_1, m^2_2, m^2_3 , m^2_4$)}
\cite{Hahn:1998yk1,vanOldenborgh:1990ycplb} 
at 
$m^2_2= m^2_3= m^2_4=0$.
Further numerical references are from the
packages K.H.P\_D0 (PHK, unpublished) %KhiemD0 (PHK, 
and MBOneLoop \cite{Usovitsch:January2018,Riemann:2019fcceejan}.
External invariants:
$(p_1^2=\pm 1,p_2^2=\pm 5,p^2_3=\pm 2,p^2_4=\pm 7,s=\pm 20,t=\pm 1)$.
Table from Ref. \cite{Phan:2018cnz}, licence: \url{https://creativecommons.org/licenses/by/4.0/}.}
\begin{center}
\begin{tabular}{ll} \hline \hline
$(p_1^2,p_2^2,p_3^2,p_4^2, s, t)$ 
                           & Four-point integral   
\\ \hline %\hline
$(-,-,-,-,-,-)$ & $d=4$, ~ $m^2_1=100$
\\
$J_4$ & $0.009\,178\,67$ 
\\
LoopTools                     &
 $0.009\,178\,670\,7$ 
%----- JU 2018-12-21 20:55
\\ MBOneLoop &  $0.009\,178\,670\,7$
\\ 
$(+,+,+,+,+,+)$       & $d=4$, ~ $m^2_1=100$
\\
$J_4$ & $-0.011\,592\,7 - 0.000\,406\,03\;\mathrm{i}$   
\\
LoopTools                      & $-0.011\,591\,7- 0.000\,406\,02\;\mathrm{i}$ 
%----- JU 2018-12-21 20:55
\\ MBOneLoop & $-0.011\,591\,736\,9 - 0.000\,406\,024\,3 %51
\;\mathrm{i}$
\\ 
$(-,-,-,-,-,-)$ & $d=5$, ~ $m^2_1=100$
\\ 
$J_4$ & $ 0.009\,268\,95  $ 
\\
K.H.P\_D0                    &
 $0.009\,268\,88 $
 %----- JU 2018-12-21 20:55
\\ MBOneLoop & $0.009\,268\,948\,8 %65
$
\\ 
$(+,+,+,+,+,+)$       & $d=5$, ~ $m^2_1=100$
\\
$J_4$ & $-0.002\,728\,89 + 0.012\,648\,8\;\mathrm{i}$
\\
K.H.P\_D0                   & \hspace*{+19.5mm}(--)
%----- JU 2018-12-21 20:55
\\ MBOneLoop & $-0.002\,728\,424\,2 +  0.012\,648\,813\,4 
\;\mathrm{i}$
\\ 
$(-,-,-,-,-,-)$ & $d=5$, ~ $m^2_1=100-10\;\mathrm{i}$
\\
$J_4$ & $ 0.009\,200\,65 + 0.000\,782\,308\;\mathrm{i}$
\\
K.H.P\_D0                    &
 $0.009\,200\,6 ~~ +0.000\,782\,301\;\mathrm{i}$
%----- JU 2018-12-21 20:55
\\ MBOneLoop & $ 0.009\,200\,648\,1 + 0.000\,782\,309\,0\;\mathrm{i}$
 \\ 
$(+,+,+,+,+,+)$       & $d=5$, ~ $m^2_1=100-10\;\mathrm{i}$
\\
$J_4$ & $-0.003\,987\,25 + 0.012\,067\;\mathrm{i}$
\\
K.H.P\_D0                   & $-0.003\,987\,23+ 0.012\,069\;\mathrm{i}$
%----- JU 2018-12-21 20:55
\\ MBOneLoop & $-0.003\,986\,770\,2 + 0.012\,067\,038\,8 
\;\mathrm{i}$
\\ \hline\hline
\end{tabular}
\end{center}
\end{table}
%%%%%%%%%%%%%%%%%%%%%%%%%%%%%%%%%%%%%%%%%%%%%%%%%%%%

% TR 2018-04-18, pages 1,4 and 2018-04-24 page 2
After applying the Cauchy theorem and summing the residues, we get
\cite{Riemann:2018LL,Phan:2018cnz}:
\bq
J_4= {J_{1234}} + J_{2341} + J_{3412} + J_{4123}
,
\eq
with $R_4=R_{1234}, R_3=R_{123}, 
R_2=R_{12}$, etc.:
\begin{align}
\label{J1234altern}
{J_{1234}} 
&=  %J_{1234} &=&
%%%%%%%%
%   2F1
  {
  \Gamma\left(2-\frac{d}{2}\right)
  }
  \frac{\partial_4 r_4}{r_4}
    \Biggl\{
\left[ 
\frac{
{
b_{123}
}
}{2}  
\left(
{-} 
R_3^{d/2-2}  
{
\Fh21\Fz{ \frac{d-3}{2}, 1}{\frac{d}{2} -1 }{\frac{R_{2}}{R_{3}} }
}
{+}  R_4^{d/2-2}  
\sqrt{\uppi} 
\frac{\Gamma\left(\frac{d}{2}-1\right) }{\Gamma\left(\frac{d}{2}-\frac{3}{2}\right) }
 % \Fh21\Fz{ 1/2, 1}{1}{ \dfrac{R_{2}}{R_{3} } } 
  {_2F_1}(d \to 4)
\right)\right]   
%%%%%%%%%%%%%%%%%%%%%%%%%%%
%   F_1
\nonumber\\                    %  ----   OK   ----
& \qquad \qquad  
+ %20190428
 \left[
{+} 
\frac{R_2^{d/2-2} }{d-3}
{
F_1 \left( \dfrac{d-3}{2}; 1, \frac{1}{2}; \dfrac{d-1}{2} ; \dfrac{R_2}{R_{4}}, \dfrac{R_2}{R_{3}}\right)
}
{-}   
R_4^{d/2-2}  
F_1 \left( d \to 4 \right)
\right]  
% % % % \nonumber\\                     %  ----   OK   ----
% % % % &&
% % % % + ~~ (1 \leftrightarrow 2)  
  \nonumber\\                   %  ----   OK   ----
& \qquad \qquad   
%%%%%%%%%%%%%%%%%%%%%%%
% F_S terms
%\Gamma\left(2-\frac{d}{2}\right)
+ %20190428
\frac{m_1^2}{8}
\frac{\Gamma\left(\frac{d}{2}-1\right) }{\Gamma\left(\frac{d}{2}-\frac{3}{2}\right) }
%\frac{\partial_4 r_4}{r_4} 
\frac{\partial_3 r_3}{r_3}   
\frac{\partial_2 r_2}{r_2} 
\frac{r_3}{r_3-m_1^2}
\frac{r_2}{r_2-m_1^2}
\nonumber\\                    %  ----   OK   ----
& \qquad \qquad 
 \Bigl[
{-} 
(m_1^2)^{d/2-2} 
\frac{\Gamma\left(\frac{d}{2}-3/2\right) }{\Gamma\left(\frac{d}{2}\right) }
{
F_S \left (\frac{d}{2}-\frac{3}{2},1,1,1,1,\frac{d}{2},\frac{d}{2},\frac{d}{2},\frac{d}{2},\frac{m_1^2}{R_4},\frac{m_1^2}{m_1^2-R_3},\frac{m_1^2}{m_1^2-R_2}
\right )
}
\nonumber\\                  %  ----   OK   ----
& \qquad \qquad 
{+} 
R_4^{d/2-2} \sqrt{\uppi} 
F_S(d \to 4)\Bigr] 
 +  (m_1^2   \leftrightarrow m_2^2 ) \Biggr\}.
\end{align}
 %endoffootnotesize%small 

For $d \to 4$, all three contributions  in square brackets approach zero, so that the massive $J_4$ gets finite in this limit, as it should do.
Table \ref{tab-D0-1} contains numerical examples.

%===========================================================================================
\subsection{The cases of vanishing Cayley determinant $\lambda_n=0$ and of vanishing Gram determinant $G_n=0$
\label{sec-tr-vanish}}
%---------------------------------------------------
% case 7.13, 7.17 of scalar-oneloop-2017-generic.pdf
We refer here to two important special cases, where the general derivations cannot be applied.

In the case of vanishing Cayley determinant, $\lambda_n=0$, we cannot introduce the inhomogeneity $R_n=-\lambda_n/G_n$ 
into the Symanzik polynomial $F_n$.
Let us assume that it is  $G_n \neq 0$, so that $r_n=0$. 
A useful alternative representation to \Eref{JNJN1} is known from the literature, see \eg Eq.~(3) in Ref. 
\cite{Fleischer:2003rm}:
\begin{eqnarray} \label{cayleyZero}
J_n(d) &=& \frac{1}{d-n-1} \sum_{k=1}^{n} \frac{\partial_k \lambda_n}{G_n} {\bf k^-} J_n(d-2).
% eqn 7.13 seems to have a factor of 1/2 wrong in scalar-oneloop-2017-generic.pdf
\end{eqnarray}

Another special case is a vanishing Gram determinant, $G_n=0$.
Here again, one may use Eq. (3) of Ref. \cite{Fleischer:2003rm} and the result is (for 
$\lambda_n\neq 0$):
\begin{eqnarray} \label{gramZero}
J_n(d) &=& - \sum_{k=1}^{n} \frac{\partial_k \lambda_n}{2 \lambda_n} {\bf k^-} J_n(d).
\end{eqnarray}
The representation was, for the special case of the vertex function, also given in Eq.\ (46) of Ref. \cite{Devaraj:1997es}.

For the vertex function, a general study of the special cases has been conducted,
as reported in Ref. \cite{Phan:2019qee}.

%===========================================================================================
\subsection{A massive four-point function with vanishing Gram determinant}
% take-overs from JU(&TR), LL2018  and the extended E.6 of  hep-ph/1809.01830v2 %t13 
As a very interesting non-trivial example, we have restudied the numerics of  a  massive four-point function with a small or 
vanishing Gram determinant 
\cite{%
Usovitsch:CERN-Jan-2018, UsovitschRiemannMinirep:2018aa,Usovitsch:April2018slides,Riemann:calc2018}.
The original example has been taken from Appendix C of Ref. \cite{Fleischer:2010sq}.

The sample outcome 
is shown in Table \ref{tab-d1111}.
The new iterative Mellin--Barnes representations deliver very precise numerical results for, \eg box functions, including cases of small or vanishing Gram determinants.
The software used is MBOneLoop \cite{Usovitsch:mboneloop}. The notational correspondences are, \eg
% From talk held at $11^{th}$ FCC-ee workshop: Theory and Experiments
% January, 8 - 11, 2019, CERN \cite{}, the Table tab-d1111.
\[
J_{4}(12-2\epsilon,1,5,1,1) \to I_{4,2222}^{[d+]^4} = D_{1111}.
\]

\begin{table}
\caption{\label{tab-d1111}
  %\it
  The Feynman integral $J_{4}(12-2\epsilon,1,5,1,1)$ 
  %as defined in \eqref{prd83-integral} 
  compared with 
numbers from Ref. \cite{Fleischer:2010sq}. 
The $I_{4,2222}^{[d+]^4}$ is the scalar integral, where propagator 2 has index $\nu_2 = 1+(1+1+1+1) = 5$, and the other propagators 
have index 1.
The integral corresponds to $D_{1111}$ in the notation of LoopTools 
\cite{Hahn:1998yk1}. For $x=0$, the Gram determinant vanishes.
We see an agreement of about 10 to 11 relevant digits. The deviations of the two calculations seem to stem from 
a limited accuracy of the Pad\'e approximations used in Ref. \cite{Fleischer:2010sq}.
Table courtesy Refs. \cite{Usovitsch:CERN-Jan-2018, UsovitschRiemannMinirep:2018aa}.
}
\center
\begin{tabular}[\linewidth]{ll}
   \hline \hline
$x$ & Value for $4! \times ~ J_{4}(12-2\epsilon,1,5,1,1)$ \\\hline
% tord 2018-05-26 line with x=0 inserted from table for D_1111
$0$       & $(2.059\,692\,897\,30 + 1.555\,949\,101\,18 \;  \mathrm{i})10^{-10}$ \;{\scriptsize \cite{Fleischer:2010sq}}
% Re D1111 2.05969289730E-10 Im D1111 1.55594910118E-10
\\
$0$      &   $(2.059\,692\,897\,30 + 1.555\,949\,101\,18\;\mathrm{i})10^{-10}$ \;
%t9 {\color{red}{???}}   
{\scriptsize MBOneLoop+Kira+MBnumerics}
\\
$10^{-8}$ &$(2.059\,692\,893\,42 + 1.555\,949\,091\,87\;\mathrm{i})10^{-10}$\;{\scriptsize \cite{Fleischer:2010sq}}
\\
$10^{-8}$ &$(2.059\,692\,893\,63 + 1.555\,949\,091\,87\;\mathrm{i})10^{-10}$\;{\scriptsize MBOneLoop+Kira+MBnumerics}
\\
$10^{-4}$ &$(2.059\,656\,094\,97 + 1.555\,856\,053\,43\;\mathrm{i})10^{-10}$\;{\scriptsize \cite{Fleischer:2010sq}}
\\
$10^{-4}$ &$(2.059\,656\,094\,89 + 1.555\,856\,053\,43\;\mathrm{i})10^{-10}$\;{\scriptsize MBOneLoop+Kira+MBnumerics}
\\\hline\hline
\end{tabular}
\end{table}
%=======================================================================================

%=========================================================================
%\begin{frame}%[allowframebreaks=0.9]
  \subsection{Calculation of Gauss hypergeometric function $_2F_1$, Appell function $F_1$, and Saran function $F_\mathrm{S}$ at arbitrary kinematics}
%--------------------------------
Little is known about the precise numerical calculation of generalised hypergeometric functions at arbitrary arguments.
Numerical calculations of specific Gauss hypergeometric functions $_2F_1$, 
Appell functions $F_1$ (Eq.\ (1) of Ref. \cite{Appell:1925}), and Lauricella--Saran functions $F_S$ (Eq.\ (2.9) of Ref. \cite{Saran:1955}) are needed for the scalar one-loop Feynman integrals:
\begin{align}
 \label{eq-2f1def}
 _2F_1(a,b;c;x)
& =
\sum_{k=0}^{\infty} \frac{(a)_k(b)_k}{ k! ~ (c)_k} ~ x^k,
\\\label{eq-f1def}
 F_1(a;b,b';c;y,z) 
%\nonumber\\ &
&=
\sum_{m,n=0}^{\infty} \frac{(a)_{m+n} (b)_m (b')_n}{ m! ~ n! ~ (c)_{m+n}} ~ y^m z^n,
\\\label{eq-fsdef}
 F_S(a_1,a_2,a_2; b_1,b_2,b_3 ; c,c,c ; x,y,z)
%\\\nonumber&
&=
\sum_{m,n,p=0}^{\infty} \frac{(a_1)_{m} (a_2)_{n+p} (b_1)_m (b_2)_n (b_3)_p}
{ m! ~ n! ~ p! ~(c)_{m+n+p}} ~ x^m y^n z^p
.
\end{align}
The  $(a)_k$ is the Pochhammer symbol.
The specific cases needed here are discussed in the appendices of Ref. \cite{Phan:2018cnz}.
Here, we repeat only few definitions.

%%%%%%%%%%%%%%%%%%%%%%%%%%%%%%%%%%%%%%%%%%%%%%%%%%%%%%%%%%%%%
%\begin{frame}{{\bf Mellin-Barnes integrals for $_2F_1$ and $F_1$ and $F_S$}}
One approach to the numerics of $_2F_1$, $F_1$, and $F_\mathrm{S}$ may be based on Mellin--Barnes representations.
For the Gauss function $_2F_1$ and the Appell function $F_1$, Mellin--Barnes representations have been known for some time.
See Eq.~(1.6.1.6) in Ref. \cite{Slater:1966},
\begin{align}\label{s-mb2f1}
& _2F_1(a,b;c;z) = 
\frac{1}{2\uppi \mathrm{i}}~
\frac{\Gamma(c)}{\Gamma(a)\Gamma(b)}
%\\\nonumber& \times 
\int_{-\mathrm{i}\infty}^{+\mathrm{i}\infty} 
\mathrm{d}s ~ (-z)^s ~ \frac{\Gamma(a+s)\Gamma(b+s)\Gamma(-s)}{\Gamma(c+s)}
,
\end{align}
and Eq.~(10) in Ref. \cite{Appell:1925}, which is a two-dimensional MB integral:
\begin{align}\label{s-mbf1}
&F_1( a;b,b';c;x,y) = 
\frac{1}{2\uppi \mathrm{i}}~
\frac{\Gamma(c)}{\Gamma(a)\Gamma(b')}
% \\\nonumber
% &
% \times
\int_{-\mathrm{i}\infty}^{+\mathrm{i}\infty} \mathrm{d}t  ~ (-y)^t ~ _2F_1(a+t,b;c+t,x)
% \\\nonumber
% &
% \times 
\frac{\Gamma(a+t)\Gamma(b'+t)\Gamma(-t)}{\Gamma(c+t)}
.
\end{align}
For the Lauricella--Saran function $F_\mathrm{S}$, the following, new, three-dimen\-sio\-nal MB integral was given in Ref. \cite{Phan:2018cnz}: 
\begin{multline}\label{s-mbfs}
F_\mathrm{S}(a_1,a_2,a_2;b_1,b_2,b_3;c,c,c;x,y,z) 
% \\\nonumber
% &
\\ = 
\frac{1}{2\uppi \mathrm{i}}
\frac{\Gamma(c)}{\Gamma(a_1)\Gamma(b_1)}
\int_{-\mathrm{i}\infty}^{+\mathrm{i}\infty} \mathrm{d}t 
(-x)^t  \frac{\Gamma(a_1+t)\Gamma(b_1+t)\Gamma(-t)}{\Gamma(c+t)}
\times  
 F_1(a_2;b_2,b_3;c+t;y,z)
.
\end{multline}

The numerics of the Gauss hypergeometric function are generally known in all detail.

For the Appell function $F_1$, the numerical mean value integration of the 
one-dimensional integral representation of Ref. \cite{Picard:1881} may be advocated,
being quoted in Eq.~(9) of Ref. \cite{Appell:1925}:
\begin{align}\label{s-f1A}
& F_1(a;b,b';c;x,y)
 =
 \frac{\Gamma(c)}{\Gamma(a)\Gamma(c-a)}
 %\\\nonumber  & \times
 \int_{0}^{1}\mathrm{d}u\frac{u^{a-1}(1-u)^{c-a-1}}{(1-xu)^b(1-yu)^{b'}}
 .
\end{align}
 
 We need three specific cases, taken at $d \geq 4$.
{For vertices,} \eg
\begin{align}\label{F1-002}
& F_1^v(d)
\equiv F_1 \left( \frac{d-2}{2};1,\frac{1}{2};\frac{d}{2};x_c,y_c \right)
%\\\nonumber&
=    
\frac{1}{2}(d-2) \int_0^1 \frac{ \mathrm{d}u~u^{\frac{d}{2}-2}}{(1-x_c u)\sqrt{1-y_c u}}
.
\end{align}
Integrability is violated at $u=0$ if not $\Re \mathrm{e}(d)>2$.
The stability of numerics is well-controlled, as exemplified in Table \ref{t-f1numer}.

%%%%%%%%%%%%%%%%%%%%%%%%%%%%%%%%%%%%%%%%%%%%%%%%%%%%%%%%%%%%
\begin{table*}
\caption[]%
{\label{t-f1numer}%\it 
The Appell function $F_1$ of the massive vertex integrals as defined in  \Eref{F1-002}.
As a proof of principle, only the constant term of the expansion in 
$d=4-2\varepsilon$ is shown, $F_1(1;1,\frac{1}{2};2;x,y)$.
Upper values from general numerics of appendices of Ref. \cite{Phan:2018cnz}; lower values from  setting $d=4$ and the use of analytical formulae.
Table courtesy of Ref. \cite{Phan:2018cnz} under licence \url{http://creativecommons.org/licenses/by/4.0/}.}
\renewcommand{\arraystretch}{1.2}
\begin{center}
\scriptsize%\footnotesize%\small %\tiny
\begin{tabular}{llll}
\hline \hline
  $x -\mathrm{i}\varepsilon_x$ & $y-\mathrm{i}\varepsilon_y$ & $F_1(1;1,\frac{1}{2};2;x,y)$ &
\\
\hline
$+11.1 - 10^{-12}\mathrm{i}$ & $+12.1 - 10^{-12} \mathrm{i}$ & 
   $-0.175\,044\,248\,073\,5 % truncated, not rounded 880657                
   $&$      - 0.054\,228\,129\,473\,2 %898100 
   \;\mathrm{i}$   
   \\ 
   & & 
   $-0.175\,044\,248\,073\,518\,778\,844\,982\,899\,12 $ & $ - 0.054\,228\,129\,473\,304\,027\,882\,097\,641\,167 \;\mathrm{i}
$
\\   %%%%%%%%%%%%%%%%%%%%%%%%%%%%%%%%%%%%%%%%
$+11.1 - 10^{-12}\mathrm{i}$   & $+12.1 + 10^{-12} \mathrm{i}$  & $ +1.710\,854\,529\,324\,4 %03295 
$&$ + 0.054\,228\,129\,473\,2 %89810
~ \mathrm{i} $
   \\  
   & & $
+1.710\,854\,529\,324\,335\,571\,348\,382\,041\,75 $&$+ 0.054\,228\,129\,471\,482\,173\,815\,892\,709\,24 \;\mathrm{i}
$
 %------------------------------------------------------------------
 \\    %%%%%%%%%%%%%%%%%%%%%%%%%%%%%%%%%%%%%%
 $+11.1 + 10^{-12}\mathrm{i}$   & $+12.1 - 10^{-12} \mathrm{i}$  & 
$ +1.710\,854\,530\,411\,4 %87717245222484052646905184  
$&$ - 0.054\,228\,129\,473\,2 %898100 
\;\mathrm{i}$
  \\   
  & & $
 +
 1.710\,854\,529\,324\,335\,571\,348\,382\,041\,75 $&$- 0.054\,228\,129\,471\,482\,173\,815\,892\,709\,24 \;\mathrm{i}$
 \\   
 %%%%%%%%%%%%%%%%%%%%%%%%%%%%%%%%%%%%%%%%%%%%%%%%%%%%%%%%
$+11.1 + 10^{-12}\;\mathrm{i}$   & $+12.1 + 10^{-12}\;\mathrm{i}$  
& 
% B
  $-0.175\,044\,248\,073\,5 %18778844982899126  
  $&$ + 0.054\,228\,129\,473\,3 ~\mathrm{i}$ %04027882097641167 ~i$
 \\   
 & & $
 -0.175\,044\,248\,073\,518\,778\,844\,982\,899\,12 $&$+ 0.054\,228\,129\,473\,304\,027\,882\,097\,641\,167 ~\mathrm{i}$
 %%%%%%%%%%%%%%%%%%%%%%%%%%%%%%%%%%%%%%%%%%%%%%%%%%%%%%%%%%%%%%%%%%%%%%%%%%%%%%%%%%%%%%%%
\\
$+12.1 - 10^{-15} ~\mathrm{i}$   & $+11.1 - 10^{-15} ~\mathrm{i}$  &
 $-0.170\,082\,716\,648\,4 %41197663 
 $&$ -0.051\,868\,484\,603\,7 %1380 
 ~\mathrm{i}$ 
 \\ $+12.1 - 10^{-10} ~\mathrm{i}$&  $+11.1 - 10^{-15} ~\mathrm{i}$ & $
 -0.170\,082\,716\,648\,000\,581\,011\,657\,492\,79 $&$ - 0.051\,868\,484\,604\,656\,749\,765\,565\,256\,21 ~\mathrm{i} $
  %---------------
 \\   
$+12.1 - 10^{-15} ~\mathrm{i}$   & $+11.1 + 10^{-15} ~\mathrm{i}$  &
 $-0.170\,082\,716\,648\,4 %411977 
 $ & $ - 1.754\,420\,290\,995\,5 %760794  
 ~\mathrm{i}$
\\  
& & $
 -0.170\,082\,716\,648\,440\,256\,472\,688\,173\,99  $ & $- 1.754\,420\,290\,995\,576\,887\,358\,425\,620\,38 ~\mathrm{i}$
 \\  
$+12.1 + 10^{-15} ~\mathrm{i}$   & $+11.1 - 10^{-15} ~\mathrm{i}$  &
 $ -0.170\,082\,716\,648\,4 %411977
 $&$  + 1.754\,420\,290\,995\,5 %760794
 ~ \mathrm{i}$
  \\ 
  & & $
  -0.170\,082\,716\,648\,440\,256\,472\,688\,173\,99 $&$+ 1.754\,420\,290\,995\,576\,887\,358\,425\,620\,38 ~\mathrm{i}$
%---------------
 \\  
$+12.1 + 10^{-15}~ \mathrm{i}$   & $+11.1 + 10^{-15} ~\mathrm{i}$  &
 $-0.170\,082\,716\,648\,4 %41197663 
 $&$ + 0.051\,868\,484\,603\,7 %1380 
 ~\mathrm{i}$ 
 \\  $+12.1 - 10^{-10} ~\mathrm{i}$ &  $+11.1 - 10^{-15} ~\mathrm{i}$ & $
 -0.170\,082\,716\,648\,000\,581\,011\,657\,492\,79 $&$ + 0.051\,868\,484\,604\,656\,749\,765\,565\,256\,21 ~\mathrm{i}$
  %%%%%%%%%%%%%%%%%%%%%%%%%%%%%%%%%%%%%%%%%%%%%%%%%%%%%%%%%%%%%%%%%%%%%%%%%%%
\\
  $+11.1 - 10^{-15}~\mathrm{i}$   & $-12.1$   
 &  $-0.053\,370\,514\,651\,8 %9944545 
 $&$ - 0.195\,769\,211\,155\,7 %3399152 
 ~\mathrm{i}$
 \\
  && 
 $-0.053\,370\,514\,651\,899\,444\,733\,494\,011\,52$ & $- 0.195\,769\,211\,155\,733\,985\,388\,920\,833\,693 ~\mathrm{i}$
  \\  
  $+11.1 + 10^{-15}~\mathrm{i}$   & $-12.1$   
 &  $-0.053\,370\,514\,651\,8 %9944545 
 $&$ + 0.195\,769\,211\,155\,7 %3399152 
 ~\mathrm{i}$
 \\
  && 
 $-0.053\,370\,514\,651\,899\,444\,733\,494\,011\,52$ & $+ 0.195\,769\,211\,155\,733\,985\,388\,920\,833\,693 ~\mathrm{i}$
 %%%%%%%%%g%%%%%%%%%%%%%%%%%%%%%%%%%%%%%%%%%%%%%%%%%%%%%%%%%%%%%%%%%%%%%%%%%%%%%%
\\
%                                  x<1<y
  $-11.1                $   & $+12.1- 10^{-12}~\mathrm{i}$  
&  $  +0.106\,086\,408\,466\,2 %1242501  
$&$  - 0.144\,744\,070\,008\,2 %18726147  
~\mathrm{i}$ 
\\  
& & $
+0.106\,086\,408\,476\,510\,642\,871\,335\,275\,99$ & $- 0.144\,744\,070\,021\,333\,407\,167\,349\,619\,088 ~\mathrm{i}$
\\
$ -11.1                $   &  $+12.1+ 10^{-12}~\mathrm{i}$  
&  $   +0.106\,086\,408\,466\,2 %1242501 
$&$  + 0.144\,744\,070\,008\,2 %18726147  
~\mathrm{i}$ 
\\ 
& & $
+0.106\,086\,408\,476\,510\,642\,871\,335\,275\,99$ & $+ 0.144\,744\,070\,021\,333\,407\,167\,349\,619\,088 ~\mathrm{i}$
%%%%%%%%%%%%%%%%%%%%%%%%%%%%%%%%%%%%%%%%%%%%%%%%%%%%%%%%%%%%%%%%%%%%%%%%%%%%%%%
\\
$ -12.1                $   &  $-11.1$  
&  $ +0.122\,456\,767\,687\,224\,028 %10327293549
$  & 
\\
 && 
$+0.122\,456\,767\,687\,224\,025\,065\,133\,951\,61$&%$ + 10^{-32} ~i$
\\\hline \hline
\end{tabular}
\end{center}
\end{table*}
For the calculation of the four-point Feynman integrals, one also needs  the Lauricella--Saran function $F_\mathrm{S}$ \cite{Saran:1955}.
Saran defines $F_\mathrm{S}$ as a three-fold sum (\Eref{eq-fsdef}), see Eq.~(2.9) in Ref. \cite{Saran:1955}. Saran derives a {three-fold integral representation  in Eq.~(2.15) and a two-fold integral  in Eq.~(2.16)}.
We recommend  use of the following representation, given on p. 304 of 
\cite{Saran:1955}: 
\begin{multline}
\label{fsgeneral}
F_\mathrm{S}(a_1,a_2,a_2;b_1,b_2,b_3;c,c,c,x,y,z)
\\
=
\frac{\Gamma(c)}{\Gamma(a_1)\Gamma(c-a_1)}
%\nonumber\\\nonumber &
\int_0^1 \mathrm{d}t \frac{t^{c-a_1-1} (1-t)^{a_1-1}}{(1-x+tx)^{b_1}}
F_1(a_2;b_2,b_3;c-a_1;ty,tz)
. 
 \end{multline}

{For the box integrals, one needs the specific case}
\begin{align}\label{fsJ4}
F_\mathrm{S}^b(d) & =
F_\mathrm{S}\left(\frac{d-3}{2},1,1;1,1,\frac{1}{2};\frac{d}{2},\frac{d}{2},\frac{d}{2},x_c,y_c,z_c\right)
\nonumber\\
& 
=
\frac{\Gamma(\frac{d}{2})}{\Gamma(\frac{d-3}{2})\Gamma(\frac{3}{2})}
%\\\nonumber & \times
\int_0^1 \mathrm{d}t \frac{\sqrt{t} (1-t)^{\frac{d-5}{2}}}{(1-x_c+x_c t)}
F_1(1;1,\frac12;\frac32;y_c t,z_c t)
.
\end{align}
Equation \eqref{fsJ4} is valid if $\Re \mathrm{e}(d)>3$.

% 2020-03-06 sectin -> subsection
\subsection*{Acknowledgement}
We would like to thank J. Gluza for a careful reading of the manuscript.
%}

%==================================
%\newpage

%====================================================================

\end{bibunit}

 \label{sec-mtools-riemann}  

\clearpage \pagestyle{empty}  \cleardoublepage
%============================================

\pagestyle{fancy}
\fancyhead[CO]{\thechapter.\thesection \hspace{1mm}  NNLO corrections in four dimensions}
\fancyhead[RO]{}
\fancyhead[LO]{}
\fancyhead[LE]{}
\fancyhead[CE]{}
\fancyhead[RE]{}
\fancyhead[CE]{R.~Pittau}
\lfoot[]{}
\cfoot{-  \thepage \hspace*{0.075mm} -}
\rfoot[]{}

\newcommand{\mur}{\mu_\mathrm{R}}
\newcommand\toGP{\,~{\to}\rule{-10.7pt}{1.93ex}^{\mbox{\tiny \rm GP}}~\,}

\begin{bibunit}[elsarticle-num]  
\section
[NNLO corrections in four dimensions \\ {\it R.~Pittau}]
{NNLO corrections in four dimensions
\label{contr:pittau}}
\noindent
{\bf Contribution\footnote{This contribution should be cited as:\\
R.~Pittau, NNLO corrections in four dimensions,  
%04 DOI:10.23731/CYRM-2020-XXX.\thepage, in:
%04 \url{http://dx.doi.org/10.23731/CYRM-2020-XXX.\thepage}, in:
DOI: \href{http://dx.doi.org/10.23731/CYRM-2020-003.\thepage}{10.23731/CYRM-2020-003.\thepage}, in:
Theory for the FCC-ee, Eds. A. Blondel, J. Gluza, S. Jadach, P. Janot and T. Riemann,\\
CERN Yellow Reports: Monographs, CERN-2020-003,
%04 \url{http://dx.doi.org/10.23731/CYRM-2020-XXX}, p. \thepage.} 
DOI: \href{http://dx.doi.org/10.23731/CYRM-2020-003}{10.23731/CYRM-2020-003},
p. \thepage.
\\ \copyright\space CERN, 2020. Published by CERN under the 
%04-2
\href{http://creativecommons.org/licenses/by/4.0/}{Creative Commons Attribution 4.0 license}.} by: R.~Pittau 
%\\Corresponding Author: Roberto~Pittau 
{[pittau@ugr.es]}}
\vspace*{.5cm}
  
%=================================================================
\subsection{Introduction}
Currently, four-dimensional techniques applied to higher-order calculations are under active investigation
\cite{Fazio:2014xea,Battistel:1998sz,Cherchiglia:2010yd,Hernandez-Pinto:2015ysa,Sborlini:2016gbr,Sborlini:2016hat,Driencourt-Mangin:2019aix,Runkel:2019yrs}. The main motivation for this is the need to simplify perturbative calculations necessary to cope with the precision requirements of the future LHC and FCC experiments.

\begin{figure}[h!]
\vspace*{-7.9cm}
\hspace*{4cm}
\centerline{
\includegraphics[width=1.6\textwidth]{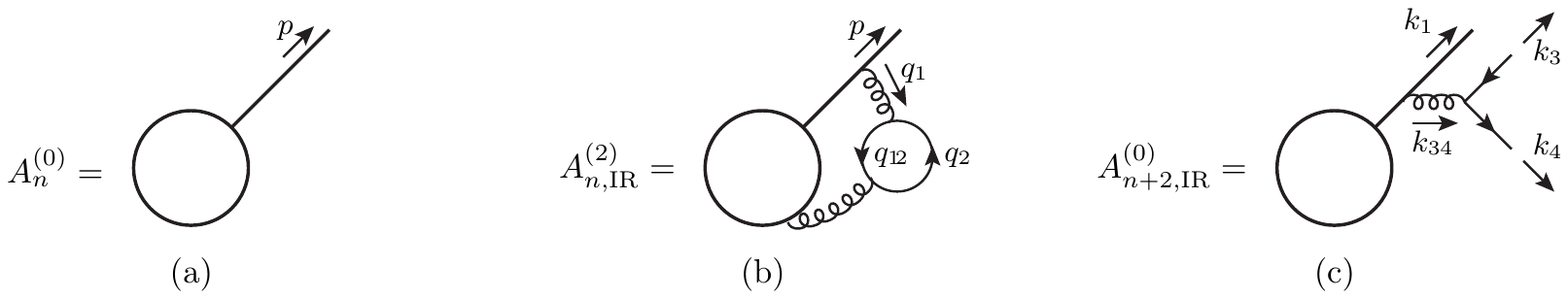}
}
\vspace*{-9.2cm}
\caption{\label{fig:BVR}
The lowest-order amplitude (a), the IR-divergent final-state virtual quark-pair correction (b), and the IR-divergent real component (c).
The empty circle stands for the emission of $n\!-\!1$ particles.
Additional IR finite corrections are created if the gluons with momenta
$q_1$ and $k_{34}$ are emitted by off-shell particles contained in the empty circle.}
\end{figure}

In this contribution, I review the four-dimensional regularisation or renormalization (FDR) approach \cite{Pittau:2012zd} to the computation of NNLO corrections in four dimensions. In particular, I describe how fully inclusive NNLO final-state quark-pair corrections \cite{Page:2018ljf}
\bea
\label{eq:sigmaBVR}
\sigma^\mathrm{NNLO} = \sigma_\mathrm{B}+\sigma_\mathrm{V}+\sigma_\mathrm{R}
\hskip 20pt {\rm with} \hskip 20pt
\left\{
\begin{tabular}{l}
$\displaystyle \sigma_\mathrm{B} = \int \mathrm{d} \Phi_n\, \sum_{\rm spin}
|A^{(0)}_{n}|^2$ \\
$\displaystyle \sigma_\mathrm{V} = \int \mathrm{d} \Phi_n\,
\sum_{\rm spin}
\left\{
 A^{(2)}_{n} (A^{(0)}_{n})^\ast 
+A^{(0)}_{n} (A^{(2)}_{n})^\ast 
\right\}$ \\
$\displaystyle \sigma_\mathrm{R} = \int \mathrm{d} \Phi_{n+2}\,
\sum_{\rm spin}
\left\{
A^{(0)}_{n+2} (A^{(0)}_{n+2})^\ast
\right\}$
\end{tabular}
\right.
\eea
are computed in FDR by directly enforcing gauge invariance and unitarity in the definition of the regularised UV- and IR-divergent integrals.
The IR-divergent parts of the amplitudes are depicted in \Fign{fig:BVR} and
$
\mathrm{d} \Phi_m :=  \updelta\left(P-\sum_{i=1}^m p_i\right) \prod^m_{i=1}
\mathrm{d}^4 p_i \updelta_+(p^2_i).
$

In \Sref{sec:fdr}, I recall the basics of FDR. The following sections deal with its use in the context of the calculation of $\sigma^\mathrm{NNLO}$ in \Eref{eq:sigmaBVR}.

\subsection{FDR integration and loop integrals}
\label{sec:fdr}
The main idea of FDR can be sketched out with the help of a simple
one-dimensional example \cite{Pittau:2019bba}. More details can be found in the relevant literature
\cite{Pittau:2012zd,Donati:2013iya,Pittau:2013qla,Donati:2013voa,Page:2015zca,Page:2018ljf,Blondel:2018mad}.
Let us assume that one needs to define the UV divergent integral
\bea
\label{eq:integral}
I = \lim_{\Lambda \to \infty}\int_0^\Lambda \mathrm{d}x \frac{x}{x+M},
\eea
where $M$ stands for a physical energy scale.
FDR identifies the UV divergent pieces in terms of integrands that do not depend on $M$, the so-called FDR vacua, and separates them by rewriting
\bea
\label{eq:integrand}
\frac{x}{x+M}= 1-\frac{M}{x}+\frac{M^2}{x(x+M)}.
\eea
The first term in the right-hand side of \Eref{eq:integrand} is the vacuum responsible for the linear ${\cal O}(\Lambda)$ UV divergence of $I$ and $1/x$ generates its $\ln \Lambda$ behaviour.
From the {\em definition} of FDR integration, both divergent contributions need to be subtracted from \Eref{eq:integral}. The subtraction of the ${\cal O}(\Lambda)$ part is performed over the full integration domain
$[0,\Lambda]$, while the logarithmic divergence is removed over the interval
$[\mur,\Lambda]$ only. The arbitrary separation scale $\mur \ne 0$ is needed to keep adimensional and finite the arguments of the logarithms appearing in the subtracted and finite parts. 
Thus
\bea
\label{eq:fdrintegral1}
I_\mathrm{FDR} := I-\lim_{\Lambda \to \infty}\left(\int_0^\Lambda \mathrm{d}x-
\int_{\mur}^\Lambda \mathrm{d}x \frac{M}{x}\right)= M \ln\frac{M}{\mur}.
\eea
The advantage of the definition in \Eref{eq:fdrintegral1} is two-fold.
\begin{itemize}
\item  The UV cut-off $\Lambda$ is traded for $\mur$, which is interpreted, straight away, as the renormalization scale.\item  Other than logarithmic UV divergences never contribute.
\end{itemize}  
The use of \Eref{eq:fdrintegral1} is inconvenient in practical calculations,
owing to the explicit appearance of $\mur$ in the integration interval.  An equivalent definition is obtained by adding an auxiliary unphysical scale $\mu$ to $x$,
\bea
\label{eq:xtoxbar}
x \to \bar x:= x+\mu,
\eea
and introducing an integral operator $\int_0^\infty [\mathrm{d}x]$, defined in such a way that it annihilates the FDR vacua before integration. Thus
\bea
\label{eq:FDR2}
I_\mathrm{FDR} = \int_0^\infty [\mathrm{d}x]  
\frac{\bar x}{\bar x+M}
= \int_0^\infty [\mathrm{d}x]  
\left(
1-\frac{M}{\bar x}
+\frac{M^2}{\bar x(\bar x+M)}
\right)
:=  M^2 \left.\lim_{\mu \to 0} \int_0^\infty \mathrm{d}x
\frac{1}{\bar x(\bar x+M)}
\right|_{\mu = \mur}\!\!\!\!\!,
\eea
where $\mu \to 0$ is an asymptotic limit.
Note that, in order to keep the structure of the subtracted terms as in \Eref{eq:integrand}, the replacement $x \to \bar x$ must be performed in
{\em both the numerator and the denominator} of the integrated function.

This strategy can be extended to more dimensions and to integrands that are rational functions of the integration variables, as is the case of multiloop integrals. For instance, typical two-loop integrals contributing
to $\sigma_\mathrm{V}(\upgamma^\ast \to \mathrm{jets})$ and $\sigma_\mathrm{V}(\mathrm{H} \to \mathrm{b \bar b}+\mathrm{jets})$ are
\bea
\label{eq:K}
K_1 := \int \left  [\mathrm{d}^4q_1 \right ] \left [\mathrm{d}^4q_2 \right ] \frac{1}{\bar q^2_1 \bar D_{1} \bar D_{2}  
\bar q^2_2  \bar q^2_{12}}, \qquad
K_2^{\rho \sigma \alpha \beta} := \int \left [\mathrm{d}^4q_1 \right ] \left  [\mathrm{d}^4q_2
\right ] \frac{q_2^\rho q_2^\sigma q_1^\alpha q_1^\beta}{\bar q^4_1 \bar D_{1} \bar D_{2}  
\bar q^2_2  \bar q^2_{12}},
\eea
where $q_{12}:= q_1+q_2$, $\bar D_{1,2}=  \bar q_1^2+2 (q_1\cdot p_{1,2})$,  $p_{1,2}^2=0$, and $\bar q_i^2:= q_i^2 -\mu^2$ ($i= 1,2,12$), in the same spirit as \Eref{eq:xtoxbar}.

FDR integration keeps shift invariance in any of the loop integration variables and the possibility of cancelling reconstructed denominators, \eg
\bea
\int \left [\mathrm{d}^4q_1 \right ] \left  [\mathrm{d}^4q_2 \right ] \frac{\bar q_1^2}{\bar q^4_1 \bar D_{1} \bar D_{2}  
\bar q^2_2  \bar q^2_{12}}= K_1.
\eea
Since, instead,
\[
 \int \left [\mathrm{d}^4q_1 \right ] \left [\mathrm{d}^4q_2 \right ] \frac{q_1^2}{\bar q^4_1 \bar D_{1} \bar D_{2} \bar q^2_2  \bar q^2_{12}} \ne K_1 \, ,
\]
 this last property is maintained only if the replacement $q_i^2 \to \bar q_i^2$ is also made in the numerator of the loop integrals whenever $q_i^2$ is generated by Feynman rules. This is called {\em global prescription} (GP), often denoted  $q_i^2 \toGP \bar q_i^2$.

GP and shift invariance guarantee results that do not depend on the chosen gauge \cite{Donati:2013iya,Donati:2013voa}. Nevertheless, unitarity should also be maintained. This requires that any given UV divergent subdiagram produce the same result when computed or manipulated separately or when embedded in the full diagram. Such a requirement is called {\em subintegration consistency} (SIC) \cite{Page:2015zca}. 
Enforcing SIC in the presence of IR-divergent integrals, such as those in \Eref{eq:K}, needs extra care. In fact, the IR treatments of $\sigma_\mathrm{V}$ and $\sigma_\mathrm{R}$ should match  each other. In the next sections, I describe how this is achieved in the computation of the observable in \Eref{eq:sigmaBVR}.

\subsection{Keeping unitarity in the virtual component}
\label{sec:virt}
Any integral contributing to $\sigma_\mathrm{V}$ has the form
\bea
\label{eq:virtint}
I_\mathrm{V}= \int \left [\mathrm{d}^4q_1 \right ] \left  [\mathrm{d}^4q_2 \right] \frac{N_\mathrm{V}}{\bar D \bar q^2_2  \bar q^2_{12}},
\eea
where $\bar D$ collects all $q_2$-independent propagators and $N_\mathrm{V}$ is the numerator of the corresponding Feynman diagram.
$I_\mathrm{V}$ can be subdivergent or globally divergent for large values of the integration momenta. For example, $K_1$ in \Eref{eq:K} only diverges when $q_2 \to \infty$, while
$K_2$ also diverges when $q_{1,2} \to \infty$. This means that FDR prescribes the subtraction of a {\em global vacuum} (GV) involving both integration variables in $K_2$, while the {\em subvacuum} (SV) developed when $q_2 \to \infty$ should be removed from both $K_1$ and $K_2$.
In addition, IR infinities are generated by the on-shell conditions $p^2_{1,2}= 0$. Even though IR divergences are automatically regulated when barring the loop denominators, a careful SIC preserving treatment is necessary in order not to spoil unitarity. Since the only possible UV subdivergence is produced by the quark loop in \Fref{fig:BVR}(b), this is accomplished as follows \cite{Page:2018ljf}.
\begin{itemize}
\item  One does not apply GP to the contractions
$g_{\rho \sigma} q_2^\rho q_2^\sigma$ when $g_{\rho \sigma}$ refers to indices external to the UV divergent subdiagram.
\item One replaces everywhere $\bar q_1^2 \to q_1^2$ {\em after} GV subtraction.
\end{itemize}

\begin{figure}
\vspace*{-8cm}
\hspace*{1.2cm}
\centerline{
\includegraphics[width=1.6\textwidth]{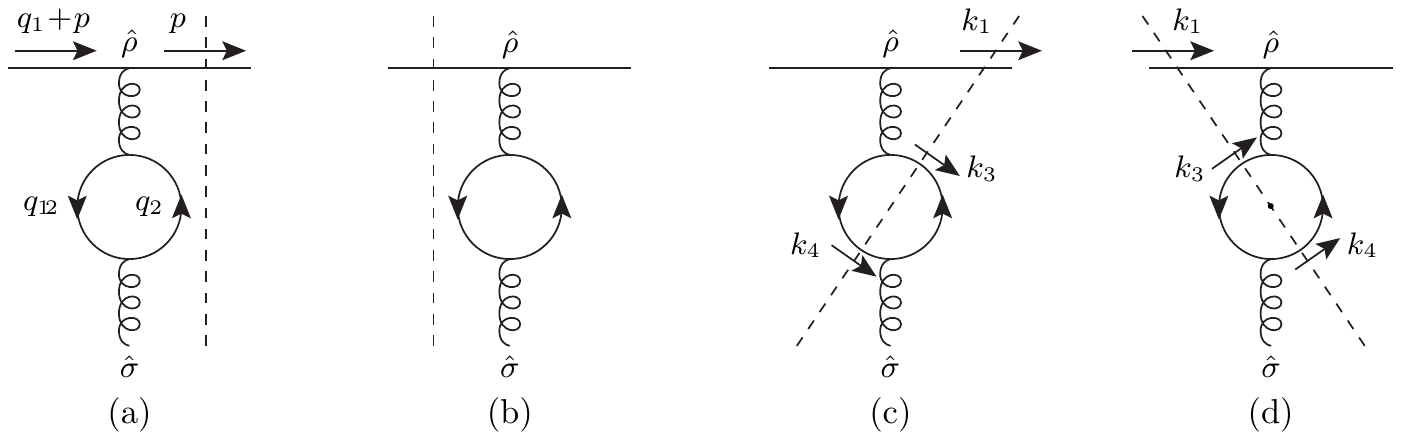}
}
\vspace*{-7.6cm}
\caption{\label{fig:cuts} Virtual and real cuts contributing to the IR-divergent parts of $\sigma_\mathrm{V}$ (a,b) and $\sigma_\mathrm{R}$ (c,d).}
\end{figure}

The external indices entering the calculation of
$\sigma_\mathrm{V}$ in \Eref{eq:sigmaBVR} are denoted  $\hat \rho$ and $\hat \sigma$
in \Fref{fig:cuts}(a,b). Using this convention, one can rephrase the first rule as follows:
$g_{\rho \sigma}q_2^\rho q_2^\sigma= q_2^2 \toGP \bar q^2_2$,
but
$g_{\hat \rho \hat \sigma}q_2^\rho q_2^\sigma := \hat q_2^2 \toGP q_2^2$,
which gives, for instance,
\begin{align}
g_{\rho \sigma}  K_2^{\rho \sigma \alpha \beta} \toGP
\bar K_2^{\alpha \beta} & =
\int \left [\mathrm{d}^4q_1 \right ]  \frac{q_1^\alpha q_1^\beta}{\bar q^4_1 \bar D_{1} \bar D_{2}}  
\!\int \left [\mathrm{d}^4q_2 \right ] \frac{1}{\bar q^2_{12}} = 0,\,
{\rm but}~\nl
g_{\hat \rho \hat \sigma}  K_2^{\rho \sigma \alpha \beta} \toGP
\hat K_2^{\alpha \beta} & = 
\int \left [\mathrm{d}^4q_1 \right ] \left  [\mathrm{d}^4q_2 \right ] \frac{q_2^2 q_1^\alpha q_1^\beta}{\bar q^4_1 \bar D_{1} \bar D_{2}  
  \bar q^2_2  \bar q^2_{12}} \ne 0,
\end{align}
where $\bar K_2^{\alpha \beta}$ vanishes because the shift $q_2 \to q_2 -q_1$ makes it proportional to the subvacuum $1/\bar q^2_2$, which is annihilated by the $\int [\mathrm{d}^4q_2]$ operator.
It can be shown \cite{Page:2015zca,Page:2018ljf} that integrals such as $\hat K_2^{\alpha \beta}$ generate the unitarity-restoring logarithms missed by $\bar K_2^{\alpha \beta}$.

As for the second rule, it states that a GV subtraction is needed first.
In the case of $\hat K_2^{\alpha \beta}$, this is achieved by rewriting
\[
\frac{1}{\bar D_1}=
\frac{1}{\bar q^2_1}-
\frac{2(q_1 \cdot p_1)}{\bar D_1 \bar q^2_1} \, .
\]
The first term gives a scaleless integral, annihilated by $\int [\mathrm{d}^4q_1] [\mathrm{d}^4q_2]$, so that
\bea
\hat K_2^{\alpha \beta}= -2
\int \left [\mathrm{d}^4q_1 \right ] \left  [\mathrm{d}^4q_2 \right ] \frac{(q_1 \cdot p_1) q_2^2 q_1^\alpha q_1^\beta}{\bar q^6_1 \bar D_{1} \bar D_{2}  
  \bar q^2_2  \bar q^2_{12}},
\eea
which is now only subdivergent when $q_2 \to \infty$, as is $K_1$
in \Eref{eq:K}. After that, the
replacement $\bar q_1^2 \to q_1^2$ produces
\bea
K_1 \to \tilde K_1 =
\int \mathrm{d}^4q_1 \left [\mathrm{d}^4q_2 \right] \frac{1}{q^2_1 D_{1} D_{2}  
  \bar q^2_2  \bar q^2_{12}},~
\hat K_2^{\alpha \beta} \to  \tilde K_2^{\alpha \beta}= -2
\int \mathrm{d}^4q_1 \left [\mathrm{d}^4q_2 \right ] \frac{(q_1 \cdot p_1) q_2^2 q_1^\alpha q_1^\beta}{q^6_1 D_{1} D_{2}  
  \bar q^2_2  \bar q^2_{12}}.
\eea
All two-loop integrals $I_\mathrm{V}$ in \Eref{eq:virtint} should be treated in this way. In the case of the $N_\mathrm{F}$ part of $\sigma_\mathrm{V}(\upgamma^\ast \to \mathrm{jets})$ and $\sigma_\mathrm{V}(\mathrm{H} \to \mathrm{b \bar b}+\mathrm{jets})$, this produces three master integrals, which can be computed as described in Appendix D of Ref. \cite{Page:2018ljf}.

After loop integration, $\sigma_\mathrm{V}$ contains logarithms of $\mu^2$ of both UV and IR origin. The former should be replaced by logarithms of $\mur^2$, as dictated by \Eref{eq:FDR2}, while the latter compensate the IR behaviour of $\sigma_\mathrm{R}$. To disentangle the two cases, it is convenient to renormalise $\sigma_\mathrm{V}$ first. This involves expressing the bare strong coupling constant
$a^0 := {\alpha^0_\mathrm{S}}/{4 \uppi}$ and the bare bottom Yukawa coupling
$y^0_\mathrm{b}$ in terms of $a := {\alpha_\mathrm{S}^{\mbox{\tiny ${\rm \overline{MS}}$}}(s)}/{4 \uppi}$ and $y_\mathrm{b}$ extracted from the the bottom pole mass $m_\mathrm{b}$. 
The relevant relations in terms of
$L := \ln {\mu^2}/{(p_1-p_2)^2}$ and $L^{\prime \prime}:= \ln {\mu^2}/{m_\mathrm{b}^2}$
are  \cite{Page:2018ljf}
\bea
\label{eq:a0}
a^0 = a \left(1+a {\delta}^{(1)}_a  \right),~\, \qquad 
y^0_\mathrm{b}= y_\mathrm{b} \left(
1+a \delta^{(1)}_y + a^2
\left(\delta^{(2)}_y+ {\delta}^{(1)}_a \delta^{(1)}_y 
\right)
\right),
\eea
with 
\bea
{\delta}^{(1)}_a =  \frac{2}{3} N_\mathrm{F} L,~\, \quad 
\delta^{(1)}_y   = -C_\mathrm{F} \left(3 L^{\prime \prime} +5  \right),~\,
\quad 
\delta^{(2)}_y   =  C_\mathrm{F} N_\mathrm{F} \left({L^{\prime \prime}}^2
+\frac{13}{3}{L^{\prime \prime}}
+ \frac{2}{3} \uppi^2+\frac{151}{18}\right).
\eea
After renormalization, the remaining $\mu^2$s are the IR ones. 

\subsection{Keeping unitarity in the real component}
\label{sec:real}
The integrands in $\sigma_\mathrm{R}$ of \Eref{eq:sigmaBVR} are represented in \Fref{fig:cuts}(c,d). They are of the form
\bea
\label{eq:Jr}
J_\mathrm{R} = \frac{N_\mathrm{R}}{S s^\alpha_{34} s^\beta_{134}}, \qquad
s_{i\cdots j} := (k_i+\dots + k_j)^2, \qquad 
~0 \le \alpha, \quad \beta\le 2,
\eea
where $N_\mathrm{R}$ is the numerator of the amplitude squared and $S$ collects the remaining propagators. Depending on the values of $\alpha$ and $\beta$, 
$J_\mathrm{R}$ becomes infrared divergent when integrated over $\Phi_{n+2}$. These IR singularities must be regulated consistently with the SIC preserving treatment of $\sigma_\mathrm{V}$ described in \Sref{sec:virt}.

The changes $q_2^2 \toGP \bar q^2_2$ and $q_{12}^2 \toGP \bar{q}_{12}^2$ in the virtual cuts of \Fref{fig:cuts}(a,b) imply the Cutkosky relation
\begin{equation}
  \label{eq:qtok}
\frac{1}{(\bar q_2^2+\mathrm{i}0^+)(\bar{q}_{12}^2+\mathrm{i}0^+)} \leftrightarrow
\left(\frac{2 \uppi}{\mathrm{i}}\right)^2
\delta_+(\bar k^2_3)
\delta_+(\bar k^2_4),
\end{equation}
with $\bar k^2_{3,4}:= k^2_{3,4}-\mu^2$.
Hence, one replaces in \Eref{eq:sigmaBVR}
$\Phi_{n+2} \to \tilde \Phi_{n+2}$, where the phase space $\tilde \Phi_{n+2}$ is such that $k^2_3=k^2_4= \mu^2$ and $k^2_i= 0$ when $i \ne 3,4$.
In Ref. \cite{Page:2018ljf}, it is proven that SV subtraction in $\sigma_\mathrm{V} $ does not alter \Eref{eq:qtok}. Analogously, the correspondence between cuts (a) and (d)
\bea
\label{eq:lastprop}
\frac{1}{(q_1+p)^2+\mathrm{i}0^+}  \leftrightarrow  \frac{2 \uppi}{\mathrm{i}} \delta_+(k^2_1)
\eea
is not altered by GV subtraction. Finally, $k^2_3$, $k^2_4$, and
$(k_3+k_4)^2= s_{34}$ in $N_\mathrm{R}$ of \Eref{eq:Jr} should be treated using the same prescriptions imposed on $q^2_2$, $q^2_{12}$, and $q^2_1$ in $N_\mathrm{V}$ of \Eref{eq:virtint}, respectively.
This means replacing
\bea
\label{eq:gprer}
k^2_{3,4} \to \bar k^2_{3,4} = 0, \qquad 
(k_3 \cdot k_4)= \frac{1}{2} \left(s_{34}-k^2_3-k^2_4 \right) \to
\frac{1}{2}(s_{34}-\bar k^2_3 -\bar k^2_4) = \frac{1}{2} s_{34}, 
\eea
where the last equalities are induced by the delta functions in \Eref{eq:qtok}.
These changes should be made everywhere in $N_\mathrm{R}$, except in contractions induced by the external indices $\hat \rho$ and $\hat \sigma$ in cuts (c,d). In this case 
\bea
\label{eq:nogpr}
g_{\hat \rho \hat \sigma } k^\rho_{3,4} k^\sigma_{3,4} \to k^2_{3,4}= \mu^2,
\qquad 
g_{\hat \rho \hat \sigma } k^\rho_3 k^\sigma_4 \to (k_3 \cdot k_4) = \frac{s_{34}-2 \mu^2}{2}.
\eea
In the case of the $N_\mathrm{F}$ part of $\sigma_\mathrm{R}(\upgamma^\ast \to \mathrm{jets})$ and $\sigma_\mathrm{R}(\mathrm{H} \to \mathrm{b} \bar {\mathrm{b}} + \mathrm{jets})$, integrating $J_\mathrm{R}$ over $\tilde \Phi_{4}$ and taking the asymptotic $\mu \to 0$ limit produces the phase space integrals reported in Appendix E of Ref. \cite{Page:2018ljf}.

\subsection{Results and conclusions}
Using the approach outlined in Sections \ref{sec:virt} and \ref{sec:real}, one reproduces the known ${\rm \overline{MS}}$ results for the $N_\mathrm{F}$ components of 
$\sigma^\mathrm{NNLO}(\mathrm{H} \to \mathrm{b \bar b} +\mathrm{jets})$ and $\sigma^\mathrm{NNLO}(\upgamma^\ast \to \mathrm{jets})$ \cite{Page:2018ljf}
\begin{align}
\label{eq:gammafin}
\sigma^\mathrm{NNLO}(\mathrm{H} \to \mathrm{b \bar b} + \mathrm{jets}) & =
\Gamma_{\mbox{\tiny BORN}}(y_\mathrm{b}^{\mbox{\tiny ${\rm \overline{MS}}$}}(M_\mathrm{H}))
\bigg\{
1+a^2 C_\mathrm{F} N_\mathrm{F} \left(8 \zeta_3
+\frac{2}{3}\uppi^2 - \frac{65}{2}
\right)
\bigg\}, \nl
\sigma^\mathrm{NNLO}(\upgamma^\ast \to \mathrm{jets}) &= \sigma_{\mbox{\tiny BORN}}
\left\{1+
 a^2 C_\mathrm{F} N_\mathrm{F} 
\left(
8 \zeta_3 -11
\right)
\right\}. 
\end{align}
This shows, for the first time,  that a fully four-dimensional framework
to compute NNLO quark-pair corrections can be constructed based on the requirement of preserving gauge invariance and unitarity.
The basic principles leading to a consistent treatment of all the parts contributing to the  NNLO results in \Eref{eq:gammafin} are also expected to remain valid  when considering more complicated environments.
A general four-dimensional NNLO procedure including initial-state IR singularities is currently under investigation.

\end{bibunit}

\label{sec-mtools-pittau}  

\clearpage \pagestyle{empty} \cleardoublepage
%============================================

\pagestyle{fancy}
\fancyhead[CO]{\thechapter.\thesection \hspace{1mm} Unsubtractions at NNLO}
\fancyhead[RO]{}
\fancyhead[LO]{}
\fancyhead[LE]{}
\fancyhead[CE]{}
\fancyhead[RE]{}
\fancyhead[CE]{
J.J. Aguilera-Verdugo,
F. Driencourt-Mangin,
J. Plenter,
S. Ram\'{\i}rez-Uribe,
G. Rodrigo,
G.F.R. Sborlini,
W.J. Torres Bobadilla,
S. Tracz}
\lfoot[]{}
\cfoot{-  \thepage \hspace*{0.075mm} -}
\rfoot[]{}

%%% definitions %%%%%%%%%%%%%%%%%%%%%%%%%%
\def\beqn{\begin{eqnarray}} 
\def\eeqn{\end{eqnarray}}
\def\beeq{\begin{eqnarray}}
\def\eeeq{\end{eqnarray}}
\def\nn{\nonumber}
\def\alphas{\alpha_{\rm S}}
\def\td#1{\tilde{\delta}\left(#1\right)}
\def\ii{\mathrm{i} 0}
    
\begin{bibunit}[elsarticle-num] 
% define the bib-style for the unit: elsarticle-num.bst
% text-1; this is the corresponding section
% \putbib[2loops] % the *.bib
% \end{bibunit}
% go-on
%--- from: bibunits.sty, adapts the font size of ``References'' to section
\let\stdthebibliography\thebibliography
\renewcommand{\thebibliography}{%
\let\section\subsection
\stdthebibliography}
%---
    
\section
[Unsubtractions at NNLO\\ {\it 
J.J.~Aguilera-Verdugo, F.~Driencourt-Mangin,
J.~Plenter,~S.~Ram\'{\i}rez-Uribe,
G.~Rodrigo, G.F.R.~Sborlini, W.J. Torres Bobadilla, S.~Tracz}]
{Unsubtractions at NNLO \label{contr:RODRIGO}}
\noindent
{\bf Contribution\footnote{This contribution should be cited as:\\
J.J.~Aguilera-Verdugo, F.~Driencourt-Mangin,
J.~Plenter,~S.~Ram\'{\i}rez-Uribe,
G.~Rodrigo, G.F.R.~Sborlini, W.J. Torres Bobadilla, S.~Tracz, Unsubtractions at NNLO,  
%04 DOI:10.23731/CYRM-2020-XXX.\thepage, in:
%04 \url{http://dx.doi.org/10.23731/CYRM-2020-XXX.\thepage}, in:
DOI: \href{http://dx.doi.org/10.23731/CYRM-2020-003.\thepage}{10.23731/CYRM-2020-003.\thepage}, in:
Theory for the FCC-ee, Eds. A. Blondel, J. Gluza, S. Jadach, P. Janot and T. Riemann,\\
CERN Yellow Reports: Monographs, CERN-2020-003,
%04 \url{http://dx.doi.org/10.23731/CYRM-2020-XXX}, p. \thepage.} 
DOI: \href{http://dx.doi.org/10.23731/CYRM-2020-003}{10.23731/CYRM-2020-003},
p. \thepage.
\\ \copyright\space CERN, 2020. Published by CERN under the 
%04-2
\href{http://creativecommons.org/licenses/by/4.0/}{Creative Commons Attribution 4.0 license}.} by: J.J. Aguilera-Verdugo, 
F. Driencourt-Mangin,
J. Plenter,
S. Ram\'{\i}rez-Uribe,
G. Rodrigo,
G.F.R. Sborlini,
W.J. Torres Bobadilla,
S. Tracz \\
Corresponding author: G. Rodrigo {[german.rodrigo@csic.es]}}
\vspace*{.5cm}

%%%%%%%%%%%%%%%%%%%%%%%%%%%%%%%%%%%%%%%%%%%%%%%%%%%%%%%%%%%%%%%%%%%%%%%%%%%%%%%%%%%%%%%%%%%%%%%%%%%%%%%%%%%%%%%%%%%%%%%%%%%
%%%%%%%%%%%%%%%%%%%%%%%%%%%%%%%%%%%%%%%%%%%%%%%%%%%%%%%%%%%%%%%%%%%%%%%%%%%%%%%%%%%%%%%%%%%%%%%%%%%%%%%%%%%%%%%%%%%%%%%%%%%

\subsection[Introduction]{Introduction}
\label{sec:Introduction:rodrigo}

Computations in perturbative quantum field theory (pQFT) feature several aspects that, although intrinsically non-physical,
are traditionally successfully eluded by modifying the dimensions of  space--time. Closed loops in pQFT implicitly extrapolate 
the validity of the Standard Model (SM) to infinite energies---equivalent to zero distance---much above the Planck scale. 
We should expect this to be a legitimate procedure if the loop scattering amplitudes that contribute to the physical observables 
are either suppressed at very high energies, or if there is a way to suppress or renormalise their contribution in this limit. 
In gauge theories like QCD, massless particles can be emitted with zero energy, and pQFT treats the quantum state with 
$N$ external partons as different from the quantum state with  emission of extra massless particles at zero energy, while 
these two states are physically identical. In addition, partons can be emitted in exactly the same direction, or, in other words, at zero distance. 
All these unphysical features have a price and lead to the emergence of infinities in the four dimensions of space--time.

In dimensional regularisation (DREG)~\cite{Bollini:1972ui,tHooft:1972tcz,Cicuta:1972jf,Ashmore:1972uj,Wilson:1972cf},
the infinities are replaced by explicit poles in $1/\varepsilon$, with $d=4-2\varepsilon$, through integration of the loop momenta 
and the phase space of real radiation. Then, the $1/\varepsilon$ ultraviolet (UV) singularities of the virtual contributions are removed by renormalization, 
and the infrared (IR) soft and collinear singularities are subtracted. 
The general idea of subtraction~\cite{Kunszt:1992tn,Frixione:1995ms,Catani:1996jh,Catani:1996vz,GehrmannDeRidder:2005cm,Catani:2007vq,Czakon:2010td,Bolzoni:2010bt,Boughezal:2015dva,Gaunt:2015pea,DelDuca:2016ily,Caola:2017dug,Magnea:2018hab}  
involves introducing counterterms that mimic the local IR behaviour
of the real components and that can easily be integrated analytically in $d$ dimensions. In this way, the integrated
form is combined with the virtual component, while the unintegrated counterterm cancels the IR
poles originated from the phase space integration of the real-radiation contribution. 

Although this procedure efficiently transforms the theory into a calculable and well-defined mathematical framework, 
a big effort needs to be invested in evaluating loop and phase space integrals in arbitrary space--time dimensions, 
which are particularly difficult at higher perturbative orders. 
In view of the highly challenging demands imposed by the expected accuracy attainable at the LHC and future colliders, like the FCC, 
there has been a recent interest in the community to define perturbative methods directly in $d=4$ space--time dimensions in order 
to avoid the complexity of working in a non-physical multidimensional space~\cite{Gnendiger:2017pys}. 
Examples of these methods are the four-dimensional formulation (FDF)~\cite{Fazio:2014xea} of the four-dimensional helicity scheme,
the six-dimensional formalism (SDF)~\cite{Bern:2002zk}, implicit regularisation (IREG)~\cite{Battistel:1998sz,Fargnoli:2010ec}, 
and four-dimensional regularisation or renormalization (FDR)~\cite{Donati:2013voa,Page:2018ljf}.\footnote{See Section~C.\ref{contr:pittau}  in this report.} 
In this section, we review the four-dimensional unsubtraction  (FDU)~\cite{Hernandez-Pinto:2015ysa,Sborlini:2016gbr,Sborlini:2016hat} method, which is based on  loop-tree duality (LTD)~\cite{Catani:2008xa,Bierenbaum:2010cy,Bierenbaum:2012th,Buchta:2014dfa,Buchta:2015wna,Driencourt-Mangin:2017gop,Driencourt-Mangin:2019aix,Aguilera-Verdugo:2019kbz}. 
The idea behind FDU is to exploit a suitable mapping of momenta between the virtual and real kinematics in such a way that the summation 
over the degenerate soft and collinear quantum states is performed locally at integrand level without the necessity of introducing IR subtractions,
whereas the UV singularities are locally suppressed at very high energies, \eg at two loops~\cite{Driencourt-Mangin:2019aix}.
The method should improve the efficiency of Monte Carlo event generators because it 
simultaneously describes real and virtual contributions. 

Finally, LTD is also a powerful framework to analyse the singular structure of scattering amplitudes directly in the loop 
momentum space, which is particularly interesting for characterizing unitarity thresholds and anomalous thresholds for specific 
kinematic configurations~\cite{Aguilera-Verdugo:2019kbz}. 

\subsection{Loop-tree duality}
\label{sec:LTD:rodrigo}

The LTD representation of a one-loop scattering amplitude is given by
\begin{eqnarray}
{\cal A}^{(1)} (\{p_n\}_N) = - \int_{\ell} {\cal N}(\ell,\{p_n\}_N) \otimes G_D(\alpha)~, 
\label{eq:LTDoneloop}
\end{eqnarray}
where $G_D(\alpha) = \sum_{i\in \alpha} \td{q_i} \prod_{j\ne i}G_D(q_i; q_j)$,
and ${\cal N} (\ell,\{p_n\}_N)$ is the numerator of the integrand, which depends on the loop momentum $\ell$ and 
the external momenta $\{p_n\}_N$.
The delta function $\td{q_i} = \mathrm{i} 2 \uppi \, \theta(q_{i,0}) \, \delta (q_i^2-m_i^2)$
sets on-shell the internal propagator with momentum $q_i=\ell+k_i$ and selects its positive energy mode, $q_{i,0}>0$.
At one loop, $\alpha=\{1, \dots, N\}$ labels all the internal momenta, and \Eref{eq:LTDoneloop} is the sum of $N$ single-cut
dual amplitudes. The dual propagators,
\begin{equation}
G_D(q_i;q_j) = \frac{1}{q_j^2-m_j^2 - \ii \, \eta \cdot k_{ji}}~,  
\end{equation}
differ from the usual Feynman propagators only by the imaginary prescription,
which now depends on 
$\eta \cdot k_{ji}$, with $k_{ji} = q_j-q_i$. The dual propagators are implicitly linear in the loop momentum, owing to the on-shell conditions.
With $\eta = (1, {\bf 0})$, which is equivalent to integrating out the energy 
component of the loop momentum, the remaining integration domain is Euclidean.  

\begin{figure}
\begin{center}
\includegraphics[width=0.35\textwidth]{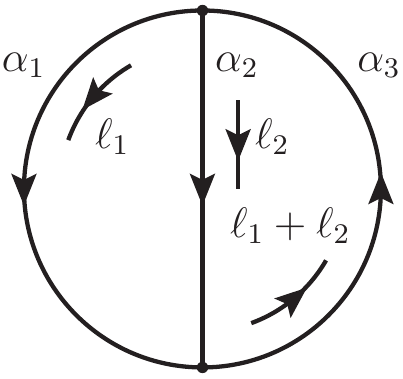}
\caption{Momentum flow of a two-loop Feynman diagram
\label{fig:twoloop}}
\end{center}
\end{figure}

At two loops, the corresponding dual representation is~\cite{Bierenbaum:2012th,Driencourt-Mangin:2019aix}
\begin{multline}
{\cal A}^{(2)}(\{p_n\}_N)  = \int_{\ell_1}\int_{\ell_2} {\cal N}(\ell_1,\ell_2,\{p_n\}_N) \otimes 
 [ G_D(\alpha_1) \, G_D(\alpha_2\cup \alpha_3) 
+ G_D(-\alpha_2\cup \alpha_1) G_D(\alpha_3)  \\
- G_D(\alpha_1) \,G_F(\alpha_2) \, G_D(\alpha_3) ]~.
\label{eq:LTDtwoloop}
\end{multline}
Now, the internal momenta are $q_i = \ell_1+k_i$, $q_j = \ell_2 + k_j$, and $q_k = \ell_1+ \ell_2 + k_k$,
and are classified into three different sets, with $i \in \alpha_1$, $j \in \alpha_2$, and $k \in \alpha_3$ (see \Fref{fig:twoloop}). 
The minus sign in front of $\alpha_2$ indicates that the momenta in $\alpha_2$ are reversed to hold
a momentum flow consistent with $\alpha_1$.
The dual representation in \Eref{eq:LTDtwoloop} spans over the sum of all possible double-cut contributions, 
with each of the two cuts belonging to a different set. 
In general, at higher orders, LTD transforms any loop integral or loop scattering 
amplitude into a sum of tree-level-like objects that are constructed by setting on-shell a number 
of internal loop propagators equal to the number of loops.

%%%%%%%%%%%%%%%%%%%%%%%%%%%%%%%%%%%%%%%%%%%%%%%%%%%%%%%%%%%%%%%%%%%%%%%%%%%%%%%%%
%%%%%%%%%%%%%%%%%%%%%%%%%%%%%%%%%%%%%%%%%%%%%%%%%%%%%%%%%%%%%%%%%%%%%%%%%%%%%%%%%

Explicit LTD representations of the scattering amplitude describing the decay $\mathrm{H}\to \upgamma \upgamma$ 
have been presented at one \cite{Driencourt-Mangin:2017gop} and two loops~\cite{Driencourt-Mangin:2019aix}.

%%%%%%%%%%%%%%%%%%%%%%%%%%%%%%%%%%%%%%%%%%%%%%%%%%%%%%%%%%%%%%%%%%%%%%%%%%%%%%%%
%%%%%%%%%%%%%%%%%%%%%%%%%%%%%%%%%%%%%%%%%%%%%%%%%%%%%%%%%%%%%%%%%%%%%%%%%%%%%%%%
\subsection{Four-dimensional unsubtraction}
\label{sec:FDU:rodrigo}

It is interesting to note that although in Eqs. (\ref{eq:LTDoneloop}) and (\ref{eq:LTDtwoloop}) the on-shell loop three-momenta are unrestricted,
all the IR and physical threshold singularities of the dual amplitudes are restricted 
to a compact region~\cite{Buchta:2014dfa,Aguilera-Verdugo:2019kbz}, as discussed in \Sref{sec:anomalous:rodrigo}.
This is essential to define the four-dimensional unsubtraction (FDU)~\cite{Hernandez-Pinto:2015ysa,Sborlini:2016gbr,Sborlini:2016hat} 
algorithm, namely, to establish a mapping between the real and virtual kinematics in order to locally cancel the IR singularities without 
the need for subtraction counterterms.

In the FDU approach, the cross-section at next-to-leading order (NLO) is constructed, as usual, 
from the renormalised one-loop virtual correction with $N$ external partons 
and the exclusive real cross-section with $N+1$ partons 
\begin{equation}
\sigma^{\rm NLO} = \int_{N} \mathrm{d}\sigma_{{\rm V}}^{(1,{\rm R})}+ \int_{N+1} \mathrm{d}\sigma_{{\rm R}}^{(1)}~,
\end{equation}
integrated over the corresponding phase space, $\int_N$ and $\int_{N+1}$. The virtual contribution is obtained from its LTD representation 
\begin{equation}
\int_N \mathrm{d}\sigma_{\rm V}^{(1,{\rm R})} = \int_{(N, \, {\vec \ell})} \, 2 \, {\rm Re} \, 
\langle {\cal M}^{(0)}_N|\bigg(\sum_i{\cal M}^{(1)}_N(\tilde \delta(q_i)) \bigg) 
- {\cal M}^{(1)}_{\rm UV} (\tilde \delta(q_{\rm UV})) \rangle  \,  \hat{\cal O}(\{p_n\}_N)~,
\label{eq:nlov}
\end{equation}
where ${\cal M}^{(0)}_N$ is the $N$-leg scattering amplitude at leading order (LO), and ${\cal M}^{(1)}_N (\tilde \delta(q_i))$ 
is the dual representation of the unrenormalised one-loop scattering amplitude with the internal momentum 
$q_i$ set on-shell. The integral is weighted with the explicit observable function $\hat{\cal O}(\{p_n\}_N)$. 
The expression includes appropriate counterterms,  ${\cal M}^{(1)}_{\rm UV} (\tilde \delta(q_{\rm UV}))$, 
that implement renormalization by subtracting the UV singularities 
locally, as discussed in Refs.~\cite{Sborlini:2016hat,Sborlini:2016gbr}, 
including UV singularities of degree higher than logarithmic that integrate to zero.

By means of an appropriate mapping between the real and virtual kinematics\cite{Sborlini:2016hat,Sborlini:2016gbr},
\begin{equation}
\{p_r'\}_{N+1} \to (q_i, \{p_n\}_N)~,
\label{mapping}
\end{equation}
the real phase space is rewritten in terms of the virtual phase space and the loop three-momentum
\begin{equation}
\int_{N+1}  =  \int_{(N, \, {\vec{\ell}})} \, \sum_i  {\cal J}_i(q_i)  \, {\cal R}_i(\{p_r'\}_{N+1})~,
\end{equation}
where ${\cal J}_i(q_i)$ is the Jacobian of the transformation with $q_i$ on-shell, and ${\cal R}_i(\{p_j'\}_{N+1})$ defines 
a complete partition of the real phase space
\begin{equation}
\sum_i {\cal R}_i (\{p_r'\}_{N+1}) = 1~.
\end{equation} 
As a result, the NLO cross-section is cast into a single integral in the Born or virtual phase space and the loop three-momentum
\begin{multline}
\sigma^{\rm NLO} =  \int_{(N, \, {\vec \ell})} \bigg[ \, 2 \, {\rm Re} \, \langle {\cal M}^{(0)}_N|\bigg(\sum_i{\cal M}^{(1)}_N(\tilde \delta(q_i)) \bigg)- {\cal M}^{(1)}_{\rm UV} (\tilde \delta(q_{\rm UV})) \rangle  \,  \hat{\cal O}(\{p_n\}_N)  \\
+ \sum_i {\cal J}_i(q_i) \, {\cal R}_i(\{p_r'\}_{N+1}) \, |{\cal M}^{(0)}_{N+1}(\{p_r'\}_{N+1})|^2 \, \hat{\cal O}(\{p_r'\}_{N+1}) \bigg]~,
\label{eq:nlovr}
\end{multline} 
and exhibits a smooth four-dimensional limit in such a way that it can be evaluated directly in four space--time dimensions. 
Explicit computations are presented in Refs.~\cite{Sborlini:2016gbr,Sborlini:2016hat}
with both massless and massive final-state quarks. More importantly, with suitable mappings in \Eref{mapping}
conveniently describing the quasi-collinear configurations,  the transition from the massive~\cite{Sborlini:2016hat}
to the massless configuration~\cite{Sborlini:2016gbr} is also smooth.

The extension of FDU to next-to-next-to-leading order (NNLO) is obvious; the total cross-section consists of three contributions 
\begin{equation}
\sigma^{\rm NNLO} = \int_{N} \mathrm{d}\sigma_{{\rm V}{\rm V}}^{(2,{\rm R})} + \int_{N+1} \mathrm{d}\sigma_{{\rm V}{\rm R}}^{(2,{\rm R})} 
+ \int_{N+2} \mathrm{d}\sigma_{{\rm R}{\rm R}}^{(2)}~,
\end{equation}
where the double virtual cross-section $\mathrm{d}\sigma_{{\rm V}{\rm V}}^{(2,{\rm R})}$ receives contributions from 
the interference of the two-loop with the Born scattering amplitudes,  the square 
of the one-loop scattering amplitude with $N$ external partons, the virtual-real cross-section 
$\mathrm{d}\sigma_{{\rm V}{\rm R}}^{(2,{\rm R})}$ includes the contributions from the interference of one-loop and tree-level scattering 
amplitudes with one extra external particle, and the double real cross-section $\mathrm{d}\sigma_{{\rm R}{\rm R}}^{(2)}$ comprises 
tree-level contributions with the emission of two extra particles. 
The LTD representation of the two-loop scattering amplitude,
$\langle {\cal M}^{(0)}_N|{\cal M}^{(2)}_N(\tilde \delta(q_i,q_j))\rangle$, is obtained from \Eref{eq:LTDtwoloop}, 
while the two-loop momenta of the squared one-loop 
amplitude are independent and generate dual contributions of the type \newline 
$\langle {\cal M}^{(1)}_N(\tilde \delta(q_i))|{\cal M}^{(1)}_N(\tilde \delta(q_j))\rangle$. In both cases, there are two independent
loop three-momenta and $N$ external momenta with which to reconstruct the 
kinematics of the tree-level corrections entering $\mathrm{d}\sigma_{{\rm R}{\rm R}}^{(2)}$
and the one-loop corrections in $\mathrm{d}\sigma_{{\rm V}{\rm R}}^{(2,{\rm R})}$:
\begin{equation}
\{p_r''\}_{N+2} \to (q_i, q_j, \{p_n\}_N)~, \qquad 
(q_k' , \{p_s'\}_{N+1} )  \to (q_i, q_j, \{p_n\}_N)~,
\end{equation}
in such a way that all the contributions to the NNLO cross-section are cast into a single phase space integral 
\begin{equation}
\sigma^{\rm NNLO} =  \int_{(N,  \, {\vec \ell_1}, \, {\vec \ell_2})}  \mathrm{d}\sigma^{\rm NNLO}~.
\end{equation}
Explicit applications of FDU at NNLO are currently under investigation. 

%%%%%%%%%%%%%%%%%%%%%%%%%%%%%%%%%%%%%%%%%%%%%%%%%%%%%%%%%%%%%%%%%%%%%%%%%%%%%%%%
%%%%%%%%%%%%%%%%%%%%%%%%%%%%%%%%%%%%%%%%%%%%%%%%%%%%%%%%%%%%%%%%%%%%%%%%%%%%%%%%
\subsection{Unitarity thresholds and anomalous thresholds}
\label{sec:anomalous:rodrigo}

An essential requirement for FDU to work is to prove that all the IR singularities of the dual amplitudes are restricted to a compact region 
of the loop three-momenta. This has recently been proven at higher orders~\cite{Aguilera-Verdugo:2019kbz},
thus extending the one-loop analysis of Ref.~\cite{Buchta:2014dfa}, as well
as analysing the case of anomalous thresholds. 
The location of the singularities of the dual amplitudes in the loop three-momentum space 
are encoded at one loop through the set of conditions
\begin{equation}
\lambda^{\pm\pm}_{ij} = \pm q_{i,0}^{(+)}  \pm q_{j,0}^{(+)}  + k_{ji,0} \to 0~, 
\label{eq:genericoneloop}
\end{equation}
where $q_{r,0}^{(+)} = \sqrt{\vec{q}_r^{~2}+m_r^2}$, with $r\in\{i,j\}$, are the on-shell loop energies. 
There are, indeed, only two independent solutions. The limit $\lambda^{++}_{ij}\to 0$ 
describes the causal unitarity threshold, and determines that $q_{r,0}^{(+)} < |k_{ji,0}|$, 
where $k_{ji,0}$ depends on the external momenta only and is therefore bounded. 
For massless partons, it also describes soft and collinear singularities. 
The other potential singularity occurs for $\lambda^{+-}_{ij}\to 0$, but this is a non-causal or unphysical threshold
and it cancels locally in the forest defined by the sum of all the on-shell dual contributions. 
For this to happen, the dual prescription of the dual propagators plays a central role. 
Finally, anomalous thresholds are determined by overlapping causal unitarity thresholds, 
\eg $\lambda^{++}_{ij}$ and $\lambda^{++}_{ik} \to 0$ simultaneously.

At two loops, the location of the singularities is determined by the set of conditions%~\cite{Driencourt-Mangin:2019aix}
\begin{equation}
\lambda^{\pm\pm\pm}_{ijk} = \pm q_{i,0}^{(+)}  \pm q_{j,0}^{(+)}  \pm q_{k,0}^{(+)} + k_{k(ij),0} \to 0~,
\label{eq:twoloopconditions}
\end{equation}
where $k_{k(ij)} = q_k-q_i-q_j$ depends on external momenta only,
with $i \in \alpha_1$, $j \in \alpha_2$, and $k \in \alpha_3$.
Now, the unitarity threshold is defined by the limit $\lambda^{+++}_{ijk}\to 0$  (or $\lambda^{---}_{ijk}\to 0$ )
with $q_{r,0}^{(+)} \le |k_{k(ij),0}|$ and $r\in\{i,j,k \}$, and the potential singularities 
at $\lambda^{++-}_{ijk}\to 0$ and $\lambda^{+--}_{ijk}\to 0$ cancel locally in the forest of all the dual contributions.
Again, the anomalous thresholds are determined by the simultaneous contribution of unitarity thresholds. 
The generalisation of \Eref{eq:twoloopconditions} to higher orders is straightforward. 

\subsection[Conclusions]{Conclusions}
\label{sec:conclusions:rodrigo}

The bottleneck in higher-order perturbative calculations is not only the evaluation of multiloop Feynman 
diagrams, but also the gathering of all the quantum corrections from different loop orders (and thus different numbers of final-state partons).  To match the expected experimental accuracy at the LHC, particularly in the high-luminosity phase, and at future colliders, 
new theoretical efforts are still needed to overcome the current precision frontier. LTD is also a powerful framework 
to analyse, comprehensively, the emergence of anomalous thresholds at higher orders.

%%%%%%%%%%%%%%%%%%%%%%%%%%%%%%%%%%%%%%%%%%%%%%%%%%%%%%%%%%%%%%%%%%%%%%%%%%%%%%%%
%%%%%%%%%%%%%%%%%%%%%%%%%%%%%%%%%%%%%%%%%%%%%%%%%%%%%%%%%%%%%%%%%%%%%%%%%%%%%%%%
%%%%%%%%%%%%%%%%%%%%%%%%%%%%%%%%%%%%%%%%%%%%%%%%%%%%%%%%%%%%%%%%%%%%%%%%%%%%%%%%
%%%%%%%%%%%%%%%%%%%%%%%%%%%%%%%%%%%%%%%%%%%%%%%%%%%%%%%%%%%%%%%%%%%%%%%%%%%%%%%%
%%%%%%%%%%%%%%%%%%%%%%%%%%%%%%%%%%%%%%%%%%%%%%%%%%%%%%%%%%%%%%%%%%%%%%%%%%%%%%%%
%%%%%%%%%%%%%%%%%%%%%%%%%%%%%%%%%%%%%%%%%%%%%%%%%%%%%%%%%%%%%%%%%%%%%%%%%%%%%%%%

\end{bibunit}

\label{sec-mtools-rodrigo}  

\clearpage \pagestyle{empty}  \cleardoublepage
%============================================

\pagestyle{fancy}
\fancyhead[CO]{\thechapter.\thesection \hspace{1mm} Numerics for elliptic Feynman integrals}
\fancyhead[RO]{}
\fancyhead[LO]{}
\fancyhead[LE]{}
\fancyhead[CE]{}
\fancyhead[RE]{}
\fancyhead[CE]{C. Bogner, I. H\"onemann, K. Tempest, A. Schweitzer, S. Weinzierl}
\lfoot[]{}
\cfoot{-  \thepage \hspace*{0.075mm} -}
\rfoot[]{}

\begin{bibunit}[elsarticle-num]

\let\stdthebibliography\thebibliography
\renewcommand{\thebibliography}{%
\let\section\subsection
\stdthebibliography}

\section
[Numerics for elliptic Feynman integrals \\ {\it C. Bogner, I. H\"onemann, K. Tempest, A. Schweitzer, S. Weinzierl}]
{Numerics for elliptic Feynman integrals
\label{contr:mtools_weinzierl}}
\noindent
{\bf Contribution\footnote{This contribution should be cited as:\\
C.~Bogner, I. H\"onemann, K. Tempest, A. Schweitzer, S. Weinzierl, Numerics for elliptic Feynman integrals,  
%04 DOI:10.23731/CYRM-2020-XXX.\thepage, in:
%04 \url{http://dx.doi.org/10.23731/CYRM-2020-XXX.\thepage}, in:
DOI: \href{http://dx.doi.org/10.23731/CYRM-2020-003.\thepage}{10.23731/CYRM-2020-003.\thepage}, in:
Theory for the FCC-ee, Eds. A. Blondel, J. Gluza, S. Jadach, P. Janot and T. Riemann,
CERN Yellow Reports: Monographs, CERN-2020-003,
%04 \url{http://dx.doi.org/10.23731/CYRM-2020-XXX}, p. \thepage.} 
DOI: \href{http://dx.doi.org/10.23731/CYRM-2020-003}{10.23731/CYRM-2020-003},
p. \thepage.
\\ \copyright\space CERN, 2020. Published by CERN under the 
%04-2
\href{http://creativecommons.org/licenses/by/4.0/}{Creative Commons Attribution 4.0 license}.} by: C. Bogner, I. H\"onemann, K. Tempest, A. Schweitzer, S. Weinzierl \\
Corresponding author: S. Weinzierl {[weinzierl@uni-mainz.de]}}
\vspace*{.5cm}

\noindent The Standard Model involves several heavy particles: 
the Z and W bosons, the Higgs boson, and the top quark.
Precision studies of these particles require, on the theoretical side, 
quantum corrections at the two-loop order and beyond.
It is a well-known fact that, starting from two loops,
Feynman integrals with massive particles can no longer be expressed in terms of multiple polylogarithms.
This  immediately raises the following question. 
What is the larger class of functions needed to express the relevant Feynman integrals?
For single-scale two-loop Feynman integrals related to a single elliptic curve we now have  the answer: 
they are expressed as iterated integrals of modular form \cite{Adams:2017ejb}.
This brings us to a second question:
is there an efficient method to evaluate these functions numerically in the full kinematic range?
In this contribution, we review how this can be done.
This review is mainly based on Refs. \cite{Bogner:2017vim,Honemann:2018mrb}.

Efficient numerical evaluation methods rely on three ingredients:
(i) an (iterated) integral rep\-re\-sen\-tation, used to transform the arguments into the region of convergence,
(ii) a (nested) sum representation, defined in the region of convergence, which can be truncated and gives a numerical approximation,
and (iii) methods to accelerate the convergence of the truncated series.
Let us illustrate this strategy for 
the numerical evaluation of the dilogarithm \cite{'tHooft:1979xw}, defined by
\begin{eqnarray}
 \mathrm{Li}_{2}(x) 
  =  \int\limits_{0}^{x} \frac{\mathrm{d}t_1}{t_1} \int\limits_0^{t_1} \frac{\mathrm{d}t_2}{1-t_2}
 = \sum\limits_{n=1}^{\infty} \frac{x^{n}}{n^{2}}.
\end{eqnarray}
The power series expansion can be evaluated numerically, provided $|x| < 1$.
Using the functional equations 
\begin{eqnarray}
 \mathrm{Li}_2(x) 
 = 
 -\mathrm{Li}_2\left(\frac{1}{x}\right) -\frac{\uppi^2}{6} -\frac{1}{2} \left( \ln(-x) \right)^2,
 & &
 \mathrm{Li}_2(x) 
 = 
 -\mbox{Li}_2(1-x) + \frac{\uppi^2}{6} -\ln(x) \ln(1-x),
 \nonumber
\end{eqnarray}
any argument of the dilogarithm can be mapped into the region
$|x| \le 1$ and $-1 \leq \mbox{Re}(x) \leq 1/2$.
The numerical computation can be accelerated  by using an expansion in $z=-\ln(1-x)$ and the
Bernoulli numbers $B_i$:
\begin{eqnarray}
\mathrm{Li}_2(x)  =  \sum\limits_{i=0}^\infty B_i \frac{z^{i+1}}{(i+1)!}.
\end{eqnarray}
Multiple polylogarithms are defined for $z_k \neq 0$ by \cite{Goncharov:1998,Borwein:2001,Moch:2001zr}
\begin{eqnarray}
 \label{mtools_weinzierl:Gfuncdef}
 G(z_1, \dots ,z_k;y)
  = 
 \int\limits_0^y \frac{\mathrm{d}y_1}{y_1-z_1}
 \int\limits_0^{y_1} \frac{\mathrm{d}y_2}{y_2-z_2} \ \dots
 \int\limits_0^{y_{k-1}} \frac{\mathrm{d}y_k}{y_k-z_k}.
\end{eqnarray}
This represents multiple polylogarithms as iterated integrals.
Alternatively, we may define multiple polylogarithms through a nested sum
\begin{eqnarray} 
\label{mtools_weinzierl:def_multiple_polylogs_sum}
 \mathrm{Li}_{m_1, \dots ,m_k}(x_1, \dots ,x_k)
   =  \sum\limits_{n_1>n_2>\dots>n_k>0}^\infty
     \frac{x_1^{n_1}}{{n_1}^{m_1}}\cdots \frac{x_k^{n_k}}{{n_k}^{m_k}}.
\end{eqnarray}
With the shorthand notation
\begin{eqnarray}
\label{mtools_weinzierl:Gshorthand}
 G_{m_1, \dots ,m_k}(z_1, \dots ,z_k;y)
  = 
 G(\underbrace{0, \dots ,0}_{m_1-1},z_1, \dots ,z_{k-1},\underbrace{0, \dots
 ,0}_{m_k-1},z_k;y),
\end{eqnarray}
where all $z_j$ for $j=1, \dots ,k$ are assumed to be non-zero, the two notations are related by
\begin{eqnarray}
\label{mtools_weinzierl:Gintrepdef}
 \mathrm{Li}_{m_1, \dots ,m_k}(x_1, \dots ,x_k)
  =  
 (-1)^k 
 G_{m_1, \dots ,m_k}\left( \frac{1}{x_1}, \frac{1}{x_1 x_2}, \dots , \frac{1}{x_1
 \cdots x_k};1 \right).
\end{eqnarray}
The numerical evaluation of multiple polylogarithms follows the same strategy \cite{Vollinga:2004sn}.
Using the integral representation, one transforms all arguments into a region, 
where the sum representation gives a converging power series expansion.
In addition, the H\"older convolution is used to accelerate the convergence of the series expansion.
The H\"older convolution reads (with $z_1 \neq 1$ and $z_k \neq 0$)
\begin{eqnarray}
\label{mtools_weinzierl:defhoelder}
G\left(z_1,\dots ,z_k; 1 \right) 
 = 
 \sum\limits_{j=0}^k \left(-1\right)^j 
  G\left(1-z_j, 1-z_{j-1}, \dots ,1-z_1; \frac{1}{2} \right)
  G\left( z_{j+1}, \dots , z_k; \frac{1}{2} \right).
\end{eqnarray}
Multiple polylogarithms are a special case of iterated integrals.
Let us briefly review Chen's definition of iterated integrals \cite{Chen:1977}:
let $M$ be an $n$-dimensional manifold and
\begin{eqnarray}
 \gamma & : & \left[0,1\right] \rightarrow M
\end{eqnarray}
a path with start point ${x}_\mathrm{i}=\gamma(0)$ and endpoint ${x}_\mathrm{f}=\gamma(1)$. 
Suppose further that $\omega_1$, \dots , $\omega_k$ are differential $1$-forms on $M$.
Let us write
\begin{eqnarray}
 f_j\left(\lambda\right) \mathrm{d}\lambda  =  \gamma^\ast \omega_j
\end{eqnarray}
for the pull-backs to the interval $[0,1]$.
For $\lambda \in [0,1]$ the $k$-fold iterated integral of $\omega_1$, \dots
, $\omega_k$ 
along the path $\gamma$ is defined by
\begin{eqnarray}
 I_{\gamma}\left(\omega_1, \dots ,\omega_k;\lambda\right)
  = 
 \int\limits_0^{\lambda} \mathrm{d}\lambda_1 f_1\left(\lambda_1\right)
 \int\limits_0^{\lambda_1} \mathrm{d}\lambda_2 f_2\left(\lambda_2\right)
 \
 \dots 
 \int\limits_0^{\lambda_{k-1}} \mathrm{d}\lambda_k f_k\left(\lambda_k\right).
\end{eqnarray}
For multiple polylogarithms, we have $\omega_j = \mathrm{d} \ln(\lambda-z_j)$.
A second special case is given by iterated integrals of modular forms \cite{Brown:2014aa}:
\begin{eqnarray}
 \omega_j  =  2 \uppi \mathrm{i} \; f_j\left(\tau\right) \; \mathrm{d}\tau,
\end{eqnarray}
where $f_j(\tau)$ is a modular form.
This type of iterated integral occurs in physics 
for the equal-mass sunrise integral \cite{Laporta:2004rb,Bloch:2013tra,Adams:2015ydq,Adams:2017ejb,Adams:2018yfj}
and the the kite integral \cite{Remiddi:2016gno,Adams:2016xah}.
A physical application is the two-loop electron self-energy in quantum electrodynamics, if the mass of the electron
is not neglected \cite{Sabry:1962,Honemann:2018mrb}.
This is a single-scale problem and we set $x=p^2/m^2$.
In all these examples, the complication is related to the equal-mass sunrise integral, which cannot be expressed in terms
of multiple polylogarithms.
This is related to the fact that the system of differential equations for this Feynman integral
contains an irreducible second-order differential operator \cite{Broadhurst:1993mw,MullerStach:2012mp,Adams:2017tga}
\begin{eqnarray}
 L  = 
 x \left(x-1\right) \left(x-9\right) \frac{\mathrm{d}^2}{\mathrm{d}x^2} 
 + \left(3x^2-20x+9\right) \frac{\mathrm{d}}{\mathrm{d}x}
 + x-3.
\end{eqnarray}
Let $\psi_1$ and $\psi_2$ be two independent solutions of the homogeneous equation
\begin{eqnarray}
\label{mtools_weinzierl:diff_eq}
 L \; \psi
  =  0.
\end{eqnarray}
$\psi_1$ and $\psi_2$ can be taken as the periods of the elliptic curve
\begin{eqnarray}
 E
 & : &
 w^2 - z
       \left(z + 4 \right) 
       \left[z^2 + 2 \left(1+x\right) z + \left(1-x\right)^2 \right]
 \; = \; 0.
\end{eqnarray}
One defines the modulus $k$ and the complementary modulus $k'$ by
\begin{eqnarray}
 k^2 
  = 
 \frac{16 \sqrt{x}}{\left(1+\sqrt{x}\right)^3 \left(3-\sqrt{x}\right)},
 \qquad 
 k'{}^2
  = 
 1 - k^2.
\end{eqnarray}
In a neighbourhood of $x=0$, the periods may be taken as
\begin{eqnarray}
\label{mtools_weinzierl:def_psi_0}
 \psi_{1,0}  =  
 \frac{4 K\left( k \right)}{\left(1+\sqrt{x}\right)^{\frac{3}{2}} \left(3-\sqrt{x}\right)^{\frac{1}{2}}},
 \qquad 
 \psi_{2,0} =  
 \frac{4 \mathrm{i} K\left( k' \right)}{\left(1+\sqrt{x}\right)^{\frac{3}{2}} \left(3-\sqrt{x}\right)^{\frac{1}{2}}}.
\end{eqnarray}
The complete elliptic integral $K(k)$ can be computed efficiently from the arithmetic-geometric mean
\begin{eqnarray}
 K\left(k\right)
  = 
 \frac{\uppi}{2 \; \mathrm{agm}\left(k',1\right)}.
\end{eqnarray}
The periods $\psi_1$ and $\psi_2$ generate a lattice. 
Any other basis of the lattice again gives
 two independent solutions of the homogeneous differential equation (\Eref{mtools_weinzierl:diff_eq}).
It is a standard convention to normalise one basis vector of the lattice to one:
$(\psi_2,\psi_1) \rightarrow (\tau, 1)$ where $\tau=\psi_2/\psi_1$ and $\mathrm{Im} \tau > 0$.
Let us now consider a change of basis:
\begin{eqnarray}
\begin{pmatrix}
 \psi_2' \\
 \psi_1' \\
 \end{pmatrix} 
 =
 \gamma
 \begin{pmatrix}
 \psi_2 \\
 \psi_1 \\
 \end{pmatrix} ,
\qquad 
 \gamma = 
 \begin{pmatrix}
 a & b \\
 c & d \\
 \end{pmatrix} .
\end{eqnarray}
The transformation should be invertible and preserve $\mathrm{Im}(\psi_2'/\psi_1') > 0$, therefore, 
$\gamma \in \mathrm{SL}_2\left({\mathbb Z}\right)$.
In terms of $\tau$ and $\tau'$, this yields
\begin{eqnarray}
 \tau' = \frac{a \tau +b}{c \tau +d}.
\end{eqnarray}
This is a modular transformation and we write $\tau'=\gamma(\tau)$.
Let us denote the complex upper half plane by $\mathbb{H}$.
A meromorphic function $f: \mathbb{H} \rightarrow \mathbb{C}$ is a modular form of modular weight $k$ 
for $\mathrm{SL}_2(\mathbb{Z})$,
if
(i) $f$ transforms under M\"obius transformations as $f\left( \tau' \right) = (c\tau+d)^k \cdot f(\tau)$ 
for all $\gamma \in \mathrm{SL}_2(\mathbb{Z})$,
(ii) $f$ is holomorphic on $\mathbb{H}$,
and (iii) $f$ is holomorphic at infinity.
Furthermore, one defines modular forms for congruence subgroups $\Gamma \subset \mathrm{SL}_2(\mathbb{Z})$ by
requiring property (i) only for $\gamma \in \Gamma$ (plus holomorphicity on ${\mathbb H}$ and at the cusps).
Relevant to us will be the congruence subgroup $\Gamma_1(6)$, defined by
\begin{eqnarray}
\Gamma_1(6) 
  = 
 \left( \begin{pmatrix}
  a & b \\ 
  c & d
 \end{pmatrix}   \in \mathrm{SL}_2(\mathbb{Z}): a,d \equiv 1\ \text{mod}\ 6, \; c \equiv 0\ \text{mod}\ 6  \right\}.
\end{eqnarray}
With $\psi_{1,0}$ and $\psi_{2,0}$ defined by \Eref{mtools_weinzierl:def_psi_0},
we set
\begin{eqnarray}
 \tau_{0}  =  \frac{\psi_{2,0}}{\psi_{1,0}},
\qquad 
 q_{0}  =  \mathrm{e}^{2 \mathrm{i} \uppi \tau_{0}}.
\end{eqnarray}
We then change the variable from $x$ to $\tau_0$ (or $q_0$) \cite{Bloch:2013tra}.
The differential equation for the master integrals $\vec{I}$ relevant to the two-loop electron self-energy then reads \begin{eqnarray}
 \frac{\mathrm{d}}{\mathrm{d}\tau_0} \vec{I}
  = 
 \varepsilon \; A(\tau_0) \; \vec{I},
\end{eqnarray}
where $A(\tau_0)$ is an $\varepsilon$-independent matrix whose entries are modular forms for $\Gamma_1(6)$ \cite{Adams:2017ejb,Adams:2018yfj}.
It follows immediately that all master integrals can be expressed in terms of iterated integrals of modular forms.

Let us now discuss how to evaluate numerically iterated integrals of modular forms in an efficient way.
The essential point is that modular forms have a $q$-expansion.
Using 
\begin{eqnarray}
 2 \uppi \mathrm{i} \; \mathrm{d}\tau_0  =  \frac{\mathrm{d}q_0}{q_0} \,
 ,
\end{eqnarray}
we may integrate term-by-term and obtain the $q_0$-expansion of the master integrals.
Truncating the $q_0$-series to the desired accuracy gives a polynomial in $q_0$.
This needs to be done only once.
The resulting polynomial can then be evaluated for different values of $q_0$ (or $x$) numerically.
Note that the conversion from $x$ to $q_0$ is also fast, since the complete elliptic integrals can be computed efficiently
with the help of the arithmetic-geometric mean.
Let us give an example.
One finds for the $\varepsilon^2$-term of the sunrise integral \cite{Honemann:2018mrb}
\begin{equation}
\label{mtools_weinzierl:expansion_sunrise}
 I_{6}^{(2)}
 = 
 3   \mathrm{Cl}_2\left(\frac{2\uppi}{3}\right)
 - 3 \sqrt{3}
   \left[
         q_0-\frac{5}{4} q_0^{2}+q_0^{3}-{\frac {11}{16}} q_0^{4}+{\frac {24}{25}} q_0^{5}
         -\frac{5}{4} q_0^{6}+{\frac {50}{49}} q_0^{7}-{\frac {53}{64}}
         q_0^{8}+q_0^{9}
   \right]
 + {\mathcal O}\left(q_0^{10}\right).
\end{equation}
We have $q_0=0$ for $x=0$ and \Eref{mtools_weinzierl:expansion_sunrise} gives a fast convergent series
in a neighbourhood of $x=0$.
We are interested in evaluating the master integrals in the full kinematic range $x \in {\mathbb R}$.
This raises the question: for which values $x \in {\mathbb R}$ do the $q_0$-series for the master integrals
converge?
Or phrased differently, for which values $x \in {\mathbb R}$ do we have $|q_0| < 1$?
It turns out that we have $|q_0| < 1$ for $x \in {\mathbb R} \backslash \{ 1,9,\infty \}$,
corresponding to $p^2 \in {\mathbb R} \backslash \{ m^2,9 m^2,\infty \}$ \cite{Bogner:2017vim}.
Thus, the $q_0$-series for the master integrals converge for all real values of $x$ except three points.
Let us stress that the $q_0$-series give the correct real and imaginary part of the master integrals,
as specified by Feynman's $\mathrm{i}\delta$ prescription.
To cover the three remaining points, $x \in \{1,9,\infty\}$, 
we recall that the periods $\psi_1$ and $\psi_2$ are not uniquely determined. 
By using four different choices for the pair of periods $(\psi_1,\psi_2)$, we may
define $q_0$, $q_1$, $q_9$ and $q_\infty$ such that
(i) the integration kernels are modular forms of $\Gamma_1(6)$
and (ii) $q_j=0$ for $x=j$ \cite{Honemann:2018mrb}.
This gives expansions around all singular points of the system of differential equations or---phrased differently---around all cusps of $\Gamma_1(6)$.
In particular, there is always a choice such that $|q_{j}| \lessapprox 0.163$ for all real values of $x$.
Truncation of the $q$-series to order ${\mathcal O}(q^{30})$ gives for the finite part 
of the two-loop electron self-energy a relative precision better than $10^{-20}$ for all real values $p^2/m^2$.

Although we focused on the two-loop electron self-energy, we expect the methods discussed here to be applicable
to any single-scale Feynman integral related to a single elliptic curve.
This is a significant step beyond Feynman integrals evaluating to multiple polylogarithms
and puts single-scale Feynman integrals related to a single elliptic curve on the same level of understanding as 
Feynman integrals evaluating to multiple polylogarithms.
With the ongoing research on Feynman integrals beyond multiple polylogarithms \cite{Laporta:2004rb,MullerStach:2011ru,Adams:2013nia,Bloch:2013tra,Remiddi:2013joa,Adams:2014vja,Bloch:2014qca,Adams:2015gva,Adams:2015ydq,Sogaard:2014jla,Bloch:2016izu,Tancredi:2015pta,Primo:2016ebd,Remiddi:2016gno,Adams:2016xah,Bonciani:2016qxi,vonManteuffel:2017hms,Adams:2017tga,Adams:2017ejb,Bogner:2017vim,Ablinger:2017bjx,Primo:2017ipr,Remiddi:2017har,Bourjaily:2017bsb,Hidding:2017jkk,Broedel:2017kkb,Broedel:2017siw,Broedel:2018iwv,Adams:2018yfj,Adams:2018bsn,Adams:2018kez,Broedel:2018qkq,Bourjaily:2018yfy,Bourjaily:2018aeq,Besier:2018jen,Mastrolia:2018uzb,Ablinger:2018zwz,Frellesvig:2019kgj,Broedel:2019hyg}
we may expect more results---in particular, on multiscale Feynman integrals beyond multiple polylogarithms---to be coming soon.

%------------------------------------------------------------------------------

\end{bibunit}

\label{sec-mtools-weinzierl}  

\clearpage \pagestyle{empty}  \cleardoublepage
%============================================

 \newcommand{\pysecdec}{{py{\textsc{SecDec}}}}
\newcommand{\secdec}{{{\textsc{SecDec}}}}
\newcommand{\secdecthree}{{{\textsc{SecDec}\ 3}}}

\pagestyle{fancy}
 \fancyhead[CO]{\thechapter \hspace{1mm} Sector decomposition and QMC integration in \pysecdec}
 \fancyhead[RO]{}
 \fancyhead[LO]{}
 \fancyhead[LE]{}
 \fancyhead[CE]{}
 \fancyhead[RE]{}
 \fancyhead[CE]{S. Borowka, G.~Heinrich, S.~Jahn, S.P.~Jones, M.~Kerner, J.~Schlenk}
 \lfoot[]{}
 \cfoot{-  \thepage \hspace*{0.075mm} -}
 \rfoot[]{}
 
 \begin{bibunit}[elsarticle-num] % define the bib-style for the unit: elsarticle-num.bst
        %  text-1; this is the corresponding section
        %\putbib[2loops] % the *.bib
        %\end{bibunit}
        % go-on
        %--- from: bibunits.sty, adapts the font size of ``References'' to section
        \let\stdthebibliography\thebibliography
        \renewcommand{\thebibliography}{%
                \let\section\subsection
                \stdthebibliography}
        %---
        
\section
[Numerical multiloop calculations: sector decomposition and QMC integration in \pysecdec \\{\it S.~Borowka, G.~Heinrich, S.~Jahn, S.P.~Jones, M.~Kerner, J.~Schlenk}]
{Numerical multiloop calculations: sector decomposition and QMC integration in \pysecdec
\label{contr:JONES}}
\noindent
{\bf 
Contribution\footnote{This contribution should be cited as:\\
S.~Borowka, G.~Heinrich, S.~Jahn, S.P.~Jones, M.~Kerner, J.~Schlenk, Numerical multiloop calculations: sector decomposition and QMC integration in \pysecdec,  
%04 DOI:10.23731/CYRM-2020-XXX.\thepage, in:
%04 \url{http://dx.doi.org/10.23731/CYRM-2020-XXX.\thepage}, in:
DOI: \href{http://dx.doi.org/10.23731/CYRM-2020-003.\thepage}{10.23731/CYRM-2020-003.\thepage}, in:
Theory for the FCC-ee, Eds. A. Blondel, J. Gluza, S. Jadach, P. Janot and T. Riemann,\\
CERN Yellow Reports: Monographs, CERN-2020-003,
%04 \url{http://dx.doi.org/10.23731/CYRM-2020-XXX}, p. \thepage.} 
DOI: \href{http://dx.doi.org/10.23731/CYRM-2020-003}{10.23731/CYRM-2020-003},
p. \thepage.
\\ \copyright\space CERN, 2020. Published by CERN under the 
%04-2
\href{http://creativecommons.org/licenses/by/4.0/}{Creative Commons Attribution 4.0 license}.} by: S.~Borowka, G.~Heinrich, S.~Jahn, S.P.~Jones, M.~Kerner, J.~Schlenk \\
Corresponding author: S.P.~Jones {[s.jones@cern.ch]}}
\vspace*{.5cm}

%\address[a]{Theoretical Physics Department, CERN, Geneva, Switzerland}
%\address[b]{Max Planck Institute for Physics, F\"ohringer Ring 6, 80805 M\"unchen, Germany}
%\address[c]{Physik-Institut, Universit{\"a}t Z{\"u}rich, Winterthurerstrasse 190, 8057 Z{\"u}rich, Switzerland}
%\address[d]{Institute for Particle Physics Phenomenology, University of Durham, Durham DH1 3LE, UK}

%=================================================================
\noindent The FCC-ee will allow the experimental uncertainties on several important observables, such
as the electroweak precision observables (EWPOs), to
be reduced by up to two orders of magnitude compared with the previous generation 
LEP and SLC experiments~\cite{Abrams:1989aw,Schael:2013ita}. To be able to best exploit this 
unprecedented boost in precision, it is also necessary for theoretical predictions to be known
with sufficient accuracy. In practice, this means that very high-order perturbative corrections
to electroweak precision observables and other processes will be required, both in the 
Standard Model (SM) and potentially also in BSM scenarios.

One of the key challenges for computing perturbative corrections is our ability to compute
the Feynman integrals that appear in these multiloop corrections. There has been very
significant progress in this direction in recent years, ranging from
purely analytic 
approaches~\cite{Badger:2017jhb,Abreu:2017hqn,Gehrmann:2018yef,Abreu:2018jgq, Abreu:2018zmy,Abreu:2018aqd,Chicherin:2018old,Chicherin:2018yne,Broedel:2019hyg,Frellesvig:2019kgj,Lee:2019zop,Bruser:2019auj,Henn:2019rmi,vonManteuffel:2019gpr,Blumlein:2018cms} 
to semi-analytical approaches based on expansions~\cite{Borowka:2018dsa,Davies:2018qvx,Mishima:2018olh,Caola:2018zye,Bonciani:2018omm,Grober:2017uho}
and also via purely numerical methods~\cite{Binoth:2000ps,Heinrich:2008si,Borowka:2018goh,Czakon:2005rk,Usovitsch:2018shx,Mandal:2018cdj,Liu:2017jxz,Driencourt-Mangin:2019aix,Runkel:2019yrs}.

So far, the method of sector decomposition has already proved to be useful for computing 
the complete electroweak two-loop corrections to Z boson production
and decay~\cite{Dubovyk:2018rlg}, 
which is of direct relevance to the FCC-ee, as well as several processes 
of significant interest at the LHC~\cite{Borowka:2016ehy,Borowka:2016ypz,Jones:2018hbb,Heinrich:2019bkc} 
and also BSM corrections~\cite{Buchalla:2018yce,Borowka:2014wla}. The latter calculations were based on \secdecthree~\cite{Borowka:2015mxa}.
Another code based on sector decomposition, {\sc Fiesta}~\cite{Smirnov:2008py,Smirnov:2013eza,Smirnov:2015mct}, has also been used successfully in various multiloop calculations, for example for numerical checks of recent evaluations of four-loop three-point functions~\cite{Bruser:2019auj,Henn:2019rmi,Lee:2019zop,vonManteuffel:2019gpr}.

In this contribution, we will briefly describe the essential aspects of this method
and provide a short update regarding some of the recent developments~\cite{Borowka:2017idc,Borowka:2018goh} that have enabled 
state-of-the-art predictions to be made using this technique.

In Section~\ref{sec:sectordecomposition}, we will introduce the method of sector decomposition
as we use it for computing Feynman integrals and describe how it leads to integrals that are suitable
for numerical evaluation. In Section~\ref{sec:qmc}, we will discuss a particular
type of quasi-Monte Carlo integration that allows us to numerically integrate the sector-decomposed loop integrals efficiently. Finally, in Section~\ref{sec:outlook}, we will give a short outlook for the 
field of numerical multiloop calculations.

\subsection{Feynman integrals and sector decomposition}
\label{sec:sectordecomposition}

A general scalar Feynman integral $I$ in $D=4-2\epsilon$ dimensions with $L$ loops and $N$ propagators $P_j$, 
each raised to a power $\nu_j$, can be written in momentum space as
\begin{align}
I = \int_{- \infty}^{\infty} \prod_{l=1}^L \left[ \mathrm{d}^D k_l \right] \frac{1}{\prod_{j=1}^N P_j^{\nu_j} },& & \mathrm{where} \quad \left[ \mathrm{d}^D k_l \right] = \frac{\mu^{4-D}}{\mathrm{i}\uppi^\frac{D}{2}} \mathrm{d}^D k_l, \quad \quad P_j = \left (q_j-m_j^2+\mathrm{i} \delta \right ),& 
\end{align}
and $q_j$ are linear combinations of external momenta $p_i$ and loop momenta $k_l$.
After introducing Feynman parameters,  the momentum integrals can be performed straightforwardly and
the integral can be recast in the form
\begin{align}
I = (-1)^{N_\nu} \frac{\Gamma(N_\nu-LD/2)}{\prod_{j=1}^N \Gamma(\nu_j)} \int_0^\infty \prod_{j=1}^N \mathrm{d}x_j \ x_j^{\nu_j-1} \delta \left (1-\sum_{i=1}^N x_i \right
)
\frac{\mathcal{U}^{N_\nu - (L+1)D/2}(\vec{x})}{\mathcal{F}^{N_\nu-LD/2}(\vec{x},s_{ij},m_j^2)},
\end{align} 
where the momentum integrals have been replaced by an $N$-fold parameter integral. Here, 
$\mathcal{U}$ and $\mathcal{F}$ are the first and second Symanzik polynomials; they are homogeneous
polynomials in the Feynman parameters of degree $L$ and $L+1$, respectively, and $N_{\nu}=\sum_j\nu_j$.
This procedure can be extended to support  Feynman integrals with tensor numerators.
There are three possibilities of poles in the dimensional regulator $\epsilon$ arising.
\begin{enumerate}
\item The overall $\Gamma(N_\nu-LD/2)$ can diverge, resulting in a single UV pole.
\item $\mathcal{U}(\vec{x})$ vanishes for some $x=0$ and has a negative exponent, resulting in a UV subdivergence.
\item $\mathcal{F}(\vec{x},s_{ij},m_j^2)$ vanishes on the boundary and has a negative exponent, giving rise to an IR divergence.
\end{enumerate}

After integrating out the $\delta$ distribution and extracting a common factor of $(-1)^{N_\nu} \Gamma(N_\nu-LD/2)$, we are faced with integrals of the form
\begin{align}
I_i = \int_0^1 \prod_{j=1}^{N-1} \mathrm{d}x_j x_j^{\nu_j-1} \frac{\mathcal{U}_i(\vec{x})^{\mathrm{expo} \mathcal{U}(\epsilon)}}{\mathcal{F}_i(\vec{x},s_{ij},m_j^2)^{ \mathrm{expo} \mathcal{F}(\epsilon)}}.
\end{align}
The sector decomposition algorithms aim to factorise, via integral transforms, the polynomials $\mathcal{U}_i$ and $\mathcal{F}_i$ (or, more generally, any product of polynomials $\mathcal{P}(\{x_j\})$) as products of a monomial and a polynomial with non-zero constant term, explicitly
\begin{align} 
\mathcal{P}(\{x_j\}) \rightarrow  \prod_j x_j^{\alpha_j} \left( c + p(\{x_j\}) \right),
\end{align}
where $\{x_j\}$ is the set of Feynman parameters, $c$ is a constant, and the polynomial $p$ has no constant term. After this procedure, singularities in $\epsilon$ resulting from the region where one or more $x_j \rightarrow 0$ can appear only from the monomials $x_j^{\alpha_j}$. In this factorised form, the integrand can now be expanded in $\epsilon$ and the coefficients of the expansion can be numerically integrated; for an overview, see Ref.~\cite{Heinrich:2008si}.

\begin{figure}
\begin{center}
\includegraphics[width=0.3\textwidth]{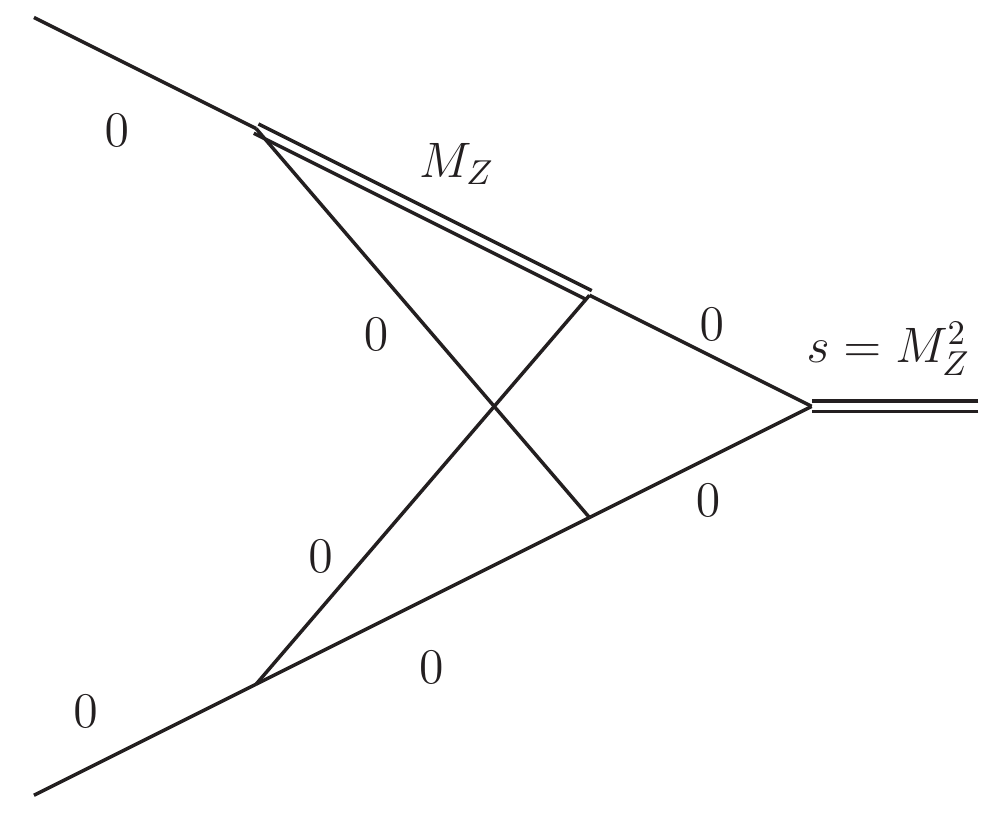}
\end{center}
\caption{A $\mathrm{Zb\bar{b}}$ vertex diagram with no Euclidean region, which can give rise to poorly convergent numerical integrals after sector decomposition. Figure taken from Ref. \cite{Dubovyk:2016zok}.}
\label{fig:zbb}
%\query{Please correct the figure labels in \Fref{fig:zbb}. Set variables
%in italic font and particle names in roman font.}
\end{figure}

If we consider only integrals for which the Mandelstam variables and masses can be chosen such that the $\mathcal{F}$ polynomial is positive semidefinite (\ie with a Euclidean region), this procedure is sufficient to render the integrals numerically integrable.\footnote{In the physical region, such integrals may still require the integration contour to be deformed into the complex
plane, in accordance with the causal $\mathrm{i}\delta$ Feynman prescription~\cite{Soper:1999xk,Binoth:2005ff,Nagy:2006xy}.} However, not all integrals of interest have a Euclidean region in this sense. Consider, for example, the three-point function depicted in \Fref{fig:zbb}, which appears in the two-loop electroweak corrections to the $\mathrm{Zb\bar{b}}$ vertex~\cite{Fleischer:1998nb,Dubovyk:2016zok}. The $\mathcal{F}$ polynomial is given by
\begin{multline}
\mathcal{F}/m_\mathrm{Z}^2 \\
= x_3^2 x_5 + x_3^2 x_4 + x_2 x_3 x_5 + x_2 x_3 x_4 + x_1 x_3 x_5 + x_1 x_3 x_4 
  + x_1 x_3^2 + x_1 x_2 x_3 + x_0 x_3 x_4 + x_0 x_3^2 + x_0 x_2 x_3 \\
  - x_1 x_2 x_4 - x_0 x_1 x_5 - x_0 x_1 x_4 - x_0 x_1 x_2  - x_0 x_1 x_3
\, , \label{eq:zbbf}
\end{multline}
where $m_\mathrm{Z}$ is the Z boson mass and $x_j$ are the Feynman parameters. Note that the massive propagator has the same mass as the external Z boson, which gives rise to terms in the $\mathcal{F}$ polynomial of differing sign, regardless of the value chosen for $m_\mathrm{Z}$. 

After sector decomposition, integrals for which the $\mathcal{F}$ polynomial is not positive semidefinite can diverge not only as some $x_j \rightarrow 0$ but also as some $x_j \rightarrow 1$. One solution for dealing with such integrals is to split the integration domain in each Feynman parameter and then map the integration boundaries back to the unit hypercube such that the divergences at $x_j \rightarrow 1$ are mapped to divergences at $x_j \rightarrow 0$. Sector decomposition can then resolve the singularities at $x_j \rightarrow 0$ as usual. Such a splitting procedure was introduced in earlier versions of \secdec~\cite{Carter:2010hi,Borowka:2012yc}, and also in {\sc Fiesta}~\cite{Smirnov:2013eza,Smirnov:2015mct}.

However, prior to \pysecdec~\cite{Borowka:2017idc}, integrals were always split at $x_j = 1/2$ and, as shown in Ref.~\cite{Jahn:2018gnp}, this can again lead to problems if the $\mathcal{F}$ polynomial vanishes at this point (which happens to be the case for the polynomial in \Eref{eq:zbbf}). The proposed solution in Ref.~\cite{Jahn:2018gnp} was therefore to split the integrals at a random point, such that, if one run produces a problematic result, it is always possible to rerun the code and avoid a problematic split.

Alternatively, it is often possible to avoid having to evaluate such problematic integrals, as well as integrals that have poor numerical convergence properties, through the use of integration by parts identities (IBPs)~\cite{Tkachov:1981wb,Chetyrkin:1981qh}. In particular, it is usually possible to express Feynman integrals in terms of a sum of (quasi-)finite integrals\footnote{Here, quasi-finite integrals are integrals for which the overall $\Gamma(N_\nu-LD/2)$ can give rise to poles in $\epsilon$ but for which no poles arise from the integration over the $\mathcal{U}$ and $\mathcal{F}$ polynomials.} with rational coefficients~\cite{Panzer:2015ida,vonManteuffel:2014qoa}. Typically, choosing a basis of (quasi-)finite integrals leads to significantly improved numerical properties; see, for example, Ref.~\cite{vonManteuffel:2017myy}. The choice of a quasi-finite basis proved advantageous for the numerical evaluation of the $\mathrm{gg} \rightarrow \mathrm{HH}$ and $\mathrm{gg} \rightarrow \mathrm{Hg}$ amplitudes~\cite{Borowka:2016ehy,Borowka:2016ypz,Jones:2018hbb}.

\subsection{Quasi-Monte
Carlo integration}
\label{sec:qmc}

Numerical integration of the sector-decomposed finite integrals can be a computationally 
intensive process. One of the most widely used tools for numerical integration is the 
{\sc Cuba} package~\cite{Hahn:2004fe,Hahn:2014fua}, which implements several different 
numerical integration routines, relying on pseudo-random sampling, quasi-random sampling, or
cubature rules. 

In the last few years, it was found that a particular type of quasi-Monte
Carlo (QMC) integration
based on rank-1 shifted lattice (R1SL) rules has particularly good convergence properties for 
the numerical integration of Feynman parametrized integrals~\cite{Li:2015foa,deDoncker:2018nqe,De_Doncker_2018}. 
An unbiased R1SL estimate $\bar{Q}_{n,m}[f]$ of the integral $I[f]$ can be obtained from the following (QMC) cubature rule~\cite{QMCActaNumerica}: 
\begin{equation}
I[f] \approx \bar{Q}_{n,m}[f] \equiv  \frac{1}{m} \sum_{k=0}^{m-1} Q_{n}^{(k)}[f], \qquad 
Q_{n}^{(k)}[f] \equiv \frac{1}{n} \sum_{i=0}^{n-1} f \left( \left\{ \frac{i \mathbf{z}}{n} + \boldsymbol{\Delta}_k \right\} \right) . \label{eq:lattice}
\end{equation}
Here, the estimate of the integral depends on the number of lattice points $n$ and the number of random shifts $m$. The generating vector $\mathbf{z} \in \mathbb{Z}^d$ is a fixed $d$-dimensional vector of integers coprime to $n$. The shift vectors $\boldsymbol{\Delta}_k \in [0,1)^d$ are $d$-dimensional vectors with components consisting of independent, uniformly distributed, random real numbers in the interval $[0,1)$. Finally, the curly brackets indicate that the fractional part of each component is taken, such that all arguments of $f$ remain in the interval $[0,1)$. An unbiased estimate of the mean-square error due to the numerical integration can be obtained by computing the variance of the different random shifts $Q_{n}^{(k)}[f]$.

The latest version of \pysecdec{} provides a public implementation of a R1SL (QMC) integrator. The 
implementation is also capable of performing numerical integration  on a
number of CUDA-compatible graphics processing units (GPUs), which can significantly
accelerate the evaluation of the integrand. The integrator, which is distributed as a header-only C++ library, can also be used as a stand-alone integration package~\cite{Borowka:2018goh}. The generating vectors distributed with the package are generated using the component-by-component  construction~\cite{nuyens2006fast}.

\subsection{Summary and outlook}
\label{sec:outlook}

We have presented new developments for the numerical calculation of multiloop integrals, focusing on the sector decomposition approach in combination with quasi-Monte Carlo (QMC) integration. We described a new feature present in \pysecdec, which allows  integrals with special (non-Euclidean)  kinematic configurations to be calculated as they occur, \eg in electroweak two-loop corrections, which previously had shown poor convergence in \secdecthree.
We also described a QMC integrator, developed in conjunction with \pysecdec{} as well as for stand-alone usage, which can lead to considerably more accurate results in a given time compared with standard Monte Carlo integration. This integrator is also capable of utilising CUDA-compatible graphics processing units (GPUs).

In view of the need for high-precision calculations with many mass scales at future colliders, as they occur, for example, in electroweak corrections, 
numerical methods are a promising approach, and are actively being  developed to best utilise recent progress in computing hardware.
Several further developments towards the automation of numerical multiloop calculations, with sector decomposition as an ingredient,  
could be envisaged. For example, to provide boundary conditions for numerical solutions to differential equations, along the lines of Refs.~\cite{Czakon:2008zk,Mandal:2018cdj},  for automated asymptotic expansions, similar to Refs.~\cite{Jantzen:2012mw,Mishima:2018olh}, or 
aiming at fully numerical evaluations of both virtual and real corrections.

\subsection*{Acknowledgements}
We  thank Tom Zirke for his contributions to earlier
versions of the code and Oliver Schulz for providing GPU computing support.

%\subsection*{References of Jones}

%\bibliographystyle{elsarticle-num}
%\bibliography{secdec}

\end{bibunit}

\label{sec-mtools-sjones}

\clearpage \pagestyle{empty}  \cleardoublepage
%============================================

\pagestyle{fancy}
\fancyhead[CO]{\thechapter.\thesection \hspace{1mm}  Analytics from numerics: five-point QCD amplitudes at two loops
}
\fancyhead[RO]{}
\fancyhead[LO]{}
\fancyhead[LE]{}
\fancyhead[CE]{}
\fancyhead[RE]{}
\fancyhead[CE]{S.~Abreu, J.~Dormans, F.~Febres~Cordero, H.~Ita, B.~Page}
\lfoot[]{}
\cfoot{-  \thepage \hspace*{0.075mm} -}
\rfoot[]{}

\begin{bibunit}[elsarticle-num]  

\section[Analytics from numerics: five-point QCD amplitudes at two loops
 \\ {\it S.~Abreu, J.~Dormans, F.~Febres~Cordero, H.~Ita, B.~Page}]
{  Analytics from numerics: five-point QCD amplitudes at two loops
\label{contr:page}}
\noindent
{\bf Contribution\footnote{This contribution should be cited as:\\
S.~Abreu, J.~Dormans, F.~Febres~Cordero, H.~Ita, B.~Page, Analytics from numerics: five-point QCD amplitudes at two loops,  
%04 DOI:10.23731/CYRM-2020-XXX.\thepage, in:
%04 \url{http://dx.doi.org/10.23731/CYRM-2020-XXX.\thepage}, in:
DOI: \href{http://dx.doi.org/10.23731/CYRM-2020-003.\thepage}{10.23731/CYRM-2020-003.\thepage}, in:
Theory for the FCC-ee, Eds. A. Blondel, J. Gluza, S. Jadach, P. Janot and T. Riemann,\\
CERN Yellow Reports: Monographs, CERN-2020-003,
%04 \url{http://dx.doi.org/10.23731/CYRM-2020-XXX}, p. \thepage.} 
DOI: \href{http://dx.doi.org/10.23731/CYRM-2020-003}{10.23731/CYRM-2020-003},
p. \thepage.
\\ \copyright\space CERN, 2020. Published by CERN under the 
%04-2
\href{http://creativecommons.org/licenses/by/4.0/}{Creative Commons Attribution 4.0 license}.} by: S.~Abreu, J.~Dormans, F.~Febres~Cordero, H.~Ita and B.~Page \\
Corresponding author: B.~Page {[bpage@ipht.fr]}}
\vspace*{.5cm}
\subsection{Introduction} \label{sec:page-intro}

The operation of the Large Hadron Collider (LHC) has been a great success, with Run 1 culminating in the
discovery of the Higgs boson by the ATLAS and CMS experiments in 2012. 
In Run 2, the LHC experiments have moved towards performing
high-precision measurements with uncertainties reaching below the percentage level for
certain observables. Looking forward to the Future Circular Collider with electron beams (FCC-ee), 
which will operate in the experimentally much cleaner environment of electron--positron initial states,
there will be an even more dramatic increase in experimental precision.
To exploit the precision measurements, the theory community will
need to provide high-precision predictions that match the experimental
uncertainties. This requires the development of efficient ways to compute these
corrections, breaking through the current computational bottlenecks.

In this section, we discuss the calculation of a key component in
making such predictions---the loop amplitude. Specifically, we discuss the
computation of an independent set of analytical two-loop five-gluon helicity
amplitudes in the leading-colour approximation. These amplitudes are an
ingredient for the phenomenologically relevant description of three-jet
production at next-to-next-to-leading order (NNLO) for hadron colliders. 
Nonetheless, the methods we present are
completely general and can also be applied to predictions for electron--positron
collisions.

The first two-loop five-gluon amplitude to be computed was the one with all
helicities positive in the leading-colour approximation, initially numerically
\cite{Badger:2013gxa} and subsequently analytically
\cite{Gehrmann:2015bfy,Dunbar:2016aux}. In the last few years, a flurry of
activity in this field led to the numerical calculation of all five-gluon
\cite{Badger:2017jhb,Abreu:2017hqn}, and then all five-parton
\cite{Badger:2018gip,Abreu:2018jgq}  amplitudes in the leading-colour approximation.
The combination of numerical frameworks with finite-field techniques, with a view to
the reconstruction of the rational functions appearing in the final results, 
was first introduced to our field in Ref.~\cite{vonManteuffel:2014ixa}, and an
algorithm applicable to multiscale calculations was presented in
Ref.~\cite{Peraro:2016wsq}. Inspired by these ideas, the four-gluon
amplitudes were analytically reconstructed from floating-point evaluations
\cite{Abreu:2017xsl}. 
The first application involving multiple scales was the single-minus 
two-loop five-gluon amplitude \cite{Badger:2018enw}. In this section,
we describe the calculation of the full set of independent five-gluon 
amplitudes in the leading-colour approximation \cite{Abreu:2018zmy}.
These results were obtained using analytical reconstruction techniques,
starting from numerical results obtained in the framework of two-loop
numerical unitarity \cite{Ita:2016oar,Abreu:2017xsl,Abreu:2017hqn,Abreu:2018jgq}. 
Recently, the
remaining five-parton amplitudes have also become available \cite{Abreu:2019odu},
and all two-loop amplitudes for three-jet production at 
NNLO in QCD are now known 
analytically in the leading-colour approximation.\footnote{The approach taken in Ref.~\cite{Abreu:2019odu} is very similar to that
described here; we refer the reader to the results presented
in Ref.~\cite{Abreu:2019odu} for a more compact expression for the five-gluon 
amplitudes and further improvements in the methodology.}

This section is organised as follows. 
In  \Sref{sec:amps}, we describe the amplitudes under consideration and the numerical 
unitarity framework employed for their evaluation. 
\Sref{sec:func_rec} describes the objects we will be computing and the simplifications 
that are made to allow for an efficient functional reconstruction.
The implementation and the results are presented in \Sref{sec:impl} and we conclude in 
 \Sref{sec:page-conclusion}.

%
%
% ----- AMPLITUDES -----------
%
%
\subsection{Amplitudes} \label{sec:amps}
We discuss the computation of the two-loop five-gluon amplitudes in QCD. 
The calculation is performed in the  
leading-colour approximation where there is a single partial amplitude.
The bare amplitude can be perturbatively expanded as
\begin{multline}
\label{eq:bareAmp}
  \mathcal{A}(\{p_i,h_i\}_{i=1,\ldots,5})\big\vert_{\textrm{leading colour}} \\
 = 
\sum_{\sigma\in S_5/Z_5} {\rm Tr}\left(
  T^{a_{\sigma(1)}} T^{a_{\sigma(2)}}
  T^{a_{\sigma(3)}} T^{a_{\sigma(4)}} T^{a_{\sigma(5)}} \right) \,
   g^{3}_0
  \left(\mathcal{A}^{(0)}
+
  \lambda
  \mathcal{A}^{(1)} +
  \lambda^2\mathcal{A}^{(2)}
+\mathcal{O}(\lambda^3)
\right).
\end{multline}
Here, $\lambda={N_c g_0^2}/{(4\uppi)^2}$, $g_0$ is the
bare QCD coupling and $S_5/Z_5$ is the set of all non-cyclic 
permutations of five indices. The amplitudes $\mathcal{A}^{(k)}$
appearing in the expansion of \Eref{eq:bareAmp} depend on 
the momenta $p_{\sigma(i)}$ and the helicities $h_{\sigma(i)}$ 
and these proceedings focus on the calculation of $\mathcal{A}^{(2)}$
in the 't Hooft--Veltman scheme of dimensional regularisation, with $D=4-2\epsilon$. 

The first step in the analytic reconstruction procedure is the numerical
evaluation of the amplitude. We evaluate the amplitudes in the framework of
two-loop numerical unitarity
\cite{Ita:2016oar,Abreu:2017xsl,Abreu:2017hqn,Abreu:2018jgq}.
The integrands of the amplitudes $\mathcal{A}^{(2)}$ are parametrized
with a decomposition in terms of master integrands and surface terms. 
On integration, the former yield the master integrals, while the latter vanish. 
Labelling the loop momenta  $\ell_{l}$, 
the parametrization we use is given by 
\begin{equation}\label{eq:AL}
  \mathcal{A}^{(2)}(\ell_l)=\sum_{\Gamma\in\Delta}
  \sum_{i\in M_\Gamma\cup S_\Gamma} c_{\Gamma,i}
  \frac{m_{\Gamma,i}(\ell_l)}{\prod_{j\in
  P_\Gamma}\rho_j},
\end{equation}
with $\Delta$ being the set of all propagator structures $\Gamma$,
$P_{\Gamma}$ the associated set of propagators, and $M_{\Gamma}$ and $S_{\Gamma}$
denoting the corresponding sets of master integrands and surface terms, respectively.
If the master integrals are known, the evaluation of the amplitude reduces to the 
determination of master coefficients $c_{\Gamma,i}$ with $i\in M_{\Gamma}$. In
numerical unitarity, this is achieved by solving a linear system, which is 
generated by sampling on-shell values of the loop momenta $\ell_l^{\Gamma}$ 
belonging to the algebraic variety of $P_{\Gamma}$.
In this limit, the leading contribution to \Eref{eq:bareAmp} factorises 
into products of tree amplitudes
\begin{equation}
  \sum_{\rm states}\prod_{i\in T_\Gamma}
  {\cal A}^{\rm tree}_i(\ell_l^\Gamma) =\!\!\!
  \sum_{\substack{\Gamma' \ge \Gamma\,,\\ 
  i\,\in\,M_{\Gamma'}\cup S_{\Gamma'}}} 
  \frac{ c_{\Gamma',i}\,m_{\Gamma',i}(\ell_l^\Gamma)}{\prod_{j\in
  (P_{\Gamma'}\setminus P_\Gamma) } \rho_j(\ell_l^\Gamma)}\,.
  \label{eq:CE}
  \end{equation}
The tree amplitudes associated with vertices in the diagram corresponding to 
$\Gamma$ are  denoted  $T_{\Gamma}$ and the sum is over the  
physical states of each internal line of $\Gamma$. 
On the right-hand side, the sum is performed over all 
propagator structures $\Gamma^{'}$, such that $P_{\Gamma} \subseteq P_{\Gamma^{'}}$. 
At two loops, subleading contributions appear, which cannot be described by a 
factorisation theorem in the on-shell limit. In practice, this complication is 
eliminated by constructing a larger system of equations, as described, for instance, 
in Ref.~\cite{Abreu:2017idw}. For a given (rational) phase space point, 
we solve the linear system in \Eref{eq:CE}  using finite-field arithmetic. 
This allows us to obtain exact results for the master integral coefficients 
in a very efficient manner. 

Once the coefficients $c_{\Gamma,i}$ are known, the amplitude
can be decomposed into a linear combination of master integrals
$\mathcal{I}_{\Gamma,i}$, according to
\begin{align}\label{eq:MI}
\mathcal{A}^{(2)} = \sum_{\Gamma \in \Delta} \sum_{i \in M_{\Gamma}} c_{\Gamma, i} \mathcal{I}_{\Gamma, i} \,,
\end{align}
with
\begin{align}\label{eq:mint}
  \mathcal{I}_{\Gamma,i} = \int \mathrm{d}^Dl_l \frac{m_{\Gamma,i}(l_l)}{\prod_{j\in P_{\Gamma}}\rho_j}\,.
\end{align}
For planar massless five-point scattering at two loops, the basis of master
integrals is known in analytical form
\cite{Papadopoulos:2015jft,Gehrmann:2018yef}.

%
%
% ----- FUNCTIONAL RECONSTRUCTION -----------
%
%
\subsection{Simplifications for functional reconstruction} \label{sec:func_rec}
Functional reconstruction techniques allow one to reconstruct rational functions
from numerical data, preferably in a finite field to avoid issues related
to loss of precision \cite{vonManteuffel:2014ixa, Peraro:2016wsq}.
By choosing an appropriate set of variables,
such as momentum twistors \cite{Hodges:2009hk}, we can guarantee that the 
coefficients $c_{\Gamma, i}$ in \Eref{eq:MI} are rational.
The specific parametrization we use is \cite{Peraro:2016wsq}
\begin{align}\label{eq:twistorVars}
  s_{12}  &=x_4,  \qquad   s_{23} =x_2 x_4,  \qquad
  s_{34} =x_4\left(\frac{(1+x_1)x_2}{x_0}+x_1(x_3-1)\right),
  \nonumber\\
  s_{45} &=x_3x_4,   \qquad
  s_{51}=x_1x_4(x_0-x_2+x_3)\,,\\
  \textrm{tr}_5 &=4\,\mathrm{i}\,\varepsilon(p_1,p_2,p_3,p_4)
  \nonumber\\
  &=x_4^2\left(x_2(1+2x_1)+x_0x_1(x_3-1)
  -\frac{x_2(1+x_1)(x_2-x_3)}{x_0}
  \right),\nonumber
\end{align}
where $s_{ij}=(p_i+p_j)^2$, with the indices defined
cyclically. 
One could, in principle, reconstruct the rational master integral
coefficients. However, the difficulty of the reconstruction is 
governed by the complexity of the function under consideration.
The amplitude $\mathcal{A}^{(2)}$ of \Eref{eq:MI} contains a lot of 
redundant information; to improve the efficiency of the reconstruction, it
is thus beneficial to remove this redundancy. Furthermore, while
\Eref{eq:MI} provides a decomposition in terms of master integrals
in dimensional regularisation, after expanding the master integrals
in $\epsilon$ there can be new linear relations between the different
terms in the Laurent expansion in $\epsilon$. We thus expect cancellations
between the different coefficients $c_{\Gamma, i}$. In this section,
we discuss how we address these issues and define the object we reconstruct.

We start by expressing the Laurent expansion of the master integrals in
\Eref{eq:mint} in terms of a basis $B$ of so-called pentagon functions $h_i
\in B$ \cite{Gehrmann:2018yef}. 
That is, we rewrite the amplitudes as
\begin{equation}\label{eq:ampAsPent}
  \mathcal{A}^{(2)}=
  \sum_{i \in B}\sum_{k=-4}^0
  \epsilon^k\,\tilde c_{k,i}(\vec x)h_i(\vec x)\, + 
  \mathcal{O}(\epsilon),
\end{equation}
where $\vec x=\{x_0,x_1,x_2,x_3,x_4\}$ and the $\tilde c_{k,i}(\vec x)$ are
rational functions of the twistor variables.
All linear relations between master integrals that appear after expansion in
$\epsilon$ are resolved in such a decomposition.

Next, we recall that the singularity structure of 
two-loop amplitudes is governed by lower-loop amplitudes 
\cite{Catani:1998bh,Sterman:2002qn, Becher:2009cu, Gardi:2009qi}. 
One can thus exploit this knowledge to subtract the pole structure 
from the amplitudes in order to obtain a finite remainder that 
contains the new two-loop information.
There is freedom in how to define the remainders, as they are only constrained
by removing the poles of the amplitudes. For
helicity amplitudes that vanish at tree level,
$\mathcal{A}^{(k)}_{\pm++++}$,
we use
\begin{equation}\label{eq:remainderHV}
  \mathcal{R}^{(2)}_{\pm++++}\!
        =\bar{\mathcal{A}}^{(2)}_{\pm++++}\!
  +S_{\epsilon}\bar{\mathcal{A}}^{(1)}_{\pm++++}
  \sum_{i=1}^5
  \frac{(-s_{i,i+1})^{-\epsilon}}{\epsilon^2}
  +\mathcal{O}(\epsilon),
\end{equation}
where $S_\epsilon=(4\uppi)^{\epsilon}\mathrm{e}^{-\epsilon\gamma_E}$.
The $\bar{\mathcal{A}}^{(k)}$ denote amplitudes
normalised to remove any ambiguity related to overall phases.
In the case of amplitudes that vanish at tree level, we
normalise to the leading order in $\epsilon$ of the (finite) 
one-loop amplitude.
For the maximally helicity violating  (MHV) amplitudes, 
$\mathcal{A}^{(k)}_{-\mp\pm++}$,
which we normalise to the corresponding tree amplitude,
we define
\begin{equation}
\label{eq:remainderMHV}
  \mathcal{R}^{(2)}_{-\mp\pm++}=
                \bar{\mathcal{A}}^{(2)}_{-\mp\pm++}
    -\left(\frac{5\,\tilde\beta_0}{2\epsilon}+{\bf I}^{
    (1)}\right)
  S_{\epsilon}\bar{\mathcal{A}}^{(1)}_{-\mp\pm++}
 +    
 \left(
  \frac{15\,\tilde\beta_0^2}{8\epsilon^2}
  +\frac{3}{2\epsilon}
  \left(\tilde\beta_0{\bf I}^{(1)}-\tilde\beta_1
  \right)
  -{\bf I}^{(2)}
  \right)
  S_{\epsilon}^2
  +\mathcal{O}(\epsilon)\,,
\end{equation}
where $\tilde\beta_i$ are the coefficients of the QCD
$\beta$ function divided by $N_c^{i+1}$ and 
${\bf I}^{(1)}$ and ${\bf I}^{(2)}$ are the standard Catani
operators at leading colour. Precise expressions for the operators in our
conventions can be found in Appendix B of Ref.~\cite{Abreu:2018jgq}.
We note that for both \Eref{eq:remainderHV} and \Eref{eq:remainderMHV} 
we require one-loop amplitudes expanded up to order $\epsilon^2$.  
By expressing the one-loop amplitudes and the Catani operators in the basis 
of pentagon functions, the remainder can be expressed in the same 
way,
\begin{equation}\label{eq:dec_as_masters}
  \mathcal{R}^{(2)}=
  \sum_{i \in B} r_i(\vec x)\,h_i(\vec x)\,.
  \end{equation}
We observe that the coefficients $r_i(\vec x)$ are rational functions of
lower total degree than the $\tilde c_{k,i}(\vec x)$ of \Eref{eq:ampAsPent}.

As a further simplification, we investigate the pole structure of the coefficients
$r_i(\vec{x})$. The alphabet determines the points in phase 
space where the pentagon functions have logarithmic singularities, and
as such provides a natural candidate to describe the pole structure of 
the coefficients. 
We use the alphabet $A$ determined in Ref.~\cite{Gehrmann:2018yef}
to build an ansatz for the denominator structure of the $r_i(\vec{x})$,
\begin{equation}\label{eq:ansatz}
  r_i(\vec x) =  
  \frac{n_i(\vec x)}{\prod_{j\in A} w_j(\vec x)^{q_{ij}}}\,.
\end{equation}
We then reconstruct the remainder on a slice 
$\vec{x}(t)=\vec{a}\cdot t + \vec{b}$, where all 
the twistor variables depend on a single parameter $t$ and $\vec{a}$ and $\vec{b}$
are random vectors of finite-field values. This reconstruction 
in one variable is drastically simpler than the full multivariate reconstruction. 
In addition, the maximal degree in $t$ on the slice corresponds to 
the total degree in $\vec{x}$. We determine the exponents $q_{ij}$ by matching 
the ansatz on the univariate slice and check its validity on a second slice. 
Having determined the denominators of the rational coefficients $r_i$, 
the reconstruction reduces to the much simpler polynomial reconstruction of 
the numerators $n_i(\vec{x})$.

The last simplification we implement is a change of basis in the space of 
pentagon functions. 
Amplitudes are expected to simplify in specific kinematic configurations
where the pentagon functions degenerate into a smaller basis,
which requires relations between the different coefficients.
This motivates the search for (helicity-dependent) bases
with  coefficients of lower total degree.
To find them, we construct linear combinations of coefficients
\begin{equation}
  \sum_{i \in B} a_{i,k} r_i(\vec x)=
  \frac{N_k(\vec x,a_{i,k})}
  {\prod_{j\in A} w_j(\vec x)^{q_{kj}'}}\,,
\end{equation}
and solve for phase space independent $a_{i,k}$ such that the 
numerators $N_k(\vec x,a_{i,k})$ factorise a subset of the
$w_j\in A$.
This can be performed on univariate slices by
only accepting solutions that are invariant over a number of
slices. The matrix $a_{i,k}$ allows us to change to a new basis 
$B'$ in the space of special functions, in which remainders
can be decomposed as in \Eref{eq:dec_as_masters}, with
coefficients $r_i'(\vec x)$ whose numerators $n_i'(\vec x)$ are
polynomials of lower total degree than those of 
\Eref{eq:ansatz}.

%
%
% ----- Implementation & Results -----------
%
%
\subsection{Implementation and results} \label{sec:impl}

The master integral coefficients of the one- and two-loop amplitudes are computed 
using numerical unitarity in a finite field.
They are combined with the corresponding 
master integrals, expressed in terms of pentagon functions, and the known pole
structure is subtracted to obtain the finite remainders 
as a linear combination of pentagon functions.
After a rotation in the space of pentagon functions and multiplication by the
predetermined denominator factors, we obtain numerical samples for the numerators 
$n_i'(\vec{x})$ in a finite field.
These samples are used to analytically reconstruct the $n_i'(\vec{x})$
with the algorithm of Ref. \cite{Peraro:2016wsq}, which we slightly modified
to allow a more efficient parallelization.
These  steps were implemented in a flexible C{}\verb!++! framework, which was
used to reconstruct the analytical form of the two-loop remainders of a basis
of five-gluon helicity  amplitudes (the other helicities can be 
obtained by parity and charge conjugation). Two finite fields of cardinality
$O(2^{31})$ were necessary for the rational reconstruction by means of the
Chinese remainder theorem.

\Tref{tab:ranks} shows the impact of the simplifications discussed in the 
previous section for each helicity. In the most complicated case, 
the $\mathrm{g}^{-}\mathrm{g}^{+}\mathrm{g}^{-}\mathrm{g}^{+}\mathrm{g}^{+}$ helicity amplitude, we must
reconstruct a polynomial of degree 53. This required 250\,000 numerical 
evaluations, with 4.5 min per evaluation.

\begin{table}
  \caption{
  Each $t^n/t^d$  denotes the total degree of
  numerator ($n$) and denominator ($d$) of the most complex
  coefficient for each helicity amplitude in the
  decomposition of \Eref{eq:ampAsPent} (second column) or 
 \Eref{eq:dec_as_masters} (third column). The fourth
  column lists the highest polynomial we reconstruct. The final
  column lists the number of letters $w_j(\vec x)$ that
  contribute in the denominator of \Eref{eq:ansatz}.}
  \label{tab:ranks}
  \begin{center}
  \begin{tabular}{lllll} \hline \hline
      Helicity  & $\tilde c_{k,i}(t)$ &   $r_i(t)$ &
      $n'_i(t)$ & $ w_j\textrm{s}\,\,\textrm{in denominator}$\\
      \hline
      $+++++$   & $t^{34}/t^{28}$ &  $t^{10}/t^{4}$ & $t^{10}$
      & \phantom{2}3   
      \\
      $-++++$   & $t^{50}/t^{42}$ &  $t^{35}/t^{28}$  & $t^{35}$
      &14 
      \\
      $--+++$  & $t^{70}/t^{65}$ &  $t^{50}/t^{45}$  & $t^{40}$
      &17
      \\
      $-+-++$  & $t^{84}/t^{82}$ &  $t^{68}/t^{66}$  & $t^{53}$
      &20
      \\ \hline\hline
  \end{tabular}
  \end{center}
  \end{table}

The results that we provide contain the one-loop amplitudes in terms of master 
integrals and the two-loop remainders in terms of pentagon functions. 
The one-loop master integrals are provided in terms of pentagon functions up to
order $\epsilon^2$. The combined size of the expressions amounts to 45\,MB without 
attempting any simplification 
(we refer the reader to Ref. \cite{Abreu:2019odu} for more compact expressions). 
These expressions can be combined to 
construct the full analytical expression for the two-loop five-gluon 
leading-colour amplitudes in the Euclidean region.
We validated our expressions by reproducing all the target 
benchmark values available in the literature 
\cite{Badger:2013gxa, Gehrmann:2015bfy, Dunbar:2016aux, Badger:2018enw, 
Badger:2017jhb, Abreu:2017hqn, Abreu:2018jgq}.

%
%
% ----- CONCLUSION -----------
%
%
\subsection{Conclusion} \label{sec:page-conclusion}

In this section, we have presented the recent computation of the analytical form of
the leading-colour contributions to the two-loop five-gluon scattering
amplitudes in pure Yang--Mills theory.
This computation was undertaken in a novel way, made possible by a collection of
mature tools. The amplitude is first numerically reduced to a basis of master
integrals with the two-loop numerical unitarity 
approach, where the coefficients take finite-field 
values~\cite{Ita:2016oar,Abreu:2017xsl,Abreu:2017hqn,Abreu:2018jgq}. 
This allows us to numerically
calculate a finite remainder, expressed in terms of pentagon
functions~\cite{Gehrmann:2018yef}. The generation of these numerical samples is
driven by a functional reconstruction algorithm~\cite{Peraro:2016wsq}, which
determines the analytical form of the pentagon-function coefficients from a series of
evaluations.
A key step in efficiently implementing this strategy was to utilise physical
information to simplify the analytical form of the objects we reconstruct, 
and hence reduce the required number of evaluations. 
First, we reconstruct the finite remainder, which
removes redundant information related to lower-loop contributions. Second,
we decompose the remainder in terms of pentagon functions to account
for relations between different master integrals after
expansion in the dimensional regulator. Next,
we exploit the knowledge of the singularity structure of the pentagon functions 
to efficiently establish the denominators of the coefficient functions. 
Finally, we find a basis of pentagon functions with coefficients of lower
degree by exploiting their reconstruction on a univariate slice.

These techniques show a great deal of potential for future calculations. Indeed,
they have very recently been used to obtain the full set of leading-colour 
contributions to the five-parton scattering amplitudes~\cite{Abreu:2019odu}. 
We foresee further applications to processes with a higher number of scales  and loops,
such as those  required for a future lepton collider in the near future.

\end{bibunit}

\label{sec-mtools-page}

\clearpage \pagestyle{empty}  \cleardoublepage
%============================================

\pagestyle{fancy}
\fancyhead[CO]{\thechapter.\thesection \hspace{1mm} Recent developments in Kira}
\fancyhead[RO]{}
\fancyhead[LO]{}
\fancyhead[LE]{}
\fancyhead[CE]{}
\fancyhead[RE]{}
\fancyhead[CE]{P. Maierh\"ofer, J. Usovitsch}
\lfoot[]{}
\cfoot{-  \thepage \hspace*{0.075mm} -}
\rfoot[]{}

\begin{bibunit}[elsarticle-num]

\let\stdthebibliography\thebibliography
\renewcommand{\thebibliography}{%
\let\section\subsection
\stdthebibliography}

\section
[Recent developments in Kira\\ {\it P. Maierh\"ofer, J.~Usovitsch}]
{Recent developments in Kira
\label{contr:mtools_usovitsch}}
\noindent
{\bf Contribution\footnote{This contribution should be cited as:\\
P. Maierh\"ofer, J.~Usovitsch, Recent developments in Kira,  
%04 DOI:10.23731/CYRM-2020-XXX.\thepage, in:
%04 \url{http://dx.doi.org/10.23731/CYRM-2020-XXX.\thepage}, in:
DOI: \href{http://dx.doi.org/10.23731/CYRM-2020-003.\thepage}{10.23731/CYRM-2020-003.\thepage}, in:
Theory for the FCC-ee, Eds. A. Blondel, J. Gluza, S. Jadach, P. Janot and T. Riemann,\\
CERN Yellow Reports: Monographs, CERN-2020-003,
%04 \url{http://dx.doi.org/10.23731/CYRM-2020-XXX}, p. \thepage.} 
DOI: \href{http://dx.doi.org/10.23731/CYRM-2020-003}{10.23731/CYRM-2020-003},
p. \thepage.
\\ \copyright\space CERN, 2020. Published by CERN under the 
%04-2
\href{http://creativecommons.org/licenses/by/4.0/}{Creative Commons Attribution 4.0 license}.} by:  P. Maierh\"ofer, J. Usovitsch \\
Corresponding author: J. Usovitsch {[usovitsj@maths.tcd.ie]}}
\vspace*{.5cm}

%------------------------------------------------------------------------------

\noindent In this section, we report on the recent progress made in the development of the
Feynman integral reduction program \kira{}.
The development is focused on algorithmic improvements that are essential to
extend the range of feasible high-precision calculations for present and future
colliders like the FCC-ee.

\subsection{Introduction}

\kira{} \cite{Maierhoefer:2017hyi} implements Laporta's algorithm
\cite{Laporta:2001dd} to reduce Feynman integrals to a
basis of master integrals.
In this approach, large systems of integration by parts \cite{Chetyrkin:1981qh}
and Lorentz invariance \cite{Gehrmann:1999as} identities, as well as symmetry
relations, are generated and solved by a variant of Gaussian elimination,
systematically expressing complicated integrals in terms of simpler integrals
with respect to\ a given complexity criterion.
Though alternative reduction techniques have been proposed and applied to
specific problems, see, \eg\ Refs.\
\cite{Smirnov:2005ky,Lee:2013mka,Larsen:2015ped,Kosower:2018obg}, to date
programs based on Laporta's algorithm
\cite{Anastasiou:2004vj,vonManteuffel:2012np,Smirnov:2019qkx} pose the only
general-purpose tools suited for large-scale applications.
Since these reduction programs constitute one of the bottlenecks of high-precision predictions, their continuous improvement is crucial to meet the
increasing demand for such calculations.

A key element of \kira{} is its equation selector to extract a linearly
independent system of equations, discarding equations that are not required to
fully reduce all integrals requested by the user. The selector is based on
Gaussian elimination using modular arithmetic on the coefficients.

\subsection{Improved symmetrization}

The detection of symmetry relations between sectors within and across topologies
received a performance boost as a result of the implementation of the algorithm
described in Ref. \cite{Pak:2011xt}.
In this approach, a canonical form of the integrand of each sector is
constructed, so that a one-to-one comparison of the representations can be made.
Additionally, the combinatorial complexity of the loop momentum shift finder to
determine the mapping prescriptions of equivalent sectors has been reduced.
Furthermore, the detection of trivial sectors received a significant speed-up by
employing \kira{}'s IBP solver instead of the less optimised previous linear
solver.

\begin{figure}
  \centering
  \includegraphics[width=0.4\linewidth]{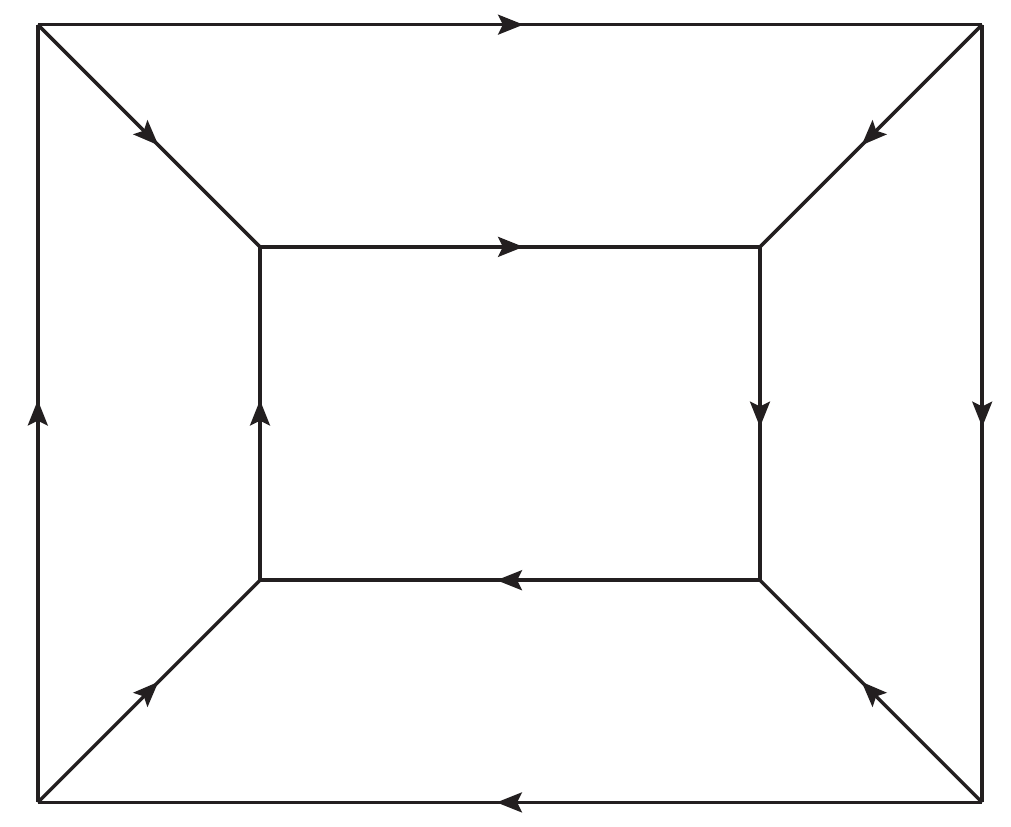}
  \caption{The cube topology is the five-loop vacuum bubble with 12 propagators of
    equal mass and octahedral symmetry. The high symmetry of 48 equivalent
    propagator permutations in the top-level sector makes this topology an ideal
    candidate for symmetrization benchmarks.}
  \label{fig:cube}
\end{figure}

As an example, the `cube topology' shown in \Fref{fig:cube}, \ie\ the
five-loop vacuum bubble with 12 propagators of equal mass and the symmetry of a
cube, can now be analysed in less than 10\,min on a state-of-the-art desktop
computer.

\subsection{Parallel simplification algorithms for coefficients}

\subsubsection{Algebraic simplifications with Fermat}

To simplify multivariate rational functions in masses and kinematic quantities,
which appear as coefficients in the Gaussian elimination steps, \kira{} relies
on the program \fermat{} \cite{Fermat}.
In almost all cases, the run time for the reduction is dominated by those
algebraic simplifications.
It turns out that, when a new coefficient is constructed from several (often
thousands of) known coefficients, combining them na\"ively and simplifying them
in one step results in an avoidable performance penalty.
Instead, \kira{} recursively combines coefficients pairwise, choosing the pairs
based on the size of their string representations.
Besides the improved performance, this strategy also offers new possibilities
for the parallelization, since the pairwise combinations can be evaluated by
different \fermat{} instances.

In the Gaussian back substitution, one can restrict a solver to calculate only
the coefficients of a specific master integral.
This allows the user to parallelize the reduction across several machines and
merge the results in a final step.

\subsubsection{Algebraic reconstruction over integers}

An alternative algorithm to simplify the coefficients is given by algebraic
reconstruction over integers, introduced in Refs.
\cite{Larsen:2015ped,Boehm:2017wjc,Boehm:2018fpv}.
This strategy is based on sampling the rational functions by setting kinematic
invariants and masses to integer values repeatedly.
Each sample can be evaluated rather quickly, but the number of samples required
to reconstruct the simplified result increases with the degree of the numerator
and denominator of the rational function, the number of invariants involved, and
the number of invariants over which it is sampled.
Of course, the sample can again be evaluated in parallel, leading to the
potential for massive parallelization on dozens of CPU cores.
An implementation of this algorithm is available in \kira{}\;\texttt{1.2} and is
continuously being improved and extended.
Furthermore, \kira{} automatically decides which simplification strategy,
\ie\ algebraic reconstruction or \fermat{}, is expected to be more efficient in
each case.
The criteria for these decisions are subject to investigation and offer room for
future improvements.

\subsubsection{Algebraic reconstruction over finite integer fields}

Instead of sampling rational functions over integers, it is also possible to
reconstruct them from samples over finite integer fields.
Mapping coefficients to a finite field limits the size of each coefficient and,
with that, the complexity of each operation.
Choosing the module as a word-size prime, numerical operations on coefficients
correspond to the native arithmetic capabilities of the employed CPU, allowing
for high performance sampling of the coefficients.
A reconstruction algorithm for multivariate rational functions was first
presented in Ref. \cite{Peraro:2016wsq}.
Recently, the library \firefly{} \cite{Klappert:2019emp} became available,
implementing a similar algorithm.
\firefly{} has been combined with \kira{} to use it for Feynman integral
reduction, calculating the samples with \kira{}'s finite integer Gaussian
elimination.
An independent implementation is available in \firesix{} \cite{Smirnov:2019qkx}.

In the sampling over (arbitrary-size) integers described here, whenever a
coefficient is required to proceed with the reduction, the solver needs to wait
until that coefficient has been reconstructed.
Using finite integers, the entire solver can be parallelized, opening the
possibility of distributing solvers over different machines.
The reconstructor can then collect the samples from the solvers and finish the
calculation when a sufficient number of samples is available.
The finite integer reconstruction is expected to become publicly available in
a future \kira{} release in combination with \firefly{}.

\subsection{Basis choice}

It is well-known that the reduction time strongly depends on the choice of the
master integrals.
In a convenient basis, the reduction coefficients tend to become much simpler
than, \eg\ in the basis that follows directly from the integral ordering.
In this respect, uniformly transcendental bases \cite{Henn:2013pwa}, finite
bases \cite{Panzer:2015ida}, or finite uniformly transcendental bases
\cite{Schabinger:2018dyi} present interesting candidates to study the impact of
the basis choice on the reduction performance.
These special choices involve linear combinations of integrals as basis elements
that we call `master equations'.

In \kira{}, integrals are represented by integer `weights' in such a way that
they obey the imposed integral ordering.
Choosing a specific basis of master integrals is already possible.
To this end, the weights are modified so that the preferred basis integrals are
regarded as simpler than all other integrals.
In the presence of master equations, a new kind of object must be introduced,
representing the master equation instead of a particular integral.
With appropriate bookkeeping, the implementation becomes straightforward and
will soon be available in a \kira{} release.

\subsection{Conclusions}

The complexity of precision calculations needed to match the accuracy of the FCC-ee experiment demands for integral reduction tools beyond the state-of-the-art capabilities.
For example, the computation of pseudo-observables at the Z boson resonance, involving reductions of three-loop Feynman diagrams with up to five scales, will be necessary to reach the accuracy that may be achieved with the FCC-ee \cite{Blondel:2018mad}.
We expect that Feynman integral reduction programs based on Laporta's algorithm will continue to play a key role in such calculations;
\eg\ by harnessing the potential of rational reconstruction, basis choices, and large-scale parallelization, we are convinced that \kira{} will keep up with the arising technical challenges.

\end{bibunit}

\label{sec-mtools-usovitsch}  

\clearpage \pagestyle{empty}  \cleardoublepage
%============================================

\pagestyle{fancy}
\fancyhead[CO]{\thechapter.\thesection \hspace{1mm} Precision Monte Carlo
  simulations with WHIZARD}
\fancyhead[RO]{}
\fancyhead[LO]{}
\fancyhead[LE]{}
\fancyhead[CE]{}
\fancyhead[RE]{}
\fancyhead[CE]{S. Bra\ss, W. Kilian, T. Ohl, J. Reuter, V. Rothe, P. Stienemeier}
\lfoot[]{}
\cfoot{-  \thepage \hspace*{0.075mm} -}
\rfoot[]{}

\begin{bibunit}[elsarticle-num]

\let\stdthebibliography\thebibliography
\renewcommand{\thebibliography}{%
\let\section\subsection
\stdthebibliography}

\section
[Precision Monte Carlo simulations with WHIZARD \\ {\it S. Bra\ss, W. Kilian, T. Ohl, J. Reuter,
  V. Rothe, P.~Stienemeier}]
{Precision Monte Carlo simulations with WHIZARD
}
\label{contr:mtools_reuter}
\noindent
{\bf Contribution\footnote{This contribution should be cited as:\\
S. Bra\ss, W. Kilian, T. Ohl, J. Reuter, V. Rothe, P.~Stienemeier, Precision Monte Carlo simulations with WHIZARD,  
%04 DOI:10.23731/CYRM-2020-XXX.\thepage, in:
%04 \url{http://dx.doi.org/10.23731/CYRM-2020-XXX.\thepage}, in:
DOI: \href{http://dx.doi.org/10.23731/CYRM-2020-003.\thepage}{10.23731/CYRM-2020-003.\thepage}, in:
Theory for the FCC-ee, Eds. A. Blondel, J. Gluza, S. Jadach, P. Janot and T. Riemann,\\
CERN Yellow Reports: Monographs, CERN-2020-003,
%04 \url{http://dx.doi.org/10.23731/CYRM-2020-XXX}, p. \thepage.} 
DOI: \href{http://dx.doi.org/10.23731/CYRM-2020-003}{10.23731/CYRM-2020-003},
p. \thepage.
\\ \copyright\space CERN, 2020. Published by CERN under the 
%04-2
\href{http://creativecommons.org/licenses/by/4.0/}{Creative Commons Attribution 4.0 license}.} by: S. Bra\ss, W. Kilian, T. Ohl, J. Reuter,
  V.~Rothe, P.~Stienemeier \\
Corresponding author: J. Reuter {[juergen.reuter@desy.de]}}
\vspace*{.5cm}
  
\noindent The precision physics programmes of the FCC-ee demand for a precise simulation of 
all Standard Model (SM) processes and possible beyond-the-SM (BSM) signals in
a state-of-the-art way by means of Monte Carlo (MC) techniques. As a standard
tool for $\mathrm{e}^+\mathrm{e}^-$ simulations, the multipurpose event generator WHIZARD~\cite{Kilian:2007gr,WHIZARD_url} has been used: this generator was  originally
developed for the TESLA project, and was later  used, \eg for the ILC 
Technical Design Report~\cite{Baer:2013cma,Behnke:2013lya}. The WHIZARD 
package has a modular structure, which serves a modern unit-test driven
software development and guarantees a high level of maintainability and 
extendability. WHIZARD comes with its own fully general tree-level matrix
element generator for the hard process, O'Mega~\cite{Moretti:2001zz}. It generates amplitudes in a recursive way, based on the graph-theoretical concepts of directed acyclical graphs, thereby avoiding all redundancies. The matrix elements are generated either as compilable modern Fortran code or as 
byte-code instructions interpreted by a virtual machine~\cite{Nejad:2014sqa}. For QCD, WHIZARD uses the colour flow formalism~\cite{Kilian:2012pz}. Matrix elements support all kinds of particles and interactions up to spin-2. A large number of BSM models are hard-coded, particularly the minimal supersymmetric Standard Model (MSSM) and next-to-MSSM (NMSSM)~\cite{Ohl:2002jp,Hagiwara:2005wg}. General BSM models can be loaded from a Lagrangian level tool, using the interface to FeynRules~\cite{Christensen:2010wz}; from  version 2.8.0 of WHIZARD on (early summer 2019) a fully fledged interface to the general UFO format is available. One of the biggest assets of WHIZARD is its general phase space parametrization, which uses a heuristic based on the dominating subprocesses, which allows  integration and simulation of processes with up to ten fermions in the final state. The integration is based on an adaptive multichannel algorithm, called VAMP~\cite{Ohl:1998jn}. Recently, this multichannel adaptive integration has been enhanced to a parallelized version using the MPI3 protocol, showing speed-ups of up to 100~\cite{brass:2018xbv}, while a first physics study using this MPI parallelized integration and event generation has been published~\cite{Ballestrero:2018anz}. 

WHIZARD allows  all the necessary ingredients for a high-precision $\mathrm{e}^+\mathrm{e}^-$ event simulation to be described: the CIRCE1/CIRCE2 modules~\cite{Ohl:1996fi} simulate the spectrum of beamstrahlung (including beam energy spectra) that comes from classical electromagnetic radiation, owing to extreme space charge densities of highly collimated bunches for high-luminosity running. This takes care of a precise description of the peaks of the luminosity spectra and a smooth mapping of the tail that does not lead to artificial spikes and kinks in differential distributions. For the beam set-up, WHIZARD furthermore allows  polarised beams to be correctly described, with arbitrary polarisation settings and fractions, asymmetric beams, and crossing angles. QED initial-state radiation (ISR) is convoluted in a collinear approximation according to a resummation of soft photons to all orders and hard-collinear photons up to third order~\cite{Skrzypek:1990qs}. While this will give a correct normalization of the cross-section to the given QED order, one explicit ISR photon per beam will be inserted into the event record. A special handler generates transverse momentum according to a physical $p_\mathrm{T}$ distribution and boosts the complete events accordingly. This treatment is also available  for the photon beam components, according to the Weizs\"acker--Williams spectrum (equivalent photon approximation, EPA). 

\begin{figure}
    \centering
    \includegraphics[width=.6\textwidth]{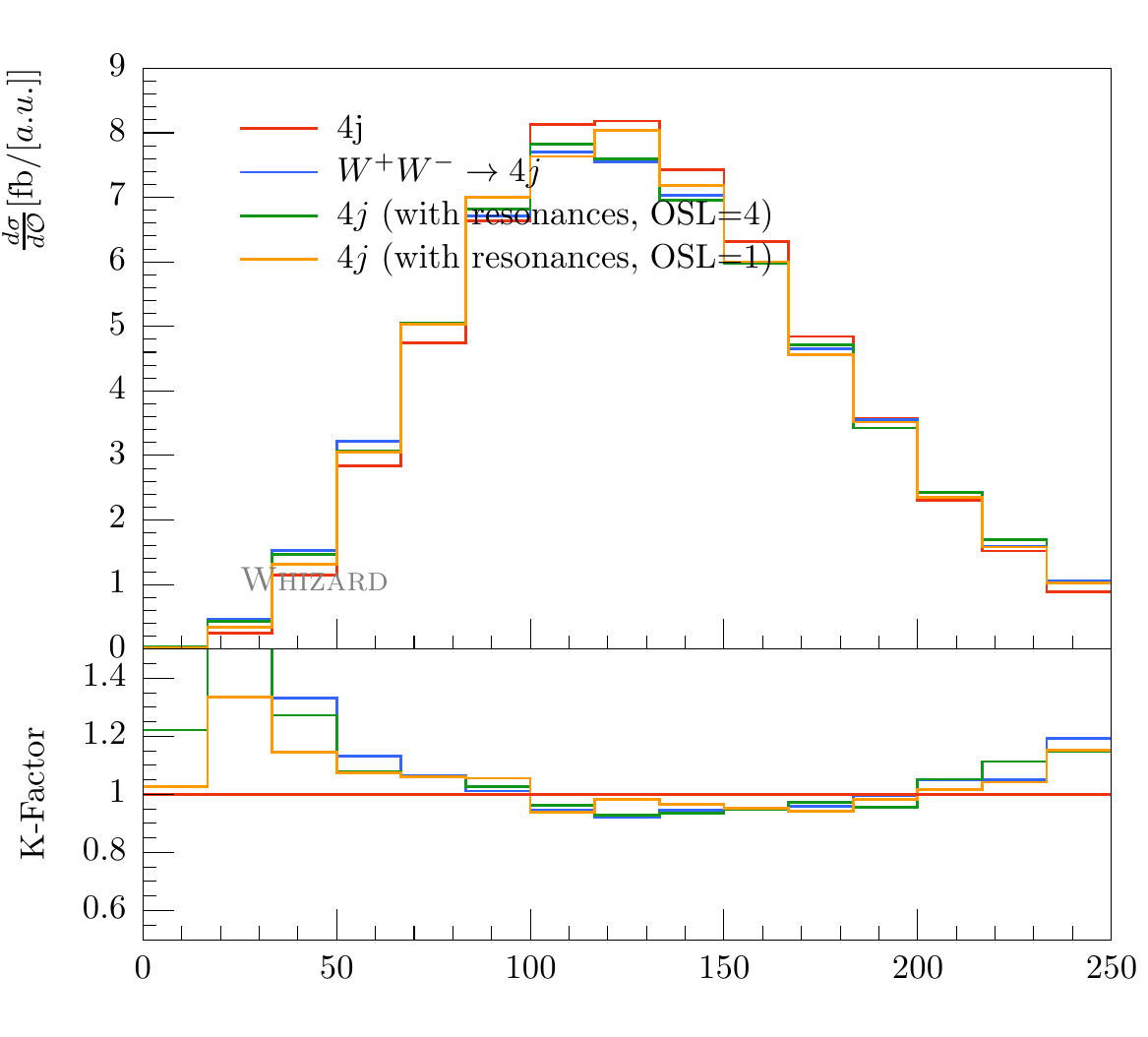}
    \caption{Energy distribution of photons in $\mathrm{e}^+\mathrm{e}^- \to \mathrm{jjjj}$ after parton
    shower and hadronization. Full amplitudes without resonance histories (red), factorised process $\mathrm{e}^+\mathrm{e}^- \to \mathrm{W}^+\mathrm{W}^- \to (\mathrm{jj})(\mathrm{jj})$ (blue), and 
    full process with resonance histories and different Breit--Wigner settings (green and orange, respectively).}
    \label{fig:whizard_res_matching}
%    \query{Please correct the figure labels in \Fref{fig:whizard_res_matching}.
%    Set variables in italic font. Set particle names, units, and the differential   d in roman font.}
\end{figure}

The MC generator WHIZARD offers a vast functionality, which cannot be given full justice here, \eg automatic generation of decays, factorised processes, including full spin correlations (which can also be switched off for case studies), specification of the helicity of decaying resonances, preset branching ratios, etc. WHIZARD supports all used HEP event formats, such as StdHEP, LHE, HepMC, LCIO, and various ASCII formats. It allows easy reweighting of event samples. WHIZARD has its own two QCD parton shower algorithms, a $k_\mathrm{T}$-ordered shower and an analytic parton shower~\cite{Kilian:2011ka}, and ships with the final version of PYTHIA6~\cite{Sjostrand:2006za} for showering and hadronization. The event records are directly interfaced and exchanged, and the framework has been validated with the full LEP dataset by the Linear Collider Generator Group in a set-up similar to the FCC-ee. Recently, we added
a corresponding interface for an externally linked PYTHIA8~\cite{Sjostrand:2007gs}, which, again,  allows direct communication between the event records of WHIZARD and PYTHIA. This offers the ability to use all the machinery for QCD jet matching and merging from PYTHIA inside WHIZARD. WHIZARD also directly interfaces Fastjet~\cite{Cacciari:2011ma} for jet clustering. 
One important feature of WHIZARD is the proper resonance matching of hadronically decaying resonances, \eg in the process $\mathrm{e}^+\mathrm{e}^- \to \mathrm{jjjj}$. This is predominantly WW production ($\sim$80\%), followed by ZZ production ($\lesssim$20\%) and the QCD four-jet continuum. When simulating full quantum theoretical amplitudes for four-jet production, the parton shower does not know intermediate resonances because of the full coherence of the process, and hence does not preserve the resonance mass of the hadronic Ws. WHIZARD allows one  to automatically determine underlying resonance histories, evaluates their approximate rates, and inserts resonance histories for final-state jets according to these rates. Figure \ref{fig:whizard_res_matching} shows, for the process $\mathrm{e}+\mathrm{e}^- \to \mathrm{jjjj}$, the photon energy distribution after hadronization and hadronic decays. The central line in the inset (red) shows the full process, which disagrees with LEP data, while the blue line shows the factorised process
$\mathrm{e}^+\mathrm{e}^- \to \mathrm{W}^+\mathrm{W}^- \to (\mathrm{jj})(\mathrm{jj})$ (where the shower program knows the resonance history) and the resonance-matched processes (green and orange). These correctly reproduce the data using full matrix elements, thereby allowing different handles on how far to take  Breit--Wigner tails of resonances into account. This type of matching has now been validated for six-jet processes, including $\mathrm{H} \to \mathrm{b}\bar{\mathrm{b}}$.

Finally, we comment on the NLO QCD capabilities of WHIZARD: WHIZARD has completed its final validation phase for lepton collider QCD NLO corrections, and version
3.0.0 will be released (approximately at the end of 2019) when proton collider processes are also completely validated. For NLO QCD corrections, WHIZARD uses the Frixione--Kunszt--Signer subtraction (FKS) formalism~\cite{Frixione:1995ms}, where  real and integrated subtraction terms are automatically  generated for all processes. WHIZARD also implements the resonance-aware variant~\cite{Jezo:2015aia}. Virtual multileg one-loop matrix elements are included from one-loop providers,
such as GoSam~\cite{Cullen:2014yla}, OpenLoops~\cite{Cascioli:2011va,Buccioni:2017yxi}, and
RECOLA~\cite{Actis:2016mpe,Denner:2017wsf}. First proof-of-principle NLO calculations have been made for electroweak corrections~\cite{Kilian:2006cj,Robens:2008sa} in lepton collisions, while NLO QCD has been implemented for LHC processes first~\cite{Binoth:2009rv,Greiner:2011mp}. 
The automated FKS subtraction has been tested and reported in a study
of off-shell $\mathrm{t}\bar{\mathrm{t}}$ and $\mathrm{t}\bar{\mathrm{t}}\mathrm{H}$ processes in lepton collisions ~\cite{Nejad:2016bci}. The complete validation of the automated NLO QCD set-up will be available after the version 3.0.0 release of WHIZARD (S. Bra\ss\ \textit{et al.}, in preparation).
WHIZARD allows  fixed-order NLO events for differential distributions to be generated at NLO QCD using weighted events, as well as automatically  POWHEG-matched and damped events~\cite{Nejad:2015opa,Reuter:2016qbi}. Decays at NLO QCD are treated in the same set-up as scattering processes. 

The scan of the top threshold is a crucial component of the FCC-ee physics programme. To determine systematic uncertainties from, \eg  event selection, WHIZARD allows  simulation at the completely exclusive final-state $\mathrm{e}^+\mathrm{e}^- \to \mathrm{W}^+ \mathrm{b W}^- \bar{\mathrm{b}}$, matching the continuum NLO QCD calculation to the NRQCD threshold NLL resummation~\cite{Bach:2017ggt}. This simulation is available via a specific top threshold model inside WHIZARD.

%------------------------------------------------------------------------------

\end{bibunit}

\label{sec-mtools-reuter}

\clearpage \pagestyle{empty}  \cleardoublepage
%============================================

\pagestyle{fancy}
\fancyhead[CO]{\thechapter.\thesection \hspace{1mm} FCC tau polarisation}
\fancyhead[RO]{}
\fancyhead[LO]{}
\fancyhead[LE]{}
\fancyhead[CE]{}
\fancyhead[RE]{}
\fancyhead[CE]{S. Banerjee, Z. Was}
\lfoot[]{}
\cfoot{-  \thepage \hspace*{0.075mm} -}
\rfoot[]{}

%%% definitions %%%%%%%%%%%%%%%%%%%%%%%%%%
\def\beqn{\begin{eqnarray}} \def\eeqn{\end{eqnarray}}
\def\beeq{\begin{eqnarray}}
\def\eeeq{\end{eqnarray}}
\def\nn{\nonumber}
\def\alphas{\alpha_{\rm S}}
    
\begin{bibunit}[elsarticle-num] % define the bib-style for the unit: elsarticle-num.bst
%  text-1; this is the corresponding section
%\putbib[2loops] % the *.bib
%\end{bibunit}
% go-on
%--- from: bibunits.sty, adapts the font size of ``References'' to section
\let\stdthebibliography\thebibliography
\renewcommand{\thebibliography}{%
\let\section\subsection
\stdthebibliography}
%---
    
\section
[FCC tau polarisation\\ {\it S.~Banerjee, Z.~Was}]
{FCC tau polarisation 
\label{contr:tau}}
\noindent
{\bf Contribution\footnote{This contribution should be cited as:\\
S.~Banerjee, Z.~Was, FCC tau polarisation,  
%04 DOI:10.23731/CYRM-2020-XXX.\thepage, in:
%04 \url{http://dx.doi.org/10.23731/CYRM-2020-XXX.\thepage}, in:
DOI: \href{http://dx.doi.org/10.23731/CYRM-2020-003.\thepage}{10.23731/CYRM-2020-003.\thepage}, in:
Theory for the FCC-ee, Eds. A. Blondel, J. Gluza, S. Jadach, P. Janot and T. Riemann,\\
CERN Yellow Reports: Monographs, CERN-2020-003,
%04 \url{http://dx.doi.org/10.23731/CYRM-2020-XXX}, p. \thepage.} 
DOI: \href{http://dx.doi.org/10.23731/CYRM-2020-003}{10.23731/CYRM-2020-003},
p. \thepage.
\\ \copyright\space CERN, 2020. Published by CERN under the 
%04-2
\href{http://creativecommons.org/licenses/by/4.0/}{Creative Commons Attribution 4.0 license}.} by: S. Banerjee, Z. Was} \\
{\bf Corresponding author:  Z.~Was {[z.was@cern.ch]}}
\vspace*{.5cm}

\noindent SM parameters, such as the $\uptau$ polarisation can be measured very precisely in $\uptau$ decays.
The phenomenology is quite similar to that of measurement of the $A_\mathrm{FB}$ parameter of the SM~\cite{Blondel:2018mad}.
Details of the $\uptau$ decay spectrum, as well as a good understanding of associated uncertainty, play an important role in this measurement of polarisation, because the spin of the $\uptau$ lepton is not measured directly.

The distribution of hadronic final-state products in decays of a $\uptau$ lepton needs to be evaluated to understand the substructure of the vertex. An important effect is related to bremsstrahlung, because the signature of every decay mode needs to take into account the final-state configurations with accompanying photons. Corresponding virtual corrections cancel the bulk of these effects and specialised programs, such as 
PHOTOS~\cite{Barberio:1993qi,Davidson:2010ew}, are useful. 

Corresponding effects can be sizeable; even during the early stages of LEP preparations, it was found~\cite{Boillot:1988re} that the 
corresponding corrections affect the slope of the $\uppi$ spectrum in $\uptau^- \to \uppi^- \upnu$, for example. This translates to a 0.013 effect on $\uptau$ idealised observable $A_\mathrm{pol}$. For more discussion and essential experimental context, see Ref.~\cite{ALEPH:2010aa}. 

However, not all of the final-state photons can be associated with bremsstrahlung. For example, in the cascade decay $\uptau^- \to \uppi^- \upomega \upnu$, a subsequent decay of $\upomega \to \uppi^0 \upgamma$ contributes to the final state $\uptau^- \to \uppi^- \uppi^0 \upgamma \upnu$, coinciding with the radiative corrections to the final state of the $\uptau^- \to \uprho^- \upnu$ decay channel. In this case, the photon originates from the $\upomega \to \uppi^0 \upgamma$ decay and is of non-QED bremsstrahlung origin.

The branching fractions for the $\uptau^- \to \uppi^- \upomega \upnu$ decay and for the $\upomega \to \uppi^0 \upgamma$ decay are 0.02 and 0.08, respectively~\cite{Tanabashi:2018oca}.  Thus, the resulting decay channel $\uptau^- \to \uppi^- \uppi^0 \upgamma \upnu$ contributes 0.0015 of all $\uptau$ decays.

Such contributions and subsequent changes of the hadronic decay energy spectrum in $\uptau$ decays need to be understood for each spin-sensitive channel. Resulting deformation of $\uptau^- \to \uprho^- \upnu$ decay spectra may mimic the contribution of the $\uptau$ polarisation that can be obtained from  future high-precision data analysis at the Belle II experiment.

This is the case when one of the $\uptau$ decay channels mimics 
bremsstrahlung correction for the other one. The dynamics of the low-energy strong interactions are difficult to obtain from 
a perturbative calculation.

Another hint of the non-point-like nature of the $\uptau$ vertex was explained in the corrections to the $\uppi$ energy spectra in the $\uptau^- \to \uppi^- \upnu$ decay channel~\cite{Decker:1993py, Decker:1994kw}. Although at the lowest order, the spectrum is fully determined by the Lorentz structure of the vertex, and the real and virtual photonic corrections play an important role in the level of precision under discussion. The dominant part of the effects of the QED bremsstrahlung from point-like sources can be seen in  Fig. 3 of Ref.~\cite{Decker:1994kw}, where the effects induced by hadronic resonances also play an important role.

The Belle II experiment is expected to collect $10^{11}$ $\uptau$ lepton decays with 50\,ab$^{-1}$ of data, and the detector is extremely well-suited to study $\uptau$ lepton physics. The backgrounds can be well-controlled in an electron--positron collider environment. We can expect that the $\uptau$ decay spectra can be measured without large degradation, owing to a highly granular electromagnetic calorimeter with large fiducial coverage, as explained in the Belle II technical design report~\cite{Abe:2010gxa}.

\end{bibunit}

\label{sec-mtools-was}

\clearpage \pagestyle{empty}  \cleardoublepage
%============================================

\pagestyle{fancy}
\fancyhead[CO]{\thechapter.\thesection \hspace{1mm} Electron--positron annihilation processes in MCSANCee}
\fancyhead[RO]{}
\fancyhead[LO]{}
\fancyhead[LE]{}
\fancyhead[CE]{}
\fancyhead[RE]{}
\fancyhead[CE]{A.~Arbuzov, S.~Bondarenko, Y.~Dydyshka, L.~Kalinovskaya, L.~Rumyantsev, R.~Sadykov, V.~Yermolchyk}
\lfoot[]{}
\cfoot{-  \thepage \hspace*{0.075mm} -}
\rfoot[]{}

%%% definitions %%%%%%%%%%%%%%%%%%%%%%%%%%
\def\beqn{\begin{eqnarray}} \def\eeqn{\end{eqnarray}}
\def\beeq{\begin{eqnarray}}
\def\eeeq{\end{eqnarray}}
\def\nn{\nonumber}
\def\alphas{\alpha_{\rm S}}
    
\begin{bibunit}[elsarticle-num] % define the bib-style for the unit: elsarticle-num.bst
%  text-1; this is the corresponding section
%\putbib[2loops] % the *.bib
%\end{bibunit}
% go-on
%--- from: bibunits.sty, adapts the font size of ``References'' to section
\let\stdthebibliography\thebibliography
\renewcommand{\thebibliography}{%
\let\section\subsection
\stdthebibliography}
%---
    
\section
[Electron--positron annihilation processes in MCSANCee \\ {\it A.~Arbuzov, S.~Bondarenko, Y.~Dydyshka, L.~Kalinovskaya, L.~Rumyantsev, R.~Sadykov, V.~Yermolchyk
}]
{Electron--positron annihilation processes in MCSANCee
\label{contr:sanc}}
\noindent
{\bf Contribution\footnote{This contribution should be cited as:\\
A.~Arbuzov, S.~Bondarenko, Y.~Dydyshka, L.~Kalinovskaya, L.~Rumyantsev, R.~Sadykov, V.~Yermolchyk, Electron--positron annihilation processes in MCSANCee,  
%04 DOI:10.23731/CYRM-2020-XXX.\thepage, in:
%04 \url{http://dx.doi.org/10.23731/CYRM-2020-XXX.\thepage}, in:
DOI: \href{http://dx.doi.org/10.23731/CYRM-2020-003.\thepage}{10.23731/CYRM-2020-003.\thepage}, in:
Theory for the FCC-ee, Eds. A. Blondel, J. Gluza, S. Jadach, P. Janot and T. Riemann,\\
CERN Yellow Reports: Monographs, CERN-2020-003,
%04 \url{http://dx.doi.org/10.23731/CYRM-2020-XXX}, p. \thepage.} 
DOI: \href{http://dx.doi.org/10.23731/CYRM-2020-003}{10.23731/CYRM-2020-003},
p. \thepage.
\\ \copyright\space CERN, 2020. Published by CERN under the 
%04-2
\href{http://creativecommons.org/licenses/by/4.0/}{Creative Commons Attribution 4.0 license}.} by: A.~Arbuzov,
S.~Bondarenko,
Y.~Dydyshka,
L.~Kalinovskaya,
L.~Rumyantsev,
R.~Sadykov,
V.~Yermolchyk} \\
{\bf Corresponding author: A.~Arbuzov  {[arbuzov@theor.jinr.ru]}}
\vspace*{.5cm}

\newcommand{\labhel}[5]{{}^{\rm{#1}}_{{#2}{#3}{#4}{#5}} } 
\newcommand{\pair}[2]{\phantom{$+0.6$}\makebox[0pt][r]{$#1$},~\phantom{$+0.6$}\makebox[0pt][r]{$#2$}}
\newcommand{\tpair}[2]{{#1},~{#2}}

\newcommand{\FCCee}{FCC-ee}
\newcommand{\SANC}{SANC\xspace}
\newcommand{\MCSANC}{\texttt{MCSANC}\xspace}
\newcommand{\MCSANCee}{\texttt{MCSANCee}\xspace}
\newcommand{\WHIZARD}{\texttt{WHIZARD}\xspace}%
\newcommand{\CalcHEP}{\texttt{CalcHEP}\xspace}%
\newcommand{\GRACE}{\texttt{GRACE}\xspace}%
\newcommand{\ROOT}{\texttt{ROOT}\xspace}%
\newcommand{\mFOAM}{\texttt{mFOAM}\xspace}%
  
\title{Electron-positron annihilation processes in MCSANCee}
  
\author{A.~Arbuzov$^{a}$, S.~Bondarenko$^{a}$, Ya.~Dydyshka$^{b}$, 
  L.~Kalinovskaya$^{b}$, L.~Rumyantsev$^{b,c}$, R.~Sadykov$^{b}$, V.~Yermolchyk$^{b}$}

%\begin{abstract}
\noindent The Monte Carlo event generator \MCSANCee is used to estimate
the significance of polarisation effects in one-loop electroweak radiative corrections.
The electron--positron annihilation processes $\mathrm{e}^+\mathrm{e}^- \to \upmu^-\upmu^+$ $(\uptau^-\uptau^+$, ZH)
are considered, taking into account conditions of future colliders. 
%\end{abstract}

\subsection{Introduction}

Radiative  corrections with effects due to polarisation of the initial particles
will play an important role in the high-precision programme at the \FCCee.
\MCSANCee is a Monte Carlo generator of unweighted events for polarised { $\mathrm{e}^+\mathrm{e}^-$}
scattering and annihilation processes with complete one-loop electroweak (EW) corrections.
The generator uses the adaptive Monte Carlo algorithm {\mFOAM} \cite{Jadach:2005ex},
which is a part of the {\ROOT}~\cite{homepageROOT} framework.

The \SANC computer system is capable of calculating cross-sections of general
Standard Model (SM) processes with up to three final-state particles~\cite{Andonov:2004hi,Arbuzov:2015yja}.
Using the \SANC system, we calculated electroweak radiative corrections
at the one-loop level to the  polarised Bhabha scattering \cite{Bardin:2017mdd,Blondel:2018mad},
which is the basic normalization process at $\mathrm{e}^+\mathrm{e}^-$ colliders.
For processes
\begin{eqnarray}
  \mathrm{e}^+ \mathrm{e}^- \to \upmu^-\upmu^+\ (\uptau^-\uptau^+,\, \mathrm{ZH})
\, ,   \label{Processes_ee}
\end{eqnarray}
we made a few upgrades of the standard procedures in the \SANC system.
We investigated the effect of the polarisation degrees of initial particles 
on the differential cross-sections. We found that the EW corrections to the total
cross-section range from $-18\%$  to $+69\%$,
when the centre-of-mass energy $\sqrt{s}$ varies in the set 250\,GeV, 500\,GeV, and 1\,TeV.

\subsection{Cross-section structure}

The cross-section of a generic {$2 \to 2(\upgamma)$}
process {$\mathrm{e}^+\mathrm{e}^- \to \mathrm{X}_3 \mathrm{X}_4 (\upgamma)$}
($\mathrm{X}_3 \mathrm{X}_4 = 
%e^-e^+,
\upmu^-\upmu^+,\uptau^-\uptau^+,\mathrm{ZH}$) reads
\begin{equation}
  \sigma_{P_{\mathrm{e}^-}P_{\mathrm{e}^+}} = \frac{1}{4}\sum_{\upchi_1,\upchi_2}(1+\upchi_1P_{\mathrm{e}^-})
  (1+\upchi_2P_{\mathrm{e}^+})\sigma_{\upchi_1\upchi_2},
\nonumber
\end{equation}
where { $\upchi_i = -1(+1)$} corresponds to a lepton with left (right) helicity state.

The cross-section at the one-loop level can be divided into four parts:
\begin{eqnarray}
\sigma^{\rm{1-loop}} = \sigma^{\rm{Born}}
+ \sigma^{\mathrm{virt}}({\lambda})
+ \sigma^{\mathrm{soft}}({\lambda},{\omega})
+ \sigma^{\mathrm{hard}}({\omega}),
\nonumber
\end{eqnarray}
where
{ $\sigma^{\mathrm{Born}}$} is the Born level cross-section, 
{ $\sigma^{\mathrm{virt}}$} is the virtual (loop) contribution,
{ $\sigma^{\mathrm{soft}}$} is the contribution due to soft-photon emission, and
{ $\sigma^{\mathrm{hard}}$} is the contribution due to hard photon emission
(with energy { $E_{\upgamma} > {\omega}$}).
Auxiliary parameters {$\lambda$} (`photon mass') and {$\omega$} cancel out after summation.

We treat all contributions using the helicity amplitudes  approach:
\begin{eqnarray} 
\sigma^{\mathrm{Part}}_{\upchi_1\upchi_2} = \dfrac{1}{2s}\sum_{\upchi_i,i\geq3}
\Bigl|\mathcal{H}\labhel{Part}{\upchi_1}{\upchi_2}{\upchi_3,\ldots}{}|^2\mathrm{d}\mathrm{LIPS},
\end{eqnarray}
where  $\mathrm{Part}\in\{\mathrm{Born}$, $\mathrm{virt}$, $\mathrm{hard}\}$, and 
$\mathrm{d}\mathrm{LIPS}$ is a volume element of the Lorentz-invariant phase space.

The soft-photon contribution is factorised in front of the Born level cross-section:
\begin{eqnarray}
\mathrm{d}\sigma\labhel{soft}{ \upchi_1}{ \upchi_2}{}{} =
\mathrm{d}\sigma\labhel{Born}{ \upchi_1}{ \upchi_2}{}{} 
\cdot 
\frac{\alpha}{2\uppi}K^\mathrm{soft}(\omega,\lambda).
\nonumber
\end{eqnarray}

\subsection{Numerical results and comparison}

The following input parameters are used for numerical estimates and comparisons:
\begin{align}
 \alpha^{-1}(0) & = 137.035\,999\,76, \nonumber\\
 M_\mathrm{W} &  = 80.451\,495\,8 \; \mbox{GeV}, & M_\mathrm{Z} &= 91.1876 \; \mbox{GeV},
& \Gamma_\mathrm{Z}& = 2.499\,77 \; \mbox{GeV}, \nonumber\\
 m_\mathrm{e} & = 0.510\,999\,07 \; \mbox{MeV}, & m_\upmu& = 0.105\,658\,389 \; \mbox{GeV},
& m_\uptau& = 1.777\,05 \; \mbox{GeV}, \nonumber\\
 m_\mathrm{d} & = 0.083 \; \mbox{GeV}, & m_\mathrm{s} &= 0.215 \; \mbox{GeV},
& m_\mathrm{b}& = 4.7 \; \mbox{GeV}, \nonumber\\
 m_\mathrm{u} & = 0.062 \; \mbox{GeV}, & m_\mathrm{c}& = 1.5 \; \mbox{GeV},
& m_\mathrm{t}& = 173.8 \; \mbox{GeV}.\nonumber
\end{align}

The following simple cuts are imposed:
\begin{align}
 |\cos{\theta}| & < 0.9,\nonumber\\
 E_{\upgamma} & > 1\UGeV \quad (\mbox{for comparison of hard bremsstrahlung}).\nonumber
\end{align}

Tuned comparison of our results for polarised Born and hard bremsstrahlung with the
results of the  {\WHIZARD}~\cite{Kilian:2007gr}
and {\CalcHEP}~\cite{Belyaev:2012qa}
programs shows an agreement within statistical errors.
The unpolarised \textit{soft + virtual} contribution agrees with
the results of Ref.~\cite{Lorca:2004fg}
for  $\mathrm{e}^+\mathrm{e}^- \to 
\upmu^+\upmu^-(\uptau^+\uptau^-)
$ and with the ones of the \GRACE system~\cite{Belanger:2003sd}.
For  $\mathrm{e}^+\mathrm{e}^- \to \mathrm{ZH}$, we found an agreement with the results of
the \GRACE system~\cite{Belanger:2003sd} 
and with those given in Ref.~\cite{Denner:1992qf}.

The integrated cross-sections of \Eref{Processes_ee} and the
relative corrections $\updelta$ are given in 
Tables \ref{mumutautau} \cite{Sadykov:2019acat}
and \ref{ZH} \cite{Bondarenko:2018sgg} for various energies and beam
polarisation degrees.

\begin{table}
\centering
\caption{Processes $\mathrm{e}^+\mathrm{e}^-\to \upmu^+\upmu^-$ and $\mathrm{e}^+\mathrm{e}^-\to\uptau^+\uptau^-$:  Born vs one loop}     
\label{mumutautau}
\begin{tabular}{lllllll}
\hline\hline
\pair{P_{\mathrm{e}^-}}{P_{\mathrm{e}^+}} &$\sigma_{\upmu^+\upmu^-}^{\mathrm{Born}}$  &$\sigma_{\upmu^+\upmu^-}^{\mathrm{1-loop}}$  &$\updelta$ 
                   &$\sigma_{\uptau^+\uptau^-}^{\mathrm{Born}}$&$\sigma_{\uptau^+\uptau^-}^{\mathrm{1-loop}}$
                   &$\updelta$ \\
& (pb) & (pb) & (\%) & (pb) & (pb ) & (\%)  \\ \hline
\multicolumn{7}{l}{$\sqrt{s} = 250$\,GeV}\\
 \pair{0}{0}      & 1.417(1)  & 2.397(1) & 69.1(1)& 1.417(1) & 2.360(1)& 66.5(1) \\%\hline
 \pair{-0.8}{0}   & 1.546(1)  & 2.614(1) & 69.1(1)& 1.546(1) & 2.575(1)& 66.5(1) \\%\hline
 \pair{-0.8}{-0.6}& 0.7690(2) & 1.301(1) & 69.2(1)& 0.7692(1)& 1.298(1)& 68.8(1) \\%\hline
 \pair{-0.8}{+0.6}& 2.323(1)  & 3.927(1) & 69.1(1)& 2.324(1) & 3.850(1)& 65.7(1) \\
\multicolumn{7}{l}{$\sqrt{s} = 500$\,GeV}\\
 \pair{0}{0}      &  0.3436(1) & 0.4696(1)& 36.7(1)& 0.3436(1)& 0.4606(1) & 34.0(3)\\%\hline
 \pair{-0.8}{0}   &0.3716(1) & 0.4953(1)  & 33.3(1)& 0.3715(1)& 0.4861(1) & 30.8(1)\\%\hline
 \pair{-0.8}{-0.6}&0.1857(1) & 0.2506(1)  & 35.0(1)& 0.1857(1)& 0.2466(1) & 32.8(1)\\%\hline
 \pair{-0.8}{+0.6}&0.5575(1) & 0.7399(1)  & 32.7(1)& 0.5575(1)& 0.7257(1) &30.1(1)\\
\multicolumn{7}{l}{$\sqrt{s} = 1000$\,GeV}\\
 \pair{0}{0}      & 0.085\,35(1) & 0.116\,3(1)  & 36.2(1)& 0.085\,34(2)& 0.113\,4(1) & 33.6(1)\\%\hline
 \pair{-0.8}{0}   & 0.092\,13(1) & 0.121\,2(1)  & 31.6(1)& 0.092\,13(1)& 0.118\,85(2)& 29.0(1)\\%\hline
 \pair{-0.8}{-0.6}& 0.046\,08(1) & 0.061\,69(1) & 33.9(1)& 0.046\,08(1)& 0.060\,67(1)& 31.7(1)\\%\hline
 \pair{-0.8}{+0.6}& 0.138\,2(1)  &  0.180\,7(1) & 30.8(1)& 0.138\,2(1) & 0.177\,0(1) & 28.1(1)\\
\hline\hline
\end{tabular}
\end{table}

\begin{table}  
  \caption{Process $\mathrm{e}^+\mathrm{e}^-\to \mathrm{ZH}$:  Born vs one
loop}
\centering  
\begin{tabular}{llll}
\hline\hline
\pair{P_{\mathrm{e}^-}}{P_{\mathrm{e}^+}} &$\sigma_\mathrm{ZH}^{\mathrm{Born}}$  &$\sigma_\mathrm{ZH}^{\mathrm{1-loop}}$
 & $\updelta$ \\
& (pb) & (pb) & (\%) \\\hline
\multicolumn{4}{l}{$\sqrt{s} = 250$\,GeV}\\
 \pair{0}{0}      & 205.64(1)& 186.6(1)& \phantom{1}$-$9.24(1) \\%\hline
 \pair{-0.8}{0}   & 242.55(1)& 201.5(1)& $-$16.94(1)\\%\hline
 \pair{-0.8}{-0.6}& 116.16(1)& 100.8(1)& $-$13.25(1)\\%\hline
 \pair{-0.8}{+0.6}& 368.93(1)& 302.2(1)& $-$18.10(1)\\
\multicolumn{4}{l}{$\sqrt{s} = 500$\,GeV}\\
 \pair{0}{0}      & 51.447(1)& 57.44(1)& 11.65(1) \\%\hline
 \pair{-0.8}{0}   & 60.680(1)& 62.71(1)& \phantom{1}3.35(2)  \\%\hline
 \pair{-0.8}{-0.6}& 29.061(1)& 31.25(1)& \phantom{1}7.54(1)  \\%\hline
 \pair{-0.8}{+0.6}& 92.299(1)& 94.17(2)& \phantom{1}2.03(2)  \\
\multicolumn{4}{l}{$\sqrt{s} = 1000$\,GeV}\\
 \pair{0}{0}      & 11.783(1)& 12.92(1)& \phantom{$-$}9.68(1)  \\%\hline
 \pair{-0.8}{0}   & 13.898(1)& 13.91(1)& \phantom{$-$}0.10(2)  \\%\hline
 \pair{-0.8}{-0.6}& \phantom{1}6.6559(1)& \phantom{1}6.995(1)& \phantom{$-$}5.09(2)  \\%\hline
 \pair{-0.8}{+0.6}& 21.140(1)& 20.83(1)& $-$1.47(2) \\
\hline\hline
\end{tabular}
\label{ZH}
\end{table}

 In these tables, we summarise the estimates of
the Born and one-loop cross-sections  and the relative corrections $\updelta$
 of the processes $\mathrm{e}^+\mathrm{e}^- \to  \upmu^+\upmu^-, (\uptau^+\uptau^-, \mathrm{ZH})$
for the set (\tpair{0}{0}; \tpair{$-$0.8}{0}; \tpair{$-$0.8}{$-$0.6}; \tpair{$-$0.8}{+0.6}) 
of longitudinal  polarisations $P_{\mathrm{e}^+}$ and $P_{\mathrm{e}^-}$
of the positron and electron beams, respectively. Values of energy $250$, $500$, and 
$1000$\,GeV were taken. The relative correction $\updelta$ is defined as
\begin{eqnarray}
\updelta = \frac{\sigma^{\mbox{1-loop}}-\sigma^{\mbox{Born}}}{\sigma^{\mbox{Born}}} \cdot 100 \%.
\end{eqnarray}
 
\subsection{Conclusion}

As can be seen from  Tables \ref{mumutautau} and \ref{ZH},
the difference between values of $\updelta$ for polarisation degrees of 
initial particles (\tpair{0}{0}) and (\tpair{$-$0.8}{0}; \tpair{$-$0.8}{$-$0.6}; \tpair{$-$0.8}{+0.6})
amounts to a significant value: 6--20\%.

In assessing theoretical uncertainties for future $\mathrm{e}^+\mathrm{e}^-$ colliders, it is
necessary to achieve an accuracy of approximately $10^{-4}$ for many observables.
Estimating the value of  $\updelta$ at different degrees of polarisation of the initial states,
we see that taking beam polarisation  into account is crucial.

Further development of the process library of the Monte Carlo generator \MCSANCee
involves $\mathrm{e}^+\mathrm{e}^- \to \upgamma\upgamma$ (plus cross-symmetric processes)
and  (`W fusion') $\mathrm{e}^+ \mathrm{e}^- \to \upnu_\mathrm{e}\upnu_\mathrm{e} \mathrm{H}$.
We have started  work on the introduction of higher-order corrections,
as well as on the implementation of multiphoton emission contributions.

%\clearpage 

%\section*{Acknowledgments}

%Results were obtained within the framework of state's task N~3.9696.2017/8.9 from Ministry of Education and Science of Russia.
 
%\subsection*{References of Rodrigo}
%

\end{bibunit}

\label{sec-mtools-sanc}

\clearpage \pagestyle{empty}  \cleardoublepage
%============================================
\pagestyle{fancy}
\fancyhead[CO]{\thechapter.\thesection \hspace{1mm} Global electroweak fit in the FCC-ee era
}
\fancyhead[RO]{}
\fancyhead[LO]{}
\fancyhead[LE]{}
\fancyhead[CE]{}
\fancyhead[RE]{}
\fancyhead[CE]{J.~Erler, M.~Schott}
\lfoot[]{}
\cfoot{-  \thepage \hspace*{0.075mm} -}
\rfoot[]{}

\begin{bibunit}[elsarticle-num]  

\section[Global electroweak fit in the FCC-ee era
 \\ {\it J.~Erler, M.~Schott}]
{Global electroweak fit in the FCC-ee era
\label{contr:erler}}
\noindent
{\bf Contribution\footnote{This contribution should be cited as:\\
J.~Erler, M.~Schott, Global electroweak fit in the FCC-ee era,  
%04 DOI:10.23731/CYRM-2020-XXX.\thepage, in:
%04 \url{http://dx.doi.org/10.23731/CYRM-2020-XXX.\thepage}, in:
DOI: \href{http://dx.doi.org/10.23731/CYRM-2020-003.\thepage}{10.23731/CYRM-2020-003.\thepage}, in:
Theory for the FCC-ee, Eds. A. Blondel, J. Gluza, S. Jadach, P. Janot and T. Riemann,\\
CERN Yellow Reports: Monographs, CERN-2020-003,
%04 \url{http://dx.doi.org/10.23731/CYRM-2020-XXX}, p. \thepage.} 
DOI: \href{http://dx.doi.org/10.23731/CYRM-2020-003}{10.23731/CYRM-2020-003},
p. \thepage.
\\ \copyright\space CERN, 2020. Published by CERN under the 
%04-2
\href{http://creativecommons.org/licenses/by/4.0/}{Creative Commons Attribution 4.0 license}.} by: J.~Erler, M.~Schott \\
Corresponding author: J.~Erler {[erler@fisica.unam.mx]}}
\vspace*{.5cm}

\noindent The top quark and Higgs boson masses have been predicted before their respective discoveries by the global fit of the Standard Model to electroweak precision data. With the Higgs boson discovery and the measurement of its mass, the last missing parameter of the Standard Model has been fixed and thus the internal consistency of the Standard Model can be probed at a new level by comparing direct measurements with the indirect predictions of the global electroweak fit. In this section, we discuss the expected precisions in the most important indirect predictions that are expected in the FCC-ee era and compare them with the state of the art.

%\subsection{Global Electroweak Fit in the FCC-ee Era}

Global electroweak analyses and fits have a long history in particle physics, starting  before the discovery of the W and Z bosons. The basic idea of the global electroweak fit is the comparison of the state-of-the-art calculations of the electroweak precision observables with the most recent experimental data to constrain the free parameters of the fit and to test the goodness
of fit. The free parameters of the SM relevant for the global electroweak analysis are the coupling constant of the electromagnetic, weak, and strong interactions, as well as the masses of the elementary fermions and bosons. This number can be reduced by fixing parameters with insignificant uncertainties compared with the sensitivity of the fit, as well as imposing the relations of the electroweak unification. The typical  floating parameters chosen in the fit are the masses of the Z and the Higgs boson, the top, the bottom, and charm quark masses, and the coupling parameters $\Delta \alpha_{5}$ and $\alpha_\mathrm{S} (m_\mathrm{Z})$. An introduction and a review of the current status of the global electroweak fit can be found in Ref. \cite{Erler:2019hds}.

Besides a global analysis of the consistency between observables and their relations, the global electroweak fit can be used to indirectly determine and hence predict the expected values of observables. Technically, this indirect parameter determination is performed by scanning the parameter in a chosen range and calculating the corresponding $\chi^2$ values. It should be noted that the value of $\chi^2_{\min}$ is not relevant for the uncertainty estimation, only its difference relative to the global minimum, $\Delta \chi^2 := \chi^2 - \chi^2_{\min}$. 

\begin{figure*}[h!]
\centering
\resizebox{0.32\textwidth}{!}{\includegraphics{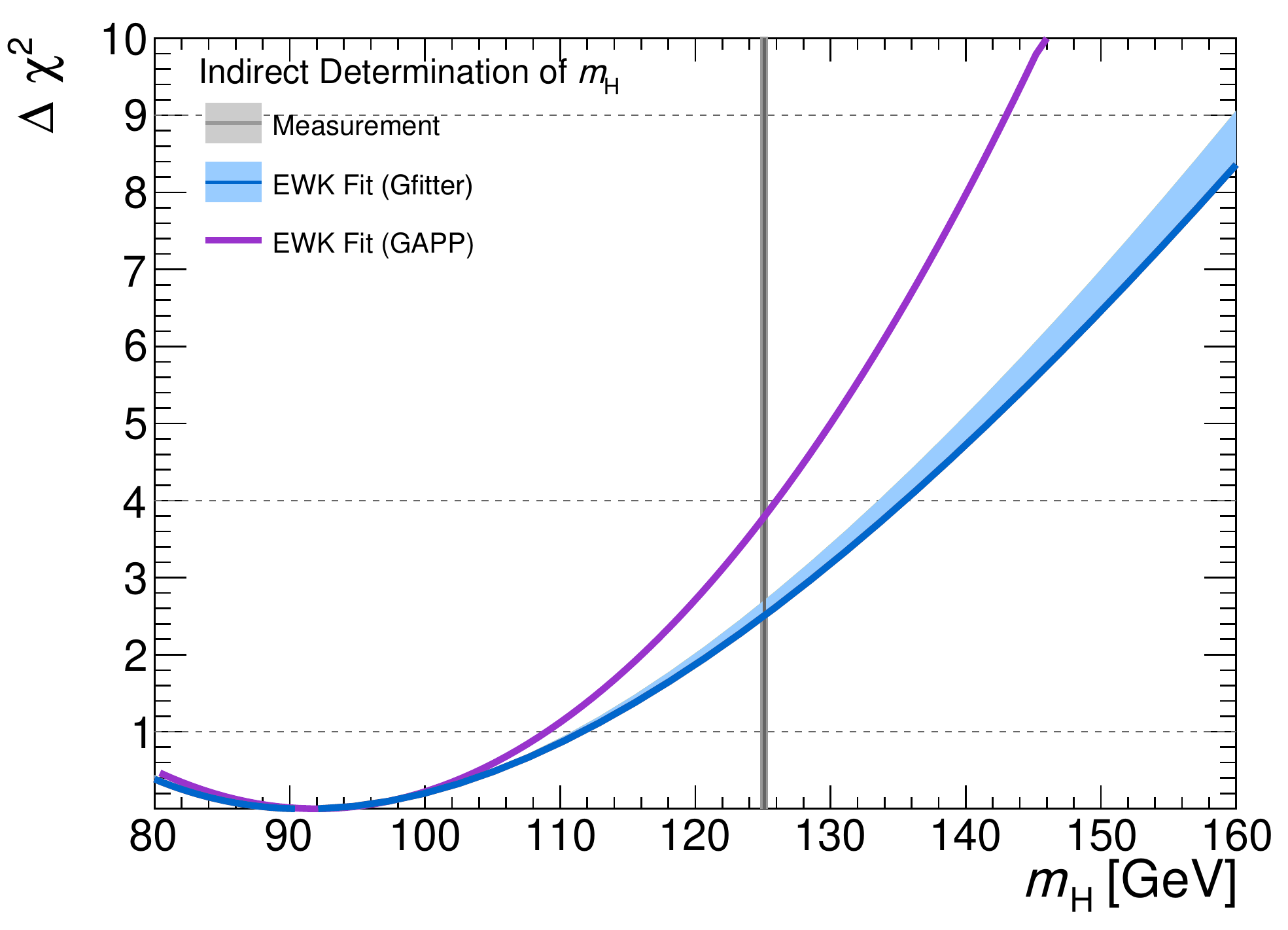}}
\resizebox{0.32\textwidth}{!}{\includegraphics{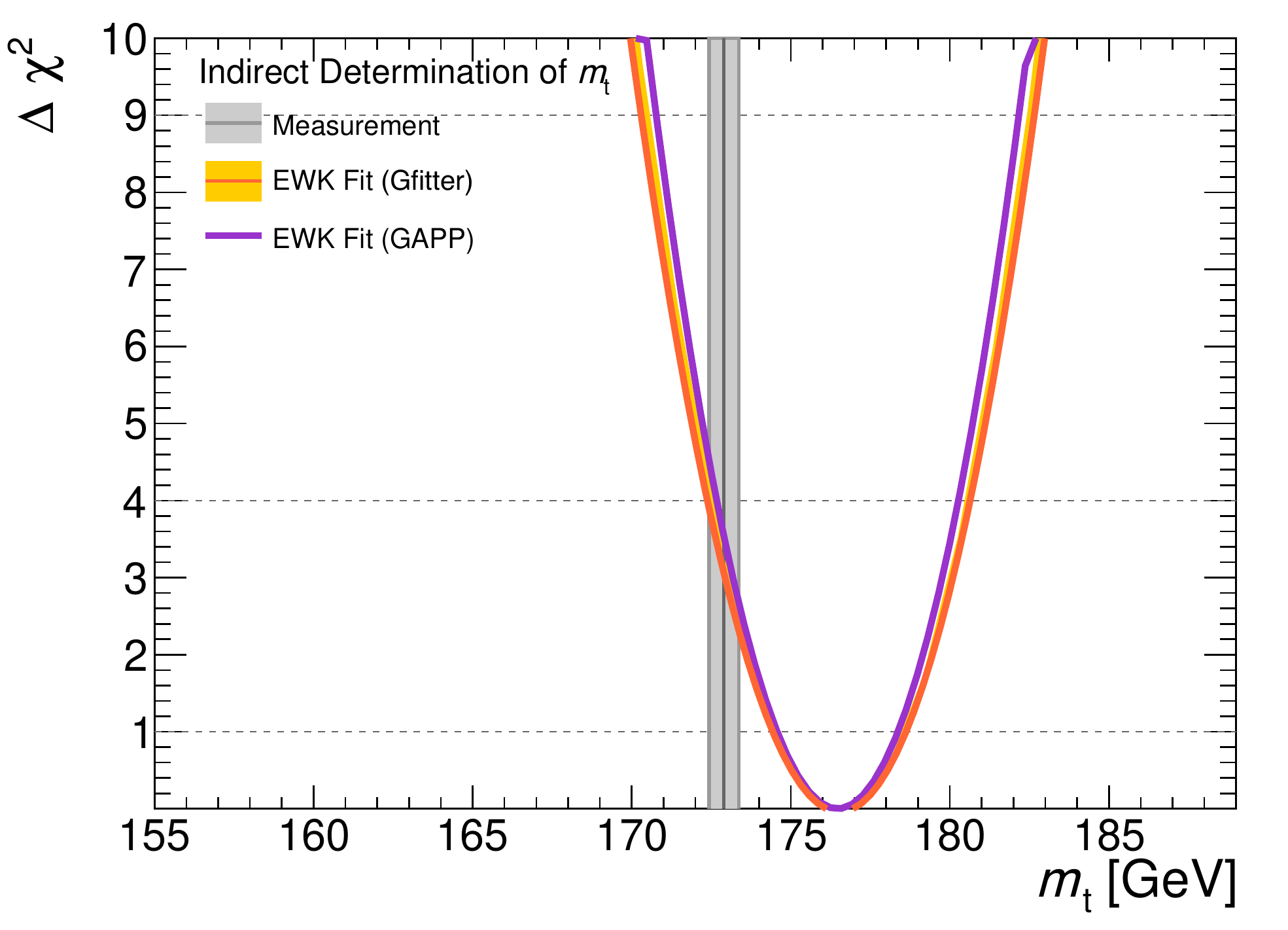}}
\resizebox{0.32\textwidth}{!}{\includegraphics{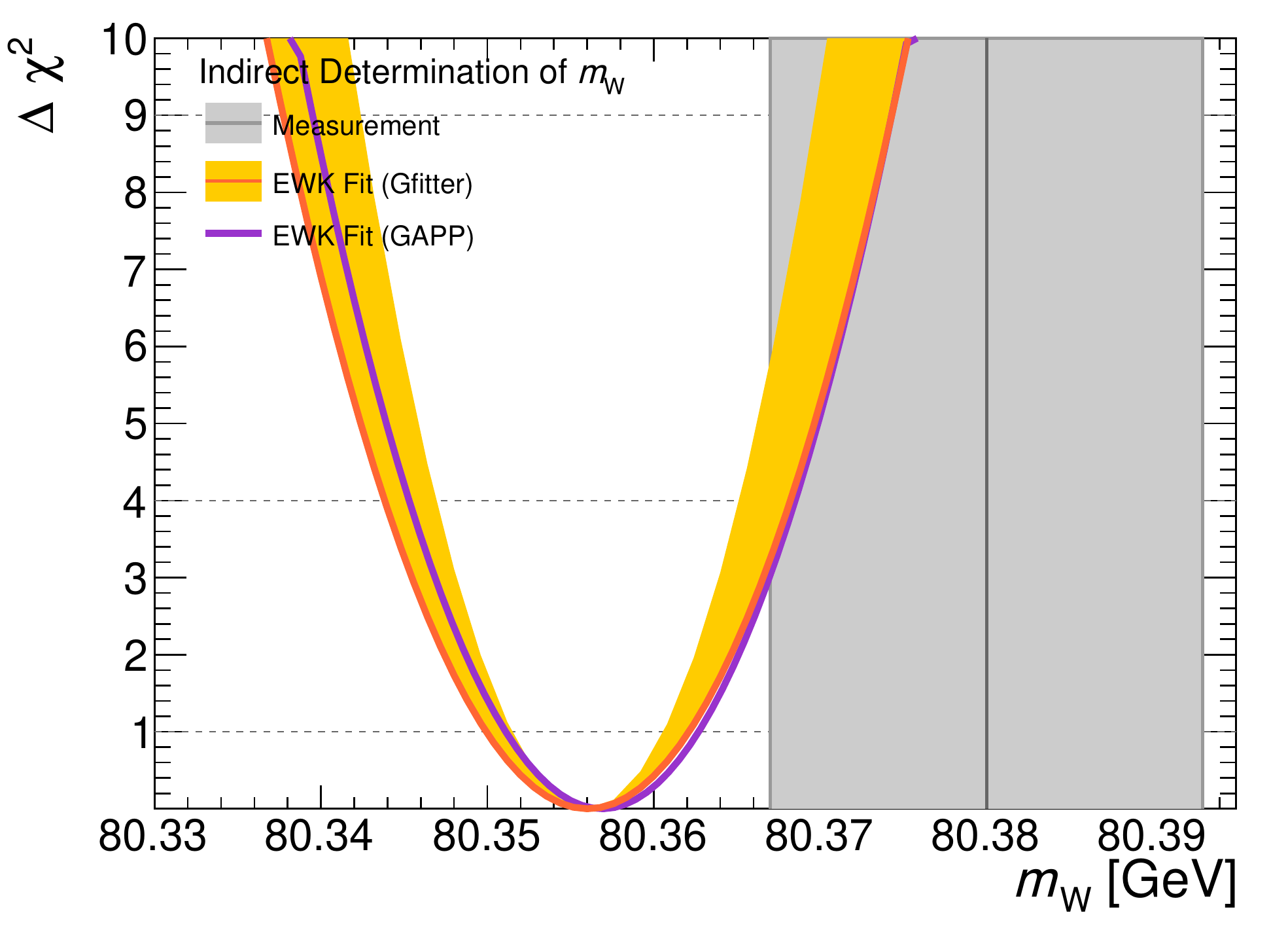}}
\caption{Comparisons of $\chi^2$ distributions for scanning different observables using the Gfitter and the GAPP, using the current experimental values and uncertainties. Theoretical uncertainties are indicated by the filled blue and yellow areas, respectively.}
\label{fig:FitNow}
%\query{Please correct the figure labels of \Fref{fig:FitNow}. Set variables
%in italic font and particle names in roman font.}
\end{figure*}

These indirect determinations have been recently performed with the latest measured values of electroweak precision observables in Ref. \cite{Erler:2019hds} and the state-of-the art fitting frameworks GAPP and Gfitter. While GAPP (Global Analysis of Particle Properties) \cite{Erler:1999ug} is a Fortran library for the evaluations of pseudo-observables, Gfitter consist of independent object-oriented C++ code \cite{Haller:2018nnx}. Both frameworks yield consistent results. Selected input parameters of the fit, including their current experimental uncertainty, are summarised in Table \ref{tab:Overview}, while the $\Delta \chi^2$ distributions for the indirect determinations of $M_\mathrm{H}$, $M_\mathrm{W}$, and $m_\mathrm{top}$ are summarised in Figure \ref{fig:FitNow}.  

\begin{table}[h!]
\caption{Overview of selected observables, their values, and current uncertainties, which are used or determined within the global electroweak fit \cite{Erler:2019hds}. The future expected FCC-ee  uncertainties are also shown \cite{Gomez-Ceballos:2013zzn, Abada:2019zxq}.}
 \label{tab:Overview}
\footnotesize
\centering
\begin{tabular}{l l  l l l l} 
\hline \hline
Parameter                                               & Current value                         & FCC-ee unc.-                    & Parameter             & Current value                                         & FCC-ee unc.-     \\      
                                                        &                                               & target                          &                               &                                                       & target          \\
\hline
$M_\mathrm{H}$                                           & $125.09 \pm 0.15$\,GeV         & $\pm0.01$\,GeV           & $M_\mathrm{Z}$                 & $91.1875 \pm 0.0021$\,GeV              & $<$$0.1$\,MeV      \\
$M_\mathrm{W}$                                           & \phantom{$1$}$80.380 \pm 0.013$\,GeV        & $\pm0.6$\,MeV            & $\Gamma_\mathrm{Z}$    & \phantom{$1$}$2.4952 \pm 0.0023$\,GeV               & 25\,keV  \\
$\Gamma_\mathrm{W}$                                      &\phantom{$10$}$2.085 \pm 0.042$\,GeV          & $\pm1.0$\,MeV            & $\sigma^0_\mathrm{had}$      & $41.540 \pm 0.037$\,nb                        & $0.004$\,nb     \\
$m_\mathrm{top}$                                               & $172.90 \pm 0.47$\,GeV             & $\pm15$\,MeV           & $R_\mathrm{b}$                 & \phantom{$1$}$0.21629 \pm 0.00066$                    & $<$$0.00006$    \\
$\Delta\alpha_\mathrm{had}[\times 10^{-5}]$    & $2758 \pm 10$                 & $\pm3$                          & $A^\mathrm{FB}_\mathrm{LR}(b)$      & \phantom{$1$}$0.0992 \pm 0.0016$                   & $\pm 0.0001$\\
\hline \hline
\end{tabular}
\end{table}

\begin{figure*}[h!]
\centering
\resizebox{0.32\textwidth}{!}{\includegraphics{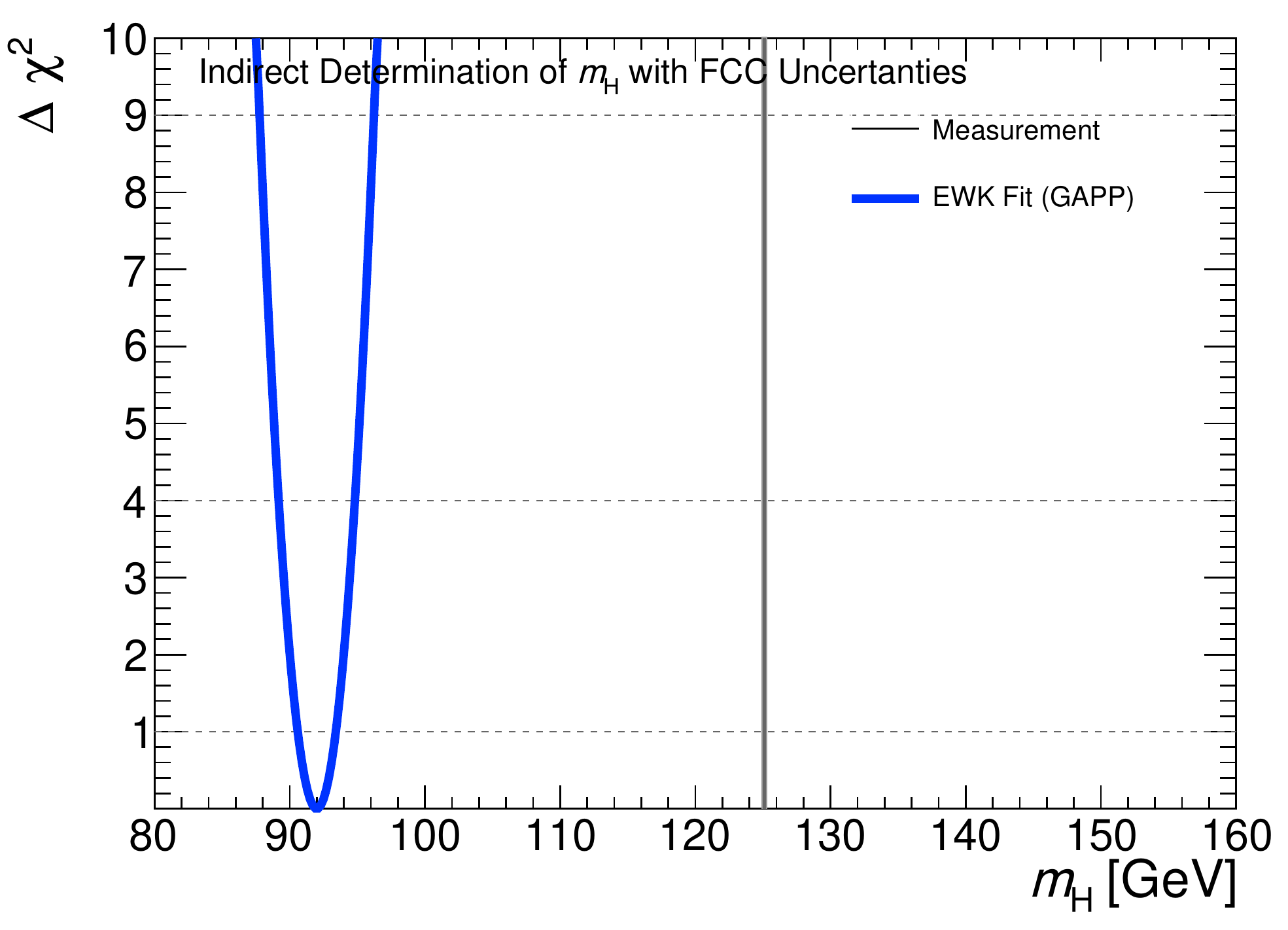}}
\resizebox{0.32\textwidth}{!}{\includegraphics{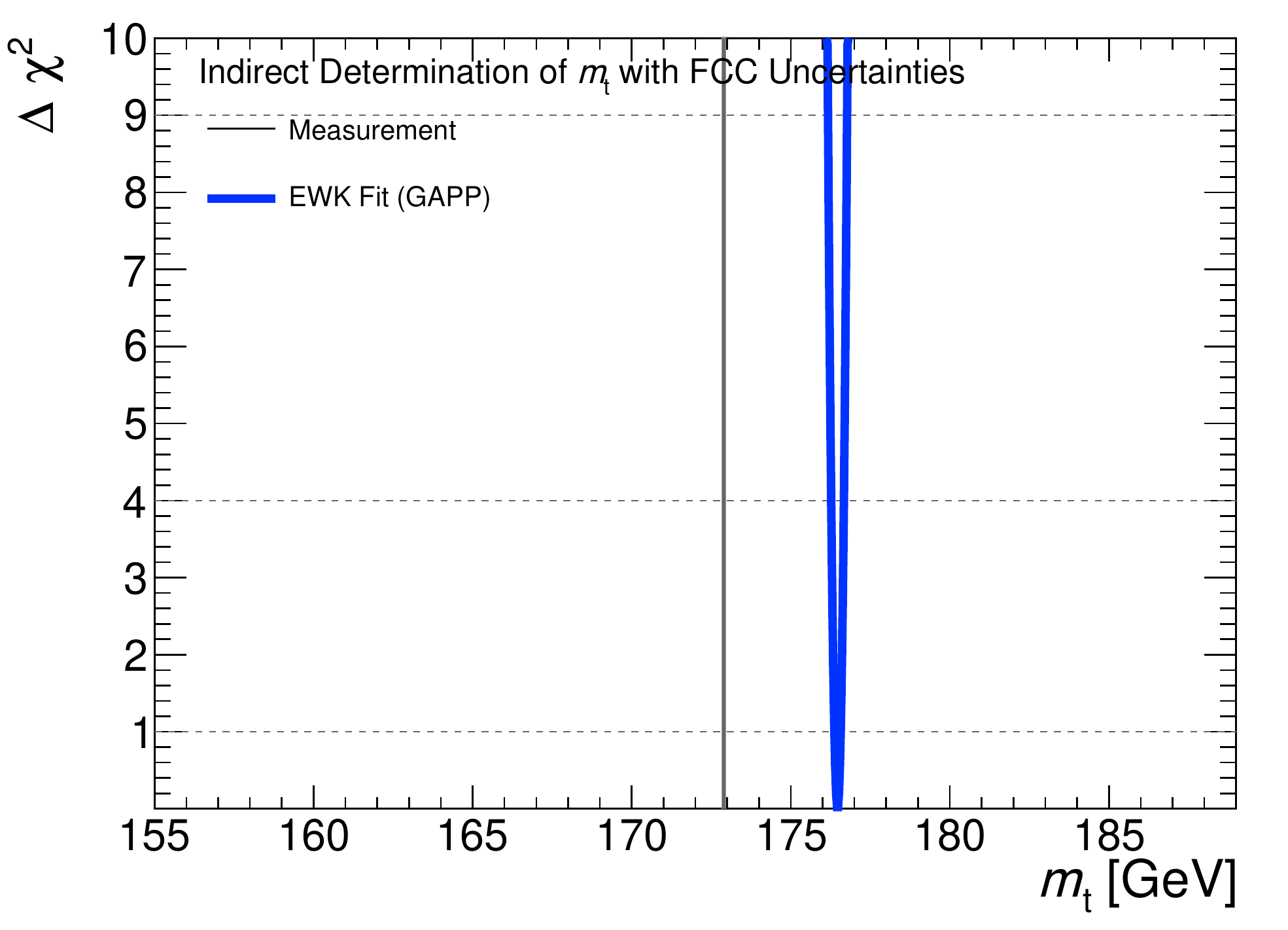}}
\resizebox{0.32\textwidth}{!}{\includegraphics{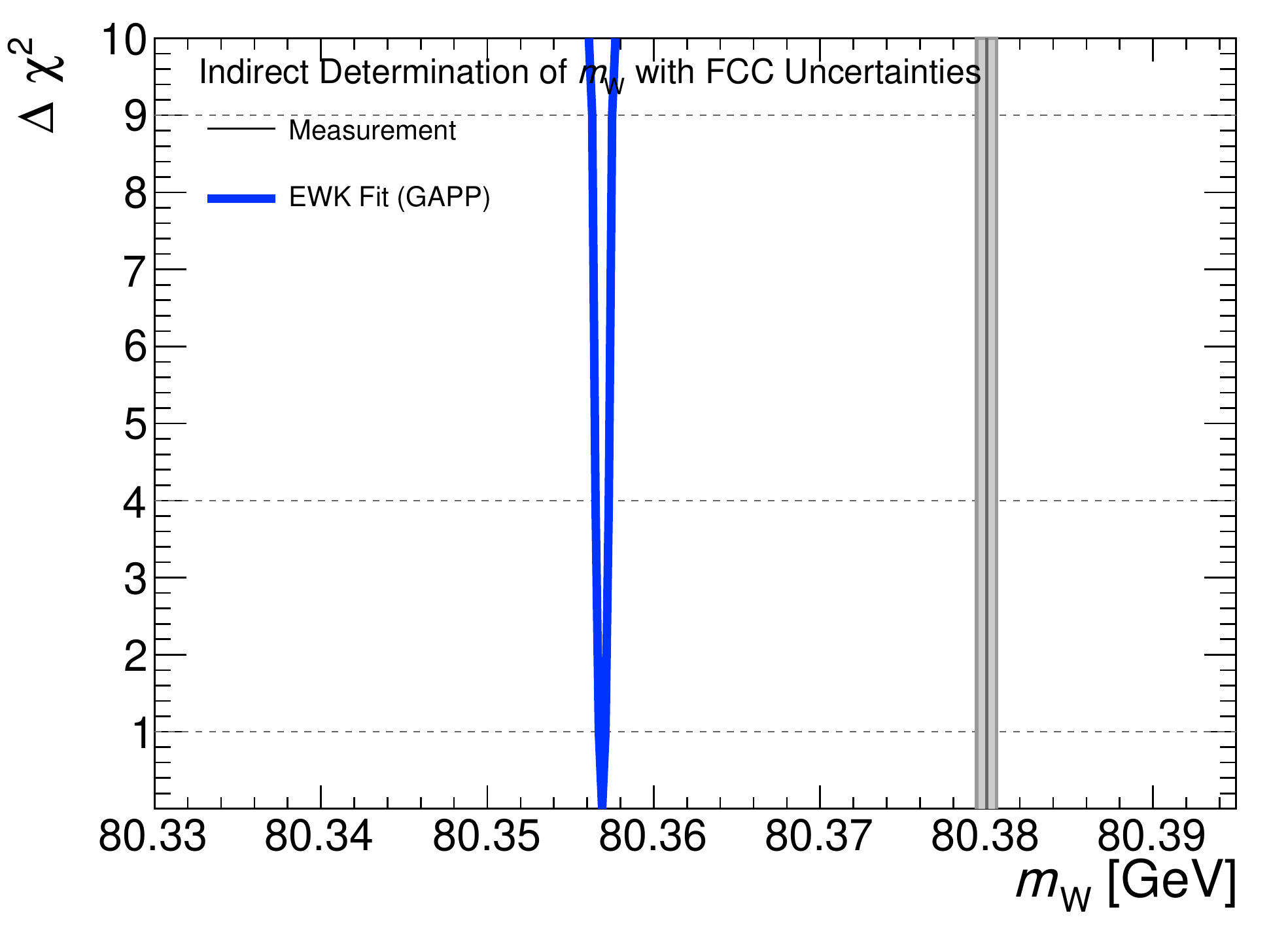}}
\caption{Comparisons of $\chi^2$ distributions for scanning different observables using GAPP with the current experimental values  but the expected uncertainties from FCC.}
\label{fig:FitFuture}
%\query{Please correct the figure labels of \Fref{fig:FitFuture}. Set variables in italic font and particle names in roman font.}
\end{figure*}

We repeat the indirect fit of these observables using the GAPP program, mainly by assuming  the FCC-ee projections and target uncertainties from Refs. \cite{Gomez-Ceballos:2013zzn, Abada:2019zxq}, as well as non-dominant theory uncertainties from unknown higher orders. It should be noted that the uncertainty in the weak mixing angle is assumed to be $\pm 5\times 10^{-6}$ during the fit.\footnote{This uncertainty combines the expected measurement precision of the asymmetry
observables, \ie it can be seen as a combination of $A^\mathrm{FB}(\upmu)$, $A^\mathrm{FB}(\mathrm{b})$ and the $\uptau$ polarisation measurements.} 

Similar studies have been previously performed  \cite{Baak:2014ora, Azzi:2017iih}. Of special importance are the significantly lower uncertainties in $m_\mathrm{Z}$, $m_\mathrm{W}$, and $m_\mathrm{top}$ (Table \ref{tab:Overview}), which could be reduced by an order of magnitude. The $\Delta \chi^2$ distributions for $M_\mathrm{H}$, $M_\mathrm{W}$, and $m_\mathrm{top}$ are summarised in Figure \ref{fig:FitFuture}, yielding precisions of the indirect determinations of $\Delta M_\mathrm{H}=\pm 1.4$\,GeV,  $\Delta M_\mathrm{W}=\pm 0.2$\,MeV, and  $\Delta m_\mathrm{top}=\pm 0.1$\,GeV. Thus, the indirect test of the internal consistency of the electroweak sector would be brought to a new level. The uncertainty in $m_\mathrm{H}$ increases from $\pm 1.4$\,GeV to $\pm 5.7$\,GeV, if no advances are made on the theory side. Likewise, the expected uncertainty in the indirectly determined value of $\Delta\alpha_\mathrm{had}$ increases from 0.05\% to 0.1\%. Last but not least, the number of active neutrinos $N_\upnu$ can be constrained at FCC-ee within $\pm 0.0006$, compared with the current result $N_\upnu = 2.992 \pm 0.007$.

\end{bibunit}

\label{sec-mtools-gfits}

\clearpage \pagestyle{empty}  \cleardoublepage
%============================================
%============================================
\chapter
%[Methods and tools]
{SMEFT} \label{chsmeft}
%============================================

\pagestyle{fancy}
\fancyhead[CO]{\thechapter.\thesection \hspace{1mm} \texttt{{CoDEx}}: BSM physics being realised as  SMEFT}
\fancyhead[RO]{}
\fancyhead[LO]{}
\fancyhead[LE]{}
\fancyhead[CE]{}
\fancyhead[RE]{}
\fancyhead[CE]{S.D. Bakshi, J. Chakrabortty, S.K. Patra}
\lfoot[]{}
\cfoot{-  \thepage \hspace*{0.075mm} -}
\rfoot[]{}

\begin{bibunit}[elsarticle-num]
\let\stdthebibliography\thebibliography
\renewcommand{\thebibliography}{%
\let\section\subsection
\stdthebibliography}

\section[\texttt{{CoDEx}}: BSM physics being realised as  SMEFT \\ {\it S.D. Bakshi, J. Chakrabortty, S.K. Patra }]
        {\texttt{{CoDEx}}: BSM physics being realised as  SMEFT
        %  \\ {\it S.D. Bakshi, J. Chakrabortty, S.K. Patra }
        }
{\bf Contribution\footnote{This contribution should be cited as:\\
S.D. Bakshi, J. Chakrabortty, S.K. Patra, \texttt{{CoDEx}}: BSM physics being realised as  SMEFT,  
%04 DOI:10.23731/CYRM-2020-XXX.\thepage,\texttt{{CoDEx}}: BSM physics being realised as  SMEFT in:
%04 \url{http://dx.doi.org/10.23731/CYRM-2020-XXX.\thepage}, in:
DOI: \href{http://dx.doi.org/10.23731/CYRM-2020-003.\thepage}{10.23731/CYRM-2020-003.\thepage}, in:
Theory for the FCC-ee, Eds. A. Blondel, J. Gluza, S. Jadach, P. Janot and T. Riemann,\\
CERN Yellow Reports: Monographs, CERN-2020-003,
%04 \url{http://dx.doi.org/10.23731/CYRM-2020-XXX}, p. \thepage.} 
DOI: \href{http://dx.doi.org/10.23731/CYRM-2020-003}{10.23731/CYRM-2020-003},
p. \thepage.
\\ \copyright\space CERN, 2020. Published by CERN under the 
%04-2
\href{http://creativecommons.org/licenses/by/4.0/}{Creative Commons Attribution 4.0 license}.} by: S.D. Bakshi, J. Chakrabortty, S.K. Patra\\
Corresponding author: S.D. Bakshi [{sdbakshi13@gmail.com}]}
\vspace*{.5cm}

        \begin{small}
                \noindent
                {\bf Program summary}                                                   \\
                {\em Program title: \texttt{CoDEx}}             \\
                {\em Version: \texttt{1.0.0}}                                   \\
                {\em Licensing provisions: \texttt{CC By 4.0}}                         \\
                {\em Programming language: \texttt{Wolfram Language$^\text{\tiny \textregistered}$}}                                   \\
                {\em Mathematica$^\text{\tiny \textregistered}$ Version: \texttt{10+}}                                   \\
                {\em URL:} \url{https://effexteam.github.io/CoDEx}                                      \\
                {\em Send BUG reports and questions:} \href{mailto:effex.package@gmail.com}{\texttt{effex.package@gmail.com}}                                   \\
        \end{small}
        
%       \newpage
        
        %% main text
        %%%%%%%%%%%%%%%%%%%%%%%%%%%%%%%%%%%%%%%%%%%%%%%%%%%%%%%%%%%%%%%%%%%%%%%%%%%%%%%
\subsection{Introduction}
        %%%%%%%%%%%%%%%%%%%%%%%%%%%%%%%%%%%%%%%%%%%%%%%%%%%%%%%%%%%%%%%%%%%%%%%%%%%%%%%
In spite of the non-observation of any new resonances after the discovery of the Standard Model (SM) Higgs-like particle, which announces the success of the SM, we have enough reason to believe the existence of theories beyond it (BSM), with the SM as a part. As any such theory will affect the electroweak and the Higgs sector, and the sensitivity of these precision observables is bound to increase in the near future, indirect estimation of the allowed room left for BSM using Standard Model effective field theory (SMEFT) is well motivated. 

Provided that the S-matrix can be expanded perturbatively in the inverse powers of the ultraviolet scale ($\Lambda^{-1}$), and the resultant series is convergent, we can integrate out heavy degrees of freedom and the higher mass dimensional operators capture their impact through -- $\sum_{i}(1/\Lambda^{d_i - 4}) C_i \mathcal{O}_i$, where $d_i$ is the operator mass dimension (>5), and $C_i$, a function of BSM parameters, is the corresponding Wilson coefficient. Among different choices of operator base, we restrict ourselves to `\mmaInlineCell{Code}{SILH}' \cite{Giudice:2007fh,Contino:2013kra} and `\mmaInlineCell{Code}{Warsaw}' \cite{Grzadkowski:2010es,Jenkins:2013zja, Jenkins:2013wua,Alonso:2013hga} bases. All WCs are computed at the cut-off scale $\Lambda$, usually identified as the mass of the heavy field. The truncation the $1/\Lambda$ series depends on the experimental precision of the observables \cite{Furnstahl:2015rha}.  %One can consult these lectures  \cite{Georgi:1994qn,Kaplan:1995uv,Manohar:1996cq,Burgess:2007pt,Rothstein:2003mp} where effective field theory has been introduced and discussed in great detail. 
Already, there has been quite good progress in building packages and libraries \cite{Gripaios:2018zrz,Falkowski:2015wza,Celis:2017hod,Criado:2017khh,Aebischer:2018bkb,Aebischer:2017ugx}. 
        
One can justifiably question the validity of choosing to use SMEFT over the full BSM Lagrangian;  the answer lies in the trade-off between the computational challenge of the full BSM and the precision of the observables. The choice of $\Lambda$ ensures the convergence of the $M_\mathrm{Z}/\Lambda$ series. Using the anomalous dimension matrix $(\gamma)$ (which is basis dependent), the SMEFT WCs  $C_i(\Lambda)$ (computed at $\Lambda$) are evolved to $C_i(M_\mathrm{Z})$, some of which are absent at the $\Lambda$ scale, as the matrix $\gamma$ contains non-zero off-diagonal elements. See Refs. \cite{Jenkins:2013zja, Jenkins:2013wua,Alonso:2013hga,Wells:2015cre} regarding the running of the SMEFT operators. We need to choose only those `complete' bases in which the precision observables are defined. 
        
\begin{figure}
        \centering
        \includegraphics[scale=.4]{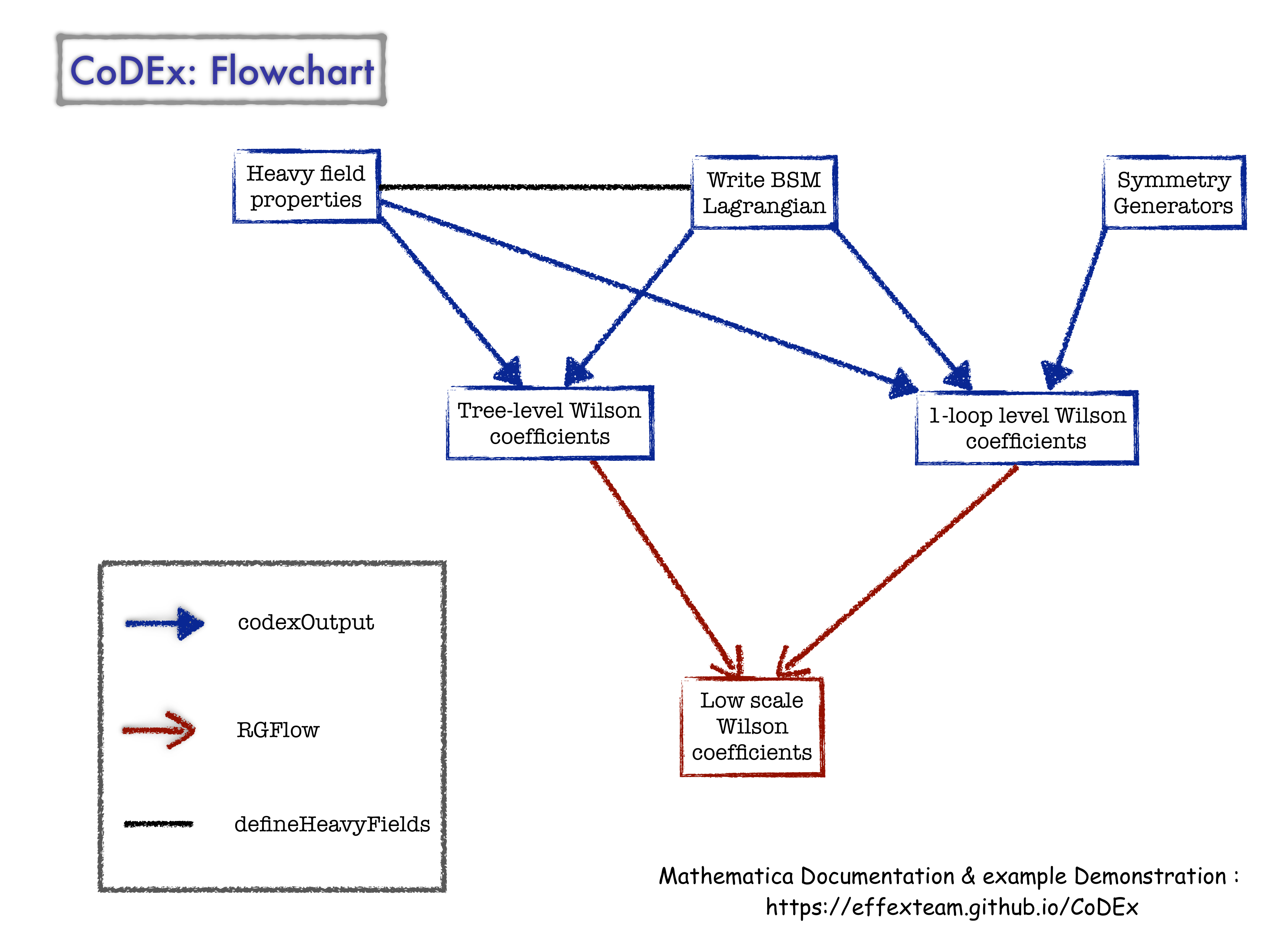}
        \caption{Flow-chart demonstrating the working principle of  \texttt{CoDEx}}
        \label{flowch}
%        \query{In \Fref{flowch}, please correct `tree level' to `tree-level' and
 %       delete the random digit 1 at the bottom of the figure.}
  %      \query{All figures must be cited in the text. Please check
   %     that the citation for \Fref{flowch} is in the correct place.}
\end{figure}

\mmaInlineCell{Code}{\mmaDef{CoDEx}}, a {\em Mathematica$^\text{\tiny \textregistered}$} package \cite{Bakshi:2018ics}, in addition to integrating out the heavy field propagators from tree and one-loop processes and generating SMEFT operators up to dimension-6, provides the WCs as a function of BSM parameters (\Fref{flowch}). In this draft, we briefly discuss the underlying principle of \mmaInlineCell{Code}{\mmaDef{CoDEx}}, and give one illustrative example of the workflow. Details about downloading, installation, and detailed documentation of the functions are available at the website \cite{CoDEx}.  

%%%%%%%%%%%%%%%%%%%%%%%%%%%%%%%%%%%%%%%%%%%%%%%%%%%%%%%%%%%%%%%%%%%%%%%%%

%%%%%%%%%%%%%%%%%%%%%%%%%%%%%%%%%%%%%%%%%%%%%%%%%%%%%%%%
\subsection{The package in detail}\label{sec:pack}
%%%%%%%%%%%%%%%%%%%%%%%%%%%%%%%%%%%%%%%%%%%%%%%%%%%%%%%
        
 \mmaInlineCell{Input}{\mmaDef{CoDEx}} is a Wilson coefficient calculator, developed in the Mathematica environment. The algorithm of this code is based on the `{\underline {co}}variant \underline{d}erivative \underline{ex}pansion'  method discussed in Refs. \cite{Gaillard:1985uh,Cheyette:1985ue,Henning:2014wua,Henning:2016lyp,Ellis:2016enq,Fuentes-Martin:2016uol,delAguila:2016zcb,Henning:2015alf,Drozd:2015rsp,Wells:2015uba,Lehman:2015coa,Huo:2015nka,Huo:2015exa,Chiang:2015ura,Lehman:2015via}.
  Each and every detail of this package can be found in Ref. \cite{CoDEx}.  The main functions provided by this program  are given in  Table \ref{tab:funclst}. Here we have demonstrated the working methodology of  \mmaInlineCell{Input}{\mmaDef{CoDEx}} with an explicit example.

\begin{table}
        \caption{Main functions provided by \texttt{CoDEx}}
        \label{tab:funclst}
               \centering
                %       \rowcolors{1}{lightgray}{lightgray}
                            \begin{tabular}{{ll}}
                        \hline \hline
                    {Function} &  {Details}  \\
                        \hline
                       \mmaInlineCell{Code}{CoDExHelp}  &  Opens the \mmaInlineCell{Code}{\mmaDef{CoDEx}} guide, with all help files listed\\ 
                       \mmaInlineCell{Code}{treeOutput}  &  Calculates WCs generated from tree-level processes\\ 
                        \mmaInlineCell{Code}{loopOutput}  &  Calculates WCs generated from one-loop processes\\ 
                        \mmaInlineCell{Code}{codexOutput}  &  Generic function for WC calculation with\\ 
                        & choices for level, bases, \etc, given with \mmaInlineCell{Code}{\mmaDef{OptionValues}}\\
                        \mmaInlineCell{Code}{defineHeavyFields}  &  Creates representation of heavy fields\\ 
                        &  Use the output to construct BSM Lagrangian\\
                        \mmaInlineCell{Code}{texTable}  &  Given a \mmaInlineCell{Code}{\mmaDef{List}}, returns the \LaTeX~output of a tabular\\ 
                        & environment, displayed or copied to clipboard$^*$\\
                        \mmaInlineCell{Code}{formPick}  &  Applied on a list of WCs from a specific operator basis,\\ 
                        & reformats the output in the specified style \\
                        \mmaInlineCell{Code}{RGFlow}  &  RG Flow of WCs of dim. 6 operators in \mmaInlineCell{Code}{`Warsaw'} basis, \\
                        &  from matching scale to a lower (arbitrary) scale  \\
                        \mmaInlineCell{Code}{initializeLoop}  &  Prepares the Isospin and colour symmetry generators \\
                        &  for a specific model with a specific heavy field content: \\
                        &  \mmaInlineCell{Code}{loopOutput} can only be run after this step is completed  \\
                        \hline \hline
                \end{tabular}\\
\raggedright{$^*${This is a simplified version of the package titled \texttt{TeXTableForm} \cite{TeXTableForm}.}}
  \end{table}

%\section{The flowchart of  \texttt{CoDex} and used functions}

%%%%%%%%%%%%%%% HOW TO RUN %%%%%%%%%%%%%%%%%%%%%%%
%%%%%%%%%%%%%%%%%%%%%%%%%%%%%%%%%%%
\subsubsection{Detailed example: electroweak $SU(2)_L$ triplet scalar with hypercharge \\$Y=1$}\label{sec:detailex}
%%%%%%%%%%%%%%%%%%%%%%%%%%%%%%%%%%%%%%%%%%%%%%%%%%%%%%%%%%%%%%%%%%%%%%%%
Here, we have demonstrated the workflow of \mmaInlineCell{Code}{\mmaDef{CoDEx}} with the help of a complete analysis of a representative model. 
%Say the Lagrangian is:
{\small 
        \begin{align}
        \label{lbsmCTS}
        \mathcal{L}_\mathrm{BSM} =  \mathcal{L}_\mathrm{SM} \ + \mathrm{Tr} [ (\mathcal{D}_{\mu} \Delta)^{\dagger} (\mathcal{D}^{\mu} \Delta ) ] - m_{\Delta}^{2} \mathrm{Tr} [ \Delta^{\dagger} \Delta ] + \mathcal{L}_\mathrm{Y} - V(H,\Delta), 
        \end{align}
}
where
        \begin{align}
      &&  V(H,\Delta) &= \zeta_{1} (H^{\dagger} H)\mathrm{Tr}[ \Delta^{\dagger} \Delta ] + \zeta_{2} (H^{\dagger} \tau^{i} H)\mathrm{Tr}[ \Delta^{\dagger} \tau^{i} \Delta ] + \left  [ \mu ( H^\mathrm{T} \mathrm{i} \sigma^{2} \Delta^{\dagger} H ) + \mathrm{h.c.} \right], \\
        \text{and} && \mathcal{L}_{Y} &= y_{\Delta} L^\mathrm{T} C \mathrm{i} \tau^2 \Delta L + \mathrm{h.c.} 
        \end{align}

Here, the heavy field is $\Delta$.  Once this heavy field, $\Delta$, is integrated out using \mmaInlineCell{Input}{\mmaDef{CoDEx}}, the effective operators up to dimension-6 for both bases are generated. The effective operators and their respective Wilson coefficients are listed in Tables \ref{tab:CTSS}
to \ref{tab:CTS5}. Next, we give the exact steps that must be followed to run the code and compute the desired results.

\begin{table*}
        \caption{Effective operators and Wilson coefficients in `SILH' basis for complex triplet scalar ($Y=1$)  model.}
\label{tab:CTSS}
%       \label{tab:CTS}
           \centering
     %       \subfloat[``SILH'' basis]{
                \begin{tabular}{ll}
                        \hline \hline
                        $O_\mathrm{2B}$  &  $\dfrac{g_\mathrm{Y}^2}{160 \uppi ^2 m_{\Delta }^2}$  \\[12pt]
                                               $O_\mathrm{2W}$  &  $\dfrac{g_\mathrm{W}^2}{240 \uppi ^2 m_{\Delta }^2}$  \\[12pt]
                        $O_\mathrm{3W}$  &  $\dfrac{g_\mathrm{W}^2}{240 \uppi ^2 m_{\Delta }^2}$  \\[12pt]
                         $O_6$  &  $-\dfrac{\zeta _1 \mu ^2}{m_{\Delta }^4}-\dfrac{\zeta _2 \mu ^2}{4 m_{\Delta }^4}-\dfrac{\zeta _1^3}{4 \uppi ^2 m_{\Delta }^2}-\dfrac{\zeta _2^2 \zeta _1}{32 \uppi ^2 m_{\Delta }^2}$  \\[12pt]
                        $O_{\text{BB}}$  &  $\dfrac{\zeta _1}{32 \uppi ^2 m_{\Delta }^2}$  \\[12pt]
                         $O_\mathrm{H}$  &  $\dfrac{\zeta _1^2}{8 \uppi ^2 m_{\Delta }^2}+\dfrac{\mu ^2}{2 m_{\Delta }^4}$  \\[12pt]
                      
                        $O_\mathrm{R}$  &  $\dfrac{\zeta _2^2}{96 \uppi ^2 m_{\Delta }^2}+\dfrac{\mu ^2}{m_{\Delta }^4}$  \\[12pt]
                                               $O_\mathrm{T}$  &  $\dfrac{\zeta _2^2}{192 \uppi ^2 m_{\Delta }^2}-\dfrac{\mu ^2}{2 m_{\Delta }^4}$  \\[12pt]
                      
                        $O_{\text{WB}}$  &  $-\dfrac{\zeta _2}{96 \uppi ^2 m_{\Delta }^2}$  \\[12pt]
                                      $O_{\text{WW}}$  &  $\dfrac{\zeta _1}{48 \uppi ^2 m_{\Delta }^2}$  \\
                        \hline \hline
                \end{tabular}
\end{table*}

\begin{table*}
        \caption{Effective operators and Wilson coefficients in `Warsaw' basis for complex triplet scalar ($Y=1$) model.}
\label{tab:CTSW}
%       \label{tab:CTS}
             \centering
    %       }%{\rule{4cm}{3cm}}
%       \subfloat[``Warsaw'' basis]{
                \begin{tabular}{ll}
                        \hline \hline
                        $Q_\mathrm{H}$  &  $-\dfrac{\zeta _1 \mu ^2}{m_{\Delta }^4}-\dfrac{\zeta _2 \mu \
                                ^2}{4 m_{\Delta }^4}-\dfrac{\zeta _1^3}{4 \uppi ^2 m_{\Delta }^2}-\dfrac{\
                                \zeta _2^2 \zeta _1}{32 \uppi ^2 m_{\Delta }^2}$  \\[12pt]
                            $Q_{\mathrm{H}\square }$  &  $\dfrac{\zeta _2^2}{192 \uppi ^2 m_{\Delta \
                                }^2}+\dfrac{\mu ^2}{2 m_{\Delta }^4}$  \\[12pt]
                         $Q_{\text{HD}}$  &  $\dfrac{\zeta _1^2}{4 \uppi ^2 m_{\Delta \
                                }^2}+\dfrac{\zeta _2^2}{96 \uppi ^2 m_{\Delta }^2}-\dfrac{2 \mu \
                                ^2}{m_{\Delta }^4}$  \\[12pt]
                         $Q_{\text{HW}}$  &  $\dfrac{\zeta _1 g_\mathrm{W}^2}{48 \uppi ^2 m_{\Delta \
                                }^2}$  \\[12pt]
                        $Q_{\text{HWB}}$  &  $-\dfrac{\zeta _2 g_\mathrm{W} g_\mathrm{Y}}{48 \uppi ^2 \
                                m_{\Delta }^2}$  \\[12pt]
                        $Q_{\text{ll}}$  &  $\dfrac{\text{y$^2_\Delta $}}{4 m_{\Delta }^2}$  \
                        \\[12pt]
                        $Q_\mathrm{W}$  &  $\dfrac{g_\mathrm{W}^3}{1440 \uppi ^2 m_{\Delta }^2}$  \\
                        \hline \hline
                \end{tabular}%\\
\end{table*}

\begin{table*}
        \caption{Mass dimension-5 effective operators and Wilson coefficients for complex triplet scalar ($Y=1$)  model.}
\label{tab:CTS5}
%       \label{tab:CTS}
            \centering
   %       }\\
%       \subfloat[``Dimension-5" basis.]{
                \begin{tabular}{ll}
                        \hline \hline
                        Dimension-5 operator & Wilson coefficient \\
                        \hline
                        llHH & $\dfrac{y_{\Delta}^{2}}{m_{\Delta}}$ \\
                        \hline \hline
                \end{tabular}
%       }
\end{table*}

%Here $\phi$ is the real singlet scalar. Once this field is integrated out, few effective operators will emerge. To obtain those effective dimension-6 operators and their respective  Wilson Coefficients using \mmaInlineCell{Code}{\mmaDef{CoDEx}}, we need to perform the following steps:

\begin{enumerate}
        \item First, load the package:
\begin{mmaCell}[index=1]{Input}
Needs["CoDEx"]
\end{mmaCell}
        \item We have to define the field $\Delta$ as:
\begin{mmaCell}[label={}]{Code}
fields =
{ 
{\mmaPat{fieldName}, \mmaPat{components}, \mmaPat{colorDim}, \mmaPat{isoDim},
\mmaPat{hyperCharge}, \mmaPat{spin}, \mmaPat{mass}}
};
\end{mmaCell}

        We follow the convention in this line.
\begin{mmaCell}[index=2]{Input}
fieldewcts=
\{
\{hf,3,1,3,1,0,\mmaSub{m}{\mmaUnd{\(\pmb{\Delta}\)}}\}
\};
\end{mmaCell}
\begin{mmaCell}{Input}
hfvecst2ss=\mmaDef{defineHeavyFields}[\mmaDef{fieldewcts]};
\end{mmaCell}

\begin{mmaCell}{Input}
\mmaUnd{\(\pmb{\delta}\)}=hfvecst2ss[[1,1]]
\end{mmaCell}
\begin{mmaCell}{Output}
\{hf[1,1]+i ihf[1,1],hf[1,2]+i ihf[1,2],hf[1,3]+i ihf[1,3]\}
\end{mmaCell}
        
\item 
        Now, we will build the Lagrangian after defining the heavy field. We need to provide only those terms that contain the heavy fields.  The kinetic terms (covariant derivative and mass terms) of the heavy field will not play any role in this construction, and thus can be ignored. The Lagrangian is written in the following way:
\begin{mmaCell}[morefunctionlocal={i}]{Input}
\mmaUnd{\(\pmb{\Delta}\)}=\mmaUnderOver{\(\pmb{\sum}\)}{i}{3}\mmaUnd{\(\pmb{\delta}\)}[[i]] PauliMatrix[i];
\end{mmaCell}

\begin{mmaCell}[moredefined={tau}]{Input}
\mmaUnd{\(\pmb{\Delta}\)c} =\mmaDef{i} tau[2].\mmaUnd{\(\pmb{\Delta}\)};
\end{mmaCell}
\begin{mmaCell}[moredefined={dag, H, tau, hermitianConjugate}]{Input}
V=\mmaUnd{\(\pmb{\zeta}\)1} dag[H].H Tr[dag[\mmaUnd{\(\pmb{\Delta}\)}].\mmaUnd{\(\pmb{\Delta}\)}]
    +\mmaUnd{\(\pmb{\zeta}\)2} \mmaUnderOver{\(\pmb{\sum}\)}{i}{3}dag[H].tau[i].H Tr[dag[\mmaUnd{\(\pmb{\Delta}\)}].tau[i].\mmaUnd{\(\pmb{\Delta}\)}]
    +\mmaUnd{\(\pmb{\mu}\)} H.(\mmaDef{i} tau[2].dag[\mmaUnd{\(\pmb{\Delta}\)}]).H 
    +\mmaUnd{\(\pmb{\mu}\)} dag[H].(-\mmaDef{i} \mmaUnd{\(\pmb{\Delta}\)}.tau[2]).hermitianConjugate[H];
\end{mmaCell}

\begin{mmaCell}[moredefined={hermitianConjugate, lepb, gamma, lep, dag}]{Input}
lyukawa=Expand[\mmaUnd{y\(\pmb{\Sigma}\)} \mmaUnderOver{\(\pmb{\sum}\)}{i}{2}\mmaUnderOver{\(\pmb{\sum}\)}{j}{2}hermitianConjugate[lepb[1][[i]]].
        gamma[0].chargeC.(\mmaUnd{\(\pmb{\Delta}\)c}[[i,j]] lep[1][[j]])
        +\mmaUnd{y\(\pmb{\Sigma}\)} \mmaUnderOver{\(\pmb{\sum}\)}{i}{2}\mmaUnderOver{\(\pmb{\sum}\)}{j}{2}lepb[1][[i]].gamma[0].
        (dag[\mmaUnd{\(\pmb{\Delta}\)c}][[i,j]] dag[chargeC].lep[1][[j]])];
\end{mmaCell}

\begin{mmaCell}{Input}
Lpotent2ss=Expand[lyukawa-V];
\end{mmaCell}

\begin{mmaCell}[moredefined={initializeLoop}]{Input}
initializeLoop["t2ss",fieldt2ss]
\end{mmaCell}

\begin{mmaCell}{Output}
Check the documentation page CoDExParafernalia for details.             
\end{mmaCell}
{\small
\begin{mmaCell}[label={\mmaShd{>>}}]{Output}
Isospin Symmetry Generators for the field `hf' are
isot2ss[1,a] = tauadj[a]
\end{mmaCell}
\begin{mmaCell}[label={\mmaShd{>>}}]{Output}
colour Symmetry Generators for the field `hf' are colt2ss[1,a] = 0
\end{mmaCell}
}
%       \item Next, we need to construct the symmetry generators:
(See the documentation of \mmaInlineCell{Input}{initializeLoop} for details.)
%       \item The last step is the computation of effective operators and associated WCs as:
        
\begin{mmaCell}[moredefined={codexOutput, model, formPick}]{Input}
wcT2SSwar=codexOutput[Lpotent2ss,fieldt2ss,model\(\pmb{\to}\)"t2ss"];
formPick["Warsaw","Detailed2",wcT2SSwar,FontSize\(\pmb{\to}\)Medium,
FontFamily\(\pmb{\to}\)"Times New Roman",Frame\(\pmb{\to}\)All]
\end{mmaCell}
        
        \item   The operators can be generated in both  `\mmaInlineCell{Code}{SILH}' and `\mmaInlineCell{Code}{Warsaw}' bases along with their respective  Wilson coefficients. This output can be exported into a \LaTeX~ format as well; see Table \ref{tab:CTSW}.
        
\begin{mmaCell}[moredefined={codexOutput, model, operBasis, formPick}]{Input}
wcT2SSsilh=codexOutput[Lpotent2ss,fieldt2ss,model\(\pmb{\to}\)"t2ss",
operBasis\(\pmb{\to}\)"SILH"];
formPick["SILH","Detailed2",wcT2SSsilh,FontSize\(\pmb{\to}\)Medium,
FontFamily\(\pmb{\to}\)"Times New Roman",Frame\(\pmb{\to}\)All]
        \end{mmaCell}
        
        The output of this is given in Table \ref{tab:CTSS}.
        
\begin{mmaCell}[moredefined={codexOutput, model, operBasis, formPick}]{Input}
wcT2SSdim5=codexOutput[Lpotent2ss,fieldt2ss,model\(\pmb{\to}\)"t2ss",
            operBasis\(\pmb{\to}\)"Dim5"];
formPick["Dim5","Detailed2",wcT2SSdim5,FontSize\(\pmb{\to}\)Medium,
FontFamily\(\pmb{\to}\)"Times New Roman",Frame\(\pmb{\to}\)All]
\end{mmaCell}

        The output of this is given in Table \ref{tab:CTS5}.
       
        \item  
         The RG evolution of these WCs can be performed only in the  `\mmaInlineCell{Code}{Warsaw}' basis, as this is the complete one using \mmaInlineCell{Code}{RGFlow}:

\begin{mmaCell}[moredefined={RGFlow, smlst1}]{Input}
RGFlow[wcT2SSwar,m,\mmaUnd{\(\pmb{\mu}\)}]
\end{mmaCell}

Let us consider that the  \mmaInlineCell{Input}{\mmaDef{CoDEx}} output, which is the WCs at the high scale, is generated, and saved as:
\begin{mmaCell}[moredefined={gW, gY}]{Input}
wcT2SSwar=\{\{"qH",-\mmaFrac{\mmaSup{\mmaUnd{\(\pmb{\zeta}\)1}}{3}}{4 \mmaSup{\mmaSub{m}{\mmaUnd{\(\pmb{\Delta}\)}}}{2} \mmaSup{\mmaDef{\(\pmb{\pi}\)}}{2}}-\mmaFrac{\mmaUnd{\(\pmb{\zeta}\)1} \mmaSup{\mmaUnd{\(\pmb{\zeta}\)2}}{2}}{32 \mmaSup{\mmaSub{m}{\mmaUnd{\(\pmb{\Delta}\)}}}{2} \mmaSup{\mmaDef{\(\pmb{\pi}\)}}{2}}-\mmaFrac{\mmaUnd{\(\pmb{\zeta}\)1} \mmaSup{\mmaUnd{\(\pmb{\mu}\)}}{2}}{\mmaSup{\mmaSub{m}{\mmaUnd{\(\pmb{\Delta}\)}}}{4}}-\mmaFrac{\mmaUnd{\(\pmb{\zeta}\)2} \mmaSup{\mmaUnd{\(\pmb{\mu}\)}}{2}}{4 \mmaSup{\mmaSub{m}{\mmaUnd{\(\pmb{\Delta}\)}}}{4}}\},
            \{"qHbox",\mmaFrac{\mmaSup{\mmaUnd{\(\pmb{\zeta}\)2}}{2}}{192 \mmaSup{\mmaSub{m}{\mmaUnd{\(\pmb{\Delta}\)}}}{2} \mmaSup{\mmaDef{\(\pmb{\pi}\)}}{2}}+\mmaFrac{\mmaSup{\mmaUnd{\(\pmb{\mu}\)}}{2}}{2 \mmaSup{\mmaSub{m}{\mmaUnd{\(\pmb{\Delta}\)}}}{4}}\},
\{"qHD",\mmaFrac{\mmaSup{\mmaUnd{\(\pmb{\zeta}\)1}}{2}}{4 \mmaSup{\mmaSub{m}{\mmaUnd{\(\pmb{\Delta}\)}}}{2} \mmaSup{\mmaDef{\(\pmb{\pi}\)}}{2}}+\mmaFrac{\mmaSup{\mmaUnd{\(\pmb{\zeta}\)2}}{2}}{96 \mmaSup{\mmaSub{m}{\mmaUnd{\(\pmb{\Delta}\)}}}{2} \mmaSup{\mmaDef{\(\pmb{\pi}\)}}{2}}-\mmaFrac{2 \mmaSup{\mmaUnd{\(\pmb{\mu}\)}}{2}}{\mmaSup{\mmaSub{m}{\mmaUnd{\(\pmb{\Delta}\)}}}{4}}\},\{"qHW",\mmaFrac{\mmaSup{gW}{2} \mmaUnd{\(\pmb{\zeta}\)1}}{48 \mmaSup{\mmaSub{m}{\mmaUnd{\(\pmb{\Delta}\)}}}{2} \mmaSup{\mmaDef{\(\pmb{\pi}\)}}{2}}\},
\{"qHWB",-\mmaFrac{gW gY \mmaUnd{\(\pmb{\zeta}\)2}}{48 \mmaSup{\mmaSub{m}{\mmaUnd{\(\pmb{\Delta}\)}}}{2} \mmaSup{\mmaDef{\(\pmb{\pi}\)}}{2}}\},\{"qll[1,1,1,1]",\mmaFrac{\mmaSup{\mmaUnd{y\(\pmb{\Sigma}\)}}{2}}{4 \mmaSup{\mmaSub{m}{\mmaUnd{\(\pmb{\Delta}\)}}}{2}}\},\{"qW",\mmaFrac{\mmaSup{gW}{3}}{1440 \mmaSup{\mmaSub{m}{\mmaUnd{\(\pmb{\Delta}\)}}}{2} \mmaSup{\mmaDef{\(\pmb{\pi}\)}}{2}}\}\}
\end{mmaCell}

        Once we declare  the matching scale (high scale) as the mass of the heavy particle (`\mmaInlineCell{Code}{\mmaUnd{m}}'), we need to recall the function \mmaInlineCell{Code}{RGFlow} to generate the WCs at low scale, as:
\begin{mmaCell}{Input}
floRes1 = \mmaDef{RGFlow}[\mmaUnd{wcT2SSwar},m,\mmaUnd{\(\pmb{\mu}\)}]
\end{mmaCell}
{\tiny
        \begin{mmaCell}{Output}
                \{\{qW,\mmaFrac{\mmaSup{gW}{3}}{1440 \mmaSup{\mmaSub{m}{\(\Delta\)}}{2} \mmaSup{\(\pi\)}{2}}+\mmaFrac{29 \mmaSup{gW}{5} Log[\mmaFrac{\(\mu\)}{m}]}{46080 \mmaSup{\mmaSub{m}{\(\Delta\)}}{2} \mmaSup{\(\pi\)}{4}}\},\{qH,-\mmaFrac{\mmaSup{\(\zeta\)1}{3}}{4 \mmaSup{\mmaSub{m}{\(\Delta\)}}{2} \mmaSup{\(\pi\)}{2}}-\mmaFrac{\(\zeta\)1 \mmaSup{\(\zeta\)2}{2}}{32 \mmaSup{\mmaSub{m}{\(\Delta\)}}{2} \mmaSup{\(\pi\)}{2}}-\mmaFrac{\(\zeta\)1 \mmaSup{\(\mu\)}{2}}{\mmaSup{\mmaSub{m}{\(\Delta\)}}{4}}-\mmaFrac{\(\zeta\)2 \mmaSup{\(\mu\)}{2}}{4 \mmaSup{\mmaSub{m}{\(\Delta\)}}{4}}-\mmaFrac{3 \mmaSup{gW}{6} \(\zeta\)1 Log[\mmaFrac{\(\mu\)}{m}]}{256 \mmaSup{\mmaSub{m}{\(\Delta\)}}{2} \mmaSup{\(\pi\)}{4}}
                -\mmaFrac{\mmaSup{gW}{4} \mmaSup{gY}{2} \(\zeta\)1 Log[\mmaFrac{\(\mu\)}{m}]}{256 \mmaSup{\mmaSub{m}{\(\Delta\)}}{2} \mmaSup{\(\pi\)}{4}}-\mmaFrac{3 \mmaSup{gW}{2} \mmaSup{\(\zeta\)1}{2} Log[\mmaFrac{\(\mu\)}{m}]}{32 \mmaSup{\mmaSub{m}{\(\Delta\)}}{2} \mmaSup{\(\pi\)}{4}}-\mmaFrac{3 \mmaSup{gW}{4} \mmaSup{\(\zeta\)1}{2} Log[\mmaFrac{\(\mu\)}{m}]}{256 \mmaSup{\mmaSub{m}{\(\Delta\)}}{2} \mmaSup{\(\pi\)}{4}}+\mmaFrac{3 \mmaSup{gY}{2} \mmaSup{\(\zeta\)1}{2} Log[\mmaFrac{\(\mu\)}{m}]}{32 \mmaSup{\mmaSub{m}{\(\Delta\)}}{2} \mmaSup{\(\pi\)}{4}}
                -\mmaFrac{3 \mmaSup{gW}{2} \mmaSup{gY}{2} \mmaSup{\(\zeta\)1}{2} Log[\mmaFrac{\(\mu\)}{m}]}{128 \mmaSup{\mmaSub{m}{\(\Delta\)}}{2} \mmaSup{\(\pi\)}{4}}-\mmaFrac{3 \mmaSup{gY}{4} \mmaSup{\(\zeta\)1}{2} Log[\mmaFrac{\(\mu\)}{m}]}{256 \mmaSup{\mmaSub{m}{\(\Delta\)}}{2} \mmaSup{\(\pi\)}{4}}+\mmaFrac{27 \mmaSup{gW}{2} \mmaSup{\(\zeta\)1}{3} Log[\mmaFrac{\(\mu\)}{m}]}{128 \mmaSup{\mmaSub{m}{\(\Delta\)}}{2} \mmaSup{\(\pi\)}{4}}+\mmaFrac{9 \mmaSup{gY}{2} \mmaSup{\(\zeta\)1}{3} Log[\mmaFrac{\(\mu\)}{m}]}{128 \mmaSup{\mmaSub{m}{\(\Delta\)}}{2} \mmaSup{\(\pi\)}{4}}
                +\mmaFrac{\mmaSup{gW}{4} \mmaSup{gY}{2} \(\zeta\)2 Log[\mmaFrac{\(\mu\)}{m}]}{256 \mmaSup{\mmaSub{m}{\(\Delta\)}}{2} \mmaSup{\(\pi\)}{4}}+\mmaFrac{\mmaSup{gW}{2} \mmaSup{gY}{4} \(\zeta\)2 Log[\mmaFrac{\(\mu\)}{m}]}{256 \mmaSup{\mmaSub{m}{\(\Delta\)}}{2} \mmaSup{\(\pi\)}{4}}-\mmaFrac{\mmaSup{gW}{2} \mmaSup{\(\zeta\)2}{2} Log[\mmaFrac{\(\mu\)}{m}]}{256 \mmaSup{\mmaSub{m}{\(\Delta\)}}{2} \mmaSup{\(\pi\)}{4}}-\mmaFrac{\mmaSup{gW}{4} \mmaSup{\(\zeta\)2}{2} Log[\mmaFrac{\(\mu\)}{m}]}{2048 \mmaSup{\mmaSub{m}{\(\Delta\)}}{2} \mmaSup{\(\pi\)}{4}}+\mmaFrac{\mmaSup{gY}{2} \mmaSup{\(\zeta\)2}{2} Log[\mmaFrac{\(\mu\)}{m}]}{256 \mmaSup{\mmaSub{m}{\(\Delta\)}}{2} \mmaSup{\(\pi\)}{4}}
                -\mmaFrac{\mmaSup{gW}{2} \mmaSup{gY}{2} \mmaSup{\(\zeta\)2}{2} Log[\mmaFrac{\(\mu\)}{m}]}{1024 \mmaSup{\mmaSub{m}{\(\Delta\)}}{2} \mmaSup{\(\pi\)}{4}}-\mmaFrac{\mmaSup{gY}{4} \mmaSup{\(\zeta\)2}{2} Log[\mmaFrac{\(\mu\)}{m}]}{2048 \mmaSup{\mmaSub{m}{\(\Delta\)}}{2} \mmaSup{\(\pi\)}{4}}+\mmaFrac{27 \mmaSup{gW}{2} \(\zeta\)1 \mmaSup{\(\zeta\)2}{2} Log[\mmaFrac{\(\mu\)}{m}]}{1024 \mmaSup{\mmaSub{m}{\(\Delta\)}}{2} \mmaSup{\(\pi\)}{4}}
                +\mmaFrac{9 \mmaSup{gY}{2} \(\zeta\)1 \mmaSup{\(\zeta\)2}{2} Log[\mmaFrac{\(\mu\)}{m}]}{1024 \mmaSup{\mmaSub{m}{\(\Delta\)}}{2} \mmaSup{\(\pi\)}{4}}+\mmaFrac{3 \mmaSup{gW}{4} \(\zeta\)1 \(\lambda\) Log[\mmaFrac{\(\mu\)}{m}]}{64 \mmaSup{\mmaSub{m}{\(\Delta\)}}{2} \mmaSup{\(\pi\)}{4}}-\mmaFrac{\mmaSup{gW}{2} \mmaSup{gY}{2} \(\zeta\)2 \(\lambda\) Log[\mmaFrac{\(\mu\)}{m}]}{64 \mmaSup{\mmaSub{m}{\(\Delta\)}}{2} \mmaSup{\(\pi\)}{4}}+\mmaFrac{5 \mmaSup{gW}{2} \mmaSup{\(\zeta\)2}{2} \(\lambda\) Log[\mmaFrac{\(\mu\)}{m}]}{1152 \mmaSup{\mmaSub{m}{\(\Delta\)}}{2} \mmaSup{\(\pi\)}{4}}
                +\mmaFrac{3 \mmaSup{gW}{2} \mmaSup{\(\mu\)}{2} Log[\mmaFrac{\(\mu\)}{m}]}{4 \mmaSup{\mmaSub{m}{\(\Delta\)}}{4} \mmaSup{\(\pi\)}{2}}+\mmaFrac{3 \mmaSup{gW}{4} \mmaSup{\(\mu\)}{2} Log[\mmaFrac{\(\mu\)}{m}]}{32 \mmaSup{\mmaSub{m}{\(\Delta\)}}{4} \mmaSup{\(\pi\)}{2}}-\mmaFrac{3 \mmaSup{gY}{2} \mmaSup{\(\mu\)}{2} Log[\mmaFrac{\(\mu\)}{m}]}{4 \mmaSup{\mmaSub{m}{\(\Delta\)}}{4} \mmaSup{\(\pi\)}{2}}
                +\mmaFrac{3 \mmaSup{gW}{2} \mmaSup{gY}{2} \mmaSup{\(\mu\)}{2} Log[\mmaFrac{\(\mu\)}{m}]}{16 \mmaSup{\mmaSub{m}{\(\Delta\)}}{4} \mmaSup{\(\pi\)}{2}}+\mmaFrac{3 \mmaSup{gY}{4} \mmaSup{\(\mu\)}{2} Log[\mmaFrac{\(\mu\)}{m}]}{32 \mmaSup{\mmaSub{m}{\(\Delta\)}}{4} \mmaSup{\(\pi\)}{2}}+\mmaFrac{27 \mmaSup{gW}{2} \(\zeta\)1 \mmaSup{\(\mu\)}{2} Log[\mmaFrac{\(\mu\)}{m}]}{32 \mmaSup{\mmaSub{m}{\(\Delta\)}}{4} \mmaSup{\(\pi\)}{2}}
                +\mmaFrac{9 \mmaSup{gY}{2} \(\zeta\)1 \mmaSup{\(\mu\)}{2} Log[\mmaFrac{\(\mu\)}{m}]}{32 \mmaSup{\mmaSub{m}{\(\Delta\)}}{4} \mmaSup{\(\pi\)}{2}}+\mmaFrac{27 \mmaSup{gW}{2} \(\zeta\)2 \mmaSup{\(\mu\)}{2} Log[\mmaFrac{\(\mu\)}{m}]}{128 \mmaSup{\mmaSub{m}{\(\Delta\)}}{4} \mmaSup{\(\pi\)}{2}}+\mmaFrac{9 \mmaSup{gY}{2} \(\zeta\)2 \mmaSup{\(\mu\)}{2} Log[\mmaFrac{\(\mu\)}{m}]}{128 \mmaSup{\mmaSub{m}{\(\Delta\)}}{4} \mmaSup{\(\pi\)}{2}}+\mmaFrac{5 \mmaSup{gW}{2} \(\lambda\) \mmaSup{\(\mu\)}{2} Log[\mmaFrac{\(\mu\)}{m}]}{12 \mmaSup{\mmaSub{m}{\(\Delta\)}}{4} \mmaSup{\(\pi\)}{2}}\},
                \{qHbox,\mmaFrac{\mmaSup{\(\zeta\)2}{2}}{192 \mmaSup{\mmaSub{m}{\(\Delta\)}}{2} \mmaSup{\(\pi\)}{2}}+\mmaFrac{\mmaSup{\(\mu\)}{2}}{2 \mmaSup{\mmaSub{m}{\(\Delta\)}}{4}}+\mmaFrac{5 \mmaSup{gY}{2} \mmaSup{\(\zeta\)1}{2} Log[\mmaFrac{\(\mu\)}{m}]}{192 \mmaSup{\mmaSub{m}{\(\Delta\)}}{2} \mmaSup{\(\pi\)}{4}}-\mmaFrac{\mmaSup{gW}{2} \mmaSup{\(\zeta\)2}{2} Log[\mmaFrac{\(\mu\)}{m}]}{768 \mmaSup{\mmaSub{m}{\(\Delta\)}}{2} \mmaSup{\(\pi\)}{4}}
                +\mmaFrac{\mmaSup{gY}{2} \mmaSup{\(\zeta\)2}{2} Log[\mmaFrac{\(\mu\)}{m}]}{1536 \mmaSup{\mmaSub{m}{\(\Delta\)}}{2} \mmaSup{\(\pi\)}{4}}-\mmaFrac{\mmaSup{gW}{2} \mmaSup{\(\mu\)}{2} Log[\mmaFrac{\(\mu\)}{m}]}{8 \mmaSup{\mmaSub{m}{\(\Delta\)}}{4} \mmaSup{\(\pi\)}{2}}-\mmaFrac{\mmaSup{gY}{2} \mmaSup{\(\mu\)}{2} Log[\mmaFrac{\(\mu\)}{m}]}{4 \mmaSup{\mmaSub{m}{\(\Delta\)}}{4} \mmaSup{\(\pi\)}{2}}\},
                \{qHD,\mmaFrac{\mmaSup{\(\zeta\)1}{2}}{4 \mmaSup{\mmaSub{m}{\(\Delta\)}}{2} \mmaSup{\(\pi\)}{2}}+\mmaFrac{\mmaSup{\(\zeta\)2}{2}}{96 \mmaSup{\mmaSub{m}{\(\Delta\)}}{2} \mmaSup{\(\pi\)}{2}}-\mmaFrac{2 \mmaSup{\(\mu\)}{2}}{\mmaSup{\mmaSub{m}{\(\Delta\)}}{4}}+\mmaFrac{9 \mmaSup{gW}{2} \mmaSup{\(\zeta\)1}{2} Log[\mmaFrac{\(\mu\)}{m}]}{128 \mmaSup{\mmaSub{m}{\(\Delta\)}}{2} \mmaSup{\(\pi\)}{4}}-\mmaFrac{5 \mmaSup{gY}{2} \mmaSup{\(\zeta\)1}{2} Log[\mmaFrac{\(\mu\)}{m}]}{384 \mmaSup{\mmaSub{m}{\(\Delta\)}}{2} \mmaSup{\(\pi\)}{4}}
                +\mmaFrac{3 \mmaSup{gW}{2} \mmaSup{\(\zeta\)2}{2} Log[\mmaFrac{\(\mu\)}{m}]}{1024 \mmaSup{\mmaSub{m}{\(\Delta\)}}{2} \mmaSup{\(\pi\)}{4}}+\mmaFrac{5 \mmaSup{gY}{2} \mmaSup{\(\zeta\)2}{2} Log[\mmaFrac{\(\mu\)}{m}]}{3072 \mmaSup{\mmaSub{m}{\(\Delta\)}}{2} \mmaSup{\(\pi\)}{4}}-\mmaFrac{9 \mmaSup{gW}{2} \mmaSup{\(\mu\)}{2} Log[\mmaFrac{\(\mu\)}{m}]}{16 \mmaSup{\mmaSub{m}{\(\Delta\)}}{4} \mmaSup{\(\pi\)}{2}}+\mmaFrac{5 \mmaSup{gY}{2} \mmaSup{\(\mu\)}{2} Log[\mmaFrac{\(\mu\)}{m}]}{16 \mmaSup{\mmaSub{m}{\(\Delta\)}}{4} \mmaSup{\(\pi\)}{2}}\},
                \{qHW,\mmaFrac{\mmaSup{gW}{2} \(\zeta\)1}{48 \mmaSup{\mmaSub{m}{\(\Delta\)}}{2} \mmaSup{\(\pi\)}{2}}-\mmaFrac{\mmaSup{gW}{6} Log[\mmaFrac{\(\mu\)}{m}]}{1536 \mmaSup{\mmaSub{m}{\(\Delta\)}}{2} \mmaSup{\(\pi\)}{4}}-\mmaFrac{53 \mmaSup{gW}{4} \(\zeta\)1 Log[\mmaFrac{\(\mu\)}{m}]}{4608 \mmaSup{\mmaSub{m}{\(\Delta\)}}{2} \mmaSup{\(\pi\)}{4}}
                -\mmaFrac{\mmaSup{gW}{2} \mmaSup{gY}{2} \(\zeta\)1 Log[\mmaFrac{\(\mu\)}{m}]}{512 \mmaSup{\mmaSub{m}{\(\Delta\)}}{2} \mmaSup{\(\pi\)}{4}}-\mmaFrac{\mmaSup{gW}{2} \mmaSup{gY}{2} \(\zeta\)2 Log[\mmaFrac{\(\mu\)}{m}]}{768 \mmaSup{\mmaSub{m}{\(\Delta\)}}{2} \mmaSup{\(\pi\)}{4}}\},\{qHB,-\mmaFrac{\mmaSup{gW}{2} \mmaSup{gY}{2} \(\zeta\)2 Log[\mmaFrac{\(\mu\)}{m}]}{256 \mmaSup{\mmaSub{m}{\(\Delta\)}}{2} \mmaSup{\(\pi\)}{4}}\},
                \{qHWB,-\mmaFrac{gW gY \(\zeta\)2}{48 \mmaSup{\mmaSub{m}{\(\Delta\)}}{2} \mmaSup{\(\pi\)}{2}}+\mmaFrac{\mmaSup{gW}{5} gY Log[\mmaFrac{\(\mu\)}{m}]}{7680 \mmaSup{\mmaSub{m}{\(\Delta\)}}{2} \mmaSup{\(\pi\)}{4}}+\mmaFrac{\mmaSup{gW}{3} gY \(\zeta\)1 Log[\mmaFrac{\(\mu\)}{m}]}{384 \mmaSup{\mmaSub{m}{\(\Delta\)}}{2} \mmaSup{\(\pi\)}{4}}-\mmaFrac{\mmaSup{gW}{3} gY \(\zeta\)2 Log[\mmaFrac{\(\mu\)}{m}]}{576 \mmaSup{\mmaSub{m}{\(\Delta\)}}{2} \mmaSup{\(\pi\)}{4}}-\mmaFrac{19 gW \mmaSup{gY}{3} \(\zeta\)2 Log[\mmaFrac{\(\mu\)}{m}]}{2304 \mmaSup{\mmaSub{m}{\(\Delta\)}}{2} \mmaSup{\(\pi\)}{4}}\},
                \{qeH[1,1],\mmaFrac{3 \mmaSup{gW}{4} \(\zeta\)1 Log[\mmaFrac{\(\mu\)}{m}] \mmaSup{Yu}{\(\dagger\)}[e]}{256 \mmaSup{\mmaSub{m}{\(\Delta\)}}{2} \mmaSup{\(\pi\)}{4}}-\mmaFrac{3 \mmaSup{gW}{2} \mmaSup{\(\zeta\)1}{2} Log[\mmaFrac{\(\mu\)}{m}] \mmaSup{Yu}{\(\dagger\)}[e]}{128 \mmaSup{\mmaSub{m}{\(\Delta\)}}{2} \mmaSup{\(\pi\)}{4}}
                +\mmaFrac{3 \mmaSup{gY}{2} \mmaSup{\(\zeta\)1}{2} Log[\mmaFrac{\(\mu\)}{m}] \mmaSup{Yu}{\(\dagger\)}[e]}{128 \mmaSup{\mmaSub{m}{\(\Delta\)}}{2} \mmaSup{\(\pi\)}{4}}+\mmaFrac{\mmaSup{gW}{2} \mmaSup{gY}{2} \(\zeta\)2 Log[\mmaFrac{\(\mu\)}{m}] \mmaSup{Yu}{\(\dagger\)}[e]}{256 \mmaSup{\mmaSub{m}{\(\Delta\)}}{2} \mmaSup{\(\pi\)}{4}}
                +\mmaFrac{\mmaSup{gW}{2} \mmaSup{\(\zeta\)2}{2} Log[\mmaFrac{\(\mu\)}{m}] \mmaSup{Yu}{\(\dagger\)}[e]}{9216 \mmaSup{\mmaSub{m}{\(\Delta\)}}{2} \mmaSup{\(\pi\)}{4}}+\mmaFrac{\mmaSup{gY}{2} \mmaSup{\(\zeta\)2}{2} Log[\mmaFrac{\(\mu\)}{m}] \mmaSup{Yu}{\(\dagger\)}[e]}{1024 \mmaSup{\mmaSub{m}{\(\Delta\)}}{2} \mmaSup{\(\pi\)}{4}}+\mmaFrac{7 \mmaSup{gW}{2} \mmaSup{\(\mu\)}{2} Log[\mmaFrac{\(\mu\)}{m}] \mmaSup{Yu}{\(\dagger\)}[e]}{24 \mmaSup{\mmaSub{m}{\(\Delta\)}}{4} \mmaSup{\(\pi\)}{2}}-\mmaFrac{3 \mmaSup{gY}{2} \mmaSup{\(\mu\)}{2} Log[\mmaFrac{\(\mu\)}{m}] \mmaSup{Yu}{\(\dagger\)}[e]}{16 \mmaSup{\mmaSub{m}{\(\Delta\)}}{4} \mmaSup{\(\pi\)}{2}}\},
                \{quH[1,1],\mmaFrac{3 \mmaSup{gW}{4} \(\zeta\)1 Log[\mmaFrac{\(\mu\)}{m}] \mmaSup{Yu}{\(\dagger\)}[u]}{512 \mmaSup{\mmaSub{m}{\(\Delta\)}}{2} \mmaSup{\(\pi\)}{4}}-\mmaFrac{3 \mmaSup{gW}{2} \mmaSup{\(\zeta\)1}{2} Log[\mmaFrac{\(\mu\)}{m}] \mmaSup{Yu}{\(\dagger\)}[u]}{128 \mmaSup{\mmaSub{m}{\(\Delta\)}}{2} \mmaSup{\(\pi\)}{4}}
                +\mmaFrac{3 \mmaSup{gY}{2} \mmaSup{\(\zeta\)1}{2} Log[\mmaFrac{\(\mu\)}{m}] \mmaSup{Yu}{\(\dagger\)}[u]}{128 \mmaSup{\mmaSub{m}{\(\Delta\)}}{2} \mmaSup{\(\pi\)}{4}}+\mmaFrac{\mmaSup{gW}{2} \mmaSup{gY}{2} \(\zeta\)2 Log[\mmaFrac{\(\mu\)}{m}] \mmaSup{Yu}{\(\dagger\)}[u]}{768 \mmaSup{\mmaSub{m}{\(\Delta\)}}{2} \mmaSup{\(\pi\)}{4}}+\mmaFrac{\mmaSup{gW}{2} \mmaSup{\(\zeta\)2}{2} Log[\mmaFrac{\(\mu\)}{m}] \mmaSup{Yu}{\(\dagger\)}[u]}{9216 \mmaSup{\mmaSub{m}{\(\Delta\)}}{2} \mmaSup{\(\pi\)}{4}}
                +\mmaFrac{\mmaSup{gY}{2} \mmaSup{\(\zeta\)2}{2} Log[\mmaFrac{\(\mu\)}{m}] \mmaSup{Yu}{\(\dagger\)}[u]}{1024 \mmaSup{\mmaSub{m}{\(\Delta\)}}{2} \mmaSup{\(\pi\)}{4}}+\mmaFrac{7 \mmaSup{gW}{2} \mmaSup{\(\mu\)}{2} Log[\mmaFrac{\(\mu\)}{m}] \mmaSup{Yu}{\(\dagger\)}[u]}{24 \mmaSup{\mmaSub{m}{\(\Delta\)}}{4} \mmaSup{\(\pi\)}{2}}-\mmaFrac{3 \mmaSup{gY}{2} \mmaSup{\(\mu\)}{2} Log[\mmaFrac{\(\mu\)}{m}] \mmaSup{Yu}{\(\dagger\)}[u]}{16 \mmaSup{\mmaSub{m}{\(\Delta\)}}{4} \mmaSup{\(\pi\)}{2}}\},
                \{qdH[1,1],\mmaFrac{3 \mmaSup{gW}{4} \(\zeta\)1 Log[\mmaFrac{\(\mu\)}{m}] \mmaSup{Yu}{\(\dagger\)}[d]}{512 \mmaSup{\mmaSub{m}{\(\Delta\)}}{2} \mmaSup{\(\pi\)}{4}}-\mmaFrac{3 \mmaSup{gW}{2} \mmaSup{\(\zeta\)1}{2} Log[\mmaFrac{\(\mu\)}{m}] \mmaSup{Yu}{\(\dagger\)}[d]}{128 \mmaSup{\mmaSub{m}{\(\Delta\)}}{2} \mmaSup{\(\pi\)}{4}}
                +\mmaFrac{3 \mmaSup{gY}{2} \mmaSup{\(\zeta\)1}{2} Log[\mmaFrac{\(\mu\)}{m}] \mmaSup{Yu}{\(\dagger\)}[d]}{128 \mmaSup{\mmaSub{m}{\(\Delta\)}}{2} \mmaSup{\(\pi\)}{4}}-\mmaFrac{\mmaSup{gW}{2} \mmaSup{gY}{2} \(\zeta\)2 Log[\mmaFrac{\(\mu\)}{m}] \mmaSup{Yu}{\(\dagger\)}[d]}{768 \mmaSup{\mmaSub{m}{\(\Delta\)}}{2} \mmaSup{\(\pi\)}{4}}+\mmaFrac{\mmaSup{gW}{2} \mmaSup{\(\zeta\)2}{2} Log[\mmaFrac{\(\mu\)}{m}] \mmaSup{Yu}{\(\dagger\)}[d]}{9216 \mmaSup{\mmaSub{m}{\(\Delta\)}}{2} \mmaSup{\(\pi\)}{4}}
                +\mmaFrac{\mmaSup{gY}{2} \mmaSup{\(\zeta\)2}{2} Log[\mmaFrac{\(\mu\)}{m}] \mmaSup{Yu}{\(\dagger\)}[d]}{1024 \mmaSup{\mmaSub{m}{\(\Delta\)}}{2} \mmaSup{\(\pi\)}{4}}+\mmaFrac{7 \mmaSup{gW}{2} \mmaSup{\(\mu\)}{2} Log[\mmaFrac{\(\mu\)}{m}] \mmaSup{Yu}{\(\dagger\)}[d]}{24 \mmaSup{\mmaSub{m}{\(\Delta\)}}{4} \mmaSup{\(\pi\)}{2}}-\mmaFrac{3 \mmaSup{gY}{2} \mmaSup{\(\mu\)}{2} Log[\mmaFrac{\(\mu\)}{m}] \mmaSup{Yu}{\(\dagger\)}[d]}{16 \mmaSup{\mmaSub{m}{\(\Delta\)}}{4} \mmaSup{\(\pi\)}{2}}\},
                \{qeW[1,1],-\mmaFrac{\mmaSup{gW}{3} \(\zeta\)1 Log[\mmaFrac{\(\mu\)}{m}] \mmaSup{Yu}{\(\dagger\)}[e]}{768 \mmaSup{\mmaSub{m}{\(\Delta\)}}{2} \mmaSup{\(\pi\)}{4}}-\mmaFrac{gW \mmaSup{gY}{2} \(\zeta\)2 Log[\mmaFrac{\(\mu\)}{m}] \mmaSup{Yu}{\(\dagger\)}[e]}{512 \mmaSup{\mmaSub{m}{\(\Delta\)}}{2} \mmaSup{\(\pi\)}{4}}\},
                \{qeB[1,1],\mmaFrac{\mmaSup{gW}{2} gY \(\zeta\)2 Log[\mmaFrac{\(\mu\)}{m}] \mmaSup{Yu}{\(\dagger\)}[e]}{512 \mmaSup{\mmaSub{m}{\(\Delta\)}}{2} \mmaSup{\(\pi\)}{4}}\},
                \{quW[1,1],-\mmaFrac{\mmaSup{gW}{3} \(\zeta\)1 Log[\mmaFrac{\(\mu\)}{m}] \mmaSup{Yu}{\(\dagger\)}[u]}{768 \mmaSup{\mmaSub{m}{\(\Delta\)}}{2} \mmaSup{\(\pi\)}{4}}-\mmaFrac{5 gW \mmaSup{gY}{2} \(\zeta\)2 Log[\mmaFrac{\(\mu\)}{m}] \mmaSup{Yu}{\(\dagger\)}[u]}{4608 \mmaSup{\mmaSub{m}{\(\Delta\)}}{2} \mmaSup{\(\pi\)}{4}}\},
                \{quB[1,1],-\mmaFrac{\mmaSup{gW}{2} gY \(\zeta\)2 Log[\mmaFrac{\(\mu\)}{m}] \mmaSup{Yu}{\(\dagger\)}[u]}{512 \mmaSup{\mmaSub{m}{\(\Delta\)}}{2} \mmaSup{\(\pi\)}{4}}\},
                \{qdW[1,1],-\mmaFrac{\mmaSup{gW}{3} \(\zeta\)1 Log[\mmaFrac{\(\mu\)}{m}] \mmaSup{Yu}{\(\dagger\)}[d]}{768 \mmaSup{\mmaSub{m}{\(\Delta\)}}{2} \mmaSup{\(\pi\)}{4}}-\mmaFrac{gW \mmaSup{gY}{2} \(\zeta\)2 Log[\mmaFrac{\(\mu\)}{m}] \mmaSup{Yu}{\(\dagger\)}[d]}{4608 \mmaSup{\mmaSub{m}{\(\Delta\)}}{2} \mmaSup{\(\pi\)}{4}}\},
                \{qdB[1,1],\mmaFrac{\mmaSup{gW}{2} gY \(\zeta\)2 Log[\mmaFrac{\(\mu\)}{m}] \mmaSup{Yu}{\(\dagger\)}[u]}{512 \mmaSup{\mmaSub{m}{\(\Delta\)}}{2} \mmaSup{\(\pi\)}{4}}\},
                \{q1Hl[1,1],-\mmaFrac{\mmaSup{gY}{2} \mmaSup{y\(\Sigma\)}{2} Log[\mmaFrac{\(\mu\)}{m}]}{96 \mmaSup{\mmaSub{m}{\(\Delta\)}}{2} \mmaSup{\(\pi\)}{2}}-\mmaFrac{\mmaSup{gY}{2} \mmaSup{\(\zeta\)1}{2} Log[\mmaFrac{\(\mu\)}{m}]}{384 \mmaSup{\mmaSub{m}{\(\Delta\)}}{2} \mmaSup{\(\pi\)}{4}}-\mmaFrac{\mmaSup{gY}{2} \mmaSup{\(\zeta\)2}{2} Log[\mmaFrac{\(\mu\)}{m}]}{6144 \mmaSup{\mmaSub{m}{\(\Delta\)}}{2} \mmaSup{\(\pi\)}{4}}+\mmaFrac{\mmaSup{gY}{2} \mmaSup{\(\mu\)}{2} Log[\mmaFrac{\(\mu\)}{m}]}{64 \mmaSup{\mmaSub{m}{\(\Delta\)}}{4} \mmaSup{\(\pi\)}{2}}\},
                \{q3Hl[1,1],\mmaFrac{\mmaSup{gW}{2} \mmaSup{y\(\Sigma\)}{2} Log[\mmaFrac{\(\mu\)}{m}]}{96 \mmaSup{\mmaSub{m}{\(\Delta\)}}{2} \mmaSup{\(\pi\)}{2}}+\mmaFrac{\mmaSup{gW}{2} \mmaSup{\(\zeta\)2}{2} Log[\mmaFrac{\(\mu\)}{m}]}{18432 \mmaSup{\mmaSub{m}{\(\Delta\)}}{2} \mmaSup{\(\pi\)}{4}}+\mmaFrac{\mmaSup{gW}{2} \mmaSup{\(\mu\)}{2} Log[\mmaFrac{\(\mu\)}{m}]}{192 \mmaSup{\mmaSub{m}{\(\Delta\)}}{4} \mmaSup{\(\pi\)}{2}}\},
                \{qHe[1,1],-\mmaFrac{\mmaSup{gY}{2} \mmaSup{\(\zeta\)1}{2} Log[\mmaFrac{\(\mu\)}{m}]}{192 \mmaSup{\mmaSub{m}{\(\Delta\)}}{2} \mmaSup{\(\pi\)}{4}}-\mmaFrac{\mmaSup{gY}{2} \mmaSup{\(\zeta\)2}{2} Log[\mmaFrac{\(\mu\)}{m}]}{3072 \mmaSup{\mmaSub{m}{\(\Delta\)}}{2} \mmaSup{\(\pi\)}{4}}+\mmaFrac{\mmaSup{gY}{2} \mmaSup{\(\mu\)}{2} Log[\mmaFrac{\(\mu\)}{m}]}{32 \mmaSup{\mmaSub{m}{\(\Delta\)}}{4} \mmaSup{\(\pi\)}{2}}\},
                \{q1Hq[1,1],\mmaFrac{\mmaSup{gY}{2} \mmaSup{\(\zeta\)1}{2} Log[\mmaFrac{\(\mu\)}{m}]}{1152 \mmaSup{\mmaSub{m}{\(\Delta\)}}{2} \mmaSup{\(\pi\)}{4}}+\mmaFrac{\mmaSup{gY}{2} \mmaSup{\(\zeta\)2}{2} Log[\mmaFrac{\(\mu\)}{m}]}{18432 \mmaSup{\mmaSub{m}{\(\Delta\)}}{2} \mmaSup{\(\pi\)}{4}}-\mmaFrac{\mmaSup{gY}{2} \mmaSup{\(\mu\)}{2} Log[\mmaFrac{\(\mu\)}{m}]}{192 \mmaSup{\mmaSub{m}{\(\Delta\)}}{4} \mmaSup{\(\pi\)}{2}}\},
                \{q3Hq[1,1],\mmaFrac{\mmaSup{gW}{2} \mmaSup{\(\zeta\)2}{2} Log[\mmaFrac{\(\mu\)}{m}]}{18432 \mmaSup{\mmaSub{m}{\(\Delta\)}}{2} \mmaSup{\(\pi\)}{4}}+\mmaFrac{\mmaSup{gW}{2} \mmaSup{\(\mu\)}{2} Log[\mmaFrac{\(\mu\)}{m}]}{192 \mmaSup{\mmaSub{m}{\(\Delta\)}}{4} \mmaSup{\(\pi\)}{2}}\},
                \{qHu[1,1],\mmaFrac{\mmaSup{gY}{2} \mmaSup{\(\zeta\)1}{2} Log[\mmaFrac{\(\mu\)}{m}]}{288 \mmaSup{\mmaSub{m}{\(\Delta\)}}{2} \mmaSup{\(\pi\)}{4}}+\mmaFrac{\mmaSup{gY}{2} \mmaSup{\(\zeta\)2}{2} Log[\mmaFrac{\(\mu\)}{m}]}{4608 \mmaSup{\mmaSub{m}{\(\Delta\)}}{2} \mmaSup{\(\pi\)}{4}}-\mmaFrac{\mmaSup{gY}{2} \mmaSup{\(\mu\)}{2} Log[\mmaFrac{\(\mu\)}{m}]}{48 \mmaSup{\mmaSub{m}{\(\Delta\)}}{4} \mmaSup{\(\pi\)}{2}}\},
                \{qHd[1,1],-\mmaFrac{\mmaSup{gY}{2} \mmaSup{\(\zeta\)1}{2} Log[\mmaFrac{\(\mu\)}{m}]}{576 \mmaSup{\mmaSub{m}{\(\Delta\)}}{2} \mmaSup{\(\pi\)}{4}}-\mmaFrac{\mmaSup{gY}{2} \mmaSup{\(\zeta\)2}{2} Log[\mmaFrac{\(\mu\)}{m}]}{9216 \mmaSup{\mmaSub{m}{\(\Delta\)}}{2} \mmaSup{\(\pi\)}{4}}+\mmaFrac{\mmaSup{gY}{2} \mmaSup{\(\mu\)}{2} Log[\mmaFrac{\(\mu\)}{m}]}{96 \mmaSup{\mmaSub{m}{\(\Delta\)}}{4} \mmaSup{\(\pi\)}{2}}\},
                \{qll[1,1,1,1],\mmaFrac{\mmaSup{y\(\Sigma\)}{2}}{4 \mmaSup{\mmaSub{m}{\(\Delta\)}}{2}}+\mmaFrac{11 \mmaSup{gW}{2} \mmaSup{y\(\Sigma\)}{2} Log[\mmaFrac{\(\mu\)}{m}]}{192 \mmaSup{\mmaSub{m}{\(\Delta\)}}{2} \mmaSup{\(\pi\)}{2}}+\mmaFrac{5 \mmaSup{gY}{2} \mmaSup{y\(\Sigma\)}{2} Log[\mmaFrac{\(\mu\)}{m}]}{64 \mmaSup{\mmaSub{m}{\(\Delta\)}}{2} \mmaSup{\(\pi\)}{2}}\},\{q1lq[1,1,1,1],-\mmaFrac{\mmaSup{gY}{2} \mmaSup{y\(\Sigma\)}{2} Log[\mmaFrac{\(\mu\)}{m}]}{96 \mmaSup{\mmaSub{m}{\(\Delta\)}}{2} \mmaSup{\(\pi\)}{2}}\},
                \{q3lq[1,1,1,1],\mmaFrac{\mmaSup{gW}{2} \mmaSup{y\(\Sigma\)}{2} Log[\mmaFrac{\(\mu\)}{m}]}{96 \mmaSup{\mmaSub{m}{\(\Delta\)}}{2} \mmaSup{\(\pi\)}{2}}\},\{qle[1,1,1,1],\mmaFrac{\mmaSup{gY}{2} \mmaSup{y\(\Sigma\)}{2} Log[\mmaFrac{\(\mu\)}{m}]}{16 \mmaSup{\mmaSub{m}{\(\Delta\)}}{2} \mmaSup{\(\pi\)}{2}}\},
                \{qlu[1,1,1,1],-\mmaFrac{\mmaSup{gY}{2} \mmaSup{y\(\Sigma\)}{2} Log[\mmaFrac{\(\mu\)}{m}]}{24 \mmaSup{\mmaSub{m}{\(\Delta\)}}{2}\mmaSup{\(\pi\)}{2}}\},\{qld[1,1,1,1],\mmaFrac{\mmaSup{gY}{2} \mmaSup{y\(\Sigma\)}{2} Log[\mmaFrac{\(\mu\)}{m}]}{48 \mmaSup{\mmaSub{m}{\(\Delta\)}}{2} \mmaSup{\(\pi\)}{2}}\}\}
        \end{mmaCell}
}

We have provided the flexibility to  users to reformat, save, or export all these WCs corresponding to the effective operators at the electroweak scale (\mmaInlineCell{Input}{\mmaUnd{\(\pmb{\mu}\)}}) to \LaTeX, using \mmaInlineCell{Input}{formPick}. We have also provided  an illustrative example:
\begin{mmaCell}{Input}
\mmaDef{formPick}["Warsaw","Detailed2",\mmaDef{floRes1},Frame\(\pmb{\to}\)All,
        FontSize\(\pmb{\to}\)\mmaDef{Medium},FontFamily\(\pmb{\to}\)"Times New Roman"]
\end{mmaCell}
\begin{mmaCell}{Output}
   
    \end{mmaCell}

                    \centering
                                    \begin{tabular}{lll}
                                \hline \hline
                                $Q_\mathrm{W}$  &  $\epsilon ^{\text{abc}}\mathrm{W}_{\uprho }{}^{\mathrm{a},\upmu }\mathrm{W}_{\upmu }{}^{\mathrm{b},\upnu }\mathrm{W}_{\upnu }{}^{\mathrm{c},\uprho }$  &  $\dfrac{29 \text{$g_\mathrm{W}$}^5 \log \left(\frac{\mu }{m_{\Delta}}\right)}{46080 \uppi ^4 m_{\Delta}^2}+\dfrac{\text{$g_\mathrm{W}$}^3}{1440 \uppi ^2 m_{\Delta}^2}$  \\
                              $\vdots$ & $\vdots$ & $\vdots$ \\
                                $Q_{\text{ld}}$  &  $\left(\bar{\mathrm{l}} \upgamma _{\mu }\text{ l)(}\bar{\mathrm{d}} \upgamma _{\mu }\text{ d)}\right.$  &  $\dfrac{{g_\mathrm{Y}}^2 \text{y$_\Sigma $}^2 \log \left(\frac{\mu }{m_{\Delta}}\right)}{48 \uppi ^2 m_{\Delta}^2}$  \\
                                \hline \hline
                        \end{tabular}
                         \end{enumerate}

\subsection*{Acknowledgement}
S.D. Bakshi thanks the organisers of the 11th FCC-ee workshop: Theory and Experiments, CERN for the invitation and providing the opportunity to present this work.

%%%%%%%%%%%%%%%%%%%%%%%%%%%%%%%%%%%%%%%%%%%%%%%%%%%%%%%%%%%%%%%%%%%%%%%%%%%%%%%

\end{bibunit}

\label{sec-smeft-SDasBakshi}

\clearpage \pagestyle{empty}  \cleardoublepage
%============================================
%============================================
\chapter
%[Methods and Tools]
{Beyond the Standard Model (BSM)} \label{chbsm}
%============================================

\pagestyle{fancy}
\fancyhead[CO]{\thechapter.\thesection \hspace{1mm} (Triple) Higgs coupling imprints at future lepton colliders}
\fancyhead[RO]{}
\fancyhead[LO]{}
\fancyhead[LE]{}
\fancyhead[CE]{}
\fancyhead[RE]{}
\fancyhead[CE]{J. Baglio, C. Weiland}
\lfoot[]{}
\cfoot{-  \thepage \hspace*{0.075mm} -}
\rfoot[]{}
%%% definitions %%%%%%%%%%%%%%%%%%%%%%%%%%
\def\be{\begin{equation}}
\def\ee{\end{equation}}
\graphicspath{{BSM_Baglio/plots/}}

\begin{bibunit}[elsarticle-num] % define the bib-style for the unit: elsarticle-num.bst
%  text-1; this is the corresponding section
%\putbib[2loops] % the *.bib
%\end{bibunit}
% go-on
%--- from: bibunits.sty, adapts the font size of ``References'' to section
\let\stdthebibliography\thebibliography
\renewcommand{\thebibliography}{%
\let\section\subsection
\stdthebibliography}
%---
    
\section
[(Triple) Higgs coupling imprints at future lepton colliders \\{\it J. Baglio, C. Weiland}]
{(Triple) Higgs coupling imprints at future lepton colliders
\label{contr:BAGLIO}}
\noindent
{\bf Contribution\footnote{This contribution should be cited as:\\
J. Baglio, C. Weiland, (Triple) Higgs coupling imprints at future lepton colliders,  
%04 DOI:10.23731/CYRM-2020-XXX.\thepage,\texttt{{CoDEx}}: BSM physics being realised as  SMEFT in:
%04 \url{http://dx.doi.org/10.23731/CYRM-2020-XXX.\thepage}, in:
DOI: \href{http://dx.doi.org/10.23731/CYRM-2020-003.\thepage}{10.23731/CYRM-2020-003.\thepage}, in:
Theory for the FCC-ee, Eds. A. Blondel, J. Gluza, S. Jadach, P. Janot and T. Riemann,\\
CERN Yellow Reports: Monographs, CERN-2020-003,
%04 \url{http://dx.doi.org/10.23731/CYRM-2020-XXX}, p. \thepage.} 
DOI: \href{http://dx.doi.org/10.23731/CYRM-2020-003}{10.23731/CYRM-2020-003},
p. \thepage.
\\ \copyright\space CERN, 2020. Published by CERN under the 
%04-2
\href{http://creativecommons.org/licenses/by/4.0/}{Creative Commons Attribution 4.0 license}.} by: J. Baglio, C. Weiland\\
Corresponding author: J. Baglio {[julien.baglio@uni-tuebingen.de]}}
\vspace*{.5cm}

\subsection{Triple Higgs coupling studies in an EFT framework}
\label{sec:hhhEFT}

The measurement of the triple Higgs coupling is one of the major goals of the future colliders. The direct measurement at lepton colliders relies on the production of Higgs boson pairs in two main channels: $\mathrm{e}^+\mathrm{e}^-\to \mathrm{ZHH}$, which is dominant at centre-of-mass energies below 1~TeV and maximal at around 500~GeV, and $\mathrm{e}^+\mathrm{e}^-\to \mathrm{HH}\upnu_\mathrm{e}\bar{\upnu}_\mathrm{e}$, which becomes dominant for high-energy colliders. This direct measurement is required to be at least at a centre-of-mass energy of 500~GeV, and is hence only possible at future linear colliders, such as the International Linear Collider (ILC), operating at 500~GeV or 1~TeV \cite{Baer:2013cma}, or the Compact Linear Collider (CLIC), operating at 1.4\,TeV (stage 2) or 3~TeV (stage 3)~\cite{CLIC:2016zwp}. The SM triple Higgs coupling sensitivity is estimated to be $\updelta\kappa_\lambda = (\lambda_\mathrm{HHH}/\lambda_\mathrm{HHH}^{\rm SM}-1) \sim 28\%$ at the 500~GeV ILC, with a luminosity of 4~ab$^{-1}$~\cite{Duerig:2016dvi,Barklow:2017awn}, and $\updelta\kappa_\lambda\sim 13\%$ at the CLIC, when combining the 1.4~TeV run, with 2.5~ab$^{-1}$ of data, and the 3~TeV run, with 5~ab$^{-1}$ of data~\cite{Abramowicz:2016zbo}.

Still, circular lepton collider projects, such as the Circular Electron--Positron Collider (CEPC)~\cite{CEPCStudyGroup:2018ghi} or the FCC-ee~\cite{Abada:2019lih,Abada:2019zxq}, which run at energies below 500~GeV (not to mention the ILC or the CLIC running at lower energies), can provide a way to constrain the triple Higgs coupling \cite{Blondel:2018aan}. Since  Ref. \cite{McCullough:2013rea},
 in which it was first proposed  to use precision measurements to constrain the triple Higgs coupling, in particular, the measurements in single Higgs production at lepton colliders, there have been studies of the combination of single and double Higgs production observables, not only at lepton but also at hadron colliders \cite{Degrassi:2017ucl,DiVita:2017eyz,DiVita:2017vrr,Maltoni:2018ttu}. The analyses use the framework of  Standard Model effective field theory (SMEFT). According to the latest ECFA report \cite{deBlas:2019rxi}, the combination of HL-LHC projections \cite{Cepeda:2019klc} with ILC exclusive single Higgs data gives $\updelta\kappa_\lambda = 26\%$ at $68\%$ CL, while with the FCC-ee (at 250 or 365~GeV) this goes down to $\updelta\kappa_\lambda = 19\%$, and with CEPC we get $\updelta\kappa_\lambda = 17\%$. We will present in more detail the results of Refs.~\cite{DiVita:2017eyz,DiVita:2017vrr}, which demonstrate how important the combination of the LHC results with an analysis at lepton colliders is, and show the potential of the FCC-ee.\footnote{Julien Baglio thanks Christophe Grojean for his very useful input to this subsection.}

\begin{figure}
  \centering
    \includegraphics[width=0.48\textwidth]{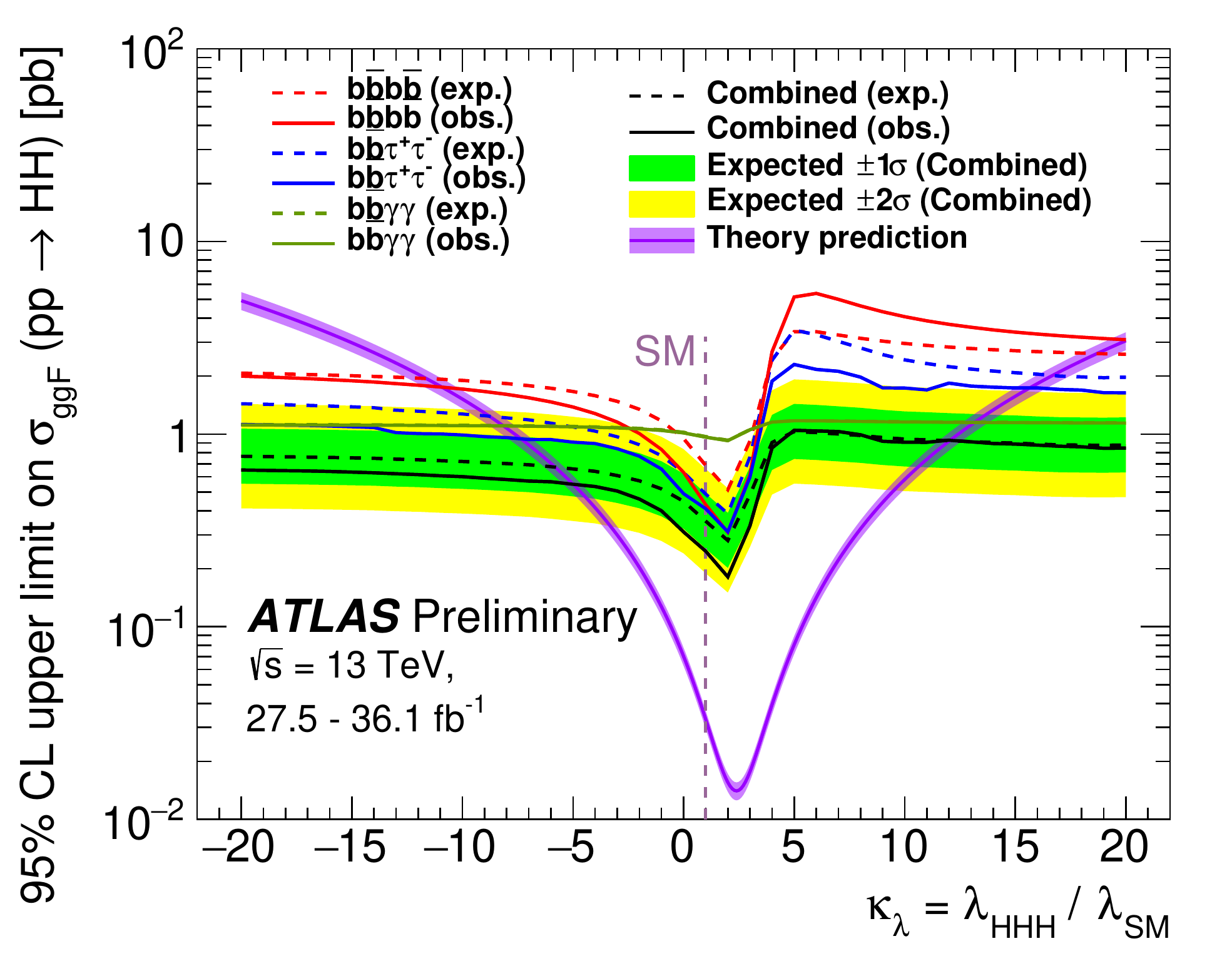}\hspace*{2mm}
    \includegraphics[width=0.44\textwidth]{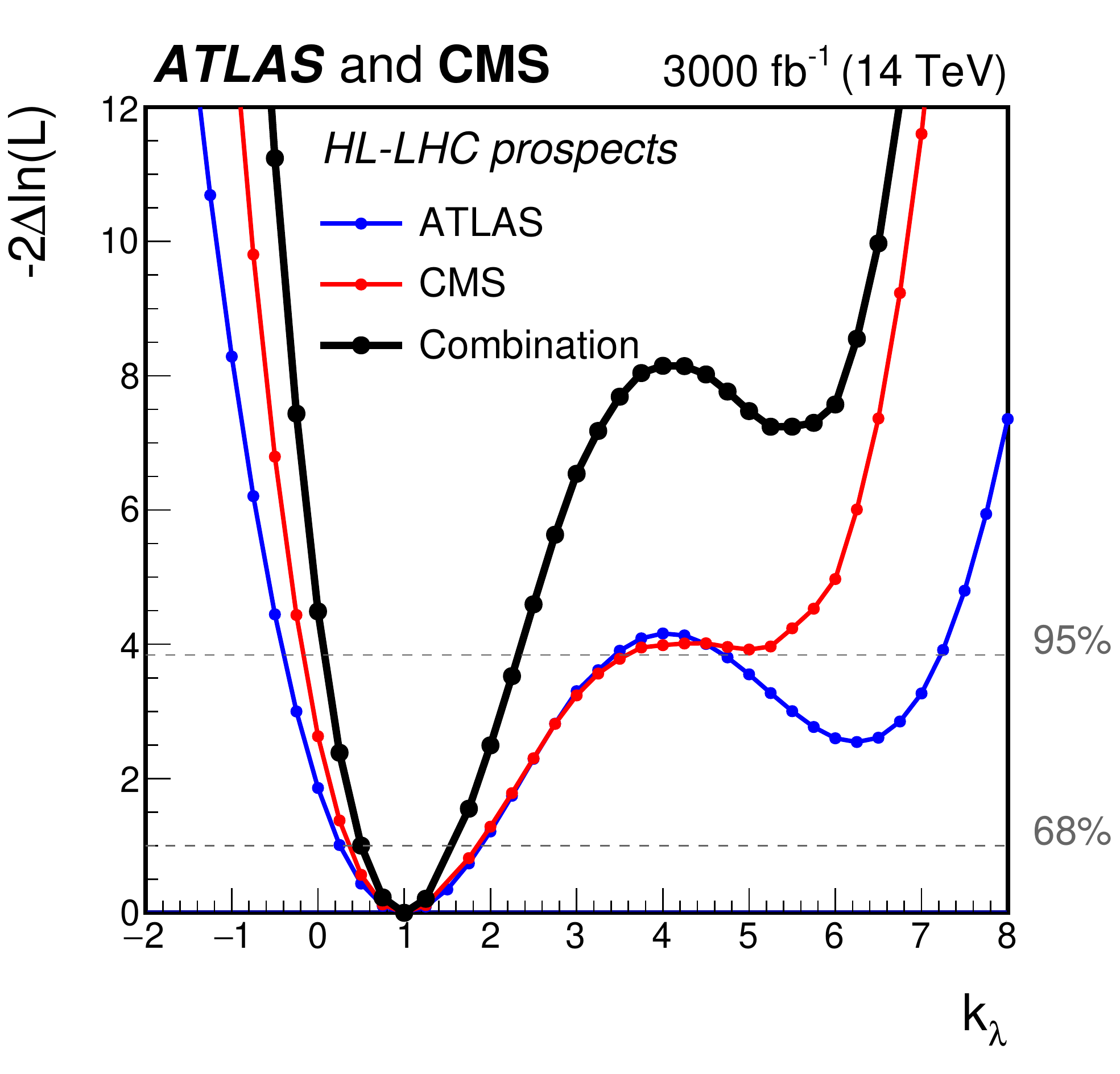}
    \caption{Left: Latest experimental bounds on the triple Higgs coupling from the ATLAS collaboration at the 13~TeV LHC, combining $\mathrm{b}\bar{\mathrm{b}}\mathrm{b}\bar{\mathrm{b}}$, $\mathrm{b}\bar{\mathrm{b}}\uptau^+\uptau^-$, and $\mathrm{b}\bar{\mathrm{b}}\upgamma\upgamma$ final states. Taken from Ref.~\cite{ATLAS:2018otd}. Right: Minimum negative-log-likelihood distribution of $\kappa_\lambda$ at the HL-LHC with 3~ab$^{-1}$ of data, including differential observables in Higgs pair production, with ATLAS (blue), CMS (red), and ATLAS+CMS (black) projected results. Figure taken from Ref.~\cite{Cepeda:2019klc}.}
\label{fig:HHLHC}
%\query{Please correct the figure labels of \Fref{fig:HHLHC}. Set variables
%in italic font (but particle names in roman font). Be %sure that your minus
%signs are minus signs  ($-$) and not hyphens (-).}
\end{figure}

Figure \ref{fig:HHLHC} (left) displays the latest experimental results available at the 13~TeV LHC for the search of non-resonance Higgs pair production and the 95\% CL limits on the triple Higgs coupling, which have been presented in Ref. \cite{ATLAS:2018otd}. The results constrain $\updelta\kappa_\lambda$ in the range $[-6.0:11.1]$. We can compare them with the projections at the HL-LHC with 3~ab$^{-1}$ presented in the HL-HE LHC report~\cite{Cepeda:2019klc} in an SMEFT framework, using a differential analysis in the channel $\mathrm{p p}\to \mathrm{HH}$. Compared with the projection in Ref. \cite{DiVita:2017eyz}, which also included single Higgs data in the channels $\mathrm{p p}\to \mathrm{W}^\pm \mathrm{H}, \mathrm{ZH}, \mathrm{t}\bar{\mathrm{t}} \mathrm{H}$, there is a substantial improvement, thanks to the experimental differential analysis. We have $-0.5\leq \updelta\kappa_\lambda\leq 0.5$ at 68\% CL and $-0.9\leq\updelta\kappa_\lambda\leq 1.3$ at 95\% CL. The degeneracy observed in Ref.~\cite{DiVita:2017eyz} with a second minimum at $\updelta\kappa_\lambda \sim 5$ is now excluded at $4\sigma$. 

%%%%
%%%%

\begin{figure}
  \centering
    \includegraphics[width=0.48\textwidth]{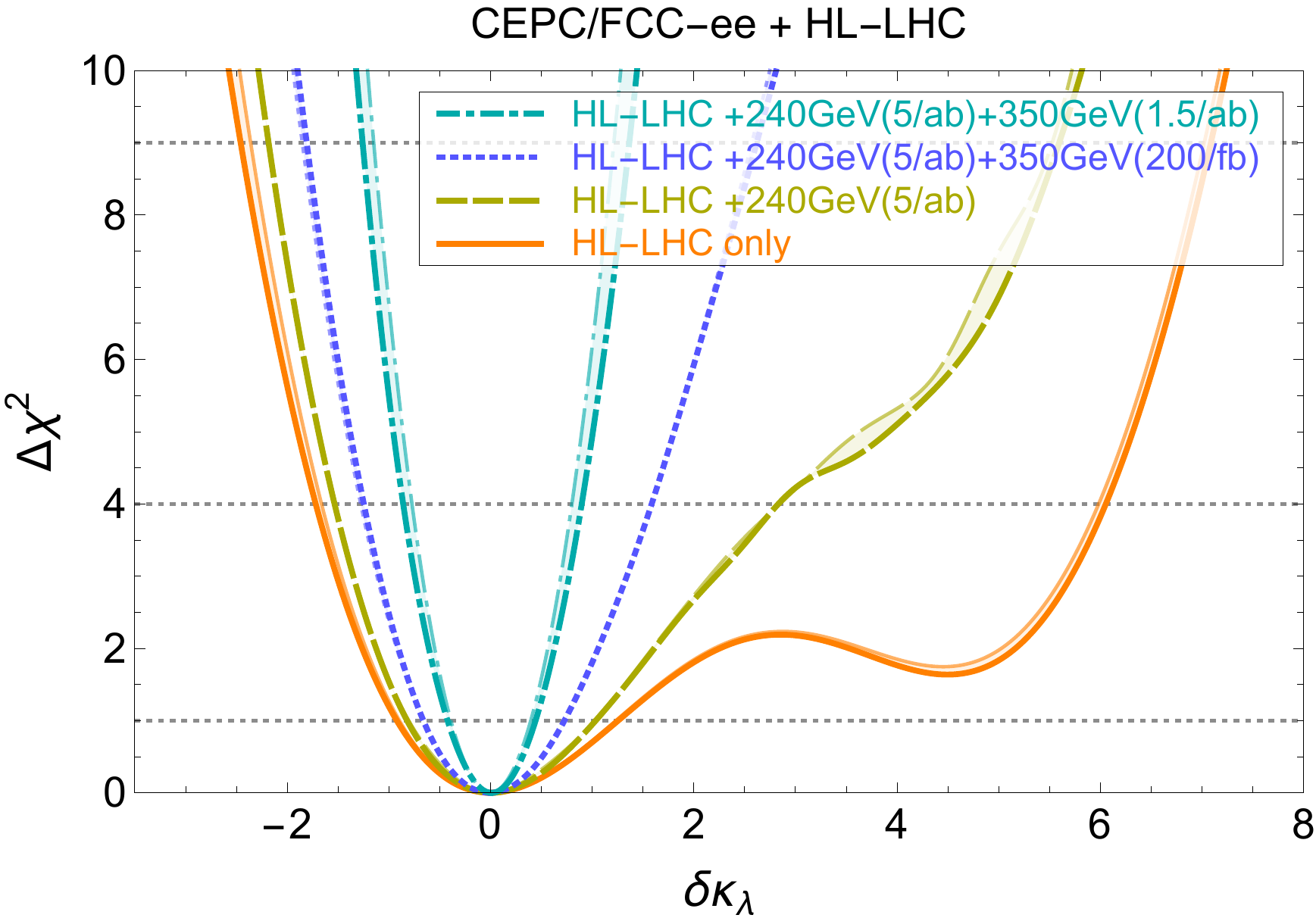}
    \includegraphics[width=0.48\textwidth]{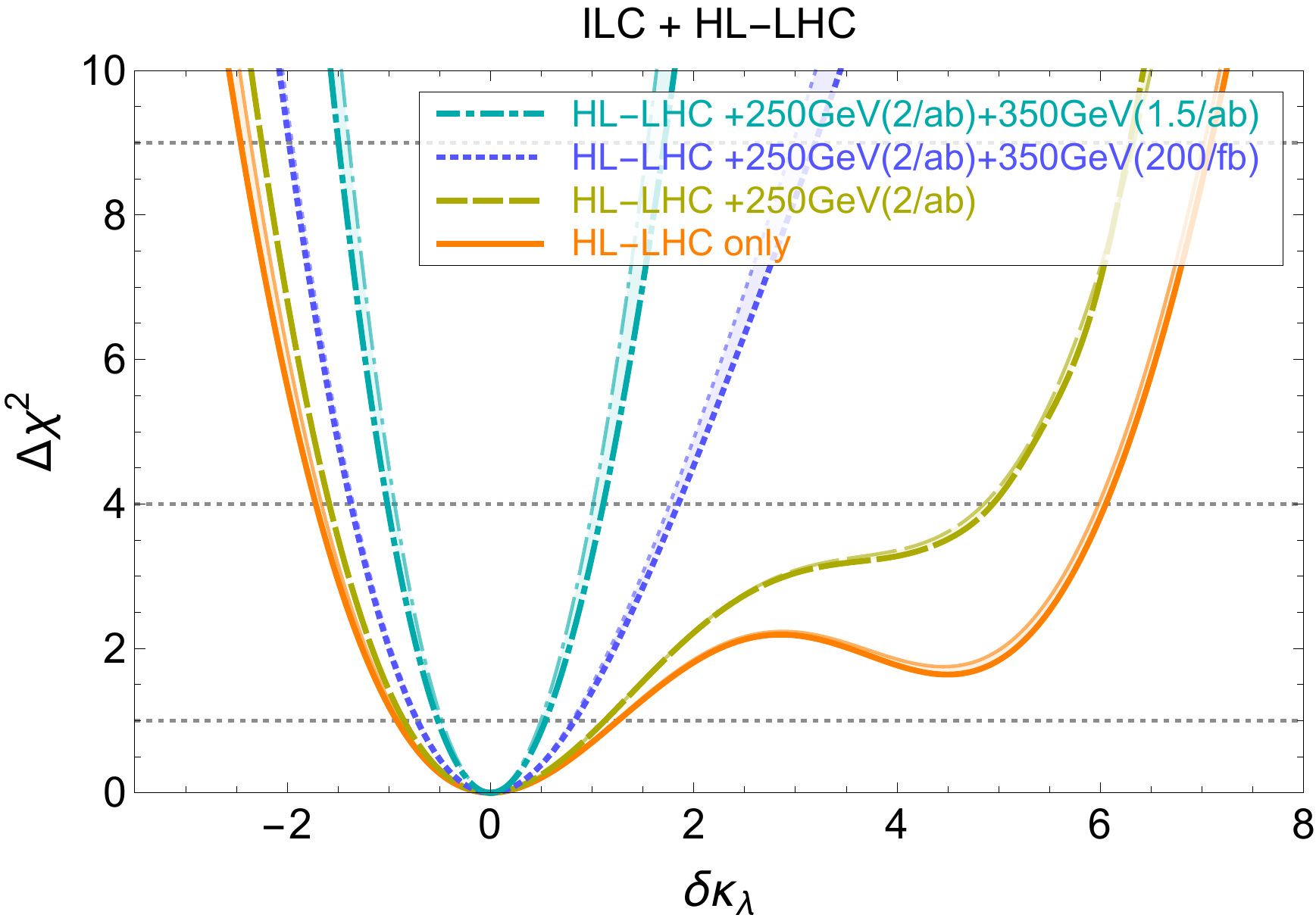}
    \caption{$\Delta\chi^2$ distributions for a global fit of the parameter $\updelta\kappa_\lambda$ at circular lepton colliders (left) or at the ILC (right), combined with HL-LHC data. The different lines compare the different centre-of-mass energies and luminosity scenarios. Figures taken from Ref.~\cite{DiVita:2017vrr}.}
\label{fig:HHHleptons}
%\query{Please correct the figure labels of \Fref{fig:HHHleptons}. Set variables
%in italic font (but particle names in roman font).}
\end{figure}

The combination with data from lepton colliders removes the second minimum even more drastically and only the SM minimum is left at $\updelta\kappa_\lambda = 0$ \cite{DiVita:2017vrr}, in particular when data from 250\,GeV and 350--365~GeV centre-of-mass energies are combined \cite{DiVita:2017vrr}. This is shown in \Fref{fig:HHHleptons}, where two set-ups are compared, the combination of HL-LHC data with circular lepton colliders (FCC-ee or CEPC) data on the left-hand side, and the combination of HL-LHC data with the ILC data on the right-hand side. In both cases, the lepton collider data consist of measurements in the channels $\mathrm{e}^+\mathrm{e}^-\to \mathrm{W}^+\mathrm{W}^-, \mathrm{ZH}, \upnu_\mathrm{e}\bar{\upnu}_\mathrm{e} \mathrm{H}$. The second minimum disappears completely even with a relatively low integrated luminosity of $\mathcal{L}=200\Ufb^{-1}$ at 350~GeV, when combined with the data at 250\,GeV. Note that the FCC-ee (or CEPC), thanks to its much higher luminosity in the 250~GeV run, is doing significantly better than the ILC.
%%%%
%%%%

\subsection{Probing heavy neutral leptons via Higgs couplings}
\label{sec:Higgsneut}

Since the confirmation of neutrino oscillations in 1998 by the Super-Kamiokande\linebreak  experiment~\cite{Fukuda:1998mi}, it has been established that at least two neutrinos have a non-zero mass~\cite{Esteban:2018azc}. This experimental fact cannot be accounted for in the SM and requires new physics. One of the simplest extensions is the addition of new heavy neutral leptons that are gauge singlets and mix with the active neutrinos to generate the light neutrino masses. An appealing model, allowing for these new fermionic states to be in the range of gigaelectronvolts to a few teraelectronvolts while having Yukawa couplings of order one, is the inverse see-saw (ISS) model~\cite{Mohapatra:1986aw,Mohapatra:1986bd,Bernabeu:1987gr}, in which a nearly conserved lepton-number symmetry~\cite{Kersten:2007vk,Moffat:2017feq} is introduced, naturally explaining the smallness of the mass of the lightest neutrino states while allowing for large couplings between the heavy neutrinos and the Higgs boson, leading to a rich phenomenology. In this view, the very precise study of the Higgs sector at lepton colliders can offer a unique opportunity to test low-scale see-saw mechanisms, such as the ISS.

\subsubsection{Heavy neutral leptons in the gigaelectronvolt regime}

\begin{figure}
    \begin{center}
    \includegraphics[width=0.6\textwidth]{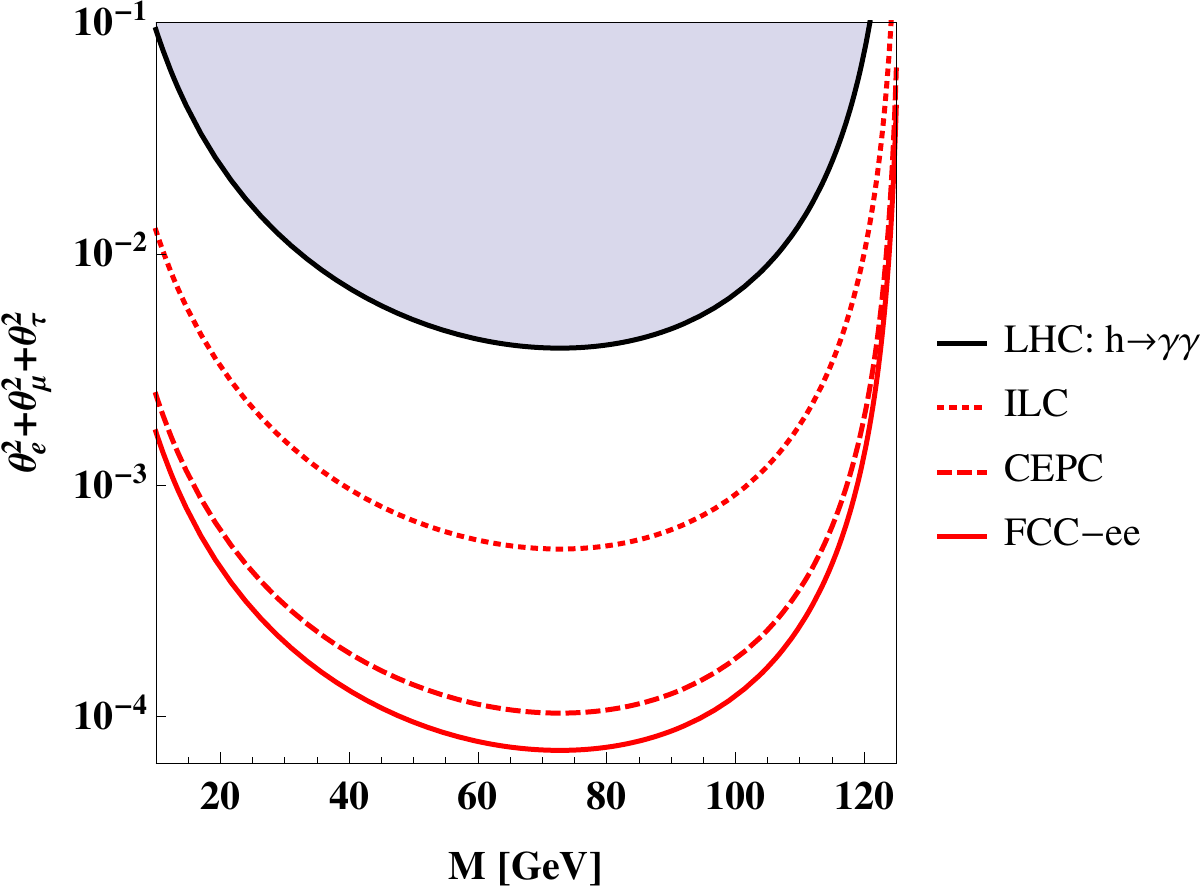}
    \end{center}
    \caption{Estimated sensitivities on the heavy sterile neutrino properties from the decay $\mathrm{H}\to \mathrm{W}^+ \mathrm{W}^-$, assuming 10 years of data collection. The black line denotes the bound from the LHC coming from $\mathrm{H}\to\upgamma\upgamma$ with up to 2015 data. Taken from Ref.~\cite{Antusch:2015mia}.}
\label{fig:HWWneut}
%\query{Please correct the figure labels of \Fref{fig:HWWneut}. Set variables
%in italic font and particle names in roman font.}
\end{figure}

We begin with the gigaelectronvolt regime. In these low-scale see-saw models, the mixing between the active and the sterile neutrinos leads to modified couplings of neutrinos to the W, Z, and Higgs bosons. This naturally leads to the idea of using precision measurements of the Higgs boson branching fractions into gauge bosons in order to test the mass range $M_\mathrm{N}<M_\mathrm{H}$, where $M_\mathrm{N}$ is the mass of the heavy neutrino states and $M_\mathrm{H}$ is the mass of the Higgs boson. As $\mathrm{H}\to \mathrm{N N}$ is allowed, the invisible Higgs decay width is modified and hence the branching fraction ${\rm BR}(\mathrm{H}\to \mathrm{W}^+\mathrm{W}^-)$ is modified via the modified total decay width $\Gamma_\mathrm{H}$. According to an analysis of 2015~\cite{Antusch:2015mia}, the FCC-ee could be the most competitive lepton collider to test this option, as demonstrated in \Fref{fig:HWWneut}. In particular, the experimental sensitivity to ${\rm BR}(\mathrm{H}\to \mathrm{W}^+\mathrm{W}^-)$ is expected to be $0.9\%$ at the FCC-ee, compared with $1.3\%$ at the CEPC, operating at 240~GeV~\cite{Ruan:2014xxa}, and $6.4\%$ at the ILC, operating at 250\,GeV~\cite{Baer:2013cma}.\footnote{The latest analysis at the ILC, using a luminosity of 500~fb$^{-1}$, states that a precision of $4.1\%$ can be achieved~\cite{Panduroviu0107:2019rms}.}
%%%%
%%%%

\subsubsection{Probing heavy neutral leptons in the multi-teraelectronvolt regime}

Since the coupling of the heavy neutral leptons to the Higgs boson can be quite large in low-scale see-saw models for masses $M_\mathrm{N}$ of a few teraelectronvolts, it is also very appealing to use, again, Higgs properties to probe a mass regime of $M_\mathrm{N}\sim \mathcal{O}(1-10\UTeV)$.

Off-diagonal couplings of the Higgs boson to heavy neutral leptons will induce charged-lepton-flavour-violating (cLFV) decays~\cite{Pilaftsis:1992st}. In particular, simplified formulae were provided in Ref.~\cite{Arganda:2014dta}, showing that cLFV Higgs decays exhibit a different functional dependence on see-saw parameters than cLFV radiative decays. They thus provide complementary observables to search for heavy neutral leptons. In a typical low-scale see-saw model like the ISS, the predicted branching fraction can be as large as $\mathrm{BR}(\mathrm{H}\to\uptau\upmu)\sim 10^{-5}$ and could even reach $\mathrm{BR}(\mathrm{H}\to\uptau\upmu)\sim 10^{-2}$ in a supersymmetric model~\cite{Arganda:2015naa}, thus being well within the reach of a Higgs factory like the FCC-ee. However, Higgs observables are also uniquely sensitive to diagonal couplings  and this was discussed in particular in Refs.~\cite{Baglio:2016ijw,Baglio:2016bop}, using the triple Higgs coupling, and in Ref.~\cite{Baglio:2017fxf}, using a direct physical observable, the production cross-section $\sigma(\mathrm{e}^+\mathrm{e}^-\to \mathrm{W}^+\mathrm{W}^- \mathrm{H})$. Taking into account all theoretical and experimental constraints that were available, the three studies have found sizeable effects.

\begin{figure}
  \centering
\includegraphics[width=0.9\textwidth]{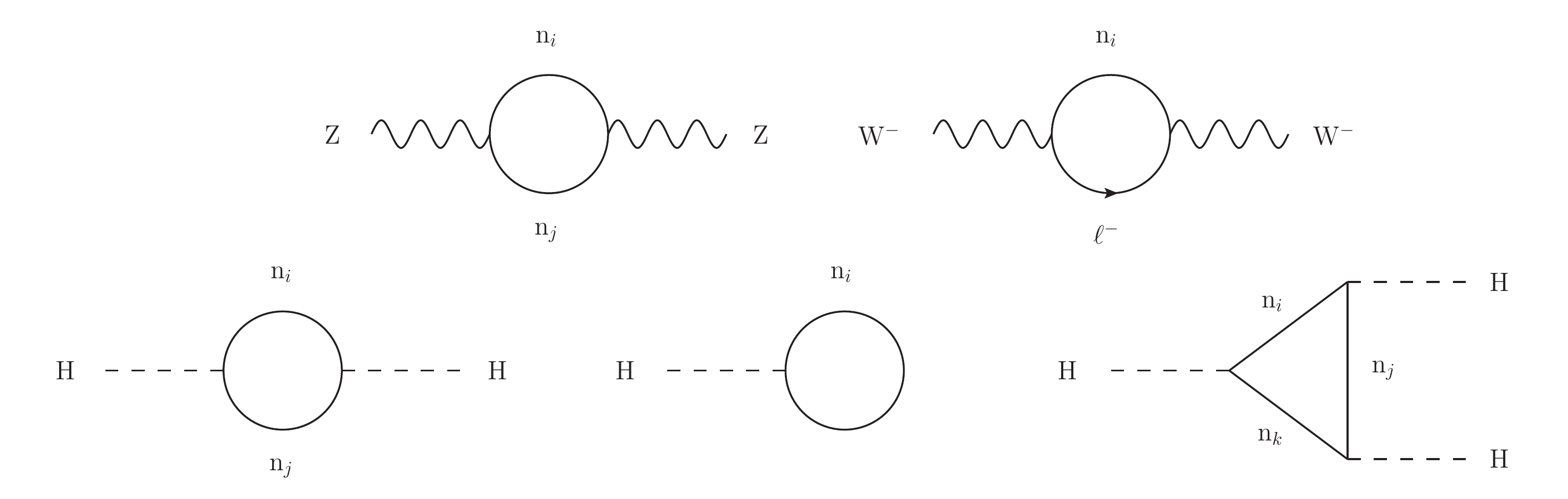}
    \caption{Representative Feynman diagrams for the one-loop corrections to $\lambda_\mathrm{HHH}$ involving the neutrinos in the ISS model.}
\label{fig:HHHdiags}
%\query{Please correct the figure labels of \Fref{fig:HHHdiags}. Set variables
%in italic font and particle names in roman font.}
\end{figure}

In the triple Higgs coupling studies, the one-loop corrections to $\lambda_\mathrm{HHH}$, defined as the physical triple Higgs coupling after electroweak symmetry breaking, are studied. The calculation is performed in the on-shell scheme and compares the SM prediction with the prediction in low-scale see-saw models (specifically the ISS presented in Ref.~\cite{Baglio:2016bop}). Representative one-loop diagrams involving the new heavy neutral leptons are given in \Fref{fig:HHHdiags} and details of the calculation and analytical formulae can be found in the original articles. The results are given  in terms of deviations with respect to the tree-level value $\lambda^{0}_\mathrm{HHH}$ and to the renormalised one-loop value in the SM $\lambda_\mathrm{HHH}^{1,{\rm SM}}$ of the triple Higgs coupling,
\begin{align}
  \Delta^{(1)} \lambda_\mathrm{HHH} & =
                               \frac{1}{\lambda^{0}}\left(\lambda_\mathrm{HHH}^{1}
                               -\lambda^0\right)\,,\,\nonumber\\
  \Delta^{\rm BSM} & = \frac{1}{\lambda_\mathrm{HHH}^{1,{\rm
                     SM}}}\left(\lambda_\mathrm{HHH}^{1}
                     -\lambda_\mathrm{HHH}^{1,{\rm SM}}\right)\,,
\end{align}
with $\lambda_\mathrm{HHH}^{1}$ being the one-loop renormalised triple Higgs coupling in the low-scale see-saw model considered. The constraints from low-energy neutrino observables are implemented via the $\mu_X^{}$ parametrization; see Ref.~\cite{Arganda:2014dta} for more details and  Appendix A of Ref.~\cite{Baglio:2016bop} for terms beyond the lowest order in the see-saw expansion. 
%All relevant theoretical and experimental bounds are taken into account; the most stringent constraint comes from the global fit to electroweak precision observables and lepton universality tests
All relevant theoretical and experimental bounds are taken into account. The most stringent constraint comes from the global fit to electroweak precision observables and lepton universality tests~\cite{Fernandez-Martinez:2016lgt}.

%%%%
%%%%

\begin{figure*}
  \centering
  \includegraphics[width=0.49\textwidth]{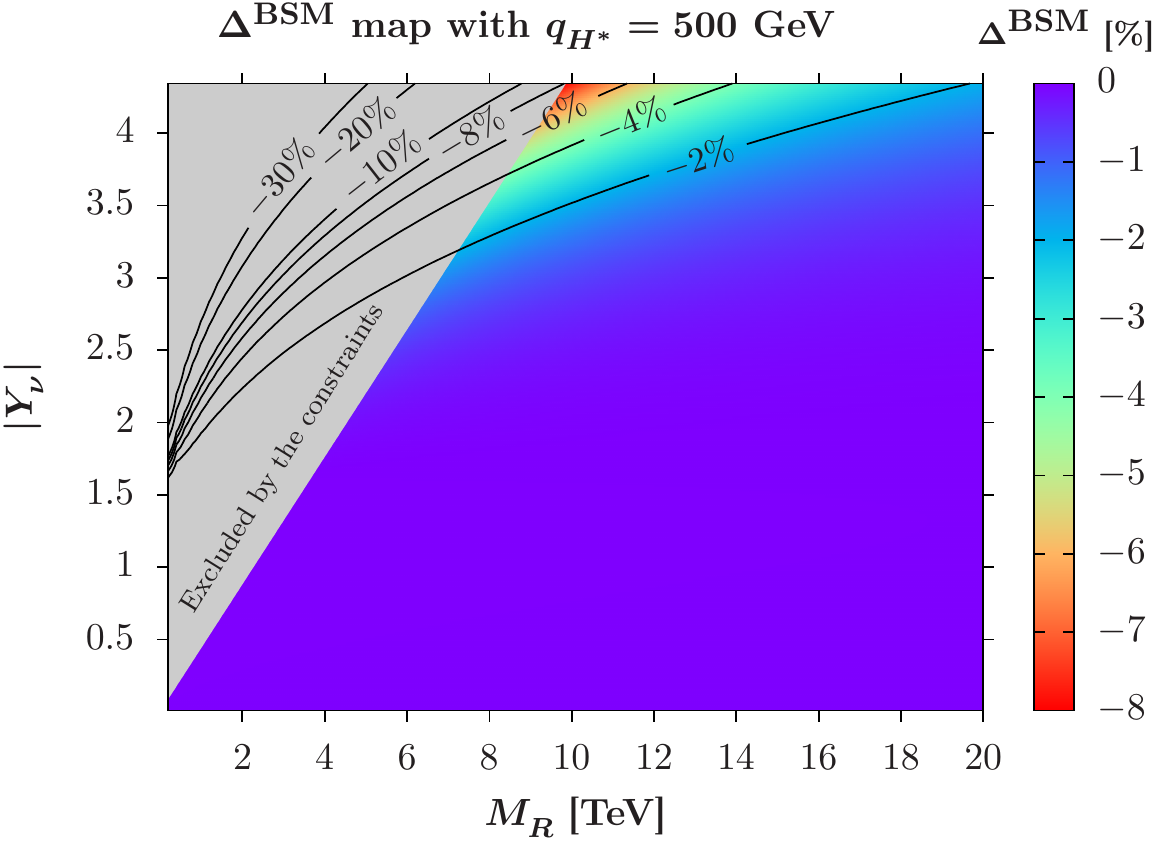}
  \includegraphics[width=0.49\textwidth]{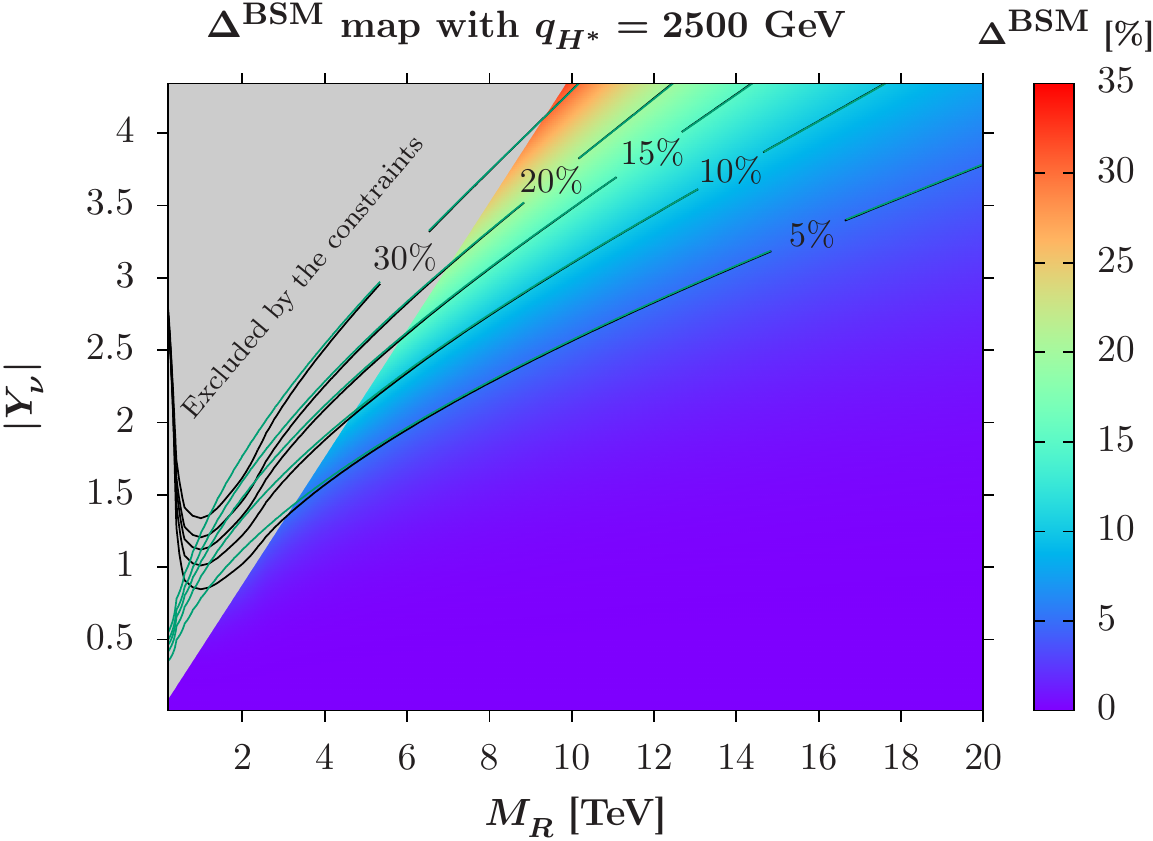}
  \caption{Contour maps of the heavy neutral lepton correction $\Delta^{\rm BSM}_{}$ to the triple Higgs coupling $\lambda_\mathrm{HHH}$ (in $\%$) as a  function of the heavy neutral lepton parameters $M_R^{}$ (in teraelectronvolts) and $|Y_\nu^{}|$ at a fixed off-shell Higgs momentum $q_{\mathrm{H}^*}^{} = 500$~GeV (left) and $q_{\mathrm{H}^*}^{}=2500$~GeV (right). The details of the spectrum are given in Ref.~\cite{Baglio:2016bop}. The grey area is excluded by the constraints on the model and the green lines on the right figure are contour lines that correspond to our approximate formula, while the black lines correspond to the full calculation.}
  \label{fig:HHHresults}
%  \query{Please correct the figure labels of \Fref{fig:HHHresults}. Set variables
%in italic font and particle names in roman font.}
\end{figure*}

Figure \ref{fig:HHHresults} displays the results of the analysis in the plane $M_R-|Y_\nu|$ where $M_R$ is the see-saw scale and $|Y_\nu|$ is the magnitude of the Yukawa coupling between the heavy neutral leptons and the Higgs boson. For an off-shell Higgs momentum of $q_{\mathrm{H}^*}=500$\,GeV splitting into two on-shell Higgs bosons, sizeable deviations can be obtained, up to $\Delta^{\rm BSM} \simeq -8\%$. Compared with the expected sensitivity of $\sim$$10\%$ at the ILC at 1\,TeV with 5\,ab$_{}^{-1}$~\cite{Fujii:2015jha} or  the FCC-hh sensitivity of $\sim$$5\%$ when two experiments were to be combined~\cite{He:2015spf}, the deviation can be probed and hence test masses of  order $\mathcal{O}(10\UTeV)$. In the case of the FCC-hh, as the hadronic centre-of-mass energy is large, the case $q_{\mathrm{H}^*}=2500$~GeV is even more interesting, with a deviation up to $\Delta^{\rm BSM} \simeq +35\%$, leading to a larger coverage of the parameter space and the possibility of testing the model at the 3\,TeV CLIC, where the sensitivity to $\lambda_\mathrm{HHH}$ is expected to be of the order of $13\%$~\cite{Abramowicz:2016zbo}. The triple Higgs coupling $\lambda_\mathrm{HHH}$ is a viable new (pseudo-)observable for the neutrino sector in order to constrain mass models, and might also be used in the context of the FCC-ee in an indirect way in $\mathrm{e}^+\mathrm{e}^-\to \mathrm{ZH}$ at the two-loop order, given the expected sensitivity the FCC-ee is supposed to reach in this channel. Studies remain to be done in this context.

\begin{figure}
  \centering
  \includegraphics[width=0.9\textwidth]{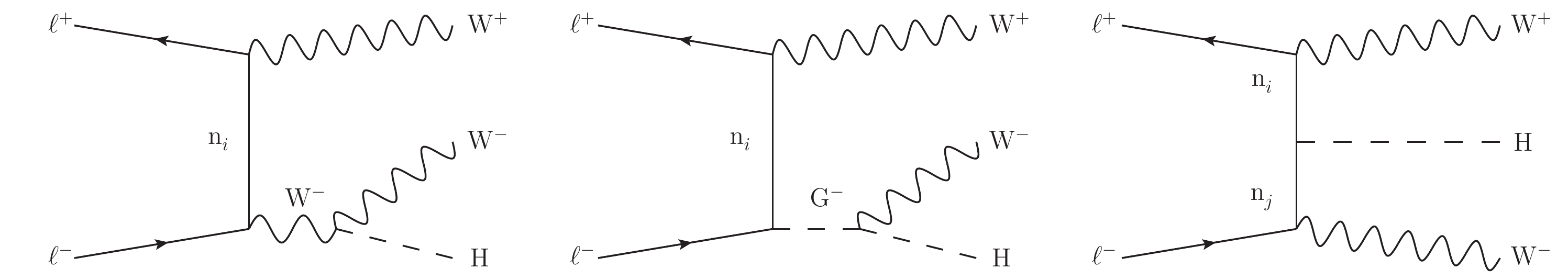}
  \caption{ISS neutrino contributions to the process $\ell^+\ell^-\to \mathrm{W}^+ \mathrm{W}^- \mathrm{H}$ in the Feynman--'t~Hooft gauge. Mirror diagrams can be obtained by flipping all the electric charges; the indices $i,j$ run from 1 to 9.}
  \label{fig:wwhdiags}
%    \query{Please correct the figure labels of \Fref{fig:wwhdiags}. Set variables
%in italic font and particle names in roman font.}
\end{figure}

The study presented in Ref.~\cite{Baglio:2017fxf} considered a more direct observable, the production cross-section $\sigma(\mathrm{e}^+\mathrm{e}^-\to \mathrm{W}^+\mathrm{W}^-\mathrm{H})$ at lepton colliders. The set-up is the same as in Ref.~\cite{Baglio:2016bop}, albeit with an updated global fit using NuFIT 3.0~\cite{Esteban:2016qun} to explain neutrino oscillations. The representative diagrams in the Feynman--'t~Hooft gauge are displayed in Fig.~\ref{fig:wwhdiags}, with the contributions of the heavy neutral leptons in the t channel.

\begin{figure}
  \centering
  \includegraphics[width=0.51\textwidth]{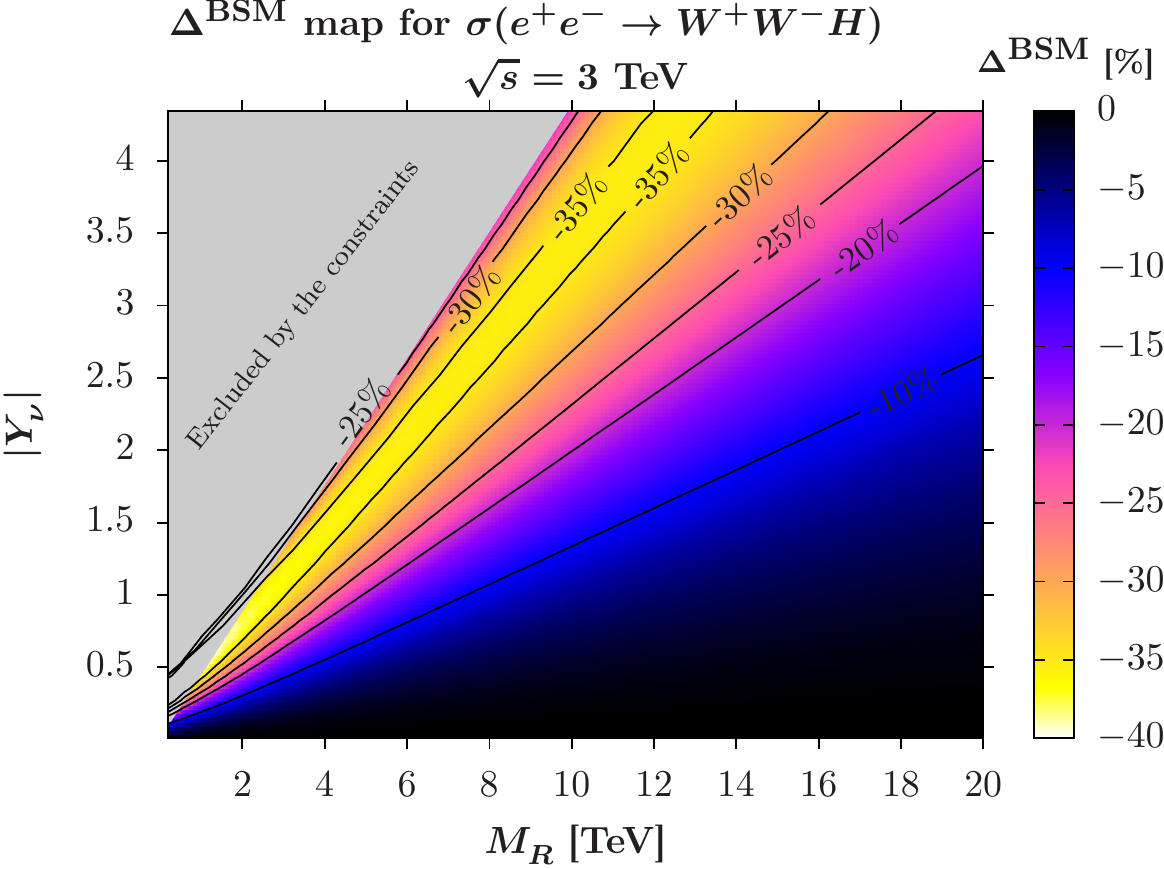}
  \includegraphics[width=0.47\textwidth]{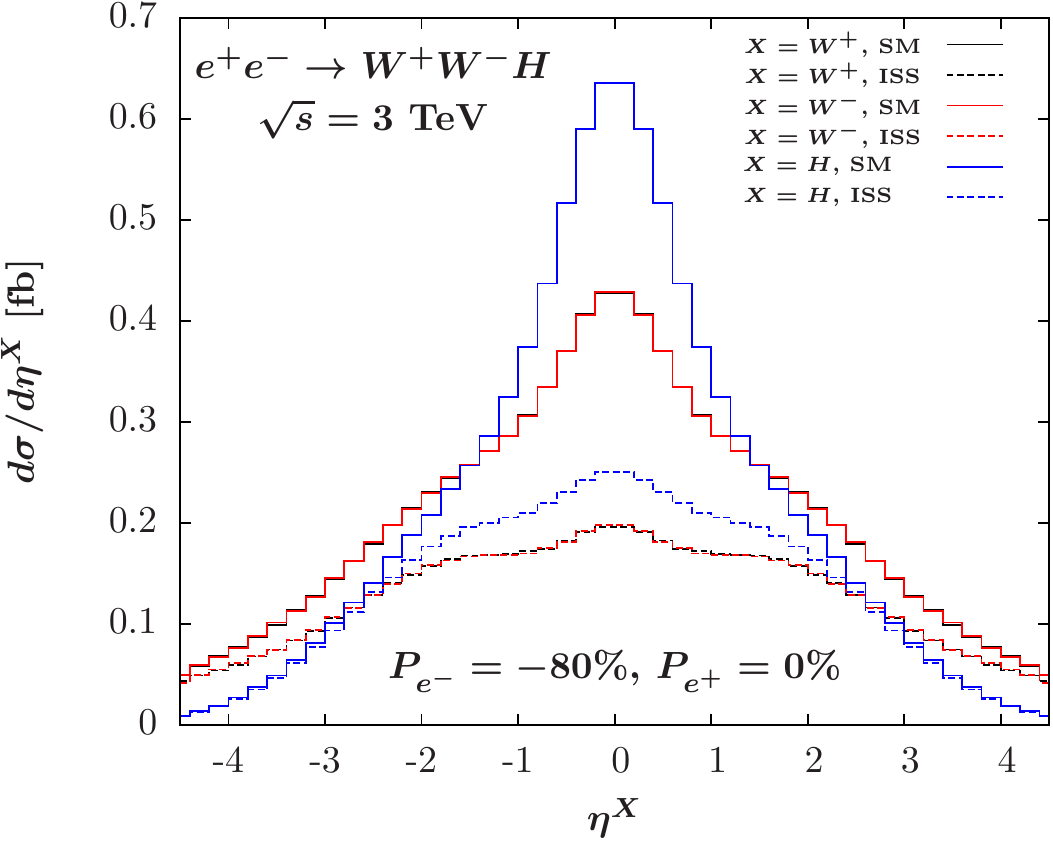}
  \caption{Left: Contour map of the neutrino corrections $\Delta_{}^{\rm BSM}$ at the 3\,TeV CLIC, using a $-80\%$ polarised electron beam, as a function of the see-saw scale $M_R^{}$ and $|Y_\nu^{}|$. Right: Pseudo-rapidity distributions of the $\mathrm{W}^+_{}$ (black), $\mathrm{W}^-_{}$ (red), and Higgs (blue) bosons. The solid curves stand for the SM predictions while the dashed curves stand for the ISS predictions, for the benchmark scenario described in the text. Figures taken from Ref.~\cite{Baglio:2017fxf}.}
  \label{fig:wwhresults}
%      \query{Please correct the figure labels of \Fref{fig:wwhresults}. Set variables
%in italic font and particle names in roman font.}
\end{figure}

The deviation $\Delta^{\rm BSM}$ now stands  for the comparison between the total cross-section\linebreak $\sigma(\mathrm{e}^+\mathrm{e}^-\to \mathrm{W}^+\mathrm{W}^- \mathrm{H})$ calculated in the ISS model and in the SM, $\Delta^{\rm BSM} = (\sigma^{\rm ISS}-\sigma^{\rm SM})/\sigma^{\rm SM}$. Using the CLIC baseline for the polarisation of the beams~\cite{CLIC:2016zwp} with an unpolarised positron beam, $P_{\mathrm{e}^+_{}}^{} = 0$, and a polarised electron beam, $P_{\mathrm{e}^-_{}}^{}=-80\%$, the contour map at 3~TeV in the same $M_R-|Y_\nu|$ plane is presented in the left-hand side of Fig.~\ref{fig:wwhresults}. Again, the grey area is excluded by the constraints that mostly originate from the global fit~\cite{Fernandez-Martinez:2016lgt}. The process $\mathrm{e}^+ \mathrm{e}^- \rightarrow \mathrm{W}^+ \mathrm{W}^- \mathrm{H}$ exhibits sizeable negative deviations, of at least $-20\%$. Note that the full results can be approximated within $1\%$ for $M_R^{} > 3$~TeV by the simple formulae presented in Ref.~\cite{Baglio:2017fxf}. Compared with the left-hand side of Fig.~\ref{fig:HHHresults}, the coverage of the parameter space is here much larger. optimised cuts can also be chosen to enhance the deviation, such as the cuts $|\eta_{\mathrm{H}/\mathrm{W}^\pm_{}}^{}| < 1$ and $E_\mathrm{H}^{} > 1$~TeV (see the right-hand side of Fig.~\ref{fig:wwhresults} for the $\eta$ distributions), which push the corrections down to $-66\%$ while keeping an ISS cross-section at a reasonable level: $0.14$~fb, as compared with $1.23$~fb before cuts. This has been studied for a benchmark scenario with $|Y_\nu| = 1$ and heavy neutrinos in the range $2.4$--$8.6$~TeV. The results means that this observable has a great potential that needs to be checked in a detailed sensitivity analysis. In the context of the FCC-ee, a similar observable could be chosen to test the effects of heavy neutral leptons in the same mass range, albeit at the one-loop level, namely the production cross-section $\sigma(\mathrm{e}^+\mathrm{e}^-\to \mathrm{ZH})$. 

%%%%
%%%%

\subsection{Conclusions}
This contribution has presented the current status of the triple Higgs coupling measurements at the LHC and the prospects for future lepton colliders. As combined studies in an EFT framework using precision measurements in single Higgs observables, as well as direct Higgs pair production, have shown, lepton colliders are able to completely remove the degeneracy in the measurement of the triple Higgs coupling beyond the $4\sigma$ level, and the combination of data collected at a centre-of-mass energy of 250~GeV with data collected at energies of at least 350~GeV is of crucial importance for very-high-precision measurements in single Higgs physics. Opportunities offered by the Higgs sector to test neutrino mass models at future lepton colliders have also been presented. The FCC-ee is very competitive to test the heavy sterile
neutrino option in the gigaelectronvolt regime. As far as the teraelectronvolt regime for the heavy neutrino scale is concerned, studies reported in the literature have shown that the CLIC and ILC at high energies could offer new avenues in the Higgs sector via precision measurements of the triple Higgs coupling, as well as of the production cross-section of a pair of W bosons in association with a Higgs boson. In the same spirit, the FCC-ee may well offer new opportunities in the same mass regime via precision calculations at one and two loops for the ZH production cross-section, which remain to be studied.

\end{bibunit}

\label{sec-bsm-baglio}
\clearpage \pagestyle{empty} \cleardoublepage
%============================================

\pagestyle{fancy}
\fancyhead[CO]{\thechapter.\thesection \hspace{1mm} Exotic Higgs decays (and long-lived particles) at future colliders}
\fancyhead[RO]{}
\fancyhead[LO]{}
\fancyhead[LE]{}
\fancyhead[CE]{}
\fancyhead[RE]{}
\fancyhead[CE]{J.F. Zurita}
\lfoot[]{}
\cfoot{-  \thepage \hspace*{0.075mm} -}
\rfoot[]{}
%%% definitions %%%%%%%%%%%%%%%%%%%%%%%%%%
\def\be{\begin{equation}}
\def\ee{\end{equation}}
\graphicspath{{BSM_Zurita/plots/}}

\begin{bibunit}[elsarticle-num]
\let\stdthebibliography\thebibliography
\renewcommand{\thebibliography}{%
\let\section\subsection
\stdthebibliography}
    
\section
[Exotic Higgs decays (and long-lived particles) at future colliders  \\{\it J.F. Zurita}]
{Exotic Higgs decays (and long-lived particles) at future colliders
\label{contr:ZURITA}}
\noindent
{\bf Contribution\footnote{This contribution should be cited as:\\
J.F. Zurita, Exotic Higgs decays (and long-lived particles) at future colliders,  
%04 DOI:10.23731/CYRM-2020-XXX.\thepage,\texttt{{CoDEx}}: BSM physics being realised as  SMEFT in:
%04 \url{http://dx.doi.org/10.23731/CYRM-2020-XXX.\thepage}, in:
DOI: \href{http://dx.doi.org/10.23731/CYRM-2020-003.\thepage}{10.23731/CYRM-2020-003.\thepage}, in:
Theory for the FCC-ee, Eds. A. Blondel, J. Gluza, S. Jadach, P. Janot and T. Riemann,\\
CERN Yellow Reports: Monographs, CERN-2020-003,
%04 \url{http://dx.doi.org/10.23731/CYRM-2020-XXX}, p. \thepage.} 
DOI: \href{http://dx.doi.org/10.23731/CYRM-2020-003}{10.23731/CYRM-2020-003},
p. \thepage.
\\ \copyright\space CERN, 2020. Published by CERN under the 
%04-2
\href{http://creativecommons.org/licenses/by/4.0/}{Creative Commons Attribution 4.0 license}.}  by: J.F. Zurita {[jose.zurita@kit.edu]}}
\vspace*{.5cm}

\subsection{Exotic Higgs decays: motivations and signatures}
\label{sec:ehd}

The theoretical motivations and the large breadth of signatures for exotic Higgs decays have been thoroughly reviewed in Ref.~\cite{Curtin:2013fra}. They were first considered as a discovery mode of new physics in the context of a hidden valley scenario~\cite{Strassler:2006im,Strassler:2006ri,Strassler:2006qa}. In the last few years, exotic Higgs decays have been revisited, as they arise ubiquitously in models of neutral naturalness, such as twin Higgs~\cite{Chacko:2005pe}, folded supersymmetry~\cite{Burdman:2006tz}, fraternal twin Higgs~\cite{Craig:2015pha}, hyperbolic Higgs~\cite{Cohen:2018mgv}, and singlet scalar top partners~\cite{Cheng:2018gvu}.

A simple proxy model for hidden valleys is obtained via a Higgs portal set-up, 
\be
{\cal L} \supset \frac{1}{2} \Bigl(\partial_\mu \phi \Bigr)^2 - \frac{1}{2} M^2 \phi^2 - A |H|^2 \phi - \frac{1}{2} \kappa |H|^2 \phi^2 - \frac{1}{6} \mu \phi^3 - \frac{1}{24} \phi^4 - \frac{1}{2} \lambda_\mathrm{H} |H|^4 \, .
\ee
The fields $H$ and $\phi$ mix, depending on $\kappa$ and $A$, giving rise to physical states $h (125)$ and $X(m_\mathrm{X})$. Note that the phenomenology is fully encapsulated by three free parameters: $m_X, c \tau (X) \equiv c \tau$, and $\mathrm{Br}(\mathrm{h} \to \mathrm{X X})$. We will assume that the $\mathrm{h} \to \mathrm{X X}$ is always kinematically open. Existing constraints on the $h(125)$ properties imply that currently the room for an exotic Higgs branching ratio, $\mathrm{Br} (\mathrm{h} \to \mathrm{XX}$) is below about 10\%. Since the mixing controls the X decay widths, a small mixing naturally gives rise to particles that travel a macroscopic distance $c \tau \gtrsim $ mm before decaying. Exotic Higgs decays are then encompassed within the larger class of `long-lived particles' (LLP) signatures.
%(See also talks by A. Thamm and W. Xue at this workshop). 
For concreteness, we review LLPs in the next subsection.

It is worth stressing that the HL-LHC will produce about $10^8$ Higgs bosons, while the CEPC and FCC-ee (240) will only give about $10^6$. Hence, there is a trade-off between the clean environment provided by the collider and the corresponding production cross-section. This already tells us that future electron--positron colliders might probe exotic Higgs branching fractions down to $10^{-5}$, while at the HL-LHC one could, in principle, go down to $10^{-6}$ or even $10^{-7}$, depending on the visibility of the target final state.

\subsection{Long-lived particles (LLPs)}

Long-lived particles are Beyond Standard Model states with macroscopic lifetimes ($\gtrsim$ nanoseconds). These are theoretically well motivated in extensions of the SM trying to solve  fundamental problems of the SM, such as dark matter
or  neutrino masses. A comprehensive overview of the theoretical motivations for LLPs can be found in Ref.~\cite{Curtin:2018mvb}, while a signature-driven document was put forward by the LLP@LHC community in Ref~\cite{Alimena:2019zri}. 

In a nutshell, to obtain a macroscopic lifetime (or a very narrow width), one is led to one of three choices: a large mass hierarchy (\eg muon decay), a compressed spectrum (\eg neutron lifetime), and feeble interactions. The latter is the one that concerns exotic Higgs decays.

In the last few years, there  several proposed detectors have been targeting neutral LLPs, such as MATHUSLA~\cite{Chou:2016lxi}, FASER~\cite{Feng:2017uoz}, CODEX-b ~\cite{Gligorov:2017nwh}, and AL3X~\cite{Gligorov:2018vkc}. Exotic Higgs decays constitute a major theoretical motivation in the design of such experiments, which can probe the difficult phase space regions where the standard triggers and object reconstruction became inefficient. These shortcomings will be detailed in the next subsection.

\subsection{Exotic Higgs decays vis-\`a-vis current LHC data}

\begin{figure}
\centering
\includegraphics[width=0.69\textwidth]{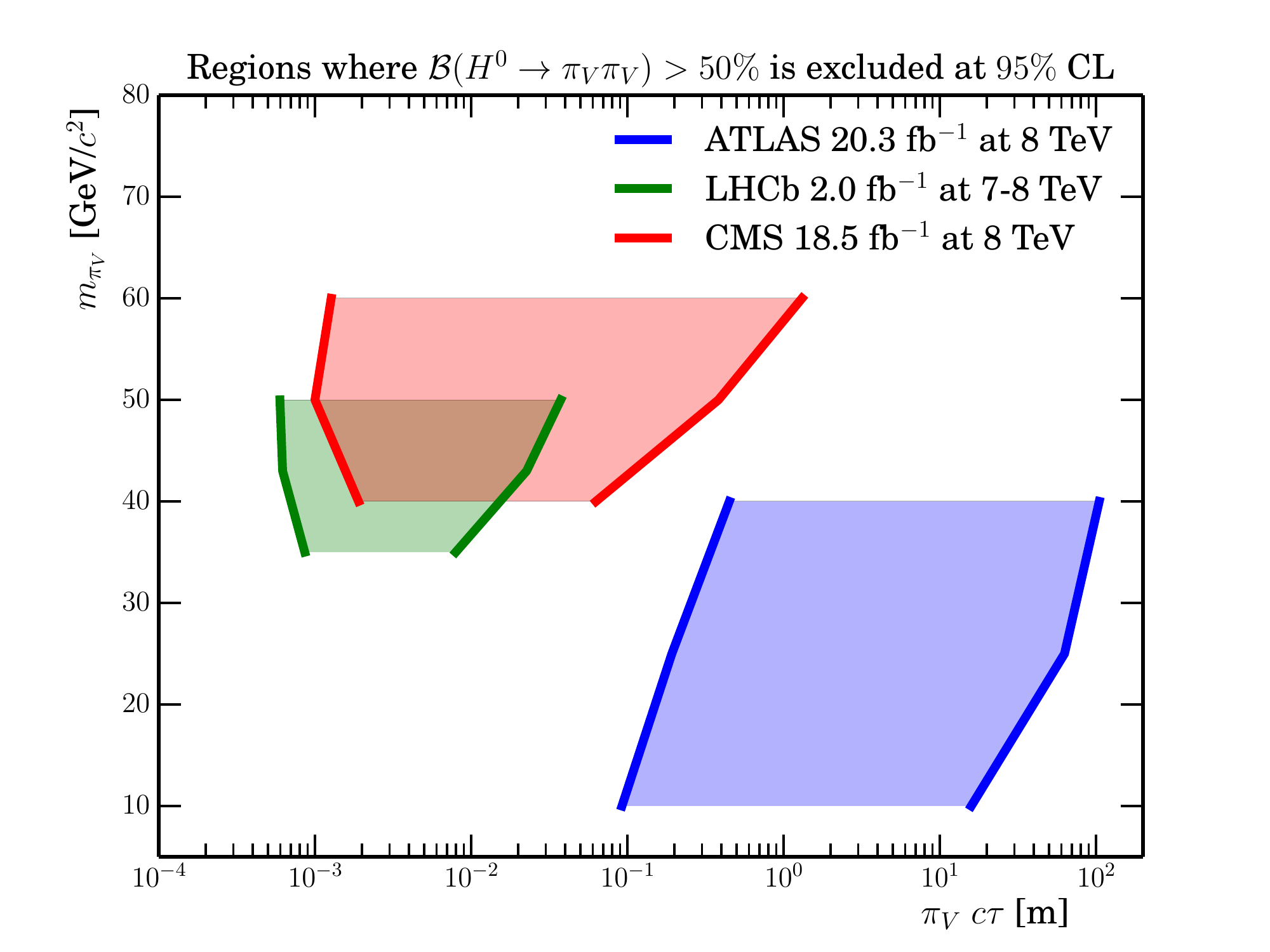}
\caption{Reach of the ATLAS~\cite{Aad:2015rba}, CMS~\cite{CMS:2014wda}, and LHCb~\cite{Aaij:2017mic} studies for $\mathrm{X} \to \mathrm{j j}$, where X is taken to be a dark pion $\uppi_\mathrm{V}$ of the hidden valley scenario. This model is in one-to-one correspondence with that described in Section~\ref{sec:ehd}. The shaded regions show where $\mathrm{Br} (\mathrm{H} \to \mathrm{X X}) > 50\%$ is excluded. Note that the area to the lower left cannot be probed by current searches. Plot taken from the supplementary material of Ref.~\cite{Aaij:2017mic}.}
\label{fig:LHCplot}
%\query{Please correct the figure labels of \Fref{fig:LHCplot}. Set variables in italic font and particle names in roman font.}
\end{figure}

Since, in the simplest scenarios, the X particle decays like a SM Higgs boson of $m_\mathrm{X}$, what occurs is that the predominant decays are into $\mathrm{b} \bar{\mathrm{b}}$ pairs, if the channel is open. In that case, the existing programme of LHC searches for displaced hadronic vertexes (see, \eg Refs.
\cite{Aad:2015rba,CMS:2014wda,Aaij:2017mic}) can cover part of the parameter space. We display the current coverage in the $c \tau$--$m_\mathrm{X}$ plane in \Fref{fig:LHCplot}. We immediately see that the current LHC data are not able to cover the region of short lifetimes ($c \tau \lesssim 10$ cm) and low masses ($m_\mathrm{X} < 35\gev$). Low masses for X imply lower boosts, so the soft jets of the event will not pass the typical $H_\mathrm{T}$ or $p_\mathrm{T}(j)$ trigger thresholds used by ATLAS and CMS.\footnote{It is worth noting that LHCb has the capability to trigger directly on displaced vertexes.} As a sample, the reported trigger efficiency of CMS for $m_\mathrm{X} = 50\gev$ and $c \tau = 30$ mm is about 2\%. The other limitation corresponds to short lifetimes, which is limited by the vertex resolution. Hence, the shortcomings of pp machines can be targeted, instead, with a collider providing better angular resolution, lower $p_\mathrm{T}$ thresholds, and more accurate vertexing, which happens at both $\mathrm{e}^- \mathrm{p}$ and $\mathrm{e}^+ \mathrm{e}^-$ machines. We stress that additional data will not alter this picture, and the low $c \tau$ and low $m_\mathrm{X}$ region would continue to be extremely hard to probe.

%%%%%%%%%%%% 
%%%%%%%%

\subsection{Future experiments: HL-LHC, FCC, CEPC, LHeC}

% 3 papers: study at HL-LHC~\cite{Curtin:2015fna}, at e-p collider~\cite{Curtin:2017bxr} and a very recent work on electron-positron machines~\cite{Alipour-Fard:2018lsf}.

\subsubsection{Proton--proton colliders}

\begin{figure}
\begin{center}
\hspace*{-1.7cm}
\begin{tabular}{c}
\includegraphics[width=17cm]{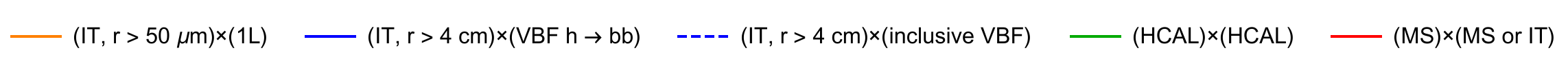}
\\
\begin{tabular}{cc}
\includegraphics[width=7.5cm]{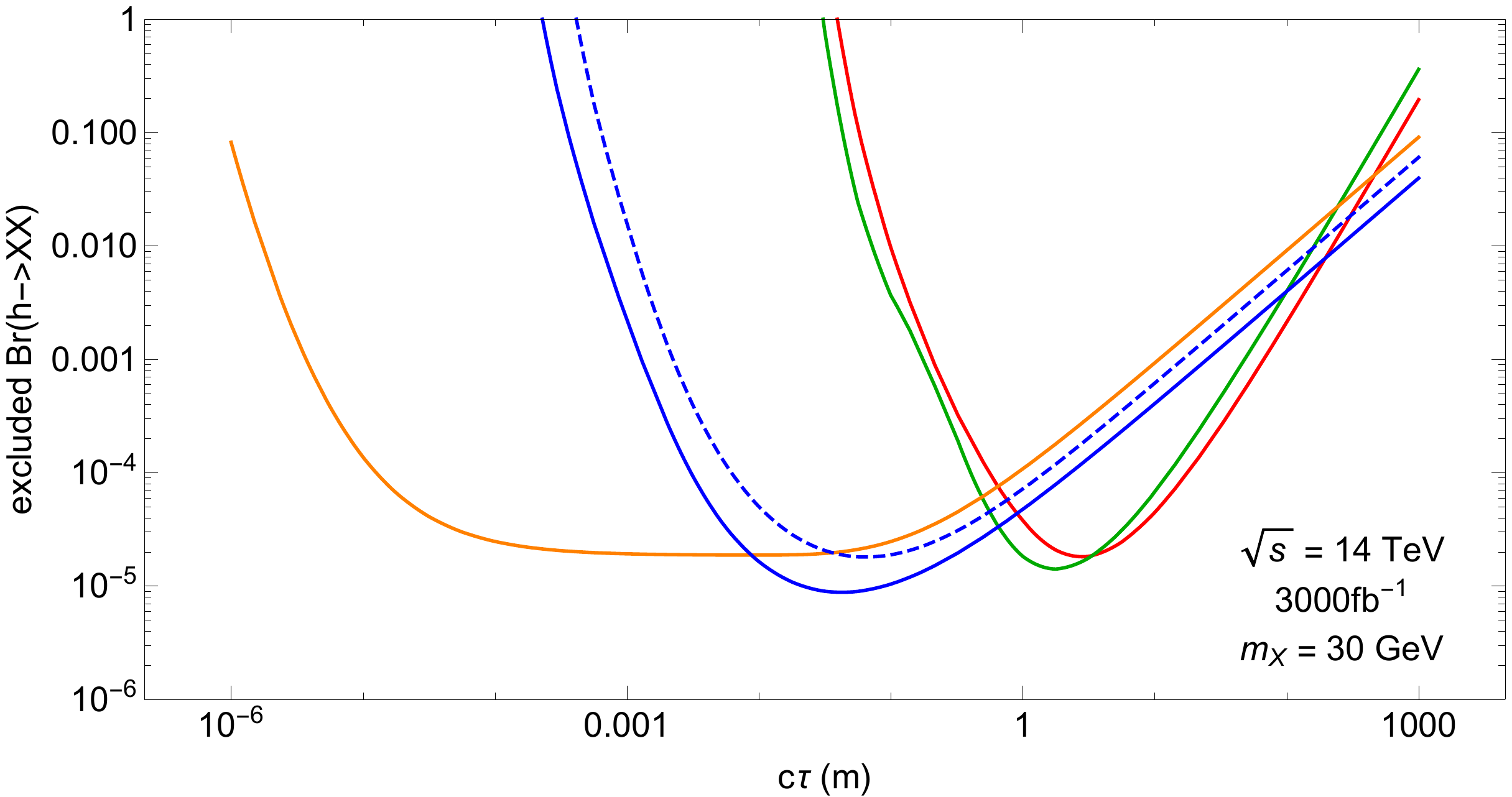}
 &
\includegraphics[width=7.5cm]{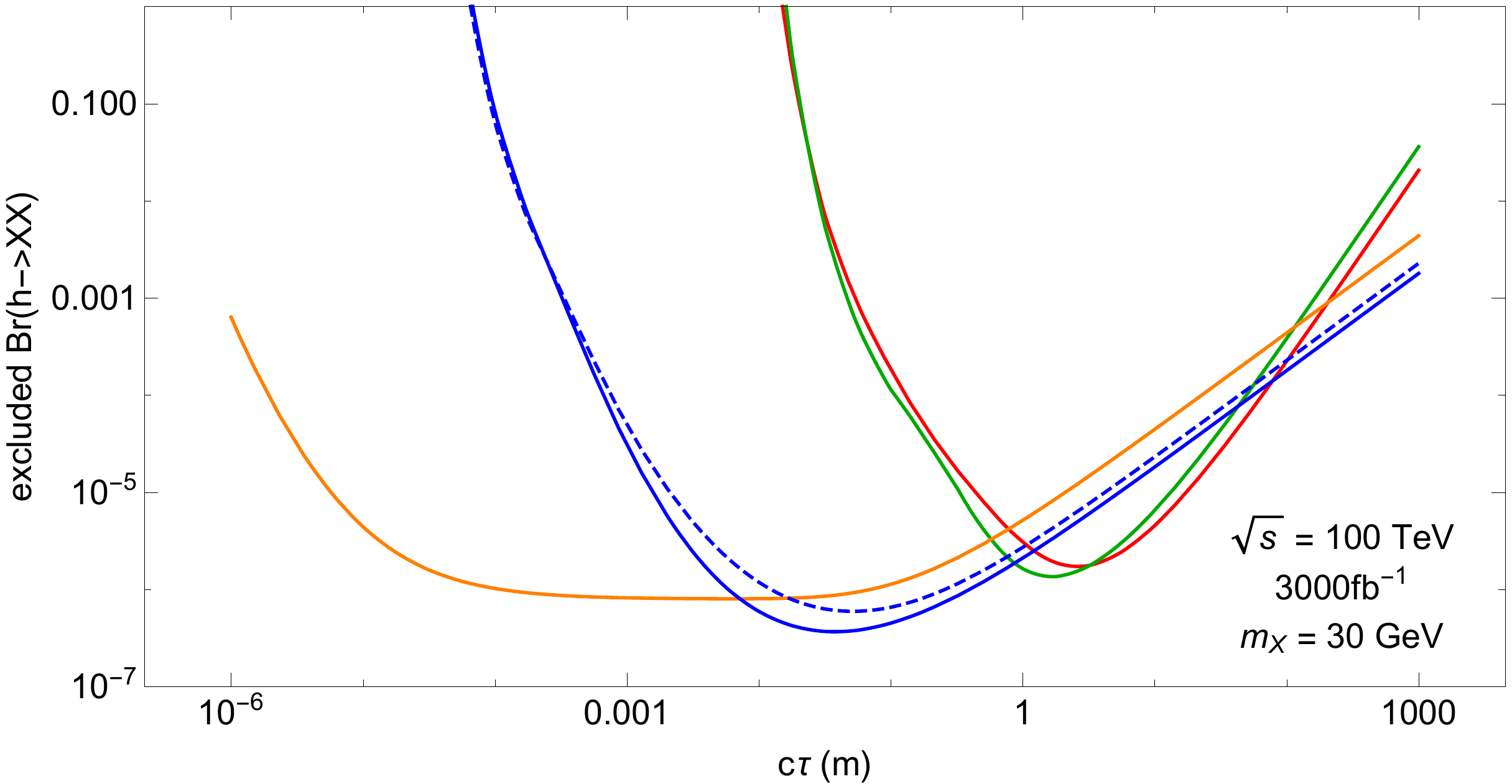}
\end{tabular}
\end{tabular}
\end{center}
\caption{
Sensitivities of the displaced searches for exotic Higgs decays at the HL-LHC (left) and FCC-hh (right), in the $c \tau$--$\mathrm{Br}(\mathrm{h} \to \mathrm{XX})$ plane, for $m_\mathrm{X}=30 \gev$. The curves  correspond to the use of different triggers and different assumptions about the reconstruction of the displaced vertexes. Plot taken from Ref~\cite{Curtin:2015fna}.}
\label{fig:pplimits}
%\query{Please correct the figure labels of \Fref{fig:pplimits}. Set variables in italic font and particle names in roman font.}
\end{figure}

We show in \Fref{fig:pplimits} (taken from Ref. \cite{Curtin:2015fna}) the expected reach at the HL-LHC ($\sqrt{s} = 14$\,TeV and total integrated luminosity of 3\,ab$^{-1}$) and at the FCC-hh ($\sqrt{s} = 100 $\,TeV and total integrated luminosity of 3\,ab$^{-1}$) for a scalar mass of $m_\mathrm{X}= 30$\,GeV. The curves indicate different choices of trigger and of reconstruction capabilities of the displaced vertex. In particular, the orange curve corresponds to one displaced vertex in the inner tracker with an impact parameter of $50\Uum$, which poses an interesting experimental challenge and thus should be regarded as an optimistic case. The blue curve corresponds to the realistic case of using VBF, $\mathrm{h} \to \mathrm{b} \bar{\mathrm{b}}$ triggers down to an impact parameter of 4\,cm.

We see that one can cover lifetimes as short as a  millimetre (or even one micrometre for the optimistic scenario), while the probed exotic branching ratios can reach down to $10^{-5}~(10^{-6})$ for the HL-LHC (FCC-hh), for the benchmark case of $m_\mathrm{X} = 30$\,GeV.  As discussed before, lower masses would suffer from a poor trigger efficiency, which opens a window of opportunity for both electron--proton and electron--positron colliders.

\subsubsection{Electron--proton colliders}
\begin{figure}
\centering
\includegraphics[width=0.79\textwidth]{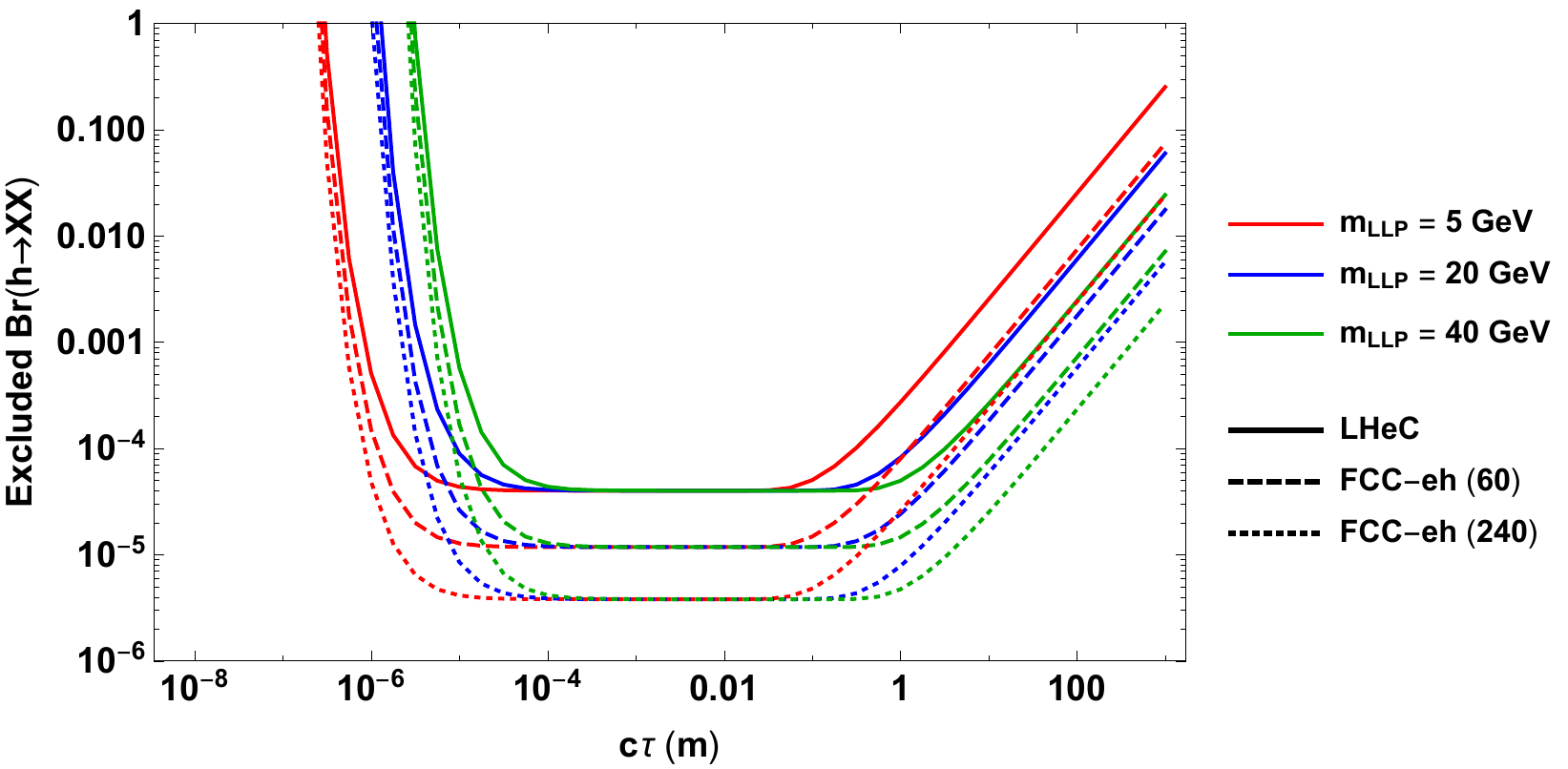}
\caption{Reach of the future electron--proton colliders: LHeC (solid), FCC-eh (60), and FCC-eh (240). The LHeC would collide a 7\,TeV proton from the LHC against a 50\,GeV electron beam, while for the FCC-eh  a 50\,TeV proton beam will collide against a 60\,GeV (design case) or 240\,GeV beam (optimistic scenario). Taken from Refs.~\cite{Curtin:2017bxr,Curtin:2018qot}.}
\label{fig:EHDep}
%\query{Please correct the figure labels of \Fref{fig:EHDep}. Set variables
%in italic font and particle names in roman font.}
\end{figure}

The reach on exotic Higgs decays for future electron--proton colliders is displayed in \Fref{fig:EHDep}. We see that the electron--proton colliders, owing to their better resolution, can test masses down to  $5\gev$ for exotic branching fractions of about $10^{-4}$. This mass range is almost impossible to probe at the LHC, because of the overwhelming multijet background. We also note that electron--proton colliders provide a smaller luminosity.\footnote{During a 25 year run period of the Future Circular Collider (FCC), the proton--proton incarnation (FCC-hh) is expected to collect 15--30\,ab$^{-1}$ while the electron--proton version will collect only 1\,$ab^{-1}$~\cite{Mangano:2651294}.} Hence, electron--proton colliders provide a window of opportunity to overcome the gaps in coverage discussed for  proton--proton colliders.
 %%%%
%%%%

\subsubsection{Electron--positron colliders}
\begin{figure}
  \centering
    \includegraphics[width=\textwidth]{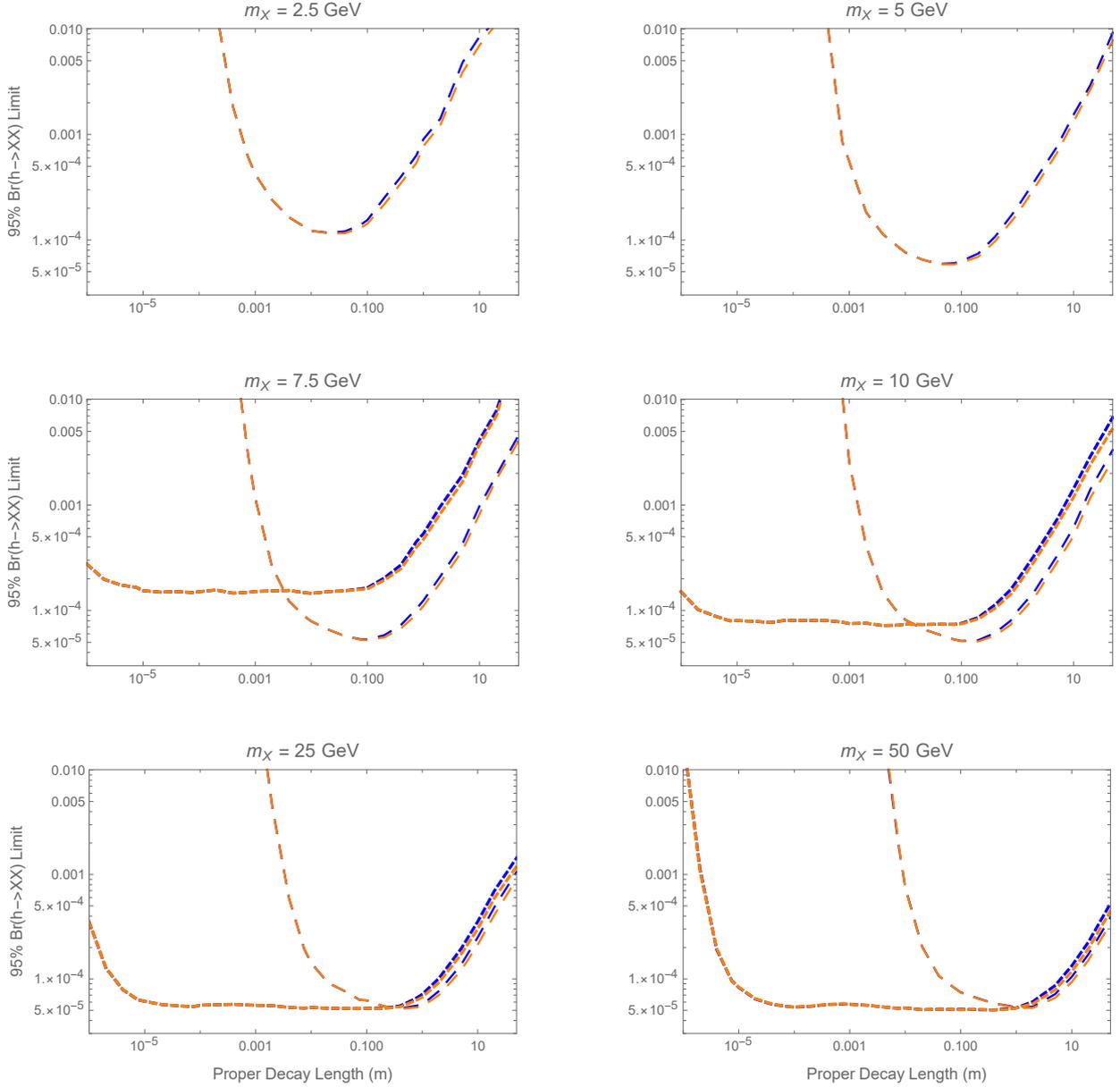}
  \caption{FCC-ee (blue) and CEPC (orange) limits on the exotic branching ratio $\mathrm{h} \to \mathrm{XX}$ at the 95\% CL. The `long-lifetime' analysis is shown with larger dashes, while smaller dashes correspond to the `large-mass' study. Taken from Ref.~\cite{Alipour-Fard:2018lsf}.}
\label{fig:EHDee}
%\query{Please correct the figure labels of \Fref{fig:EHDee}. Set variables
%in italic font and particle names in roman font.}
\end{figure}

Finally, we take a look at the $\mathrm{e}^+$--$\mathrm{e}^-$ case. A detailed analysis is reported in Ref~\cite{Alipour-Fard:2018lsf}; here, we briefly summarise the most salient points. This study considers the Higgs-strahlung process $\mathrm{e}^+ \mathrm{e}^- \to \mathrm{h Z}$ with leptonic decays of the Z boson for both the FCC-ee~\cite{Gomez-Ceballos:2013zzn} and the CEPC~\cite{CEPCStudyGroup:2018ghi,CEPCStudyGroup:2018rmc}. A set of basic selection cuts allows us to achieve a zero-background regime for the irreducible SM processes.\footnote{Backgrounds from particles originating away from the interaction point (\eg beam halo, cosmic muons, cavern radiation) are not considered.} Two different strategies are pursued: the large mass and the long-lifetime regime. The main difference between the two is in the requirements on the minimal distance between the displaced vertexes. The results are shown in \Fref{fig:EHDee}.

%%%%
%%%%

One immediately sees that the $\mathrm{e}^+ \mathrm{e}^-$ colliders can test exotic branching fractions down to $5 \times 10^{-5}$. Moreover,  they can go low in mass, down to a few gigaelectronvolts, and they can also probe decay lengths down to micrometres, where the proton--proton colliders would be ineffective. 

\subsection{Conclusions}
In this contribution, I have summarised the existing studies on exotic Higgs decays at current and future colliders. While the proton--proton machines would, in principle, be the best option, owing to their larger energies and luminosities, we have also seen that the phase space regions where the LHC and FCC-hh lose steam, namely, low X masses and short lifetimes provide a unique window of opportunity for both $\mathrm{e}^- \mathrm{p}$ and $\mathrm{e}^+ \mathrm{e}^-$ colliders. The latter two types of machine have only recently
been  studied, and thus there is naturally much room for improvement. It
should also be stressed that these kinds of study can help to optimise the detector design of future colliders.

\end{bibunit}

\label{sec-bsm-zurita}  
\clearpage \pagestyle{empty} \cleardoublepage
%============================================

\pagestyle{fancy}
\fancyhead[CO]{\thechapter.\thesection \hspace{1mm} Precision predictions for Higgs decays in the (N)MSSM}
\fancyhead[RO]{}
\fancyhead[LO]{}
\fancyhead[LE]{}
\fancyhead[CE]{}
\fancyhead[RE]{}
\fancyhead[CE]{F. Domingo, S. Heinemeyer, S. Pa\ss ehr, G. Weiglein}
\lfoot[]{}
\cfoot{-  \thepage \hspace*{0.075mm} -}
\rfoot[]{}
%%% definitions %%%%%%%%%%%%%%%%%%%%%%%%%%
\def\be{\begin{equation}}
\def\ee{\end{equation}}
\graphicspath{{BSM_Heinemeyer/plots/}}

\begin{bibunit}[elsarticle-num] % define the bib-style for the unit: elsarticle-num.bst
%  text-1; this is the corresponding section
%\putbib[2loops] % the *.bib
%\end{bibunit}
% go-on
%--- from: bibunits.sty, adapts the font size of ``References'' to section
\let\stdthebibliography\thebibliography
\renewcommand{\thebibliography}{%
\let\section\subsection
\stdthebibliography}
%---

{

\section
[Precision predictions for Higgs decays in the (N)MSSM \\{\it F. Domingo, S. Heinemeyer, S. Pa\ss ehr, G. Weiglein}]
{Precision predictions for Higgs decays in the (N)MSSM
\label{contr:HEINEMEYER}}
\noindent
{\bf Contribution\footnote{This contribution should be cited as:\\
F. Domingo, S. Heinemeyer, S. Pa\ss ehr, G. Weiglein, Precision predictions for Higgs decays in the (N)MSSM,  
%04 DOI:10.23731/CYRM-2020-XXX.\thepage,\texttt{{CoDEx}}: BSM physics being realised as  SMEFT in:
%04 \url{http://dx.doi.org/10.23731/CYRM-2020-XXX.\thepage}, in:
DOI: \href{http://dx.doi.org/10.23731/CYRM-2020-003.\thepage}{10.23731/CYRM-2020-003.\thepage}, in:
Theory for the FCC-ee, Eds. A. Blondel, J. Gluza, S. Jadach, P. Janot and T. Riemann,\\
CERN Yellow Reports: Monographs, CERN-2020-003,
%04 \url{http://dx.doi.org/10.23731/CYRM-2020-XXX}, p. \thepage.} 
DOI: \href{http://dx.doi.org/10.23731/CYRM-2020-003}{10.23731/CYRM-2020-003},
p. \thepage.
\\ \copyright\space CERN, 2020. Published by CERN under the 
%04-2
\href{http://creativecommons.org/licenses/by/4.0/}{Creative Commons Attribution 4.0 license}.} by: F. Domingo, S. Heinemeyer, S. Pa\ss ehr, G. Weiglein\\
Corresponding author: S. Heinemeyer {[Sven.Heinemeyer@cern.ch]}}
\vspace*{.5cm}

\subsection{Introduction}
\label{heinemeyer-sec:intro}

The signal that was discovered in the Higgs searches at ATLAS and~CMS
at a mass of $\sim$$125$\,GeV\,\cite{Aad:2012tfa,Chatrchyan:2012xdj,Khachatryan:2016vau}
is, within current theoretical and experimental uncertainties,
compatible with the properties of the Higgs boson predicted within
the Standard Model~(SM) of particle physics. No conclusive signs of
physics beyond the~SM have been reported so far. However, the
measurements of Higgs signal strengths for the various channels leave
considerable room for Beyond Standard Model~(BSM)
interpretations. Consequently, the investigation of the precise
properties of the discovered Higgs boson will be one of the prime
goals at the~LHC and beyond.  While the mass of the observed particle
is already known with excellent
accuracy\,\cite{Aad:2015zhl,Sirunyan:2017exp}, significant
improvements of the information about the couplings of the observed
state are expected from the upcoming runs of
the~LHC\,\cite{CMS-HL,ATLAS-HL,Khachatryan:2016vau,Testa:2017,Cepeda:2017}
and even more so from the high-precision measurements at a future
$\mathrm{e}^+\mathrm{e}^-$~collider\,\cite{Baer:2013cma,Fujii:2015jha,Fujii:2017ekh,Moortgat-Picka:2015yla,Gomez-Ceballos:2013zzn,An:2018dwb,Abada:2019zxq,Abada:2019lih,FCC-CDR}.
For the accurate study of the properties of the Higgs boson, precise
predictions for the various partial decay widths, the branching ratios
(BRs), and the Higgs boson production cross-sections, 
along with their theoretical uncertainties, are indispensable.

Motivated by the `hierarchy problem', supersymmetry (SUSY) inspired
extensions of the SM play a prominent role in the investigations of
possible new physics. As such, the minimal supersymmetric Standard
Model (MSSM) \cite{Nilles:1983ge,Haber:1984rc} or its singlet
extension,
the~next-to-MSSM~(NMSSM) \cite{Ellwanger:2009dp,Maniatis:2009re},
have been the object of many studies in the last decades. Despite this
attention, these models are not yet prepared for an era of precision
tests, as the uncertainties at the level of the Higgs mass
calculation\,\cite{Degrassi:2002fi,Staub:2015aea,Drechsel:2016htw} are
about one order of magnitude larger than the experimental
uncertainty. At the level of the decays, the theoretical uncertainty
arising from unknown higher-order corrections has been estimated for
the case of the Higgs boson of the~SM (where the Higgs mass is treated
as a free input parameter) in
Refs.\,\cite{Denner:2011mq,Heinemeyer:2013tqa} and updated in
Ref.\,\cite{deFlorian:2016spz}: depending on the channel and the Higgs
mass, it typically falls in the range of $\sim$$0.5$--$5\%$. To our
knowledge, no similar analysis has been performed in SUSY-inspired
models (or other BSM models),
but one can expect the uncertainties from missing higher-order
corrections to be larger in general---with many nuances, depending on
the characteristics of the Higgs state and the considered point in
parameter space: we provide some discussion of this issue at the end
of this section. In addition, parametric uncertainties that are induced
by the experimental errors of the input parameters should be taken
into account. For the case of the~SM decays, those parametric
uncertainties have been discussed in the references cited. In
the~SUSY~case, the parametric uncertainties induced by the (known)~SM
input parameters can be determined in the same way as for the~SM,
while the dependence on unknown~SUSY parameters can be utilised in
setting constraints on those parameters. While still competitive
today, the level of accuracy of the theoretical predictions of
Higgs boson decays in~SUSY~models should soon become outclassed by the
achieved experimental precision (in particular at future $\mathrm{e}^+\mathrm{e}^-$ colliders)
on the decays of the observed Higgs
signal. Without comparable accuracy of the theoretical predictions,
the impact of the exploitation of the precision data will be
diminished---either in terms of further constraining the parameter
space or of interpreting deviations from the~SM results. Further
efforts towards improving the theoretical accuracy are therefore
necessary in order to enable a thorough investigation of the
phenomenology of these models. Besides the decays of the~SM-like state
at 125\,GeV of a~SUSY~model---where the goal is clearly to reach an
accuracy that is comparable to the case of the~SM---it is also of
interest to obtain reliable and accurate predictions for the decays of
the other Higgs bosons in the spectrum. The decays of the non-SM-like
Higgs bosons can be affected by large higher-order corrections as a
consequence of either large enhancement factors or a suppression of
the lowest-order contribution. Confronting accurate predictions with
the available search limits yields important constraints on the
parameter space.
Here, we review the evaluation of the decays of the neutral
Higgs bosons of the $\mathbb{Z}_3$-conserving~NMSSM into
SM~particles, as presented in Ref.~\cite{Domingo:2018uim}.

Current work focusing on NMSSM Higgs decays is part of the effort
for developing a version
of \texttt{FeynHiggs} \cite{Heinemeyer:1998np,Heinemeyer:1998yj,
  Degrassi:2002fi, Frank:2006yh, Hahn:2013ria, Bahl:2016brp,
  Bahl:2017aev, Bahl:2018qog, FH-www} dedicated to
the NMSSM \cite{Drechsel:2016jdg,Domingo:2017rhb}. The general
methodology relies on a Feynman-diagrammatic calculation of radiative
corrections, which
employs \texttt{FeynArts} \cite{Kublbeck:1990xc,Hahn:2000kx},
\texttt{FormCalc} \cite{Hahn:1998yk},
and \texttt{LoopTools} ,\cite{Hahn:1998yk}. The 
renormalization scheme has been implemented within the NMSSM \cite{Domingo:2017rhb}  in such a way that the result in the MSSM limit of the NMSSM
exactly coincides with the~MSSM result obtained from
\texttt{FeynHiggs} without any further adjustments of parameters.

%%%%%%%%%%%%%%%%%%%%%%%%%%%%%%%%%%%%%%%%%%%%%%%%%%%%%%%%%%%%%%%%%%%%%%%%%%%%%%
%%%%%%%%%%%%%%%%%%%%%%%%%%%%%%%%%%%%%%%%%%%%%%%%%%%%%%%%%%%%%%%%%%%%%%%%%%%%%%

\subsection{Higgs decays to SM particles in the \cp-violating NMSSM}
\label{heinemeyer-sec:theory}

In this section, we review the technical aspects of our calculation
of the Higgs decays. Our notation and the renormalization scheme that
we employ for the~$\mathbb{Z}_3$-conserving~NMSSM in the general case
of complex parameters are presented in
Section 2 of Ref. 
\cite{Domingo:2017rhb}, and we refer the reader to that article for
further details.

\subsubsection{Decay amplitudes for a physical (on-shell) Higgs state---generalities}
\label{heinemeyer-sec:general}

\paragraph{On-shell external Higgs leg}

In this section, we consider the decays of a physical Higgs state,
\IE\ an eigenstate of the inverse propagator matrix for the Higgs
fields, evaluated at the corresponding pole eigenvalue. The connection
between such a physical state and the tree-level Higgs~fields entering
the Feynman diagrams is non-trivial in general since the higher-order
contributions induce mixing among the Higgs states and between the
Higgs states and the gauge bosons (as well as the associated Goldstone
bosons). The LSZ reduction fully determines the (non-unitary)
transition matrix $\mathbf{Z}^{\mbox{\tiny mix}}$ between the
loop-corrected mass eigenstates and the lowest-order states. Then, the
amplitude describing the decay of the physical state $h_i^{\mbox{\tiny
    phys}}$ (we shall omit the superscript `phys' later on), into
\EG~a fermion pair $\mathrm{f\bar{f}}$, relates to the amplitudes in terms of
the tree-level states $h_j^0$ according to (see the following for the mixing
with gauge bosons and Goldstone bosons):
\begin{align}
  \label{eq:RelationPhysAmplitude}
  \Amp{}{}{h_i^{\mbox{\tiny  phys}}\to \mathrm{f\bar{f}}\,}&=
  Z^{\mbox{\tiny mix}}_{ij}\,\Amp{}{}{h_j^0\to \mathrm{f\bar{f}}\,}\,.
\end{align}
Here, we characterize the physical Higgs states according to the
procedure outlined in Ref. \cite{Domingo:2017rhb} (see also Refs.
\cite{Frank:2006yh,Williams:2011bu,Fuchs:2016swt}).\begin{enumerate}
\item The Higgs self-energies include full one-loop and
  leading $\mathcal{O}{(\alpha_\mathrm{t}\alpha_\mathrm{s},\alpha_\mathrm{t}^2)}$ two-loop
  corrections (with two-loop effects obtained in
  the MSSM approximation via the publicly available
  code \texttt{Feyn\-Higgs}\footnote{The Higgs masses
  in \texttt{FeynHiggs} could be computed with additional improvements,
  such as additional fixed-order
  results \cite{Passehr:2017ufr,Borowka:2018anu} or the resummation
  of large logarithms for very heavy SUSY
  particles \cite{Hahn:2013ria,Bahl:2016brp,Bahl:2017aev}; for
  simplicity, we do not take such refinements into account in this section.}).
\item The pole masses correspond to the zeros of the determinant of
  the inverse propagator matrix.
\item The~$(5 \times 5)$~matrix $\mathbf{Z}^{\mbox{\tiny mix}}$ is
  obtained in terms of the solutions of the eigenvector equation for
  the effective mass matrix evaluated at the poles, and satisfying the
  appropriate normalization conditions (see
  Section \,\href{https://arxiv.org/pdf/1706.00437.pdf#subsection.2.6}{$2.6$}
  of Ref., \cite{Domingo:2017rhb}).
\end{enumerate}
In correcting the external Higgs legs by the full
matrix~$\mathbf{Z}^{\mbox{\tiny mix}}$---instead of employing a simple
diagrammatic expansion---we resum contributions to the transition
amplitudes that are formally of higher-loop order. This resummation is
convenient for taking into account numerically relevant leading
higher-order contributions. It can, in fact, be crucial for the frequent
case where radiative corrections mix states that are almost
mass-degenerate in order to properly describe the resonance-type
effects that are induced by the mixing. Conversely, care needs
to be taken to avoid the occurrence of non-decoupling terms when Higgs
states are well-separated in mass, since higher-order effects can
spoil the order-by-order cancellations with vertex corrections.

We stress that all public tools, with the exception of \FH, neglect
the full effect of the transition to the physical Higgs states encoded
within~$\mathbf{Z}^{\mbox{\tiny mix}}$, and instead employ the unitary
approximation~$\mathbf{U}^0$ neglecting external momenta (which is in
accordance with leading-order or QCD-improved leading-order
predictions). We refer the reader to Refs.
\cite{Frank:2006yh,Fuchs:2016swt,Domingo:2017rhb} for the details
of the definition of~$\mathbf{U}^0$ or~$\mathbf{U}^m$ (another unitary
approximation), as well as a discussion of their impact at the level of
Higgs decay widths.

\paragraph{Higgs--electroweak mixing}

For the mass~determination, we do not take into account contributions
arising from the mixing of the Higgs fields with the neutral Goldstone
or Z bosons, since these corrections enter at the subdominant
two-loop level (contributions of this kind can also be compensated by
appropriate field-renormalization
conditions \cite{Hollik:2002mv}). We note that, in the \cp-conserving
case, only external \cp-odd Higgs components are affected by such a
mixing. Yet, at the level of the decay amplitudes, the Higgs mixing
with the Goldstone and Z bosons already enters  at the one-loop order
(even if the corresponding self-energies are cancelled by an
appropriate field-renormalization condition, this procedure will
still provide a contribution to
the $h_i\mathrm{f\bar{f}}$ counterterm). Therefore, for a complete one-loop
result of the decay amplitudes, it is, in general, necessary to
incorporate Higgs--Goldstone and Higgs--Z self-energy transition
diagrams\,\cite{Williams:2007dc,Fowler:2009ay,Williams:2011bu}. In the
following, we evaluate such contributions to the decay amplitudes in
the usual diagrammatic fashion (as prescribed by the~LSZ~reduction),
with the help of the \texttt{FeynArts} model file for
the~\cp-violating~NMSSM\,\cite{Domingo:2017rhb}. The corresponding
one-loop amplitudes (including the associated counterterms) will be
symbolically denoted as~$\AmpE{1L}{\mathrm{G/Z}}{}$. These amplitudes can be
written in terms of the self-energies $\Sigma_{h_i\mathrm{G/Z}}$ with Higgs and
Goldstone or Z bosons in the external legs. In turn, these
self-energies are connected by a Slavnov--Taylor identity (see
\EG~Appendix \href{https://arxiv.org/pdf/0807.4668.pdf#page=29}{A} of Ref.
\cite{Baro:2008bg}):\footnote{We denote the imaginary unit by $\I$.}
\begin{subequations}
\begin{align}\label{eq:hG-hZ}
  \begin{split}
    0 &=
    M_\mathrm{Z}\,\Sigma_{h_i\mathrm{G}}{\left(p^2\right)} + \I\,p^2\,\Sigma_{h_i\mathrm{Z}}{\left(p^2\right)}
    + M_\mathrm{Z}\left(p^2-m_{h_i}^2\right)f{\left(p^2\right)}\\
    &\quad -\frac{e}{2\,s_{\text{w}}\,c_{\text{w}}}\sum_j\left[
      \Un{i1}\Un{j4}-\Un{i2}\Un{j5}-\Un{j1}\Un{i4}+\Un{j2}\Un{i5}
      \right]T_{h_j}\,,
  \end{split}\\
  \begin{split}
    f{\left(p^2\right)} &\equiv
    \begin{aligned}[t]
      -\frac{\alpha}{16\,\uppi\,s_{\text{w}}\,c_{\text{w}}}\sum_j &
      \left[\Un{i1}\Un{j4}-\Un{i2}\Un{j5}-\Un{j1}\Un{i4}+\Un{j2}\Un{i5}\right]\\
      &\times\left[c_\beta\,\Un{j1} + s_\beta\,\Un{j2}\right]\Bnull{p^2,m_{h_j}^2,M_Z^2},
    \end{aligned}
  \end{split}
\end{align}
\end{subequations}
where the~$T_{h_i}$ correspond to the tadpole terms of the Higgs
potential and~$\Un{ij}$ are the elements of the transition matrix
between the gauge- and tree-level mass-eigenstate bases of the Higgs
bosons---the notation is introduced in
Section\,\href{https://arxiv.org/pdf/1706.00437.pdf#subsection.2.1}{2.1}
of Ref. \cite{Domingo:2017rhb}. Similar relations in the~MSSM are also
provided in
Eq. (\href{https://arxiv.org/pdf/1103.1335.pdf#page=69}{127}) of Ref.
\cite{Williams:2011bu}. We checked this identity at the numerical
level.

\paragraph{Inclusion of one-loop contributions}

The wave function normalization factors contained
in $\mathbf{Z}^{\mbox{\tiny mix}}$, together with the described
treatment of the mixing with the Goldstone and Z bosons, ensure the
correct on-shell properties of the external Higgs leg in the decay
amplitude, so that no further diagrams correcting this external leg
are needed. Moreover, the SM~fermions and gauge bosons are also
treated as on-shell particles in our renormalization scheme. Beyond
the transition to the loop-corrected states incorporated
by~$\mathbf{Z}^{\mbox{\tiny mix}}$, we thus compute the decay
amplitudes at the one-loop order as the sum of the tree-level
contribution~$\AmpE{tree}{}$ (possibly equal to zero), the
Higgs--electroweak one-loop mixing $\AmpE{1L}{\mathrm{G/Z}}{}$ and the
(renormalised) one-loop vertex corrections~$\AmpE{1L}{{\rm vert}}{}$
(including counterterm contributions)---we note that each of these
pieces of the full amplitude is separately ultraviolet-finite. In the
example of the $\mathrm{f\bar{f}}$ decay, the amplitudes with a tree-level
external Higgs field~$h_j^0$---on the right-hand side of
\Eref{eq:RelationPhysAmplitude}---thus symbolically read
\begin{align}
  \label{eq:1LAmplitude}
  \Amp{}{}{h_j^0\to \mathrm{f\bar{f}}\,} &=
  \Amp{tree}{}{h_j^0\to \mathrm{f\bar{f}}\,}+
  \Amp{1L}{\mathrm{G/Z}}{h_j^0\to \mathrm{f\bar{f}}\,}+
  \Amp{1L}{{\rm vert}}{h_j^0\to \mathrm{f\bar{f}}\,}\,.
\end{align}
All the pieces on the right-hand side of this equation are computed
with the help of
\texttt{FeynArts} \cite{Kublbeck:1990xc,Hahn:2000kx},
\texttt{FormCalc} \cite{Hahn:1998yk}, and
\texttt{LoopTools} \cite{Hahn:1998yk}, according to the prescriptions
that are encoded in the model file for
the \cp-violating~NMSSM. However, we use a specific treatment for some
of the contributions, such as~QED and~QCD~one-loop corrections to
Higgs decays into final-state particles that are electrically or
colour~charged, or include certain higher-order corrections. We
describe these channel-specific modifications in the following
subsections.

\paragraph{Goldstone-boson couplings}

The cubic Higgs--Goldstone-boson vertices can be expressed as
\begin{multline}\label{eq:treelevelG}
  \mathcal{L}\ni-\frac{1}{\sqrt{2}\,v}\left\{
  \sum_j m^2_{h_j}\left[\cos\beta\,\Un{j1}+\sin\beta\,\Un{j2}\right]
  h_j^0\left[G^+G^-+\tfrac{1}{2}\left(G^{0}\right)^2\right]\right.\\
 \shoveright {} + \Big[\sum_j\left(m^2_{\mathrm{H}^{\pm}}-m^2_{h_j}\right)
    \left(\sin\beta\left[\Un{j1}+\I\,\Un{j4}\right]
    -\cos\beta\left[\Un{j2}-\I\,\Un{j5}\right]\right)
    h_j^0H^+G^-+\text{h.\,c.}\Big]\\
  \left.+\frac{1}{2}\sum_{j,\,k}\left(m^2_{h_k}-m^2_{h_j}\right)
  \left[\Un{j1}\Un{k4}-\Un{j2}\Un{k5}-\left(j\leftrightarrow k\right)\right]
  h_j^0h_k^0G^0\right\}.
\end{multline}
The doublet vacuum expectation
value~(VEV),~\mbox{$v=\,M_\mathrm{W}\,s_{\text{w}}/\sqrt{2\,\uppi\,\alpha}$}, is
expressed in terms of the gauge-boson masses $M_\mathrm{W}$ and $M_\mathrm{Z}$
\mbox{$\big(s_{\text{w}}=\sqrt{1-M_\mathrm{W}^2/M_\mathrm{Z}^2}\big)$}, as well as the
electromagnetic coupling $\alpha$. The symbol $m_{h_j}^2$,
($j=1,\ldots,5$), represents the tree-level mass squared of the
neutral Higgs state $h_j^0$, and $m^2_{\mathrm{H}^{\pm}}$ represents the mass squared of
the charged Higgs state.

The use of the tree-level couplings of \Eref{eq:treelevelG}, together
with a physical (loop-corrected) external Higgs
leg~\mbox{$h_i=\sum_jZ^{\mbox{\tiny mix}}_{ij}\,h^0_j$}, is potentially
problematic regarding the gauge properties of the matrix elements. The
structure of the gauge theory and its renormalization indeed guarantee
that the gauge identities are observed at the order of the calculation
(one loop). However, the evaluation of Feynman amplitudes is not
protected against a violation of the gauge identities at the
(incomplete) two-loop order. We detected such gauge-violating effects
of two-loop order at several points in our calculation of the
neutral Higgs decays.
\begin{enumerate}
\item The Ward identity in \mbox{$h_i\to\upgamma\upgamma$} is not
  satisfied (see also Ref.\,\cite{Benbrik:2012rm}).
\item Infrared (IR) divergences of the virtual corrections
  in \mbox{$h_i\to \mathrm{W}^+\mathrm{W}^-$} do not cancel their counterparts in the
  bremsstrahlung process \mbox{$h_i\to \mathrm{W}^+\mathrm{W}^-\upgamma$} (see also
  Ref.\,\cite{Gonzalez:2012mq}).
\item Computing \mbox{$h_i\to \mathrm{f\bar{f}}$} in an $R_{\xi}$~gauge entails
  non-vanishing dependence of the amplitudes on the electroweak
  gauge-fixing parameters $\xi_\mathrm{Z}$ and $\xi_\mathrm{W}$.
\end{enumerate}
As these gauge-breaking effects could intervene with sizeable and
uncontrolled numerical impact, it is desirable to add two-loop order
terms, restoring the gauge identities at the level of the matrix
elements. Technically, there are different possible procedures to
achieve this: one would amount to replacing the kinematic Higgs masses
that appear in Higgs--gauge-boson couplings with tree-level Higgs
masses; we prefer the alternative procedure, which involves changing the
Higgs--Goldstone-boson couplings of \Eref{eq:treelevelG}: for the
Higgs mass associated to the external Higgs leg, the loop-corrected
Higgs mass $M_{h_i}$ is used instead of the tree-level one. This is
actually the form of the Higgs--Goldstone-boson coupling that would
be expected in an effective field theory of the physical Higgs
boson $h_i$. Using the definition of~$Z^{\mbox{\tiny mix}}_{ij}$ as an
eigenvector of the loop-corrected mass matrix for the
eigenvalue $M^2_{h_i}$---see
Section \href{https://arxiv.org/pdf/1706.00437.pdf#subsection.2.6}{$2.6$}
of  Ref. \cite{Domingo:2017rhb}---one can verify that the effective
Higgs--Goldstone-boson vertices employing the physical Higgs mass
differ from their tree-level counterparts by a term of one-loop order
(proportional to the Higgs self-energies) so that the alteration of
the one-loop amplitudes is indeed of two-loop order. Employing this
shift of the Higgs--Goldstone couplings cures the gauge-related issues
that we mentioned earlier.

Another issue with gauge invariance appears in connection with the
amplitudes $\AmpE{1L}{\mathrm{G/Z}}{}$. The Goldstone and Z boson
propagators generate denominators with pole $M_\mathrm{Z}^2$ (or $\xi_\mathrm{Z}\,M_\mathrm{Z}^2$
in an $R_{\xi}$~gauge): in virtue of the Slavnov--Taylor identity of
 \Eref{eq:hG-hZ}, these terms should cancel one another in the
total amplitude at the one-loop order---we refer the reader to
Section \href{https://arxiv.org/pdf/1103.1335.pdf#page=11}{4.3} of Ref. 
\cite{Williams:2011bu} for a detailed discussion.  However, the
term \mbox{$\left(p^2-M_\mathrm{Z}^2\right)^{-1}$}
multiplying $f{\left(p^2\right)}$ of \Eref{eq:hG-hZ} only
vanishes if~\mbox{$p^2=m^2_{h_i}$}: if we
employ~\mbox{$p^2=M^2_{h_i}$} (the loop-corrected Higgs mass), the
cancellation is spoilt by a term of two-loop order. To
address this problem, we redefine $\AmpE{1L}{\mathrm{G/Z}}{}$ by adding a two-loop term:
\begin{align}\label{EWHmix}
  \tAmp{1L}{\mathrm{G/Z}}{h_i\to \mathrm{f\bar{f}}\,} &\equiv
  Z^{\text{mix}}_{ij}\cdot\Amp{1L}{\mathrm{G/Z}}{h_j^0\to \mathrm{f\bar{f}}\,}
  +\frac{\Gamma^{\text{tree}}_{G\mathrm{f\bar{f}}}}{M_{h_i}^2}\sum_{j,\,k}
  \hat{\Sigma}_{h_jh_k}{\left(M_{h_i}^2\right)}\cdot Z^{\text{mix}}_{ik}\,
  \frac{f{\left(M_{h_i}^2\right)}\,\xi_\mathrm{Z}\,M_\mathrm{Z}^2}{M_{h_i}^2-\xi_\mathrm{Z}\,M_\mathrm{Z}^2} ,
\end{align}
where $\Gamma^{\text{tree}}_{\mathrm{Gf\bar{f}}}$ represents the tree-level
vertex of the neutral Goldstone boson with the fermion f (in the
particular example of a Higgs decay into $\mathrm{f\bar{f}}$). Then it is
straightforward to check that $\tAmp{1L}{\mathrm{G/Z}}{}$
is gauge-invariant. The transformation of \Eref{EWHmix} can also
be interpreted as a two-loop shift redefining $\Sigma_{h_i\mathrm{Z}}$, so
that it satisfies a generalised Slavnov--Taylor identity of the form
of \Eref{eq:hG-hZ}, but applied to a physical (loop-corrected)
Higgs field, with the
term $\left(p^2-m^2_{h_i}\right)f{\left(p^2\right)}$ of
\Eref{eq:hG-hZ} replaced
with $\left(p^2-M^2_{h_i}\right)f{\left(p^2\right)}$.

\paragraph{Numerical input in the one-loop corrections}

As usual, the numerical values of the input parameters need to reflect
the adopted renormalization scheme, and the input parameters
corresponding to different schemes differ from each other by shifts of
the appropriate loop order (at the loop level, there exists some
freedom to use a numerical value of an input parameter that differs
from the tree-level value by a one-loop shift, since the difference
induced in this way is of higher order). Concerning the input values
of the relevant light quark masses, we follow in our evaluation the
choice of \texttt{FeynHiggs} and employ $\overline{\mbox{MS}}$ quark
masses with three-loop QCD corrections evaluated at the scale of the
mass of the decaying Higgs, $m_\mathrm{q}^{\overline{\mbox{\tiny
      MS}}}(M_{h_i})$, in the loop functions and the definition of the
Yukawa couplings. In addition, the input value for the pole top mass
is converted to $m_\mathrm{t}^{\overline{\mbox{\tiny MS}}}(m_\mathrm{t})$ using up to
two-loop~QCD and one-loop top Yukawa or electroweak corrections
(corresponding to the higher-order corrections included in the
Higgs boson mass calculation). Furthermore, the $\tan\beta$-enhanced
contributions are always included in the defining relation between the
bottom Yukawa coupling and the bottom mass (and similarly for all
other down-type quarks). Concerning the Higgs VEV appearing in the
relation between the Yukawa couplings and the fermion masses, we
parametrize it in terms of $\alpha(M_\mathrm{Z})$. Finally, the strong coupling
constant employed in SUSY-QCD diagrams is set to the scale of the
supersymmetric particles entering the loop. We will comment on
deviations from these settings if needed.%
\footnote{Possibly large contributions by electroweak double-logarithms of the Sudakov
type as well as the corresponding counterparts in fermionic Higgs decays
with additional real radiation of gauge bosons are investigated in a
separate article \cite{H13:inprep}.}

\subsubsection{Higgs decays into SM fermions}
\label{heinemeyer-sec:Higgs decays into SM fermions}

Our calculation of the Higgs decay amplitudes into SM~fermions closely
follows the procedure outlined in the previous subsection. However, we
include the QCD and QED corrections separately, making use of
analytical formulae that are well-documented in the
literature \cite{Braaten:1980yq,Drees:1990dq}. We also employ an
effective description of the Higgs--$\mathrm{b\bar{b}}$ interactions in order
to resum potentially large effects for large values
of $\tan\beta$. Next, we comment on these two issues and discuss
further the derivation of the decay widths for this class of channel.

\paragraph{Tree-level amplitude}

At the tree level, the decay $h^0_j\to \mathrm{f\bar{f}}$ is determined by the
Yukawa coupling $Y_\mathrm{f}$ and the decomposition of the tree-level
state $h^0_j$ in terms of the Higgs-doublet components:
\begin{align}
  \label{eq:hfftree}
  \Amp{tree}{}{h^0_j\to \mathrm{f\bar{f}}} &=
  -\I \frac{Y_\mathrm{f}}{\sqrt{2}} \bar{u}_\mathrm{f}{\left(p_\mathrm{f}\right)} \left \{
  \delta_{\mathrm{f}, d_k/e_k}\Un{j1} + \delta_{\mathrm{f}, u_k}\Un{j2}
  -\I \gamma_5\left[\delta_{\mathrm{f}, d_k/e_k}\Un{j4}+\delta_{\mathrm{f},
  u_k}\Un{j5}\right]
  \right \} v_\mathrm{f}{\left(p_{\bar{\mathrm{f}}}\right)}
\\
  &\equiv -\I \bar{u}_\mathrm{f}{\left(p_\mathrm{f}\right)}
  \left\{g_{h_j\mathrm{ff}}^S-\gamma_5 g_{h_j\mathrm{ff}}^P\right\}v_\mathrm{f}{\left(p_{\bar{\mathrm{f}}}\right)}\,.
\end{align}
The $\delta$s are Kronecker symbols selecting the appropriate Higgs
matrix element for the fermionic final state, $u_k=\mathrm{u,c,t}$,
\mbox{$d_k=\mathrm{d,s,b}$}, or $e_k=\mathrm{e},\upmu,\uptau$. We have written the amplitude
in the Dirac-fermion convention, separating the scalar
part $g_{h_j\mathrm{ff}}^S$ (first two terms between curly brackets in the
first line) from the pseudo-scalar one $g_{h_j\mathrm{ff}}^P$ (last two
terms). The fermion and antifermion spinors are denoted
$\bar{u}_\mathrm{f}(p_\mathrm{f})$ and $v_\mathrm{f}(p_{\bar{\mathrm{f}}})$, respectively.

\paragraph{Case of the $\mathrm{b\bar{b}}$ final state: $\tan\beta$-enhanced corrections}

In the case of a decay to $\mathrm{b\bar{b}}$ (and analogously for down-type
quarks of first and second generation, but with smaller numerical
impact), the loop contributions that receive a $\tan\beta$ enhancement
may have a sizeable impact, thus justifying an effective description of
the Higgs--$\mathrm{b\bar{b}}$ vertex that provides a resummation of large
contributions \cite{Banks:1987iu, Hall:1993gn, Hempfling:1993kv,
  Carena:1994bv, Carena:1999py, Eberl:1999he, Williams:2011bu,
  Baglio:2013iia}. We denote the neutral components of $\mathcal{H}_1$
and $\mathcal{H}_2$ from
Eq. (\href{https://arxiv.org/pdf/1706.00437.pdf#equation.2.2}{2.2})
of Ref. \cite{Domingo:2017rhb} by $H_\mathrm{d}^0$ and $H_\mathrm{u}^0$, respectively. The
large $\tan\beta$-enhanced effects arise from contributions to
the $\left(H_u^{0}\right)^*\,\bar{b}\,P_\mathrm{L}\,{b}$ operator---$P_\mathrm{L,R}$
are the left- and right-handed projectors in the Dirac description of
the $b$ spinors---and can be parametrized in the following fashion:
\begin{align}
\label{eq:Leffbb}
  \mathcal{L}^{\text{eff}} = -Y_\mathrm{b}\,\bar{\mathrm{b}}\left[H_\mathrm{d}^0 +
    \frac{\Delta_\mathrm{b}}{\tan\beta}\left(\frac{\lambda}{\mueff}\,S\,H_\mathrm{u}^0\right)^*
    \right] P_\mathrm{L}\,b + \text{h.c.}
  \equiv -\sum_j g_{h_j\mathrm{bb}}^{L\,\text{eff}}\,h^0_j\,\bar{b}\,P_\mathrm{L}\,b + \text{h.\,c.}
\end{align}
Here, $\Delta_\mathrm{b}$ is a coefficient that is determined via the
calculation of the relevant ($\tan\beta$-enhanced) one-loop diagrams
to the Higgs--$\mathrm{b\bar{b}}$ vertex, involving gluino--sbottom,
chargino--stop, and\\ neutralino--sbottom loops.\footnote{Two-loop
  corrections to $\Delta_\mathrm{b}$ have also been reported in Refs.
  \cite{Noth:2008tw,Noth:2010jy}.} The symbol~$\mu_{\text{eff}}$
represents the effective $\mu$ term that is generated when the singlet
field acquires a VEV. The specific form of the
operator, $\left(S\,H_\mathrm{u}^0\right)^*\,\bar{b}\,P_\mathrm{L}\,b$, is designed so
as to preserve the $\mathbb{Z}_3$ symmetry, and it can be shown that
this operator is the one that gives rise to leading contributions to
the $\tan\beta$-enhanced effects. We evaluate $\Delta_\mathrm{b}$ at a scale
corresponding to the arithmetic mean of the masses of the
contributing SUSY particles: this choice is consistent with the
definition of $\Delta_\mathrm{b}$ employed for the Higgs mass calculation.

From the parametrization of \Eref{eq:Leffbb}, one can derive the
non-trivial relation between the `genuine' Yukawa coupling $Y_\mathrm{b}$ and
the effective bottom~mass $m_\mathrm{b}$: $Y_\mathrm{b}= {m_\mathrm{b}}
/ ({v_1\left(1 +
    \Delta_\mathrm{b}\right))}$. Then, the effective couplings of the neutral
Higgs fields to $\mathrm{b\bar{b}}$ read:
\begin{align}
  \label{eq:bbeffcoup}
  g_{h_j\mathrm{bb}}^{L\,\text{eff}} &= 
  \frac{m_\mathrm{b}}{\sqrt{2}\,v_1\left(1 + \Delta_b\right)}\left\{
  \Un{j1} + \I\,\Un{j4}
  +\frac{\Delta_\mathrm{b}}{\tan\beta}\left(
  \Un{j2} - \I\,\Un{j5} +
  \tfrac{\lambda^*\,v_2}{\mueff^*}\left[\Un{j3} - \I\,\Un{j6}\right]
  \right)\right\}.
\end{align}
This can be used to substitute $\Amp{tree}{}{h^0_j\to
  \mathrm{b\bar{b}}\,}$ in \Eref{eq:1LAmplitude} for:
\begin{align}\label{eq:effHbb}
  \Amp{eff}{}{h^0_j\to \mathrm{b\bar{b}}\,} =-\I\,\bar{u}_\mathrm{b}{\left(p_\mathrm{b}\right)}\left[
    g_{h_j\mathrm{bb}}^{L\,\text{eff}}\,P_\mathrm{L} + g_{h_j\mathrm{bb}}^{L\,\text{eff}\,*}\,P_\mathrm{R}
    \right] v_\mathrm{b}{\left(p_{\bar{\mathrm{b}}}\right)}\,,
\end{align}
where this expression resums the effect of $\tan\beta$-enhanced
corrections to the $h^0_j\mathrm{b\bar{b}}$ vertex. However, if one now adds
the one-loop amplitude $\AmpE{1L}{{\rm vert}}{}$, the one-loop effects
associated with the $\tan\beta$-enhanced contributions would be
included twice.  To avoid this double counting, the terms that are
linear in $\Delta_\mathrm{b}$ in \Eref{eq:bbeffcoup} need to be subtracted.
Employing the `subtraction' couplings
\begin{align}
  g_{h_j\mathrm{bb}}^{L\,\text{sub}} = \frac{m_\mathrm{b}\,\Delta_\mathrm{b}}{\sqrt{2}\,v_1}\left\{
    \Un{j1} + \I\,\Un{j4} -
    \frac{1}{\tan\beta}\left(\Un{j2} - \I\,\Un{j5} +
    \frac{\lambda^*\,v_\mathrm{u}}{\mueff^*}\left[\Un{j3} - \I\,\Un{j6}\right]
    \right)\right\} \, ,
    \end{align}
we define the following `tree-level' amplitude for the Higgs decays
into bottom quarks:
\begin{subequations}
\begin{align}
  \Amp{tree}{}{h^0_j\to \mathrm{b\bar{b}}\,} &=
  \Amp{eff}{}{h^0_j\to \mathrm{b\bar{b}}\,}+\Amp{sub}{}{h^0_j\to \mathrm{b\bar{b}}\,}\,,\\
  \Amp{sub}{}{h^0_j\to \mathrm{b\bar{b}}\,} &\equiv
  -\I\,\bar{u}_\mathrm{b}{\left(p_\mathrm{b}\right)}\left[
    g_{h_j\mathrm{bb}}^{L\,\text{sub}}\,P_\mathrm{L} + g_{h_j\mathrm{bb}}^{L\,\text{sub}\,*}\,P_\mathrm{R}
    \right]v_\mathrm{b}{\left(p_{\bar{\mathrm{b}}}\right)}\,.
\end{align}
\end{subequations}

\paragraph{QCD and QED corrections}

The inclusion of QCD and QED corrections requires a proper treatment
of~IR~effects in the decay amplitudes. The IR-divergent parts of the
virtual contributions by gluons or photons in $\AmpE{1L}{{\rm vert}}{}$ are
cancelled by their counterparts in processes with radiated photons or
gluons. We directly employ the QCD and~QED~correction factors that are
well-known analytically (see next) and therefore omit the Feynman
diagrams involving a photon or gluon propagator when computing, with
\texttt{FeynArts} and \texttt{FormCalc}, the one-loop corrections to
the $h^0_j\mathrm{f\bar{f}}$ vertex and to the fermion mass and wave function
counterterms.  The QCD and~QED correction factors applying to the
fermionic decays of a \cp-even Higgs state are given in Ref.
\cite{Braaten:1980yq}. The \cp-odd case was addressed later in Ref.
\cite{Drees:1990dq}. In the \cp-violating case, it is useful to
observe that the $h_j\mathrm{f\bar{f}}$ scalar and pseudo-scalar operators do
not interfere, so that the \cp-even and \cp-odd correction factors can
be applied directly at the level of the amplitudes---although they
were obtained at the level of the squared amplitudes:
\begin{subequations}
\label{eq:hffQCD}
\begin{align}
  \Amp{tree+QCD/QED}{}{h^0_j\to \mathrm{f\bar{f}}\,} &=
  -\I\,\frac{m_\mathrm{f}^{\overline{\mbox{\tiny MS}}}{\left(M_{h_i}\right)}}{m_\mathrm{f}}\,
  \bar{u}_\mathrm{f}{\left(p_\mathrm{f}\right)}\left\{
  g_{h_j\mathrm{ff}}^S\,c_S-\gamma_5\,g_{h_j\mathrm{ff}}^P\,c_P
  \right\}v_\mathrm{f}{\left(p_{\bar{\mathrm{f}}}\right)}\,,\\[-.2ex]
  c_{S,P} &= \sqrt{1+c_{S,P}^{\mbox{\tiny QED}}+c_{S,P}^{\mbox{\tiny QCD}}}\,,\\[-.3ex]
  c_{S,P}^{\mbox{\tiny QED}} &\equiv
  \frac{\alpha}{\uppi}\,Q_\mathrm{f}^2\,\Delta_{S,P}{\left(\sqrt{1-\tfrac{4\,m^2_\mathrm{f}}{M^2_{h_i}}}\right)},\\[-.2ex]
  c_{S,P}^{\mbox{\tiny QCD}} &\equiv
  \frac{\alpha_s{\left(M_{h_i}\right)}}{\uppi}\,C_2{(f)}\left[
    \Delta_{S,P}{\left(\sqrt{1-\tfrac{4\,m^2_\mathrm{f}}{M^2_{h_i}}}\right)}+
    2+3\log{\left(\frac{M_{h_i}}{m_\mathrm{f}}\right)}
  \right].
\end{align}
\end{subequations}
Here, $Q_\mathrm{f}$ is the electric charge of the fermion $f$, $C_2(f)$ is
equal to~$4/3$ for quarks and equal to~$0$ for leptons, $M_{h_i}$
corresponds to the kinematic (pole) mass in the Higgs decay under
consideration and the functions~$\Delta_{S,P}$ are explicated in
\EG\ Section \href{https://arxiv.org/pdf/hep-ph/9503443.pdf#page=9}{$4$}
of Ref. \cite{Dabelstein:1995js}. In the limit of~$M_{h_i}\gg m_\mathrm{f}$,
both $\Delta_{S,P}$ reduce
to $\left[-3\log{\left(M_{h_i}/m_\mathrm{f}\right)}+\frac{9}{4}\right]$. As
noted  in Ref. \cite{Braaten:1980yq}, the leading logarithm in
the QCD correction factor can be absorbed by the introduction of a
running \MSbar\ fermion mass in the definition of the Yukawa
coupling $Y_\mathrm{f}$. Therefore, it is motivated to
factorise $m_\mathrm{f}^{\overline{\mbox{\tiny MS}}}(M_{h_i})$, with higher
orders included in the definition of the~QCD~beta~function.

The QCD (and QED) correction factors generally induce a sizeable shift
of the tree-level width of as much as $\simord$$50\%$. While these
effects were formally derived at the one-loop order, we apply them
over the full amplitudes (without the~QCD and~QED corrections), \IE~we
include the one-loop vertex amplitude
without QCD$/$QED~corrections $\AmpE{1L wo.\ QCD/QED}{{\rm vert}}{}$
and $\AmpE{1L}{\mathrm{G/Z}}{}$ in the definitions of the
couplings $g_{h_j\mathrm{ff}}^{S,P}$ that are employed in
\Eref{eq:hffQCD}---we will use the
notation $g_{h_j\mathrm{ff}}^{S,P\,\text{1L}}$ in the following. The adopted factorisation
corresponds to a particular choice of the higher-order contributions
beyond the ones that have been explicitly calculated.

\paragraph{Decay width}

Putting together the various pieces discussed before, we can express
the decay amplitude at the one-loop order as
\begin{subequations}
\begin{align}
  \Amp{}{}{h_i\to \mathrm{f\bar{f}}\,} & =
  -\I\,\frac{m_\mathrm{f}^{\overline{\mbox{\tiny MS}}}{\left(M_{h_i}\right)}}{m_\mathrm{f}}\,
  Z^{\mbox{\tiny mix}}_{ij}\,\bar{u}_\mathrm{f}{\left(p_\mathrm{f}\right)}\left\{
  g_{h_j\mathrm{ff}}^{S\,\text{1L}}\,c_S-\gamma_5\,g_{h_j\mathrm{ff}}^{P\,\text{1L}}\,c_P
  \right\}v_\mathrm{f}{\left(p_{\bar{\mathrm{f}}}\right)}\,,\\
  - \I\,\bar{u}_\mathrm{f}{\left(p_\mathrm{f}\right)}\left\{g_{h_j\mathrm{ff}}^{S\,\text{1L}}-\gamma_5\,g_{h_j\mathrm{ff}}^{P\,\text{1L}}\right\}v_\mathrm{f}{\left(p_{\bar{\mathrm{f}}}\right)}  & \equiv
  \left(\AmpE{tree}{}+\Amp{1L wo.\ QCD/QED}{{\rm vert}}{}+\AmpE{1L}{\mathrm{G/Z}}{}\right)
  \!\!{\left[h_j\to \mathrm{f\bar{f}}\,\right]}\,. %\span\omit
\end{align}
\end{subequations}
Summing over spinor and colour degrees of freedom, the decay width is
then obtained as
\begin{align}
  \Gamma{\left[h_i\to \mathrm{f\bar{f}}\,\right]} &=
  \frac{1}{16\,\uppi\,M_{h_i}}\sqrt{1-\frac{4m_\mathrm{f}^2}{M_{h_i}^2}}
  \sum_{\parbox{\widthof{\tiny polarisation,}}{\centering\tiny polarisation,\\\tiny colour}}
  \left|\Amp{}{}{h_i^{\mbox{\tiny phys.}}\to \mathrm{f\bar{f}}\,}\right|^2\,.
\end{align}
At the considered order, we could dismiss the one-loop squared terms
in $|\Amp{}{}{h_i\to \mathrm{f\bar{f}}\,}|^2$. However, to
tackle the case where the contributions from irreducible one-loop
diagrams are numerically larger than the tree-level amplitude, we keep
the corresponding squared terms in the expression  (it should be
noted that the QCD and QED~corrections have been stripped off from the
one-loop amplitude, which  gets squared).  The approach of incorporating
the squared terms should give a reliable result in a situation where
the tree-level result is significantly suppressed, since the other
missing contribution at this order, consisting of the tree-level
amplitude times the two-loop amplitude, would be suppressed, owing to the
small tree-level result. In such a case, however, the higher-order
uncertainties are expected to be comparatively larger than in the case
where one-loop effects are subdominant to the tree level.

The kinematic masses of the fermions are easily identified in the
leptonic case. For decays into top quarks, the `pole' mass $m_\mathrm{t}$ is
used, while for all other decays into quarks we employ
the $\overline{\mbox{MS}}$ masses evaluated at the scale of the Higgs
mass $m_\mathrm{q}^{\overline{\mbox{\tiny MS}}}(M_{h_i})$. We note that these
kinematic masses have little impact on the decay widths, as long as
the Higgs state is much heavier. In the~NMSSM, however, singlet-like
Higgs states can be very light, in which case the choice of
an $\overline{\mbox{MS}}$ mass is problematic. Yet, in this case, the
Higgs state is typically near threshold so that the free-parton
approximation in the final state is not expected to be reliable. Our
current code is not properly equipped to address decays directly at
threshold independently of the issue of running kinematic
masses. Improved descriptions of the hadronic decays of Higgs states
close to the $\mathrm{b\bar{b}}$ threshold or in the chiral limit have been
presented in,
\EG\ Refs. \cite{Drees:1989du,Fullana:2007uq,McKeen:2008gd,Domingo:2011rn,Dolan:2014ska,Domingo:2016yih}.

\subsubsection{Decays into SM gauge bosons}
\label{heinemeyer-sec:Decays into SM gauge bosons}
 
Now we consider Higgs decays into the gauge bosons of the~SM. Almost
each of these channels requires a specific processing in order to
include higher-order corrections consistently or to deal with
off-shell effects.

\paragraph{Decays into electroweak gauge bosons}

Higgs decays into on-shell Ws and Zs can easily be included at
the one-loop order in comparable fashion to the fermionic
decays. However, the notion of WW or ZZ final states usually
includes contributions from off-shell gauge bosons as well,
encompassing a wide range of four-fermion final states. Such off-shell
effects mostly impact the decays of Higgs bosons with a mass below
the WW or ZZ thresholds. Instead of a full processing of the
off-shell decays at one-loop order, we pursue two distinct evaluations
of the decay widths in these channels.

Our first approach is that already employed in \texttt{FeynHiggs} for
the corresponding decays in the ~MSSM. It involves exploiting the
precise one-loop results of \texttt{Prophecy4f} for the SM Higgs
decays into four
fermions\,\cite{Bredenstein:2006rh,Bredenstein:2006nk,Bredenstein:2006ha}. For
an (N)MSSM~Higgs boson $h_i$, the~SM~decay width is thus evaluated at
the mass $M_{h_i}$ and then rescaled by the squared ratio of the
tree-level couplings to gauge bosons for $h_i$ and an~SM~Higgs
boson $H_{\mbox{\tiny SM}}$~($V=\mathrm{W,Z}$):
\begin{subequations}
\begin{align}
  \Gamma{\left[h_i\to VV\right]} &=
  \Gamma^{\mbox{\tiny SM}}{\left[H_{\mbox{\tiny SM}}(M_{h_i})\to VV\right]}
  \left|\mathcal{R}_{ij}\cdot
  \frac{g_{h_jVV}^{\mbox{\tiny NMSSM}}}{g_{\mathrm{H}VV}^{\mbox{\tiny SM}}}\right|^2,
  \label{ratio_NMSSM_SM}\\
  \frac{g_{h_jVV}^{\mbox{\tiny NMSSM}}}{g_{\mathrm{H}VV}^{\mbox{\tiny SM}}} &\equiv
  \cos\beta\,\Un{j1}+\sin\beta\,\Un{j2}\,,
\end{align}
\end{subequations}
where \mbox{$\Gamma{\left[h_i\to VV\right]}$} represents the decay
width of the physical Higgs state $h_i$ in the~NMSSM,
while \mbox{$\Gamma^{\mbox{\tiny SM}}{\left[H_{\mbox{\tiny
          SM}}{\left(M_{h_i}\right)}\to VV\right]}$} denotes the decay
width of an SM Higgs boson with  mass $M_{h_i}$. The matrix
elements $\mathcal{R}_{ij}$ reflect the connection between the
tree-level Higgs states and the physical states. This role is similar
to $\mathbf{Z}^{\mbox{\tiny mix}}$. However, decoupling in
the SM limit of the model yields the additional condition that the
ratio in \Eref{ratio_NMSSM_SM} reduces to $1$ in this limit for
the SM-like Higgs boson of the NMSSM. For this reason,
\texttt{FeynHiggs} employs the matrix $\mathbf{U}^m$
(or $\mathbf{U}^0$) as a unitary approximation
of $\mathbf{Z}^{\mbox{\tiny mix}}$---see
Section \href{https://arxiv.org/pdf/1706.00437.pdf#subsection.2.6}{$2.6$}
of Ref. \cite{Domingo:2017rhb}. An alternative choice involves
using~\mbox{$X_{ij}\equiv \left.Z^{\mbox{\tiny
      mix}}_{ij}\middle/\sqrt{\sum_k |Z^{\mbox{\tiny
        mix}}_{ik}|^2}\right.$}. However, the difference of the widths
when employing $\mathbf{U}^0$, $\mathbf{U}^m$,
$\mathbf{Z}^{\mbox{\tiny mix}}$, or $\mathbf{X} \equiv
\left(X_{ij}\right)$ corresponds to effects of higher order, which
should be regarded as part of the higher-order uncertainty. The
rescaling of the one-loop SM width should only be applied for
the SM-like Higgs of the NMSSM, where this implementation of
the $h_i\to VV$ widths is expected to provide an approximation that is
relatively close to a full one-loop result incorporating all NMSSM
contributions. However, for the other Higgs states of the NMSSM,
one-loop contributions beyond the SM may well be dominant. Actually,
the farther the
quantity \mbox{$[\mathcal{R}_{ij}\cdot\Un{j2}]\big/[\mathcal{R}_{ij}\cdot\Un{j1}]$}
departs from~$\tan\beta$, the more inaccurate the prediction based
on SM-like radiative corrections becomes.

Our second approach involves a one-loop calculation of the Higgs
decay widths into on-shell gauge bosons (see Ref. \cite{Gonzalez:2012mq}
for the MSSM case), including tree-level off-shell effects. This
evaluation is meant to address the case of heavy Higgs bosons at the
full one-loop order. The restriction to on-shell kinematics is
justified above the threshold for electroweak gauge-boson production
(off-shell effects at the one-loop level could be included via a
numerical integration over the squared momenta of the gauge bosons in
the final state---see Refs.\cite{Hollik:2010ji,Hollik:2011xd} for a
discussion in the~MSSM). For details of our implementation,  see Ref. \cite{Domingo:2018uim}, with the noteworthy feature
that contributions from Higgs--electroweak mixing $\AmpE{1L}{\mathrm{G/Z}}{}$
vanish. In the case of the $\mathrm{W}^+\mathrm{W}^-$ final state,
the QED~IR divergences are regularised with a photon~mass and cancel
with bremsstrahlung corrections: soft and hard bremsstrahlung are
included according to Refs. \cite{Kniehl:1991xe,Kniehl:1993ay} (see
also Ref.\,\cite{Gonzalez:2012mq}). We stress that the exact cancellation
of the IR divergences is only achieved through the replacement of
the $h_i\mathrm{G}^+\mathrm{G}^-$ coupling with the expression in terms of the kinematic
Higgs mass (see Ref. \cite{Domingo:2018uim} for more details). This fact had
already been observed in Ref. \cite{Gonzalez:2012mq}. To extend
the validity of the calculation below the threshold, we process the
Born-order term separately, applying an off-shell kinematic
integration over the squared external momentum of the gauge
bosons---see,
\EG~Eq. (\href{https://arxiv.org/pdf/1612.07651.pdf#equation.2.37}{37})
in Ref. \cite{Spira:2016ztx}. Thus, this evaluation is performed at tree
level below threshold and at full one-loop order (for the on-shell
case) above threshold. The vanishing on-shell kinematic factor
multiplying the contributions of one-loop order ensures the continuity
of the prediction at threshold. Finally, we include the one-loop
squared term in the calculation. Indeed, as we will discuss later,
the tree-level contribution vanishes for a decoupling doublet, meaning
that the Higgs decays to WW/ZZ  can be dominated by one-loop
effects. To this end, the infrared divergences of two-loop order are
regularised in an ad-hoc fashion---which appears compulsory as long as
the two-loop order is incomplete---making use of the one-loop real
radiation and estimating the logarithmic term in the imaginary part of
the one-loop amplitude.

\paragraph{Radiative decays into gauge bosons}

Higgs decays into photon pairs, gluon pairs, or $\upgamma \mathrm{Z}$ appear at
the one-loop level---\IE~\mbox{$\AmpE{tree}{}=0$} for all these
channels. We compute the one-loop order using the \texttt{FeynArts}
model file, although the results are well-known analytically in the
literature---see, \EG\ Ref.~\cite{Benbrik:2012rm} or
Section \href{https://arxiv.org/pdf/1402.3522.pdf#section*.3}{III} of Ref.
\cite{Belanger:2014roa} (Ref. \cite{Spira:2016ztx} for the~MSSM). The
electromagnetic coupling in these channels is set to the
value $\alpha(0)$, corresponding to the Thomson limit.

The use of tree-level Higgs--Goldstone couplings together with
loop-corrected kinematic Higgs masses $M_{h_i}$ in our calculation
would induce an effective violation of Ward~identities by two-loop
order terms in the amplitude: we choose to restore the proper gauge
structure by redefining the Higgs--Goldstone couplings in terms of
the kinematic Higgs mass $M_{h_i}$ (see Ref. \cite{Domingo:2018uim} for more
details). Since our calculation is
restricted to the leading---here, one-loop---order, the transition of
the amplitude from tree-level to physical Higgs states is performed
via $\mathbf{U}^m$ or $\mathbf{X}$ instead of $\mathbf{Z}^{\mbox{\tiny
    mix}}$ in order to ensure the appropriate behaviour in the
decoupling limit.

Leading QCD corrections to the diphoton Higgs decays have received
substantial attention in the literature. A frequently used
approximation for this channel involves multiplying the amplitudes
driven by quark and squark loops by the
factors $\left[1-\alpha_\mathrm{s}(M_{h_i})/\uppi\right]$
and $\left[1+8\,\alpha_\mathrm{s}(M_{h_i})/(3\,\uppi)\right]$, respectively---see,
\EG\ Ref. \cite{Lee:2003nta}. However, these simple factors are only valid
in the limit of heavy quarks and squarks (compared with the mass of the
decaying Higgs boson). More general analytical expressions can be
found in, \EG~Ref. \cite{Aglietti:2006tp}. In our calculation, we apply
the correction
factors $[1+C^S{(\tau_\mathrm{q})}\,\alpha_\mathrm{s}(M_{h_i})/\uppi ]$
and $[1+C^P{(\tau_\mathrm{q})}\,\alpha_\mathrm{s}(M_{h_i})/\uppi ]$ to the
contributions of the quark q to the \cp-even and
the \cp-odd $h_i\upgamma\upgamma$ operators, respectively,
and $[1+C{(\tau_{\tilde{Q}})}\,\alpha_\mathrm{s}(M_{h_i})/\uppi ]$ to
the contributions of the squark~$\tilde{Q}$ (to the \cp-even
operator).  Here, $\tau_X$ denotes the
ratio $\left[4\,m^2_X{(M_{h_i}/2)}/M^2_{h_i}\right]$. The
coefficients $C^{S,P}$ and $C$ are extracted from Refs.
\cite{Spira:1995rr} and \cite{Muhlleitner:2006wx}. To
obtain a consistent inclusion of
the $\mathcal{O}{(\alpha_\mathrm{s})}$ corrections, the quark and squark
masses $m_X$ entering the one-loop amplitudes or the correction
factors are chosen as defined in
Eq. (\href{https://arxiv.org/pdf/hep-ph/9504378.pdf#page=7}{5}) of Ref.
\cite{Spira:1995rr} and 
Eq. (\href{https://arxiv.org/pdf/hep-ph/0612254.pdf#equation.2.12}{12})
of Ref. \cite{Muhlleitner:2006wx} (rather than \MSbar\ running masses).

The QCD corrections to the digluon decays include virtual corrections
but also gluon and light quark radiation. They are thus technically
defined at the level of the squared amplitudes. In the limit of heavy
quarks and squarks, the corrections are known beyond~NLO---see the
discussion in Ref. \cite{Spira:2016ztx} for a list of references. The
full dependence in mass was derived at NLO in Refs.
\cite{Spira:1995rr,Muhlleitner:2006wx}, for both quark and squark
loops. In our implementation, we follow the prescriptions of
Eqs. (\href{https://arxiv.org/pdf/1612.07651.pdf#equation.2.51}{51}),
(\href{https://arxiv.org/pdf/1612.07651.pdf#equation.2.63}{63}),
and (\href{https://arxiv.org/pdf/1612.07651.pdf#equation.2.67}{67}) of Ref.
\cite{Spira:2016ztx} in the limit of light radiated quarks and heavy
particles in the loop. For consistency, the masses of the particles in
the one-loop amplitude are taken as pole masses. Effects beyond this
approximation can be sizeable, as evidenced by
Fig. \href{https://arxiv.org/PS_cache/hep-ph/ps/9504/9504378v1.fig1-22.png}{20}
of Ref. \cite{Spira:1995rr} and
Fig. \href{https://arxiv.org/pdf/hep-ph/0612254.pdf#page.21}{12} of Ref.
\cite{Muhlleitner:2006wx}. As the \cp-even and \cp-odd
Higgs--gg operators do not interfere, it is straightforward to
include both correction factors in the \cp-violating case. Finally, we
note that parts of the leading QCD corrections to $h_i\to \mathrm{gg}$ are
induced by the real radiation of quark--antiquark pairs. In the case
of the heavier quark flavours (top, bottom, and possibly charm), the
channels are experimentally  easily distinguishable from gluonic
decays. Therefore, the partial widths related to these corrections
could be attached to the Higgs decays into quarks
instead \cite{Djouadi:1995gt}. The resolution of this ambiguity would
involve a dedicated experimental analysis of the kinematics of the
gluon radiation in $h_i\to \mathrm{gq\bar{q}}$ (collinear or back-to-back
emission).

The QCD corrections to the quark loops of an~SM Higgs decay
into $\upgamma \mathrm{Z}$ have been studied in Refs.
\cite{Spira:1991tj,Bonciani:2015eua,Gehrmann:2015dua}, but we do
not consider them here.

%%%%%%%%%%%%%%%%%%%%%%%%%%%%%%%%%%%%%%%%%%%%%%%%%%%%%%%%%%%%%%%%%%%%%%%%%%%%%%
%%%%%%%%%%%%%%%%%%%%%%%%%%%%%%%%%%%%%%%%%%%%%%%%%%%%%%%%%%%%%%%%%%%%%%%%%%%%%%

\subsection{Discussion concerning the remaining theoretical uncertainties}

Next, we provide a summary of the main sources of theoretical
uncertainties from unknown higher-order corrections applying to our
calculation of the NMSSM Higgs decays. We do not discuss here the
parametric theoretical uncertainties arising from the experimental
errors of the input parameters. For the experimentally
known SM-type parameters, the induced uncertainties can be determined
in the same way as for the SM case (see,
\EG~Ref.\,\cite{Denner:2011mq}). The dependence on the
unknown SUSY parameters, however, is usually not treated as
a theoretical uncertainty but rather exploited for setting indirect
constraints on those parameters.

\subsubsection{Higgs decays into quarks (\boldmath$h_i\to \mathrm{q}\bar{\mathrm{q}}$, $\mathrm{q}=\mathrm{c,b,t}$)}
In our evaluation, these decays have been implemented at full one-loop
order, \IE~at QCD, electroweak, and SUSY next-to-leading
order (NLO). In addition, leading QCD logarithmic effects have been
resummed within the parametrization of the Yukawa couplings in terms
of a running quark mass at the scale of the Higgs mass. The Higgs
propagator-type corrections determining the mass of the considered
Higgs particle, as well as the wave function normalization at the
external Higgs leg of the process, contain full one-loop and dominant
two-loop contributions.

For an estimate of the remaining theoretical uncertainties, several
higher-order effects should be taken into account.
\begin{enumerate}
\item First, we should assess the magnitude of the
  missing QCD NNLO (two-loop) effects. We stress that there should be
  no large logarithms associated with these corrections, since these are
  already resummed through the choice of running parameters and the
  renormalization scale. For the remaining QCD pieces, we can directly
  consider the situation in the SM. In the case of the light quarks,
  the QCD contributions of higher order have been evaluated and amount
  to $\simord$$4\%$ at \mbox{$m_\mathrm{H}=120$}\,GeV (see,
  \EG~Ref.\,\cite{Baikov:2005rw}). For the top quark, the uncertainty
  due to missing QCD NNLO effects was estimated
  at $5\%$\,\cite{Denner:2011mq}.
\item Concerning the electroweak corrections,
  the numerical analysis in Ref. \cite{Domingo:2018uim} 
  suggests that the one-loop contribution is small---at the percentage
  level---for an SM-like Higgs, which is consistent with earlier
  estimates in the~SM\,\cite{Denner:2011mq}. For the heavy Higgs
  states, the numerical analysis in Ref. \cite{Domingo:2018uim}
  indicates a larger impact of such
  effects---at the level of $\simord$$10\%$ in the considered
  scenario. Assuming that the electroweak NNLO corrections are
  comparable to the squared one-loop effects, our estimate for pure
  electroweak higher orders in decays of heavy Higgs states reaches
  the percentage level. In fact, for \mbox{multiteraelectronvolt} Higgs bosons, the
  electroweak Sudakov logarithms may require a
  resummation (see Ref. \cite{H13:inprep}).
  Furthermore, mixed electroweak--QCD contributions are
  expected to be larger than the pure electroweak NNLO corrections,
  adding a few more percent to the uncertainty budget. For light Higgs
  states, the electroweak effects are much smaller, since the
  Sudakov logarithms remain of comparatively modest size.
\item Finally, the variations with the squark masses in
  the numerical analysis in Ref. \cite{Domingo:2018uim}
  for the heavy doublet states show that the
  one-loop SUSY effects could amount to~$5$--$10\%$ for a
  sub\-teraelectronvolt stop or sbottom spectrum. In such a case, the two-loop~SUSY and
  the mixed QCD or electroweak--SUSY corrections may reach the percentage
  level. Conversely, for very heavy squark spectra, we expect
  to recover an effective singlet-extended two-Higgs-doublet model (an
  effective SM if the heavy doublet and singlet states also decouple)
  at low energy. However, all the parameters of this low-energy
  effective field theory implicitly depend on the SUSY radiative
  effects, since unsuppressed logarithms of SUSY origin generate terms
  of dimension $\le$4---\EG~in the Higgs potential or the Higgs
  couplings to SM fermions. Conversely, the explicit dependence
  of the Higgs decay widths on SUSY higher-order corrections is
  suppressed for a large SUSY scale. In this case, the uncertainty
  from~SUSY~corrections reduces to a parametric effect, that of the
  matching between the~NMSSM and the low-energy Lagrangian---\EG~in
  the SM limit, the uncertainty on the mass prediction for the SM-like
  Higgs continues to depend on~SUSY~logarithms and would indirectly
  affect the uncertainty on the decay widths.
\end{enumerate}
Considering all these higher-order effects together, we conclude that
the decay widths of the SM-like Higgs should be relatively well-controlled (up to $\simord$$5\%$), while those of a heavy Higgs state
could receive sizeable higher-order contributions, possibly adding up
to the level of $\simord$$10\%$.

\subsubsection{Higgs decays into leptons}
Here, QCD corrections appear only at two-loop order in the Higgs
propagator-type corrections, as well as in the counterterms of the
electroweak parameters, and only from three-loop order onwards in the
genuine vertex corrections. Thus, the theory uncertainty is expected
to be substantially smaller than in the case of quark~final
states. For an SM-like Higgs, associated uncertainties were estimated
to be below the percentage level \cite{deFlorian:2016spz}. For heavy
Higgs states, however, electroweak one-loop corrections are enhanced
by Sudakov logarithms (see Ref. \cite{H13:inprep})
and reach the $\simord$$10\%$ level for Higgs
masses of the order of $1$\,TeV, so that the two-loop effects could
amount to a few percent. In addition, light status may generate a
sizeable contribution of SUSY origin, where the unknown corrections are
of two-loop electroweak order.

\subsubsection{Higgs decays into \boldmath $\mathrm{WW/ZZ}$}
The complexity of these channels is illustrated by our presentation of
two separate estimates, expected to perform differently in various
regimes.
\begin{enumerate}
\item In the SM, the uncertainty of \texttt{Prophecy4f} in the
  evaluation of these channels was assessed at the subpercentage level
  below $500$\,GeV, but up to $\simord$$15\%$
  at $1$\,TeV\,\cite{Denner:2011mq}. For an SM-like Higgs, our
  numerical analysis in Ref. \cite{Domingo:2018uim}
  shows that the one-loop electroweak corrections
  are somewhat below $10\%$, making plausible a
  subpercentage uncertainty in the results employing
  \texttt{Prophecy4f}. Conversely, the assumption that the
  decay widths for an NMSSM Higgs boson can be obtained through a
  simple rescaling of the result for the width in the SM by tree-level
  couplings is, in itself, a source of uncertainties. We expect this
  approximation to be accurate only in the limit of a
  decoupling SM-like composition of the NMSSM Higgs boson. If
  these SM-like characteristics are altered through radiative
  corrections of SUSY origins or NMSSM Higgs mixing effects---both of
  which may still reach the level of several~percentage in a
  phenomenologically realistic set-up---the uncertainty in the
  rescaling procedure for the decay widths should be of corresponding
  magnitude.
\item In the case of heavier states, our numerical analysis in Ref.
 \cite{Domingo:2018uim} indicates that the previous procedure is
  unreliable in the mass range $\gsim$$500$\,GeV. In particular, for
  heavy doublets in the decoupling limit, radiative corrections
  dominate over the---then vanishing---tree-level amplitude, shifting
  the widths by orders of magnitude. In such a case, our one-loop
  calculation captures only the leading order and one can expect
  sizeable contributions at the two-loop level: as discussed
  in the numerical analysis in Ref. \cite{Domingo:2018uim}, 
  shifting the quark masses
  between pole and \MSbar~values---two legitimate choices at the
  one-loop order that differ in the treatment of QCD two-loop
  contributions---results in modifications of the widths of
  order $\simord$$50\%$. Conversely, one expects the decays of a
  decoupling heavy doublet into electroweak gauge bosons to remain a
  subdominant channel, so that a less accurate prediction may be
  tolerable. It should be noted, however, that the magnitude of the
  corresponding widths is sizeably enhanced by the effects of one-loop
  order; this may be of interest regarding their phenomenological
  impact.
\end{enumerate}

\subsubsection{Radiative decays into gauge bosons}
As these channels appear at the one-loop order, our (QCD-corrected)
results represent (only) an improved leading-order evaluation. Yet the
situation is contrasted.
\begin{enumerate}
\item In the SM, the uncertainty for a Higgs decay into $\upgamma\upgamma$
  was estimated at the level of $1\%$ in Ref. \cite{Denner:2011mq};
  however, the corresponding calculation includes both QCD NLO and
  electroweak NLO corrections. In our case, only QCD NLO corrections
  (with full mass dependence) are taken into account. The comparison
  with \texttt{NMSSMCALC} in Ref. \cite{Domingo:2018uim} provides us with a
  lower bound on the magnitude of electroweak~NLO
  and QCD NNLO effects: both evaluations are of the same order but
  differ by a few~percent. The uncertainty in the SUSY contribution
  should be considered separately, as light charginos or sfermions
  could have a sizeable impact. In any case, we expect the accuracy of
  our calculation to perform at the level of $\gsim$$4\%$.
\item In the case of the Higgs decays into gluons, for
  the SM prediction---including QCD corrections with full mass
  dependence and electroweak two-loop effects---an uncertainty
  of $3\%$ from QCD effects and~$1\%$ from electroweak effects was
  estimated in Ref. \cite{Denner:2011mq}. In our case,
  the QCD corrections are only included in the heavy-loop
  approximation, and NLO electroweak contributions have not been
  considered. Consequently, the uncertainty budget should settle above
  the corresponding estimate for the SM quoted here. In the case of
  heavy Higgs bosons, the squark spectrum could have a significant
  impact on the QCD two-loop corrections, as exemplified in
  Fig. \href{https://arxiv.org/pdf/hep-ph/0612254.pdf#page.11}{5} of Ref.
  \cite{Muhlleitner:2006wx}.
\item For \mbox{$h_i\to \upgamma \mathrm{Z}$}, QCD~corrections are not yet
available, so  the uncertainty should be above the $\simord$$5\%$
  estimated in the SM \cite{Denner:2011mq}.
\end{enumerate}

\subsubsection{Additional sources of uncertainty from higher orders}
For an uncertainty estimate, the following effects apply to
essentially all channels and should be considered as well.
\begin{enumerate}
\item The mixing in the Higgs sector plays a central role in the
  determination of the decay widths. Following the treatment in \FH,
  we have considered $\mathbf{Z}^{\mbox{\tiny mix}}$ in all our
  one-loop evaluations, as prescribed by the~LSZ~reduction. Most
  public codes consider a unitary approximation in the limit of the
  effective scalar potential ($\mathbf{U}^0$, in our notation). The
  analysis of Ref. \cite{Domingo:2017rhb} and our most recent analysis in
Ref.   \cite{Domingo:2018uim}---employing $\mathbf{U}^m$, a more reliable
  unitary approximation than $\mathbf{U}^0$---indicate that the
  different choices of mixing matrices may affect the Higgs decays by
  a few percent (and far more in contrived cases). However, even the
  use of $\mathbf{Z}^{\mbox{\tiny mix}}$ is, of course, subject to
  uncertainties from unknown higher-order corrections. While the Higgs
  propagator-type corrections determining the mass of the considered
  Higgs boson and the wave function normalization contain corrections
  up to the two-loop order, the corresponding prediction for the mass
  of the SM-like Higgs still has an uncertainty at the level of
  about $2\%$, depending on the SUSY spectrum.
\item In this section, we confined ourselves to the evaluation of the
  Higgs decay widths into SM~particles and did not consider the
  branching ratios. For the latter, an implementation at the full
  one-loop order of many other two-body decays, relevant, in particular,
  for the heavy Higgs states, would be desirable, but goes beyond
  the scope of the present analysis. Furthermore,  to consider
  the Higgs branching ratios at the one-loop order, we would have to
  consider three-body widths at the tree level, for
  instance \mbox{$h_i\to \mathrm{b\bar{b}Z}$}, since these are formally of the
  same magnitude as the one-loop effects for two-body decays \cite{H13:inprep}.
  In addition, these three-body~decays---typically real~radiation of
  electroweak and Higgs bosons---exhibit Sudakov~logarithms that would
  require resummation in the limit of heavy Higgs states \cite{H13:inprep}.
\item At decay thresholds, the approximation of free particles in the
  final state is not sufficient, and a more accurate treatment would
  require the evaluation of final-state interactions. Several cases
  have been discussed in,
  \EG\ Refs.~\cite{Domingo:2011rn,Domingo:2016yih,Haisch:2018kqx}.
\end{enumerate}

In this discussion, we did not attempt to provide a quantitative
estimate of the remaining theoretical uncertainties from unknown
higher-order corrections, as such an estimate would, in any case,
sensitively depend on the considered region in parameter space.
Instead, we have pointed out the various sources of higher-order
uncertainties remaining at the level of our
state-of-the-art~evaluation of the Higgs decays into SM particles in
the NMSSM. For a decoupling SM-like Higgs boson, one would ideally
expect that the level of accuracy of the predictions approaches that achieved in the SM. However, even in this limit,
missing NNLO pieces---which are known for the SM, but not for
the NMSSM---give rise to a somewhat larger theoretical uncertainty in
the NMSSM. Furthermore, uncertainties of parametric nature (for
instance, from the theoretical prediction of the Higgs boson mass) need
to be taken into account as well. For heavy Higgs states, the impact
of electroweak Sudakov logarithms and SUSY corrections add to the
theoretical uncertainty to an extent that is strongly dependent on the
details of the spectrum and the characteristics of the Higgs
state (see Ref. \cite{H13:inprep}).
For a decoupling doublet at $\simord$$1$\,TeV, an uncertainty
of \mbox{$\simord$$5$--$15\%$} may be used as a guideline for the
fermionic and radiative decays, while the uncertainty may be as large
as $\simord$$50\%$ in \mbox{$h_i\to \mathrm{WW/ZZ}$}.

}

\end{bibunit}

\label{sec-bsm-heinemeyer}
\clearpage \pagestyle{empty} \cleardoublepage 

%\input{MTools/mt_intro.tex} %\label{sec-mtintro}
%%%%%%%%%%%%%%%%%%%%%%%%%%%%%%%%%%%%%%%%%%%%%%%%%
%\input{BSM_your1/bsm_your1.tex} \label{sec-bsm-your1}  
%\clearpage \pagestyle{empty}  \cleardoublepage

%################################################
\clearpage
%%%%%%%%%%%%%%%%%%%%%%%%%%%%%%%%%%%%%%%%%%%%%%%%% 
 
%%%%%%%%%%%%%%%%%%%%%%%%%%%%%%%%%%
% 2020-03-06 one line added
\phantomsection
\addcontentsline{toc}{chapter}{Acknowledgements}
\chapter*{Acknowledgements}

%\vspace{10mm} \noindent
%{\bf\Large Acknowledgments}

\vspace{2mm}\noindent
 
\noindent
The work of \textit{J.J. Aguilera-Verdugo, F. Driencourt-Mangin, J. Plenter, S. Ram\'{\i}rez-Uribe, G. Rodrigo, G.F.R. Sborlini, W.J. Torres Bobadilla}, and {\it S. Tracz} was supported by the Spanish Government (Agencia Estatal de Investigacion) and ERDF funds from European Commission (grant numbers FPA2017-84445-P and SEV-2014-0398), by Generalitat Valenciana (grant number PROMETEO/2017/ 053), and by Consejo Superior de Investigaciones Cient\'ificas (grant number PIE-201750E021).
\\
\textit{J.J.~Aguilera-Verdugo}  acknowledges support
from Generalitat Valenciana (GRISOLIAP/2018/101).
\\
Results by \textit{A. Arbuzov, S. Bondarenko, Y. Dydyshka, L. Kalinovskaya, L. Rumyantsev, R. Sadykov}, and \textit{V. Yermolchyk} are obtained in
the framework of state’s task  N 3.9696.2017/8.9 of the Ministry of Education and Science of Russia.
\\
The work of \textit{J. Baglio} is supported by the Institutional Strategy of the University of T\"ubingen (DFG, ZUK 63) and the Carl-Zeiss foundation.
\\
{\textit{S.D. Bakshi}} acknowledges the financial support of IIT Kanpur and 
an
Arepalli-Karumuri travel grant for attending
this conference at CERN.
\\
The work of \textit{M.~Beneke, C.~Bobeth}, and \textit{R.~Szafron}\ was supported by the DFG Son\-derfor\-schungs\-bereich/Transregio 110 `Symmetries and the Emergence of Structure in QCD'.
\\
\textit{S.~Borowka} gratefully acknowledges the financial support of the ERC Starting Grant `MathAm' (39568).
\\
The work of \textit{J.~Chakrabortty} is supported by the Department of Science and
Technology, Government of India, under grant number IFA12/PH/34 (INSPIRE Faculty Award),  
the Science and Engineering Research Board, Government of India, under agreement
number SERB/PHY/2016348 (Early Career Research Award), and
 an Initiation Research Grant, agreement number IITK/PHY/2015077, of IIT Kanpur.
\\
The work of \textit{M.~Chrzaszcz, Z.~Was}, and \textit{J.~Zaremba} is partly supported by the Polish National Science Center, grant number 2016/23/B/ST2/03927,
 and the CERN FCC Design Study Programme.
 \\
The work of \textit{J.~Gluza} is supported in part by the Polish National Science Centre, grant number 2017/25/B/ST2/01987 and by international mobilities for research activities of the University of Hradec Kr\'alov\'e, CZ.02.2.69/0.0/0.0/16\_027/0008487.
\\
The work of \textit{J.A.~Gracey} was supported by a DFG Mercator Fellowship.
\\
The work of \textit{S.~Heinemeyer} is supported
in part by the MEINCOP Spain under contract 
FPA2016-78022-P, in part by the Spanish Agencia
Estatal de Investigacion (AEI) and the EU
Fondo Europeo de Desarrollo Regional (FEDER)
through the project FPA2016-78645-P, in part by
the AEI through the grant IFT Centro de Excelencia
Severo Ochoa SEV-2016-0597, and by the
`Spanish Red Consolider Multidark' FPA2017-90566-REDC.
\\
The work of \textit{S.~Jadach, M.~Skrzypek}, and \textit{Z.~Was} is partly supported by the Polish National Science Center, grant number 2016/23/B/ST2/03927,
and the CERN FCC Design Study Programme.
\\
 \textit{A.~Kardos} acknowledges financial support from the Premium Postdoctoral Fellowship
programme of the Hungarian Academy of Sciences. This work was supported by grant 
K 125105 of the National Research, Development and Innovation Fund in Hungary.
\\
\textit{M.~Kerner} acknowledges supported by the Swiss National Science Foundation (SNF) under grant number 200020-175595.
\\
\textit{P.~Maierh\"ofer} acknowledges support by the state of Baden-W\"urttemberg through bwHPC and
the German Research Foundation (DFG) through grant number INST 39/963-1 FUGG.
\\
The work of \textit{R.~Pittau}\ was supported by the MECD project FPA2016-78220-C3-3-P.
\\ 
\textit{J.~Plenter} acknowledges support from 
the `la Caixa' Foundation (grant numbers ID 100010434 and  LCF/BQ/IN17/11620037) and from the European Union's H2020-MSCA 
Grant (agreement number 713673).  
\\
\textit{S. Ram\'{\i}rez-Uribe}  acknowledges support from CONACYT.
\\
The work of \textit{T.~Riemann} is 
funded by Deutsche Rentenversicherung Bund. He is 
supported in part by a 2015 Alexander von Humboldt Honorary Research Scholarship of the Foundation for Polish Sciences (FNP) 
and by the Polish National Science Centre (NCN) under  grant agreement 
2017/25/B/ST2/01987.
Support of participation at the workshop from the FCC group is acknowledged.
\\
The research of \textit{J.~Schlenk} was supported by the European Union through the ERC Advanced Grant MC@NNLO (340983).
\\
{\textit{C.~Schwinn}} acknowledges support by the Heisenberg Programme of the DFG and a fellowship of the Collaborative Research Centre SFB 676 `Particles, Strings, and the Early Universe' at Hamburg University. \\
\textit{W.J.~Torres Bobadilla} acknowledges support from the Spanish Government (grant number FJCI-2017-32128).
\\
\textit{J.~Usovitsch} has received funding from the European Research Council (ERC) under
the European Union's Horizon 2020 research and innovation programme (grant
agreement  647356, CutLoops).
\\
\textit{C. Weiland} received financial support from the European Research Council under the European Union's Seventh Framework Programme (FP/2007-2013)/ERC Grant NuMass, agreement number 617143, and  is also supported in part by the US Department of Energy under contract DE-FG02-95ER40896 and in part by the PITT PACC. His work was  done in collaboration with S. Pascoli.
\\
This report was partly supported by COST (European Cooperation in Science and Technology) Action CA16201 PARTICLEFACE and the CERN FCC design study programme.

%%%%%%%%%%%%%% Bibliography %%%%%%%%%%%%%%%%%%%%%
\bibliographystyle{elsarticle-num}

%\input{all2.bbl}
 
%references are made in each section  separately,  

\clearpage
\pagestyle{empty}

\cleardoublepage
\pagestyle{empty}
%\pagecolor{amber}
\vphantom{Some Text}

\clearpage %\newpage
\vphantom{Some Text}

%\pagecolor{amber}

\end{document}